\documentclass[10pt]{article}
 \usepackage[super,compress]{cite}
\usepackage{graphicx}
\usepackage{graphicx}
\usepackage{amsfonts}
\usepackage{amsmath}
\usepackage{breqn}

\usepackage{latexsym}
\begin{document}

\pagestyle{plain}



\title{The analytic structure of the BFKL equation and reflection identities of harmonic sums at weight five.
}



\author{ Mohammad Joubat\\ Department of Applied Mathematics \\   Ariel University\\  Ariel, 40700, Israel\\ mohammad.joubat@msmail.ariel.ac.il \vspace{1cm} \\
Alexander Prygarin\\Department of Physics \\ Ariel University\\  Ariel, 40700, Israel \\               alexanderp@ariel.ac.il }

\maketitle
\vspace{-1cm}
\begin{abstract}
We analyze the  structure of the  eigenvalue of the color-singlet Balitsky-Fadin-Kuraev-Lipatov~(BFKL) equation in N=4 SYM in terms of the meromorphic   functions  obtained by the analytic continuation of harmonic sums from positive even integer values of the argument to the complex plane. The meromorphic functions we discuss  have  pole singularities at negative integers and take finite values at all other points. 
We derive  the reflection identities for harmonic sums at weight five decomposing a product of two harmonic sums with mixed pole structure into a linear combination of terms each having a pole at either negative or non-negative values of the argument. The pole decomposition demonstrates how the product of two simpler harmonic sums can build more complicated harmonic sums at higher weight. We list a minimal irreducible  set of bilinear reflection identities at weight five which presents the main result of the paper.
We show how the reflection identities can be used to restore the functional form of  the  next-to-leading eigenvalue of the color-singlet BFKL equation in N=4 SYM , i.e. we argue that it is possible to restore the full functional form on the entire complex plane provided one has information how the function looks like on just two lines on the complex plane.  
Finally we discuss how  non-linear reflection identities can be   constructed from our result with the use of well known quasi-shuffle relations for harmonic sums. 
\end{abstract}


\section{Introduction}
The Balitsky-Fadin-Kuraev-Lipatov~(BFKL) equation  has been derived  about four decades ago,\cite{BFKL1,BFKL2,BFKL3,BFKL4,BFKL5}  the leading Regge trajectory (Pomeron) in the framework of the perturbative gauge theory, in particular the Quantum Chromodynamics~(QCD). The BFKL approach is based on identifying the leading contributions in the perturbative expansion, the terms accompanied by the  large logarithm of center-of-mass energy. This way one  separates the dynamics of the longitudinal and  transverse degrees of freedom, where the longitudinal momentum contributes to the large parameter~(logarithm of center-of-mass energy ) and plays a role of the "time" parameter~\footnote{More precisely the logarithm of the  center-of-mass energy is similar to the imaginary time in the Schroedinger-like equation. } of the evolution, while the transverse momenta defines    the time-independent Hamiltonian.  The BFKL equation is derived by considering all possible Feynman diagrams in the perturbative QCD selecting those that give  the leading order~(LO)  power of  logarithm of the center-of-mass energy. The resulting LO BFKL equation can be viewed also as a propagation  of a bound state of two Reggeized gluons in the $t$-channel. The BFKL equation initiated  a lot of activity in the field of analytic perturbative calculations as well phenomenological studies and comparison to the experimental data.  For more details on the BFKL equation, its  derivation, self-consistency, bootstrap, the issue of unitarity violation etc. the reader is referred to a profound review book by Ioffe, Fadin and Lipatov,\cite{Ioffe:2010zz}. For the purpose of the present discussion we only want to emphasize the following important aspects of this approach. The BFKL equation is formulated through the perturbative expansion that provides a solid background and good control of the calculations, but it also makes it extremely difficult to go to higher orders of the perturbative expansion. In fact, at the moment we only know the leading~(LO) and next-to-leading~(NLO) order of the BFKL equation and the  full solution  in both QCD and its supersymmetric extensions. The exact analytic solution was found due an interesting feature of the BFKL equation, namely, it can be reformulated in terms of the Heisenberg spin chain and thus integrable,\cite{int}\hspace{0.05cm}. The color singlet BFKL equation is of particular interest because it is related to the pomeron in the Regge theory and we do not know that much about it beyond  next-to-leading order due to involved calculations in the old-fashioned diagramatic expansion. Some good news may come from maximally super-symmetric Yang-Mills gauge theory, called $N=4$ SYM, where the calculation may be done by exploiting spin chain analogy of the BFKL dynamics. 

In this paper we continue discussion of our previous study,\cite{Prygarin:2018tng, Prygarin:2018cog, Prygarin:2019ruv,Bondarenko:2015tba,Bondarenko:2016tws}
aiming at generalizing recent results  by Gromov, Levkovich-Maslyuk and Sizov,\cite{Gromov:2015vua},  Caron Huot and Herraren,\cite{Caron-Huot:2016tzz} and Alfimov, Gromov and Sizov,\cite{Alfimov:2018cms} on the color singlet next-to-next-to-leading~(NLO) BFKL eigenvalue for a case of arbitrary values of  conformal spin and anomalous dimension. Our approach is to consider all possible non-linear combinations of the functions, which are obtained from the analytic continuation of the harmonic sums to the complex plane. The resulting meromorphic functions enjoy the reflection identities, which we derive in this paper.

In this section we introduce all necessary definitions and concepts and then in Section~\ref{motivation} we show some examples of use of the reflection identities derived below. This puts the present discussion in the broader context of well known state-of-art results. 

 The reflection identities for the analytically continued harmonic sums represent a product of two harmonic sums of argument $z$ and $-1-z$ is expressed through a linear combination of other harmonic sums of the same arguments, i.e.
\begin{eqnarray}
S_{a_1,a_2,...}(z)S_{b_1, b_2, ...}(-1-z)=S_{c_1,c_2,...}(z)+...+S_{d_1,d_2,...}(-1-z)+... 
\end{eqnarray}

We consider an analytic continuation of harmonic sums from positive even integer values of the argument to the complex plane except for isolated pole singularities at negative integers. This analytic continuation is done using the generalization of Carlson's theorem in terms of the Mellin transform of the Harmonic Polylogarithms~(HPL).
.   
We derive  the reflection identities for  the resulting meromorphic functions~(analytically continued harmonic sums)  at weight five  decomposing a product of two harmonic sums with mixed pole structure into a linear combination of terms each having a pole at either negative or non-negative values of the argument.

The reflection identities at weight two are not new and were known long time ago in the context of functions related to the Euler Gamma function. To the best of our knowledge they appear the  earliest in  Chapter 20 of the book by Nielsen,\cite{handbuch}
.
 At weight three and four they were recently calculated by one of the  authors,\cite{Prygarin:2018tng,Prygarin:2018cog}. This paper deals with weight five. 

 After the analytic continuation  of the harmonic sums to the complex plane the resulting meromorphic functions  have pole singularities at negative integers. The reflection identities present a pole separation for a product of two sums with mixed pole structure. 
 We call those functional relations the reflection identities because the argument of the harmonic sums is reflected with respect to the point $\frac{z+(-1-z)}{2}=-\frac{1}{2}$. The reflection identities up to  weight four were published in our previous study,\cite{Prygarin:2018tng,Prygarin:2018cog} and here we present them at weight five. 
 
The harmonic sums are defined through a nested summation with their argument being the upper limit in the outermost sum,\cite{HS1,Vermaseren:1998uu,Blumlein:1998if,Remiddi:1999ew}
\begin{eqnarray}\label{defS}
S_{a_1,a_2,...,a_k}(n)=  \sum_{n \geq i_1 \geq i_2 \geq ... \geq i_k \geq 1 }   \frac{\mathtt{sign}(a_1)^{i_1}}{i_1^{|a_1|}}... \frac{\mathtt{sign}(a_k)^{i_k}}{i_k^{|a_k|}},\;\; n \in \mathbb{N}^*
\end{eqnarray}
In this paper we consider the harmonic sums with only real integer values of $a_i$, which build the alphabet of the possible negative and positive indices.  
In Eq.~(\ref{defS}) $k$ is  the depth and $w=\sum_{i=1}^{k}|a_i|$ is the weight of the harmonic sum $S_{a_1,a_2,...,a_k}(n)$. 

The indices of harmonic sums $a_1,a_2,...,a_k$ can be either positive or negative integers and label uniquely $S_{a_1,a_2,...,a_k}(n)$ for any given 
weight. However there is no unique way of building the functional basis for a given weight because the harmonic sums are subject to so called quasi-shuffle relations, where a linear combination of $S_{a_1,a_2,...,a_k}(n)$ with the same argument but all possible permutations of indices can be expressed through 
a non-linear combinations of harmonic sums at lower weight.
 There is also some freedom in choosing the irreducible minimal set of $S_{a_1,a_2,...,a_k}(n)$ that builds those non-linear combinations.  
The quasi-shuffle relations make a connection between the linear and non-linear combinations of the harmonic sums of the same argument. For example,  the quasi-shuffle relation at depth two reads
\begin{eqnarray}\label{shuffle}
S_{a,b}(z)+S_{b,a}(z)= S_{a}(z) S_{b}(z)+S_{\textrm{ sign}(a) \textrm{ sign}(b)(|a|+|b|)}(z)
\end{eqnarray}
The quasi-shuffle relations of the harmonic sums is closely connected to the quasi-shuffle algebra of the  harmonic polylogarithms,\cite{Remiddi:1999ew}.

There is another type of identity called the duplication identities where a combination of harmonic sums of argument $n$ can be expressed through 
a harmonic sum of the argument $2 n$. The duplication identities introduce additional freedom in choosing the functional basis.

  In this paper we consider the analytic continuation of the harmonic sums from positive integer values of the argument to the complex plane.
  The resulting meromorphic functions is defined on the complex plane except for isolated poles at negative integer values of the argument. 
  The analytic continuation is done in terms of the Mellin transform of corresponding Harmonic Polylogarithms and was recently used by Gromov, Levkovich-Maslyuk and Sizov,\cite{Gromov:2015vua,Alfimov:2018cms} and Caron Huot and Herraren,\cite{Caron-Huot:2016tzz} for expressing the  eigenvalue Balitsky-Fadin-Kuraev-Lipatov~(BFKL)  equation using the principle of Maximal Transcedentality,\cite{Kotikov:2006ts} in super Yang-Mills $\mathcal{N}=4$ field theory. We plan to use their results together with analysis done by one of the authors and collaborators,\cite{Bondarenko:2015tba,Bondarenko:2016tws} to understand the general structure of the   BFKL equation in QCD and beyond.

 The Mellin transform allows to make the analytic continuation to the complex plane. This method is well known and implemented in the dedicated Mathematica package by   Gromov, Levkovich-Maslyuk and Sizov,\cite{Gromov:2015vua,Alfimov:2018cms} which we use for our calculations. Here we present a simple example that illustrates the general idea.
  We follow the lines of papers of  Kotikov, Lipatov, Onishchenko and Velizhanin,\cite{Kotikov:2004er} as well as   Kotikov and Velizhanin,\cite{Kotikov:2005gr} introducing the analytic continuation of harmonic sums from positive integer values of the argument to the complex plane. 
  We consider the harmonic sum 
 \begin{eqnarray}\label{s1}
 S_{-1}(z)=\sum_{k=1}^{z} \frac{(-1)^k}{k}, \;\; z \in \mathbb{N}^*
\end{eqnarray}  
The corresponding Mellin transform reads
\begin{eqnarray}
\int_0^1 \frac{1}{1+x} x^z= (-1)^z \left(S_{-1}(z)+\ln 2\right)
\end{eqnarray}
One can see that $S_{-1}(z)$ on its own is not an analytic function because of the term $(-1)^z $ and we impose that we start from even integer values of the argument $z$. In this case we define its analytic continuation from even positive integers to all positive integers through,\cite{Kotikov:2004er, Kotikov:2005gr}
\begin{eqnarray}
\bar{S}_{-1}^{+}(z)=(-1)^z S_{-1}(z)+((-1)^z-1) \ln 2
\end{eqnarray} 
 and thus we can write 
 \begin{eqnarray}
 \bar{S}_{-1}^{+}(z)=\int_0^1 \frac{1}{1+x} x^z-\ln 2
\end{eqnarray}
This way we defined  $\bar{S}_{-1}^{+}(z)$ using the Mellin transform of ratio function $ \frac{1}{1+x}$. In more complicated cases of other harmonic sums one includes also Harmonic Polylogarithms on top of the ratio functions, but the general procedure is very similar and largely covered in  a number of publications,\cite{AblingerThesis,Ablinger:2011te,Blumlein:2009ta,Blumlein:1998if,Blumlein:2009fz}\hspace{0.02cm}.

 It is worth mentioning that there is another analytic continuation for the harmonic sum, from odd positive integer values of the argument, which is different for harmonic sums with at least one negative index and  denoted by  $\bar{S}_{a_1,a_2,...}^{-}(z)$. 
Both analytic continuations are equally valid. 
 Our goal is to find a closed expression of the BFKL eigenvalue for all possible values of anomalous dimension and conformal spin, so that  we follow the notation of Gromov, Levkovich-Maslyuk and Sizov,\cite{Gromov:2015vua}  and use $\bar{S}^{+}(z)$ throughout the text. For simplicity of presentation in this paper we write everywhere $S_{a_1,a_2,...}(z)$ instead of $\bar{S}_{a_1,a_2,...}^{+}(z)$.   

 As it was already mentioned there is no unique way in defining a minimal irreducible set of harmonic sums due to the functional relations between them. 
 For example, one can use the quasi-shuffle relations and then the minimal irreducible basis would include quadratic terms $S_{a_1}(z)S_{a_2}(z)$ in place of either $S_{a_1,a_2}(z)$ or $S_{a_2,a_1}(z)$. It is convenient to use quasi-shuffle relations to remove from the minimal basis the harmonic sums with the first index being equal $1$, because those are divergent as $z \to \infty$. Then, the remaining harmonic sums give transcendental constants at $z \to \infty$. Most of constants are reducible and one is free to choose an irreducible set of transcendental  constants at any given weight. We use the set implemented in the HarmonicSums package,\cite{AblingerThesis,Ablinger:2011te}
 
\begin{eqnarray}\label{C1}
C_1=\{\log (2)\}
\end{eqnarray}
and 
\begin{eqnarray}\label{C2}
C_2=\left\{\pi ^2,\log ^2(2)\right\}
\end{eqnarray}
and 
\begin{eqnarray}\label{C3}
C_3=\left\{\pi ^2 \log (2),\log ^3(2),\zeta _3\right\}
\end{eqnarray}
and 
\begin{eqnarray}\label{C4}
C_4=\left\{\pi ^4,\pi ^2 \log ^2(2),\log ^4(2),\text{Li}_4\left(\frac{1}{2}\right),\zeta _3 \log
   (2)\right\}
\end{eqnarray}
as well as 
\begin{eqnarray}\label{C5}
C_5&=&\left\{\pi ^4 \log(2),\pi ^2 \log^3(2),\log^5(2),\text{Li}_4\left(\frac{1}{2}\right) \log(2),
\right.
\nonumber
\\
&& 
\left.
\hspace{2cm}\text{Li}_5\left(\frac{1}{2}\right),\pi ^2 \zeta _3,\zeta _3 \log^2(2),\zeta
   _5\right\}
\end{eqnarray}

 where $C_w$ stands for a minimal set of irreducible constants at given weight $w$. There is only one of those  at $w=1$, two at $w=2$, three at $w=3$, five at $w=4$ and eight irreducible constants at weight $w=5$.

We choose to use a linear minimal set of the harmonic sums to represent our results. In this set we do not apply  quasi-shuffle relations and thus all the terms of the basis are linear in $S_{a_1,a_2,...}(z)$. This choice is dictated mostly by a convenience and was also used by Caron Huot and Herraren,\cite{Caron-Huot:2016tzz} on which we would like to rely   in  our future calculations. The minimal linear set of harmonic sums we use is  as follows 
\begin{eqnarray}\label{B_1}
B_1=\left\{S_{-1},S_1\right\}
\end{eqnarray}
  and  
\begin{eqnarray}\label{B_2}
B_2 = \left\{S_{-2},S_2,S_{-1,1},S_{1,-1},S_{1,1},S_{-1,-1}\right\}
\end{eqnarray}
and 
\begin{eqnarray}\label{B_3}
B_3&=&\left\{S_{-3},S_3,S_{-2,-1},S_{-2,1},S_{2,-1},S_{2,1},S_{-1,1,-1},S_{-1,1,1}, S_{1,-2},S_{1,2},S_{1
   ,-1,-1},
   \right.
\nonumber
\\
&& 
\left.
S_{1,-1,1},
S_{1,1,-1},S_{1,1,1},S_{-1,-2},S_{-1,2},S_{-1,-1,-1},S_{-1,-1,1}\right\}
\end{eqnarray}
and
\begin{eqnarray}\label{B_4}
B_4&=&\left\{S_{-4},S_4,S_{-3,-1},S_{-3,1},S_{2,-2},S_{3,-1},S_{3,1},S_{-2,-1,-1},S_{-2,-1,1},S_{-2,1,-1},\right.
\nonumber
\\
&& 
\left.
S_{-2,1,1},S_{2,-1,-1},S
   _{2,-1,1},S_{2,1,-1},S_{2,1,1},S_{-1,1,-1,-1},S_{-1,1,-1,1},S_{-1,1,1,1},
   \right.
\nonumber
\\
&& 
\left.
S_{1,-3},S_{1,3},S_{1,-2,-1},S_{1,-2,1},S_{1,-1,
   -2},S_{1,-1,2},S_{1,1,-2},S_{1,1,2},S_{1,2,-1},S_{1,2,1},
   \right.
\nonumber
\\
&& 
\left.
S_{1,-1,-1,-1},S_{1,-1,-1,1},S_{1,-1,1,-1},S_{1,-1,1,1},S_{1,1,-
   1,-1},S_{1,1,-1,1},S_{1,1,1,-1},
   \right.
\nonumber
\\
&& 
\left.
S_{1,1,1,1},S_{-1,-3},S_{-1,3},S_{-1,-2,-1},S_{-1,-2,1},S_{-1,-1,-2},S_{-1,-1,2},S_{-1,1,
   -2},
   \right.
\nonumber
\\
&& 
\left.
S_{-1,1,2},S_{-1,2,-1},S_{-1,2,1},S_{-1,-1,-1,-1},S_{-1,-1,-1,1},S_{-1,-1,1,-1},S_{-1,-1,1,1},
\right.
\nonumber
\\
&& 
\left.
S_{-1,1,1,-1},S_{-2,-2}
   ,S_{-2,2},S_{2,2}\right\}
\end{eqnarray}
as well as 
\begin{eqnarray}\label{B_5}
B_5&=&\left\{S_{-5},S_5,S_{-4,-1},S_{-4,1},S_{-3,-2},S_{-3,2},S_{3,-2},S_{3,2},S_{4,-1},S_{4,1},S_{-3,-1,-1},
 \right.
\nonumber
\\
&& 
\left.
S_{-3,-1,1},S_{-3,1,-
   1},S_{-3,1,1},S_{-2,-1,-2},S_{-2,1,-2},S_{2,-2,-1},S_{2,-2,1},S_{2,-1,-2},
    \right.
\nonumber
\\
&& 
\left.
S_{2,1,-2},S_{2,2,-1},S_{2,2,1},S_{3,-1,-1},S_{
   3,-1,1},S_{3,1,-1},S_{3,1,1},S_{-2,-1,-1,-1},
    \right.
\nonumber
\\
&& 
\left.
S_{-2,-1,-1,1},S_{-2,-1,1,-1},S_{-2,-1,1,1},S_{-2,1,-1,-1},S_{-2,1,-1,1},S_{
   -2,1,1,-1},
    \right.
\nonumber
\\
&& 
\left.
S_{-2,1,1,1},S_{2,-1,-1,-1},S_{2,-1,-1,1},S_{2,-1,1,-1},S_{2,-1,1,1},S_{2,1,-1,-1},
 \right.
\nonumber
\\
&& 
\left.
S_{2,1,-1,1},S_{2,1,1,-1},S
   _{2,1,1,1},S_{-1,1,-1,-1,-1},S_{-1,1,-1,-1,1},S_{-1,1,-1,1,1},
    \right.
\nonumber
\\
&& 
\left.
S_{-1,1,1,-1,-1},S_{-1,1,1,1,-1},S_{-1,1,1,1,1},S_{1,-4},S_
   {1,4},S_{1,-3,-1},S_{1,-3,1},S_{1,-2,-2},
    \right.
\nonumber
\\
&& 
\left.
S_{1,-2,2},S_{1,-1,-3},S_{1,-1,3},S_{1,1,-3},S_{1,1,3},S_{1,2,-2},S_{1,2,2},S_{1
   ,3,-1},S_{1,3,1},
    \right.
\nonumber
\\
&& 
\left.
S_{1,-2,-1,-1},S_{1,-2,-1,1},S_{1,-2,1,-1},S_{1,-2,1,1},S_{1,-1,-2,-1},S_{1,-1,-2,1},
 \right.
\nonumber
\\
&& 
\left.
S_{1,-1,-1,-2},S_{1
   ,-1,-1,2},S_{1,-1,1,-2},S_{1,-1,1,2},S_{1,-1,2,-1},S_{1,-1,2,1},
    \right.
\nonumber
\\
&& 
\left.
S_{1,1,-2,-1},S_{1,1,-2,1},S_{1,1,-1,-2},S_{1,1,-1,2},S_{
   1,1,1,-2},S_{1,1,1,2},S_{1,1,2,-1},S_{1,1,2,1},
    \right.
\nonumber
\\
&& 
\left.
S_{1,2,-1,-1},S_{1,2,-1,1},S_{1,2,1,-1},S_{1,2,1,1},S_{1,-1,-1,-1,-1},S_{1
   ,-1,-1,-1,1},
    \right.
\nonumber
\\
&& 
\left.
S_{1,-1,-1,1,-1},S_{1,-1,-1,1,1},S_{1,-1,1,-1,-1},S_{1,-1,1,-1,1},S_{1,-1,1,1,-1},S_{1,-1,1,1,1},
 \right.
\nonumber
\\
&& 
\left.
S_{1,1,-1,-
   1,-1},S_{1,1,-1,-1,1},S_{1,1,-1,1,-1},S_{1,1,-1,1,1},S_{1,1,1,-1,-1},S_{1,1,1,-1,1},
    \right.
\nonumber
\\
&& 
\left.
S_{1,1,1,1,-1},S_{1,1,1,1,1},S_{-1,-4
   },S_{-1,4},S_{-1,-3,-1},S_{-1,-3,1},S_{-1,-2,-2},S_{-1,-2,2},
    \right.
\nonumber
\\
&& 
\left.
S_{-1,-1,-3},S_{-1,-1,3},S_{-1,1,-3},S_{-1,1,3},S_{-1,2,-2},
   S_{-1,2,2},S_{-1,3,-1},S_{-1,3,1},
    \right.
\nonumber
\\
&& 
\left.
S_{-1,-2,-1,-1},S_{-1,-2,-1,1},S_{-1,-2,1,-1},S_{-1,-2,1,1},S_{-1,-1,-2,-1},S_{-1,-1,-2
   ,1},
    \right.
\nonumber
\\
&& 
\left.
S_{-1,-1,-1,-2},S_{-1,-1,-1,2},S_{-1,-1,1,-2},S_{-1,-1,1,2},S_{-1,-1,2,-1},S_{-1,-1,2,1},
 \right.
\nonumber
\\
&& 
\left.
S_{-1,1,-2,-1},S_{-1,1,-2,1}
   ,S_{-1,1,-1,-2},S_{-1,1,-1,2},S_{-1,1,1,-2},S_{-1,1,1,2},S_{-1,1,2,-1},
    \right.
\nonumber
\\
&& 
\left.
S_{-1,1,2,1},S_{-1,2,-1,-1},S_{-1,2,-1,1},S_{-1,2,
   1,-1},S_{-1,2,1,1},S_{-1,-1,-1,-1,-1},
    \right.
\nonumber
\\
&& 
\left.
S_{-1,-1,-1,-1,1},S_{-1,-1,-1,1,-1},S_{-1,-1,-1,1,1},S_{-1,-1,1,-1,-1},S_{-1,-1,1,-
   1,1},
    \right.
\nonumber
\\
&& 
\left.
S_{-1,-1,1,1,-1},S_{-1,-1,1,1,1},S_{-1,1,-1,1,-1},S_{-1,1,1,-1,1},S_{-2,-3},S_{-2,3},S_{2,-3},S_{2,3},
 \right.
\nonumber
\\
&& 
\left.
S_{-2,-2,-1},S
   _{-2,-2,1},S_{-2,-1,2},S_{-2,1,2},S_{-2,2,-1},S_{-2,2,1},S_{2,-1,2},S_{2,1,2}\right\}
\end{eqnarray}

A comprehensive discussion on harmonic sums,  irreducible constants, functional identities and possible choice of the minimal set of functions at given weight is presented by J.~Ablinger,\cite{AblingerThesis}\hspace{0.02cm}. 
 In this paper we focus only at the reflection identities for harmonic sums at weight $w=5$ analytically continued from even positive points to complex plane.  In the next Section we discuss them in more details along the method we use in our calculations. 
 
We also discuss the physical motivation for the reflection identities focusing on the well known color singlet NLO BFKL eigenvalue in $\mathcal{N}=4$ SYM. We demonstrate how its full functional form can be restored just from two specific values of the conformal spin while keeping the dependence on the anomalous dimension.

\section{Motivation: restoring NLO eigenvalue}\label{motivation}

In this section we illustrate how the reflection identities derived in this paper can be used for restoring the full functional dependence on the conformal spin of BFKL eigenvalue. The BFKL eigenvalue is a function of two variables, the anomalous dimension $\nu$ which takes real continuous values and the conformal spin $n$ which is defined at discreet integers. In some specific applications one can consider analytic continuation to the complex plane, but those cases are beyond the scope of the present paper. 

As it was already mentioned it is a very non-trivial task to calculate higher order corrections to the BFKL eigenvalue using a well established perturbation theory. Instead one can consider analogous systems based on the modern integrability approaches and have exact analytic results in some regions of the $(\nu, n)$ plane. This was done by by N.~ Gromov, F.~Levkovich-Maslyuk  and G.~Sizov,\cite{Gromov:2015vua} using Quantum Spectral Curves and the resulting singlet NNLO BFKL eigenvalue in $\mathcal{N}=4$ SYM was calculated for a specific value of the conformal spin $n=0$ while keeping the full dependence on the anomalous dimension $\nu$. 

 Caron Huot and Herraren,\cite{Caron-Huot:2016tzz} extended this analysis to    other  specific values of the conformal spin $n=1,2,...$ providing infinitely many lines at the $(\nu, n)$ plane. Despite plenty of data we are still lacking the full functional form  of the singlet NNLO BFKL eigenvalue in $\mathcal{N}=4$ SYM mainly because we do not know the space of functions it is built of.  A natural candidate would be the harmonic sums of one variable considered in the present paper, but the problem is that we need to express the function of two variables in terms of the function of one variable and there are many ways to do that. 

In both cases of above mentioned   Refs.~\cite{Gromov:2015vua,Caron-Huot:2016tzz}  the authors considered the pseudo-holomorphic separable form of the singlet NNLO BFKL eigenvalue for any specific value of the conformal spin. To the best of our understanding  this pseudo-holomorphic separable form differs from value to value of the conformal spin and it seem to be not suitable for writing a closed expression valid for all values of the conformal spin. 

It follows from the form of NLO BFKL eigenvalue that  one needs to relax the condition of pseudo-holomorphic separability and consider bilinear forms of mixed functions of  holomorphic and antiholomorphic variables. This approach clearly works at the NLO level, where we know the exact closed expression for arbitrary values of the conformal spin while preserving  the full dependence on the anomalous dimension. In the following we illustrate this idea for the case of the singlet NLO BFKL eigenvalue in $\mathcal{N}=4$ SYM using the reflection identities for harmonic sums at weight $w=3$ derived by one of the authors,\cite{Prygarin:2018tng, Prygarin:2018cog}\hspace{0.02cm}. In the rest of this section we use the expressions and  the notation of Kotikov and Lipatov,\cite{Kotikov:2002ab} for the case of  $\mathcal{N}=4$ SYM. 

For simplicity of presentation let us define a complex variable
\begin{eqnarray}\label{zzbar}
z= -\frac{1}{2}+ i\nu +\frac{|n|}{2}, \;\; 
\bar{z}= -\frac{1}{2}- i\nu +\frac{|n|}{2}. 
\end{eqnarray}
Then the leading-order~(LO) and the next-to-leading~(NLO)  BFKL eigenvalue in the color singlet channel reads~(see Section 3 of the paper by Kotikov and Lipatov,\cite{Kotikov:2002ab})
\[ \omega =4\,\hat{a}\,\biggl[\chi (z,\bar{z} )+\delta
(z,\bar{z} )\,\hat{a}\, \biggr]\,.\,
\]
The LO function is given by  
\begin{eqnarray}\label{chi}
\chi(z,\bar{z} )=2 \psi(1)- \psi(1+z)-\psi(1+\bar{z})=-S_1(z)-S_1(\bar{z}),  
\end{eqnarray}
where  the digamma function defined by 
\begin{eqnarray}
\psi(z)=\frac{d \ln \Gamma(z)}{ dz },
\end{eqnarray}
for $\Gamma(z)$ being  Euler gamma function. In Eq.~(\ref{chi}) we used the fact the digamma function is the analytic continuation of the harmonic sum $S_1(z)$.

The function  that determines the NLO part of the BFKL eigenvalue reads 
\begin{eqnarray}\label{delta}
\delta (z,\bar{z} )=\phi (1+z)+\phi (1+\bar{z})-2 \chi(z, \bar{z})
\biggl(\beta' (1+z)+\beta' (1+\bar{z})+\zeta_2 \biggr),
\end{eqnarray}
where 
\begin{eqnarray}\label{phi}
\phi (z) &=&3\zeta (3)+\Psi ^{^{\prime \prime }}(z)-2\Phi _{2}(z)+2\beta
^{^{\prime }}(z)\Bigl(\Psi (1)-\Psi (z)\Bigr)\,.  
\end{eqnarray}
The most complicated term appearing in Eq.~(\ref{phi}) can be written as follows
\begin{eqnarray}
\Phi _2(z )= \sum_{k=0}^\infty \frac{\left( \beta ^{\prime
}(k+1)+(-1)^k\Psi ^{\prime }(k+1)\right) }{k+z }
-\sum_{k=0}^\infty \frac{(-1)^k\Bigl( \Psi (k+1)-\Psi (1)\Bigr) } {(k+
z)^2 }
\end{eqnarray}
for the derivative of  the Dirichlet beta function
\begin{eqnarray}\label{betap}
\beta ^{\prime }(z)=\frac{1}{4}\Biggl[\Psi ^{\prime }\Bigl(\frac{z+1}{2}%
\Bigr)-\Psi ^{\prime }\Bigl(\frac{z}{2}\Bigr)\Biggr]= -\sum _{r=0}^{\infty}%
\frac{(-1)^r}{(z+r)^2}= - S_{-2}(z-1)-\frac{\zeta_2}{2}\,.  
\end{eqnarray}
In Eq.~(\ref{betap}) we used the fact the derivative of  the Dirichlet beta function is related to the analytic continuation of the harmonic sum $ S_{-2}(z-1)$ from even positive values of the argument to the complex plane.  In a similar way we write the expression in Eq.~(\ref{phi}) in terms of the analytically continued harmonic sums as follows
\begin{eqnarray}\label{phiS}
\phi(z)=4 S_{1,-2}(z-1)-2 S_{-3}(z-1)+\frac{1}{3} \pi ^2 S_{1}(z-1)+2 S_{3}(z-1).
\end{eqnarray}
Then the NLO function in Eq.~(\ref{delta}) reads
\begin{eqnarray}\label{deltaS}
\delta (z,\bar{z} )=\phi (1+z)+\phi (1+\bar{z})-2 
\biggl(S_1(z)+S_1(\bar{z}) \biggr)
\biggl(S_{-2}(z)+S_{-2}(\bar{z}) \biggr)
\end{eqnarray}

At this point we have the exact expression for the LO and NLO BFKL eigenvalue in the color singlet state in $\mathcal{N}=4$ SYM expressed in terms of the analytically continued harmonic sums. It is no different from the one given in the  paper of Kotikov and Lipatov,\cite{Kotikov:2002ab}. 

Now we consider a special case of zero conformal spin for expression in  Eq.~(\ref{deltaS}). For  $n=0$ the holomorphic variables $z$ and $\bar{z}$ are not  independent anymore and related by  $\bar{z}=-1-z$ suggesting that it would be more natural to write $\delta (z,\bar{z} )$ in  Eq.~(\ref{deltaS}) as a sum of two parts 
\begin{eqnarray}\label{deltanzero}
\delta (z,\bar{z} )= F_2(z)+F_2(-1-z), 
\end{eqnarray}
where 
\begin{eqnarray}
\frac{F_2(z)}{4} =-\frac{3}{2}\zeta_3+\pi^2 \ln 2 +\frac{\pi^2}{3}S_1(z)
+2 S_{3}(z)+\pi^2 S_{-1}(z)-4 S_{-2,1}(z),
\end{eqnarray}
using the reflection identity derived by one of the  authors,\cite{Prygarin:2018tng} 
\begin{eqnarray}
&&  S_1(z) S_{-2}(-1-z) = -\frac{1}{2} \pi ^2 \log (2)+\frac{3 \zeta (3)}{4}-\frac{1}{4} \pi ^2
   S_{-1}(-1-z)
   -\frac{1}{4} \pi ^2 S_{-1}(z)
   \nonumber
    \\
 && 
 \hspace{1.9cm}+\frac{1}{12} \pi ^2 S_1(-1-z)-\frac{1}{12} \pi ^2
   S_1(z)+S_{-2,1}(z)+S_{1,-2}(-1-z)
\end{eqnarray}
for the mixed term $S_1(z) S_{-2}(\bar{z})$  appearing in the last term of   Eq.~(\ref{deltaS}). 

In fact,  this example    reproduces the starting point of studies of  Gromov, Levkovich-Maslyuk, Sizov,\cite{Gromov:2015vua} and  Caron Huot and Herraren,\cite{Caron-Huot:2016tzz}  which laid a path to the corresponding expression for the NNLO case expressed in terms of the harmonic sums at weight $w=5$. The only difference is that is those studies the authors  did not use the reflection identities rather a direct pole decomposition. 
The same pole decomposition we used for deriving the reflection identities so effectively there is no difference between the two approaches at this point.

The need for the reflection identities becomes obvious when we try to go back from the expression for $n=0$  in Eq.~(\ref{deltanzero}) in attempt to restore the full functional dependence on the conformal spin in Eq.~(\ref{deltaS}).  In this case  we need to consider all possible combinations of
 $S_{\{a\}}(z)S_{\{b\}}(\bar{z})$ appearing as a direct product of harmonic sums at weight $w=1$ and weight $w=2$ and then use the corresponding reflection identities $S_{\{a\}}(z)S_{\{b\}}(-1-z)$ in the following way. 
 
 Let us assume we do not know a closed expression valid for an arbitrary value of the conformal spin, i.e. we do not know the exact form of the expression in Eq.~(\ref{phiS}). We do know that this expression must be built of the harmonic sums of the uniform weight $w=3$ because this corresponds to the highest  transcendentality principle formulated for  $\mathcal{N}=4$ SYM  by Kotikov and Lipatov,\cite{Kotikov:2006ts}\hspace{0.02cm}. Then we  can consider the following terms that should  build our ansatz 
 \begin{eqnarray}\label{guess}
 \beta_i \biggl(S_{\{a_i\}}(z)S_{\{b_i\}}(\bar{z}) - S_{\{a_i\}}(z)S_{\{b_i\}}(-1-z) \biggr), 
 \end{eqnarray}
 where $\beta_i$ is the unknown parameter to be fixed and the index $i$ runs over all possible terms in the  direct product of harmonic sums at weight $w=1$ and weight $w=2$. 
 
 Obviously, at $n=0$ the expression Eq.~(\ref{guess}) vanishes and  we cannot fix the coefficients $\beta_i$ using only information from Eq.~(\ref{deltanzero}), but if we add to it all possible harmonic sums of either $z$ or $\bar{z}$ as well as  constants of uniform weight $w=3$ we can fix the coefficients of those, merely based on the pole expansion of Eq.~(\ref{deltanzero}).  So we are left with only a limited set of unknown parameters $\beta_i$ and to fix those we need to consider only one additional value of the conformal spin, say $n=1$. Provided we have this expression, it makes it possible to  calculate $\beta_i$ using pole expansion of the harmonic sums and comparing them for the two expressions. 
 
 Fortunately, the analytic expressions in terms of harmonic sums for both $n=0$ and $n=1$ as well as many others are available in  Refs.~\cite{Gromov:2015vua,Caron-Huot:2016tzz} and we only need to consider all possible reflection identities up to weight $w=5$ following the steps we have outlined at the example of the NLO BFKL eigenvalue. In this  paper we derived the reflection identities at weight $w=5$, which adds to our previous studies for $w=2$, $w=3$ and $w=4$,\cite{Prygarin:2018tng, Prygarin:2018cog}\hspace{0.02cm}. The missing part is the reflection identities at weight $w=5$  of the type $S_{\{a_i\}}(z)S_{\{b_i\}}(-z)$, which  are  still to be calculated. Note,  the missing part is related to the case $n=1$, where $z+\bar{z}=0$.

In the next section we discuss methods and results of our calculations focusing on some technical aspects of building the proper space of functions, performing the pole expansion of the harmonic sums and building coefficient equations for the reflection identities.

\newpage

\section{Methods and Results}

The reflection identities at weight $w=5$ are obtained by taking a product of harmonic sums of argument $z$ at weight $w=1$ and harmonic sums of 
argument $-1-z$
at weight $w=4$, i.e. $B_1 \otimes \bar{B}_4$, and also by taking a product  of harmonic sums of argument $z$ at weight $w=2$ and $-1-z$ at weight $w=3$, 
i.e. $B_2 \otimes \bar{B}_3$.

The number of basis harmonic sums in $B_1$, $B_2$ and $B_3$ is given by 
\begin{eqnarray}\label{L1L2L3L4}
\text{Length}(B_1)=2, \;\;\; \text{Length}(B_2)=6,  \;\;\;  \text{Length}(B_3)=18,
\;\; \text{Length}(B_4)=54
\end{eqnarray}
so that the number of elements in the products $B_1 \otimes \bar{B}_4$ and  $B_2 \otimes \bar{B}_3$ reads 
\begin{eqnarray}
B_1 \otimes \bar{B}_4=2 \times 54=108
\end{eqnarray}
and 
\begin{eqnarray}
B_2 \otimes \bar{B}_3=6 \times 18=108,
\end{eqnarray}
resulting in the total number of irreducible reflections identities at weight $w=5$ being equal to $108+108=216$.

In order to calculate the reflection identities at weight $w=5$ we use the  basis harmonic sums at $w=5$ listed in Eq.~(\ref{B_5}) together with basis 
 harmonic sums at lower weight listed in the work of J.~Ablinger~(\ref{B_1})-(\ref{B_4}) multiplied by irreducible constants at corresponding weight listed in Eqs.~(\ref{C1})-(\ref{C4}).  This should be supplemented by irreducible constants at weight $w=5$ listed in Eqs.~(\ref{C5}).
The number of basis sums in $B_5$ equals
\begin{eqnarray}\label{L5}
 \text{Length}(B_5)=162
\end{eqnarray}
so that   the total number of terms in the expansion ansatz at $w=5$ 
\begin{eqnarray}
 B_5 + B_4 \otimes C_1+B_3 \otimes C_2+B_2 \otimes C_3+B_1 \otimes C_4+C_5
\end{eqnarray}
 is given by
 \begin{eqnarray}\label{Lanz5}
\text{Length}(\text{ANZ}_5)=162 + 54 \times 1+18 \times 2+6  \times 3
+2  \times 5 + 8=288
\end{eqnarray}
\newpage
The full expansion ansatz at $w=5$ is given by 
\begin{eqnarray}\label{anz5}
\text{ANZ}_5&=&
\left\{\pi ^4 \log (2),\pi ^2 \log ^3(2),\log ^5(2),\log (2) \text{Li}_4\left(\frac{1}{2}\right),\text{Li}_5\left(\frac{1}{2}\right),S_{-5},
\log (2) S_{-4},
    \right.
\nonumber 
\\
&& 
\left.
\pi ^2
   S_{-3},\log ^2(2) S_{-3},
\pi ^2 \log (2) S_{-2},\log ^3(2) S_{-2},\pi ^4 S_{-1},\pi ^2 \log ^2(2) S_{-1},
    \right.
\nonumber 
\\
&& 
\left.
\log ^4(2)
   S_{-1},\text{Li}_4\left(\frac{1}{2}\right) S_{-1},\pi ^4 S_1,\pi ^2 \log ^2(2) S_1,\log ^4(2) S_1,\text{Li}_4\left(\frac{1}{2}\right) S_1,
       \right.
\nonumber 
\\
&& 
\left.
\pi ^2 \log (2)
   S_2,\log ^3(2) S_2,\pi ^2 S_3,\log ^2(2) S_3,\log (2) S_4,S_5,S_{-4,-1},S_{-4,1},
       \right.
\nonumber 
\\
&& 
\left.
S_{-3,-2},\log (2) S_{-3,-1},\log (2)
   S_{-3,1},S_{-3,2},S_{-2,-3},\log (2) S_{-2,-2},\pi ^2 S_{-2,-1},
       \right.
\nonumber 
\\
&& 
\left.
\log ^2(2) S_{-2,-1},\pi ^2 S_{-2,1},\log ^2(2)
   S_{-2,1},\log (2) S_{-2,2},S_{-2,3},S_{-1,-4},
       \right.
\nonumber 
\\
&& 
\left.
\log (2) S_{-1,-3},\pi ^2 S_{-1,-2},\log ^2(2) S_{-1,-2},\pi ^2 \log (2)
   S_{-1,-1},\log ^3(2) S_{-1,-1},
       \right.
\nonumber 
\\
&& 
\left.
\pi ^2 \log (2) S_{-1,1},\log ^3(2) S_{-1,1},\pi ^2 S_{-1,2},\log ^2(2) S_{-1,2},\log (2)
   S_{-1,3},S_{-1,4},
       \right.
\nonumber 
\\
&& 
\left.
S_{1,-4},\log (2) S_{1,-3},\pi ^2 S_{1,-2},\log ^2(2) S_{1,-2},\pi ^2 \log (2) S_{1,-1},
    \right.
\nonumber 
\\
&& 
\left.
\log ^3(2)
   S_{1,-1},\pi ^2 \log (2) S_{1,1},
   \log ^3(2) S_{1,1},\pi ^2 S_{1,2},\log ^2(2) S_{1,2},\log (2)
   S_{1,3},
       \right.
\nonumber 
\\
&& 
\left.
S_{1,4},S_{2,-3},\log (2) S_{2,-2},\pi ^2 S_{2,-1},\log ^2(2) S_{2,-1},\pi ^2 S_{2,1},\log ^2(2) S_{2,1},
    \right.
\nonumber 
\\
&& 
\left.
\log (2)
   S_{2,2},S_{2,3},S_{3,-2},\log (2) S_{3,-1},\log (2)
   S_{3,1},S_{3,2},S_{4,-1},S_{4,1},S_{-3,-1,-1},
       \right.
\nonumber 
\\
&& 
\left.
S_{-3,-1,1},S_{-3,1,-1},S_{-3,1,1},S_{-2,-2,-1},S_{-2,-2,1},S_{-2,-1,-2},
\log (2) S_{-2,-1,-1},
    \right.
\nonumber 
\\
&& 
\left.
\log (2) S_{-2,-1,1},S_{-2,-1,2},S_{-2,1,-2},\log (2) S_{-2,1,-1},\log (2)
   S_{-2,1,1},S_{-2,1,2},
       \right.
\nonumber 
\\
&& 
\left.
S_{-2,2,-1},S_{-2,2,1},S_{-1,-3,-1},S_{-1,-3,1},S_{-1,-2,-2},\log (2) S_{-1,-2,-1},
    \right.
\nonumber 
\\
&& 
\left.
\log (2)
   S_{-1,-2,1},S_{-1,-2,2},S_{-1,-1,-3},\log (2) S_{-1,-1,-2},\pi ^2 S_{-1,-1,-1},
          \right.
\nonumber 
\\
&& 
\left.
\log ^2(2) S_{-1,-1,-1},\pi ^2
   S_{-1,-1,1},\log ^2(2) S_{-1,-1,1},\log (2) S_{-1,-1,2},S_{-1,-1,3},
       \right.
\nonumber 
\\
&& 
\left.
S_{-1,1,-3},\log (2) S_{-1,1,-2},\pi ^2
   S_{-1,1,-1},\log ^2(2) S_{-1,1,-1},\pi ^2 S_{-1,1,1},
       \right.
\nonumber 
\\
&& 
\left.
\log ^2(2) S_{-1,1,1},\log (2)
   S_{-1,1,2},S_{-1,1,3},S_{-1,2,-2},\log (2) S_{-1,2,-1},
       \right.
\nonumber 
\\
&& 
\left.
\log (2)
   S_{-1,2,1},S_{-1,2,2},S_{-1,3,-1},S_{-1,3,1},S_{1,-3,-1},S_{1,-3,1},
   S_{1,-2,-2},
       \right.
\nonumber 
\\
&& 
\left.
\log (2) S_{1,-2,-1},\log (2)
   S_{1,-2,1},S_{1,-2,2},S_{1,-1,-3},\log (2) S_{1,-1,-2},\pi ^2 S_{1,-1,-1},
       \right.
\nonumber 
\\
&& 
\left.
\log ^2(2) S_{1,-1,-1},\pi ^2 S_{1,-1,1},
   \log
   ^2(2) S_{1,-1,1},\log (2) S_{1,-1,2},S_{1,-1,3},
       \right.
\nonumber 
\\
&& 
\left.
S_{1,1,-3},\log (2) S_{1,1,-2},\pi ^2 S_{1,1,-1},\log ^2(2)
   S_{1,1,-1},\pi ^2 S_{1,1,1},
   \log ^2(2) S_{1,1,1},
       \right.
\nonumber 
\\
&& 
\left.
\log (2) S_{1,1,2},S_{1,1,3},S_{1,2,-2},\log (2) S_{1,2,-1},\log (2)
   S_{1,2,1},S_{1,2,2},S_{1,3,-1},S_{1,3,1},
       \right.
\nonumber 
\\
&& 
\left.
S_{2,-2,-1},S_{2,-2,1},S_{2,-1,-2},\log (2) S_{2,-1,-1},\log (2)
   S_{2,-1,1},S_{2,-1,2},S_{2,1,-2},
       \right.
\nonumber 
\\
&& 
\left.
\log (2) S_{2,1,-1},\log (2)
   S_{2,1,1},S_{2,1,2},S_{2,2,-1},S_{2,2,1},S_{3,-1,-1},S_{3,-1,1},S_{3,1,-1},
       \right.
\nonumber 
\\
&& 
\left.
S_{3,1,1},S_{-2,-1,-1,-1},S_{-2,-1,-1,1},S_{-2
   ,-1,1,-1},S_{-2,-1,1,1},S_{-2,1,-1,-1},
       \right.
\nonumber 
\\
&& 
\left.
S_{-2,1,-1,1},S_{-2,1,1,-1},S_{-2,1,1,1},S_{-1,-2,-1,-1},S_{-1,-2,-1,1},S_{-1,-2,1
   ,-1},
       \right.
\nonumber 
\\
&& 
\left.
S_{-1,-2,1,1},S_{-1,-1,-2,-1},S_{-1,-1,-2,1},S_{-1,-1,-1,-2},\log (2) S_{-1,-1,-1,-1},
    \right.
\nonumber 
\\
&& 
\left.
\log (2)
   S_{-1,-1,-1,1},
   S_{-1,-1,-1,2},S_{-1,-1,1,-2},\log (2) S_{-1,-1,1,-1},\log (2)
   S_{-1,-1,1,1},  \right. \nonumber 
\end{eqnarray}
   \newpage 
\begin{eqnarray}
   && \left. S_{-1,-1,1,2},S_{-1,-1,2,-1},S_{-1,-1,2,1},S_{-1,1,-2,-1},S_{-1,1,-2,1},
    S_{-1,1,-1,-2},
        \right.
\nonumber 
\\
&& 
\left.
\log (2)
   S_{-1,1,-1,-1},\log (2) S_{-1,1,-1,1},S_{-1,1,-1,2},S_{-1,1,1,-2},\log (2) S_{-1,1,1,-1},
       \right.
\nonumber 
\\
&& 
\left.
\log (2)
   S_{-1,1,1,1},S_{-1,1,1,2},S_{-1,1,2,-1},S_{-1,1,2,1},
   S_{-1,2,-1,-1},S_{-1,2,-1,1},S_{-1,2,1,-1},
       \right.
\nonumber 
\\
&& 
\left.
S_{-1,2,1,1},S_{1,-2,-1,-
   1},S_{1,-2,-1,1},S_{1,-2,1,-1},S_{1,-2,1,1},S_{1,-1,-2,-1},
   S_{1,-1,-2,1},
       \right.
\nonumber 
\\
&& 
\left.
S_{1,-1,-1,-2},\log (2) S_{1,-1,-1,-1},\log (2)
   S_{1,-1,-1,1},S_{1,-1,-1,2},S_{1,-1,1,-2},
       \right.
\nonumber 
\\
&& 
\left.
\log (2) S_{1,-1,1,-1},\log (2)
   S_{1,-1,1,1},S_{1,-1,1,2},S_{1,-1,2,-1},S_{1,-1,2,1},
   S_{1,1,-2,-1},S_{1,1,-2,1},
       \right.
\nonumber 
\\
&& 
\left.
S_{1,1,-1,-2},\log (2) S_{1,1,-1,-1},\log
   (2) S_{1,1,-1,1},S_{1,1,-1,2},S_{1,1,1,-2},\log (2) S_{1,1,1,-1},
       \right.
\nonumber 
\\
&& 
\left.
\log (2)
   S_{1,1,1,1},S_{1,1,1,2},S_{1,1,2,-1},
   S_{1,1,2,1},S_{1,2,-1,-1},S_{1,2,-1,1},S_{1,2,1,-1},S_{1,2,1,1},
       \right.
\nonumber 
\\
&& 
\left.
S_{2,-1,-1,-1},S_{2,
   -1,-1,1},S_{2,-1,1,-1},S_{2,-1,1,1},S_{2,1,-1,-1},
   S_{2,1,-1,1},S_{2,1,1,-1},S_{2,1,1,1},
       \right.
\nonumber 
\\
&& 
\left.
S_{-1,-1,-1,-1,-1},S_{-1,-1,-1,-1
   ,1},S_{-1,-1,-1,1,-1},S_{-1,-1,-1,1,1},S_{-1,-1,1,-1,-1},
       \right.
\nonumber 
\\
&& 
\left.
S_{-1,-1,1,-1,1},S_{-1,-1,1,1,-1},S_{-1,-1,1,1,1},S_{-1,1,-1,-1,
   -1},S_{-1,1,-1,-1,1},
       \right.
\nonumber 
\\
&& 
\left.
S_{-1,1,-1,1,-1},S_{-1,1,-1,1,1},S_{-1,1,1,-1,-1},
S_{-1,1,1,-1,1},S_{-1,1,1,1,-1},S_{-1,1,1,1,1},
    \right.
\nonumber 
\\
&& 
\left.
S_{
   1,-1,-1,-1,-1},S_{1,-1,-1,-1,1},S_{1,-1,-1,1,-1},S_{1,-1,-1,1,1},
   S_{1,-1,1,-1,-1},S_{1,-1,1,-1,1},
       \right.
\nonumber 
\\
&& 
\left.
S_{1,-1,1,1,-1},S_{1,-1
   ,1,1,1},S_{1,1,-1,-1,-1},S_{1,1,-1,-1,1},S_{1,1,-1,1,-1},S_{1,1,-1,1,1},
       \right.
\nonumber 
\\
&& 
\left.
S_{1,1,1,-1,-1},S_{1,1,1,-1,1},S_{1,1,1,1,-1},S_{
   1,1,1,1,1},\pi ^2 \zeta _3,\log ^2(2) \zeta _3,S_{-2} \zeta _3,\log (2) S_{-1} \zeta _3,
       \right.
\nonumber 
\\
&& 
\left.
\log (2) S_1 \zeta _3,
   S_2 \zeta
   _3,\zeta _3 S_{-1,-1},\zeta _3 S_{-1,1},\zeta _3 S_{1,-1},\zeta _3 S_{1,1},\zeta _5\frac{}{}\right\}
\end{eqnarray}
  
It is worth emphasizing that that we \emph{do not reduce}  $\text{ANZ}_5$  using quasi-shuffle relations because we want to stick to the \emph{linear basis} used by  Caron Huot and Herraren,\cite{Caron-Huot:2016tzz} for calculations of the color singlet  NNLO BFKL eigenvalue. 
 It should also  be  mentioned that the  total number of the basis elements in either linear  $\text{ANZ}_5$ or non-linear basis  obtained with the use of  quasi-shuffle relations is \emph{the same}. The choice of the basis does not effects the final result and  is solely a matter of convenience.

The expansion of the product of two functions of argument $z$ and argument $-1-z$ we search in terms of two sets of $\text{ANZ}_5$, one of argument $z$ and another one of argument $-1-z$. The total number of elements in this expression equals $288 \times 2-8=568$, where we remove redundant five 
constants at $w=5$ because they are the same for both arguments. We fix the $185$ free coefficients using pole expansion of the product
 $s_{a_1,a_2,..}(z)s_{b_1,b_2,..}(-1-z)$ around negative integers of $z=-5,...,-1$ and expanding to the second order of the expansion parameter. It turns out
 that to fix all $568$ free coefficients we need only expansion up to first order and we use the second order of the expansion to double check our results.
 We checked our results listed in the Appendix by a direct numerical calculation at the accuracy  $10^{-10}$. 

Below we give two examples of reflection identities, $ S_1(z)S_{-2,2}(-1-z)$ for harmonic sums with positive indices and 
 $S_{1}(z)    S_{2,2}(-1-z)$ for harmonic sums with negative indices. 
  All $216$  irreducible 
  reflection identities at weight five are listed in the Appendix. The two examples   are 
\begin{eqnarray} \label{S1Sm22}
  && S_1(z) S_{-2,2}(-1-z) =  -\zeta _2 S_{-2,1}(-1-z)-S_{-3,2}(-1-z)+S_{-2,1,2}(-1-z) 
    \nonumber
   \\
     && \hspace{1cm}
     +S_{1,-2,2}(-1-z)-4 \text{Li}_4\left(\frac{1}{2}\right) S_{-1}(-1-z)+4 \text{Li}_4\left(\frac{1}{2}\right) S_1(-1-z)
      \nonumber
   \\
     && \hspace{1cm}
     +2 \zeta _3 S_{-2}(-1-z)+\frac{23}{40} \zeta _2^2 S_{-1}(-1-z)-\frac{51}{40} \zeta _2^2 S_1(-1-z)
      \nonumber
   \\
     && \hspace{1cm}
     +\zeta _2 \log^2(2) S_{-1}(-1-z)-\zeta _2 \log^2(2) S_1(-1-z)-\frac{7}{2} \zeta _3 \log(2)S_{-1}(-1-z)
      \nonumber
   \\
     && \hspace{1cm}
     +\frac{7}{2} \zeta _3 \log(2)S_1(-1-z)-\frac{1}{6} \log^4(2) S_{-1}(-1-z)+\frac{1}{6} \log^4(2) S_1(-1-z)
      \nonumber
   \\
     && \hspace{1cm}
     +\zeta _2 S_{-2,1}(z)+S_{-4,1}(z)-S_{-2,2,1}(z)-\frac{5 \zeta _2 \zeta _3}{2}-\frac{51 \zeta _5}{32}+\frac{2}{3} \zeta _2 \log^3(2)
      \nonumber
   \\
     && \hspace{1cm}
     -\frac{7}{2} \zeta _3 \log^2(2)+\frac{23}{20} \zeta _2^2 \log(2)-4 \text{Li}_4\left(\frac{1}{2}\right) S_{-1}(z)-4 \text{Li}_4\left(\frac{1}{2}\right) S_1(z)
      \nonumber
   \\
     && \hspace{1cm}
     +8 \text{Li}_5\left(\frac{1}{2}\right)+\frac{23}{40} \zeta _2^2 S_{-1}(z)+\frac{51}{40} \zeta _2^2 S_1(z)-\zeta _2 S_{-3}(z)+\zeta _3 S_{-2}(z)
      \nonumber
   \\
     && \hspace{1cm}
     +\zeta _2 S_{-1}(z) \log^2(2)+\zeta _2 S_1(z) \log^2(2)-\frac{7}{2} \zeta _3 S_{-1}(z) \ln _2-\frac{7}{2} \zeta _3 S_1(z) \ln _2
      \nonumber
   \\
     && \hspace{1cm}
     -\frac{1}{6} S_{-1}(z) \log^4(2)-\frac{1}{6} S_1(z) \log^4(2)-\frac{\log^5(2)}{15} 
   \end{eqnarray}
listed   in Eq.~(\ref{s1scm22}) of the Appendix and 
  \begin{eqnarray}
    && S_1(z) S_{2,2}(-1-z)  =   -\zeta _2 S_{2,1}(-1-z)-S_{3,2}(-1-z)+S_{1,2,2}(-1-z)
     \nonumber
   \\
     && \hspace{1cm}
    +S_{2,1,2}(-1-z)-\frac{7}{10} \zeta _2^2 S_1(-1-z)+2 \zeta _3 S_2(-1-z)-\zeta _2 S_{2,1}(z)-S_{4,1}(z)
     \nonumber
   \\
     && \hspace{1cm}
     +S_{2,2,1}(z)+3 \zeta _3 \zeta _2-\frac{5 \zeta _5}{2}+\frac{7}{10} \zeta _2^2 S_1(z)+\zeta _2 S_3(z)-\zeta _3 S_2(z)  
\end{eqnarray}
given in   Eq.~(\ref{s1sc22}). 
  
One can see that  the reflection identities for harmonic sums with negative indices are  more complicated than those with only positive indices and this happens mostly due to appearance of constant $\ln ( 2)$, which originates from sign alternating summation in $S_{-1}(z)$   absent for positive indices.

In the present paper we consider only bilinear reflection identities expressing a product of two harmonic sums of argument $z$ and $-1-z$ in terms of a linear combination of other sums of the same arguments. One can consider also non-linear (trilinear, quadlinear etc.) identities, but whose reducible and  form   a linear combination of the bilinear identities presented in this paper. For example, we can consider a trilinear term
$S_1(z)S_{-2}(-1-z)S_{2}(-1-z)$ and write it as 
\begin{eqnarray}\label{triples1scm2sc2}
S_1(z)S_{-2}(-1-z)S_{2}(-1-z)&=&S_1(z)S_{-2,2}(-1-z)+S_1(z)S_{2,-2}(-1-z)
\nonumber 
\\
&& -S_1(z)S_{-4}(-1-z) \hspace{1cm}
\end{eqnarray}
where we used a quasi-shuffle identity from Eq.~(\ref{shuffle})
\begin{eqnarray}
S_{-2,2}(z)+S_{2,-2}(z)-S_{-4}(z)=S_{-2}(z)S_{2}(z)
\end{eqnarray}
The expression for $S_1(z)S_{2,-2}(-1-z)$ is given in Eq.~(\ref{s1sc2m2}) and for $S_1(z)S_{-4}(-1-z)$  in Eq.~(\ref{s1scm4}). Plugging those together with  Eq.~(\ref{S1Sm22}) into Eq.~(\ref{triples1scm2sc2}) we get 
 \begin{eqnarray}
 S_1(z)& S_{-2}(-1-z)&S_{2}(-1-z) =
-\zeta _2 S_{-2,1}(-1-z)+\zeta _2 S_{-2,1}(z)
 \nonumber
   \\
     &&
     -\frac{3}{2} \zeta _2
   S_{2,-1}(-1-z)
   +\frac{3}{2} \zeta _2 S_{2,-1}(z)+\frac{1}{2} \zeta _2
   S_{2,1}(-1-z)
    \nonumber
   \\
     &&
     +\frac{1}{2} \zeta _2
   S_{2,1}(z)+S_{-4,1}(z)-S_{-3,2}(-1-z)-S_{1,-4}(-1-z)
    \nonumber
   \\
     &&
     -S_{3,-2}(-1-z)+S_{-2,1,2}(
   -1-z)-S_{-2,2,1}(z)
    \nonumber
   \\
     &&
     +S_{1,-2,2}(-1-z)+S_{1,2,-2}(-1-z)-S_{2,-2,1}(z)
      \nonumber
   \\
     &&
     +S_{2,1,-2}(
   -1-z)-\frac{9 \zeta _3 \zeta _2}{8}+\frac{85 \zeta _5}{32}-\frac{3}{4} \zeta
   _2^2 \log (2)
    \nonumber
   \\
     &&
     -\frac{3}{8} \zeta _2^2 S_{-1}(-1-z)-\frac{3}{8} \zeta _2^2
   S_{-1}(z)+\frac{1}{2} \zeta _2^2 S_1(-1-z)
    \nonumber
   \\
     &&
     -\frac{1}{2} \zeta _2^2
   S_1(z)+\frac{3}{2} \zeta _2 S_{-3}(-1-z)-\zeta _2 S_{-3}(z)-\frac{1}{2} \zeta
   _2 S_3(-1-z)
    \nonumber
   \\
     &&
     +\frac{11}{8} \zeta _3 S_{-2}(-1-z)
   +\frac{13}{8} \zeta _3
   S_{-2}(z)+\frac{1}{8} \zeta _3 S_2(-1-z)-\frac{5}{8} \zeta _3
   S_2(z)
    \nonumber
   \\
     &&
     +\frac{3}{2} \zeta _2 \log (2) S_{-2}(-1-z)-\frac{3}{2} \zeta _2 \log (2)
   S_{-2}(z)
    \nonumber
   \\
     &&
     -\frac{3}{2} \zeta _2 \log (2) S_2(-1-z)+\frac{3}{2} \zeta _2 \log (2)
   S_2(z)
\end{eqnarray}  

In a similar way one can can build any non-linear  reflection identity using quasi-shuffle relations for harmonic sums and the bilinear reflection identities listed in the Appendix of this paper. All possible quasi-shuffle relations 
 required for the present discussion  are available in the HarmonicSums package by J.~Ablinger. The  quasi-shuffle relations before and after analytic continuation of the harmonic sums to the complex plane are the same.

 
\section{Discussion and Conclusions}

We consider meromorphic functions obtained by analytic continuation of the harmonic sums from positive  integers  to the complex plane except for isolated pole singularities at negative integer values of the argument. We call those functions the analytically continued harmonic sums or simply the harmonic sums for clarity of the presentation. 
We discuss the reflection identities for harmonic sums of weight five. There are $216$ irreducible bilinear identities listed in the Appendix. All other bilinear reflection identities are easily obtained by a trivial change of argument $z \leftrightarrow -1-z$. The  non-linear (trilinear, quadlinear etc.) identities for a product of three and four harmonic sums are obtained from the identities listed in the Appendix using the quasi-shuffle relations for harmonic sums. In our analysis we use the linear basis for harmonic sums and limit ourselves to harmonic sums analytically continued from even integer values of the argument to the complex plane. The analytic continuation from odd integers is beyond  the scope of the present study. 

In deriving the reflection identities presented in this paper we used Harmonic Sums package by J.~Ablinger,\cite{AblingerThesis}  HPL package by D.~ Maitre,\cite{Maitre:2005uu} and dedicated Mathematica package for pomeron NNLO eigenvalue by N.~ Gromov, F.~Levkovich-Maslyuk and G.~Sizov,\cite{Gromov:2015vua}\hspace{0.02cm}. 

We expanded around positive and negative integer points  the  product of two harmonic sums $S_{a_1,a_2,..}(z)S_{b_1,b_2,...}(-1-z)$ and the functional basis built of pure Harmonic Sums with constants of relevant weight. Then we compared the coefficients of the irreducible constants of a given weight and solved the resulting set of coefficient equations. We used higher order expansion to cross check our results.   The bilinear reflection identities presented here are derived from the pole expansion based on the Mellin transform and then checked them  against the quasi-shuffle identities and numerical calculations of the corresponding harmonic sums on the complex plane. 

We attach a Mathematica notebook with our results.

\section{Acknowledgments}
We would like to thank Fedor Levkovich-Maslyuk and Mikhail Alfimov for fruitful discussions on details of their calculations of NNLO BFKL eigenvalue. We are grateful to
Simon Caron-Huot for explaining us the structure of his result on NNLO BFKL eigenvalue
and his calculation techniques. 

We are thankful to Victor Fadin for sharing with us his knowledge of the perturbative calculations of the  higher order 
corrections to the BFKL equation. 

We are deeply indebted to Jochen Bartels for his hospitality and enlightening discussions on BFKL physics during
our stay at University of Hamburg where this project was initiated.

This paper is dedicated to memory of Lev Lipatov and based on numerous discussions 
with him on that topic.

\newpage
\renewcommand{\theequation}{A-\arabic{equation}}    
  \setcounter{equation}{0}  
  \section*{APPENDIX}  

We present the irreducible reflection identities at weight $w=5$ using a compact and convenient notation as follows. 
The harmonic sum  of argument $z$ we denote by $s_{a_1, a_2, ...,a_n}$, whereas for  harmonic sums of argument $-1-z$ we use a bar notation and write it  as $\bar{s}_{a_1, a_2, ...,a_n}$. 

The symbol $\zeta_n$ stands for Riemann zeta function, $\ln_2$ stands for $\ln 2 \simeq 0.693147$, $\text{LiHalf}_n$ stands for the polylogarithm of argument $1/2$, i.e. $\text{Li}_n \left(\frac{1}{2}\right)$. 

Here we list the smallest irreducible set of the reflection identities, the rest of the reflection identities can be obtained by a change of the argument $z \leftrightarrow -1-z$.

   \begin{dmath}[style={\small}]     
  s_{-2} \bar{s}_{-3}  =   \bar{s}_{-2,-3}-\frac{1}{2} \zeta _2 \bar{s}_{-3}-\frac{1}{2} \zeta _2 \bar{s}_3+\frac{3}{4} \zeta _3 \bar{s}_{-2}-\frac{9}{4} \zeta _3 \bar{s}_2-s_{-3,-2}-\frac{3 \zeta _3 \zeta _2}{4}+\frac{75 \zeta _5}{8}-\frac{1}{2} \zeta _2 s_{-3}+\frac{\zeta _2 s_3}{2}-\frac{3}{4} \zeta _3 s_{-2}-\frac{9 \zeta _3 s_2}{4}  \end{dmath}
   \begin{dmath}[style={\small}]     
  s_{-2} \bar{s}_3  =   \bar{s}_{-2,3}-\frac{9}{4} \zeta _2^2 \bar{s}_{-1}-\frac{1}{2} \zeta _2 \bar{s}_{-3}-\frac{1}{2} \zeta _2 \bar{s}_3-\frac{3}{2} \zeta _3 \bar{s}_{-2}+s_{3,-2}-\frac{3 \zeta _3 \zeta _2}{4}+\frac{15 \zeta _5}{16}-\frac{9}{2} \zeta _2^2 \ln _2-\frac{9}{4} \zeta _2^2 s_{-1}-\frac{1}{2} \zeta _2 s_{-3}+\frac{\zeta _2 s_3}{2}+\frac{3}{2} \zeta _3 s_{-2}  \end{dmath}
   \begin{dmath}[style={\small}]     s_{-2} \bar{s}_{-2,-1}  =   -\frac{1}{2} \zeta _2 \bar{s}_{-2,-1}-\bar{s}_{4,-1}+2 \bar{s}_{-2,-2,-1}+\ln _2 \bar{s}_{-2,-2}-\ln _2 \bar{s}_{-2,2}+8 \text{Li}_4\left(\frac{1}{2}\right) \bar{s}_{-1}-\frac{1}{2} \zeta _2 \bar{s}_{-3}+\frac{3}{4} \zeta _3 \bar{s}_{-2}-\frac{53}{20} \zeta _2^2 \bar{s}_{-1}-2 \zeta _2 \ln _2^2 \bar{s}_{-1}-\zeta _2 \ln _2 \bar{s}_{-2}+\frac{7}{2} \zeta _3 \ln _2 \bar{s}_{-1}+\zeta _2 \ln _2 \bar{s}_2+\frac{1}{3} \ln _2^4 \bar{s}_{-1}+\ln _2 \bar{s}_{-4}-\ln _2 \bar{s}_4-\frac{1}{2} \zeta _2 s_{-2,-1}+s_{3,-2}-s_{-2,-1,-2}-\ln _2 s_{-2,-2}-\ln _2 s_{-2,2}-\frac{15 \zeta _2 \zeta _3}{8}+\frac{281 \zeta _5}{8}-\frac{2}{3} \zeta _2 \ln _2^3-\frac{49}{5} \zeta _2^2 \ln _2+8 \text{Li}_4\left(\frac{1}{2}\right) s_{-1}-32 \text{Li}_5\left(\frac{1}{2}\right)-8 \text{Li}_4\left(\frac{1}{2}\right) \ln _2-\frac{53}{20} \zeta _2^2 s_{-1}-\frac{1}{2} \zeta _2 s_{-3}+\frac{\zeta _2 s_3}{2}+\frac{3}{2} \zeta _3 s_{-2}-2 \zeta _2 s_{-1} \ln _2^2+\zeta _2 s_2 \ln _2+\frac{7}{2} \zeta _3 s_{-1} \ln _2+\frac{1}{3} s_{-1} \ln _2^4-\frac{\ln _2^5}{15}  \end{dmath}
   \begin{dmath}[style={\small}]     s_{-2} \bar{s}_{-2,1}  =   -\frac{3}{2} \zeta _2 \bar{s}_{-2,-1}-\frac{1}{2} \zeta _2 \bar{s}_{-2,1}-\bar{s}_{4,1}+2 \bar{s}_{-2,-2,1}+\zeta _2 \bar{s}_3+\frac{5}{4} \zeta _3 \bar{s}_{-2}-\frac{9}{4} \zeta _3 \bar{s}_2-\frac{3}{2} \zeta _2 \ln _2 \bar{s}_{-2}+\frac{3}{2} \zeta _2 \ln _2 \bar{s}_2-\frac{3}{2} \zeta _2 s_{-2,-1}+\frac{1}{2} \zeta _2 s_{-2,1}-s_{-3,-2}+s_{-2,1,-2}-\frac{15 \zeta _3 \zeta _2}{4}+\frac{75 \zeta _5}{8}-\frac{1}{2} \zeta _2 s_{-3}+\frac{\zeta _2 s_3}{2}-\frac{1}{2} \zeta _3 s_{-2}-\frac{9 \zeta _3 s_2}{4}-\frac{3}{2} \zeta _2 s_{-2} \ln _2+\frac{3}{2} \zeta _2 s_2 \ln _2  \end{dmath}
   \begin{dmath}[style={\small}]     s_{-2} \bar{s}_{2,-1}  =   -\frac{1}{2} \zeta _2 \bar{s}_{2,-1}-\bar{s}_{-4,-1}+\bar{s}_{-2,2,-1}+\bar{s}_{2,-2,-1}+\ln _2 \bar{s}_{2,-2}-\ln _2 \bar{s}_{2,2}-\frac{1}{4} \zeta _3 \bar{s}_{-2}+\zeta _3 \bar{s}_2-\frac{1}{2} \zeta _2 \bar{s}_3+\zeta _2 \ln _2 \bar{s}_{-2}+\frac{7}{2} \zeta _3 \ln _2 \bar{s}_{-1}-\zeta _2 \ln _2 \bar{s}_2-\ln _2 \bar{s}_{-4}+\ln _2 \bar{s}_4+\frac{1}{2} \zeta _2 s_{2,-1}-s_{-3,-2}+s_{2,-1,-2}+\ln _2 s_{2,-2}+\ln _2 s_{2,2}-\frac{3 \zeta _2 \zeta _3}{8}-\frac{5 \zeta _5}{16}+2 \zeta _2 \ln _2^3+\frac{9}{2} \zeta _2^2 \ln _2-8 \text{Li}_4\left(\frac{1}{2}\right) \ln _2-\frac{1}{2} \zeta _2 s_{-3}+\frac{\zeta _2 s_3}{2}+\frac{1}{4} \zeta _3 s_{-2}-\frac{5 \zeta _3 s_2}{4}-\zeta _2 s_{-2} \ln _2+\frac{7}{2} \zeta _3 s_{-1} \ln _2-\frac{\ln _2^5}{3}  \end{dmath}
   \begin{dmath}[style={\small}]     s_{-2} \bar{s}_{2,1}  =   -\frac{3}{2} \zeta _2 \bar{s}_{2,-1}-\frac{1}{2} \zeta _2 \bar{s}_{2,1}-\bar{s}_{-4,1}+\bar{s}_{-2,2,1}+\bar{s}_{2,-2,1}-\frac{9}{8} \zeta _2^2 \bar{s}_{-1}+\zeta _2 \bar{s}_{-3}-\frac{13}{8} \zeta _3 \bar{s}_{-2}+\frac{5}{8} \zeta _3 \bar{s}_2+\frac{3}{2} \zeta _2 \ln _2 \bar{s}_{-2}-\frac{3}{2} \zeta _2 \ln _2 \bar{s}_2+\frac{3}{2} \zeta _2 s_{2,-1}-\frac{1}{2} \zeta _2 s_{2,1}+s_{3,-2}-s_{2,1,-2}+\frac{15 \zeta _3 \zeta _2}{8}-\frac{85 \zeta _5}{32}-\frac{9}{4} \zeta _2^2 \ln _2-\frac{9}{8} \zeta _2^2 s_{-1}-\frac{1}{2} \zeta _2 s_{-3}+\frac{\zeta _2 s_3}{2}+\frac{13}{8} \zeta _3 s_{-2}-\frac{\zeta _3 s_2}{8}-\frac{3}{2} \zeta _2 s_{-2} \ln _2+\frac{3}{2} \zeta _2 s_2 \ln _2  \end{dmath}
   \begin{dmath}[style={\small}]     s_{-2} \bar{s}_{-1,1,-1}  =   \frac{5}{8} \zeta _3 \bar{s}_{-1,1}-\frac{1}{2} \zeta _2 \bar{s}_{-1,2}-\frac{1}{2} \zeta _2 \bar{s}_{-1,1,-1}+\frac{3}{2} \zeta _2 \ln _2 \bar{s}_{-1,-1}-\bar{s}_{-1,-3,-1}-\bar{s}_{3,1,-1}+\bar{s}_{-2,-1,1,-1}+\bar{s}_{-1,-2,1,-1}+\bar{s}_{-1,1,-2,-1}+\frac{1}{2} \ln _2^2 \bar{s}_{-1,-2}-\frac{1}{2} \ln _2^2 \bar{s}_{-1,2}-\ln _2 \bar{s}_{-1,-3}+\ln _2 \bar{s}_{-1,3}+\ln _2 \bar{s}_{-1,1,-2}-\ln _2 \bar{s}_{-1,1,2}+\frac{1}{8} \zeta _3 \bar{s}_{-2}-\frac{1}{8} \zeta _2^2 \bar{s}_{-1}+\frac{3}{2} \zeta _2 \ln _2^2 \bar{s}_{-1}-\zeta _2 \ln _2 \bar{s}_{-2}-\frac{1}{2} \zeta _2 \ln _2 \bar{s}_2+\frac{1}{2} \ln _2^2 \bar{s}_{-3}-\frac{1}{2} \ln _2^2 \bar{s}_3-\frac{1}{2} \zeta _2 s_{-2,-1}-\frac{1}{2} \zeta _2 s_{-1,-2}-\frac{13}{8} \zeta _3 s_{-1,1}+\frac{1}{2} \zeta _2 s_{-1,2}+\frac{1}{2} \zeta _2 s_{-1,1,-1}-\frac{3}{2} \zeta _2 \ln _2 s_{-1,-1}+\zeta _2 \ln _2 s_{-1,1}+s_{3,-2}-s_{-2,-1,-2}-s_{-1,-2,-2}+s_{-1,1,-1,-2}+\frac{1}{2} \ln _2^2 s_{-1,-2}+\frac{1}{2} \ln _2^2 s_{-1,2}-\ln _2 s_{-2,-2}-\ln _2 s_{-2,2}+\ln _2 s_{-1,1,-2}+\ln _2 s_{-1,1,2}-\frac{3 \zeta _2 \zeta _3}{8}+\frac{685 \zeta _5}{64}+\frac{2}{3} \zeta _2 \ln _2^3-\frac{3}{4} \zeta _3 \ln _2^2-\frac{21}{10} \zeta _2^2 \ln _2+6 \text{Li}_4\left(\frac{1}{2}\right) s_{-1}-10 \text{Li}_5\left(\frac{1}{2}\right)-6 \text{Li}_4\left(\frac{1}{2}\right) \ln _2-\frac{9}{4} \zeta _2^2 s_{-1}-\frac{1}{2} \zeta _2 s_{-3}+\frac{\zeta _2 s_3}{2}+\frac{3}{2} \zeta _3 s_{-2}-\frac{5}{2} \zeta _2 s_{-1} \ln _2^2+\zeta _2 s_2 \ln _2+\frac{15}{4} \zeta _3 s_{-1} \ln _2+\frac{1}{4} s_{-1} \ln _2^4-\frac{\ln _2^5}{6}  \end{dmath}
   \begin{dmath}[style={\small}]     s_{-2} \bar{s}_{-1,1,1}  =   \zeta _2 \bar{s}_{-1,-2}-\frac{21}{8} \zeta _3 \bar{s}_{-1,-1}+\frac{5}{8} \zeta _3 \bar{s}_{-1,1}-\frac{3}{2} \zeta _2 \bar{s}_{-1,1,-1}-\frac{1}{2} \zeta _2 \bar{s}_{-1,1,1}+\frac{3}{2} \zeta _2 \ln _2 \bar{s}_{-1,-1}-\frac{3}{2} \zeta _2 \ln _2 \bar{s}_{-1,1}-\bar{s}_{-1,-3,1}-\bar{s}_{3,1,1}+\bar{s}_{-2,-1,1,1}+\bar{s}_{-1,-2,1,1}+\bar{s}_{-1,1,-2,1}+3 \text{Li}_4\left(\frac{1}{2}\right) \bar{s}_{-1}+\frac{7}{8} \zeta _3 \bar{s}_{-2}-\frac{13}{20} \zeta _2^2 \bar{s}_{-1}+\frac{1}{8} \zeta _3 \bar{s}_2-\frac{1}{2} \zeta _2 \ln _2 \bar{s}_{-2}+\frac{1}{2} \zeta _2 \ln _2 \bar{s}_2+\frac{1}{8} \ln _2^4 \bar{s}_{-1}+\frac{1}{6} \ln _2^3 \bar{s}_{-2}-\frac{1}{6} \ln _2^3 \bar{s}_2-\frac{3}{2} \zeta _2 s_{-2,-1}+\frac{1}{2} \zeta _2 s_{-2,1}-\frac{1}{2} \zeta _2 s_{-1,-2}+\frac{21}{8} \zeta _3 s_{-1,-1}-\frac{1}{8} \zeta _3 s_{-1,1}+\frac{1}{2} \zeta _2 s_{-1,2}+\frac{3}{2} \zeta _2 s_{-1,1,-1}-\frac{1}{2} \zeta _2 s_{-1,1,1}-\frac{3}{2} \zeta _2 \ln _2 s_{-1,-1}+\frac{3}{2} \zeta _2 \ln _2 s_{-1,1}-s_{-3,-2}+s_{-2,1,-2}+s_{-1,2,-2}-s_{-1,1,1,-2}-\frac{43 \zeta _2 \zeta _3}{16}+\frac{101 \zeta _5}{32}+\frac{3}{5} \zeta _2^2 \ln _2-3 \text{Li}_4\left(\frac{1}{2}\right) s_{-1}+2 \text{Li}_5\left(\frac{1}{2}\right)+\frac{39}{40} \zeta _2^2 s_{-1}-\frac{1}{2} \zeta _2 s_{-3}+\frac{\zeta _2 s_3}{2}-\frac{3}{4} \zeta _3 s_{-2}-\frac{5 \zeta _3 s_2}{2}-\zeta _2 s_{-2} \ln _2+2 \zeta _2 s_2 \ln _2-\frac{1}{8} s_{-1} \ln _2^4-\frac{1}{6} s_{-2} \ln _2^3-\frac{1}{6} s_2 \ln _2^3-\frac{\ln _2^5}{60}  \end{dmath}
   \begin{dmath}[style={\small}]     s_{-2} \bar{s}_{1,-2}  =   -\zeta _2 \bar{s}_{1,2}-\bar{s}_{-3,-2}+\bar{s}_{-2,1,-2}+\bar{s}_{1,-2,-2}+\frac{3}{4} \zeta _2^2 \bar{s}_{-1}+\frac{37}{40} \zeta _2^2 \bar{s}_1-\frac{1}{2} \zeta _2 \bar{s}_{-3}+\frac{1}{2} \zeta _2 \bar{s}_3+\frac{1}{8} \zeta _3 \bar{s}_{-2}-\frac{13}{8} \zeta _3 \bar{s}_2-\zeta _2 s_{1,2}-s_{-3,-2}+s_{1,-2,-2}-\frac{15 \zeta _3 \zeta _2}{4}+\frac{115 \zeta _5}{16}+\frac{3}{2} \zeta _2^2 \ln _2+\frac{3}{4} \zeta _2^2 s_{-1}+\frac{27}{40} \zeta _2^2 s_1-\frac{1}{2} \zeta _2 s_{-3}+\frac{\zeta _2 s_3}{2}-\frac{1}{8} \zeta _3 s_{-2}-\frac{13 \zeta _3 s_2}{8}  \end{dmath}
   \begin{dmath}[style={\small}]     s_{-2} \bar{s}_{1,2}  =   -\frac{1}{2} \zeta _2 \bar{s}_{1,-2}-\frac{7}{2} \zeta _3 \bar{s}_{1,-1}-\frac{1}{2} \zeta _2 \bar{s}_{1,2}-\bar{s}_{-3,2}+\bar{s}_{-2,1,2}+\bar{s}_{1,-2,2}-4 \text{Li}_4\left(\frac{1}{2}\right) \bar{s}_{-1}+4 \text{Li}_4\left(\frac{1}{2}\right) \bar{s}_1+2 \zeta _3 \bar{s}_{-2}-\frac{7}{40} \zeta _2^2 \bar{s}_{-1}-\frac{61}{40} \zeta _2^2 \bar{s}_1+\zeta _2 \ln _2^2 \bar{s}_{-1}-\zeta _2 \ln _2^2 \bar{s}_1-\frac{1}{6} \ln _2^4 \bar{s}_{-1}+\frac{1}{6} \ln _2^4 \bar{s}_1+\frac{1}{2} \zeta _2 s_{1,-2}-\frac{7}{2} \zeta _3 s_{1,-1}-\frac{1}{2} \zeta _2 s_{1,2}+s_{3,-2}-s_{1,2,-2}-\frac{15 \zeta _2 \zeta _3}{8}-\frac{51 \zeta _5}{32}+\frac{2}{3} \zeta _2 \ln _2^3-\frac{7}{20} \zeta _2^2 \ln _2-4 \text{Li}_4\left(\frac{1}{2}\right) s_{-1}+4 \text{Li}_4\left(\frac{1}{2}\right) s_1+8 \text{Li}_5\left(\frac{1}{2}\right)-\frac{7}{40} \zeta _2^2 s_{-1}-\frac{11}{8} \zeta _2^2 s_1-\frac{1}{2} \zeta _2 s_{-3}+\frac{\zeta _2 s_3}{2}+\frac{3}{2} \zeta _3 s_{-2}+\zeta _2 s_{-1} \ln _2^2-\zeta _2 s_1 \ln _2^2-\frac{1}{6} s_{-1} \ln _2^4+\frac{1}{6} s_1 \ln _2^4-\frac{\ln _2^5}{15}  \end{dmath}
   \begin{dmath}[style={\small}]     s_{-2} \bar{s}_{1,-1,-1}  =   -\frac{1}{2} \zeta _2 \bar{s}_{1,-2}-\frac{1}{4} \zeta _3 \bar{s}_{1,-1}-\frac{1}{2} \zeta _2 \bar{s}_{1,-1,-1}-\bar{s}_{-3,-1,-1}-\bar{s}_{1,3,-1}+\bar{s}_{-2,1,-1,-1}+\bar{s}_{1,-2,-1,-1}+\bar{s}_{1,-1,-2,-1}-\ln _2^2 \bar{s}_{1,-2}+\ln _2^2 \bar{s}_{1,2}+\ln _2 \bar{s}_{1,-3}-\ln _2 \bar{s}_{1,3}+\ln _2 \bar{s}_{1,-1,-2}-\ln _2 \bar{s}_{1,-1,2}+\text{Li}_4\left(\frac{1}{2}\right) \bar{s}_{-1}+\zeta _3 \bar{s}_{-2}-\frac{4}{5} \zeta _2^2 \bar{s}_{-1}-\frac{1}{8} \zeta _2^2 \bar{s}_1-\zeta _2 \ln _2^2 \bar{s}_{-1}-\frac{1}{2} \zeta _2 \ln _2 \bar{s}_{-2}+\frac{7}{8} \zeta _3 \ln _2 \bar{s}_{-1}-\frac{7}{8} \zeta _3 \ln _2 \bar{s}_1+\frac{1}{2} \zeta _2 \ln _2 \bar{s}_2+\frac{1}{24} \ln _2^4 \bar{s}_{-1}-\frac{1}{2} \ln _2^3 \bar{s}_{-2}+\frac{1}{2} \ln _2^3 \bar{s}_2+\ln _2^2 \bar{s}_{-3}-\ln _2^2 \bar{s}_3-\frac{1}{2} \zeta _2 s_{-2,-1}+\frac{1}{2} \zeta _2 s_{1,-2}-\frac{5}{2} \zeta _3 s_{1,-1}-\frac{1}{2} \zeta _2 s_{1,2}+\frac{1}{2} \zeta _2 s_{1,-1,-1}+\zeta _2 \ln _2 s_{1,-1}+s_{3,-2}-s_{-2,-1,-2}-s_{1,2,-2}+s_{1,-1,-1,-2}+\ln _2^2 s_{1,-2}+\ln _2^2 s_{1,2}-\ln _2 s_{-2,-2}-\ln _2 s_{-2,2}+\ln _2 s_{1,-1,-2}+\ln _2 s_{1,-1,2}-\frac{27 \zeta _2 \zeta _3}{16}+\frac{747 \zeta _5}{64}-\frac{8}{3} \zeta _2 \ln _2^3+\frac{15}{4} \zeta _3 \ln _2^2-\frac{39}{8} \zeta _2^2 \ln _2+\text{Li}_4\left(\frac{1}{2}\right) s_{-1}+4 \text{Li}_4\left(\frac{1}{2}\right) s_1-8 \text{Li}_5\left(\frac{1}{2}\right)+4 \text{Li}_4\left(\frac{1}{2}\right) \ln _2-\frac{4}{5} \zeta _2^2 s_{-1}-\frac{41}{40} \zeta _2^2 s_1-\frac{1}{2} \zeta _2 s_{-3}+\frac{\zeta _2 s_3}{2}+\frac{3}{2} \zeta _3 s_{-2}-\zeta _2 s_{-1} \ln _2^2-\frac{1}{2} \zeta _2 s_{-2} \ln _2+\frac{1}{2} \zeta _2 s_2 \ln _2+\frac{7}{8} \zeta _3 s_{-1} \ln _2-\frac{5}{8} \zeta _3 s_1 \ln _2+\frac{1}{24} s_{-1} \ln _2^4+\frac{1}{6} s_1 \ln _2^4+\frac{1}{2} s_{-2} \ln _2^3+\frac{1}{2} s_2 \ln _2^3+\frac{7 \ln _2^5}{30}  \end{dmath}
   \begin{dmath}[style={\small}]     s_{-2} \bar{s}_{1,-1,1}  =   \frac{1}{2} \zeta _2 \bar{s}_{1,-2}+\frac{5}{8} \zeta _3 \bar{s}_{1,-1}+\frac{1}{2} \zeta _2 \bar{s}_{1,2}-\frac{3}{2} \zeta _2 \bar{s}_{1,-1,-1}-\frac{1}{2} \zeta _2 \bar{s}_{1,-1,1}-\frac{3}{2} \zeta _2 \ln _2 \bar{s}_{1,-1}-\bar{s}_{-3,-1,1}-\bar{s}_{1,3,1}+\bar{s}_{-2,1,-1,1}+\bar{s}_{1,-2,-1,1}+\bar{s}_{1,-1,-2,1}-\frac{1}{2} \ln _2^2 \bar{s}_{1,-2}+\frac{1}{2} \ln _2^2 \bar{s}_{1,2}-\frac{1}{2} \zeta _2 \bar{s}_{-3}-\frac{1}{8} \zeta _3 \bar{s}_{-2}+\frac{3}{4} \zeta _2^2 \bar{s}_{-1}+\frac{1}{4} \zeta _2^2 \bar{s}_1-\frac{3}{2} \zeta _3 \bar{s}_2+\frac{1}{2} \zeta _2 \bar{s}_3-\frac{3}{4} \zeta _2 \ln _2^2 \bar{s}_{-1}-\frac{3}{4} \zeta _2 \ln _2^2 \bar{s}_1+\frac{1}{2} \zeta _2 \ln _2 \bar{s}_{-2}+\zeta _2 \ln _2 \bar{s}_2-\frac{1}{3} \ln _2^3 \bar{s}_{-2}+\frac{1}{3} \ln _2^3 \bar{s}_2+\frac{1}{2} \ln _2^2 \bar{s}_{-3}-\frac{1}{2} \ln _2^2 \bar{s}_3-\frac{3}{2} \zeta _2 s_{-2,-1}+\frac{1}{2} \zeta _2 s_{-2,1}-\frac{1}{8} \zeta _3 s_{1,-1}-\zeta _2 s_{1,2}+\frac{3}{2} \zeta _2 s_{1,-1,-1}-\frac{1}{2} \zeta _2 s_{1,-1,1}+\frac{3}{2} \zeta _2 \ln _2 s_{1,-1}-s_{-3,-2}+s_{-2,1,-2}+s_{1,-2,-2}-s_{1,-1,1,-2}+\frac{1}{2} \ln _2^2 s_{1,-2}+\frac{1}{2} \ln _2^2 s_{1,2}-\frac{43 \zeta _2 \zeta _3}{16}+\frac{411 \zeta _5}{32}-\frac{11}{6} \zeta _2 \ln _2^3+\frac{15}{8} \zeta _3 \ln _2^2-\frac{9}{10} \zeta _2^2 \ln _2+2 \text{Li}_4\left(\frac{1}{2}\right) s_1-8 \text{Li}_5\left(\frac{1}{2}\right)+\frac{3}{4} \zeta _2^2 s_{-1}-\frac{21}{40} \zeta _2^2 s_1-\frac{1}{2} \zeta _2 s_{-3}+\frac{\zeta _2 s_3}{2}+\frac{1}{4} \zeta _3 s_{-2}-\frac{3 \zeta _3 s_2}{2}-\frac{3}{4} \zeta _2 s_{-1} \ln _2^2+\frac{3}{4} \zeta _2 s_1 \ln _2^2-2 \zeta _2 s_{-2} \ln _2+\zeta _2 s_2 \ln _2+\frac{1}{12} s_1 \ln _2^4+\frac{1}{3} s_{-2} \ln _2^3+\frac{1}{3} s_2 \ln _2^3+\frac{\ln _2^5}{15}  \end{dmath}
   \begin{dmath}[style={\small}]     s_{-2} \bar{s}_{1,1,-1}  =   \frac{5}{8} \zeta _3 \bar{s}_{1,1}-\frac{1}{2} \zeta _2 \bar{s}_{1,2}-\frac{1}{2} \zeta _2 \bar{s}_{1,1,-1}+\frac{3}{2} \zeta _2 \ln _2 \bar{s}_{1,-1}-\bar{s}_{-3,1,-1}-\bar{s}_{1,-3,-1}+\bar{s}_{-2,1,1,-1}+\bar{s}_{1,-2,1,-1}+\bar{s}_{1,1,-2,-1}+\frac{1}{2} \ln _2^2 \bar{s}_{1,-2}-\frac{1}{2} \ln _2^2 \bar{s}_{1,2}-\ln _2 \bar{s}_{1,-3}+\ln _2 \bar{s}_{1,3}+\ln _2 \bar{s}_{1,1,-2}-\ln _2 \bar{s}_{1,1,2}-3 \text{Li}_4\left(\frac{1}{2}\right) \bar{s}_1+\frac{23}{20} \zeta _2^2 \bar{s}_1+\frac{1}{8} \zeta _3 \bar{s}_2-\frac{3}{4} \zeta _2 \ln _2^2 \bar{s}_{-1}+\frac{9}{4} \zeta _2 \ln _2^2 \bar{s}_1-\zeta _2 \ln _2 \bar{s}_{-2}+\frac{21}{8} \zeta _3 \ln _2 \bar{s}_{-1}-\frac{21}{8} \zeta _3 \ln _2 \bar{s}_1-\frac{1}{2} \zeta _2 \ln _2 \bar{s}_2-\frac{1}{8} \ln _2^4 \bar{s}_1+\frac{1}{6} \ln _2^3 \bar{s}_{-2}-\frac{1}{6} \ln _2^3 \bar{s}_2-\frac{1}{2} \ln _2^2 \bar{s}_{-3}+\frac{1}{2} \ln _2^2 \bar{s}_3+\frac{1}{2} \zeta _2 s_{1,-2}+\frac{13}{8} \zeta _3 s_{1,1}-\frac{1}{2} \zeta _2 s_{1,2}+\frac{1}{2} \zeta _2 s_{2,-1}-\frac{1}{2} \zeta _2 s_{1,1,-1}+\frac{3}{2} \zeta _2 \ln _2 s_{1,-1}-\zeta _2 \ln _2 s_{1,1}-s_{-3,-2}+s_{1,-2,-2}+s_{2,-1,-2}-s_{1,1,-1,-2}-\frac{1}{2} \ln _2^2 s_{1,-2}-\frac{1}{2} \ln _2^2 s_{1,2}+\ln _2 s_{2,-2}+\ln _2 s_{2,2}-\ln _2 s_{1,1,-2}-\ln _2 s_{1,1,2}-\frac{31 \zeta _2 \zeta _3}{16}-\frac{139 \zeta _5}{64}+\frac{11}{6} \zeta _2 \ln _2^3-\frac{15}{8} \zeta _3 \ln _2^2+\frac{15}{4} \zeta _2^2 \ln _2-3 \text{Li}_4\left(\frac{1}{2}\right) s_1+6 \text{Li}_5\left(\frac{1}{2}\right)-6 \text{Li}_4\left(\frac{1}{2}\right) \ln _2+\frac{39}{40} \zeta _2^2 s_1-\frac{1}{2} \zeta _2 s_{-3}+\frac{\zeta _2 s_3}{2}-\frac{3 \zeta _3 s_2}{2}-\frac{3}{4} \zeta _2 s_{-1} \ln _2^2+\frac{7}{4} \zeta _2 s_1 \ln _2^2-\frac{1}{2} \zeta _2 s_{-2} \ln _2+\frac{1}{2} \zeta _2 s_2 \ln _2+\frac{21}{8} \zeta _3 s_{-1} \ln _2-\frac{9}{8} \zeta _3 s_1 \ln _2-\frac{1}{8} s_1 \ln _2^4-\frac{1}{6} s_{-2} \ln _2^3-\frac{1}{6} s_2 \ln _2^3-\frac{3 \ln _2^5}{10}  \end{dmath}
   \begin{dmath}[style={\small}]     s_{-2} \bar{s}_{1,1,1}  =   \zeta _2 \bar{s}_{1,-2}-\frac{21}{8} \zeta _3 \bar{s}_{1,-1}+\frac{5}{8} \zeta _3 \bar{s}_{1,1}-\frac{3}{2} \zeta _2 \bar{s}_{1,1,-1}-\frac{1}{2} \zeta _2 \bar{s}_{1,1,1}+\frac{3}{2} \zeta _2 \ln _2 \bar{s}_{1,-1}-\frac{3}{2} \zeta _2 \ln _2 \bar{s}_{1,1}-\bar{s}_{-3,1,1}-\bar{s}_{1,-3,1}+\bar{s}_{-2,1,1,1}+\bar{s}_{1,-2,1,1}+\bar{s}_{1,1,-2,1}-3 \text{Li}_4\left(\frac{1}{2}\right) \bar{s}_{-1}+3 \text{Li}_4\left(\frac{1}{2}\right) \bar{s}_1+\zeta _3 \bar{s}_{-2}+\frac{3}{20} \zeta _2^2 \bar{s}_{-1}-\frac{13}{20} \zeta _2^2 \bar{s}_1-\frac{1}{8} \ln _2^4 \bar{s}_{-1}+\frac{1}{8} \ln _2^4 \bar{s}_1+\frac{1}{2} \zeta _2 s_{1,-2}-\frac{21}{8} \zeta _3 s_{1,-1}+\frac{1}{8} \zeta _3 s_{1,1}-\frac{1}{2} \zeta _2 s_{1,2}+\frac{3}{2} \zeta _2 s_{2,-1}-\frac{1}{2} \zeta _2 s_{2,1}-\frac{3}{2} \zeta _2 s_{1,1,-1}+\frac{1}{2} \zeta _2 s_{1,1,1}+\frac{3}{2} \zeta _2 \ln _2 s_{1,-1}-\frac{3}{2} \zeta _2 \ln _2 s_{1,1}+s_{3,-2}-s_{1,2,-2}-s_{2,1,-2}+s_{1,1,1,-2}-\frac{9 \zeta _2 \zeta _3}{16}-\frac{107 \zeta _5}{32}+\frac{3}{10} \zeta _2^2 \ln _2-3 \text{Li}_4\left(\frac{1}{2}\right) s_{-1}+3 \text{Li}_4\left(\frac{1}{2}\right) s_1+6 \text{Li}_5\left(\frac{1}{2}\right)+\frac{3}{20} \zeta _2^2 s_{-1}-\frac{39}{40} \zeta _2^2 s_1-\frac{1}{2} \zeta _2 s_{-3}+\frac{\zeta _2 s_3}{2}+\frac{13}{8} \zeta _3 s_{-2}-\frac{\zeta _3 s_2}{8}-\frac{3}{2} \zeta _2 s_{-2} \ln _2+\frac{3}{2} \zeta _2 s_2 \ln _2-\frac{1}{8} s_{-1} \ln _2^4+\frac{1}{8} s_1 \ln _2^4-\frac{\ln _2^5}{20}  \end{dmath}
   \begin{dmath}[style={\small}]     s_{-2} \bar{s}_{-1,-2}  =   -\zeta _2 \bar{s}_{-1,2}-\bar{s}_{3,-2}+\bar{s}_{-2,-1,-2}+\bar{s}_{-1,-2,-2}-4 \text{Li}_4\left(\frac{1}{2}\right) \bar{s}_{-1}+\frac{1}{2} \zeta _2 \bar{s}_{-3}-\frac{3}{2} \zeta _3 \bar{s}_{-2}+\frac{9}{8} \zeta _2^2 \bar{s}_{-1}-\frac{1}{2} \zeta _2 \bar{s}_3+\zeta _2 \ln _2^2 \bar{s}_{-1}-\frac{1}{2} \zeta _2 \ln _2 \bar{s}_{-2}+\frac{1}{2} \zeta _2 \ln _2 \bar{s}_2-\frac{1}{6} \ln _2^4 \bar{s}_{-1}+\zeta _2 s_{-1,2}+s_{3,-2}-s_{-1,-2,-2}-\frac{547 \zeta _5}{32}+\frac{4}{3} \zeta _2 \ln _2^3+\frac{71}{20} \zeta _2^2 \ln _2-4 \text{Li}_4\left(\frac{1}{2}\right) s_{-1}+16 \text{Li}_5\left(\frac{1}{2}\right)-\frac{19}{40} \zeta _2^2 s_{-1}-\frac{1}{2} \zeta _2 s_{-3}+\frac{\zeta _2 s_3}{2}+\frac{3}{2} \zeta _3 s_{-2}+\zeta _2 s_{-1} \ln _2^2+\frac{1}{2} \zeta _2 s_{-2} \ln _2+\frac{1}{2} \zeta _2 s_2 \ln _2-\frac{1}{6} s_{-1} \ln _2^4-\frac{2 \ln _2^5}{15}  \end{dmath}
   \begin{dmath}[style={\small}]     s_{-2} \bar{s}_{-1,2}  =   -\frac{1}{2} \zeta _2 \bar{s}_{-1,-2}-\frac{7}{2} \zeta _3 \bar{s}_{-1,-1}-\frac{1}{2} \zeta _2 \bar{s}_{-1,2}-\bar{s}_{3,2}+\bar{s}_{-2,-1,2}+\bar{s}_{-1,-2,2}+4 \text{Li}_4\left(\frac{1}{2}\right) \bar{s}_{-1}+\zeta _3 \bar{s}_{-2}-\frac{61}{40} \zeta _2^2 \bar{s}_{-1}+\zeta _3 \bar{s}_2-\zeta _2 \ln _2^2 \bar{s}_{-1}-\frac{1}{2} \zeta _2 \ln _2 \bar{s}_{-2}+\frac{1}{2} \zeta _2 \ln _2 \bar{s}_2+\frac{1}{6} \ln _2^4 \bar{s}_{-1}-\frac{1}{2} \zeta _2 s_{-1,-2}+\frac{7}{2} \zeta _3 s_{-1,-1}+\frac{1}{2} \zeta _2 s_{-1,2}-s_{-3,-2}+s_{-1,2,-2}+\frac{3 \zeta _2 \zeta _3}{4}+\frac{75 \zeta _5}{32}-\frac{7}{4} \zeta _2^2 \ln _2-4 \text{Li}_4\left(\frac{1}{2}\right) s_{-1}+\frac{11}{8} \zeta _2^2 s_{-1}-\frac{1}{2} \zeta _2 s_{-3}+\frac{\zeta _2 s_3}{2}-\zeta _3 s_{-2}-\frac{5 \zeta _3 s_2}{2}+\zeta _2 s_{-1} \ln _2^2+\frac{1}{2} \zeta _2 s_{-2} \ln _2+\frac{1}{2} \zeta _2 s_2 \ln _2-\frac{1}{6} s_{-1} \ln _2^4  \end{dmath}
   \begin{dmath}[style={\small}]     s_{-2} \bar{s}_{-1,-1,-1}  =   -\frac{1}{2} \zeta _2 \bar{s}_{-1,-2}-\frac{1}{4} \zeta _3 \bar{s}_{-1,-1}-\frac{1}{2} \zeta _2 \bar{s}_{-1,-1,-1}-\bar{s}_{-1,3,-1}-\bar{s}_{3,-1,-1}+\bar{s}_{-2,-1,-1,-1}+\bar{s}_{-1,-2,-1,-1}+\bar{s}_{-1,-1,-2,-1}-\ln _2^2 \bar{s}_{-1,-2}+\ln _2^2 \bar{s}_{-1,2}+\ln _2 \bar{s}_{-1,-3}-\ln _2 \bar{s}_{-1,3}+\ln _2 \bar{s}_{-1,-1,-2}-\ln _2 \bar{s}_{-1,-1,2}+\frac{1}{4} \zeta _3 \bar{s}_{-2}-\frac{1}{8} \zeta _2^2 \bar{s}_{-1}+\frac{3}{4} \zeta _3 \bar{s}_2+\frac{1}{2} \zeta _2 \ln _2 \bar{s}_{-2}-\frac{1}{2} \zeta _2 \ln _2 \bar{s}_2+\frac{2}{3} \ln _2^3 \bar{s}_{-2}-\frac{2}{3} \ln _2^3 \bar{s}_2-\ln _2^2 \bar{s}_{-3}+\ln _2^2 \bar{s}_3-\frac{1}{2} \zeta _2 s_{-1,-2}+\frac{5}{2} \zeta _3 s_{-1,-1}+\frac{1}{2} \zeta _2 s_{-1,2}+\frac{1}{2} \zeta _2 s_{2,-1}-\frac{1}{2} \zeta _2 s_{-1,-1,-1}-\zeta _2 \ln _2 s_{-1,-1}-s_{-3,-2}+s_{-1,2,-2}+s_{2,-1,-2}-s_{-1,-1,-1,-2}-\ln _2^2 s_{-1,-2}-\ln _2^2 s_{-1,2}+\ln _2 s_{2,-2}+\ln _2 s_{2,2}-\ln _2 s_{-1,-1,-2}-\ln _2 s_{-1,-1,2}+\frac{7 \zeta _2 \zeta _3}{8}-\frac{271 \zeta _5}{32}+\frac{3}{2} \zeta _3 \ln _2^2+\frac{41}{20} \zeta _2^2 \ln _2-4 \text{Li}_4\left(\frac{1}{2}\right) s_{-1}+8 \text{Li}_5\left(\frac{1}{2}\right)+\frac{41}{40} \zeta _2^2 s_{-1}-\frac{1}{2} \zeta _2 s_{-3}+\frac{\zeta _2 s_3}{2}-\frac{1}{4} \zeta _3 s_{-2}-\frac{7 \zeta _3 s_2}{4}-\frac{1}{2} \zeta _2 s_{-2} \ln _2+\frac{1}{2} \zeta _2 s_2 \ln _2+\frac{3}{2} \zeta _3 s_{-1} \ln _2-\frac{1}{6} s_{-1} \ln _2^4-\frac{2}{3} s_{-2} \ln _2^3-\frac{2}{3} s_2 \ln _2^3-\frac{\ln _2^5}{15}  \end{dmath}
   \begin{dmath}[style={\small}]     s_{-2} \bar{s}_{-1,-1,1}  =   \frac{1}{2} \zeta _2 \bar{s}_{-1,-2}+\frac{5}{8} \zeta _3 \bar{s}_{-1,-1}+\frac{1}{2} \zeta _2 \bar{s}_{-1,2}-\frac{3}{2} \zeta _2 \bar{s}_{-1,-1,-1}-\frac{1}{2} \zeta _2 \bar{s}_{-1,-1,1}-\frac{3}{2} \zeta _2 \ln _2 \bar{s}_{-1,-1}-\bar{s}_{-1,3,1}-\bar{s}_{3,-1,1}+\bar{s}_{-2,-1,-1,1}+\bar{s}_{-1,-2,-1,1}+\bar{s}_{-1,-1,-2,1}-\frac{1}{2} \ln _2^2 \bar{s}_{-1,-2}+\frac{1}{2} \ln _2^2 \bar{s}_{-1,2}-\text{Li}_4\left(\frac{1}{2}\right) \bar{s}_{-1}+\frac{1}{2} \zeta _2 \bar{s}_{-3}-\frac{13}{8} \zeta _3 \bar{s}_{-2}-\frac{3}{40} \zeta _2^2 \bar{s}_{-1}-\frac{1}{2} \zeta _2 \bar{s}_3-\frac{1}{2} \zeta _2 \ln _2^2 \bar{s}_{-1}+\zeta _2 \ln _2 \bar{s}_{-2}+\frac{1}{2} \zeta _2 \ln _2 \bar{s}_2-\frac{1}{24} \ln _2^4 \bar{s}_{-1}+\frac{1}{2} \ln _2^3 \bar{s}_{-2}-\frac{1}{2} \ln _2^3 \bar{s}_2-\frac{1}{2} \ln _2^2 \bar{s}_{-3}+\frac{1}{2} \ln _2^2 \bar{s}_3+\frac{1}{8} \zeta _3 s_{-1,-1}+\zeta _2 s_{-1,2}+\frac{3}{2} \zeta _2 s_{2,-1}-\frac{1}{2} \zeta _2 s_{2,1}-\frac{3}{2} \zeta _2 s_{-1,-1,-1}+\frac{1}{2} \zeta _2 s_{-1,-1,1}-\frac{3}{2} \zeta _2 \ln _2 s_{-1,-1}+s_{3,-2}-s_{-1,-2,-2}-s_{2,1,-2}+s_{-1,-1,1,-2}-\frac{1}{2} \ln _2^2 s_{-1,-2}-\frac{1}{2} \ln _2^2 s_{-1,2}+\frac{9 \zeta _2 \zeta _3}{16}-\frac{617 \zeta _5}{64}-\frac{1}{3} \zeta _2 \ln _2^3+\frac{3}{4} \zeta _3 \ln _2^2+\frac{51}{40} \zeta _2^2 \ln _2-3 \text{Li}_4\left(\frac{1}{2}\right) s_{-1}+8 \text{Li}_5\left(\frac{1}{2}\right)+\frac{1}{5} \zeta _2^2 s_{-1}-\frac{1}{2} \zeta _2 s_{-3}+\frac{\zeta _2 s_3}{2}+\frac{13}{8} \zeta _3 s_{-2}-\frac{\zeta _3 s_2}{8}-\frac{1}{2} \zeta _2 s_{-1} \ln _2^2-\zeta _2 s_{-2} \ln _2+2 \zeta _2 s_2 \ln _2-\frac{1}{8} s_{-1} \ln _2^4-\frac{1}{2} s_{-2} \ln _2^3-\frac{1}{2} s_2 \ln _2^3-\frac{\ln _2^5}{15}  \end{dmath}
   \begin{dmath}[style={\small}]     s_2 \bar{s}_{-3}  =   -\bar{s}_{2,-3}-\frac{9}{4} \zeta _2^2 \bar{s}_{-1}-\frac{5}{4} \zeta _3 \bar{s}_{-2}-\frac{3}{4} \zeta _3 \bar{s}_2-s_{-3,2}-\frac{15 \zeta _5}{16}-\frac{9}{2} \zeta _2^2 \ln _2-\frac{9}{4} \zeta _2^2 s_{-1}+\frac{5}{4} \zeta _3 s_{-2}-\frac{3 \zeta _3 s_2}{4}  \end{dmath}
   \begin{dmath}[style={\small}]     s_2 \bar{s}_3  =   -\bar{s}_{2,3}-2 \zeta _3 \bar{s}_2+s_{3,2}+10 \zeta _5-2 \zeta _3 s_2  \end{dmath}
   \begin{dmath}[style={\small}]     s_2 \bar{s}_{-2,-1}  =   -\frac{1}{2} \zeta _2 \bar{s}_{-2,-1}+\bar{s}_{-4,-1}-\bar{s}_{-2,2,-1}-\bar{s}_{2,-2,-1}+\ln _2 \bar{s}_{-2,-2}-\ln _2 \bar{s}_{-2,2}+\frac{1}{4} \zeta _3 \bar{s}_{-2}-\zeta _3 \bar{s}_2+\frac{1}{2} \zeta _2 \bar{s}_3-\zeta _2 \ln _2 \bar{s}_{-2}+\frac{7}{2} \zeta _3 \ln _2 \bar{s}_{-1}+\zeta _2 \ln _2 \bar{s}_2+\ln _2 \bar{s}_{-4}-\ln _2 \bar{s}_4-\frac{1}{2} \zeta _2 s_{-2,-1}+s_{3,2}-s_{-2,-1,2}-\ln _2 s_{-2,-2}-\ln _2 s_{-2,2}-\frac{9 \zeta _2 \zeta _3}{8}+\frac{5 \zeta _5}{16}+2 \zeta _2 \ln _2^3+\frac{53}{10} \zeta _2^2 \ln _2-8 \text{Li}_4\left(\frac{1}{2}\right) \ln _2-\zeta _3 s_{-2}-\zeta _3 s_2+\zeta _2 s_2 \ln _2+\frac{7}{2} \zeta _3 s_{-1} \ln _2-\frac{\ln _2^5}{3}  \end{dmath}
   \begin{dmath}[style={\small}]     s_2 \bar{s}_{-2,1}  =   \zeta _2 \bar{s}_{-2,1}+\bar{s}_{-4,1}-\bar{s}_{-2,2,1}-\bar{s}_{2,-2,1}-\frac{9}{8} \zeta _2^2 \bar{s}_{-1}-\zeta _2 \bar{s}_{-3}+\frac{13}{8} \zeta _3 \bar{s}_{-2}-\frac{5}{8} \zeta _3 \bar{s}_2-\zeta _2 s_{-2,1}-s_{-3,2}+s_{-2,1,2}-\frac{3 \zeta _3 \zeta _2}{4}+\frac{85 \zeta _5}{32}-\frac{9}{4} \zeta _2^2 \ln _2-\frac{9}{8} \zeta _2^2 s_{-1}+\frac{11}{8} \zeta _3 s_{-2}-\frac{5 \zeta _3 s_2}{8}  \end{dmath}
   \begin{dmath}[style={\small}]     s_2 \bar{s}_{2,-1}  =   -\frac{1}{2} \zeta _2 \bar{s}_{2,-1}+\bar{s}_{4,-1}-2 \bar{s}_{2,2,-1}+\ln _2 \bar{s}_{2,-2}-\ln _2 \bar{s}_{2,2}+8 \text{Li}_4\left(\frac{1}{2}\right) \bar{s}_{-1}+\frac{1}{2} \zeta _2 \bar{s}_{-3}-\frac{5}{4} \zeta _3 \bar{s}_{-2}-\frac{53}{20} \zeta _2^2 \bar{s}_{-1}+\frac{1}{2} \zeta _3 \bar{s}_2-2 \zeta _2 \ln _2^2 \bar{s}_{-1}+\zeta _2 \ln _2 \bar{s}_{-2}+\frac{7}{2} \zeta _3 \ln _2 \bar{s}_{-1}-\zeta _2 \ln _2 \bar{s}_2+\frac{1}{3} \ln _2^4 \bar{s}_{-1}-\ln _2 \bar{s}_{-4}+\ln _2 \bar{s}_4+\frac{1}{2} \zeta _2 s_{2,-1}-s_{-3,2}+s_{2,-1,2}+\ln _2 s_{2,-2}+\ln _2 s_{2,2}+\frac{3 \zeta _2 \zeta _3}{8}+\frac{123 \zeta _5}{4}-\frac{2}{3} \zeta _2 \ln _2^3-\frac{53}{5} \zeta _2^2 \ln _2+8 \text{Li}_4\left(\frac{1}{2}\right) s_{-1}-32 \text{Li}_5\left(\frac{1}{2}\right)-8 \text{Li}_4\left(\frac{1}{2}\right) \ln _2-\frac{53}{20} \zeta _2^2 s_{-1}+\frac{5}{4} \zeta _3 s_{-2}+\frac{5 \zeta _3 s_2}{4}-2 \zeta _2 s_{-1} \ln _2^2-\zeta _2 s_{-2} \ln _2+\frac{7}{2} \zeta _3 s_{-1} \ln _2+\frac{1}{3} s_{-1} \ln _2^4-\frac{\ln _2^5}{15}  \end{dmath}
   \begin{dmath}[style={\small}]     s_2 \bar{s}_{2,1}  =   \zeta _2 \bar{s}_{2,1}+\bar{s}_{4,1}-2 \bar{s}_{2,2,1}+\zeta _3 \bar{s}_2-\zeta _2 \bar{s}_3+\zeta _2 s_{2,1}+s_{3,2}-s_{2,1,2}-3 \zeta _2 \zeta _3+10 \zeta _5-2 \zeta _3 s_2  \end{dmath}
   \begin{dmath}[style={\small}]     s_2 \bar{s}_{-1,1,-1}  =   \frac{1}{2} \zeta _2 \bar{s}_{-1,-2}-\frac{7}{8} \zeta _3 \bar{s}_{-1,-1}+\frac{1}{4} \zeta _3 \bar{s}_{-1,1}-\frac{1}{2} \zeta _2 \bar{s}_{-1,1,-1}-\frac{3}{2} \zeta _2 \ln _2 \bar{s}_{-1,1}+\bar{s}_{-3,1,-1}+\bar{s}_{-1,3,-1}-\bar{s}_{-1,1,2,-1}-\bar{s}_{-1,2,1,-1}-\bar{s}_{2,-1,1,-1}+\frac{1}{2} \ln _2^2 \bar{s}_{-1,-2}-\frac{1}{2} \ln _2^2 \bar{s}_{-1,2}-\ln _2 \bar{s}_{-1,-3}+\ln _2 \bar{s}_{-1,3}+\ln _2 \bar{s}_{-1,1,-2}-\ln _2 \bar{s}_{-1,1,2}+\frac{1}{8} \zeta _2^2 \bar{s}_{-1}-\frac{1}{8} \zeta _3 \bar{s}_2-\frac{3}{2} \zeta _2 \ln _2^2 \bar{s}_{-1}+\frac{1}{2} \zeta _2 \ln _2 \bar{s}_{-2}+\zeta _2 \ln _2 \bar{s}_2+\frac{1}{2} \ln _2^2 \bar{s}_{-3}-\frac{1}{2} \ln _2^2 \bar{s}_3-\frac{1}{2} \zeta _2 s_{-2,-1}+\frac{7}{8} \zeta _3 s_{-1,-1}+\zeta _3 s_{-1,1}+\frac{1}{2} \zeta _2 s_{-1,1,-1}-\frac{1}{2} \zeta _2 \ln _2 s_{-1,1}+s_{3,2}-s_{-2,-1,2}-s_{-1,-2,2}+s_{-1,1,-1,2}+\frac{1}{2} \ln _2^2 s_{-1,-2}+\frac{1}{2} \ln _2^2 s_{-1,2}-\ln _2 s_{-2,-2}-\ln _2 s_{-2,2}+\ln _2 s_{-1,1,-2}+\ln _2 s_{-1,1,2}-\frac{9 \zeta _2 \zeta _3}{8}+\frac{85 \zeta _5}{32}-\frac{3}{4} \zeta _3 \ln _2^2+\frac{71}{40} \zeta _2^2 \ln _2-2 \text{Li}_4\left(\frac{1}{2}\right) s_{-1}-6 \text{Li}_4\left(\frac{1}{2}\right) \ln _2+\frac{2}{5} \zeta _2^2 s_{-1}-\zeta _3 s_{-2}-\zeta _3 s_2-\frac{1}{2} \zeta _2 s_{-1} \ln _2^2+\zeta _2 s_2 \ln _2+\frac{15}{4} \zeta _3 s_{-1} \ln _2-\frac{1}{12} s_{-1} \ln _2^4-\frac{\ln _2^5}{4}  \end{dmath}
   \begin{dmath}[style={\small}]     s_2 \bar{s}_{-1,1,1}  =   2 \zeta _3 \bar{s}_{-1,1}-\zeta _2 \bar{s}_{-1,2}+\zeta _2 \bar{s}_{-1,1,1}+\bar{s}_{-3,1,1}+\bar{s}_{-1,3,1}-\bar{s}_{-1,1,2,1}-\bar{s}_{-1,2,1,1}-\bar{s}_{2,-1,1,1}-3 \text{Li}_4\left(\frac{1}{2}\right) \bar{s}_{-1}-\frac{1}{8} \zeta _3 \bar{s}_{-2}+\frac{13}{20} \zeta _2^2 \bar{s}_{-1}-\frac{7}{8} \zeta _3 \bar{s}_2-\frac{1}{2} \zeta _2 \ln _2 \bar{s}_{-2}+\frac{1}{2} \zeta _2 \ln _2 \bar{s}_2-\frac{1}{8} \ln _2^4 \bar{s}_{-1}+\frac{1}{6} \ln _2^3 \bar{s}_{-2}-\frac{1}{6} \ln _2^3 \bar{s}_2-\zeta _2 s_{-2,1}-\zeta _3 s_{-1,1}+\zeta _2 s_{-1,1,1}-s_{-3,2}+s_{-2,1,2}+s_{-1,2,2}-s_{-1,1,1,2}+\frac{\zeta _2 \zeta _3}{8}-\frac{265 \zeta _5}{32}-\frac{1}{3} \zeta _2 \ln _2^3+\frac{7}{4} \zeta _2^2 \ln _2-3 \text{Li}_4\left(\frac{1}{2}\right) s_{-1}+10 \text{Li}_5\left(\frac{1}{2}\right)-\frac{3}{20} \zeta _2^2 s_{-1}+\frac{9}{8} \zeta _3 s_{-2}-\frac{7 \zeta _3 s_2}{8}+\frac{1}{2} \zeta _2 s_{-2} \ln _2+\frac{1}{2} \zeta _2 s_2 \ln _2-\frac{1}{8} s_{-1} \ln _2^4-\frac{1}{6} s_{-2} \ln _2^3-\frac{1}{6} s_2 \ln _2^3-\frac{\ln _2^5}{12}  \end{dmath}
   \begin{dmath}[style={\small}]     s_2 \bar{s}_{1,-2}  =   \frac{1}{2} \zeta _2 \bar{s}_{1,-2}-\frac{7}{2} \zeta _3 \bar{s}_{1,-1}-\frac{1}{2} \zeta _2 \bar{s}_{1,2}+\bar{s}_{3,-2}-\bar{s}_{1,2,-2}-\bar{s}_{2,1,-2}-4 \text{Li}_4\left(\frac{1}{2}\right) \bar{s}_{-1}+4 \text{Li}_4\left(\frac{1}{2}\right) \bar{s}_1-\frac{1}{2} \zeta _2 \bar{s}_{-3}+\frac{13}{8} \zeta _3 \bar{s}_{-2}+\frac{1}{5} \zeta _2^2 \bar{s}_{-1}-\frac{15}{8} \zeta _2^2 \bar{s}_1-\frac{1}{8} \zeta _3 \bar{s}_2+\frac{1}{2} \zeta _2 \bar{s}_3+\zeta _2 \ln _2^2 \bar{s}_{-1}-\zeta _2 \ln _2^2 \bar{s}_1-\frac{1}{6} \ln _2^4 \bar{s}_{-1}+\frac{1}{6} \ln _2^4 \bar{s}_1-\frac{1}{2} \zeta _2 s_{1,-2}-\frac{7}{2} \zeta _3 s_{1,-1}-\frac{1}{2} \zeta _2 s_{1,2}-s_{-3,2}+s_{1,-2,2}-\frac{15 \zeta _2 \zeta _3}{8}-\frac{83 \zeta _5}{16}+\frac{2}{3} \zeta _2 \ln _2^3+\frac{2}{5} \zeta _2^2 \ln _2-4 \text{Li}_4\left(\frac{1}{2}\right) s_{-1}+4 \text{Li}_4\left(\frac{1}{2}\right) s_1+8 \text{Li}_5\left(\frac{1}{2}\right)+\frac{1}{5} \zeta _2^2 s_{-1}-\frac{41}{40} \zeta _2^2 s_1+\frac{15}{8} \zeta _3 s_{-2}-\frac{\zeta _3 s_2}{8}+\zeta _2 s_{-1} \ln _2^2-\zeta _2 s_1 \ln _2^2-\frac{1}{6} s_{-1} \ln _2^4+\frac{1}{6} s_1 \ln _2^4-\frac{\ln _2^5}{15}  \end{dmath}
   \begin{dmath}[style={\small}]     s_2 \bar{s}_{1,2}  =   \bar{s}_{3,2}-\bar{s}_{1,2,2}-\bar{s}_{2,1,2}+\frac{17}{10} \zeta _2^2 \bar{s}_1-2 \zeta _3 \bar{s}_2+s_{3,2}-s_{1,2,2}+\frac{5 \zeta _5}{2}+\frac{7}{10} \zeta _2^2 s_1-2 \zeta _3 s_2  \end{dmath}
   \begin{dmath}[style={\small}]     s_2 \bar{s}_{1,-1,-1}  =   \frac{1}{4} \zeta _3 \bar{s}_{1,-1}+\frac{1}{2} \zeta _2 \bar{s}_{1,2}-\frac{1}{2} \zeta _2 \bar{s}_{1,-1,-1}+\bar{s}_{1,-3,-1}+\bar{s}_{3,-1,-1}-\bar{s}_{1,-1,2,-1}-\bar{s}_{1,2,-1,-1}-\bar{s}_{2,1,-1,-1}-\ln _2^2 \bar{s}_{1,-2}+\ln _2^2 \bar{s}_{1,2}+\ln _2 \bar{s}_{1,-3}-\ln _2 \bar{s}_{1,3}+\ln _2 \bar{s}_{1,-1,-2}-\ln _2 \bar{s}_{1,-1,2}-\text{Li}_4\left(\frac{1}{2}\right) \bar{s}_1+\frac{37}{40} \zeta _2^2 \bar{s}_1-\zeta _3 \bar{s}_2+\zeta _2 \ln _2^2 \bar{s}_1-\frac{1}{2} \zeta _2 \ln _2 \bar{s}_{-2}+\frac{7}{8} \zeta _3 \ln _2 \bar{s}_{-1}-\frac{7}{8} \zeta _3 \ln _2 \bar{s}_1+\frac{1}{2} \zeta _2 \ln _2 \bar{s}_2-\frac{1}{24} \ln _2^4 \bar{s}_1-\frac{1}{2} \ln _2^3 \bar{s}_{-2}+\frac{1}{2} \ln _2^3 \bar{s}_2+\ln _2^2 \bar{s}_{-3}-\ln _2^2 \bar{s}_3-\frac{1}{2} \zeta _2 s_{-2,-1}+\zeta _3 s_{1,-1}+\frac{1}{2} \zeta _2 s_{1,-1,-1}+\zeta _2 \ln _2 s_{1,-1}+s_{3,2}-s_{-2,-1,2}-s_{1,2,2}+s_{1,-1,-1,2}+\ln _2^2 s_{1,-2}+\ln _2^2 s_{1,2}-\ln _2 s_{-2,-2}-\ln _2 s_{-2,2}+\ln _2 s_{1,-1,-2}+\ln _2 s_{1,-1,2}+\frac{3 \zeta _2 \zeta _3}{16}-\frac{605 \zeta _5}{64}+\frac{1}{3} \zeta _2 \ln _2^3+\frac{15}{4} \zeta _3 \ln _2^2+\frac{81}{40} \zeta _2^2 \ln _2-3 \text{Li}_4\left(\frac{1}{2}\right) s_1+10 \text{Li}_5\left(\frac{1}{2}\right)+4 \text{Li}_4\left(\frac{1}{2}\right) \ln _2+\frac{33}{40} \zeta _2^2 s_1-\zeta _3 s_{-2}-\zeta _3 s_2+\zeta _2 s_1 \ln _2^2-\frac{1}{2} \zeta _2 s_{-2} \ln _2+\frac{1}{2} \zeta _2 s_2 \ln _2+\frac{7}{8} \zeta _3 s_{-1} \ln _2-\frac{5}{8} \zeta _3 s_1 \ln _2-\frac{1}{8} s_1 \ln _2^4+\frac{1}{2} s_{-2} \ln _2^3+\frac{1}{2} s_2 \ln _2^3+\frac{\ln _2^5}{12}  \end{dmath}
   \begin{dmath}[style={\small}]     s_2 \bar{s}_{1,-1,1}  =   -\frac{1}{2} \zeta _2 \bar{s}_{1,-2}-\frac{5}{8} \zeta _3 \bar{s}_{1,-1}-\frac{1}{2} \zeta _2 \bar{s}_{1,2}+\zeta _2 \bar{s}_{1,-1,1}+\frac{3}{2} \zeta _2 \ln _2 \bar{s}_{1,-1}+\bar{s}_{1,-3,1}+\bar{s}_{3,-1,1}-\bar{s}_{1,-1,2,1}-\bar{s}_{1,2,-1,1}-\bar{s}_{2,1,-1,1}-\frac{1}{2} \ln _2^2 \bar{s}_{1,-2}+\frac{1}{2} \ln _2^2 \bar{s}_{1,2}-\text{Li}_4\left(\frac{1}{2}\right) \bar{s}_{-1}+\text{Li}_4\left(\frac{1}{2}\right) \bar{s}_1-\frac{1}{2} \zeta _2 \bar{s}_{-3}+\frac{3}{2} \zeta _3 \bar{s}_{-2}-\frac{13}{40} \zeta _2^2 \bar{s}_{-1}-\frac{27}{40} \zeta _2^2 \bar{s}_1+\frac{1}{8} \zeta _3 \bar{s}_2+\frac{1}{2} \zeta _2 \bar{s}_3+\frac{1}{4} \zeta _2 \ln _2^2 \bar{s}_{-1}+\frac{5}{4} \zeta _2 \ln _2^2 \bar{s}_1-\zeta _2 \ln _2 \bar{s}_{-2}-\frac{1}{2} \zeta _2 \ln _2 \bar{s}_2-\frac{1}{24} \ln _2^4 \bar{s}_{-1}+\frac{1}{24} \ln _2^4 \bar{s}_1-\frac{1}{3} \ln _2^3 \bar{s}_{-2}+\frac{1}{3} \ln _2^3 \bar{s}_2+\frac{1}{2} \ln _2^2 \bar{s}_{-3}-\frac{1}{2} \ln _2^2 \bar{s}_3-\zeta _2 s_{-2,1}-\frac{1}{2} \zeta _2 s_{1,-2}-\frac{29}{8} \zeta _3 s_{1,-1}-\frac{1}{2} \zeta _2 s_{1,2}+\zeta _2 s_{1,-1,1}+\frac{3}{2} \zeta _2 \ln _2 s_{1,-1}-s_{-3,2}+s_{-2,1,2}+s_{1,-2,2}-s_{1,-1,1,2}+\frac{1}{2} \ln _2^2 s_{1,-2}+\frac{1}{2} \ln _2^2 s_{1,2}-\frac{23 \zeta _2 \zeta _3}{8}+\frac{245 \zeta _5}{64}+\frac{5}{6} \zeta _2 \ln _2^3+\frac{15}{8} \zeta _3 \ln _2^2-\frac{13}{8} \zeta _2^2 \ln _2-\text{Li}_4\left(\frac{1}{2}\right) s_{-1}+3 \text{Li}_4\left(\frac{1}{2}\right) s_1-\frac{13}{40} \zeta _2^2 s_{-1}-\frac{23}{40} \zeta _2^2 s_1+\frac{17}{8} \zeta _3 s_{-2}+\frac{\zeta _3 s_2}{8}+\frac{1}{4} \zeta _2 s_{-1} \ln _2^2-\frac{1}{4} \zeta _2 s_1 \ln _2^2-\frac{1}{2} \zeta _2 s_{-2} \ln _2-\frac{1}{2} \zeta _2 s_2 \ln _2-\frac{1}{24} s_{-1} \ln _2^4+\frac{1}{8} s_1 \ln _2^4+\frac{1}{3} s_{-2} \ln _2^3+\frac{1}{3} s_2 \ln _2^3  \end{dmath}
   \begin{dmath}[style={\small}]     s_2 \bar{s}_{1,1,-1}  =   \frac{1}{2} \zeta _2 \bar{s}_{1,-2}-\frac{7}{8} \zeta _3 \bar{s}_{1,-1}+\frac{1}{4} \zeta _3 \bar{s}_{1,1}-\frac{1}{2} \zeta _2 \bar{s}_{1,1,-1}-\frac{3}{2} \zeta _2 \ln _2 \bar{s}_{1,1}+\bar{s}_{1,3,-1}+\bar{s}_{3,1,-1}-\bar{s}_{1,1,2,-1}-\bar{s}_{1,2,1,-1}-\bar{s}_{2,1,1,-1}+\frac{1}{2} \ln _2^2 \bar{s}_{1,-2}-\frac{1}{2} \ln _2^2 \bar{s}_{1,2}-\ln _2 \bar{s}_{1,-3}+\ln _2 \bar{s}_{1,3}+\ln _2 \bar{s}_{1,1,-2}-\ln _2 \bar{s}_{1,1,2}+3 \text{Li}_4\left(\frac{1}{2}\right) \bar{s}_{-1}-\frac{1}{8} \zeta _3 \bar{s}_{-2}-\frac{51}{40} \zeta _2^2 \bar{s}_{-1}+\frac{1}{8} \zeta _2^2 \bar{s}_1-\frac{3}{4} \zeta _2 \ln _2^2 \bar{s}_{-1}-\frac{3}{4} \zeta _2 \ln _2^2 \bar{s}_1+\frac{1}{2} \zeta _2 \ln _2 \bar{s}_{-2}+\frac{21}{8} \zeta _3 \ln _2 \bar{s}_{-1}-\frac{21}{8} \zeta _3 \ln _2 \bar{s}_1+\zeta _2 \ln _2 \bar{s}_2+\frac{1}{8} \ln _2^4 \bar{s}_{-1}+\frac{1}{6} \ln _2^3 \bar{s}_{-2}-\frac{1}{6} \ln _2^3 \bar{s}_2-\frac{1}{2} \ln _2^2 \bar{s}_{-3}+\frac{1}{2} \ln _2^2 \bar{s}_3-\frac{7}{8} \zeta _3 s_{1,-1}-\zeta _3 s_{1,1}+\frac{1}{2} \zeta _2 s_{2,-1}-\frac{1}{2} \zeta _2 s_{1,1,-1}+\frac{1}{2} \zeta _2 \ln _2 s_{1,1}-s_{-3,2}+s_{1,-2,2}+s_{2,-1,2}-s_{1,1,-1,2}-\frac{1}{2} \ln _2^2 s_{1,-2}-\frac{1}{2} \ln _2^2 s_{1,2}+\ln _2 s_{2,-2}+\ln _2 s_{2,2}-\ln _2 s_{1,1,-2}-\ln _2 s_{1,1,2}-\frac{\zeta _2 \zeta _3}{16}+14 \zeta _5+\frac{1}{6} \zeta _2 \ln _2^3-\frac{15}{8} \zeta _3 \ln _2^2-\frac{29}{8} \zeta _2^2 \ln _2+3 \text{Li}_4\left(\frac{1}{2}\right) s_{-1}+2 \text{Li}_4\left(\frac{1}{2}\right) s_1-14 \text{Li}_5\left(\frac{1}{2}\right)-6 \text{Li}_4\left(\frac{1}{2}\right) \ln _2-\frac{51}{40} \zeta _2^2 s_{-1}-\frac{2}{5} \zeta _2^2 s_1+\zeta _3 s_{-2}+\zeta _3 s_2-\frac{3}{4} \zeta _2 s_{-1} \ln _2^2-\frac{1}{4} \zeta _2 s_1 \ln _2^2-\frac{1}{2} \zeta _2 s_{-2} \ln _2+\frac{1}{2} \zeta _2 s_2 \ln _2+\frac{21}{8} \zeta _3 s_{-1} \ln _2-\frac{9}{8} \zeta _3 s_1 \ln _2+\frac{1}{8} s_{-1} \ln _2^4+\frac{1}{12} s_1 \ln _2^4-\frac{1}{6} s_{-2} \ln _2^3-\frac{1}{6} s_2 \ln _2^3-\frac{2 \ln _2^5}{15}  \end{dmath}
   \begin{dmath}[style={\small}]     s_2 \bar{s}_{1,1,1}  =   -\zeta _2 \bar{s}_{1,2}+\zeta _2 \bar{s}_{1,1,1}+2 \zeta _3 \bar{s}_{1,1}+\bar{s}_{1,3,1}+\bar{s}_{3,1,1}-\bar{s}_{1,1,2,1}-\bar{s}_{1,2,1,1}-\bar{s}_{2,1,1,1}+\frac{1}{2} \zeta _2^2 \bar{s}_1-\zeta _3 \bar{s}_2+\zeta _2 s_{2,1}-\zeta _2 s_{1,1,1}+\zeta _3 s_{1,1}+s_{3,2}-s_{1,2,2}-s_{2,1,2}+s_{1,1,1,2}-3 \zeta _3 \zeta _2+\frac{15 \zeta _5}{2}+\frac{3}{10} \zeta _2^2 s_1-2 \zeta _3 s_2  \end{dmath}
   \begin{dmath}[style={\small}]     s_2 \bar{s}_{-1,-2}  =   \frac{1}{2} \zeta _2 \bar{s}_{-1,-2}-\frac{7}{2} \zeta _3 \bar{s}_{-1,-1}-\frac{1}{2} \zeta _2 \bar{s}_{-1,2}+\bar{s}_{-3,-2}-\bar{s}_{-1,2,-2}-\bar{s}_{2,-1,-2}+4 \text{Li}_4\left(\frac{1}{2}\right) \bar{s}_{-1}+\frac{1}{2} \zeta _2 \bar{s}_{-3}-\frac{9}{8} \zeta _2^2 \bar{s}_{-1}+\frac{3}{2} \zeta _3 \bar{s}_2-\frac{1}{2} \zeta _2 \bar{s}_3-\zeta _2 \ln _2^2 \bar{s}_{-1}-\frac{1}{2} \zeta _2 \ln _2 \bar{s}_{-2}+\frac{1}{2} \zeta _2 \ln _2 \bar{s}_2+\frac{1}{6} \ln _2^4 \bar{s}_{-1}+\frac{1}{2} \zeta _2 s_{-1,-2}+\frac{7}{2} \zeta _3 s_{-1,-1}+\frac{1}{2} \zeta _2 s_{-1,2}+s_{3,2}-s_{-1,-2,2}-\frac{3 \zeta _2 \zeta _3}{4}+\frac{235 \zeta _5}{32}-\frac{7}{4} \zeta _2^2 \ln _2-4 \text{Li}_4\left(\frac{1}{2}\right) s_{-1}+\frac{71}{40} \zeta _2^2 s_{-1}-2 \zeta _3 s_2+\zeta _2 s_{-1} \ln _2^2+\frac{1}{2} \zeta _2 s_{-2} \ln _2+\frac{1}{2} \zeta _2 s_2 \ln _2-\frac{1}{6} s_{-1} \ln _2^4  \end{dmath}
   \begin{dmath}[style={\small}]     s_2 \bar{s}_{-1,2}  =   \bar{s}_{-3,2}-\bar{s}_{-1,2,2}-\bar{s}_{2,-1,2}-4 \text{Li}_4\left(\frac{1}{2}\right) \bar{s}_{-1}-\zeta _3 \bar{s}_{-2}+\frac{61}{40} \zeta _2^2 \bar{s}_{-1}-\zeta _3 \bar{s}_2+\zeta _2 \ln _2^2 \bar{s}_{-1}-\frac{1}{2} \zeta _2 \ln _2 \bar{s}_{-2}+\frac{1}{2} \zeta _2 \ln _2 \bar{s}_2-\frac{1}{6} \ln _2^4 \bar{s}_{-1}-s_{-3,2}+s_{-1,2,2}-\frac{507 \zeta _5}{32}+\frac{4}{3} \zeta _2 \ln _2^3+\frac{71}{20} \zeta _2^2 \ln _2-4 \text{Li}_4\left(\frac{1}{2}\right) s_{-1}+16 \text{Li}_5\left(\frac{1}{2}\right)-\frac{7}{8} \zeta _2^2 s_{-1}+\zeta _3 s_{-2}-\zeta _3 s_2+\zeta _2 s_{-1} \ln _2^2+\frac{1}{2} \zeta _2 s_{-2} \ln _2+\frac{1}{2} \zeta _2 s_2 \ln _2-\frac{1}{6} s_{-1} \ln _2^4-\frac{2 \ln _2^5}{15}  \end{dmath}
   \begin{dmath}[style={\small}]     s_2 \bar{s}_{-1,-1,-1}  =   \frac{1}{4} \zeta _3 \bar{s}_{-1,-1}+\frac{1}{2} \zeta _2 \bar{s}_{-1,2}-\frac{1}{2} \zeta _2 \bar{s}_{-1,-1,-1}+\bar{s}_{-3,-1,-1}+\bar{s}_{-1,-3,-1}-\bar{s}_{-1,-1,2,-1}-\bar{s}_{-1,2,-1,-1}-\bar{s}_{2,-1,-1,-1}-\ln _2^2 \bar{s}_{-1,-2}+\ln _2^2 \bar{s}_{-1,2}+\ln _2 \bar{s}_{-1,-3}-\ln _2 \bar{s}_{-1,3}+\ln _2 \bar{s}_{-1,-1,-2}-\ln _2 \bar{s}_{-1,-1,2}-\frac{3}{4} \zeta _3 \bar{s}_{-2}+\frac{1}{8} \zeta _2^2 \bar{s}_{-1}-\frac{1}{4} \zeta _3 \bar{s}_2+\frac{1}{2} \zeta _2 \ln _2 \bar{s}_{-2}-\frac{1}{2} \zeta _2 \ln _2 \bar{s}_2+\frac{2}{3} \ln _2^3 \bar{s}_{-2}-\frac{2}{3} \ln _2^3 \bar{s}_2-\ln _2^2 \bar{s}_{-3}+\ln _2^2 \bar{s}_3-\zeta _3 s_{-1,-1}+\frac{1}{2} \zeta _2 s_{2,-1}-\frac{1}{2} \zeta _2 s_{-1,-1,-1}-\zeta _2 \ln _2 s_{-1,-1}-s_{-3,2}+s_{-1,2,2}+s_{2,-1,2}-s_{-1,-1,-1,2}-\ln _2^2 s_{-1,-2}-\ln _2^2 s_{-1,2}+\ln _2 s_{2,-2}+\ln _2 s_{2,2}-\ln _2 s_{-1,-1,-2}-\ln _2 s_{-1,-1,2}+\frac{\zeta _2 \zeta _3}{8}+\frac{231 \zeta _5}{32}-\frac{4}{3} \zeta _2 \ln _2^3+\frac{3}{2} \zeta _3 \ln _2^2-\frac{13}{4} \zeta _2^2 \ln _2+4 \text{Li}_4\left(\frac{1}{2}\right) s_{-1}-8 \text{Li}_5\left(\frac{1}{2}\right)-\frac{13}{8} \zeta _2^2 s_{-1}+\frac{3}{4} \zeta _3 s_{-2}+\frac{3 \zeta _3 s_2}{4}-2 \zeta _2 s_{-1} \ln _2^2-\frac{1}{2} \zeta _2 s_{-2} \ln _2+\frac{1}{2} \zeta _2 s_2 \ln _2+\frac{3}{2} \zeta _3 s_{-1} \ln _2+\frac{1}{6} s_{-1} \ln _2^4-\frac{2}{3} s_{-2} \ln _2^3-\frac{2}{3} s_2 \ln _2^3+\frac{\ln _2^5}{15}  \end{dmath}
   \begin{dmath}[style={\small}]     s_2 \bar{s}_{-1,-1,1}  =   -\frac{1}{2} \zeta _2 \bar{s}_{-1,-2}-\frac{5}{8} \zeta _3 \bar{s}_{-1,-1}-\frac{1}{2} \zeta _2 \bar{s}_{-1,2}+\zeta _2 \bar{s}_{-1,-1,1}+\frac{3}{2} \zeta _2 \ln _2 \bar{s}_{-1,-1}+\bar{s}_{-3,-1,1}+\bar{s}_{-1,-3,1}-\bar{s}_{-1,-1,2,1}-\bar{s}_{-1,2,-1,1}-\bar{s}_{2,-1,-1,1}-\frac{1}{2} \ln _2^2 \bar{s}_{-1,-2}+\frac{1}{2} \ln _2^2 \bar{s}_{-1,2}+\text{Li}_4\left(\frac{1}{2}\right) \bar{s}_{-1}+\frac{1}{2} \zeta _2 \bar{s}_{-3}+\frac{3}{40} \zeta _2^2 \bar{s}_{-1}+\frac{13}{8} \zeta _3 \bar{s}_2-\frac{1}{2} \zeta _2 \bar{s}_3+\frac{1}{2} \zeta _2 \ln _2^2 \bar{s}_{-1}-\frac{1}{2} \zeta _2 \ln _2 \bar{s}_{-2}-\zeta _2 \ln _2 \bar{s}_2+\frac{1}{24} \ln _2^4 \bar{s}_{-1}+\frac{1}{2} \ln _2^3 \bar{s}_{-2}-\frac{1}{2} \ln _2^3 \bar{s}_2-\frac{1}{2} \ln _2^2 \bar{s}_{-3}+\frac{1}{2} \ln _2^2 \bar{s}_3+\frac{1}{2} \zeta _2 s_{-1,-2}+\frac{29}{8} \zeta _3 s_{-1,-1}+\frac{1}{2} \zeta _2 s_{-1,2}+\zeta _2 s_{2,1}-\zeta _2 s_{-1,-1,1}-\frac{3}{2} \zeta _2 \ln _2 s_{-1,-1}+s_{3,2}-s_{-1,-2,2}-s_{2,1,2}+s_{-1,-1,1,2}-\frac{1}{2} \ln _2^2 s_{-1,-2}-\frac{1}{2} \ln _2^2 s_{-1,2}-\frac{9 \zeta _2 \zeta _3}{8}+\frac{209 \zeta _5}{32}-\frac{2}{3} \zeta _2 \ln _2^3+\frac{3}{4} \zeta _3 \ln _2^2-\frac{7}{20} \zeta _2^2 \ln _2-3 \text{Li}_4\left(\frac{1}{2}\right) s_{-1}-2 \text{Li}_5\left(\frac{1}{2}\right)+\frac{53}{40} \zeta _2^2 s_{-1}-2 \zeta _3 s_2-\frac{1}{2} \zeta _2 s_{-1} \ln _2^2+\frac{1}{2} \zeta _2 s_{-2} \ln _2+\frac{1}{2} \zeta _2 s_2 \ln _2-\frac{1}{8} s_{-1} \ln _2^4-\frac{1}{2} s_{-2} \ln _2^3-\frac{1}{2} s_2 \ln _2^3+\frac{\ln _2^5}{60}  \end{dmath}
   \begin{dmath}[style={\small}]     \bar{s}_{-3} s_{-1,1}  =   \zeta _2 \bar{s}_{-1,-2}+\frac{7}{4} \zeta _3 \bar{s}_{-1,-1}-\frac{3}{4} \zeta _3 \bar{s}_{-1,1}+\bar{s}_{-2,-3}-\bar{s}_{-1,1,-3}-2 \text{Li}_4\left(\frac{1}{2}\right) \bar{s}_{-1}-\frac{1}{2} \zeta _2 \bar{s}_{-3}+\frac{3}{4} \zeta _3 \bar{s}_{-2}+\frac{5}{4} \zeta _2^2 \bar{s}_{-1}-\frac{19}{8} \zeta _3 \bar{s}_2-\frac{1}{2} \zeta _2 \bar{s}_3+\frac{1}{2} \zeta _2 \ln _2^2 \bar{s}_{-1}+\frac{3}{2} \zeta _2 \ln _2 \bar{s}_2-\frac{1}{12} \ln _2^4 \bar{s}_{-1}+\frac{1}{2} \ln _2^2 \bar{s}_{-3}+\frac{1}{2} \ln _2^2 \bar{s}_3+\zeta _2 s_{-1,-2}-\frac{7}{4} \zeta _3 s_{-1,-1}-\frac{3}{4} \zeta _3 s_{-1,1}+s_{4,1}-s_{-3,-1,1}-s_{-1,-3,1}-3 \zeta _2 \zeta _3+\frac{225 \zeta _5}{32}+\frac{3}{4} \zeta _3 \ln _2^2-\frac{3}{4} \zeta _2^2 \ln _2+2 \text{Li}_4\left(\frac{1}{2}\right) s_{-1}-\frac{3}{5} \zeta _2^2 s_{-1}-\frac{1}{2} \zeta _2 s_{-3}-\frac{\zeta _2 s_3}{2}-\frac{5 \zeta _3 s_2}{8}-\frac{1}{2} \zeta _2 s_{-1} \ln _2^2+\frac{3}{2} \zeta _2 s_2 \ln _2+\frac{1}{12} s_{-1} \ln _2^4+\frac{1}{2} s_{-3} \ln _2^2-\frac{1}{2} s_3 \ln _2^2  \end{dmath}
   \begin{dmath}[style={\small}]     \bar{s}_3 s_{-1,1}  =   \zeta _3 \bar{s}_{-1,1}+\zeta _2 \bar{s}_{-1,2}+\bar{s}_{-2,3}-\bar{s}_{-1,1,3}-2 \text{Li}_4\left(\frac{1}{2}\right) \bar{s}_{-1}-\frac{1}{2} \zeta _2 \bar{s}_{-3}-\frac{13}{8} \zeta _3 \bar{s}_{-2}-\zeta _2^2 \bar{s}_{-1}-\frac{1}{2} \zeta _2 \bar{s}_3+\frac{1}{2} \zeta _2 \ln _2^2 \bar{s}_{-1}+\frac{3}{2} \zeta _2 \ln _2 \bar{s}_{-2}-\frac{1}{12} \ln _2^4 \bar{s}_{-1}+\frac{1}{2} \ln _2^2 \bar{s}_{-3}+\frac{1}{2} \ln _2^2 \bar{s}_3+\zeta _3 s_{-1,1}-\zeta _2 s_{-1,2}-s_{-4,1}+s_{-1,3,1}+s_{3,-1,1}-\frac{27 \zeta _5}{8}+\frac{1}{3} \zeta _2 \ln _2^3+\frac{3}{4} \zeta _3 \ln _2^2+\frac{1}{5} \zeta _2^2 \ln _2-2 \text{Li}_4\left(\frac{1}{2}\right) s_{-1}+4 \text{Li}_5\left(\frac{1}{2}\right)+\frac{3}{5} \zeta _2^2 s_{-1}+\frac{1}{2} \zeta _2 s_{-3}+\frac{\zeta _2 s_3}{2}+\frac{5}{8} \zeta _3 s_{-2}+\frac{1}{2} \zeta _2 s_{-1} \ln _2^2-\frac{3}{2} \zeta _2 s_{-2} \ln _2-\frac{1}{12} s_{-1} \ln _2^4+\frac{1}{2} s_{-3} \ln _2^2-\frac{1}{2} s_3 \ln _2^2-\frac{\ln _2^5}{30}  \end{dmath}
   \begin{dmath}[style={\small}]     s_{-1,1} \bar{s}_{-2,-1}  =   -\frac{\ln _2^5}{15}+\frac{1}{12} s_{-1} \ln _2^4+\frac{1}{3} \bar{s}_{-1} \ln _2^4+\frac{1}{2} s_{-2} \ln _2^3-\frac{1}{2} s_2 \ln _2^3+\frac{7}{6} \zeta _2 \ln _2^3+\frac{1}{2} \bar{s}_{-2} \ln _2^3-\frac{1}{2} \bar{s}_2 \ln _2^3+\frac{5}{2} s_{-1} \zeta _2 \ln _2^2+\frac{3}{2} \zeta _3 \ln _2^2+\frac{1}{2} \bar{s}_{-3} \ln _2^2-2 \zeta _2 \bar{s}_{-1} \ln _2^2-\frac{1}{2} \bar{s}_3 \ln _2^2-\frac{1}{2} s_{-1,-2} \ln _2^2+\frac{1}{2} s_{-1,2} \ln _2^2+\bar{s}_{-2,-1} \ln _2^2+\frac{1}{2} \bar{s}_{-1,-2} \ln _2^2-\frac{1}{2} \bar{s}_{-1,2} \ln _2^2-\frac{29}{5} \zeta _2^2 \ln _2-6 \text{Li}_4\left(\frac{1}{2}\right) \ln _2-3 s_{-2} \zeta _2 \ln _2-\frac{1}{2} s_2 \zeta _2 \ln _2-\frac{3}{8} s_{-1} \zeta _3 \ln _2+\bar{s}_{-4} \ln _2-\zeta _2 \bar{s}_{-2} \ln _2+\frac{21}{8} \zeta _3 \bar{s}_{-1} \ln _2+\zeta _2 \bar{s}_2 \ln _2-\bar{s}_4 \ln _2+\frac{3}{2} \zeta _2 s_{-1,-1} \ln _2+\frac{3}{2} \zeta _2 s_{-1,1} \ln _2+s_{3,-1} \ln _2+s_{3,1} \ln _2+\bar{s}_{-2,-2} \ln _2-\bar{s}_{-2,2} \ln _2+\bar{s}_{-1,-3} \ln _2-\frac{3}{2} \zeta _2 \bar{s}_{-1,-1} \ln _2+\frac{3}{2} \zeta _2 \bar{s}_{-1,1} \ln _2-\bar{s}_{-1,3} \ln _2-\bar{s}_{3,-1} \ln _2+\bar{s}_{3,1} \ln _2-s_{-2,-1,-1} \ln _2-s_{-2,-1,1} \ln _2-s_{-1,-2,-1} \ln _2-s_{-1,-2,1} \ln _2+\bar{s}_{-2,-1,-1} \ln _2-\bar{s}_{-2,-1,1} \ln _2+\bar{s}_{-1,-2,-1} \ln _2-\bar{s}_{-1,-2,1} \ln _2-\frac{13}{20} s_{-1} \zeta _2^2-22 \text{Li}_5\left(\frac{1}{2}\right)+2 \text{Li}_4\left(\frac{1}{2}\right) s_{-1}+s_3 \zeta _2+\frac{9}{4} s_{-2} \zeta _3-\frac{3 \zeta _2 \zeta _3}{2}+\frac{1563 \zeta _5}{64}-\frac{1}{2} \zeta _2 \bar{s}_{-3}+\frac{5}{8} \zeta _3 \bar{s}_{-2}-\frac{101}{40} \zeta _2^2 \bar{s}_{-1}+8 \text{Li}_4\left(\frac{1}{2}\right) \bar{s}_{-1}-s_{-4,1}-\zeta _2 s_{-2,-1}-\frac{1}{2} \zeta _2 s_{-1,-2}-\frac{5}{8} \zeta _3 s_{-1,1}-\frac{1}{2} \zeta _2 s_{-1,2}-\frac{1}{2} \zeta _2 \bar{s}_{-2,-1}-\frac{5}{8} \zeta _3 \bar{s}_{-1,1}+\frac{1}{2} \zeta _2 \bar{s}_{-1,2}-\bar{s}_{4,-1}+s_{-2,2,1}+s_{-1,3,1}+2 s_{3,-1,1}+2 \bar{s}_{-2,-2,-1}+\bar{s}_{-1,-3,-1}+\bar{s}_{3,1,-1}-2 s_{-2,-1,-1,1}-s_{-1,-2,-1,1}-\bar{s}_{-2,-1,1,-1}-\bar{s}_{-1,-2,1,-1}-\bar{s}_{-1,1,-2,-1}  \end{dmath}
   \begin{dmath}[style={\small}]     s_{-1,1} \bar{s}_{-2,1}  =   -\frac{3}{2} \zeta _2 \bar{s}_{-2,-1}-\frac{1}{2} \zeta _2 \bar{s}_{-2,1}-\zeta _2 \bar{s}_{-1,-2}+\frac{21}{8} \zeta _3 \bar{s}_{-1,-1}-\frac{5}{8} \zeta _3 \bar{s}_{-1,1}-\bar{s}_{4,1}+2 \bar{s}_{-2,-2,1}+\bar{s}_{-1,-3,1}+\bar{s}_{3,1,1}-\bar{s}_{-2,-1,1,1}-\bar{s}_{-1,-2,1,1}-\bar{s}_{-1,1,-2,1}+\frac{1}{2} \ln _2^2 \bar{s}_{-2,-1}+\frac{1}{2} \ln _2^2 \bar{s}_{-2,1}-3 \text{Li}_4\left(\frac{1}{2}\right) \bar{s}_{-1}+\frac{3}{8} \zeta _3 \bar{s}_{-2}+\frac{13}{20} \zeta _2^2 \bar{s}_{-1}-\frac{19}{8} \zeta _3 \bar{s}_2+\zeta _2 \bar{s}_3+\frac{3}{4} \zeta _2 \ln _2^2 \bar{s}_{-1}-\zeta _2 \ln _2 \bar{s}_{-2}+\zeta _2 \ln _2 \bar{s}_2-\frac{1}{8} \ln _2^4 \bar{s}_{-1}+\frac{1}{3} \ln _2^3 \bar{s}_{-2}-\frac{1}{3} \ln _2^3 \bar{s}_2+\frac{1}{2} \zeta _2 s_{-2,-1}+\frac{1}{2} \zeta _2 s_{-2,1}+\zeta _2 s_{-1,-2}-\frac{21}{8} \zeta _3 s_{-1,-1}-\frac{5}{8} \zeta _3 s_{-1,1}+s_{4,1}-s_{-3,-1,1}-s_{-2,-2,1}-s_{-1,-3,1}-s_{3,1,1}+s_{-2,-1,1,1}+s_{-2,1,-1,1}+s_{-1,-2,1,1}+\frac{1}{2} \ln _2^2 s_{-2,-1}-\frac{1}{2} \ln _2^2 s_{-2,1}-\frac{3 \zeta _2 \zeta _3}{2}+\frac{199 \zeta _5}{32}-\frac{1}{6} \zeta _2 \ln _2^3+\frac{3}{2} \zeta _3 \ln _2^2-\frac{8}{5} \zeta _2^2 \ln _2+3 \text{Li}_4\left(\frac{1}{2}\right) s_{-1}-2 \text{Li}_5\left(\frac{1}{2}\right)-\frac{29}{40} \zeta _2^2 s_{-1}-\frac{1}{2} \zeta _2 s_{-3}-\frac{\zeta _2 s_3}{2}+\frac{1}{8} \zeta _3 s_{-2}+\frac{\zeta _3 s_2}{4}-\frac{3}{4} \zeta _2 s_{-1} \ln _2^2+\frac{1}{2} \zeta _2 s_{-2} \ln _2+\zeta _2 s_2 \ln _2+\frac{1}{8} s_{-1} \ln _2^4+\frac{1}{3} s_{-2} \ln _2^3-\frac{1}{3} s_2 \ln _2^3+\frac{1}{2} s_{-3} \ln _2^2-\frac{1}{2} s_3 \ln _2^2+\frac{\ln _2^5}{60}  \end{dmath}
   \begin{dmath}[style={\small}]     s_{-1,1} \bar{s}_{2,-1}  =   -\frac{1}{12} s_{-1} \ln _2^4+\frac{1}{2} s_{-2} \ln _2^3-\frac{1}{2} s_2 \ln _2^3-\frac{3}{2} \zeta _2 \ln _2^3-\frac{1}{2} \bar{s}_{-2} \ln _2^3+\frac{1}{2} \bar{s}_2 \ln _2^3-\frac{5}{2} s_{-1} \zeta _2 \ln _2^2+\frac{3}{2} \zeta _3 \ln _2^2-\frac{1}{2} \bar{s}_{-3} \ln _2^2+\frac{1}{2} \bar{s}_3 \ln _2^2-\frac{1}{2} s_{-1,-2} \ln _2^2+\frac{1}{2} s_{-1,2} \ln _2^2-\frac{1}{2} \bar{s}_{-1,-2} \ln _2^2+\frac{1}{2} \bar{s}_{-1,2} \ln _2^2+\bar{s}_{2,-1} \ln _2^2+\frac{1}{2} \zeta _2^2 \ln _2+\frac{1}{2} s_{-2} \zeta _2 \ln _2+3 s_2 \zeta _2 \ln _2+\frac{3}{8} s_{-1} \zeta _3 \ln _2-\bar{s}_{-4} \ln _2+\zeta _2 \bar{s}_{-2} \ln _2+\frac{21}{8} \zeta _3 \bar{s}_{-1} \ln _2-\zeta _2 \bar{s}_2 \ln _2+\bar{s}_4 \ln _2-s_{-3,-1} \ln _2-s_{-3,1} \ln _2-\frac{3}{2} \zeta _2 s_{-1,-1} \ln _2-\frac{3}{2} \zeta _2 s_{-1,1} \ln _2-\bar{s}_{-3,-1} \ln _2+\bar{s}_{-3,1} \ln _2-\bar{s}_{-1,-3} \ln _2+\frac{3}{2} \zeta _2 \bar{s}_{-1,-1} \ln _2-\frac{3}{2} \zeta _2 \bar{s}_{-1,1} \ln _2+\bar{s}_{-1,3} \ln _2+\bar{s}_{2,-2} \ln _2-\bar{s}_{2,2} \ln _2+s_{-1,2,-1} \ln _2+s_{-1,2,1} \ln _2+s_{2,-1,-1} \ln _2+s_{2,-1,1} \ln _2+\bar{s}_{-1,2,-1} \ln _2-\bar{s}_{-1,2,1} \ln _2+\bar{s}_{2,-1,-1} \ln _2-\bar{s}_{2,-1,1} \ln _2+\frac{13}{20} s_{-1} \zeta _2^2-2 \text{Li}_4\left(\frac{1}{2}\right) s_{-1}-s_{-3} \zeta _2-\frac{9 s_2 \zeta _3}{4}-\frac{3 \zeta _2 \zeta _3}{2}+\frac{75 \zeta _5}{32}-\frac{1}{4} \zeta _3 \bar{s}_{-2}+\frac{1}{8} \zeta _2^2 \bar{s}_{-1}+\frac{7}{8} \zeta _3 \bar{s}_2-\frac{1}{2} \zeta _2 \bar{s}_3+\frac{1}{2} \zeta _2 s_{-1,-2}+\frac{7}{8} \zeta _3 s_{-1,-1}+\frac{1}{4} \zeta _3 s_{-1,1}+\frac{1}{2} \zeta _2 s_{-1,2}+\zeta _2 s_{2,-1}+s_{4,1}-\bar{s}_{-4,-1}+\frac{1}{2} \zeta _2 \bar{s}_{-1,-2}-\frac{7}{8} \zeta _3 \bar{s}_{-1,-1}+\frac{1}{4} \zeta _3 \bar{s}_{-1,1}-\frac{1}{2} \zeta _2 \bar{s}_{2,-1}-2 s_{-3,-1,1}-s_{-1,-3,1}-s_{2,2,1}+\bar{s}_{-3,1,-1}+\bar{s}_{-2,2,-1}+\bar{s}_{-1,3,-1}+\bar{s}_{2,-2,-1}+s_{-1,2,-1,1}+2 s_{2,-1,-1,1}-\bar{s}_{-1,1,2,-1}-\bar{s}_{-1,2,1,-1}-\bar{s}_{2,-1,1,-1}  \end{dmath}
   \begin{dmath}[style={\small}]     s_{-1,1} \bar{s}_{2,1}  =   2 \zeta _3 \bar{s}_{-1,1}-\zeta _2 \bar{s}_{-1,2}-\frac{3}{2} \zeta _2 \bar{s}_{2,-1}-\frac{1}{2} \zeta _2 \bar{s}_{2,1}-\bar{s}_{-4,1}+\bar{s}_{-3,1,1}+\bar{s}_{-2,2,1}+\bar{s}_{-1,3,1}+\bar{s}_{2,-2,1}-\bar{s}_{-1,1,2,1}-\bar{s}_{-1,2,1,1}-\bar{s}_{2,-1,1,1}+\frac{1}{2} \ln _2^2 \bar{s}_{2,-1}+\frac{1}{2} \ln _2^2 \bar{s}_{2,1}-3 \text{Li}_4\left(\frac{1}{2}\right) \bar{s}_{-1}+\zeta _2 \bar{s}_{-3}-\frac{7}{4} \zeta _3 \bar{s}_{-2}-\frac{19}{40} \zeta _2^2 \bar{s}_{-1}-\frac{1}{4} \zeta _3 \bar{s}_2+\frac{3}{4} \zeta _2 \ln _2^2 \bar{s}_{-1}+\zeta _2 \ln _2 \bar{s}_{-2}-\zeta _2 \ln _2 \bar{s}_2-\frac{1}{8} \ln _2^4 \bar{s}_{-1}-\frac{1}{3} \ln _2^3 \bar{s}_{-2}+\frac{1}{3} \ln _2^3 \bar{s}_2+2 \zeta _3 s_{-1,1}-\zeta _2 s_{-1,2}-\frac{1}{2} \zeta _2 s_{2,-1}-\frac{1}{2} \zeta _2 s_{2,1}-s_{-4,1}+s_{-3,1,1}+s_{-1,3,1}+s_{2,-2,1}+s_{3,-1,1}-s_{-1,2,1,1}-s_{2,-1,1,1}-s_{2,1,-1,1}-\frac{1}{2} \ln _2^2 s_{2,-1}+\frac{1}{2} \ln _2^2 s_{2,1}+\frac{9 \zeta _2 \zeta _3}{8}-\frac{175 \zeta _5}{16}+\frac{5}{6} \zeta _2 \ln _2^3-\frac{9}{8} \zeta _3 \ln _2^2+2 \zeta _2^2 \ln _2-3 \text{Li}_4\left(\frac{1}{2}\right) s_{-1}+10 \text{Li}_5\left(\frac{1}{2}\right)+\frac{29}{40} \zeta _2^2 s_{-1}+\frac{1}{2} \zeta _2 s_{-3}+\frac{\zeta _2 s_3}{2}-\frac{1}{4} \zeta _3 s_{-2}-\frac{\zeta _3 s_2}{8}+\frac{3}{4} \zeta _2 s_{-1} \ln _2^2-\zeta _2 s_{-2} \ln _2-\frac{1}{2} \zeta _2 s_2 \ln _2-\frac{1}{8} s_{-1} \ln _2^4+\frac{1}{3} s_{-2} \ln _2^3-\frac{1}{3} s_2 \ln _2^3+\frac{1}{2} s_{-3} \ln _2^2-\frac{1}{2} s_3 \ln _2^2-\frac{\ln _2^5}{12}  \end{dmath}
   \begin{dmath}[style={\small}]     s_{-1,1} \bar{s}_{-1,1,-1}  =   -\frac{17 \ln _2^5}{120}+\frac{1}{6} s_{-1} \ln _2^4+\frac{1}{12} \bar{s}_{-1} \ln _2^4+\frac{2}{3} s_{-2} \ln _2^3-\frac{2}{3} s_2 \ln _2^3+\frac{9}{4} \zeta _2 \ln _2^3+\frac{1}{6} \bar{s}_{-2} \ln _2^3-\frac{1}{6} \bar{s}_2 \ln _2^3+\frac{5}{6} s_{-1,-1} \ln _2^3-\frac{5}{6} s_{-1,1} \ln _2^3-\frac{1}{6} \bar{s}_{-1,-1} \ln _2^3+\frac{1}{6} \bar{s}_{-1,1} \ln _2^3+3 s_{-1} \zeta _2 \ln _2^2-\frac{3}{4} \zeta _3 \ln _2^2+\frac{1}{2} \bar{s}_{-3} \ln _2^2+\zeta _2 \bar{s}_{-1} \ln _2^2-\frac{1}{2} \bar{s}_3 \ln _2^2-\frac{1}{2} s_{-1,-2} \ln _2^2+\frac{1}{2} s_{-1,2} \ln _2^2-s_{2,1} \ln _2^2+\frac{1}{2} \bar{s}_{-1,-2} \ln _2^2-\frac{1}{2} \bar{s}_{-1,2} \ln _2^2+2 s_{-1,-1,1} \ln _2^2+\bar{s}_{-1,1,-1} \ln _2^2-\frac{3}{2} \zeta _2^2 \ln _2-9 \text{Li}_4\left(\frac{1}{2}\right) \ln _2-2 s_{-2} \zeta _2 \ln _2-\frac{3}{2} s_2 \zeta _2 \ln _2+\frac{3}{4} s_{-1} \zeta _3 \ln _2-\frac{1}{2} \zeta _2 \bar{s}_{-2} \ln _2+\frac{21}{8} \zeta _3 \bar{s}_{-1} \ln _2-\zeta _2 \bar{s}_2 \ln _2+4 \zeta _2 s_{-1,-1} \ln _2+\frac{5}{2} \zeta _2 s_{-1,1} \ln _2+s_{3,-1} \ln _2+s_{3,1} \ln _2-\bar{s}_{-1,-3} \ln _2+3 \zeta _2 \bar{s}_{-1,-1} \ln _2+\bar{s}_{-1,3} \ln _2+\bar{s}_{2,-2} \ln _2-\bar{s}_{2,2} \ln _2-s_{-2,-1,-1} \ln _2-s_{-2,-1,1} \ln _2-2 s_{-1,-2,-1} \ln _2-2 s_{-1,-2,1} \ln _2-s_{2,1,-1} \ln _2-s_{2,1,1} \ln _2-\bar{s}_{-1,-2,-1} \ln _2+\bar{s}_{-1,-2,1} \ln _2-2 \bar{s}_{-1,-1,-2} \ln _2+2 \bar{s}_{-1,-1,2} \ln _2+\bar{s}_{-1,1,-2} \ln _2-\bar{s}_{-1,1,2} \ln _2-\bar{s}_{2,1,-1} \ln _2+\bar{s}_{2,1,1} \ln _2+2 s_{-1,-1,1,-1} \ln _2+2 s_{-1,-1,1,1} \ln _2+s_{-1,1,-1,-1} \ln _2+s_{-1,1,-1,1} \ln _2+2 \bar{s}_{-1,-1,1,-1} \ln _2-2 \bar{s}_{-1,-1,1,1} \ln _2+\bar{s}_{-1,1,-1,-1} \ln _2-\bar{s}_{-1,1,-1,1} \ln _2-\frac{47}{40} s_{-1} \zeta _2^2-10 \text{Li}_5\left(\frac{1}{2}\right)+4 \text{Li}_4\left(\frac{1}{2}\right) s_{-1}+s_3 \zeta _2+\frac{3}{2} s_{-2} \zeta _3+\frac{3 s_2 \zeta _3}{4}-\frac{\zeta _2 \zeta _3}{4}+\frac{661 \zeta _5}{64}-\frac{1}{8} \zeta _3 \bar{s}_{-2}-\frac{19}{20} \zeta _2^2 \bar{s}_{-1}+2 \text{Li}_4\left(\frac{1}{2}\right) \bar{s}_{-1}+\frac{1}{4} \zeta _3 \bar{s}_2-s_{-4,1}-\zeta _2 s_{-2,-1}-\frac{3}{2} \zeta _2 s_{-1,-2}-\frac{3}{2} \zeta _3 s_{-1,-1}-\frac{11}{8} \zeta _3 s_{-1,1}-\frac{1}{2} \zeta _2 s_{-1,2}-\frac{1}{2} \zeta _2 s_{2,-1}-\frac{1}{2} \zeta _2 s_{2,1}-\frac{1}{2} \zeta _3 \bar{s}_{-1,-1}+\frac{7}{8} \zeta _3 \bar{s}_{-1,1}-\frac{1}{2} \zeta _2 \bar{s}_{-1,2}-\frac{1}{2} \zeta _2 \bar{s}_{2,-1}+s_{-2,2,1}+\zeta _2 s_{-1,-1,-1}+\zeta _2 s_{-1,-1,1}+\zeta _2 s_{-1,1,-1}+2 s_{-1,3,1}+s_{2,-2,1}+2 s_{3,-1,1}-\bar{s}_{-1,-3,-1}+\zeta _2 \bar{s}_{-1,-1,-1}-\frac{1}{2} \zeta _2 \bar{s}_{-1,1,-1}-\bar{s}_{2,2,-1}-\bar{s}_{3,1,-1}-2 s_{-2,-1,-1,1}-3 s_{-1,-2,-1,1}-2 s_{-1,-1,-2,1}-s_{-1,1,2,1}-s_{2,1,-1,1}+\bar{s}_{-2,-1,1,-1}+3 \bar{s}_{-1,-2,1,-1}+2 \bar{s}_{-1,-1,2,-1}+\bar{s}_{-1,1,-2,-1}+2 \bar{s}_{2,1,1,-1}+2 s_{-1,-1,1,-1,1}+2 s_{-1,1,-1,-1,1}-4 \bar{s}_{-1,-1,1,1,-1}-2 \bar{s}_{-1,1,-1,1,-1}  \end{dmath}
   \begin{dmath}[style={\small}]     s_{-1,1} \bar{s}_{-1,1,1}  =   \frac{2 \ln _2^5}{15}+\frac{1}{8} s_{-1} \ln _2^4+\frac{1}{8} \bar{s}_{-1} \ln _2^4+\frac{1}{3} s_{-2} \ln _2^3-\frac{1}{3} s_2 \ln _2^3-\frac{1}{3} \zeta _2 \ln _2^3+\frac{1}{6} \bar{s}_{-2} \ln _2^3-\frac{1}{6} \bar{s}_2 \ln _2^3+\frac{1}{6} s_{-1,-1} \ln _2^3-\frac{1}{2} s_{-1,1} \ln _2^3-\frac{1}{6} \bar{s}_{-1,-1} \ln _2^3+\frac{1}{6} \bar{s}_{-1,1} \ln _2^3+\frac{1}{2} s_{-3} \ln _2^2-\frac{1}{2} s_3 \ln _2^2-\frac{3}{4} s_{-1} \zeta _2 \ln _2^2+\frac{7}{16} \zeta _3 \ln _2^2+\frac{1}{2} s_{-2,-1} \ln _2^2-\frac{1}{2} s_{-2,1} \ln _2^2+\frac{1}{2} s_{-1,-2} \ln _2^2-\frac{1}{2} s_{-1,2} \ln _2^2-\frac{1}{2} s_{-1,1,-1} \ln _2^2+\frac{1}{2} s_{-1,1,1} \ln _2^2+\frac{1}{2} \bar{s}_{-1,1,-1} \ln _2^2+\frac{1}{2} \bar{s}_{-1,1,1} \ln _2^2-\frac{31}{20} \zeta _2^2 \ln _2+\frac{1}{2} s_{-2} \zeta _2 \ln _2+s_2 \zeta _2 \ln _2-\frac{1}{2} \zeta _2 \bar{s}_{-2} \ln _2+\frac{1}{2} \zeta _2 \bar{s}_2 \ln _2-\frac{1}{2} \zeta _2 s_{-1,-1} \ln _2+\frac{1}{2} \zeta _2 \bar{s}_{-1,-1} \ln _2-\frac{1}{2} \zeta _2 \bar{s}_{-1,1} \ln _2-\frac{57}{40} s_{-1} \zeta _2^2+4 \text{Li}_5\left(\frac{1}{2}\right)+3 \text{Li}_4\left(\frac{1}{2}\right) s_{-1}-\frac{1}{2} s_{-3} \zeta _2-\frac{s_3 \zeta _2}{2}+\frac{1}{8} s_{-2} \zeta _3+\frac{s_2 \zeta _3}{4}-\frac{7 \zeta _2 \zeta _3}{16}+\frac{13 \zeta _5}{8}+\frac{7}{8} \zeta _3 \bar{s}_{-2}-\frac{7}{4} \zeta _2^2 \bar{s}_{-1}+3 \text{Li}_4\left(\frac{1}{2}\right) \bar{s}_{-1}+\frac{1}{8} \zeta _3 \bar{s}_2+\frac{1}{2} \zeta _2 s_{-2,-1}+\frac{1}{2} \zeta _2 s_{-2,1}+\frac{3}{2} \zeta _2 s_{-1,-2}-\frac{25}{8} \zeta _3 s_{-1,-1}-\zeta _3 s_{-1,1}+\frac{1}{2} \zeta _2 s_{-1,2}+\zeta _2 s_{2,1}+s_{4,1}+\zeta _2 \bar{s}_{-1,-2}-\frac{23}{8} \zeta _3 \bar{s}_{-1,-1}-\frac{9}{8} \zeta _3 \bar{s}_{-1,1}+\zeta _2 \bar{s}_{2,1}-s_{-3,-1,1}-s_{-2,-2,1}-2 s_{-1,-3,1}-2 \zeta _2 s_{-1,-1,1}-\frac{1}{2} \zeta _2 s_{-1,1,-1}-\frac{1}{2} \zeta _2 s_{-1,1,1}-s_{2,2,1}-s_{3,1,1}-\bar{s}_{-1,-3,1}-2 \zeta _2 \bar{s}_{-1,-1,1}-\frac{3}{2} \zeta _2 \bar{s}_{-1,1,-1}-\frac{1}{2} \zeta _2 \bar{s}_{-1,1,1}-\bar{s}_{2,2,1}-\bar{s}_{3,1,1}+s_{-2,-1,1,1}+s_{-2,1,-1,1}+2 s_{-1,-2,1,1}+2 s_{-1,-1,2,1}+s_{-1,1,-2,1}+s_{-1,2,-1,1}+s_{2,1,1,1}+\bar{s}_{-2,-1,1,1}+3 \bar{s}_{-1,-2,1,1}+2 \bar{s}_{-1,-1,2,1}+\bar{s}_{-1,1,-2,1}+2 \bar{s}_{2,1,1,1}-2 s_{-1,-1,1,1,1}-s_{-1,1,-1,1,1}-s_{-1,1,1,-1,1}-4 \bar{s}_{-1,-1,1,1,1}-2 \bar{s}_{-1,1,-1,1,1}  \end{dmath}
   \begin{dmath}[style={\small}]     s_{-1,1} \bar{s}_{1,-2}  =   \frac{\ln _2^5}{30}+\frac{1}{12} s_{-1} \ln _2^4-\frac{1}{12} \bar{s}_{-1} \ln _2^4+\frac{1}{12} \bar{s}_1 \ln _2^4-\frac{2}{3} \zeta _2 \ln _2^3+\frac{1}{2} s_{-3} \ln _2^2-\frac{1}{2} s_3 \ln _2^2-\frac{1}{2} s_1 \zeta _2 \ln _2^2+\frac{15}{8} \zeta _3 \ln _2^2-\frac{1}{2} \zeta _2 \bar{s}_{-1} \ln _2^2+\frac{1}{2} \zeta _2 \bar{s}_1 \ln _2^2-\frac{1}{2} s_{1,-2} \ln _2^2+\frac{1}{2} s_{1,2} \ln _2^2+\frac{1}{2} \bar{s}_{1,-2} \ln _2^2+\frac{1}{2} \bar{s}_{1,2} \ln _2^2-\frac{29}{20} \zeta _2^2 \ln _2+\frac{3}{2} s_{-2} \zeta _2 \ln _2-\frac{3}{2} \zeta _2 \bar{s}_{-2} \ln _2+\frac{3}{2} \zeta _2 \bar{s}_2 \ln _2+\frac{3}{2} \zeta _2 s_{-1,-1} \ln _2-\frac{3}{2} \zeta _2 s_{-1,1} \ln _2-\frac{3}{2} \zeta _2 s_{1,-1} \ln _2-\frac{3}{2} \zeta _2 \bar{s}_{-1,-1} \ln _2+\frac{3}{2} \zeta _2 \bar{s}_{-1,1} \ln _2+\frac{3}{2} \zeta _2 \bar{s}_{1,-1} \ln _2-\frac{31}{40} s_{-1} \zeta _2^2-4 \text{Li}_5\left(\frac{1}{2}\right)+2 \text{Li}_4\left(\frac{1}{2}\right) s_{-1}-\frac{1}{2} s_{-3} \zeta _2-\frac{s_3 \zeta _2}{2}-\frac{5}{8} s_{-2} \zeta _3-\frac{13 \zeta _2 \zeta _3}{8}+\frac{119 \zeta _5}{16}-\frac{1}{2} \zeta _2 \bar{s}_{-3}+\frac{1}{4} \zeta _3 \bar{s}_{-2}+\frac{53}{40} \zeta _2^2 \bar{s}_{-1}-2 \text{Li}_4\left(\frac{1}{2}\right) \bar{s}_{-1}-\frac{11}{40} \zeta _2^2 \bar{s}_1+2 \text{Li}_4\left(\frac{1}{2}\right) \bar{s}_1-\frac{7}{4} \zeta _3 \bar{s}_2+\frac{1}{2} \zeta _2 \bar{s}_3+\frac{3}{2} \zeta _2 s_{-2,-1}+\frac{1}{2} \zeta _2 s_{-2,1}+\zeta _2 s_{-1,-2}-\frac{7}{4} \zeta _3 s_{-1,-1}+\frac{1}{2} \zeta _3 s_{-1,1}+\frac{1}{2} \zeta _2 s_{1,-2}+\frac{5}{8} \zeta _3 s_{1,-1}+\frac{1}{2} \zeta _2 s_{1,2}+s_{4,1}-\bar{s}_{-3,-2}-\frac{3}{2} \zeta _2 \bar{s}_{-2,-1}+\frac{1}{2} \zeta _2 \bar{s}_{-2,1}-\frac{1}{2} \zeta _2 \bar{s}_{-1,-2}+\frac{7}{4} \zeta _3 \bar{s}_{-1,-1}-\frac{1}{4} \zeta _3 \bar{s}_{-1,1}+\frac{1}{2} \zeta _2 \bar{s}_{-1,2}-\frac{1}{8} \zeta _3 \bar{s}_{1,-1}-\zeta _2 \bar{s}_{1,2}-s_{-3,-1,1}-s_{-2,-2,1}-s_{-1,-3,1}-\frac{3}{2} \zeta _2 s_{-1,1,-1}-\frac{1}{2} \zeta _2 s_{-1,1,1}-\frac{3}{2} \zeta _2 s_{1,-1,-1}-\frac{1}{2} \zeta _2 s_{1,-1,1}-s_{1,3,1}+2 \bar{s}_{-2,1,-2}+\frac{3}{2} \zeta _2 \bar{s}_{-1,1,-1}-\frac{1}{2} \zeta _2 \bar{s}_{-1,1,1}+\bar{s}_{-1,2,-2}+\bar{s}_{1,-2,-2}+\frac{3}{2} \zeta _2 \bar{s}_{1,-1,-1}-\frac{1}{2} \zeta _2 \bar{s}_{1,-1,1}+s_{-1,1,-2,1}+s_{1,-2,-1,1}+s_{1,-1,-2,1}-2 \bar{s}_{-1,1,1,-2}-\bar{s}_{1,-1,1,-2}  \end{dmath}

   \begin{dmath}[style={\small}]     s_{-1,1} \bar{s}_{1,-1,-1}  =   \frac{11 \ln _2^5}{60}+\frac{1}{2} s_{-1} \ln _2^4-\frac{7}{24} s_1 \ln _2^4+\frac{1}{6} \bar{s}_{-1} \ln _2^4-\frac{1}{6} \bar{s}_1 \ln _2^4+\frac{5}{6} \zeta _2 \ln _2^3+s_{-1,1} \ln _2^3+\frac{3}{2} s_{-1} \zeta _2 \ln _2^2+\frac{9}{4} s_1 \zeta _2 \ln _2^2+2 \zeta _3 \ln _2^2+\bar{s}_{-3} \ln _2^2-\frac{5}{4} \zeta _2 \bar{s}_{-1} \ln _2^2+\frac{1}{2} \zeta _2 \bar{s}_1 \ln _2^2-\bar{s}_3 \ln _2^2-\frac{3}{2} s_{-2,-1} \ln _2^2-\frac{1}{2} s_{-2,1} \ln _2^2-\frac{1}{2} s_{-1,-2} \ln _2^2+\frac{1}{2} s_{-1,2} \ln _2^2+\frac{1}{2} \bar{s}_{-2,-1} \ln _2^2-\frac{1}{2} \bar{s}_{-2,1} \ln _2^2+\bar{s}_{-1,-2} \ln _2^2-\bar{s}_{-1,2} \ln _2^2-\frac{1}{2} \bar{s}_{1,-2} \ln _2^2+\frac{1}{2} \bar{s}_{1,2} \ln _2^2+\frac{3}{2} s_{-1,1,-1} \ln _2^2+\frac{1}{2} s_{-1,1,1} \ln _2^2+\frac{3}{2} s_{1,-1,-1} \ln _2^2+\frac{1}{2} s_{1,-1,1} \ln _2^2-\frac{1}{2} \bar{s}_{-1,1,-1} \ln _2^2+\frac{1}{2} \bar{s}_{-1,1,1} \ln _2^2+\frac{1}{2} \bar{s}_{1,-1,-1} \ln _2^2+\frac{1}{2} \bar{s}_{1,-1,1} \ln _2^2-\frac{163}{40} \zeta _2^2 \ln _2+3 \text{Li}_4\left(\frac{1}{2}\right) \ln _2-\frac{7}{2} s_{-2} \zeta _2 \ln _2-\frac{1}{4} s_{-1} \zeta _3 \ln _2-\frac{3}{4} s_1 \zeta _3 \ln _2-\frac{1}{2} \zeta _2 \bar{s}_{-2} \ln _2+\frac{3}{8} \zeta _3 \bar{s}_{-1} \ln _2-\frac{5}{4} \zeta _3 \bar{s}_1 \ln _2+\frac{1}{2} \zeta _2 \bar{s}_2 \ln _2+\frac{1}{2} \zeta _2 s_{-1,-1} \ln _2+\frac{3}{2} \zeta _2 s_{-1,1} \ln _2+\frac{7}{2} \zeta _2 s_{1,-1} \ln _2+s_{3,-1} \ln _2+s_{3,1} \ln _2-\bar{s}_{-2,-2} \ln _2+\bar{s}_{-2,2} \ln _2-\frac{1}{2} \zeta _2 \bar{s}_{-1,-1} \ln _2+\frac{1}{2} \zeta _2 \bar{s}_{-1,1} \ln _2+\bar{s}_{1,-3} \ln _2-\bar{s}_{1,3} \ln _2-2 s_{-2,-1,-1} \ln _2-2 s_{-2,-1,1} \ln _2-s_{-1,-2,-1} \ln _2-s_{-1,-2,1} \ln _2-s_{1,2,-1} \ln _2-s_{1,2,1} \ln _2-\bar{s}_{-2,-1,-1} \ln _2+\bar{s}_{-2,-1,1} \ln _2+\bar{s}_{-1,1,-2} \ln _2-\bar{s}_{-1,1,2} \ln _2+2 \bar{s}_{1,-1,-2} \ln _2-2 \bar{s}_{1,-1,2} \ln _2-\bar{s}_{1,2,-1} \ln _2+\bar{s}_{1,2,1} \ln _2+s_{-1,1,-1,-1} \ln _2+s_{-1,1,-1,1} \ln _2+2 s_{1,-1,-1,-1} \ln _2+2 s_{1,-1,-1,1} \ln _2+\bar{s}_{-1,1,-1,-1} \ln _2-\bar{s}_{-1,1,-1,1} \ln _2+2 \bar{s}_{1,-1,-1,-1} \ln _2-2 \bar{s}_{1,-1,-1,1} \ln _2-\frac{1}{5} s_{-1} \zeta _2^2-\frac{41}{40} s_1 \zeta _2^2-9 \text{Li}_5\left(\frac{1}{2}\right)+\text{Li}_4\left(\frac{1}{2}\right) s_{-1}+4 \text{Li}_4\left(\frac{1}{2}\right) s_1+s_3 \zeta _2+\frac{9}{4} s_{-2} \zeta _3-\frac{7 \zeta _2 \zeta _3}{4}+\frac{193 \zeta _5}{16}+\zeta _3 \bar{s}_{-2}-\frac{27}{40} \zeta _2^2 \bar{s}_{-1}+\text{Li}_4\left(\frac{1}{2}\right) \bar{s}_{-1}+\frac{1}{5} \zeta _2^2 \bar{s}_1-\text{Li}_4\left(\frac{1}{2}\right) \bar{s}_1-s_{-4,1}-\frac{3}{2} \zeta _2 s_{-2,-1}-\frac{1}{2} \zeta _2 s_{-2,1}-\frac{1}{2} \zeta _2 s_{-1,-2}-\frac{5}{8} \zeta _3 s_{-1,1}-\frac{1}{2} \zeta _2 s_{-1,2}-\frac{9}{4} \zeta _3 s_{1,-1}-\zeta _2 s_{1,2}-\frac{1}{2} \zeta _2 \bar{s}_{-2,1}-\zeta _3 \bar{s}_{-1,1}-\frac{1}{2} \zeta _2 \bar{s}_{1,-2}-\frac{3}{8} \zeta _3 \bar{s}_{1,-1}+2 s_{-2,2,1}+\frac{1}{2} \zeta _2 s_{-1,1,-1}+\frac{1}{2} \zeta _2 s_{-1,1,1}+s_{-1,3,1}+s_{1,-3,1}+\frac{3}{2} \zeta _2 s_{1,-1,-1}+\frac{1}{2} \zeta _2 s_{1,-1,1}+2 s_{3,-1,1}-\bar{s}_{-3,-1,-1}-\bar{s}_{-2,-2,-1}+\frac{1}{2} \zeta _2 \bar{s}_{-1,1,1}-\frac{1}{2} \zeta _2 \bar{s}_{1,-1,-1}+\frac{1}{2} \zeta _2 \bar{s}_{1,-1,1}-\bar{s}_{1,3,-1}-3 s_{-2,-1,-1,1}-s_{-1,-2,-1,1}-s_{-1,1,2,1}-2 s_{1,-1,2,1}-2 s_{1,2,-1,1}+\bar{s}_{-2,-1,1,-1}+2 \bar{s}_{-2,1,-1,-1}+\bar{s}_{-1,1,-2,-1}+\bar{s}_{-1,2,-1,-1}+\bar{s}_{1,-2,-1,-1}+2 \bar{s}_{1,-1,-2,-1}+\bar{s}_{1,2,1,-1}+s_{-1,1,-1,-1,1}+3 s_{1,-1,-1,-1,1}-\bar{s}_{-1,1,-1,1,-1}-2 \bar{s}_{-1,1,1,-1,-1}-2 \bar{s}_{1,-1,-1,1,-1}-\bar{s}_{1,-1,1,-1,-1}  \end{dmath}
   \begin{dmath}[style={\small}]     s_{-1,1} \bar{s}_{1,-1,1}  =   \frac{17 \ln _2^5}{120}+\frac{1}{2} s_{-1} \ln _2^4-\frac{1}{4} s_1 \ln _2^4+\frac{1}{24} \bar{s}_{-1} \ln _2^4-\frac{5}{12} \zeta _2 \ln _2^3+\frac{2}{3} s_{-1,1} \ln _2^3+\frac{1}{2} s_{-3} \ln _2^2-\frac{1}{2} s_3 \ln _2^2-\frac{1}{2} s_{-1} \zeta _2 \ln _2^2+\frac{21}{16} \zeta _3 \ln _2^2+\frac{1}{2} \bar{s}_{-3} \ln _2^2-\frac{5}{4} \zeta _2 \bar{s}_{-1} \ln _2^2-\frac{1}{2} \zeta _2 \bar{s}_1 \ln _2^2-\frac{1}{2} \bar{s}_3 \ln _2^2-s_{-2,1} \ln _2^2-\frac{1}{2} s_{1,-2} \ln _2^2+\frac{1}{2} s_{1,2} \ln _2^2+\frac{1}{2} \bar{s}_{-2,-1} \ln _2^2-\frac{1}{2} \bar{s}_{-2,1} \ln _2^2+\frac{1}{2} \bar{s}_{-1,-2} \ln _2^2-\frac{1}{2} \bar{s}_{-1,2} \ln _2^2-\frac{1}{2} \bar{s}_{1,-2} \ln _2^2+\frac{1}{2} \bar{s}_{1,2} \ln _2^2+\frac{1}{2} s_{-1,1,-1} \ln _2^2+\frac{1}{2} s_{-1,1,1} \ln _2^2+s_{1,-1,1} \ln _2^2-\frac{1}{2} \bar{s}_{-1,1,-1} \ln _2^2+\frac{1}{2} \bar{s}_{-1,1,1} \ln _2^2+\bar{s}_{1,-1,1} \ln _2^2-\frac{41}{10} \zeta _2^2 \ln _2+\frac{3}{2} s_{-2} \zeta _2 \ln _2+\zeta _2 \bar{s}_{-2} \ln _2+\frac{1}{2} \zeta _2 \bar{s}_2 \ln _2+\frac{1}{2} \zeta _2 s_{-1,-1} \ln _2-\frac{3}{2} \zeta _2 s_{-1,1} \ln _2-\frac{3}{2} \zeta _2 s_{1,-1} \ln _2-\frac{1}{2} \zeta _2 \bar{s}_{-1,-1} \ln _2-\zeta _2 \bar{s}_{-1,1} \ln _2-\frac{3}{2} \zeta _2 \bar{s}_{1,-1} \ln _2-\frac{17}{20} s_{-1} \zeta _2^2-\frac{3}{5} s_1 \zeta _2^2-17 \text{Li}_5\left(\frac{1}{2}\right)+3 \text{Li}_4\left(\frac{1}{2}\right) s_{-1}+3 \text{Li}_4\left(\frac{1}{2}\right) s_1-\frac{1}{2} s_{-3} \zeta _2-\frac{s_3 \zeta _2}{2}+\frac{3}{8} s_{-2} \zeta _3-\frac{21 \zeta _2 \zeta _3}{16}+\frac{579 \zeta _5}{32}-\frac{1}{2} \zeta _2 \bar{s}_{-3}+\frac{53}{40} \zeta _2^2 \bar{s}_{-1}-2 \text{Li}_4\left(\frac{1}{2}\right) \bar{s}_{-1}-\frac{7}{20} \zeta _2^2 \bar{s}_1+3 \text{Li}_4\left(\frac{1}{2}\right) \bar{s}_1-\frac{13}{8} \zeta _3 \bar{s}_2+\frac{1}{2} \zeta _2 \bar{s}_3+2 \zeta _2 s_{-2,-1}+\zeta _2 s_{-2,1}+\zeta _2 s_{-1,-2}-\frac{13}{8} \zeta _3 s_{-1,-1}-\frac{1}{8} \zeta _3 s_{-1,1}+\frac{1}{2} \zeta _2 s_{1,-2}-\frac{3}{8} \zeta _3 s_{1,-1}+\frac{1}{2} \zeta _2 s_{1,2}+s_{4,1}+\frac{1}{2} \zeta _2 \bar{s}_{-2,-1}+\frac{1}{2} \zeta _2 \bar{s}_{-2,1}-\frac{1}{2} \zeta _2 \bar{s}_{-1,-2}+\frac{13}{8} \zeta _3 \bar{s}_{-1,-1}+\frac{1}{2} \zeta _2 \bar{s}_{-1,2}+\frac{1}{2} \zeta _2 \bar{s}_{1,-2}-\frac{3}{8} \zeta _3 \bar{s}_{1,-1}+\frac{1}{2} \zeta _2 \bar{s}_{1,2}-s_{-3,-1,1}-2 s_{-2,-2,1}-s_{-1,-3,1}-\frac{3}{2} \zeta _2 s_{-1,1,-1}-\frac{1}{2} \zeta _2 s_{-1,1,1}-2 \zeta _2 s_{1,-1,-1}-\zeta _2 s_{1,-1,1}-s_{1,3,1}-s_{3,1,1}-\bar{s}_{-3,-1,1}-\bar{s}_{-2,-2,1}-\frac{1}{2} \zeta _2 \bar{s}_{-1,1,-1}-\frac{1}{2} \zeta _2 \bar{s}_{-1,1,1}-2 \zeta _2 \bar{s}_{1,-1,-1}-\zeta _2 \bar{s}_{1,-1,1}-\bar{s}_{1,3,1}+2 s_{-2,-1,1,1}+s_{-2,1,-1,1}+s_{-1,-2,1,1}+s_{-1,1,-2,1}+s_{1,-2,-1,1}+2 s_{1,-1,-2,1}+s_{1,2,1,1}+\bar{s}_{-2,-1,1,1}+2 \bar{s}_{-2,1,-1,1}+\bar{s}_{-1,1,-2,1}+\bar{s}_{-1,2,-1,1}+\bar{s}_{1,-2,-1,1}+2 \bar{s}_{1,-1,-2,1}+\bar{s}_{1,2,1,1}-s_{-1,1,-1,1,1}-2 s_{1,-1,-1,1,1}-s_{1,-1,1,-1,1}-\bar{s}_{-1,1,-1,1,1}-2 \bar{s}_{-1,1,1,-1,1}-2 \bar{s}_{1,-1,-1,1,1}-\bar{s}_{1,-1,1,-1,1}  \end{dmath}
   \begin{dmath}[style={\small}]     s_{-1,1} \bar{s}_{1,1,-1}  =   -\frac{\ln _2^5}{20}-\frac{1}{6} s_1 \ln _2^4+\frac{1}{8} \bar{s}_{-1} \ln _2^4-\frac{1}{6} \bar{s}_1 \ln _2^4+\frac{2}{3} s_{-2} \ln _2^3-\frac{2}{3} s_2 \ln _2^3-\frac{4}{3} \zeta _2 \ln _2^3-\frac{1}{3} s_{-1,1} \ln _2^3-\frac{2}{3} s_{1,-1} \ln _2^3+\frac{2}{3} s_{1,1} \ln _2^3-\frac{1}{3} \bar{s}_{1,-1} \ln _2^3+\frac{1}{3} \bar{s}_{1,1} \ln _2^3-\frac{3}{2} s_{-1} \zeta _2 \ln _2^2-\frac{3}{2} s_1 \zeta _2 \ln _2^2+\frac{13}{8} \zeta _3 \ln _2^2-\frac{1}{2} \bar{s}_{-3} \ln _2^2+\zeta _2 \bar{s}_1 \ln _2^2+\frac{1}{2} \bar{s}_3 \ln _2^2+s_{-2,1} \ln _2^2-\frac{1}{2} s_{-1,-2} \ln _2^2+\frac{1}{2} s_{-1,2} \ln _2^2-\frac{1}{2} \bar{s}_{-1,-2} \ln _2^2+\frac{1}{2} \bar{s}_{-1,2} \ln _2^2-s_{-1,1,1} \ln _2^2-s_{1,-1,1} \ln _2^2+\bar{s}_{1,1,-1} \ln _2^2+\frac{7}{8} \zeta _2^2 \ln _2-3 \text{Li}_4\left(\frac{1}{2}\right) \ln _2+\frac{3}{2} s_{-2} \zeta _2 \ln _2+2 s_2 \zeta _2 \ln _2-\frac{1}{2} s_{-1} \zeta _3 \ln _2-\frac{1}{4} s_1 \zeta _3 \ln _2-\frac{3}{2} \zeta _2 \bar{s}_{-2} \ln _2+\frac{21}{8} \zeta _3 \bar{s}_{-1} \ln _2-s_{-3,-1} \ln _2-s_{-3,1} \ln _2-2 \zeta _2 s_{-1,1} \ln _2-\frac{3}{2} \zeta _2 s_{1,-1} \ln _2-2 \zeta _2 s_{1,1} \ln _2+\bar{s}_{-2,-2} \ln _2-\bar{s}_{-2,2} \ln _2+\frac{3}{2} \zeta _2 \bar{s}_{-1,1} \ln _2-\bar{s}_{1,-3} \ln _2+2 \zeta _2 \bar{s}_{1,-1} \ln _2-\frac{1}{2} \zeta _2 \bar{s}_{1,1} \ln _2+\bar{s}_{1,3} \ln _2+s_{-2,1,-1} \ln _2+s_{-2,1,1} \ln _2+s_{-1,2,-1} \ln _2+s_{-1,2,1} \ln _2+s_{1,-2,-1} \ln _2+s_{1,-2,1} \ln _2+s_{2,-1,-1} \ln _2+s_{2,-1,1} \ln _2-\bar{s}_{-2,1,-1} \ln _2+\bar{s}_{-2,1,1} \ln _2-\bar{s}_{-1,1,-2} \ln _2+\bar{s}_{-1,1,2} \ln _2-\bar{s}_{1,-2,-1} \ln _2+\bar{s}_{1,-2,1} \ln _2-\bar{s}_{1,-1,-2} \ln _2+\bar{s}_{1,-1,2} \ln _2+\bar{s}_{1,1,-2} \ln _2-\bar{s}_{1,1,2} \ln _2-s_{-1,1,1,-1} \ln _2-s_{-1,1,1,1} \ln _2-s_{1,-1,1,-1} \ln _2-s_{1,-1,1,1} \ln _2-s_{1,1,-1,-1} \ln _2-s_{1,1,-1,1} \ln _2+\bar{s}_{-1,1,1,-1} \ln _2-\bar{s}_{-1,1,1,1} \ln _2+\bar{s}_{1,-1,1,-1} \ln _2-\bar{s}_{1,-1,1,1} \ln _2+\bar{s}_{1,1,-1,-1} \ln _2-\bar{s}_{1,1,-1,1} \ln _2+\frac{13}{40} s_{-1} \zeta _2^2+\frac{17}{20} s_1 \zeta _2^2+\text{Li}_5\left(\frac{1}{2}\right)-\text{Li}_4\left(\frac{1}{2}\right) s_{-1}-3 \text{Li}_4\left(\frac{1}{2}\right) s_1-s_{-3} \zeta _2-\frac{3}{4} s_{-2} \zeta _3-\frac{3 s_2 \zeta _3}{2}-\frac{7 \zeta _2 \zeta _3}{4}+\frac{83 \zeta _5}{32}+\frac{1}{4} \zeta _3 \bar{s}_{-2}+\frac{13}{40} \zeta _2^2 \bar{s}_1-\text{Li}_4\left(\frac{1}{2}\right) \bar{s}_1-\frac{1}{8} \zeta _3 \bar{s}_2+\frac{1}{2} \zeta _2 s_{-2,-1}+\frac{1}{2} \zeta _2 s_{-2,1}+\frac{1}{2} \zeta _2 s_{-1,-2}-\frac{1}{8} \zeta _3 s_{-1,-1}+\frac{3}{4} \zeta _3 s_{-1,1}+\frac{1}{2} \zeta _2 s_{-1,2}+\zeta _2 s_{1,-2}+\frac{3}{4} \zeta _3 s_{1,-1}+\frac{3}{2} \zeta _3 s_{1,1}+\zeta _2 s_{2,-1}+s_{4,1}-\frac{1}{2} \zeta _2 \bar{s}_{-2,-1}+\frac{1}{8} \zeta _3 \bar{s}_{-1,-1}-\frac{1}{4} \zeta _3 \bar{s}_{-1,1}-\frac{1}{4} \zeta _3 \bar{s}_{1,-1}+\frac{3}{4} \zeta _3 \bar{s}_{1,1}-\frac{1}{2} \zeta _2 \bar{s}_{1,2}-2 s_{-3,-1,1}-s_{-2,-2,1}-s_{-1,-3,1}-\frac{1}{2} \zeta _2 s_{-1,1,-1}-\frac{1}{2} \zeta _2 s_{-1,1,1}-\frac{1}{2} \zeta _2 s_{1,-1,-1}-\frac{1}{2} \zeta _2 s_{1,-1,1}-\zeta _2 s_{1,1,-1}-s_{1,3,1}-s_{2,2,1}-\bar{s}_{-3,1,-1}-\bar{s}_{-2,2,-1}+\frac{1}{2} \zeta _2 \bar{s}_{-1,1,-1}-\bar{s}_{1,-3,-1}+\frac{1}{2} \zeta _2 \bar{s}_{1,-1,-1}-\frac{1}{2} \zeta _2 \bar{s}_{1,1,-1}+s_{-2,1,-1,1}+s_{-1,1,-2,1}+s_{-1,2,-1,1}+2 s_{1,-2,-1,1}+s_{1,-1,-2,1}+s_{1,1,2,1}+2 s_{2,-1,-1,1}+3 \bar{s}_{-2,1,1,-1}+\bar{s}_{-1,1,2,-1}+\bar{s}_{-1,2,1,-1}+2 \bar{s}_{1,-2,1,-1}+\bar{s}_{1,-1,2,-1}+\bar{s}_{1,1,-2,-1}-s_{-1,1,1,-1,1}-s_{1,-1,1,-1,1}-2 s_{1,1,-1,-1,1}-3 \bar{s}_{-1,1,1,1,-1}-2 \bar{s}_{1,-1,1,1,-1}-\bar{s}_{1,1,-1,1,-1}  \end{dmath}
   \begin{dmath}[style={\small}]     s_{-1,1} \bar{s}_{1,1,1}  =   \frac{\ln _2^5}{120}-\frac{1}{8} s_1 \ln _2^4+\frac{1}{3} s_{-2} \ln _2^3-\frac{1}{3} s_2 \ln _2^3-\frac{1}{12} \zeta _2 \ln _2^3-\frac{1}{3} s_{1,-1} \ln _2^3+\frac{1}{3} s_{1,1} \ln _2^3-\frac{1}{3} \bar{s}_{1,-1} \ln _2^3+\frac{1}{3} \bar{s}_{1,1} \ln _2^3+\frac{1}{2} s_{-3} \ln _2^2-\frac{1}{2} s_3 \ln _2^2+\frac{3}{4} s_1 \zeta _2 \ln _2^2+\frac{7}{16} \zeta _3 \ln _2^2-\frac{1}{2} s_{1,-2} \ln _2^2+\frac{1}{2} s_{1,2} \ln _2^2-\frac{1}{2} s_{2,-1} \ln _2^2+\frac{1}{2} s_{2,1} \ln _2^2+\frac{1}{2} s_{1,1,-1} \ln _2^2-\frac{1}{2} s_{1,1,1} \ln _2^2+\frac{1}{2} \bar{s}_{1,1,-1} \ln _2^2+\frac{1}{2} \bar{s}_{1,1,1} \ln _2^2+\frac{43}{40} \zeta _2^2 \ln _2-s_{-2} \zeta _2 \ln _2-\frac{1}{2} s_2 \zeta _2 \ln _2+\zeta _2 s_{1,-1} \ln _2+\frac{1}{2} \zeta _2 s_{1,1} \ln _2+\zeta _2 \bar{s}_{1,-1} \ln _2-\zeta _2 \bar{s}_{1,1} \ln _2+\frac{6}{5} s_{-1} \zeta _2^2+\frac{9}{40} s_1 \zeta _2^2+7 \text{Li}_5\left(\frac{1}{2}\right)-3 \text{Li}_4\left(\frac{1}{2}\right) s_{-1}+\frac{1}{2} s_{-3} \zeta _2+\frac{s_3 \zeta _2}{2}-\frac{1}{4} s_{-2} \zeta _3-\frac{s_2 \zeta _3}{8}-\frac{7 \zeta _2 \zeta _3}{16}-\frac{337 \zeta _5}{64}+\zeta _3 \bar{s}_{-2}-\frac{2}{5} \zeta _2^2 \bar{s}_{-1}-3 \text{Li}_4\left(\frac{1}{2}\right) \bar{s}_{-1}-\frac{6}{5} \zeta _2^2 \bar{s}_1+3 \text{Li}_4\left(\frac{1}{2}\right) \bar{s}_1-s_{-4,1}-\zeta _2 s_{-2,1}+2 \zeta _3 s_{-1,1}-\zeta _2 s_{-1,2}-\frac{1}{2} \zeta _2 s_{1,-2}+\frac{1}{4} \zeta _3 s_{1,-1}+\frac{1}{8} \zeta _3 s_{1,1}-\frac{1}{2} \zeta _2 s_{1,2}-\frac{1}{2} \zeta _2 s_{2,-1}-\frac{1}{2} \zeta _2 s_{2,1}+\zeta _2 \bar{s}_{-2,1}-\zeta _3 \bar{s}_{-1,1}+\zeta _2 \bar{s}_{1,-2}-\frac{11}{4} \zeta _3 \bar{s}_{1,-1}-\frac{1}{4} \zeta _3 \bar{s}_{1,1}+s_{-3,1,1}+s_{-2,2,1}+\zeta _2 s_{-1,1,1}+s_{-1,3,1}+s_{1,-3,1}+\zeta _2 s_{1,-1,1}+\frac{1}{2} \zeta _2 s_{1,1,-1}+\frac{1}{2} \zeta _2 s_{1,1,1}+s_{2,-2,1}+s_{3,-1,1}-\bar{s}_{-3,1,1}-\bar{s}_{-2,2,1}-\zeta _2 \bar{s}_{-1,1,1}-\bar{s}_{1,-3,1}-\zeta _2 \bar{s}_{1,-1,1}-\frac{3}{2} \zeta _2 \bar{s}_{1,1,-1}-\frac{1}{2} \zeta _2 \bar{s}_{1,1,1}-s_{-2,1,1,1}-s_{-1,1,2,1}-s_{-1,2,1,1}-s_{1,-2,1,1}-s_{1,-1,2,1}-s_{1,1,-2,1}-s_{1,2,-1,1}-s_{2,-1,1,1}-s_{2,1,-1,1}+3 \bar{s}_{-2,1,1,1}+\bar{s}_{-1,1,2,1}+\bar{s}_{-1,2,1,1}+2 \bar{s}_{1,-2,1,1}+\bar{s}_{1,-1,2,1}+\bar{s}_{1,1,-2,1}+s_{-1,1,1,1,1}+s_{1,-1,1,1,1}+s_{1,1,-1,1,1}+s_{1,1,1,-1,1}-3 \bar{s}_{-1,1,1,1,1}-2 \bar{s}_{1,-1,1,1,1}-\bar{s}_{1,1,-1,1,1}  \end{dmath}
   \begin{dmath}[style={\small}]     s_{-1,1} \bar{s}_{-1,-2}  =   \frac{1}{2} \zeta _2 \bar{s}_{-1,-2}-\frac{1}{4} \zeta _3 \bar{s}_{-1,-1}+\frac{13}{8} \zeta _3 \bar{s}_{-1,1}-\frac{3}{2} \zeta _2 \bar{s}_{-1,2}-\frac{3}{2} \zeta _2 \bar{s}_{2,-1}+\frac{1}{2} \zeta _2 \bar{s}_{2,1}+3 \zeta _2 \bar{s}_{-1,-1,-1}-\zeta _2 \bar{s}_{-1,-1,1}+\frac{5}{2} \zeta _2 \ln _2 \bar{s}_{-1,-1}-\zeta _2 \ln _2 \bar{s}_{-1,1}-\bar{s}_{3,-2}+\bar{s}_{-2,-1,-2}+2 \bar{s}_{-1,-2,-2}+\bar{s}_{2,1,-2}-2 \bar{s}_{-1,-1,1,-2}-\bar{s}_{-1,1,-1,-2}+\frac{1}{2} \ln _2^2 \bar{s}_{-1,-2}+\frac{1}{2} \ln _2^2 \bar{s}_{-1,2}-4 \text{Li}_4\left(\frac{1}{2}\right) \bar{s}_{-1}+\frac{1}{2} \zeta _2 \bar{s}_{-3}-\frac{13}{8} \zeta _3 \bar{s}_{-2}+\frac{1}{2} \zeta _2^2 \bar{s}_{-1}+\frac{1}{8} \zeta _3 \bar{s}_2-\frac{1}{2} \zeta _2 \bar{s}_3+\zeta _2 \ln _2^2 \bar{s}_{-1}+\zeta _2 \ln _2 \bar{s}_{-2}-\zeta _2 \ln _2 \bar{s}_2-\frac{1}{6} \ln _2^4 \bar{s}_{-1}-\frac{1}{2} \zeta _2 s_{-1,-2}-\frac{5}{4} \zeta _3 s_{-1,-1}+\frac{13}{8} \zeta _3 s_{-1,1}-\frac{3}{2} \zeta _2 s_{-1,2}-\frac{3}{2} \zeta _2 s_{2,-1}-\frac{1}{2} \zeta _2 s_{2,1}+3 \zeta _2 s_{-1,-1,-1}+\zeta _2 s_{-1,-1,1}+\frac{7}{2} \zeta _2 \ln _2 s_{-1,-1}-\zeta _2 \ln _2 s_{-1,1}-s_{-4,1}+2 s_{-1,3,1}+s_{2,-2,1}+s_{3,-1,1}-s_{-1,-2,-1,1}-2 s_{-1,-1,-2,1}+\frac{1}{2} \ln _2^2 s_{-1,-2}-\frac{1}{2} \ln _2^2 s_{-1,2}-\frac{715 \zeta _5}{64}-\frac{1}{6} \zeta _2 \ln _2^3-\frac{3}{4} \zeta _3 \ln _2^2+\frac{29}{8} \zeta _2^2 \ln _2-2 \text{Li}_4\left(\frac{1}{2}\right) s_{-1}+10 \text{Li}_5\left(\frac{1}{2}\right)+\frac{31}{40} \zeta _2^2 s_{-1}+\frac{1}{2} \zeta _2 s_{-3}+\frac{\zeta _2 s_3}{2}+\frac{5 \zeta _3 s_2}{8}+\frac{1}{2} \zeta _2 s_{-1} \ln _2^2-\frac{3}{2} \zeta _2 s_2 \ln _2-\frac{1}{12} s_{-1} \ln _2^4+\frac{1}{2} s_{-3} \ln _2^2-\frac{1}{2} s_3 \ln _2^2-\frac{\ln _2^5}{12}  \end{dmath}
   \begin{dmath}[style={\small}]     s_{-1,1} \bar{s}_{-1,2}  =   -\frac{1}{2} \zeta _2 \bar{s}_{-1,-2}-\frac{37}{8} \zeta _3 \bar{s}_{-1,-1}-\zeta _3 \bar{s}_{-1,1}-\frac{1}{2} \zeta _2 \bar{s}_{-1,2}-\zeta _2 \bar{s}_{2,1}+2 \zeta _2 \bar{s}_{-1,-1,1}+\zeta _2 \ln _2 \bar{s}_{-1,-1}+\frac{1}{2} \zeta _2 \ln _2 \bar{s}_{-1,1}-\bar{s}_{3,2}+\bar{s}_{-2,-1,2}+2 \bar{s}_{-1,-2,2}+\bar{s}_{2,1,2}-2 \bar{s}_{-1,-1,1,2}-\bar{s}_{-1,1,-1,2}+\frac{1}{2} \ln _2^2 \bar{s}_{-1,-2}+\frac{1}{2} \ln _2^2 \bar{s}_{-1,2}+4 \text{Li}_4\left(\frac{1}{2}\right) \bar{s}_{-1}+\zeta _3 \bar{s}_{-2}-\frac{41}{40} \zeta _2^2 \bar{s}_{-1}+\zeta _3 \bar{s}_2-\zeta _2 \ln _2^2 \bar{s}_{-1}-\frac{1}{2} \zeta _2 \ln _2 \bar{s}_{-2}+\frac{1}{2} \zeta _2 \ln _2 \bar{s}_2+\frac{1}{6} \ln _2^4 \bar{s}_{-1}+\frac{3}{2} \zeta _2 s_{-1,-2}-\frac{11}{8} \zeta _3 s_{-1,-1}-\zeta _3 s_{-1,1}+\frac{1}{2} \zeta _2 s_{-1,2}+\zeta _2 s_{2,1}-2 \zeta _2 s_{-1,-1,1}-\zeta _2 \ln _2 s_{-1,-1}+\frac{1}{2} \zeta _2 \ln _2 s_{-1,1}+s_{4,1}-s_{-3,-1,1}-2 s_{-1,-3,1}-s_{2,2,1}+2 s_{-1,-1,2,1}+s_{-1,2,-1,1}+\frac{1}{2} \ln _2^2 s_{-1,-2}-\frac{1}{2} \ln _2^2 s_{-1,2}-\frac{3 \zeta _2 \zeta _3}{2}+\frac{13 \zeta _5}{8}-\frac{1}{6} \zeta _2 \ln _2^3-\frac{3}{4} \zeta _3 \ln _2^2-\frac{1}{20} \zeta _2^2 \ln _2+2 \text{Li}_4\left(\frac{1}{2}\right) s_{-1}+4 \text{Li}_5\left(\frac{1}{2}\right)-\frac{9}{10} \zeta _2^2 s_{-1}-\frac{1}{2} \zeta _2 s_{-3}-\frac{\zeta _2 s_3}{2}-\frac{5 \zeta _3 s_2}{8}-\frac{1}{2} \zeta _2 s_{-1} \ln _2^2+\frac{3}{2} \zeta _2 s_2 \ln _2+\frac{1}{12} s_{-1} \ln _2^4+\frac{1}{2} s_{-3} \ln _2^2-\frac{1}{2} s_3 \ln _2^2-\frac{\ln _2^5}{30}  \end{dmath}
   \begin{dmath}[style={\small}]     s_{-1,1} \bar{s}_{-1,-1,-1}  =   \frac{37 \ln _2^5}{120}-\frac{5}{24} s_{-1} \ln _2^4-\frac{1}{24} \bar{s}_{-1} \ln _2^4-\frac{35}{12} \zeta _2 \ln _2^3+\frac{1}{6} \bar{s}_{-2} \ln _2^3-\frac{1}{6} \bar{s}_2 \ln _2^3-\frac{7}{6} s_{-1,-1} \ln _2^3-\frac{1}{6} s_{-1,1} \ln _2^3+\frac{1}{6} \bar{s}_{-1,-1} \ln _2^3-\frac{1}{6} \bar{s}_{-1,1} \ln _2^3-\frac{15}{4} s_{-1} \zeta _2 \ln _2^2+\frac{9}{4} \zeta _3 \ln _2^2-\bar{s}_{-3} \ln _2^2+\frac{1}{4} \zeta _2 \bar{s}_{-1} \ln _2^2+\bar{s}_3 \ln _2^2-\frac{1}{2} s_{-1,-2} \ln _2^2+\frac{1}{2} s_{-1,2} \ln _2^2+\frac{3}{2} s_{2,-1} \ln _2^2+\frac{1}{2} s_{2,1} \ln _2^2-\frac{3}{2} \bar{s}_{-1,-2} \ln _2^2+\frac{3}{2} \bar{s}_{-1,2} \ln _2^2+\frac{1}{2} \bar{s}_{2,-1} \ln _2^2-\frac{1}{2} \bar{s}_{2,1} \ln _2^2-3 s_{-1,-1,-1} \ln _2^2-s_{-1,-1,1} \ln _2^2+\bar{s}_{-1,-1,1} \ln _2^2+\frac{137}{40} \zeta _2^2 \ln _2+5 \text{Li}_4\left(\frac{1}{2}\right) \ln _2+\frac{7}{2} s_2 \zeta _2 \ln _2+s_{-1} \zeta _3 \ln _2+\frac{1}{2} \zeta _2 \bar{s}_{-2} \ln _2-\frac{7}{8} \zeta _3 \bar{s}_{-1} \ln _2-\frac{1}{2} \zeta _2 \bar{s}_2 \ln _2-s_{-3,-1} \ln _2-s_{-3,1} \ln _2-5 \zeta _2 s_{-1,-1} \ln _2-\frac{1}{2} \zeta _2 s_{-1,1} \ln _2+\bar{s}_{-1,-3} \ln _2+\frac{1}{2} \zeta _2 \bar{s}_{-1,-1} \ln _2-\frac{1}{2} \zeta _2 \bar{s}_{-1,1} \ln _2-\bar{s}_{-1,3} \ln _2-\bar{s}_{2,-2} \ln _2+\bar{s}_{2,2} \ln _2+2 s_{-1,2,-1} \ln _2+2 s_{-1,2,1} \ln _2+2 s_{2,-1,-1} \ln _2+2 s_{2,-1,1} \ln _2+3 \bar{s}_{-1,-1,-2} \ln _2-3 \bar{s}_{-1,-1,2} \ln _2-\bar{s}_{-1,2,-1} \ln _2+\bar{s}_{-1,2,1} \ln _2-\bar{s}_{2,-1,-1} \ln _2+\bar{s}_{2,-1,1} \ln _2-3 s_{-1,-1,-1,-1} \ln _2-3 s_{-1,-1,-1,1} \ln _2+3 \bar{s}_{-1,-1,-1,-1} \ln _2-3 \bar{s}_{-1,-1,-1,1} \ln _2+\frac{49}{40} s_{-1} \zeta _2^2+18 \text{Li}_5\left(\frac{1}{2}\right)-5 \text{Li}_4\left(\frac{1}{2}\right) s_{-1}-s_{-3} \zeta _2-\frac{9 s_2 \zeta _3}{4}-\frac{\zeta _2 \zeta _3}{4}-\frac{253 \zeta _5}{16}+\frac{1}{4} \zeta _3 \bar{s}_{-2}+\frac{13}{40} \zeta _2^2 \bar{s}_{-1}-\text{Li}_4\left(\frac{1}{2}\right) \bar{s}_{-1}+\frac{3}{4} \zeta _3 \bar{s}_2+\frac{1}{2} \zeta _2 s_{-1,-2}+\frac{21}{8} \zeta _3 s_{-1,-1}-\frac{1}{4} \zeta _3 s_{-1,1}+\frac{3}{2} \zeta _2 s_{-1,2}+\frac{3}{2} \zeta _2 s_{2,-1}+\frac{1}{2} \zeta _2 s_{2,1}+s_{4,1}-\frac{1}{2} \zeta _2 \bar{s}_{-1,-2}-\frac{9}{8} \zeta _3 \bar{s}_{-1,-1}-\frac{1}{4} \zeta _3 \bar{s}_{-1,1}-\frac{1}{2} \zeta _2 \bar{s}_{2,1}-2 s_{-3,-1,1}-2 s_{-1,-3,1}-2 \zeta _2 s_{-1,-1,-1}-\zeta _2 s_{-1,-1,1}-2 s_{2,2,1}-\frac{1}{2} \zeta _2 \bar{s}_{-1,-1,-1}+\zeta _2 \bar{s}_{-1,-1,1}-\bar{s}_{-1,3,-1}-\bar{s}_{2,-2,-1}-\bar{s}_{3,-1,-1}+3 s_{-1,-1,2,1}+3 s_{-1,2,-1,1}+3 s_{2,-1,-1,1}+\bar{s}_{-2,-1,-1,-1}+2 \bar{s}_{-1,-2,-1,-1}+3 \bar{s}_{-1,-1,-2,-1}+\bar{s}_{-1,2,1,-1}+\bar{s}_{2,-1,1,-1}+\bar{s}_{2,1,-1,-1}-4 s_{-1,-1,-1,-1,1}-3 \bar{s}_{-1,-1,-1,1,-1}-2 \bar{s}_{-1,-1,1,-1,-1}-\bar{s}_{-1,1,-1,-1,-1}  \end{dmath}
   \begin{dmath}[style={\small}]     s_{-1,1} \bar{s}_{-1,-1,1}  =   \frac{3 \ln _2^5}{20}-\frac{1}{4} s_{-1} \ln _2^4-\zeta _2 \ln _2^3+\frac{1}{6} \bar{s}_{-2} \ln _2^3-\frac{1}{6} \bar{s}_2 \ln _2^3-\frac{5}{6} s_{-1,-1} \ln _2^3-\frac{1}{6} s_{-1,1} \ln _2^3+\frac{1}{6} \bar{s}_{-1,-1} \ln _2^3-\frac{1}{6} \bar{s}_{-1,1} \ln _2^3+\frac{1}{2} s_{-3} \ln _2^2-\frac{1}{2} s_3 \ln _2^2+\frac{1}{2} s_{-1} \zeta _2 \ln _2^2-\frac{1}{16} \zeta _3 \ln _2^2-\frac{1}{2} \bar{s}_{-3} \ln _2^2-\frac{3}{4} \zeta _2 \bar{s}_{-1} \ln _2^2+\frac{1}{2} \bar{s}_3 \ln _2^2+\frac{1}{2} s_{-1,-2} \ln _2^2-\frac{1}{2} s_{-1,2} \ln _2^2+s_{2,1} \ln _2^2-\bar{s}_{-1,-2} \ln _2^2+\bar{s}_{-1,2} \ln _2^2+\frac{1}{2} \bar{s}_{2,-1} \ln _2^2-\frac{1}{2} \bar{s}_{2,1} \ln _2^2-\frac{1}{2} s_{-1,-1,-1} \ln _2^2-\frac{3}{2} s_{-1,-1,1} \ln _2^2-\frac{1}{2} \bar{s}_{-1,-1,-1} \ln _2^2+\frac{3}{2} \bar{s}_{-1,-1,1} \ln _2^2+\frac{67}{20} \zeta _2^2 \ln _2-\frac{3}{2} s_2 \zeta _2 \ln _2+\frac{1}{2} \zeta _2 \bar{s}_{-2} \ln _2+\zeta _2 \bar{s}_2 \ln _2+3 \zeta _2 s_{-1,-1} \ln _2-\frac{1}{2} \zeta _2 s_{-1,1} \ln _2-\frac{5}{2} \zeta _2 \bar{s}_{-1,-1} \ln _2-\frac{1}{2} \zeta _2 \bar{s}_{-1,1} \ln _2+\frac{29}{20} s_{-1} \zeta _2^2+10 \text{Li}_5\left(\frac{1}{2}\right)-6 \text{Li}_4\left(\frac{1}{2}\right) s_{-1}+\frac{1}{2} s_{-3} \zeta _2+\frac{s_3 \zeta _2}{2}-\frac{3 s_2 \zeta _3}{8}+\frac{\zeta _2 \zeta _3}{16}-\frac{347 \zeta _5}{32}+\frac{1}{2} \zeta _2 \bar{s}_{-3}-\frac{7}{4} \zeta _3 \bar{s}_{-2}-\frac{1}{10} \zeta _2^2 \bar{s}_{-1}+\frac{1}{8} \zeta _3 \bar{s}_2-\frac{1}{2} \zeta _2 \bar{s}_3-s_{-4,1}-\frac{1}{2} \zeta _2 s_{-1,-2}+\frac{5}{8} \zeta _3 s_{-1,-1}+\frac{7}{4} \zeta _3 s_{-1,1}-\frac{3}{2} \zeta _2 s_{-1,2}-2 \zeta _2 s_{2,-1}-\zeta _2 s_{2,1}+\zeta _2 \bar{s}_{-1,-2}-\frac{1}{2} \zeta _3 \bar{s}_{-1,-1}+\frac{7}{4} \zeta _3 \bar{s}_{-1,1}+\frac{1}{2} \zeta _2 \bar{s}_{2,-1}+\frac{1}{2} \zeta _2 \bar{s}_{2,1}+s_{-3,1,1}+\frac{7}{2} \zeta _2 s_{-1,-1,-1}+\frac{3}{2} \zeta _2 s_{-1,-1,1}+2 s_{-1,3,1}+2 s_{2,-2,1}+s_{3,-1,1}-\frac{5}{2} \zeta _2 \bar{s}_{-1,-1,-1}-\frac{3}{2} \zeta _2 \bar{s}_{-1,-1,1}-\bar{s}_{-1,3,1}-\bar{s}_{2,-2,1}-\bar{s}_{3,-1,1}-s_{-1,-2,-1,1}-3 s_{-1,-1,-2,1}-2 s_{-1,2,1,1}-2 s_{2,-1,1,1}-s_{2,1,-1,1}+\bar{s}_{-2,-1,-1,1}+2 \bar{s}_{-1,-2,-1,1}+3 \bar{s}_{-1,-1,-2,1}+\bar{s}_{-1,2,1,1}+\bar{s}_{2,-1,1,1}+\bar{s}_{2,1,-1,1}+3 s_{-1,-1,-1,1,1}+s_{-1,-1,1,-1,1}-3 \bar{s}_{-1,-1,-1,1,1}-2 \bar{s}_{-1,-1,1,-1,1}-\bar{s}_{-1,1,-1,-1,1}  \end{dmath}
   \begin{dmath}[style={\small}]     \bar{s}_{-3} s_{1,-1}  =   -\frac{3}{4} \zeta _3 \bar{s}_{1,-1}-\frac{1}{2} \zeta _2 \bar{s}_{1,2}+\bar{s}_{-2,-3}-\bar{s}_{1,-1,-3}-\ln _2 \bar{s}_{1,-3}-\ln _2 \bar{s}_{1,3}+2 \text{Li}_4\left(\frac{1}{2}\right) \bar{s}_1+\frac{3}{4} \zeta _3 \bar{s}_{-2}+\frac{3}{10} \zeta _2^2 \bar{s}_1-\frac{5}{8} \zeta _3 \bar{s}_2-\frac{1}{2} \zeta _2 \ln _2^2 \bar{s}_1+\zeta _2 \ln _2 \bar{s}_{-2}+\frac{7}{4} \zeta _3 \ln _2 \bar{s}_{-1}+\frac{1}{4} \zeta _3 \ln _2 \bar{s}_1-\frac{1}{2} \zeta _2 \ln _2 \bar{s}_2+\frac{1}{12} \ln _2^4 \bar{s}_1-\frac{1}{2} \ln _2^2 \bar{s}_{-3}-\frac{1}{2} \ln _2^2 \bar{s}_3-\frac{3}{4} \zeta _3 s_{1,-1}-\frac{1}{2} \zeta _2 s_{1,2}+s_{-4,-1}-s_{-3,1,-1}-s_{1,-3,-1}-\ln _2 s_{1,-3}+\ln _2 s_{1,3}-\frac{3 \zeta _2 \zeta _3}{2}+\frac{35 \zeta _5}{16}+\zeta _2 \ln _2^3-\frac{3}{4} \zeta _3 \ln _2^2+\frac{13}{4} \zeta _2^2 \ln _2-2 \text{Li}_4\left(\frac{1}{2}\right) s_1-4 \text{Li}_4\left(\frac{1}{2}\right) \ln _2+\frac{11}{10} \zeta _2^2 s_1+\frac{\zeta _2 s_3}{2}-\frac{5 \zeta _3 s_2}{8}+\frac{1}{2} \zeta _2 s_1 \ln _2^2-\zeta _2 s_{-2} \ln _2-\frac{1}{2} \zeta _2 s_2 \ln _2+\frac{7}{4} \zeta _3 s_{-1} \ln _2-\frac{7}{4} \zeta _3 s_1 \ln _2-\frac{1}{12} s_1 \ln _2^4-\frac{1}{2} s_{-3} \ln _2^2+\frac{1}{2} s_3 \ln _2^2+s_{-4} \ln _2-s_4 \ln _2-\frac{\ln _2^5}{6}  \end{dmath}
   \begin{dmath}[style={\small}]     \bar{s}_3 s_{1,-1}  =   -\frac{1}{2} \zeta _2 \bar{s}_{1,-2}-\frac{3}{4} \zeta _3 \bar{s}_{1,-1}+\bar{s}_{-2,3}-\bar{s}_{1,-1,3}-\ln _2 \bar{s}_{1,-3}-\ln _2 \bar{s}_{1,3}+2 \text{Li}_4\left(\frac{1}{2}\right) \bar{s}_{-1}+\frac{1}{8} \zeta _3 \bar{s}_{-2}-\frac{49}{40} \zeta _2^2 \bar{s}_{-1}-\frac{29}{40} \zeta _2^2 \bar{s}_1-\frac{1}{2} \zeta _2 \ln _2^2 \bar{s}_{-1}-\frac{1}{2} \zeta _2 \ln _2 \bar{s}_{-2}+\frac{7}{4} \zeta _3 \ln _2 \bar{s}_{-1}+\frac{1}{4} \zeta _3 \ln _2 \bar{s}_1+\zeta _2 \ln _2 \bar{s}_2+\frac{1}{12} \ln _2^4 \bar{s}_{-1}-\frac{1}{2} \ln _2^2 \bar{s}_{-3}-\frac{1}{2} \ln _2^2 \bar{s}_3+\frac{1}{2} \zeta _2 s_{1,-2}-\frac{3}{4} \zeta _3 s_{1,-1}-s_{4,-1}+s_{1,3,-1}+s_{3,1,-1}-\ln _2 s_{1,-3}+\ln _2 s_{1,3}-\frac{3 \zeta _2 \zeta _3}{8}+\frac{163 \zeta _5}{16}+\frac{1}{3} \zeta _2 \ln _2^3-\frac{3}{4} \zeta _3 \ln _2^2-\frac{93}{20} \zeta _2^2 \ln _2+2 \text{Li}_4\left(\frac{1}{2}\right) s_{-1}-8 \text{Li}_5\left(\frac{1}{2}\right)-4 \text{Li}_4\left(\frac{1}{2}\right) \ln _2-\frac{49}{40} \zeta _2^2 s_{-1}+\frac{1}{8} \zeta _2^2 s_1-\frac{1}{2} \zeta _2 s_{-3}+\frac{5}{8} \zeta _3 s_{-2}-\frac{1}{2} \zeta _2 s_{-1} \ln _2^2+\frac{1}{2} \zeta _2 s_{-2} \ln _2+\zeta _2 s_2 \ln _2+\frac{7}{4} \zeta _3 s_{-1} \ln _2-\frac{7}{4} \zeta _3 s_1 \ln _2+\frac{1}{12} s_{-1} \ln _2^4-\frac{1}{2} s_{-3} \ln _2^2+\frac{1}{2} s_3 \ln _2^2+s_{-4} \ln _2-s_4 \ln _2-\frac{\ln _2^5}{10}  \end{dmath}
   \begin{dmath}[style={\small}]     s_{1,-1} \bar{s}_{-2,-1}  =   \frac{\ln _2^5}{5}+\frac{7}{24} s_{-1} \ln _2^4-\frac{1}{6} s_1 \ln _2^4+\frac{7}{24} \bar{s}_{-1} \ln _2^4-5 \zeta _2 \ln _2^3+\frac{1}{2} s_{-3} \ln _2^2-\frac{1}{2} s_3 \ln _2^2-\frac{13}{4} s_{-1} \zeta _2 \ln _2^2-\frac{1}{2} s_1 \zeta _2 \ln _2^2+\frac{15}{4} \zeta _3 \ln _2^2+\bar{s}_{-3} \ln _2^2-\frac{13}{4} \zeta _2 \bar{s}_{-1} \ln _2^2+\frac{3}{2} \zeta _2 \bar{s}_1 \ln _2^2-\bar{s}_3 \ln _2^2+\frac{1}{2} s_{-2,-1} \ln _2^2-\frac{1}{2} s_{-2,1} \ln _2^2-s_{1,-2} \ln _2^2+s_{1,2} \ln _2^2+\frac{1}{2} \bar{s}_{-2,-1} \ln _2^2-\frac{3}{2} \bar{s}_{-2,1} \ln _2^2-\bar{s}_{1,-2} \ln _2^2+\bar{s}_{1,2} \ln _2^2-\frac{313}{40} \zeta _2^2 \ln _2+s_{-4} \ln _2-s_4 \ln _2+\frac{3}{2} s_{-2} \zeta _2 \ln _2+s_2 \zeta _2 \ln _2+\frac{21}{4} s_{-1} \zeta _3 \ln _2-3 s_1 \zeta _3 \ln _2+\bar{s}_{-4} \ln _2+\frac{21}{4} \zeta _3 \bar{s}_{-1} \ln _2-3 \zeta _3 \bar{s}_1 \ln _2+\zeta _2 \bar{s}_2 \ln _2-\bar{s}_4 \ln _2+2 s_{-3,-1} \ln _2+s_{-2,-2} \ln _2-s_{-2,2} \ln _2-s_{1,-3} \ln _2+s_{1,3} \ln _2+2 \bar{s}_{-3,-1} \ln _2+\bar{s}_{-2,-2} \ln _2-\bar{s}_{-2,2} \ln _2-\bar{s}_{1,-3} \ln _2+\bar{s}_{1,3} \ln _2-2 s_{-2,1,-1} \ln _2-2 s_{1,-2,-1} \ln _2-2 \bar{s}_{-2,1,-1} \ln _2-2 \bar{s}_{1,-2,-1} \ln _2-\frac{13}{5} s_{-1} \zeta _2^2+\frac{61}{40} s_1 \zeta _2^2-24 \text{Li}_5\left(\frac{1}{2}\right)+7 \text{Li}_4\left(\frac{1}{2}\right) s_{-1}-4 \text{Li}_4\left(\frac{1}{2}\right) s_1-\frac{1}{2} s_{-3} \zeta _2-\frac{3}{8} s_{-2} \zeta _3+\frac{1501 \zeta _5}{64}-\frac{1}{2} \zeta _2 \bar{s}_{-3}-\frac{1}{4} \zeta _3 \bar{s}_{-2}-\frac{13}{5} \zeta _2^2 \bar{s}_{-1}+7 \text{Li}_4\left(\frac{1}{2}\right) \bar{s}_{-1}+\frac{1}{8} \zeta _2^2 \bar{s}_1+\frac{1}{2} \zeta _2 s_{-2,1}+\frac{1}{2} \zeta _2 s_{1,-2}+\frac{1}{4} \zeta _3 s_{1,-1}-s_{4,-1}+\frac{1}{2} \zeta _2 \bar{s}_{-2,1}+\frac{1}{2} \zeta _2 \bar{s}_{1,-2}+\frac{1}{4} \zeta _3 \bar{s}_{1,-1}-\bar{s}_{4,-1}+s_{-3,-1,-1}+s_{-2,-2,-1}+s_{1,3,-1}+s_{3,1,-1}+\bar{s}_{-3,-1,-1}+2 \bar{s}_{-2,-2,-1}+\bar{s}_{1,3,-1}-s_{-2,-1,1,-1}-s_{-2,1,-1,-1}-s_{1,-2,-1,-1}-\bar{s}_{-2,1,-1,-1}-\bar{s}_{1,-2,-1,-1}-\bar{s}_{1,-1,-2,-1}  \end{dmath}
   \begin{dmath}[style={\small}]     s_{1,-1} \bar{s}_{-2,1}  =   -\frac{1}{2} \zeta _2 \bar{s}_{-2,-1}-\frac{1}{2} \zeta _2 \bar{s}_{-2,1}-\frac{1}{2} \zeta _2 \bar{s}_{1,-2}-\frac{5}{8} \zeta _3 \bar{s}_{1,-1}-\frac{1}{2} \zeta _2 \bar{s}_{1,2}-\bar{s}_{4,1}+\bar{s}_{-3,-1,1}+2 \bar{s}_{-2,-2,1}+\bar{s}_{1,3,1}-\bar{s}_{-2,1,-1,1}-\bar{s}_{1,-2,-1,1}-\bar{s}_{1,-1,-2,1}-\ln _2^2 \bar{s}_{-2,1}-\frac{1}{2} \ln _2^2 \bar{s}_{1,-2}+\frac{1}{2} \ln _2^2 \bar{s}_{1,2}+\ln _2 \bar{s}_{-3,-1}+\ln _2 \bar{s}_{-3,1}-\ln _2 \bar{s}_{-2,1,-1}-\ln _2 \bar{s}_{-2,1,1}-\ln _2 \bar{s}_{1,-2,-1}-\ln _2 \bar{s}_{1,-2,1}+\frac{1}{2} \zeta _2 \bar{s}_{-3}+\frac{11}{8} \zeta _3 \bar{s}_{-2}+\frac{1}{2} \zeta _2^2 \bar{s}_1-\frac{3}{4} \zeta _3 \bar{s}_2+\frac{1}{2} \zeta _2 \bar{s}_3-2 \zeta _2 \ln _2 \bar{s}_{-2}+\frac{21}{8} \zeta _3 \ln _2 \bar{s}_{-1}-\frac{5}{4} \zeta _3 \ln _2 \bar{s}_1-\frac{1}{6} \ln _2^3 \bar{s}_{-2}+\frac{1}{6} \ln _2^3 \bar{s}_2+\frac{1}{2} \ln _2^2 \bar{s}_{-3}-\frac{1}{2} \ln _2^2 \bar{s}_3-\frac{1}{2} \zeta _2 s_{-2,-1}-\frac{5}{8} \zeta _3 s_{1,-1}-\frac{1}{2} \zeta _2 s_{1,2}+s_{-4,-1}-2 s_{-3,1,-1}-s_{-2,2,-1}-s_{1,-3,-1}+2 s_{-2,1,1,-1}+s_{1,-2,1,-1}-\frac{1}{2} \ln _2^2 s_{1,-2}+\frac{1}{2} \ln _2^2 s_{1,2}+\ln _2 s_{-3,-1}-\ln _2 s_{-3,1}+\ln _2 s_{-2,-2}-\ln _2 s_{-2,2}-\ln _2 s_{1,-3}+\ln _2 s_{1,3}-\ln _2 s_{-2,1,-1}+\ln _2 s_{-2,1,1}-\ln _2 s_{1,-2,-1}+\ln _2 s_{1,-2,1}-\frac{21 \zeta _2 \zeta _3}{8}-\frac{111 \zeta _5}{32}+\frac{2}{3} \zeta _2 \ln _2^3+\frac{9}{8} \zeta _3 \ln _2^2+\frac{29}{10} \zeta _2^2 \ln _2-2 \text{Li}_4\left(\frac{1}{2}\right) s_1+8 \text{Li}_5\left(\frac{1}{2}\right)+\frac{41}{40} \zeta _2^2 s_1+\frac{\zeta _2 s_3}{2}+\frac{1}{4} \zeta _3 s_{-2}-\frac{3 \zeta _3 s_2}{4}+\frac{1}{2} \zeta _2 s_1 \ln _2^2-\frac{3}{2} \zeta _2 s_{-2} \ln _2+\frac{21}{8} \zeta _3 s_{-1} \ln _2-\frac{7}{4} \zeta _3 s_1 \ln _2-\frac{1}{12} s_1 \ln _2^4-\frac{1}{6} s_{-2} \ln _2^3+\frac{1}{6} s_2 \ln _2^3+s_{-4} \ln _2-s_4 \ln _2-\frac{\ln _2^5}{15}  \end{dmath}
   \begin{dmath}[style={\small}]     s_{1,-1} \bar{s}_{2,-1}  =   \frac{1}{4} \zeta _3 \bar{s}_{1,-1}+\frac{1}{2} \zeta _2 \bar{s}_{1,2}+\frac{1}{2} \zeta _2 \bar{s}_{2,1}-\bar{s}_{-4,-1}+\bar{s}_{-2,2,-1}+\bar{s}_{1,-3,-1}+\bar{s}_{2,-2,-1}+\bar{s}_{3,-1,-1}-\bar{s}_{1,-1,2,-1}-\bar{s}_{1,2,-1,-1}-\bar{s}_{2,1,-1,-1}+\ln _2^2 \bar{s}_{1,-2}-\ln _2^2 \bar{s}_{1,2}+\frac{1}{2} \ln _2^2 \bar{s}_{2,-1}-\frac{3}{2} \ln _2^2 \bar{s}_{2,1}+\ln _2 \bar{s}_{1,-3}-\ln _2 \bar{s}_{1,3}+\ln _2 \bar{s}_{2,-2}-\ln _2 \bar{s}_{2,2}+2 \ln _2 \bar{s}_{3,-1}-2 \ln _2 \bar{s}_{1,2,-1}-2 \ln _2 \bar{s}_{2,1,-1}-\text{Li}_4\left(\frac{1}{2}\right) \bar{s}_1-\frac{1}{4} \zeta _3 \bar{s}_{-2}+\frac{7}{40} \zeta _2^2 \bar{s}_1-\frac{1}{2} \zeta _2 \bar{s}_3+\frac{3}{2} \zeta _2 \ln _2^2 \bar{s}_{-1}-\frac{5}{4} \zeta _2 \ln _2^2 \bar{s}_1+\zeta _2 \ln _2 \bar{s}_{-2}+\frac{9}{4} \zeta _3 \ln _2 \bar{s}_1-\frac{1}{24} \ln _2^4 \bar{s}_1-\ln _2^2 \bar{s}_{-3}+\ln _2^2 \bar{s}_3-\ln _2 \bar{s}_{-4}+\ln _2 \bar{s}_4+\frac{1}{4} \zeta _3 s_{1,-1}-\frac{1}{2} \zeta _2 s_{1,2}-\frac{1}{2} \zeta _2 s_{2,1}+s_{-4,-1}-s_{-3,1,-1}-s_{1,-3,-1}-s_{2,-2,-1}-s_{3,-1,-1}+s_{1,2,-1,-1}+s_{2,-1,1,-1}+s_{2,1,-1,-1}-\ln _2^2 s_{1,-2}+\ln _2^2 s_{1,2}-\frac{1}{2} \ln _2^2 s_{2,-1}+\frac{1}{2} \ln _2^2 s_{2,1}-\ln _2 s_{1,-3}+\ln _2 s_{1,3}-\ln _2 s_{2,-2}+\ln _2 s_{2,2}-2 \ln _2 s_{3,-1}+2 \ln _2 s_{1,2,-1}+2 \ln _2 s_{2,1,-1}-\frac{625 \zeta _5}{64}+\frac{23}{6} \zeta _2 \ln _2^3+\frac{9}{8} \zeta _3 \ln _2^2+\frac{25}{8} \zeta _2^2 \ln _2-3 \text{Li}_4\left(\frac{1}{2}\right) s_1+10 \text{Li}_5\left(\frac{1}{2}\right)+\frac{43}{40} \zeta _2^2 s_1+\frac{\zeta _2 s_3}{2}+\frac{3 \zeta _3 s_2}{8}+\frac{3}{2} \zeta _2 s_{-1} \ln _2^2+\frac{9}{4} \zeta _2 s_1 \ln _2^2-\zeta _2 s_{-2} \ln _2-\frac{3}{2} \zeta _2 s_2 \ln _2-\frac{9}{4} \zeta _3 s_1 \ln _2-\frac{1}{8} s_1 \ln _2^4+\frac{1}{2} s_{-3} \ln _2^2-\frac{1}{2} s_3 \ln _2^2+s_{-4} \ln _2-s_4 \ln _2-\frac{\ln _2^5}{12}  \end{dmath}

   \begin{dmath}[style={\small}]     s_{1,-1} \bar{s}_{-1,1,-1}  =   \frac{19 \ln _2^5}{120}+\frac{7}{24} s_{-1} \ln _2^4-\frac{1}{24} s_1 \ln _2^4+\frac{5}{24} \bar{s}_{-1} \ln _2^4-\frac{1}{4} \bar{s}_1 \ln _2^4-\frac{5}{12} \zeta _2 \ln _2^3+\frac{1}{2} s_{-3} \ln _2^2-\frac{1}{2} s_3 \ln _2^2-5 s_{-1} \zeta _2 \ln _2^2+\frac{37}{16} \zeta _3 \ln _2^2+\frac{1}{2} \bar{s}_{-3} \ln _2^2+\frac{5}{2} \zeta _2 \bar{s}_{-1} \ln _2^2+\frac{9}{4} \zeta _2 \bar{s}_1 \ln _2^2-\frac{1}{2} \bar{s}_3 \ln _2^2+s_{-2,-1} \ln _2^2-2 s_{-2,1} \ln _2^2+\frac{1}{2} s_{-1,-2} \ln _2^2-\frac{1}{2} s_{-1,2} \ln _2^2-s_{1,-2} \ln _2^2+s_{1,2} \ln _2^2-\frac{1}{2} \bar{s}_{-2,-1} \ln _2^2+\frac{1}{2} \bar{s}_{-2,1} \ln _2^2-\frac{1}{2} \bar{s}_{-1,-2} \ln _2^2+\frac{1}{2} \bar{s}_{-1,2} \ln _2^2-\frac{1}{2} \bar{s}_{1,-2} \ln _2^2+\frac{1}{2} \bar{s}_{1,2} \ln _2^2-s_{-1,1,-1} \ln _2^2+2 s_{-1,1,1} \ln _2^2-\frac{1}{2} s_{1,-1,-1} \ln _2^2+\frac{3}{2} s_{1,-1,1} \ln _2^2+\bar{s}_{-1,1,-1} \ln _2^2-2 \bar{s}_{-1,1,1} \ln _2^2+\frac{1}{2} \bar{s}_{1,-1,-1} \ln _2^2-\frac{1}{2} \bar{s}_{1,-1,1} \ln _2^2-\frac{21}{5} \zeta _2^2 \ln _2+s_{-4} \ln _2-s_4 \ln _2+\frac{3}{2} s_{-2} \zeta _2 \ln _2+s_2 \zeta _2 \ln _2+\frac{53}{8} s_{-1} \zeta _3 \ln _2-\frac{5}{8} s_1 \zeta _3 \ln _2-\frac{3}{2} \zeta _2 \bar{s}_{-2} \ln _2-\frac{11}{8} \zeta _3 \bar{s}_{-1} \ln _2-\frac{19}{8} \zeta _3 \bar{s}_1 \ln _2+2 s_{-3,-1} \ln _2+2 s_{-2,-2} \ln _2-2 s_{-2,2} \ln _2+s_{-1,-3} \ln _2-\zeta _2 s_{-1,-1} \ln _2-\frac{3}{2} \zeta _2 s_{-1,1} \ln _2-s_{-1,3} \ln _2-s_{1,-3} \ln _2+s_{1,3} \ln _2-\bar{s}_{-2,-2} \ln _2+\bar{s}_{-2,2} \ln _2-\bar{s}_{-1,-3} \ln _2+\zeta _2 \bar{s}_{-1,-1} \ln _2+\frac{3}{2} \zeta _2 \bar{s}_{-1,1} \ln _2+\bar{s}_{-1,3} \ln _2+\frac{3}{2} \zeta _2 \bar{s}_{1,-1} \ln _2-4 s_{-2,1,-1} \ln _2-2 s_{-1,1,-2} \ln _2+2 s_{-1,1,2} \ln _2-2 s_{-1,2,-1} \ln _2-2 s_{1,-2,-1} \ln _2-s_{1,-1,-2} \ln _2+s_{1,-1,2} \ln _2+2 \bar{s}_{-2,1,-1} \ln _2+2 \bar{s}_{-1,1,-2} \ln _2-2 \bar{s}_{-1,1,2} \ln _2+2 \bar{s}_{-1,2,-1} \ln _2+\bar{s}_{1,-1,-2} \ln _2-\bar{s}_{1,-1,2} \ln _2+4 s_{-1,1,1,-1} \ln _2+2 s_{1,-1,1,-1} \ln _2-4 \bar{s}_{-1,1,1,-1} \ln _2-2 \bar{s}_{1,-1,1,-1} \ln _2-\frac{12}{5} s_{-1} \zeta _2^2-11 \text{Li}_5\left(\frac{1}{2}\right)+6 \text{Li}_4\left(\frac{1}{2}\right) s_{-1}-\frac{1}{2} s_{-3} \zeta _2-\frac{3}{8} s_{-2} \zeta _3+\frac{355 \zeta _5}{32}+\frac{1}{8} \zeta _3 \bar{s}_{-2}+\frac{13}{40} \zeta _2^2 \bar{s}_1-\text{Li}_4\left(\frac{1}{2}\right) \bar{s}_1+\zeta _2 s_{-2,1}+\frac{3}{8} \zeta _3 s_{-1,1}+\frac{1}{2} \zeta _2 s_{-1,2}+\frac{1}{2} \zeta _2 s_{1,-2}+\frac{1}{4} \zeta _3 s_{1,-1}-s_{4,-1}-\frac{1}{2} \zeta _2 \bar{s}_{-2,1}-\frac{3}{8} \zeta _3 \bar{s}_{-1,1}-\frac{1}{2} \zeta _2 \bar{s}_{-1,2}-\frac{1}{8} \zeta _3 \bar{s}_{1,-1}+s_{-3,-1,-1}+2 s_{-2,-2,-1}+s_{-1,-3,-1}-\zeta _2 s_{-1,1,1}-\frac{1}{2} \zeta _2 s_{1,-1,1}+s_{1,3,-1}+s_{3,1,-1}-\bar{s}_{-2,-2,-1}-\bar{s}_{-1,-3,-1}+\zeta _2 \bar{s}_{-1,1,1}+\frac{1}{2} \zeta _2 \bar{s}_{1,-1,1}-\bar{s}_{3,1,-1}-s_{-2,-1,1,-1}-2 s_{-2,1,-1,-1}-s_{-1,-2,1,-1}-2 s_{-1,1,-2,-1}-s_{-1,2,-1,-1}-s_{1,-2,-1,-1}-s_{1,-1,-2,-1}+2 \bar{s}_{-2,-1,1,-1}+\bar{s}_{-2,1,-1,-1}+\bar{s}_{-1,-2,1,-1}+2 \bar{s}_{-1,1,-2,-1}+\bar{s}_{-1,2,-1,-1}+\bar{s}_{1,-1,-2,-1}+\bar{s}_{1,2,1,-1}+s_{-1,1,-1,1,-1}+2 s_{-1,1,1,-1,-1}+s_{1,-1,1,-1,-1}-\bar{s}_{-1,1,-1,1,-1}-2 \bar{s}_{-1,1,1,-1,-1}-2 \bar{s}_{1,-1,-1,1,-1}-\bar{s}_{1,-1,1,-1,-1}  \end{dmath}
   \begin{dmath}[style={\small}]     s_{1,-1} \bar{s}_{-1,1,1}  =   \frac{\ln _2^5}{10}+\frac{1}{8} s_{-1} \ln _2^4-\frac{1}{12} s_1 \ln _2^4+\frac{1}{4} \bar{s}_{-1} \ln _2^4-\frac{1}{12} \bar{s}_1 \ln _2^4+\frac{1}{6} \zeta _2 \ln _2^3-\frac{1}{3} s_{1,-1} \ln _2^3-\frac{1}{2} s_{-1} \zeta _2 \ln _2^2+\frac{1}{4} s_1 \zeta _2 \ln _2^2+\frac{7}{8} \zeta _3 \ln _2^2-\frac{3}{4} \zeta _2 \bar{s}_{-1} \ln _2^2+\frac{1}{2} \zeta _2 \bar{s}_1 \ln _2^2+\frac{1}{2} s_{-2,-1} \ln _2^2-\frac{1}{2} s_{-2,1} \ln _2^2-\frac{1}{2} s_{1,-2} \ln _2^2+\frac{1}{2} s_{1,2} \ln _2^2-\frac{1}{2} \bar{s}_{-2,-1} \ln _2^2+\frac{1}{2} \bar{s}_{-2,1} \ln _2^2-\frac{1}{2} \bar{s}_{-1,-2} \ln _2^2+\frac{1}{2} \bar{s}_{-1,2} \ln _2^2-\frac{1}{2} s_{-1,1,-1} \ln _2^2+\frac{1}{2} s_{-1,1,1} \ln _2^2-\frac{1}{2} s_{1,-1,-1} \ln _2^2+\frac{1}{2} s_{1,-1,1} \ln _2^2+\frac{1}{2} \bar{s}_{-1,1,-1} \ln _2^2-\frac{3}{2} \bar{s}_{-1,1,1} \ln _2^2+\frac{1}{2} \bar{s}_{1,-1,-1} \ln _2^2-\frac{1}{2} \bar{s}_{1,-1,1} \ln _2^2-\frac{7}{5} \zeta _2^2 \ln _2-\text{Li}_4\left(\frac{1}{2}\right) \ln _2+s_{-4} \ln _2-s_4 \ln _2-2 s_{-2} \zeta _2 \ln _2+\frac{1}{2} s_2 \zeta _2 \ln _2+\frac{29}{8} s_{-1} \zeta _3 \ln _2-\frac{1}{8} s_1 \zeta _3 \ln _2-\frac{3}{8} \zeta _3 \bar{s}_{-1} \ln _2-\frac{3}{4} \zeta _3 \bar{s}_1 \ln _2+s_{-3,-1} \ln _2-s_{-3,1} \ln _2+2 s_{-2,-2} \ln _2-2 s_{-2,2} \ln _2+s_{-1,-3} \ln _2-\frac{1}{2} \zeta _2 s_{-1,-1} \ln _2+2 \zeta _2 s_{-1,1} \ln _2-s_{-1,3} \ln _2-s_{1,-3} \ln _2+\zeta _2 s_{1,-1} \ln _2+s_{1,3} \ln _2+\frac{1}{2} \zeta _2 \bar{s}_{-1,-1} \ln _2-\frac{5}{2} \zeta _2 \bar{s}_{-1,1} \ln _2-2 s_{-2,1,-1} \ln _2+2 s_{-2,1,1} \ln _2-2 s_{-1,1,-2} \ln _2+2 s_{-1,1,2} \ln _2-s_{-1,2,-1} \ln _2+s_{-1,2,1} \ln _2-s_{1,-2,-1} \ln _2+s_{1,-2,1} \ln _2-s_{1,-1,-2} \ln _2+s_{1,-1,2} \ln _2+\bar{s}_{-2,1,-1} \ln _2+\bar{s}_{-2,1,1} \ln _2+\bar{s}_{-1,2,-1} \ln _2+\bar{s}_{-1,2,1} \ln _2+2 s_{-1,1,1,-1} \ln _2-2 s_{-1,1,1,1} \ln _2+s_{1,-1,1,-1} \ln _2-s_{1,-1,1,1} \ln _2-2 \bar{s}_{-1,1,1,-1} \ln _2-2 \bar{s}_{-1,1,1,1} \ln _2-\bar{s}_{1,-1,1,-1} \ln _2-\bar{s}_{1,-1,1,1} \ln _2-\frac{1}{8} s_{-1} \zeta _2^2-\frac{3}{40} s_1 \zeta _2^2-7 \text{Li}_5\left(\frac{1}{2}\right)+\text{Li}_4\left(\frac{1}{2}\right) s_1+\frac{s_3 \zeta _2}{2}+\frac{1}{2} s_{-2} \zeta _3-s_2 \zeta _3-\frac{21 \zeta _2 \zeta _3}{16}+\frac{269 \zeta _5}{32}+\zeta _3 \bar{s}_{-2}-\frac{3}{40} \zeta _2^2 \bar{s}_{-1}+\text{Li}_4\left(\frac{1}{2}\right) \bar{s}_{-1}-\frac{3}{5} \zeta _2^2 \bar{s}_1+3 \text{Li}_4\left(\frac{1}{2}\right) \bar{s}_1+s_{-4,-1}-\zeta _2 s_{-2,-1}-\frac{1}{2} \zeta _2 s_{-1,-2}+\zeta _3 s_{-1,-1}-\frac{1}{2} \zeta _3 s_{-1,1}-\frac{9}{8} \zeta _3 s_{1,-1}-\frac{1}{2} \zeta _2 s_{1,2}+\frac{1}{2} \zeta _2 \bar{s}_{-2,-1}+\frac{1}{2} \zeta _2 \bar{s}_{-2,1}+\frac{1}{2} \zeta _2 \bar{s}_{-1,-2}-\zeta _3 \bar{s}_{-1,-1}+\frac{5}{8} \zeta _3 \bar{s}_{-1,1}+\frac{1}{2} \zeta _2 \bar{s}_{-1,2}-\zeta _3 \bar{s}_{1,-1}-2 s_{-3,1,-1}-2 s_{-2,2,-1}+\zeta _2 s_{-1,1,-1}-s_{-1,3,-1}-s_{1,-3,-1}+\frac{1}{2} \zeta _2 s_{1,-1,-1}-\bar{s}_{-2,-2,1}-\bar{s}_{-1,-3,1}-\zeta _2 \bar{s}_{-1,1,-1}-\zeta _2 \bar{s}_{-1,1,1}-\frac{1}{2} \zeta _2 \bar{s}_{1,-1,-1}-\frac{1}{2} \zeta _2 \bar{s}_{1,-1,1}-\bar{s}_{3,1,1}+3 s_{-2,1,1,-1}+2 s_{-1,1,2,-1}+2 s_{-1,2,1,-1}+s_{1,-2,1,-1}+s_{1,-1,2,-1}+2 \bar{s}_{-2,-1,1,1}+\bar{s}_{-2,1,-1,1}+\bar{s}_{-1,-2,1,1}+2 \bar{s}_{-1,1,-2,1}+\bar{s}_{-1,2,-1,1}+\bar{s}_{1,-1,-2,1}+\bar{s}_{1,2,1,1}-3 s_{-1,1,1,1,-1}-s_{1,-1,1,1,-1}-\bar{s}_{-1,1,-1,1,1}-2 \bar{s}_{-1,1,1,-1,1}-2 \bar{s}_{1,-1,-1,1,1}-\bar{s}_{1,-1,1,-1,1}  \end{dmath}
   \begin{dmath}[style={\small}]     s_{1,-1} \bar{s}_{1,-2}  =   -\frac{\ln _2^5}{5}-\frac{1}{4} s_1 \ln _2^4-\frac{1}{6} \bar{s}_1 \ln _2^4-\frac{1}{2} s_{-3} \ln _2^2+\frac{1}{2} s_3 \ln _2^2-\frac{3}{2} s_{-1} \zeta _2 \ln _2^2+3 s_1 \zeta _2 \ln _2^2-\frac{15}{8} \zeta _3 \ln _2^2-\frac{3}{2} \zeta _2 \bar{s}_{-1} \ln _2^2+\frac{5}{2} \zeta _2 \bar{s}_1 \ln _2^2+\frac{1}{2} s_{1,-2} \ln _2^2-\frac{1}{2} s_{1,2} \ln _2^2-\frac{1}{2} \bar{s}_{1,-2} \ln _2^2-\frac{1}{2} \bar{s}_{1,2} \ln _2^2+\frac{23}{5} \zeta _2^2 \ln _2-4 \text{Li}_4\left(\frac{1}{2}\right) \ln _2+s_{-4} \ln _2-s_4 \ln _2-\frac{3}{2} s_{-2} \zeta _2 \ln _2+\frac{7}{4} s_{-1} \zeta _3 \ln _2-\frac{21}{4} s_1 \zeta _3 \ln _2-\frac{1}{2} \zeta _2 \bar{s}_{-2} \ln _2+\frac{7}{4} \zeta _3 \bar{s}_{-1} \ln _2-\frac{9}{4} \zeta _3 \bar{s}_1 \ln _2+\frac{1}{2} \zeta _2 \bar{s}_2 \ln _2-2 s_{1,-3} \ln _2+2 \zeta _2 s_{1,-1} \ln _2-\frac{1}{2} \zeta _2 s_{1,1} \ln _2+2 s_{1,3} \ln _2-s_{2,-2} \ln _2+s_{2,2} \ln _2+2 \zeta _2 \bar{s}_{1,-1} \ln _2-\frac{3}{2} \zeta _2 \bar{s}_{1,1} \ln _2+\bar{s}_{2,-2} \ln _2+\bar{s}_{2,2} \ln _2+2 s_{1,1,-2} \ln _2-2 s_{1,1,2} \ln _2-2 \bar{s}_{1,1,-2} \ln _2-2 \bar{s}_{1,1,2} \ln _2+\frac{1}{4} s_{-1} \zeta _2^2+\frac{19}{8} s_1 \zeta _2^2+4 \text{Li}_5\left(\frac{1}{2}\right)-6 \text{Li}_4\left(\frac{1}{2}\right) s_1+\frac{s_3 \zeta _2}{2}-\frac{5 s_2 \zeta _3}{8}-\frac{7 \zeta _2 \zeta _3}{2}+\frac{13 \zeta _5}{8}-\frac{1}{2} \zeta _2 \bar{s}_{-3}+\frac{1}{8} \zeta _3 \bar{s}_{-2}+\frac{1}{4} \zeta _2^2 \bar{s}_{-1}+\frac{7}{4} \zeta _2^2 \bar{s}_1-4 \text{Li}_4\left(\frac{1}{2}\right) \bar{s}_1-\frac{13}{8} \zeta _3 \bar{s}_2+\frac{1}{2} \zeta _2 \bar{s}_3+s_{-4,-1}-\frac{1}{8} \zeta _3 s_{1,-1}+\frac{5}{4} \zeta _3 s_{1,1}-\zeta _2 s_{1,2}+\frac{1}{2} \zeta _2 s_{2,-1}-\bar{s}_{-3,-2}+\zeta _2 \bar{s}_{1,-2}-\frac{1}{8} \zeta _3 \bar{s}_{1,-1}+\frac{13}{4} \zeta _3 \bar{s}_{1,1}-\zeta _2 \bar{s}_{1,2}+\frac{1}{2} \zeta _2 \bar{s}_{2,-1}-s_{-3,1,-1}-2 s_{1,-3,-1}-\zeta _2 s_{1,1,-1}-s_{2,-2,-1}+\bar{s}_{-2,1,-2}+2 \bar{s}_{1,-2,-2}-\zeta _2 \bar{s}_{1,1,-1}+\bar{s}_{2,-1,-2}+s_{1,-2,1,-1}+2 s_{1,1,-2,-1}-\bar{s}_{1,-1,1,-2}-2 \bar{s}_{1,1,-1,-2}  \end{dmath}
   \begin{dmath}[style={\small}]     s_{1,-1} \bar{s}_{1,2}  =   -\frac{2 \ln _2^5}{15}+\frac{1}{24} s_{-1} \ln _2^4-\frac{1}{8} s_1 \ln _2^4+\frac{1}{24} \bar{s}_{-1} \ln _2^4+\frac{1}{8} \bar{s}_1 \ln _2^4+\frac{5}{6} \zeta _2 \ln _2^3-\frac{1}{2} s_{-3} \ln _2^2+\frac{1}{2} s_3 \ln _2^2-\frac{1}{4} s_{-1} \zeta _2 \ln _2^2+\frac{3}{4} s_1 \zeta _2 \ln _2^2-\frac{15}{8} \zeta _3 \ln _2^2-\frac{1}{4} \zeta _2 \bar{s}_{-1} \ln _2^2-\frac{3}{4} \zeta _2 \bar{s}_1 \ln _2^2+\frac{1}{2} s_{1,-2} \ln _2^2-\frac{1}{2} s_{1,2} \ln _2^2-\frac{1}{2} \bar{s}_{1,-2} \ln _2^2-\frac{1}{2} \bar{s}_{1,2} \ln _2^2-\frac{87}{40} \zeta _2^2 \ln _2-4 \text{Li}_4\left(\frac{1}{2}\right) \ln _2+s_{-4} \ln _2-s_4 \ln _2+\frac{3}{2} s_2 \zeta _2 \ln _2+\frac{7}{4} s_{-1} \zeta _3 \ln _2-\frac{21}{4} s_1 \zeta _3 \ln _2-\frac{1}{2} \zeta _2 \bar{s}_{-2} \ln _2+\frac{7}{4} \zeta _3 \bar{s}_{-1} \ln _2-\frac{9}{4} \zeta _3 \bar{s}_1 \ln _2+\frac{1}{2} \zeta _2 \bar{s}_2 \ln _2-2 s_{1,-3} \ln _2+\frac{1}{2} \zeta _2 s_{1,-1} \ln _2-2 \zeta _2 s_{1,1} \ln _2+2 s_{1,3} \ln _2-s_{2,-2} \ln _2+s_{2,2} \ln _2+\frac{1}{2} \zeta _2 \bar{s}_{1,-1} \ln _2+\bar{s}_{2,-2} \ln _2+\bar{s}_{2,2} \ln _2+2 s_{1,1,-2} \ln _2-2 s_{1,1,2} \ln _2-2 \bar{s}_{1,1,-2} \ln _2-2 \bar{s}_{1,1,2} \ln _2-\frac{37}{40} s_{-1} \zeta _2^2+\frac{29}{20} s_1 \zeta _2^2-4 \text{Li}_5\left(\frac{1}{2}\right)+\text{Li}_4\left(\frac{1}{2}\right) s_{-1}-3 \text{Li}_4\left(\frac{1}{2}\right) s_1-\frac{1}{2} s_{-3} \zeta _2+\frac{7}{8} s_{-2} \zeta _3-\frac{s_2 \zeta _3}{4}-\frac{\zeta _2 \zeta _3}{2}+\frac{401 \zeta _5}{64}+\zeta _3 \bar{s}_{-2}-\frac{37}{40} \zeta _2^2 \bar{s}_{-1}+\text{Li}_4\left(\frac{1}{2}\right) \bar{s}_{-1}-\frac{17}{20} \zeta _2^2 \bar{s}_1+3 \text{Li}_4\left(\frac{1}{2}\right) \bar{s}_1+\zeta _3 \bar{s}_2+\zeta _2 s_{1,-2}-\frac{15}{8} \zeta _3 s_{1,-1}+\frac{1}{2} \zeta _3 s_{1,1}+\frac{1}{2} \zeta _2 s_{2,-1}-s_{4,-1}-\bar{s}_{-3,2}-\frac{15}{8} \zeta _3 \bar{s}_{1,-1}-2 \zeta _3 \bar{s}_{1,1}+\frac{1}{2} \zeta _2 \bar{s}_{2,-1}-\zeta _2 s_{1,1,-1}+2 s_{1,3,-1}+s_{2,2,-1}+s_{3,1,-1}+\bar{s}_{-2,1,2}+2 \bar{s}_{1,-2,2}-\zeta _2 \bar{s}_{1,1,-1}+\bar{s}_{2,-1,2}-2 s_{1,1,2,-1}-s_{1,2,1,-1}-\bar{s}_{1,-1,1,2}-2 \bar{s}_{1,1,-1,2}  \end{dmath}
   \begin{dmath}[style={\small}]     s_{1,-1} \bar{s}_{1,-1,-1}  =   -\frac{\ln _2^5}{60}+\frac{1}{8} s_{-1} \ln _2^4-\frac{1}{6} s_1 \ln _2^4+\frac{1}{8} \bar{s}_{-1} \ln _2^4+\frac{1}{12} \bar{s}_1 \ln _2^4+\frac{2}{3} s_{-2} \ln _2^3-\frac{2}{3} s_2 \ln _2^3-\frac{16}{3} \zeta _2 \ln _2^3+\frac{1}{6} \bar{s}_{-2} \ln _2^3-\frac{1}{6} \bar{s}_2 \ln _2^3-\frac{5}{6} s_{1,-1} \ln _2^3+\frac{11}{6} s_{1,1} \ln _2^3-\frac{5}{6} \bar{s}_{1,-1} \ln _2^3+\frac{5}{6} \bar{s}_{1,1} \ln _2^3+\frac{1}{2} s_{-3} \ln _2^2-\frac{1}{2} s_3 \ln _2^2-\frac{3}{2} s_{-1} \zeta _2 \ln _2^2-\frac{9}{4} s_1 \zeta _2 \ln _2^2+\frac{87}{16} \zeta _3 \ln _2^2+\bar{s}_{-3} \ln _2^2-\frac{3}{2} \zeta _2 \bar{s}_{-1} \ln _2^2-\frac{1}{2} \zeta _2 \bar{s}_1 \ln _2^2-\bar{s}_3 \ln _2^2+\frac{1}{2} s_{-2,-1} \ln _2^2-\frac{1}{2} s_{-2,1} \ln _2^2-\frac{3}{2} s_{1,-2} \ln _2^2+\frac{3}{2} s_{1,2} \ln _2^2-2 s_{2,-1} \ln _2^2-\bar{s}_{1,-2} \ln _2^2+\bar{s}_{1,2} \ln _2^2-\frac{1}{2} s_{1,-1,-1} \ln _2^2+\frac{1}{2} s_{1,-1,1} \ln _2^2+4 s_{1,1,-1} \ln _2^2+\frac{1}{2} \bar{s}_{1,-1,-1} \ln _2^2-\frac{3}{2} \bar{s}_{1,-1,1} \ln _2^2-\frac{181}{40} \zeta _2^2 \ln _2+s_{-4} \ln _2-s_4 \ln _2+s_{-2} \zeta _2 \ln _2+\frac{3}{2} s_2 \zeta _2 \ln _2+\frac{21}{8} s_{-1} \zeta _3 \ln _2-\frac{15}{8} s_1 \zeta _3 \ln _2+\frac{21}{8} \zeta _3 \bar{s}_{-1} \ln _2-\frac{9}{8} \zeta _3 \bar{s}_1 \ln _2+2 s_{-3,-1} \ln _2+s_{-2,-2} \ln _2-s_{-2,2} \ln _2-2 s_{1,-3} \ln _2-\zeta _2 s_{1,-1} \ln _2-\frac{5}{2} \zeta _2 s_{1,1} \ln _2+2 s_{1,3} \ln _2-s_{2,-2} \ln _2+s_{2,2} \ln _2+\bar{s}_{1,-3} \ln _2+\frac{1}{2} \zeta _2 \bar{s}_{1,-1} \ln _2+\frac{1}{2} \zeta _2 \bar{s}_{1,1} \ln _2-\bar{s}_{1,3} \ln _2+\bar{s}_{2,-2} \ln _2-\bar{s}_{2,2} \ln _2-2 s_{-2,1,-1} \ln _2-4 s_{1,-2,-1} \ln _2-s_{1,-1,-2} \ln _2+s_{1,-1,2} \ln _2+2 s_{1,1,-2} \ln _2-2 s_{1,1,2} \ln _2-2 s_{2,-1,-1} \ln _2+2 \bar{s}_{1,-2,-1} \ln _2+\bar{s}_{1,-1,-2} \ln _2-\bar{s}_{1,-1,2} \ln _2-2 \bar{s}_{1,1,-2} \ln _2+2 \bar{s}_{1,1,2} \ln _2+2 \bar{s}_{2,-1,-1} \ln _2+2 s_{1,-1,1,-1} \ln _2+4 s_{1,1,-1,-1} \ln _2-2 \bar{s}_{1,-1,1,-1} \ln _2-4 \bar{s}_{1,1,-1,-1} \ln _2-\frac{53}{40} s_{-1} \zeta _2^2+\frac{61}{40} s_1 \zeta _2^2-10 \text{Li}_5\left(\frac{1}{2}\right)+3 \text{Li}_4\left(\frac{1}{2}\right) s_{-1}-4 \text{Li}_4\left(\frac{1}{2}\right) s_1-\frac{1}{2} s_{-3} \zeta _2+\frac{1}{8} s_{-2} \zeta _3-\frac{s_2 \zeta _3}{2}+\frac{661 \zeta _5}{64}+\frac{1}{4} \zeta _3 \bar{s}_{-2}-\frac{53}{40} \zeta _2^2 \bar{s}_{-1}+3 \text{Li}_4\left(\frac{1}{2}\right) \bar{s}_{-1}-\frac{21}{40} \zeta _2^2 \bar{s}_1+2 \text{Li}_4\left(\frac{1}{2}\right) \bar{s}_1+\frac{3}{4} \zeta _3 \bar{s}_2+\frac{1}{2} \zeta _2 s_{-2,1}+\zeta _2 s_{1,-2}-\frac{3}{8} \zeta _3 s_{1,-1}+\zeta _3 s_{1,1}+\frac{1}{2} \zeta _2 s_{2,-1}-s_{4,-1}-\frac{1}{2} \zeta _2 \bar{s}_{1,-2}-\frac{3}{4} \zeta _3 \bar{s}_{1,-1}-\frac{3}{2} \zeta _3 \bar{s}_{1,1}-\frac{1}{2} \zeta _2 \bar{s}_{2,-1}+s_{-3,-1,-1}+s_{-2,-2,-1}-\frac{1}{2} \zeta _2 s_{1,-1,1}-\zeta _2 s_{1,1,-1}+2 s_{1,3,-1}+s_{2,2,-1}+s_{3,1,-1}-\bar{s}_{-3,-1,-1}+\frac{1}{2} \zeta _2 \bar{s}_{1,-1,1}+\zeta _2 \bar{s}_{1,1,-1}-\bar{s}_{1,3,-1}-\bar{s}_{2,2,-1}-s_{-2,-1,1,-1}-s_{-2,1,-1,-1}-2 s_{1,-2,-1,-1}-s_{1,-1,-2,-1}-2 s_{1,1,2,-1}-s_{1,2,1,-1}-s_{2,-1,-1,-1}+\bar{s}_{-2,1,-1,-1}+3 \bar{s}_{1,-2,-1,-1}+\bar{s}_{1,-1,-2,-1}+2 \bar{s}_{1,1,2,-1}+2 \bar{s}_{2,-1,-1,-1}+s_{1,-1,-1,1,-1}+s_{1,-1,1,-1,-1}+2 s_{1,1,-1,-1,-1}-2 \bar{s}_{1,-1,1,-1,-1}-4 \bar{s}_{1,1,-1,-1,-1}  \end{dmath}
   \begin{dmath}[style={\small}]     s_{1,-1} \bar{s}_{1,-1,1}  =   -\frac{13 \ln _2^5}{40}-\frac{1}{6} s_1 \ln _2^4-\frac{1}{4} \bar{s}_1 \ln _2^4+\frac{1}{3} s_{-2} \ln _2^3-\frac{1}{3} s_2 \ln _2^3-\frac{11}{4} \zeta _2 \ln _2^3+\frac{1}{6} \bar{s}_{-2} \ln _2^3-\frac{1}{6} \bar{s}_2 \ln _2^3-\frac{1}{2} s_{1,-1} \ln _2^3+\frac{7}{6} s_{1,1} \ln _2^3-\frac{5}{6} \bar{s}_{1,-1} \ln _2^3+\frac{5}{6} \bar{s}_{1,1} \ln _2^3-\frac{3}{4} s_{-1} \zeta _2 \ln _2^2+\frac{5}{4} s_1 \zeta _2 \ln _2^2+\frac{17}{8} \zeta _3 \ln _2^2+\frac{1}{2} \bar{s}_{-3} \ln _2^2-\frac{3}{4} \zeta _2 \bar{s}_{-1} \ln _2^2-\frac{9}{4} \zeta _2 \bar{s}_1 \ln _2^2-\frac{1}{2} \bar{s}_3 \ln _2^2-\frac{1}{2} s_{1,-2} \ln _2^2+\frac{1}{2} s_{1,2} \ln _2^2-s_{2,-1} \ln _2^2-\frac{1}{2} \bar{s}_{1,-2} \ln _2^2+\frac{1}{2} \bar{s}_{1,2} \ln _2^2+2 s_{1,1,-1} \ln _2^2-\bar{s}_{1,-1,1} \ln _2^2+\frac{39}{10} \zeta _2^2 \ln _2-5 \text{Li}_4\left(\frac{1}{2}\right) \ln _2+s_{-4} \ln _2-s_4 \ln _2-2 s_{-2} \zeta _2 \ln _2+\frac{1}{2} s_2 \zeta _2 \ln _2+\frac{13}{8} s_{-1} \zeta _3 \ln _2-\frac{13}{8} s_1 \zeta _3 \ln _2+\frac{13}{8} \zeta _3 \bar{s}_{-1} \ln _2-\zeta _3 \bar{s}_1 \ln _2+\frac{3}{2} \zeta _2 \bar{s}_2 \ln _2+s_{-3,-1} \ln _2-s_{-3,1} \ln _2+s_{-2,-2} \ln _2-s_{-2,2} \ln _2-2 s_{1,-3} \ln _2+2 \zeta _2 s_{1,-1} \ln _2-\frac{3}{2} \zeta _2 s_{1,1} \ln _2+2 s_{1,3} \ln _2-s_{2,-2} \ln _2+s_{2,2} \ln _2-\frac{3}{2} \zeta _2 \bar{s}_{1,-1} \ln _2-\frac{7}{2} \zeta _2 \bar{s}_{1,1} \ln _2-s_{-2,1,-1} \ln _2+s_{-2,1,1} \ln _2-2 s_{1,-2,-1} \ln _2+2 s_{1,-2,1} \ln _2-s_{1,-1,-2} \ln _2+s_{1,-1,2} \ln _2+2 s_{1,1,-2} \ln _2-2 s_{1,1,2} \ln _2-s_{2,-1,-1} \ln _2+s_{2,-1,1} \ln _2+\bar{s}_{1,-2,-1} \ln _2+\bar{s}_{1,-2,1} \ln _2+\bar{s}_{2,-1,-1} \ln _2+\bar{s}_{2,-1,1} \ln _2+s_{1,-1,1,-1} \ln _2-s_{1,-1,1,1} \ln _2+2 s_{1,1,-1,-1} \ln _2-2 s_{1,1,-1,1} \ln _2-\bar{s}_{1,-1,1,-1} \ln _2-\bar{s}_{1,-1,1,1} \ln _2-2 \bar{s}_{1,1,-1,-1} \ln _2-2 \bar{s}_{1,1,-1,1} \ln _2+\frac{1}{4} s_{-1} \zeta _2^2+\frac{57}{40} s_1 \zeta _2^2+4 \text{Li}_5\left(\frac{1}{2}\right)-4 \text{Li}_4\left(\frac{1}{2}\right) s_1+\frac{s_3 \zeta _2}{2}+\frac{1}{4} s_{-2} \zeta _3-\frac{3 s_2 \zeta _3}{4}-\frac{53 \zeta _2 \zeta _3}{16}+\frac{13 \zeta _5}{8}-\frac{1}{2} \zeta _2 \bar{s}_{-3}-\frac{1}{8} \zeta _3 \bar{s}_{-2}+\frac{1}{4} \zeta _2^2 \bar{s}_{-1}+\frac{11}{5} \zeta _2^2 \bar{s}_1-6 \text{Li}_4\left(\frac{1}{2}\right) \bar{s}_1-\frac{3}{2} \zeta _3 \bar{s}_2+\frac{1}{2} \zeta _2 \bar{s}_3+s_{-4,-1}-\frac{1}{2} \zeta _2 s_{-2,-1}-\frac{1}{8} \zeta _3 s_{1,-1}+\frac{3}{2} \zeta _3 s_{1,1}-\zeta _2 s_{1,2}+\frac{1}{2} \zeta _2 s_{2,-1}+\frac{3}{2} \zeta _2 \bar{s}_{1,-2}+\frac{7}{8} \zeta _3 \bar{s}_{1,-1}+3 \zeta _3 \bar{s}_{1,1}-\frac{1}{2} \zeta _2 \bar{s}_{1,2}+\zeta _2 \bar{s}_{2,-1}-2 s_{-3,1,-1}-s_{-2,2,-1}-2 s_{1,-3,-1}+\frac{1}{2} \zeta _2 s_{1,-1,-1}-\zeta _2 s_{1,1,-1}-s_{2,-2,-1}-\bar{s}_{-3,-1,1}-\frac{1}{2} \zeta _2 \bar{s}_{1,-1,-1}-\frac{1}{2} \zeta _2 \bar{s}_{1,-1,1}-2 \zeta _2 \bar{s}_{1,1,-1}-\bar{s}_{1,3,1}-\bar{s}_{2,2,1}+2 s_{-2,1,1,-1}+3 s_{1,-2,1,-1}+s_{1,-1,2,-1}+2 s_{1,1,-2,-1}+s_{2,-1,1,-1}+\bar{s}_{-2,1,-1,1}+3 \bar{s}_{1,-2,-1,1}+\bar{s}_{1,-1,-2,1}+2 \bar{s}_{1,1,2,1}+2 \bar{s}_{2,-1,-1,1}-2 s_{1,-1,1,1,-1}-2 s_{1,1,-1,1,-1}-2 \bar{s}_{1,-1,1,-1,1}-4 \bar{s}_{1,1,-1,-1,1}  \end{dmath}
   \begin{dmath}[style={\small}]     s_{1,-1} \bar{s}_{1,1,-1}  =   -\frac{\ln _2^5}{12}-\frac{1}{4} s_1 \ln _2^4-\frac{1}{6} \bar{s}_1 \ln _2^4+\frac{23}{6} \zeta _2 \ln _2^3+\frac{1}{6} \bar{s}_{-2} \ln _2^3-\frac{1}{6} \bar{s}_2 \ln _2^3-\frac{1}{6} s_{1,-1} \ln _2^3-\frac{1}{6} s_{1,1} \ln _2^3-\frac{1}{6} \bar{s}_{1,-1} \ln _2^3+\frac{1}{6} \bar{s}_{1,1} \ln _2^3+\frac{1}{2} s_{-3} \ln _2^2-\frac{1}{2} s_3 \ln _2^2-\frac{1}{4} s_{-1} \zeta _2 \ln _2^2+\frac{21}{4} s_1 \zeta _2 \ln _2^2-\frac{21}{16} \zeta _3 \ln _2^2-\frac{1}{2} \bar{s}_{-3} \ln _2^2-\frac{1}{4} \zeta _2 \bar{s}_{-1} \ln _2^2+5 \zeta _2 \bar{s}_1 \ln _2^2+\frac{1}{2} \bar{s}_3 \ln _2^2-\frac{3}{2} s_{1,-2} \ln _2^2+\frac{3}{2} s_{1,2} \ln _2^2-s_{2,-1} \ln _2^2+2 s_{2,1} \ln _2^2-\frac{1}{2} \bar{s}_{2,-1} \ln _2^2+\frac{1}{2} \bar{s}_{2,1} \ln _2^2+\frac{3}{2} s_{1,1,-1} \ln _2^2-\frac{7}{2} s_{1,1,1} \ln _2^2+\frac{3}{2} \bar{s}_{1,1,-1} \ln _2^2-\frac{5}{2} \bar{s}_{1,1,1} \ln _2^2+\frac{29}{8} \zeta _2^2 \ln _2+s_{-4} \ln _2-s_4 \ln _2-s_{-2} \zeta _2 \ln _2-\frac{3}{2} s_2 \zeta _2 \ln _2+\frac{9}{8} s_{-1} \zeta _3 \ln _2-\frac{57}{8} s_1 \zeta _3 \ln _2-\frac{1}{2} \zeta _2 \bar{s}_{-2} \ln _2+\frac{9}{8} \zeta _3 \bar{s}_{-1} \ln _2-\frac{39}{8} \zeta _3 \bar{s}_1 \ln _2-\zeta _2 \bar{s}_2 \ln _2-2 s_{1,-3} \ln _2+\frac{3}{2} \zeta _2 s_{1,-1} \ln _2+2 \zeta _2 s_{1,1} \ln _2+2 s_{1,3} \ln _2-2 s_{2,-2} \ln _2+2 s_{2,2} \ln _2-2 s_{3,-1} \ln _2-\bar{s}_{1,-3} \ln _2+\frac{3}{2} \zeta _2 \bar{s}_{1,-1} \ln _2+\frac{5}{2} \zeta _2 \bar{s}_{1,1} \ln _2+\bar{s}_{1,3} \ln _2-\bar{s}_{2,-2} \ln _2+\bar{s}_{2,2} \ln _2+3 s_{1,1,-2} \ln _2-3 s_{1,1,2} \ln _2+4 s_{1,2,-1} \ln _2+4 s_{2,1,-1} \ln _2+3 \bar{s}_{1,1,-2} \ln _2-3 \bar{s}_{1,1,2} \ln _2+2 \bar{s}_{1,2,-1} \ln _2+2 \bar{s}_{2,1,-1} \ln _2-6 s_{1,1,1,-1} \ln _2-6 \bar{s}_{1,1,1,-1} \ln _2+\frac{12}{5} s_1 \zeta _2^2+10 \text{Li}_5\left(\frac{1}{2}\right)-6 \text{Li}_4\left(\frac{1}{2}\right) s_1+\frac{s_3 \zeta _2}{2}+\frac{3 s_2 \zeta _3}{8}-10 \zeta _5+\frac{8}{5} \zeta _2^2 \bar{s}_1-4 \text{Li}_4\left(\frac{1}{2}\right) \bar{s}_1+\frac{1}{8} \zeta _3 \bar{s}_2+s_{-4,-1}-\frac{5}{8} \zeta _3 s_{1,1}-\zeta _2 s_{1,2}-\zeta _2 s_{2,1}-\frac{1}{2} \zeta _3 \bar{s}_{1,1}-\frac{1}{2} \zeta _2 \bar{s}_{1,2}-\frac{1}{2} \zeta _2 \bar{s}_{2,1}-s_{-3,1,-1}-2 s_{1,-3,-1}+\frac{3}{2} \zeta _2 s_{1,1,1}-2 s_{2,-2,-1}-s_{3,-1,-1}-\bar{s}_{-3,1,-1}-\bar{s}_{1,-3,-1}+\frac{3}{2} \zeta _2 \bar{s}_{1,1,1}-\bar{s}_{2,-2,-1}+s_{1,-2,1,-1}+3 s_{1,1,-2,-1}+2 s_{1,2,-1,-1}+s_{2,-1,1,-1}+2 s_{2,1,-1,-1}+\bar{s}_{-2,1,1,-1}+2 \bar{s}_{1,-2,1,-1}+3 \bar{s}_{1,1,-2,-1}+\bar{s}_{1,2,-1,-1}+\bar{s}_{2,-1,1,-1}+\bar{s}_{2,1,-1,-1}-s_{1,1,-1,1,-1}-3 s_{1,1,1,-1,-1}-\bar{s}_{1,-1,1,1,-1}-2 \bar{s}_{1,1,-1,1,-1}-3 \bar{s}_{1,1,1,-1,-1}  \end{dmath}
   \begin{dmath}[style={\small}]     s_{1,-1} \bar{s}_{1,1,1}  =   \frac{\ln _2^5}{120}-\frac{1}{24} s_{-1} \ln _2^4-\frac{1}{24} \bar{s}_{-1} \ln _2^4+\frac{1}{12} \bar{s}_1 \ln _2^4-\frac{1}{12} \zeta _2 \ln _2^3+\frac{1}{6} \bar{s}_{-2} \ln _2^3-\frac{1}{6} \bar{s}_2 \ln _2^3-\frac{1}{6} s_{1,-1} \ln _2^3+\frac{1}{6} s_{1,1} \ln _2^3-\frac{1}{6} \bar{s}_{1,-1} \ln _2^3+\frac{1}{6} \bar{s}_{1,1} \ln _2^3+\frac{1}{4} s_1 \zeta _2 \ln _2^2+\frac{1}{8} \zeta _3 \ln _2^2-\frac{1}{4} \zeta _2 \bar{s}_1 \ln _2^2-\frac{1}{2} s_{1,-2} \ln _2^2+\frac{1}{2} s_{1,2} \ln _2^2-\frac{1}{2} s_{2,-1} \ln _2^2+\frac{1}{2} s_{2,1} \ln _2^2-\frac{1}{2} \bar{s}_{1,-2} \ln _2^2+\frac{1}{2} \bar{s}_{1,2} \ln _2^2-\frac{1}{2} \bar{s}_{2,-1} \ln _2^2+\frac{1}{2} \bar{s}_{2,1} \ln _2^2+s_{1,1,-1} \ln _2^2-s_{1,1,1} \ln _2^2+\bar{s}_{1,1,-1} \ln _2^2-2 \bar{s}_{1,1,1} \ln _2^2-\frac{79}{40} \zeta _2^2 \ln _2+\text{Li}_4\left(\frac{1}{2}\right) \ln _2+s_{-4} \ln _2-s_4 \ln _2-\frac{1}{2} s_{-2} \zeta _2 \ln _2+2 s_2 \zeta _2 \ln _2+\frac{1}{8} s_{-1} \zeta _3 \ln _2-\frac{29}{8} s_1 \zeta _3 \ln _2-\frac{1}{2} \zeta _2 \bar{s}_{-2} \ln _2+\frac{1}{8} \zeta _3 \bar{s}_{-1} \ln _2-\frac{5}{4} \zeta _3 \bar{s}_1 \ln _2+\frac{1}{2} \zeta _2 \bar{s}_2 \ln _2-2 s_{1,-3} \ln _2+\zeta _2 s_{1,-1} \ln _2-\frac{5}{2} \zeta _2 s_{1,1} \ln _2+2 s_{1,3} \ln _2-2 s_{2,-2} \ln _2+2 s_{2,2} \ln _2-s_{3,-1} \ln _2+s_{3,1} \ln _2+\zeta _2 \bar{s}_{1,-1} \ln _2-3 \zeta _2 \bar{s}_{1,1} \ln _2+3 s_{1,1,-2} \ln _2-3 s_{1,1,2} \ln _2+2 s_{1,2,-1} \ln _2-2 s_{1,2,1} \ln _2+2 s_{2,1,-1} \ln _2-2 s_{2,1,1} \ln _2+\bar{s}_{1,2,-1} \ln _2+\bar{s}_{1,2,1} \ln _2+\bar{s}_{2,1,-1} \ln _2+\bar{s}_{2,1,1} \ln _2-3 s_{1,1,1,-1} \ln _2+3 s_{1,1,1,1} \ln _2-3 \bar{s}_{1,1,1,-1} \ln _2-3 \bar{s}_{1,1,1,1} \ln _2-\frac{1}{20} s_{-1} \zeta _2^2+\frac{1}{4} s_1 \zeta _2^2+2 \text{Li}_5\left(\frac{1}{2}\right)-\text{Li}_4\left(\frac{1}{2}\right) s_{-1}-\frac{1}{2} s_{-3} \zeta _2+s_{-2} \zeta _3-\frac{s_2 \zeta _3}{2}-\frac{9 \zeta _2 \zeta _3}{16}+\frac{\zeta _5}{8}+\frac{7}{8} \zeta _3 \bar{s}_{-2}-\frac{1}{20} \zeta _2^2 \bar{s}_{-1}-\text{Li}_4\left(\frac{1}{2}\right) \bar{s}_{-1}-\frac{19}{40} \zeta _2^2 \bar{s}_1+2 \text{Li}_4\left(\frac{1}{2}\right) \bar{s}_1+\frac{1}{8} \zeta _3 \bar{s}_2+\zeta _2 s_{1,-2}-\frac{15}{8} \zeta _3 s_{1,-1}+\frac{3}{4} \zeta _3 s_{1,1}+\zeta _2 s_{2,-1}-s_{4,-1}+\frac{1}{2} \zeta _2 \bar{s}_{1,-2}-\frac{15}{8} \zeta _3 \bar{s}_{1,-1}+\frac{1}{2} \zeta _3 \bar{s}_{1,1}+\frac{1}{2} \zeta _2 \bar{s}_{1,2}+\frac{1}{2} \zeta _2 \bar{s}_{2,-1}+\frac{1}{2} \zeta _2 \bar{s}_{2,1}-\frac{3}{2} \zeta _2 s_{1,1,-1}+2 s_{1,3,-1}+2 s_{2,2,-1}+2 s_{3,1,-1}-\bar{s}_{-3,1,1}-\bar{s}_{1,-3,1}-\frac{3}{2} \zeta _2 \bar{s}_{1,1,-1}-\frac{3}{2} \zeta _2 \bar{s}_{1,1,1}-\bar{s}_{2,-2,1}-3 s_{1,1,2,-1}-3 s_{1,2,1,-1}-3 s_{2,1,1,-1}+\bar{s}_{-2,1,1,1}+2 \bar{s}_{1,-2,1,1}+3 \bar{s}_{1,1,-2,1}+\bar{s}_{1,2,-1,1}+\bar{s}_{2,-1,1,1}+\bar{s}_{2,1,-1,1}+4 s_{1,1,1,1,-1}-\bar{s}_{1,-1,1,1,1}-2 \bar{s}_{1,1,-1,1,1}-3 \bar{s}_{1,1,1,-1,1}  \end{dmath}

   \begin{dmath}[style={\small}]     s_{1,-1} \bar{s}_{-1,-1,-1}  =   \frac{5 \ln _2^5}{24}+\frac{11}{24} s_{-1} \ln _2^4-\frac{5}{12} s_1 \ln _2^4+\frac{1}{4} \bar{s}_{-1} \ln _2^4-\frac{1}{12} \bar{s}_1 \ln _2^4+\frac{2}{3} s_{-2} \ln _2^3-\frac{2}{3} s_2 \ln _2^3+\frac{25}{12} \zeta _2 \ln _2^3+\frac{2}{3} s_{-1,-1} \ln _2^3-\frac{2}{3} s_{-1,1} \ln _2^3-\frac{4}{3} s_{1,-1} \ln _2^3-\frac{2}{3} \bar{s}_{-1,-1} \ln _2^3+\frac{2}{3} \bar{s}_{-1,1} \ln _2^3+\frac{1}{2} s_{-3} \ln _2^2-\frac{1}{2} s_3 \ln _2^2+\frac{5}{2} s_{-1} \zeta _2 \ln _2^2+\frac{5}{4} s_1 \zeta _2 \ln _2^2+\frac{7}{16} \zeta _3 \ln _2^2-\bar{s}_{-3} \ln _2^2+\frac{1}{4} \zeta _2 \bar{s}_{-1} \ln _2^2-\frac{5}{4} \zeta _2 \bar{s}_1 \ln _2^2+\bar{s}_3 \ln _2^2+2 s_{-2,-1} \ln _2^2+\frac{1}{2} s_{-1,-2} \ln _2^2-\frac{1}{2} s_{-1,2} \ln _2^2-s_{1,-2} \ln _2^2+s_{1,2} \ln _2^2-\frac{1}{2} s_{2,-1} \ln _2^2+\frac{1}{2} s_{2,1} \ln _2^2+\bar{s}_{1,-2} \ln _2^2-\bar{s}_{1,2} \ln _2^2+\frac{1}{2} s_{-1,-1,-1} \ln _2^2-\frac{1}{2} s_{-1,-1,1} \ln _2^2-2 s_{-1,1,-1} \ln _2^2-2 s_{1,-1,-1} \ln _2^2+\frac{1}{2} \bar{s}_{-1,-1,-1} \ln _2^2-\frac{3}{2} \bar{s}_{-1,-1,1} \ln _2^2-\frac{9}{8} \zeta _2^2 \ln _2+s_{-4} \ln _2-s_4 \ln _2-\frac{3}{2} s_{-2} \zeta _2 \ln _2-s_2 \zeta _2 \ln _2-\frac{5}{8} s_{-1} \zeta _3 \ln _2-\frac{1}{8} s_1 \zeta _3 \ln _2-\frac{5}{8} \zeta _3 \bar{s}_{-1} \ln _2+\frac{17}{8} \zeta _3 \bar{s}_1 \ln _2+s_{-2,-2} \ln _2-s_{-2,2} \ln _2+s_{-1,-3} \ln _2+\zeta _2 s_{-1,-1} \ln _2+\frac{3}{2} \zeta _2 s_{-1,1} \ln _2-s_{-1,3} \ln _2-s_{1,-3} \ln _2+\zeta _2 s_{1,-1} \ln _2+s_{1,3} \ln _2-s_{2,-2} \ln _2+s_{2,2} \ln _2-2 s_{3,-1} \ln _2+\bar{s}_{-2,-2} \ln _2-\bar{s}_{-2,2} \ln _2+\bar{s}_{-1,-3} \ln _2+\frac{1}{2} \zeta _2 \bar{s}_{-1,-1} \ln _2+\frac{1}{2} \zeta _2 \bar{s}_{-1,1} \ln _2-\bar{s}_{-1,3} \ln _2+2 s_{-2,-1,-1} \ln _2+2 s_{-1,-2,-1} \ln _2+s_{-1,-1,-2} \ln _2-s_{-1,-1,2} \ln _2-s_{-1,1,-2} \ln _2+s_{-1,1,2} \ln _2-s_{1,-1,-2} \ln _2+s_{1,-1,2} \ln _2+2 s_{1,2,-1} \ln _2+2 s_{2,1,-1} \ln _2+2 \bar{s}_{-2,-1,-1} \ln _2+2 \bar{s}_{-1,-2,-1} \ln _2+\bar{s}_{-1,-1,-2} \ln _2-\bar{s}_{-1,-1,2} \ln _2-\bar{s}_{-1,1,-2} \ln _2+\bar{s}_{-1,1,2} \ln _2-\bar{s}_{1,-1,-2} \ln _2+\bar{s}_{1,-1,2} \ln _2-2 s_{-1,-1,1,-1} \ln _2-2 s_{-1,1,-1,-1} \ln _2-2 s_{1,-1,-1,-1} \ln _2-2 \bar{s}_{-1,-1,1,-1} \ln _2-2 \bar{s}_{-1,1,-1,-1} \ln _2-2 \bar{s}_{1,-1,-1,-1} \ln _2-\frac{1}{5} s_1 \zeta _2^2-5 \text{Li}_5\left(\frac{1}{2}\right)+\text{Li}_4\left(\frac{1}{2}\right) s_1+\frac{s_3 \zeta _2}{2}+\frac{1}{2} s_{-2} \zeta _3-\frac{s_2 \zeta _3}{8}+\frac{145 \zeta _5}{32}+\zeta _3 \bar{s}_{-2}-\frac{1}{5} \zeta _2^2 \bar{s}_{-1}+\text{Li}_4\left(\frac{1}{2}\right) \bar{s}_{-1}-\frac{17}{20} \zeta _2^2 \bar{s}_1+3 \text{Li}_4\left(\frac{1}{2}\right) \bar{s}_1+s_{-4,-1}-\frac{1}{2} \zeta _2 s_{-2,-1}-\frac{1}{2} \zeta _2 s_{-1,-2}+\frac{1}{8} \zeta _3 s_{-1,-1}-\frac{1}{2} \zeta _3 s_{-1,1}-\frac{3}{4} \zeta _3 s_{1,-1}-\frac{1}{2} \zeta _2 s_{1,2}-\frac{1}{2} \zeta _2 s_{2,1}-\frac{1}{2} \zeta _2 \bar{s}_{-2,-1}-\frac{1}{2} \zeta _2 \bar{s}_{-1,-2}-\frac{1}{2} \zeta _3 \bar{s}_{-1,-1}-\frac{3}{4} \zeta _3 \bar{s}_{-1,1}-\zeta _3 \bar{s}_{1,-1}-s_{-3,1,-1}-s_{-2,2,-1}+\frac{1}{2} \zeta _2 s_{-1,-1,1}+\frac{1}{2} \zeta _2 s_{-1,1,-1}-s_{-1,3,-1}-s_{1,-3,-1}+\frac{1}{2} \zeta _2 s_{1,-1,-1}-s_{2,-2,-1}-s_{3,-1,-1}-\bar{s}_{-2,2,-1}+\frac{1}{2} \zeta _2 \bar{s}_{-1,-1,1}+\frac{1}{2} \zeta _2 \bar{s}_{-1,1,-1}-\bar{s}_{-1,3,-1}+\frac{1}{2} \zeta _2 \bar{s}_{1,-1,-1}-\bar{s}_{3,-1,-1}+s_{-2,-1,-1,-1}+s_{-1,-2,-1,-1}+s_{-1,-1,-2,-1}+s_{-1,1,2,-1}+s_{-1,2,1,-1}+s_{1,-1,2,-1}+s_{1,2,-1,-1}+s_{2,-1,1,-1}+s_{2,1,-1,-1}+3 \bar{s}_{-2,-1,-1,-1}+2 \bar{s}_{-1,-2,-1,-1}+\bar{s}_{-1,-1,-2,-1}+\bar{s}_{-1,1,2,-1}+\bar{s}_{1,-1,2,-1}+\bar{s}_{1,2,-1,-1}-s_{-1,-1,-1,1,-1}-s_{-1,-1,1,-1,-1}-s_{-1,1,-1,-1,-1}-s_{1,-1,-1,-1,-1}-\bar{s}_{-1,-1,1,-1,-1}-2 \bar{s}_{-1,1,-1,-1,-1}-3 \bar{s}_{1,-1,-1,-1,-1}  \end{dmath}
   \begin{dmath}[style={\small}]     s_{1,-1} \bar{s}_{-1,-1,1}  =   \frac{3 \ln _2^5}{20}+\frac{3}{8} s_{-1} \ln _2^4-\frac{5}{24} s_1 \ln _2^4+\frac{1}{24} \bar{s}_{-1} \ln _2^4-\frac{1}{3} \bar{s}_1 \ln _2^4+\frac{1}{3} s_{-2} \ln _2^3-\frac{1}{3} s_2 \ln _2^3-\frac{1}{2} \zeta _2 \ln _2^3+\frac{1}{3} s_{-1,-1} \ln _2^3-\frac{1}{3} s_{-1,1} \ln _2^3-s_{1,-1} \ln _2^3-\frac{2}{3} \bar{s}_{-1,-1} \ln _2^3+\frac{2}{3} \bar{s}_{-1,1} \ln _2^3-s_{-1} \zeta _2 \ln _2^2+\frac{1}{2} s_1 \zeta _2 \ln _2^2+\frac{3}{4} \zeta _3 \ln _2^2-\frac{1}{2} \bar{s}_{-3} \ln _2^2-\frac{3}{2} \zeta _2 \bar{s}_{-1} \ln _2^2-\frac{1}{2} \zeta _2 \bar{s}_1 \ln _2^2+\frac{1}{2} \bar{s}_3 \ln _2^2+s_{-2,-1} \ln _2^2-\frac{1}{2} s_{1,-2} \ln _2^2+\frac{1}{2} s_{1,2} \ln _2^2+\frac{1}{2} \bar{s}_{1,-2} \ln _2^2-\frac{1}{2} \bar{s}_{1,2} \ln _2^2-s_{-1,1,-1} \ln _2^2-s_{1,-1,-1} \ln _2^2-\bar{s}_{-1,-1,1} \ln _2^2-\frac{7}{10} \zeta _2^2 \ln _2+\text{Li}_4\left(\frac{1}{2}\right) \ln _2+s_{-4} \ln _2-s_4 \ln _2-\frac{1}{2} s_{-2} \zeta _2 \ln _2+2 s_2 \zeta _2 \ln _2-\frac{5}{8} s_{-1} \zeta _3 \ln _2+\frac{5}{8} s_1 \zeta _3 \ln _2+\frac{3}{2} \zeta _2 \bar{s}_{-2} \ln _2-\frac{7}{8} \zeta _3 \bar{s}_{-1} \ln _2+\frac{3}{2} \zeta _3 \bar{s}_1 \ln _2+s_{-2,-2} \ln _2-s_{-2,2} \ln _2+s_{-1,-3} \ln _2-2 \zeta _2 s_{-1,-1} \ln _2+\frac{1}{2} \zeta _2 s_{-1,1} \ln _2-s_{-1,3} \ln _2-s_{1,-3} \ln _2+\zeta _2 s_{1,-1} \ln _2+s_{1,3} \ln _2-s_{2,-2} \ln _2+s_{2,2} \ln _2-s_{3,-1} \ln _2+s_{3,1} \ln _2-\frac{3}{2} \zeta _2 \bar{s}_{-1,-1} \ln _2-2 \zeta _2 \bar{s}_{-1,1} \ln _2-\frac{3}{2} \zeta _2 \bar{s}_{1,-1} \ln _2+s_{-2,-1,-1} \ln _2-s_{-2,-1,1} \ln _2+s_{-1,-2,-1} \ln _2-s_{-1,-2,1} \ln _2+s_{-1,-1,-2} \ln _2-s_{-1,-1,2} \ln _2-s_{-1,1,-2} \ln _2+s_{-1,1,2} \ln _2-s_{1,-1,-2} \ln _2+s_{1,-1,2} \ln _2+s_{1,2,-1} \ln _2-s_{1,2,1} \ln _2+s_{2,1,-1} \ln _2-s_{2,1,1} \ln _2+\bar{s}_{-2,-1,-1} \ln _2+\bar{s}_{-2,-1,1} \ln _2+\bar{s}_{-1,-2,-1} \ln _2+\bar{s}_{-1,-2,1} \ln _2-s_{-1,-1,1,-1} \ln _2+s_{-1,-1,1,1} \ln _2-s_{-1,1,-1,-1} \ln _2+s_{-1,1,-1,1} \ln _2-s_{1,-1,-1,-1} \ln _2+s_{1,-1,-1,1} \ln _2-\bar{s}_{-1,-1,1,-1} \ln _2-\bar{s}_{-1,-1,1,1} \ln _2-\bar{s}_{-1,1,-1,-1} \ln _2-\bar{s}_{-1,1,-1,1} \ln _2-\bar{s}_{1,-1,-1,-1} \ln _2-\bar{s}_{1,-1,-1,1} \ln _2-\frac{3}{20} s_{-1} \zeta _2^2-\frac{61}{40} s_1 \zeta _2^2+5 \text{Li}_5\left(\frac{1}{2}\right)+4 \text{Li}_4\left(\frac{1}{2}\right) s_1-\frac{1}{2} s_{-3} \zeta _2+\frac{3}{4} s_{-2} \zeta _3-\frac{s_2 \zeta _3}{4}+\frac{5 \zeta _2 \zeta _3}{16}-\frac{269 \zeta _5}{64}+\frac{1}{2} \zeta _2 \bar{s}_{-3}-\frac{13}{8} \zeta _3 \bar{s}_{-2}+\frac{41}{40} \zeta _2^2 \bar{s}_{-1}-4 \text{Li}_4\left(\frac{1}{2}\right) \bar{s}_{-1}+\frac{7}{20} \zeta _2^2 \bar{s}_1-3 \text{Li}_4\left(\frac{1}{2}\right) \bar{s}_1-\frac{1}{2} \zeta _2 \bar{s}_3-\frac{1}{2} \zeta _2 s_{-2,-1}+\frac{1}{4} \zeta _3 s_{-1,-1}-\frac{3}{4} \zeta _3 s_{-1,1}+\frac{1}{2} \zeta _2 s_{-1,2}+\frac{1}{2} \zeta _2 s_{1,-2}-\frac{5}{8} \zeta _3 s_{1,-1}+\frac{1}{2} \zeta _2 s_{2,-1}-s_{4,-1}+\zeta _2 \bar{s}_{-2,-1}+\zeta _2 \bar{s}_{-1,-2}+\frac{3}{4} \zeta _3 \bar{s}_{-1,-1}+\frac{3}{2} \zeta _3 \bar{s}_{-1,1}-\frac{1}{2} \zeta _2 \bar{s}_{1,-2}+\frac{13}{8} \zeta _3 \bar{s}_{1,-1}+\frac{1}{2} \zeta _2 \bar{s}_{1,2}+s_{-2,-2,-1}+s_{-1,-3,-1}-\frac{1}{2} \zeta _2 s_{-1,-1,-1}+\frac{1}{2} \zeta _2 s_{-1,1,-1}+\frac{1}{2} \zeta _2 s_{1,-1,-1}+s_{1,3,-1}+s_{2,2,-1}+2 s_{3,1,-1}-\bar{s}_{-2,2,1}-\frac{1}{2} \zeta _2 \bar{s}_{-1,-1,-1}-\frac{1}{2} \zeta _2 \bar{s}_{-1,-1,1}-\zeta _2 \bar{s}_{-1,1,-1}-\bar{s}_{-1,3,1}-\zeta _2 \bar{s}_{1,-1,-1}-\bar{s}_{3,-1,1}-s_{-2,-1,1,-1}-2 s_{-1,-2,1,-1}-s_{-1,-1,2,-1}-s_{-1,1,-2,-1}-s_{1,-1,-2,-1}-s_{1,2,1,-1}-2 s_{2,1,1,-1}+3 \bar{s}_{-2,-1,-1,1}+2 \bar{s}_{-1,-2,-1,1}+\bar{s}_{-1,-1,-2,1}+\bar{s}_{-1,1,2,1}+\bar{s}_{1,-1,2,1}+\bar{s}_{1,2,-1,1}+2 s_{-1,-1,1,1,-1}+s_{-1,1,-1,1,-1}+s_{1,-1,-1,1,-1}-\bar{s}_{-1,-1,1,-1,1}-2 \bar{s}_{-1,1,-1,-1,1}-3 \bar{s}_{1,-1,-1,-1,1}  \end{dmath}
   \begin{dmath}[style={\small}]     \bar{s}_{-3} s_{1,1}  =   -\zeta _2 \bar{s}_{1,-2}-\frac{7}{4} \zeta _3 \bar{s}_{1,-1}+\frac{3}{4} \zeta _3 \bar{s}_{1,1}-\bar{s}_{2,-3}+\bar{s}_{1,1,-3}-2 \text{Li}_4\left(\frac{1}{2}\right) \bar{s}_{-1}+2 \text{Li}_4\left(\frac{1}{2}\right) \bar{s}_1-\frac{1}{4} \zeta _3 \bar{s}_{-2}+\frac{1}{10} \zeta _2^2 \bar{s}_{-1}-\frac{5}{4} \zeta _2^2 \bar{s}_1-\frac{3}{4} \zeta _3 \bar{s}_2+\frac{1}{2} \zeta _2 \ln _2^2 \bar{s}_{-1}-\frac{1}{2} \zeta _2 \ln _2^2 \bar{s}_1-\frac{1}{12} \ln _2^4 \bar{s}_{-1}+\frac{1}{12} \ln _2^4 \bar{s}_1+\zeta _2 s_{1,-2}-\frac{7}{4} \zeta _3 s_{1,-1}-\frac{3}{4} \zeta _3 s_{1,1}+s_{-4,1}-s_{-3,1,1}-s_{1,-3,1}-\frac{3 \zeta _2 \zeta _3}{4}-\frac{39 \zeta _5}{16}+\frac{1}{3} \zeta _2 \ln _2^3+\frac{1}{5} \zeta _2^2 \ln _2-2 \text{Li}_4\left(\frac{1}{2}\right) s_{-1}+2 \text{Li}_4\left(\frac{1}{2}\right) s_1+4 \text{Li}_5\left(\frac{1}{2}\right)+\frac{1}{10} \zeta _2^2 s_{-1}-\frac{3}{5} \zeta _2^2 s_1-\zeta _2 s_{-3}+2 \zeta _3 s_{-2}+\frac{1}{2} \zeta _2 s_{-1} \ln _2^2-\frac{1}{2} \zeta _2 s_1 \ln _2^2-\frac{1}{12} s_{-1} \ln _2^4+\frac{1}{12} s_1 \ln _2^4-\frac{\ln _2^5}{30}  \end{dmath}
   \begin{dmath}[style={\small}]     \bar{s}_3 s_{1,1}  =   -\zeta _2 \bar{s}_{1,2}-\zeta _3 \bar{s}_{1,1}-\bar{s}_{2,3}+\bar{s}_{1,1,3}+\frac{11}{10} \zeta _2^2 \bar{s}_1-\zeta _3 \bar{s}_2-\zeta _2 s_{1,2}+\zeta _3 s_{1,1}-s_{4,1}+s_{1,3,1}+s_{3,1,1}-3 \zeta _3 \zeta _2+\frac{15 \zeta _5}{2}+\frac{1}{2} \zeta _2^2 s_1+\zeta _2 s_3-2 \zeta _3 s_2  \end{dmath}
   \begin{dmath}[style={\small}]     s_{1,1} \bar{s}_{-2,-1}  =   -\frac{\ln _2^5}{20}-\frac{1}{8} s_1 \ln _2^4-\frac{1}{8} \bar{s}_1 \ln _2^4+\frac{1}{6} s_{-2} \ln _2^3-\frac{1}{6} s_2 \ln _2^3-\zeta _2 \ln _2^3+\frac{1}{6} \bar{s}_{-2} \ln _2^3-\frac{1}{6} \bar{s}_2 \ln _2^3+\frac{1}{2} s_{-3} \ln _2^2-\frac{1}{2} s_3 \ln _2^2-\frac{3}{2} s_{-1} \zeta _2 \ln _2^2+\frac{9}{4} s_1 \zeta _2 \ln _2^2+\frac{21}{8} \zeta _3 \ln _2^2+\frac{1}{2} \bar{s}_{-3} \ln _2^2-\frac{3}{2} \zeta _2 \bar{s}_{-1} \ln _2^2+\frac{9}{4} \zeta _2 \bar{s}_1 \ln _2^2-\frac{1}{2} \bar{s}_3 \ln _2^2+\frac{1}{2} s_{-2,-1} \ln _2^2-\frac{1}{2} s_{-2,1} \ln _2^2-\frac{1}{2} s_{1,-2} \ln _2^2+\frac{1}{2} s_{1,2} \ln _2^2+\frac{1}{2} \bar{s}_{-2,-1} \ln _2^2-\frac{1}{2} \bar{s}_{-2,1} \ln _2^2-\frac{1}{2} \bar{s}_{1,-2} \ln _2^2+\frac{1}{2} \bar{s}_{1,2} \ln _2^2+\frac{97}{40} \zeta _2^2 \ln _2-\frac{3}{2} s_{-2} \zeta _2 \ln _2-\frac{1}{2} s_2 \zeta _2 \ln _2+\frac{21}{8} s_{-1} \zeta _3 \ln _2-\frac{5}{4} s_1 \zeta _3 \ln _2+\bar{s}_{-4} \ln _2-\zeta _2 \bar{s}_{-2} \ln _2+\frac{21}{8} \zeta _3 \bar{s}_{-1} \ln _2-\frac{7}{4} \zeta _3 \bar{s}_1 \ln _2+\zeta _2 \bar{s}_2 \ln _2-\bar{s}_4 \ln _2+s_{-3,-1} \ln _2+s_{-3,1} \ln _2+\frac{3}{2} \zeta _2 s_{1,-1} \ln _2+\frac{3}{2} \zeta _2 s_{1,1} \ln _2+\bar{s}_{-3,-1} \ln _2-\bar{s}_{-3,1} \ln _2+\bar{s}_{-2,-2} \ln _2-\bar{s}_{-2,2} \ln _2-\bar{s}_{1,-3} \ln _2+\frac{3}{2} \zeta _2 \bar{s}_{1,-1} \ln _2-\frac{3}{2} \zeta _2 \bar{s}_{1,1} \ln _2+\bar{s}_{1,3} \ln _2-s_{-2,1,-1} \ln _2-s_{-2,1,1} \ln _2-s_{1,-2,-1} \ln _2-s_{1,-2,1} \ln _2-\bar{s}_{-2,1,-1} \ln _2+\bar{s}_{-2,1,1} \ln _2-\bar{s}_{1,-2,-1} \ln _2+\bar{s}_{1,-2,1} \ln _2+\frac{29}{40} s_1 \zeta _2^2+6 \text{Li}_5\left(\frac{1}{2}\right)-3 \text{Li}_4\left(\frac{1}{2}\right) s_1+\frac{1}{2} s_{-3} \zeta _2+\frac{s_3 \zeta _2}{2}-\frac{1}{8} s_{-2} \zeta _3-\frac{s_2 \zeta _3}{4}-\frac{21 \zeta _2 \zeta _3}{8}-\frac{119 \zeta _5}{64}+\frac{1}{4} \zeta _3 \bar{s}_{-2}+\frac{23}{20} \zeta _2^2 \bar{s}_1-3 \text{Li}_4\left(\frac{1}{2}\right) \bar{s}_1-\frac{7}{8} \zeta _3 \bar{s}_2+\frac{1}{2} \zeta _2 \bar{s}_3-\frac{1}{2} \zeta _2 s_{-2,-1}-\frac{1}{2} \zeta _2 s_{-2,1}-\frac{1}{2} \zeta _2 s_{1,-2}-\frac{5}{8} \zeta _3 s_{1,1}-\frac{1}{2} \zeta _2 s_{1,2}-s_{4,1}+\bar{s}_{-4,-1}-\frac{1}{2} \zeta _2 \bar{s}_{-2,-1}+\frac{5}{8} \zeta _3 \bar{s}_{1,1}-\frac{1}{2} \zeta _2 \bar{s}_{1,2}+s_{-3,-1,1}+s_{-2,-2,1}+s_{1,3,1}+s_{3,1,1}-\bar{s}_{-3,1,-1}-\bar{s}_{-2,2,-1}-\bar{s}_{1,-3,-1}-\bar{s}_{2,-2,-1}-s_{-2,-1,1,1}-s_{-2,1,-1,1}-s_{1,-2,-1,1}+\bar{s}_{-2,1,1,-1}+\bar{s}_{1,-2,1,-1}+\bar{s}_{1,1,-2,-1}  \end{dmath}
   \begin{dmath}[style={\small}]     s_{1,1} \bar{s}_{-2,1}  =   \zeta _2 \bar{s}_{-2,1}+\zeta _2 \bar{s}_{1,-2}-\frac{21}{8} \zeta _3 \bar{s}_{1,-1}+\frac{5}{8} \zeta _3 \bar{s}_{1,1}+\bar{s}_{-4,1}-\bar{s}_{-3,1,1}-\bar{s}_{-2,2,1}-\bar{s}_{1,-3,1}-\bar{s}_{2,-2,1}+\bar{s}_{-2,1,1,1}+\bar{s}_{1,-2,1,1}+\bar{s}_{1,1,-2,1}-3 \text{Li}_4\left(\frac{1}{2}\right) \bar{s}_{-1}+3 \text{Li}_4\left(\frac{1}{2}\right) \bar{s}_1-\zeta _2 \bar{s}_{-3}+\frac{21}{8} \zeta _3 \bar{s}_{-2}-\frac{39}{40} \zeta _2^2 \bar{s}_{-1}-\frac{13}{20} \zeta _2^2 \bar{s}_1-\frac{5}{8} \zeta _3 \bar{s}_2+\frac{3}{4} \zeta _2 \ln _2^2 \bar{s}_{-1}-\frac{3}{4} \zeta _2 \ln _2^2 \bar{s}_1-\frac{1}{8} \ln _2^4 \bar{s}_{-1}+\frac{1}{8} \ln _2^4 \bar{s}_1+\zeta _2 s_{-2,1}+\zeta _2 s_{1,-2}-\frac{21}{8} \zeta _3 s_{1,-1}-\frac{5}{8} \zeta _3 s_{1,1}+s_{-4,1}-2 s_{-3,1,1}-s_{-2,2,1}-s_{1,-3,1}+2 s_{-2,1,1,1}+s_{1,-2,1,1}-\frac{11 \zeta _5}{16}+\frac{1}{2} \zeta _2 \ln _2^3-\frac{39}{20} \zeta _2^2 \ln _2-3 \text{Li}_4\left(\frac{1}{2}\right) s_{-1}+3 \text{Li}_4\left(\frac{1}{2}\right) s_1+6 \text{Li}_5\left(\frac{1}{2}\right)-\frac{39}{40} \zeta _2^2 s_{-1}-\frac{29}{40} \zeta _2^2 s_1-\zeta _2 s_{-3}+3 \zeta _3 s_{-2}+\frac{3}{4} \zeta _2 s_{-1} \ln _2^2-\frac{3}{4} \zeta _2 s_1 \ln _2^2-\frac{1}{8} s_{-1} \ln _2^4+\frac{1}{8} s_1 \ln _2^4-\frac{\ln _2^5}{20}  \end{dmath}
   \begin{dmath}[style={\small}]     s_{1,1} \bar{s}_{2,-1}  =   \frac{\ln _2^5}{15}+\frac{5}{24} s_{-1} \ln _2^4-\frac{1}{12} s_1 \ln _2^4+\frac{5}{24} \bar{s}_{-1} \ln _2^4+\frac{1}{6} s_{-2} \ln _2^3-\frac{1}{6} s_2 \ln _2^3+\frac{1}{2} \zeta _2 \ln _2^3-\frac{1}{6} \bar{s}_{-2} \ln _2^3+\frac{1}{6} \bar{s}_2 \ln _2^3+\frac{1}{2} s_{-3} \ln _2^2-\frac{1}{2} s_3 \ln _2^2+\frac{1}{4} s_{-1} \zeta _2 \ln _2^2-s_1 \zeta _2 \ln _2^2-\frac{1}{2} \bar{s}_{-3} \ln _2^2+\frac{1}{4} \zeta _2 \bar{s}_{-1} \ln _2^2-\frac{3}{2} \zeta _2 \bar{s}_1 \ln _2^2+\frac{1}{2} \bar{s}_3 \ln _2^2-\frac{1}{2} s_{1,-2} \ln _2^2+\frac{1}{2} s_{1,2} \ln _2^2-\frac{1}{2} s_{2,-1} \ln _2^2+\frac{1}{2} s_{2,1} \ln _2^2+\frac{1}{2} \bar{s}_{1,-2} \ln _2^2-\frac{1}{2} \bar{s}_{1,2} \ln _2^2+\frac{1}{2} \bar{s}_{2,-1} \ln _2^2-\frac{1}{2} \bar{s}_{2,1} \ln _2^2-\frac{207}{40} \zeta _2^2 \ln _2-2 \text{Li}_4\left(\frac{1}{2}\right) \ln _2+\frac{1}{2} s_{-2} \zeta _2 \ln _2+\frac{3}{2} s_2 \zeta _2 \ln _2+\frac{7}{8} s_{-1} \zeta _3 \ln _2-\frac{9}{4} s_1 \zeta _3 \ln _2-\bar{s}_{-4} \ln _2+\zeta _2 \bar{s}_{-2} \ln _2+\frac{7}{8} \zeta _3 \bar{s}_{-1} \ln _2-\zeta _2 \bar{s}_2 \ln _2+\bar{s}_4 \ln _2-\frac{3}{2} \zeta _2 s_{1,-1} \ln _2-\frac{3}{2} \zeta _2 s_{1,1} \ln _2-s_{3,-1} \ln _2-s_{3,1} \ln _2+\bar{s}_{1,-3} \ln _2-\frac{3}{2} \zeta _2 \bar{s}_{1,-1} \ln _2+\frac{3}{2} \zeta _2 \bar{s}_{1,1} \ln _2-\bar{s}_{1,3} \ln _2+\bar{s}_{2,-2} \ln _2-\bar{s}_{2,2} \ln _2+\bar{s}_{3,-1} \ln _2-\bar{s}_{3,1} \ln _2+s_{1,2,-1} \ln _2+s_{1,2,1} \ln _2+s_{2,1,-1} \ln _2+s_{2,1,1} \ln _2-\bar{s}_{1,2,-1} \ln _2+\bar{s}_{1,2,1} \ln _2-\bar{s}_{2,1,-1} \ln _2+\bar{s}_{2,1,1} \ln _2-\frac{11}{8} s_{-1} \zeta _2^2+\frac{13}{20} s_1 \zeta _2^2-18 \text{Li}_5\left(\frac{1}{2}\right)+5 \text{Li}_4\left(\frac{1}{2}\right) s_{-1}-2 \text{Li}_4\left(\frac{1}{2}\right) s_1-\frac{1}{2} s_{-3} \zeta _2-\frac{s_3 \zeta _2}{2}+\frac{1}{4} s_{-2} \zeta _3+\frac{s_2 \zeta _3}{8}+\frac{67 \zeta _5}{4}+\frac{1}{2} \zeta _2 \bar{s}_{-3}-\frac{9}{8} \zeta _3 \bar{s}_{-2}-\frac{11}{8} \zeta _2^2 \bar{s}_{-1}+5 \text{Li}_4\left(\frac{1}{2}\right) \bar{s}_{-1}-\frac{1}{8} \zeta _2^2 \bar{s}_1+\frac{1}{2} \zeta _3 \bar{s}_2+s_{-4,1}+\frac{1}{2} \zeta _2 s_{1,-2}+\frac{7}{8} \zeta _3 s_{1,-1}+\frac{1}{4} \zeta _3 s_{1,1}+\frac{1}{2} \zeta _2 s_{1,2}+\frac{1}{2} \zeta _2 s_{2,-1}+\frac{1}{2} \zeta _2 s_{2,1}-\frac{1}{2} \zeta _2 \bar{s}_{1,-2}+\frac{7}{8} \zeta _3 \bar{s}_{1,-1}-\frac{1}{4} \zeta _3 \bar{s}_{1,1}-\frac{1}{2} \zeta _2 \bar{s}_{2,-1}+\bar{s}_{4,-1}-s_{-3,1,1}-s_{1,-3,1}-s_{2,-2,1}-s_{3,-1,1}-\bar{s}_{1,3,-1}-2 \bar{s}_{2,2,-1}-\bar{s}_{3,1,-1}+s_{1,2,-1,1}+s_{2,-1,1,1}+s_{2,1,-1,1}+\bar{s}_{1,1,2,-1}+\bar{s}_{1,2,1,-1}+\bar{s}_{2,1,1,-1}  \end{dmath}
   \begin{dmath}[style={\small}]     s_{1,1} \bar{s}_{2,1}  =   \zeta _2 \bar{s}_{1,2}+\zeta _2 \bar{s}_{2,1}-2 \zeta _3 \bar{s}_{1,1}+\bar{s}_{4,1}-\bar{s}_{1,3,1}-2 \bar{s}_{2,2,1}-\bar{s}_{3,1,1}+\bar{s}_{1,1,2,1}+\bar{s}_{1,2,1,1}+\bar{s}_{2,1,1,1}-\frac{1}{2} \zeta _2^2 \bar{s}_1-\zeta _2 \bar{s}_3+2 \zeta _3 \bar{s}_2-\zeta _2 s_{1,2}-\zeta _2 s_{2,1}+2 \zeta _3 s_{1,1}-s_{4,1}+s_{1,3,1}+s_{2,2,1}+2 s_{3,1,1}-s_{1,2,1,1}-2 s_{2,1,1,1}+\frac{5 \zeta _5}{2}+\frac{17}{10} \zeta _2^2 s_1+\zeta _2 s_3-3 \zeta _3 s_2  \end{dmath}
   \begin{dmath}[style={\small}]     s_{1,1} \bar{s}_{-1,1,-1}  =   \frac{\ln _2^5}{12}-\frac{1}{24} s_1 \ln _2^4+\frac{1}{8} \bar{s}_{-1} \ln _2^4-\frac{5}{24} \bar{s}_1 \ln _2^4-\frac{5}{6} \zeta _2 \ln _2^3+\frac{1}{6} \bar{s}_{-2} \ln _2^3-\frac{1}{6} \bar{s}_2 \ln _2^3+\frac{1}{6} s_{1,-1} \ln _2^3-\frac{1}{6} s_{1,1} \ln _2^3-\frac{1}{6} \bar{s}_{1,-1} \ln _2^3+\frac{1}{6} \bar{s}_{1,1} \ln _2^3+\frac{1}{2} s_{-3} \ln _2^2-\frac{1}{2} s_3 \ln _2^2-\frac{1}{2} s_{-1} \zeta _2 \ln _2^2+\frac{7}{4} s_1 \zeta _2 \ln _2^2+\frac{23}{16} \zeta _3 \ln _2^2+\frac{1}{2} \bar{s}_{-3} \ln _2^2-\frac{3}{4} \zeta _2 \bar{s}_{-1} \ln _2^2-\frac{1}{4} \zeta _2 \bar{s}_1 \ln _2^2-\frac{1}{2} \bar{s}_3 \ln _2^2+\frac{1}{2} s_{-2,-1} \ln _2^2-\frac{3}{2} s_{-2,1} \ln _2^2+\frac{1}{2} s_{-1,-2} \ln _2^2-\frac{1}{2} s_{-1,2} \ln _2^2-\frac{1}{2} s_{1,-2} \ln _2^2+\frac{1}{2} s_{1,2} \ln _2^2-\frac{1}{2} \bar{s}_{1,-2} \ln _2^2+\frac{1}{2} \bar{s}_{1,2} \ln _2^2-\frac{1}{2} s_{-1,1,-1} \ln _2^2+\frac{3}{2} s_{-1,1,1} \ln _2^2+s_{1,-1,1} \ln _2^2+\frac{1}{2} \bar{s}_{-1,1,-1} \ln _2^2-\frac{1}{2} \bar{s}_{-1,1,1} \ln _2^2+\frac{47}{40} \zeta _2^2 \ln _2+\text{Li}_4\left(\frac{1}{2}\right) \ln _2-\frac{5}{2} s_{-2} \zeta _2 \ln _2+\frac{1}{2} s_2 \zeta _2 \ln _2+3 s_{-1} \zeta _3 \ln _2-\frac{1}{2} s_1 \zeta _3 \ln _2+\zeta _2 \bar{s}_{-2} \ln _2-\zeta _3 \bar{s}_{-1} \ln _2-\frac{13}{8} \zeta _3 \bar{s}_1 \ln _2+\frac{1}{2} \zeta _2 \bar{s}_2 \ln _2+s_{-3,-1} \ln _2+s_{-3,1} \ln _2-\frac{1}{2} \zeta _2 s_{-1,-1} \ln _2+\frac{5}{2} \zeta _2 s_{-1,1} \ln _2+\frac{5}{2} \zeta _2 s_{1,-1} \ln _2+\frac{1}{2} \zeta _2 s_{1,1} \ln _2-\bar{s}_{-2,-2} \ln _2+\bar{s}_{-2,2} \ln _2-\bar{s}_{-1,-3} \ln _2+\frac{1}{2} \zeta _2 \bar{s}_{-1,-1} \ln _2-2 \zeta _2 \bar{s}_{-1,1} \ln _2+\bar{s}_{-1,3} \ln _2-\zeta _2 \bar{s}_{1,-1} \ln _2-\frac{1}{2} \zeta _2 \bar{s}_{1,1} \ln _2-2 s_{-2,1,-1} \ln _2-2 s_{-2,1,1} \ln _2-s_{-1,2,-1} \ln _2-s_{-1,2,1} \ln _2-s_{1,-2,-1} \ln _2-s_{1,-2,1} \ln _2+\bar{s}_{-2,1,-1} \ln _2-\bar{s}_{-2,1,1} \ln _2+2 \bar{s}_{-1,1,-2} \ln _2-2 \bar{s}_{-1,1,2} \ln _2+\bar{s}_{-1,2,-1} \ln _2-\bar{s}_{-1,2,1} \ln _2+\bar{s}_{1,-1,-2} \ln _2-\bar{s}_{1,-1,2} \ln _2+2 s_{-1,1,1,-1} \ln _2+2 s_{-1,1,1,1} \ln _2+s_{1,-1,1,-1} \ln _2+s_{1,-1,1,1} \ln _2-2 \bar{s}_{-1,1,1,-1} \ln _2+2 \bar{s}_{-1,1,1,1} \ln _2-\bar{s}_{1,-1,1,-1} \ln _2+\bar{s}_{1,-1,1,1} \ln _2+\frac{3}{40} s_{-1} \zeta _2^2+\frac{1}{8} s_1 \zeta _2^2+5 \text{Li}_5\left(\frac{1}{2}\right)-\text{Li}_4\left(\frac{1}{2}\right) s_{-1}+\frac{1}{2} s_{-3} \zeta _2+\frac{s_3 \zeta _2}{2}+\frac{5}{8} s_{-2} \zeta _3-s_2 \zeta _3-\frac{21 \zeta _2 \zeta _3}{16}-\frac{135 \zeta _5}{64}-\frac{1}{4} \zeta _3 \bar{s}_{-2}+\frac{1}{8} \zeta _2^2 \bar{s}_{-1}+\frac{33}{40} \zeta _2^2 \bar{s}_1-2 \text{Li}_4\left(\frac{1}{2}\right) \bar{s}_1+\frac{1}{8} \zeta _3 \bar{s}_2-\zeta _2 s_{-2,-1}-\zeta _2 s_{-2,1}-\frac{1}{2} \zeta _2 s_{-1,-2}+\zeta _3 s_{-1,-1}-\frac{5}{8} \zeta _3 s_{-1,1}-\frac{1}{2} \zeta _2 s_{-1,2}-\frac{1}{2} \zeta _2 s_{1,-2}-\frac{3}{4} \zeta _3 s_{1,-1}+\frac{1}{8} \zeta _3 s_{1,1}-\frac{1}{2} \zeta _2 s_{1,2}-s_{4,1}+\frac{1}{2} \zeta _2 \bar{s}_{-2,-1}+\frac{1}{2} \zeta _2 \bar{s}_{-1,-2}-\zeta _3 \bar{s}_{-1,-1}+\frac{1}{2} \zeta _3 \bar{s}_{-1,1}+\frac{1}{4} \zeta _3 \bar{s}_{1,-1}-\frac{1}{8} \zeta _3 \bar{s}_{1,1}+s_{-3,-1,1}+2 s_{-2,-2,1}+s_{-1,-3,1}+\zeta _2 s_{-1,1,-1}+\zeta _2 s_{-1,1,1}+\frac{1}{2} \zeta _2 s_{1,-1,-1}+\frac{1}{2} \zeta _2 s_{1,-1,1}+s_{1,3,1}+s_{3,1,1}+\bar{s}_{-3,1,-1}+\bar{s}_{-2,2,-1}-\zeta _2 \bar{s}_{-1,1,-1}+\bar{s}_{-1,3,-1}-\frac{1}{2} \zeta _2 \bar{s}_{1,-1,-1}-s_{-2,-1,1,1}-2 s_{-2,1,-1,1}-s_{-1,-2,1,1}-2 s_{-1,1,-2,1}-s_{-1,2,-1,1}-s_{1,-2,-1,1}-s_{1,-1,-2,1}-2 \bar{s}_{-2,1,1,-1}-2 \bar{s}_{-1,1,2,-1}-2 \bar{s}_{-1,2,1,-1}-\bar{s}_{1,-2,1,-1}-\bar{s}_{1,-1,2,-1}-\bar{s}_{2,-1,1,-1}+s_{-1,1,-1,1,1}+2 s_{-1,1,1,-1,1}+s_{1,-1,1,-1,1}+3 \bar{s}_{-1,1,1,1,-1}+2 \bar{s}_{1,-1,1,1,-1}+\bar{s}_{1,1,-1,1,-1}  \end{dmath}
   \begin{dmath}[style={\small}]     s_{1,1} \bar{s}_{-1,1,1}  =   -\zeta _2 \bar{s}_{-2,1}+3 \zeta _3 \bar{s}_{-1,1}-\zeta _2 \bar{s}_{-1,2}+\frac{1}{8} \zeta _3 \bar{s}_{1,-1}+\frac{7}{8} \zeta _3 \bar{s}_{1,1}+2 \zeta _2 \bar{s}_{-1,1,1}+\zeta _2 \bar{s}_{1,-1,1}+\frac{1}{2} \zeta _2 \ln _2 \bar{s}_{1,-1}-\frac{1}{2} \zeta _2 \ln _2 \bar{s}_{1,1}+\bar{s}_{-3,1,1}+\bar{s}_{-2,2,1}+\bar{s}_{-1,3,1}-2 \bar{s}_{-2,1,1,1}-2 \bar{s}_{-1,1,2,1}-2 \bar{s}_{-1,2,1,1}-\bar{s}_{1,-2,1,1}-\bar{s}_{1,-1,2,1}-\bar{s}_{2,-1,1,1}+3 \bar{s}_{-1,1,1,1,1}+2 \bar{s}_{1,-1,1,1,1}+\bar{s}_{1,1,-1,1,1}-\frac{1}{6} \ln _2^3 \bar{s}_{1,-1}+\frac{1}{6} \ln _2^3 \bar{s}_{1,1}-3 \text{Li}_4\left(\frac{1}{2}\right) \bar{s}_{-1}-\frac{1}{8} \zeta _3 \bar{s}_{-2}+\frac{6}{5} \zeta _2^2 \bar{s}_{-1}+\frac{11}{20} \zeta _2^2 \bar{s}_1-\frac{7}{8} \zeta _3 \bar{s}_2-\frac{1}{2} \zeta _2 \ln _2 \bar{s}_{-2}+\frac{1}{2} \zeta _2 \ln _2 \bar{s}_2-\frac{1}{8} \ln _2^4 \bar{s}_1+\frac{1}{6} \ln _2^3 \bar{s}_{-2}-\frac{1}{6} \ln _2^3 \bar{s}_2+2 \zeta _2 s_{-2,1}-3 \zeta _3 s_{-1,1}+\zeta _2 s_{-1,2}+\zeta _2 s_{1,-2}-\frac{23}{8} \zeta _3 s_{1,-1}-\frac{7}{8} \zeta _3 s_{1,1}-2 \zeta _2 s_{-1,1,1}-\zeta _2 s_{1,-1,1}+\frac{1}{2} \zeta _2 \ln _2 s_{1,-1}+\frac{1}{2} \zeta _2 \ln _2 s_{1,1}+s_{-4,1}-2 s_{-3,1,1}-2 s_{-2,2,1}-s_{-1,3,1}-s_{1,-3,1}+3 s_{-2,1,1,1}+2 s_{-1,1,2,1}+2 s_{-1,2,1,1}+s_{1,-2,1,1}+s_{1,-1,2,1}-3 s_{-1,1,1,1,1}-s_{1,-1,1,1,1}-\frac{1}{6} \ln _2^3 s_{1,-1}-\frac{1}{6} \ln _2^3 s_{1,1}-\frac{407 \zeta _5}{64}-\frac{1}{4} \zeta _2 \ln _2^3+\frac{39}{40} \zeta _2^2 \ln _2-3 \text{Li}_4\left(\frac{1}{2}\right) s_{-1}+3 \text{Li}_4\left(\frac{1}{2}\right) s_1+9 \text{Li}_5\left(\frac{1}{2}\right)-\frac{6}{5} \zeta _2^2 s_{-1}-\frac{6}{5} \zeta _2^2 s_1-\zeta _2 s_{-3}+3 \zeta _3 s_{-2}+\frac{\ln _2^5}{40}  \end{dmath}
   \begin{dmath}[style={\small}]     s_{1,1} \bar{s}_{1,-2}  =   \zeta _2 \bar{s}_{1,-2}-\frac{35}{8} \zeta _3 \bar{s}_{1,-1}+\frac{3}{8} \zeta _3 \bar{s}_{1,1}-\zeta _2 \bar{s}_{1,2}+\frac{3}{2} \zeta _2 \bar{s}_{2,-1}-\frac{1}{2} \zeta _2 \bar{s}_{2,1}-3 \zeta _2 \bar{s}_{1,1,-1}+\zeta _2 \bar{s}_{1,1,1}+3 \zeta _2 \ln _2 \bar{s}_{1,-1}-3 \zeta _2 \ln _2 \bar{s}_{1,1}+\bar{s}_{3,-2}-2 \bar{s}_{1,2,-2}-2 \bar{s}_{2,1,-2}+3 \bar{s}_{1,1,1,-2}-5 \text{Li}_4\left(\frac{1}{2}\right) \bar{s}_{-1}+5 \text{Li}_4\left(\frac{1}{2}\right) \bar{s}_1-\frac{1}{2} \zeta _2 \bar{s}_{-3}+\frac{7}{4} \zeta _3 \bar{s}_{-2}+\zeta _2^2 \bar{s}_{-1}-\frac{41}{20} \zeta _2^2 \bar{s}_1-\frac{1}{4} \zeta _3 \bar{s}_2+\frac{1}{2} \zeta _2 \bar{s}_3-\frac{1}{4} \zeta _2 \ln _2^2 \bar{s}_{-1}+\frac{1}{4} \zeta _2 \ln _2^2 \bar{s}_1-\frac{3}{2} \zeta _2 \ln _2 \bar{s}_{-2}+\frac{3}{2} \zeta _2 \ln _2 \bar{s}_2-\frac{5}{24} \ln _2^4 \bar{s}_{-1}+\frac{5}{24} \ln _2^4 \bar{s}_1+2 \zeta _2 s_{1,-2}-\frac{35}{8} \zeta _3 s_{1,-1}+\frac{9}{8} \zeta _3 s_{1,1}+\frac{3}{2} \zeta _2 s_{2,-1}+\frac{1}{2} \zeta _2 s_{2,1}-3 \zeta _2 s_{1,1,-1}-\zeta _2 s_{1,1,1}+3 \zeta _2 \ln _2 s_{1,-1}-3 \zeta _2 \ln _2 s_{1,1}+s_{-4,1}-s_{-3,1,1}-2 s_{1,-3,1}-s_{2,-2,1}+s_{1,-2,1,1}+2 s_{1,1,-2,1}-3 \zeta _2 \zeta _3-\frac{185 \zeta _5}{32}-\frac{1}{6} \zeta _2 \ln _2^3+2 \zeta _2^2 \ln _2-5 \text{Li}_4\left(\frac{1}{2}\right) s_{-1}+5 \text{Li}_4\left(\frac{1}{2}\right) s_1+10 \text{Li}_5\left(\frac{1}{2}\right)+\zeta _2^2 s_{-1}-\frac{67}{40} \zeta _2^2 s_1-\zeta _2 s_{-3}+\frac{21}{8} \zeta _3 s_{-2}-\frac{5 \zeta _3 s_2}{8}-\frac{1}{4} \zeta _2 s_{-1} \ln _2^2+\frac{1}{4} \zeta _2 s_1 \ln _2^2-\frac{3}{2} \zeta _2 s_{-2} \ln _2+\frac{3}{2} \zeta _2 s_2 \ln _2-\frac{5}{24} s_{-1} \ln _2^4+\frac{5}{24} s_1 \ln _2^4-\frac{\ln _2^5}{12}  \end{dmath}
   \begin{dmath}[style={\small}]     s_{1,1} \bar{s}_{1,2}  =   \zeta _2 \bar{s}_{2,1}-2 \zeta _2 \bar{s}_{1,1,1}+3 \zeta _3 \bar{s}_{1,1}+\bar{s}_{3,2}-2 \bar{s}_{1,2,2}-2 \bar{s}_{2,1,2}+3 \bar{s}_{1,1,1,2}+\frac{6}{5} \zeta _2^2 \bar{s}_1-2 \zeta _3 \bar{s}_2-2 \zeta _2 s_{1,2}-\zeta _2 s_{2,1}+2 \zeta _2 s_{1,1,1}+3 \zeta _3 s_{1,1}-s_{4,1}+2 s_{1,3,1}+s_{2,2,1}+s_{3,1,1}-2 s_{1,1,2,1}-s_{1,2,1,1}-3 \zeta _3 \zeta _2+10 \zeta _5+\frac{4}{5} \zeta _2^2 s_1+\zeta _2 s_3-2 \zeta _3 s_2  \end{dmath}
   \begin{dmath}[style={\small}]     s_{1,1} \bar{s}_{1,-1,-1}  =   \frac{3 \ln _2^5}{40}+\frac{2}{3} s_{-2} \ln _2^3-\frac{2}{3} s_2 \ln _2^3-\frac{3}{4} \zeta _2 \ln _2^3-\frac{2}{3} s_{1,-1} \ln _2^3+\frac{5}{3} s_{1,1} \ln _2^3-\frac{1}{3} \bar{s}_{1,-1} \ln _2^3+\frac{1}{3} \bar{s}_{1,1} \ln _2^3+\frac{1}{2} s_{-3} \ln _2^2-\frac{1}{2} s_3 \ln _2^2-\frac{1}{2} s_{-1} \zeta _2 \ln _2^2+\frac{5}{2} s_1 \zeta _2 \ln _2^2+\frac{85}{16} \zeta _3 \ln _2^2+\bar{s}_{-3} \ln _2^2-\frac{1}{2} \zeta _2 \bar{s}_{-1} \ln _2^2+\frac{5}{4} \zeta _2 \bar{s}_1 \ln _2^2-\bar{s}_3 \ln _2^2+\frac{1}{2} s_{-2,-1} \ln _2^2-\frac{1}{2} s_{-2,1} \ln _2^2-s_{1,-2} \ln _2^2+s_{1,2} \ln _2^2-\frac{3}{2} s_{2,-1} \ln _2^2-\frac{1}{2} s_{2,1} \ln _2^2-\frac{3}{2} \bar{s}_{1,-2} \ln _2^2+\frac{3}{2} \bar{s}_{1,2} \ln _2^2-\frac{1}{2} \bar{s}_{2,-1} \ln _2^2+\frac{1}{2} \bar{s}_{2,1} \ln _2^2-\frac{1}{2} s_{1,-1,-1} \ln _2^2+\frac{1}{2} s_{1,-1,1} \ln _2^2+3 s_{1,1,-1} \ln _2^2+s_{1,1,1} \ln _2^2+\frac{1}{2} \bar{s}_{1,-1,-1} \ln _2^2-\frac{1}{2} \bar{s}_{1,-1,1} \ln _2^2+\bar{s}_{1,1,-1} \ln _2^2-\bar{s}_{1,1,1} \ln _2^2+\frac{1}{10} \zeta _2^2 \ln _2+5 \text{Li}_4\left(\frac{1}{2}\right) \ln _2-s_{-2} \zeta _2 \ln _2-s_2 \zeta _2 \ln _2+s_{-1} \zeta _3 \ln _2-\frac{1}{4} s_1 \zeta _3 \ln _2-\frac{1}{2} \zeta _2 \bar{s}_{-2} \ln _2+\zeta _3 \bar{s}_{-1} \ln _2-\frac{1}{8} \zeta _3 \bar{s}_1 \ln _2+\frac{1}{2} \zeta _2 \bar{s}_2 \ln _2+s_{-3,-1} \ln _2+s_{-3,1} \ln _2+\frac{3}{2} \zeta _2 s_{1,-1} \ln _2+\frac{5}{2} \zeta _2 s_{1,1} \ln _2+\bar{s}_{1,-3} \ln _2+\frac{1}{2} \zeta _2 \bar{s}_{1,-1} \ln _2-\frac{1}{2} \zeta _2 \bar{s}_{1,1} \ln _2-\bar{s}_{1,3} \ln _2+\bar{s}_{2,-2} \ln _2-\bar{s}_{2,2} \ln _2-s_{-2,1,-1} \ln _2-s_{-2,1,1} \ln _2-2 s_{1,-2,-1} \ln _2-2 s_{1,-2,1} \ln _2-s_{2,-1,-1} \ln _2-s_{2,-1,1} \ln _2+\bar{s}_{1,-2,-1} \ln _2-\bar{s}_{1,-2,1} \ln _2+\bar{s}_{1,-1,-2} \ln _2-\bar{s}_{1,-1,2} \ln _2-2 \bar{s}_{1,1,-2} \ln _2+2 \bar{s}_{1,1,2} \ln _2+\bar{s}_{2,-1,-1} \ln _2-\bar{s}_{2,-1,1} \ln _2+s_{1,-1,1,-1} \ln _2+s_{1,-1,1,1} \ln _2+2 s_{1,1,-1,-1} \ln _2+2 s_{1,1,-1,1} \ln _2-\bar{s}_{1,-1,1,-1} \ln _2+\bar{s}_{1,-1,1,1} \ln _2-2 \bar{s}_{1,1,-1,-1} \ln _2+2 \bar{s}_{1,1,-1,1} \ln _2+\frac{3}{20} s_1 \zeta _2^2+6 \text{Li}_5\left(\frac{1}{2}\right)+\frac{1}{2} s_{-3} \zeta _2+\frac{s_3 \zeta _2}{2}-\frac{1}{8} s_{-2} \zeta _3-\frac{s_2 \zeta _3}{4}-\frac{21 \zeta _2 \zeta _3}{16}-\frac{13 \zeta _5}{8}+\frac{19}{40} \zeta _2^2 \bar{s}_1-\zeta _3 \bar{s}_2-\frac{1}{2} \zeta _2 s_{-2,-1}-\frac{1}{2} \zeta _2 s_{-2,1}-\zeta _2 s_{1,-2}+\frac{1}{8} \zeta _3 s_{1,-1}-\frac{3}{8} \zeta _3 s_{1,1}-\zeta _2 s_{1,2}-\frac{1}{2} \zeta _2 s_{2,-1}-\frac{1}{2} \zeta _2 s_{2,1}-s_{4,1}+\frac{1}{4} \zeta _3 \bar{s}_{1,-1}+\frac{9}{8} \zeta _3 \bar{s}_{1,1}+\frac{1}{2} \zeta _2 \bar{s}_{1,2}+\frac{1}{2} \zeta _2 \bar{s}_{2,1}+s_{-3,-1,1}+s_{-2,-2,1}+\frac{1}{2} \zeta _2 s_{1,-1,-1}+\frac{1}{2} \zeta _2 s_{1,-1,1}+\zeta _2 s_{1,1,-1}+\zeta _2 s_{1,1,1}+2 s_{1,3,1}+s_{2,2,1}+s_{3,1,1}+\bar{s}_{1,-3,-1}-\frac{1}{2} \zeta _2 \bar{s}_{1,-1,-1}-\zeta _2 \bar{s}_{1,1,1}+\bar{s}_{2,-2,-1}+\bar{s}_{3,-1,-1}-s_{-2,-1,1,1}-s_{-2,1,-1,1}-2 s_{1,-2,-1,1}-s_{1,-1,-2,1}-2 s_{1,1,2,1}-s_{1,2,1,1}-s_{2,-1,-1,1}-\bar{s}_{1,-2,1,-1}-\bar{s}_{1,-1,2,-1}-2 \bar{s}_{1,1,-2,-1}-2 \bar{s}_{1,2,-1,-1}-\bar{s}_{2,-1,1,-1}-2 \bar{s}_{2,1,-1,-1}+s_{1,-1,-1,1,1}+s_{1,-1,1,-1,1}+2 s_{1,1,-1,-1,1}+\bar{s}_{1,-1,1,1,-1}+2 \bar{s}_{1,1,-1,1,-1}+3 \bar{s}_{1,1,1,-1,-1}  \end{dmath}

   \begin{dmath}[style={\small}]     s_{1,1} \bar{s}_{1,1,-1}  =   \frac{\ln _2^5}{120}+\frac{1}{24} s_{-1} \ln _2^4+\frac{1}{24} \bar{s}_{-1} \ln _2^4-\frac{1}{12} \zeta _2 \ln _2^3-\frac{1}{3} s_{1,1} \ln _2^3+\frac{1}{2} s_{-3} \ln _2^2-\frac{1}{2} s_3 \ln _2^2-\frac{1}{4} s_{-1} \zeta _2 \ln _2^2-s_1 \zeta _2 \ln _2^2-\frac{23}{16} \zeta _3 \ln _2^2-\frac{1}{2} \bar{s}_{-3} \ln _2^2-\frac{1}{4} \zeta _2 \bar{s}_{-1} \ln _2^2-\frac{3}{4} \zeta _2 \bar{s}_1 \ln _2^2+\frac{1}{2} \bar{s}_3 \ln _2^2-s_{1,-2} \ln _2^2+s_{1,2} \ln _2^2-\frac{1}{2} s_{2,-1} \ln _2^2+\frac{3}{2} s_{2,1} \ln _2^2+\frac{1}{2} \bar{s}_{1,-2} \ln _2^2-\frac{1}{2} \bar{s}_{1,2} \ln _2^2+\frac{1}{2} s_{1,1,-1} \ln _2^2-\frac{5}{2} s_{1,1,1} \ln _2^2+\frac{1}{2} \bar{s}_{1,1,-1} \ln _2^2-\frac{1}{2} \bar{s}_{1,1,1} \ln _2^2-\frac{13}{4} \zeta _2^2 \ln _2-\text{Li}_4\left(\frac{1}{2}\right) \ln _2-\frac{1}{2} s_{-2} \zeta _2 \ln _2+\frac{5}{2} s_2 \zeta _2 \ln _2+s_{-1} \zeta _3 \ln _2-\frac{7}{2} s_1 \zeta _3 \ln _2+\zeta _3 \bar{s}_{-1} \ln _2-\frac{29}{8} \zeta _3 \bar{s}_1 \ln _2+\frac{3}{2} \zeta _2 \bar{s}_2 \ln _2+\frac{1}{2} \zeta _2 s_{1,-1} \ln _2-\frac{9}{2} \zeta _2 s_{1,1} \ln _2-s_{3,-1} \ln _2-s_{3,1} \ln _2-\bar{s}_{1,-3} \ln _2+\frac{1}{2} \zeta _2 \bar{s}_{1,-1} \ln _2-\frac{7}{2} \zeta _2 \bar{s}_{1,1} \ln _2+\bar{s}_{1,3} \ln _2-\bar{s}_{2,-2} \ln _2+\bar{s}_{2,2} \ln _2+2 s_{1,2,-1} \ln _2+2 s_{1,2,1} \ln _2+2 s_{2,1,-1} \ln _2+2 s_{2,1,1} \ln _2+3 \bar{s}_{1,1,-2} \ln _2-3 \bar{s}_{1,1,2} \ln _2+\bar{s}_{1,2,-1} \ln _2-\bar{s}_{1,2,1} \ln _2+\bar{s}_{2,1,-1} \ln _2-\bar{s}_{2,1,1} \ln _2-3 s_{1,1,1,-1} \ln _2-3 s_{1,1,1,1} \ln _2-3 \bar{s}_{1,1,1,-1} \ln _2+3 \bar{s}_{1,1,1,1} \ln _2-\frac{9}{20} s_{-1} \zeta _2^2+\frac{1}{4} s_1 \zeta _2^2-4 \text{Li}_5\left(\frac{1}{2}\right)+\text{Li}_4\left(\frac{1}{2}\right) s_{-1}-\frac{1}{2} s_{-3} \zeta _2-\frac{s_3 \zeta _2}{2}+s_{-2} \zeta _3-\frac{5 s_2 \zeta _3}{8}-\frac{9 \zeta _2 \zeta _3}{16}+\frac{155 \zeta _5}{32}+\frac{1}{8} \zeta _3 \bar{s}_{-2}-\frac{9}{20} \zeta _2^2 \bar{s}_{-1}+\text{Li}_4\left(\frac{1}{2}\right) \bar{s}_{-1}+\frac{1}{8} \zeta _2^2 \bar{s}_1-\frac{1}{4} \zeta _3 \bar{s}_2+s_{-4,1}+\zeta _2 s_{1,-2}-\frac{9}{8} \zeta _3 s_{1,-1}+\frac{11}{8} \zeta _3 s_{1,1}+\zeta _2 s_{1,2}+\zeta _2 s_{2,-1}+\zeta _2 s_{2,1}+\frac{1}{2} \zeta _2 \bar{s}_{1,-2}-\frac{9}{8} \zeta _3 \bar{s}_{1,-1}+\frac{3}{4} \zeta _3 \bar{s}_{1,1}+\frac{1}{2} \zeta _2 \bar{s}_{2,-1}-s_{-3,1,1}-2 s_{1,-3,1}-\frac{3}{2} \zeta _2 s_{1,1,-1}-\frac{3}{2} \zeta _2 s_{1,1,1}-2 s_{2,-2,1}-s_{3,-1,1}-\frac{3}{2} \zeta _2 \bar{s}_{1,1,-1}+\bar{s}_{1,3,-1}+\bar{s}_{2,2,-1}+\bar{s}_{3,1,-1}+s_{1,-2,1,1}+3 s_{1,1,-2,1}+2 s_{1,2,-1,1}+s_{2,-1,1,1}+2 s_{2,1,-1,1}-3 \bar{s}_{1,1,2,-1}-3 \bar{s}_{1,2,1,-1}-3 \bar{s}_{2,1,1,-1}-s_{1,1,-1,1,1}-3 s_{1,1,1,-1,1}+6 \bar{s}_{1,1,1,1,-1}  \end{dmath}
   \begin{dmath}[style={\small}]     s_{1,1} \bar{s}_{1,1,1}  =   -\zeta _2 \bar{s}_{1,2}-\zeta _2 \bar{s}_{2,1}+3 \zeta _2 \bar{s}_{1,1,1}+4 \zeta _3 \bar{s}_{1,1}+\bar{s}_{1,3,1}+\bar{s}_{2,2,1}+\bar{s}_{3,1,1}-3 \bar{s}_{1,1,2,1}-3 \bar{s}_{1,2,1,1}-3 \bar{s}_{2,1,1,1}+6 \bar{s}_{1,1,1,1,1}+\frac{8}{5} \zeta _2^2 \bar{s}_1-\zeta _3 \bar{s}_2-2 \zeta _2 s_{1,2}-2 \zeta _2 s_{2,1}+3 \zeta _2 s_{1,1,1}+5 \zeta _3 s_{1,1}-s_{4,1}+2 s_{1,3,1}+2 s_{2,2,1}+2 s_{3,1,1}-3 s_{1,1,2,1}-3 s_{1,2,1,1}-3 s_{2,1,1,1}+4 s_{1,1,1,1,1}+10 \zeta _5+\frac{12}{5} \zeta _2^2 s_1+\zeta _2 s_3-3 \zeta _3 s_2  \end{dmath}
   \begin{dmath}[style={\small}]     s_{1,1} \bar{s}_{-1,-2}  =   -\frac{1}{8} s_{-1} \ln _2^4+\frac{1}{8} s_1 \ln _2^4+\frac{1}{8} \bar{s}_{-1} \ln _2^4+\frac{1}{24} \bar{s}_1 \ln _2^4-\frac{1}{6} \zeta _2 \ln _2^3-\frac{1}{4} s_{-1} \zeta _2 \ln _2^2+\frac{1}{4} s_1 \zeta _2 \ln _2^2-\frac{1}{4} \zeta _2 \bar{s}_{-1} \ln _2^2-\frac{3}{4} \zeta _2 \bar{s}_1 \ln _2^2-\frac{5}{8} \zeta _2^2 \ln _2-\frac{3}{2} s_{-2} \zeta _2 \ln _2+\frac{3}{2} s_2 \zeta _2 \ln _2+\zeta _2 \bar{s}_{-2} \ln _2-\zeta _2 \bar{s}_2 \ln _2-\frac{3}{2} \zeta _2 s_{-1,-1} \ln _2+\frac{3}{2} \zeta _2 s_{-1,1} \ln _2+2 \zeta _2 s_{1,-1} \ln _2-\zeta _2 s_{1,1} \ln _2+\frac{3}{2} \zeta _2 \bar{s}_{-1,-1} \ln _2-\frac{3}{2} \zeta _2 \bar{s}_{-1,1} \ln _2-\zeta _2 \bar{s}_{1,-1} \ln _2+\zeta _2 \bar{s}_{1,1} \ln _2+\frac{9}{10} s_{-1} \zeta _2^2-\frac{9}{40} s_1 \zeta _2^2-3 \text{Li}_4\left(\frac{1}{2}\right) s_{-1}+3 \text{Li}_4\left(\frac{1}{2}\right) s_1+s_3 \zeta _2+\frac{5}{8} s_{-2} \zeta _3-\frac{21 s_2 \zeta _3}{8}-\frac{\zeta _2 \zeta _3}{4}+\frac{175 \zeta _5}{64}+\frac{1}{2} \zeta _2 \bar{s}_{-3}-\frac{1}{8} \zeta _3 \bar{s}_{-2}-\frac{29}{40} \zeta _2^2 \bar{s}_{-1}+3 \text{Li}_4\left(\frac{1}{2}\right) \bar{s}_{-1}+\frac{9}{40} \zeta _2^2 \bar{s}_1+\text{Li}_4\left(\frac{1}{2}\right) \bar{s}_1+\frac{13}{8} \zeta _3 \bar{s}_2-\frac{1}{2} \zeta _2 \bar{s}_3-\frac{3}{2} \zeta _2 s_{-2,-1}-\frac{1}{2} \zeta _2 s_{-2,1}-\zeta _2 s_{-1,-2}+\frac{21}{8} \zeta _3 s_{-1,-1}-\frac{5}{8} \zeta _3 s_{-1,1}-\frac{5}{8} \zeta _3 s_{1,-1}+\frac{13}{8} \zeta _3 s_{1,1}-\zeta _2 s_{1,2}-s_{4,1}+\bar{s}_{-3,-2}+\frac{3}{2} \zeta _2 \bar{s}_{-2,-1}-\frac{1}{2} \zeta _2 \bar{s}_{-2,1}+\frac{1}{2} \zeta _2 \bar{s}_{-1,-2}-\frac{21}{8} \zeta _3 \bar{s}_{-1,-1}+\frac{1}{8} \zeta _3 \bar{s}_{-1,1}-\frac{1}{2} \zeta _2 \bar{s}_{-1,2}-\frac{1}{2} \zeta _2 \bar{s}_{1,-2}+\frac{1}{8} \zeta _3 \bar{s}_{1,-1}-\frac{13}{8} \zeta _3 \bar{s}_{1,1}+\frac{1}{2} \zeta _2 \bar{s}_{1,2}+s_{-2,-2,1}+s_{-1,-3,1}+\frac{3}{2} \zeta _2 s_{-1,1,-1}+\frac{1}{2} \zeta _2 s_{-1,1,1}+\frac{3}{2} \zeta _2 s_{1,-1,-1}+\frac{1}{2} \zeta _2 s_{1,-1,1}+s_{1,3,1}+s_{3,1,1}-\bar{s}_{-2,1,-2}-\frac{3}{2} \zeta _2 \bar{s}_{-1,1,-1}+\frac{1}{2} \zeta _2 \bar{s}_{-1,1,1}-\bar{s}_{-1,2,-2}-\bar{s}_{1,-2,-2}-\frac{3}{2} \zeta _2 \bar{s}_{1,-1,-1}+\frac{1}{2} \zeta _2 \bar{s}_{1,-1,1}-\bar{s}_{2,-1,-2}-s_{-1,-2,1,1}-s_{-1,1,-2,1}-s_{1,-1,-2,1}+\bar{s}_{-1,1,1,-2}+\bar{s}_{1,-1,1,-2}+\bar{s}_{1,1,-1,-2}  \end{dmath}
   \begin{dmath}[style={\small}]     s_{1,1} \bar{s}_{-1,2}  =   \zeta _2 \bar{s}_{-2,1}+\zeta _3 \bar{s}_{-1,1}+\zeta _3 \bar{s}_{1,-1}+\zeta _3 \bar{s}_{1,1}-\zeta _2 \bar{s}_{-1,1,1}-\zeta _2 \bar{s}_{1,-1,1}+\frac{1}{2} \zeta _2 \ln _2 \bar{s}_{1,-1}-\frac{1}{2} \zeta _2 \ln _2 \bar{s}_{1,1}+\bar{s}_{-3,2}-\bar{s}_{-2,1,2}-\bar{s}_{-1,2,2}-\bar{s}_{1,-2,2}-\bar{s}_{2,-1,2}+\bar{s}_{-1,1,1,2}+\bar{s}_{1,-1,1,2}+\bar{s}_{1,1,-1,2}-3 \text{Li}_4\left(\frac{1}{2}\right) \bar{s}_{-1}-\text{Li}_4\left(\frac{1}{2}\right) \bar{s}_1-\zeta _3 \bar{s}_{-2}+\frac{43}{40} \zeta _2^2 \bar{s}_{-1}-\frac{1}{20} \zeta _2^2 \bar{s}_1-\zeta _3 \bar{s}_2-\frac{1}{4} \zeta _2 \ln _2^2 \bar{s}_{-1}+\frac{5}{4} \zeta _2 \ln _2^2 \bar{s}_1-\frac{1}{2} \zeta _2 \ln _2 \bar{s}_{-2}+\frac{1}{2} \zeta _2 \ln _2 \bar{s}_2-\frac{1}{8} \ln _2^4 \bar{s}_{-1}-\frac{1}{24} \ln _2^4 \bar{s}_1+\zeta _2 s_{-2,1}-2 \zeta _3 s_{-1,1}+\zeta _2 s_{-1,2}+\zeta _2 s_{1,-2}-2 \zeta _3 s_{1,-1}-\zeta _3 s_{1,1}-\zeta _2 s_{-1,1,1}-\zeta _2 s_{1,-1,1}+\frac{1}{2} \zeta _2 \ln _2 s_{1,-1}+\frac{1}{2} \zeta _2 \ln _2 s_{1,1}+s_{-4,1}-s_{-3,1,1}-s_{-2,2,1}-s_{-1,3,1}-s_{1,-3,1}+s_{-1,1,2,1}+s_{-1,2,1,1}+s_{1,-1,2,1}+\frac{\zeta _2 \zeta _3}{8}-\frac{773 \zeta _5}{64}-\frac{1}{6} \zeta _2 \ln _2^3+\frac{119}{40} \zeta _2^2 \ln _2-3 \text{Li}_4\left(\frac{1}{2}\right) s_{-1}+3 \text{Li}_4\left(\frac{1}{2}\right) s_1+12 \text{Li}_5\left(\frac{1}{2}\right)+\frac{11}{40} \zeta _2^2 s_{-1}-\frac{43}{40} \zeta _2^2 s_1-\zeta _2 s_{-3}+2 \zeta _3 s_{-2}-\frac{1}{4} \zeta _2 s_{-1} \ln _2^2+\frac{1}{4} \zeta _2 s_1 \ln _2^2-\frac{1}{8} s_{-1} \ln _2^4+\frac{1}{8} s_1 \ln _2^4-\frac{\ln _2^5}{10}  \end{dmath}
   \begin{dmath}[style={\small}]     s_{1,1} \bar{s}_{-1,-1,-1}  =   \frac{\ln _2^5}{6}+\frac{1}{2} s_{-1} \ln _2^4-\frac{1}{2} s_1 \ln _2^4+\frac{1}{6} \bar{s}_{-1} \ln _2^4-\frac{1}{8} \bar{s}_1 \ln _2^4+\frac{2}{3} s_{-2} \ln _2^3-\frac{2}{3} s_2 \ln _2^3-\frac{1}{2} \zeta _2 \ln _2^3+\frac{1}{6} \bar{s}_{-2} \ln _2^3-\frac{1}{6} \bar{s}_2 \ln _2^3+\frac{2}{3} s_{-1,-1} \ln _2^3-\frac{2}{3} s_{-1,1} \ln _2^3-\frac{7}{6} s_{1,-1} \ln _2^3-\frac{1}{6} s_{1,1} \ln _2^3-\frac{1}{3} \bar{s}_{-1,-1} \ln _2^3+\frac{1}{3} \bar{s}_{-1,1} \ln _2^3-\frac{1}{6} \bar{s}_{1,-1} \ln _2^3+\frac{1}{6} \bar{s}_{1,1} \ln _2^3+\frac{1}{2} s_{-3} \ln _2^2-\frac{1}{2} s_3 \ln _2^2-s_{-1} \zeta _2 \ln _2^2-s_1 \zeta _2 \ln _2^2-\frac{7}{16} \zeta _3 \ln _2^2-\bar{s}_{-3} \ln _2^2-\frac{1}{4} \zeta _2 \bar{s}_1 \ln _2^2+\bar{s}_3 \ln _2^2+\frac{3}{2} s_{-2,-1} \ln _2^2+\frac{1}{2} s_{-2,1} \ln _2^2+\frac{1}{2} s_{-1,-2} \ln _2^2-\frac{1}{2} s_{-1,2} \ln _2^2-\frac{1}{2} s_{1,-2} \ln _2^2+\frac{1}{2} s_{1,2} \ln _2^2-\frac{1}{2} s_{2,-1} \ln _2^2+\frac{1}{2} s_{2,1} \ln _2^2-\frac{1}{2} \bar{s}_{-2,-1} \ln _2^2+\frac{1}{2} \bar{s}_{-2,1} \ln _2^2-\frac{1}{2} \bar{s}_{-1,-2} \ln _2^2+\frac{1}{2} \bar{s}_{-1,2} \ln _2^2+\bar{s}_{1,-2} \ln _2^2-\bar{s}_{1,2} \ln _2^2+\frac{1}{2} s_{-1,-1,-1} \ln _2^2-\frac{1}{2} s_{-1,-1,1} \ln _2^2-\frac{3}{2} s_{-1,1,-1} \ln _2^2-\frac{1}{2} s_{-1,1,1} \ln _2^2-\frac{3}{2} s_{1,-1,-1} \ln _2^2-\frac{1}{2} s_{1,-1,1} \ln _2^2+\frac{1}{2} \bar{s}_{-1,-1,-1} \ln _2^2-\frac{1}{2} \bar{s}_{-1,-1,1} \ln _2^2+\frac{1}{2} \bar{s}_{-1,1,-1} \ln _2^2-\frac{1}{2} \bar{s}_{-1,1,1} \ln _2^2+\frac{1}{2} \bar{s}_{1,-1,-1} \ln _2^2-\frac{1}{2} \bar{s}_{1,-1,1} \ln _2^2-\frac{7}{10} \zeta _2^2 \ln _2-\text{Li}_4\left(\frac{1}{2}\right) \ln _2+s_{-2} \zeta _2 \ln _2+s_2 \zeta _2 \ln _2-\frac{3}{4} s_1 \zeta _3 \ln _2+\frac{1}{2} \zeta _2 \bar{s}_{-2} \ln _2+\frac{1}{4} \zeta _3 \bar{s}_{-1} \ln _2+\frac{5}{8} \zeta _3 \bar{s}_1 \ln _2-\frac{1}{2} \zeta _2 \bar{s}_2 \ln _2-\zeta _2 s_{-1,-1} \ln _2-\zeta _2 s_{-1,1} \ln _2-\frac{3}{2} \zeta _2 s_{1,-1} \ln _2-\frac{1}{2} \zeta _2 s_{1,1} \ln _2-s_{3,-1} \ln _2-s_{3,1} \ln _2+\bar{s}_{-2,-2} \ln _2-\bar{s}_{-2,2} \ln _2+\bar{s}_{-1,-3} \ln _2-\bar{s}_{-1,3} \ln _2-\frac{1}{2} \zeta _2 \bar{s}_{1,-1} \ln _2+\frac{1}{2} \zeta _2 \bar{s}_{1,1} \ln _2+s_{-2,-1,-1} \ln _2+s_{-2,-1,1} \ln _2+s_{-1,-2,-1} \ln _2+s_{-1,-2,1} \ln _2+s_{1,2,-1} \ln _2+s_{1,2,1} \ln _2+s_{2,1,-1} \ln _2+s_{2,1,1} \ln _2+\bar{s}_{-2,-1,-1} \ln _2-\bar{s}_{-2,-1,1} \ln _2+\bar{s}_{-1,-2,-1} \ln _2-\bar{s}_{-1,-2,1} \ln _2+\bar{s}_{-1,-1,-2} \ln _2-\bar{s}_{-1,-1,2} \ln _2-\bar{s}_{-1,1,-2} \ln _2+\bar{s}_{-1,1,2} \ln _2-\bar{s}_{1,-1,-2} \ln _2+\bar{s}_{1,-1,2} \ln _2-s_{-1,-1,1,-1} \ln _2-s_{-1,-1,1,1} \ln _2-s_{-1,1,-1,-1} \ln _2-s_{-1,1,-1,1} \ln _2-s_{1,-1,-1,-1} \ln _2-s_{1,-1,-1,1} \ln _2-\bar{s}_{-1,-1,1,-1} \ln _2+\bar{s}_{-1,-1,1,1} \ln _2-\bar{s}_{-1,1,-1,-1} \ln _2+\bar{s}_{-1,1,-1,1} \ln _2-\bar{s}_{1,-1,-1,-1} \ln _2+\bar{s}_{1,-1,-1,1} \ln _2-\frac{7}{20} s_{-1} \zeta _2^2+\frac{1}{5} s_1 \zeta _2^2-3 \text{Li}_5\left(\frac{1}{2}\right)+\text{Li}_4\left(\frac{1}{2}\right) s_{-1}-\text{Li}_4\left(\frac{1}{2}\right) s_1-\frac{1}{2} s_{-3} \zeta _2-\frac{s_3 \zeta _2}{2}+\frac{1}{4} s_{-2} \zeta _3+\frac{s_2 \zeta _3}{8}+\frac{3 \zeta _2 \zeta _3}{16}+\frac{17 \zeta _5}{8}-\frac{3}{4} \zeta _3 \bar{s}_{-2}-\frac{1}{5} \zeta _2^2 \bar{s}_{-1}+\text{Li}_4\left(\frac{1}{2}\right) \bar{s}_{-1}-\frac{1}{8} \zeta _2^2 \bar{s}_1-\frac{1}{4} \zeta _3 \bar{s}_2+s_{-4,1}+\frac{1}{2} \zeta _2 s_{-2,-1}+\frac{1}{2} \zeta _2 s_{-2,1}+\frac{1}{2} \zeta _2 s_{-1,-2}-\frac{1}{8} \zeta _3 s_{-1,-1}-\frac{1}{4} \zeta _3 s_{-1,1}+\frac{1}{2} \zeta _2 s_{-1,2}+\frac{1}{2} \zeta _2 s_{1,-2}+\frac{3}{8} \zeta _3 s_{1,-1}-\frac{1}{4} \zeta _3 s_{1,1}+\frac{1}{2} \zeta _2 s_{1,2}+\frac{1}{2} \zeta _2 s_{2,-1}+\frac{1}{2} \zeta _2 s_{2,1}+\frac{1}{2} \zeta _2 \bar{s}_{-2,1}+\frac{1}{4} \zeta _3 \bar{s}_{-1,-1}+\frac{1}{8} \zeta _3 \bar{s}_{-1,1}+\frac{1}{2} \zeta _2 \bar{s}_{-1,2}+\frac{3}{4} \zeta _3 \bar{s}_{1,-1}+\frac{1}{4} \zeta _3 \bar{s}_{1,1}-s_{-3,1,1}-s_{-2,2,1}-\frac{1}{2} \zeta _2 s_{-1,-1,-1}-\frac{1}{2} \zeta _2 s_{-1,-1,1}-\frac{1}{2} \zeta _2 s_{-1,1,-1}-\frac{1}{2} \zeta _2 s_{-1,1,1}-s_{-1,3,1}-s_{1,-3,1}-\frac{1}{2} \zeta _2 s_{1,-1,-1}-\frac{1}{2} \zeta _2 s_{1,-1,1}-s_{2,-2,1}-s_{3,-1,1}+\bar{s}_{-3,-1,-1}+\bar{s}_{-2,-2,-1}+\bar{s}_{-1,-3,-1}-\frac{1}{2} \zeta _2 \bar{s}_{-1,-1,-1}-\frac{1}{2} \zeta _2 \bar{s}_{-1,1,1}-\frac{1}{2} \zeta _2 \bar{s}_{1,-1,1}+s_{-2,-1,-1,1}+s_{-1,-2,-1,1}+s_{-1,-1,-2,1}+s_{-1,1,2,1}+s_{-1,2,1,1}+s_{1,-1,2,1}+s_{1,2,-1,1}+s_{2,-1,1,1}+s_{2,1,-1,1}-\bar{s}_{-2,-1,1,-1}-\bar{s}_{-2,1,-1,-1}-\bar{s}_{-1,-2,1,-1}-\bar{s}_{-1,-1,2,-1}-\bar{s}_{-1,1,-2,-1}-\bar{s}_{-1,2,-1,-1}-\bar{s}_{1,-2,-1,-1}-\bar{s}_{1,-1,-2,-1}-\bar{s}_{2,-1,-1,-1}-s_{-1,-1,-1,1,1}-s_{-1,-1,1,-1,1}-s_{-1,1,-1,-1,1}-s_{1,-1,-1,-1,1}+\bar{s}_{-1,-1,1,1,-1}+\bar{s}_{-1,1,-1,1,-1}+\bar{s}_{-1,1,1,-1,-1}+\bar{s}_{1,-1,-1,1,-1}+\bar{s}_{1,-1,1,-1,-1}+\bar{s}_{1,1,-1,-1,-1}  \end{dmath}
   \begin{dmath}[style={\small}]     s_{1,1} \bar{s}_{-1,-1,1}  =   \frac{17 \ln _2^5}{120}+\frac{1}{4} s_{-1} \ln _2^4-\frac{3}{8} s_1 \ln _2^4+\frac{1}{8} \bar{s}_{-1} \ln _2^4-\frac{1}{8} \bar{s}_1 \ln _2^4+\frac{1}{3} s_{-2} \ln _2^3-\frac{1}{3} s_2 \ln _2^3-\frac{5}{12} \zeta _2 \ln _2^3+\frac{1}{6} \bar{s}_{-2} \ln _2^3-\frac{1}{6} \bar{s}_2 \ln _2^3+\frac{1}{3} s_{-1,-1} \ln _2^3-\frac{1}{3} s_{-1,1} \ln _2^3-\frac{5}{6} s_{1,-1} \ln _2^3-\frac{1}{6} s_{1,1} \ln _2^3-\frac{1}{3} \bar{s}_{-1,-1} \ln _2^3+\frac{1}{3} \bar{s}_{-1,1} \ln _2^3-\frac{1}{6} \bar{s}_{1,-1} \ln _2^3+\frac{1}{6} \bar{s}_{1,1} \ln _2^3-\frac{1}{2} s_{-1} \zeta _2 \ln _2^2+\frac{1}{4} s_1 \zeta _2 \ln _2^2-\frac{1}{8} \zeta _3 \ln _2^2-\frac{1}{2} \bar{s}_{-3} \ln _2^2+\frac{1}{4} \zeta _2 \bar{s}_{-1} \ln _2^2+\frac{1}{2} \zeta _2 \bar{s}_1 \ln _2^2+\frac{1}{2} \bar{s}_3 \ln _2^2+\frac{1}{2} s_{-2,-1} \ln _2^2+\frac{1}{2} s_{-2,1} \ln _2^2-\frac{1}{2} \bar{s}_{-2,-1} \ln _2^2+\frac{1}{2} \bar{s}_{-2,1} \ln _2^2-\frac{1}{2} \bar{s}_{-1,-2} \ln _2^2+\frac{1}{2} \bar{s}_{-1,2} \ln _2^2+\frac{1}{2} \bar{s}_{1,-2} \ln _2^2-\frac{1}{2} \bar{s}_{1,2} \ln _2^2-\frac{1}{2} s_{-1,1,-1} \ln _2^2-\frac{1}{2} s_{-1,1,1} \ln _2^2-\frac{1}{2} s_{1,-1,-1} \ln _2^2-\frac{1}{2} s_{1,-1,1} \ln _2^2+\frac{1}{2} \bar{s}_{-1,1,-1} \ln _2^2-\frac{1}{2} \bar{s}_{-1,1,1} \ln _2^2+\frac{1}{2} \bar{s}_{1,-1,-1} \ln _2^2-\frac{1}{2} \bar{s}_{1,-1,1} \ln _2^2+\frac{71}{40} \zeta _2^2 \ln _2-s_{-2} \zeta _2 \ln _2+s_2 \zeta _2 \ln _2-\zeta _2 \bar{s}_{-2} \ln _2-\frac{1}{2} \zeta _2 \bar{s}_2 \ln _2-\zeta _2 s_{-1,-1} \ln _2+\zeta _2 s_{-1,1} \ln _2+\frac{3}{2} \zeta _2 s_{1,-1} \ln _2-\frac{1}{2} \zeta _2 s_{1,1} \ln _2+\zeta _2 \bar{s}_{-1,-1} \ln _2+\frac{1}{2} \zeta _2 \bar{s}_{-1,1} \ln _2+\zeta _2 \bar{s}_{1,-1} \ln _2+\frac{1}{2} \zeta _2 \bar{s}_{1,1} \ln _2+\frac{29}{20} s_{-1} \zeta _2^2+\frac{39}{40} s_1 \zeta _2^2+3 \text{Li}_5\left(\frac{1}{2}\right)-3 \text{Li}_4\left(\frac{1}{2}\right) s_{-1}+s_3 \zeta _2-\frac{1}{4} s_{-2} \zeta _3-\frac{11 s_2 \zeta _3}{4}+\frac{\zeta _2 \zeta _3}{8}-\frac{197 \zeta _5}{64}+\frac{1}{2} \zeta _2 \bar{s}_{-3}-\frac{1}{8} \zeta _3 \bar{s}_{-2}+\frac{19}{40} \zeta _2^2 \bar{s}_{-1}-\frac{3}{8} \zeta _2^2 \bar{s}_1+\frac{7}{4} \zeta _3 \bar{s}_2-\frac{1}{2} \zeta _2 \bar{s}_3-\frac{3}{2} \zeta _2 s_{-2,-1}-\frac{1}{2} \zeta _2 s_{-2,1}-\zeta _2 s_{-1,-2}+\frac{11}{4} \zeta _3 s_{-1,-1}+\frac{1}{4} \zeta _3 s_{-1,1}+\frac{1}{4} \zeta _3 s_{1,-1}+\frac{7}{4} \zeta _3 s_{1,1}-\zeta _2 s_{1,2}-\zeta _2 s_{2,1}-s_{4,1}-\frac{1}{2} \zeta _2 \bar{s}_{-2,-1}-\frac{1}{2} \zeta _2 \bar{s}_{-2,1}-\frac{1}{2} \zeta _2 \bar{s}_{-1,-2}+\frac{1}{4} \zeta _3 \bar{s}_{-1,-1}+\frac{1}{8} \zeta _3 \bar{s}_{-1,1}-\frac{1}{2} \zeta _2 \bar{s}_{-1,2}-\frac{1}{2} \zeta _2 \bar{s}_{1,-2}+\frac{1}{8} \zeta _3 \bar{s}_{1,-1}-\frac{7}{4} \zeta _3 \bar{s}_{1,1}+\frac{1}{2} \zeta _2 \bar{s}_{1,2}+s_{-2,-2,1}+s_{-1,-3,1}+\zeta _2 s_{-1,-1,1}+\frac{3}{2} \zeta _2 s_{-1,1,-1}+\frac{1}{2} \zeta _2 s_{-1,1,1}+\frac{3}{2} \zeta _2 s_{1,-1,-1}+\frac{1}{2} \zeta _2 s_{1,-1,1}+s_{1,3,1}+s_{2,2,1}+2 s_{3,1,1}+\bar{s}_{-3,-1,1}+\bar{s}_{-2,-2,1}+\bar{s}_{-1,-3,1}+\zeta _2 \bar{s}_{-1,-1,1}+\frac{1}{2} \zeta _2 \bar{s}_{-1,1,-1}+\frac{1}{2} \zeta _2 \bar{s}_{-1,1,1}+\frac{1}{2} \zeta _2 \bar{s}_{1,-1,-1}+\frac{1}{2} \zeta _2 \bar{s}_{1,-1,1}-s_{-2,-1,1,1}-2 s_{-1,-2,1,1}-s_{-1,-1,2,1}-s_{-1,1,-2,1}-s_{1,-1,-2,1}-s_{1,2,1,1}-2 s_{2,1,1,1}-\bar{s}_{-2,-1,1,1}-\bar{s}_{-2,1,-1,1}-\bar{s}_{-1,-2,1,1}-\bar{s}_{-1,-1,2,1}-\bar{s}_{-1,1,-2,1}-\bar{s}_{-1,2,-1,1}-\bar{s}_{1,-2,-1,1}-\bar{s}_{1,-1,-2,1}-\bar{s}_{2,-1,-1,1}+2 s_{-1,-1,1,1,1}+s_{-1,1,-1,1,1}+s_{1,-1,-1,1,1}+\bar{s}_{-1,-1,1,1,1}+\bar{s}_{-1,1,-1,1,1}+\bar{s}_{-1,1,1,-1,1}+\bar{s}_{1,-1,-1,1,1}+\bar{s}_{1,-1,1,-1,1}+\bar{s}_{1,1,-1,-1,1}  \end{dmath}
   \begin{dmath}[style={\small}]     \bar{s}_{-3} s_{-1,-1}  =   \frac{3}{4} \zeta _3 \bar{s}_{-1,-1}+\frac{1}{2} \zeta _2 \bar{s}_{-1,2}-\bar{s}_{2,-3}+\bar{s}_{-1,-1,-3}+\ln _2 \bar{s}_{-1,-3}+\ln _2 \bar{s}_{-1,3}-\frac{1}{4} \zeta _3 \bar{s}_{-2}-\frac{61}{40} \zeta _2^2 \bar{s}_{-1}-\frac{3}{4} \zeta _3 \bar{s}_2+\frac{1}{2} \zeta _2 \ln _2 \bar{s}_{-2}+\frac{3}{2} \zeta _3 \ln _2 \bar{s}_{-1}+\frac{1}{2} \zeta _2 \ln _2 \bar{s}_2+\ln _2^2 \bar{s}_{-3}+\ln _2^2 \bar{s}_3-\frac{3}{4} \zeta _3 s_{-1,-1}-\frac{1}{2} \zeta _2 s_{-1,2}+s_{4,-1}-s_{-3,-1,-1}-s_{-1,-3,-1}-\ln _2 s_{-1,-3}+\ln _2 s_{-1,3}+\frac{3 \zeta _2 \zeta _3}{8}+\frac{15 \zeta _5}{32}+\frac{3}{2} \zeta _3 \ln _2^2-\frac{61}{20} \zeta _2^2 \ln _2-\frac{1}{8} \zeta _2^2 s_{-1}+\frac{1}{2} \zeta _2 s_{-3}+\frac{1}{4} \zeta _3 s_{-2}-\frac{1}{2} \zeta _2 s_{-2} \ln _2+\frac{1}{2} \zeta _2 s_2 \ln _2+s_{-3} \ln _2^2-s_3 \ln _2^2-s_{-4} \ln _2+s_4 \ln _2  \end{dmath}
   \begin{dmath}[style={\small}]     \bar{s}_3 s_{-1,-1}  =   \frac{1}{2} \zeta _2 \bar{s}_{-1,-2}+\frac{3}{4} \zeta _3 \bar{s}_{-1,-1}-\bar{s}_{2,3}+\bar{s}_{-1,-1,3}+\ln _2 \bar{s}_{-1,-3}+\ln _2 \bar{s}_{-1,3}+\frac{29}{40} \zeta _2^2 \bar{s}_{-1}-\zeta _3 \bar{s}_2+\frac{1}{2} \zeta _2 \ln _2 \bar{s}_{-2}+\frac{3}{2} \zeta _3 \ln _2 \bar{s}_{-1}+\frac{1}{2} \zeta _2 \ln _2 \bar{s}_2+\ln _2^2 \bar{s}_{-3}+\ln _2^2 \bar{s}_3+\frac{1}{2} \zeta _2 s_{-1,-2}-\frac{3}{4} \zeta _3 s_{-1,-1}-s_{-4,-1}+s_{-1,3,-1}+s_{3,-1,-1}-\ln _2 s_{-1,-3}+\ln _2 s_{-1,3}-\frac{9 \zeta _2 \zeta _3}{8}+\frac{85 \zeta _5}{32}+\frac{3}{2} \zeta _3 \ln _2^2+\frac{29}{20} \zeta _2^2 \ln _2+\frac{1}{8} \zeta _2^2 s_{-1}-\frac{\zeta _2 s_3}{2}-\frac{\zeta _3 s_2}{4}-\frac{1}{2} \zeta _2 s_{-2} \ln _2+\frac{1}{2} \zeta _2 s_2 \ln _2+s_{-3} \ln _2^2-s_3 \ln _2^2-s_{-4} \ln _2+s_4 \ln _2  \end{dmath}
   \begin{dmath}[style={\small}]     s_{-1,-1} \bar{s}_{-2,-1}  =   -\frac{1}{2} \zeta _2 \bar{s}_{-2,-1}-\frac{1}{2} \zeta _2 \bar{s}_{-1,-2}-\frac{1}{4} \zeta _3 \bar{s}_{-1,-1}+\bar{s}_{-4,-1}-\bar{s}_{-2,2,-1}-\bar{s}_{-1,3,-1}-\bar{s}_{2,-2,-1}-\bar{s}_{3,-1,-1}+\bar{s}_{-2,-1,-1,-1}+\bar{s}_{-1,-2,-1,-1}+\bar{s}_{-1,-1,-2,-1}+2 \ln _2^2 \bar{s}_{-2,-1}+\ln _2^2 \bar{s}_{-1,-2}-\ln _2^2 \bar{s}_{-1,2}+\ln _2 \bar{s}_{-2,-2}-\ln _2 \bar{s}_{-2,2}+\ln _2 \bar{s}_{-1,-3}-\ln _2 \bar{s}_{-1,3}-2 \ln _2 \bar{s}_{3,-1}+2 \ln _2 \bar{s}_{-2,-1,-1}+2 \ln _2 \bar{s}_{-1,-2,-1}+\frac{1}{2} \zeta _3 \bar{s}_{-2}-\frac{1}{8} \zeta _2^2 \bar{s}_{-1}-\frac{1}{4} \zeta _3 \bar{s}_2+\frac{1}{2} \zeta _2 \bar{s}_3-\frac{3}{2} \zeta _2 \ln _2 \bar{s}_{-2}+3 \zeta _3 \ln _2 \bar{s}_{-1}+\frac{1}{2} \zeta _2 \ln _2 \bar{s}_2+\frac{2}{3} \ln _2^3 \bar{s}_{-2}-\frac{2}{3} \ln _2^3 \bar{s}_2+\ln _2^2 \bar{s}_{-3}-\ln _2^2 \bar{s}_3+\ln _2 \bar{s}_{-4}-\ln _2 \bar{s}_4+\frac{1}{2} \zeta _2 s_{-2,-1}+\frac{1}{2} \zeta _2 s_{-1,-2}+\frac{1}{4} \zeta _3 s_{-1,-1}-s_{-4,-1}+s_{-2,2,-1}+s_{-1,3,-1}+2 s_{3,-1,-1}-2 s_{-2,-1,-1,-1}-s_{-1,-2,-1,-1}-\ln _2^2 s_{-1,-2}+\ln _2^2 s_{-1,2}-\ln _2 s_{-2,-2}+\ln _2 s_{-2,2}-\ln _2 s_{-1,-3}+\ln _2 s_{-1,3}+2 \ln _2 s_{3,-1}-2 \ln _2 s_{-2,-1,-1}-2 \ln _2 s_{-1,-2,-1}-\frac{261 \zeta _5}{32}+\frac{2}{3} \zeta _2 \ln _2^3+3 \zeta _3 \ln _2^2+\frac{53}{20} \zeta _2^2 \ln _2-4 \text{Li}_4\left(\frac{1}{2}\right) s_{-1}+8 \text{Li}_5\left(\frac{1}{2}\right)+\frac{61}{40} \zeta _2^2 s_{-1}-\frac{\zeta _2 s_3}{2}-\frac{3}{4} \zeta _3 s_{-2}-\frac{\zeta _3 s_2}{2}+\zeta _2 s_{-1} \ln _2^2+\frac{1}{2} \zeta _2 s_{-2} \ln _2+\frac{1}{2} \zeta _2 s_2 \ln _2-3 \zeta _3 s_{-1} \ln _2-\frac{1}{6} s_{-1} \ln _2^4+\frac{2}{3} s_{-2} \ln _2^3-\frac{2}{3} s_2 \ln _2^3-s_{-4} \ln _2+s_4 \ln _2-\frac{\ln _2^5}{15}  \end{dmath}
   \begin{dmath}[style={\small}]     s_{-1,-1} \bar{s}_{-2,1}  =   \frac{11 \ln _2^5}{60}-\frac{1}{8} s_{-1} \ln _2^4-\frac{1}{24} \bar{s}_{-1} \ln _2^4+\frac{1}{2} s_{-2} \ln _2^3-\frac{1}{2} s_2 \ln _2^3-\frac{5}{6} \zeta _2 \ln _2^3+\frac{1}{2} \bar{s}_{-2} \ln _2^3-\frac{1}{2} \bar{s}_2 \ln _2^3+\frac{1}{2} s_{-3} \ln _2^2-\frac{1}{2} s_3 \ln _2^2+\frac{3}{4} s_{-1} \zeta _2 \ln _2^2+3 \zeta _3 \ln _2^2+\frac{1}{2} \bar{s}_{-3} \ln _2^2+\frac{1}{4} \zeta _2 \bar{s}_{-1} \ln _2^2-\frac{1}{2} \bar{s}_3 \ln _2^2+\frac{1}{2} s_{-2,-1} \ln _2^2-\frac{1}{2} s_{-2,1} \ln _2^2-\frac{1}{2} s_{-1,-2} \ln _2^2+\frac{1}{2} s_{-1,2} \ln _2^2+\frac{3}{2} \bar{s}_{-2,-1} \ln _2^2+\frac{1}{2} \bar{s}_{-2,1} \ln _2^2+\frac{1}{2} \bar{s}_{-1,-2} \ln _2^2-\frac{1}{2} \bar{s}_{-1,2} \ln _2^2-\frac{9}{10} \zeta _2^2 \ln _2+6 \text{Li}_4\left(\frac{1}{2}\right) \ln _2-s_{-4} \ln _2+s_4 \ln _2-\frac{1}{2} s_{-2} \zeta _2 \ln _2+\frac{1}{2} s_2 \zeta _2 \ln _2-\frac{21}{8} s_{-1} \zeta _3 \ln _2+\frac{3}{2} \zeta _2 \bar{s}_{-2} \ln _2+\frac{3}{8} \zeta _3 \bar{s}_{-1} \ln _2+\frac{1}{2} \zeta _2 \bar{s}_2 \ln _2-s_{-2,-2} \ln _2+s_{-2,2} \ln _2-s_{-1,-3} \ln _2+s_{-1,3} \ln _2+s_{3,-1} \ln _2-s_{3,1} \ln _2-\bar{s}_{3,-1} \ln _2-\bar{s}_{3,1} \ln _2-s_{-2,-1,-1} \ln _2+s_{-2,-1,1} \ln _2-s_{-1,-2,-1} \ln _2+s_{-1,-2,1} \ln _2+\bar{s}_{-2,-1,-1} \ln _2+\bar{s}_{-2,-1,1} \ln _2+\bar{s}_{-1,-2,-1} \ln _2+\bar{s}_{-1,-2,1} \ln _2+\frac{43}{40} s_{-1} \zeta _2^2+8 \text{Li}_5\left(\frac{1}{2}\right)-3 \text{Li}_4\left(\frac{1}{2}\right) s_{-1}+\frac{1}{2} s_{-3} \zeta _2+\frac{3}{8} s_{-2} \zeta _3-\frac{447 \zeta _5}{64}-\frac{1}{2} \zeta _2 \bar{s}_{-3}-\frac{9}{20} \zeta _2^2 \bar{s}_{-1}-\text{Li}_4\left(\frac{1}{2}\right) \bar{s}_{-1}-\frac{5}{8} \zeta _3 \bar{s}_2-\frac{1}{2} \zeta _2 \bar{s}_3-\frac{1}{2} \zeta _2 s_{-2,1}-\frac{5}{8} \zeta _3 s_{-1,-1}-\frac{1}{2} \zeta _2 s_{-1,2}+s_{4,-1}+\bar{s}_{-4,1}+\frac{1}{2} \zeta _2 \bar{s}_{-2,-1}+\frac{1}{2} \zeta _2 \bar{s}_{-2,1}+\frac{1}{2} \zeta _2 \bar{s}_{-1,-2}+\frac{5}{8} \zeta _3 \bar{s}_{-1,-1}+\frac{1}{2} \zeta _2 \bar{s}_{-1,2}-s_{-3,-1,-1}-s_{-2,-2,-1}-s_{-1,-3,-1}-s_{3,1,-1}-\bar{s}_{-2,2,1}-\bar{s}_{-1,3,1}-\bar{s}_{2,-2,1}-\bar{s}_{3,-1,1}+s_{-2,-1,1,-1}+s_{-2,1,-1,-1}+s_{-1,-2,1,-1}+\bar{s}_{-2,-1,-1,1}+\bar{s}_{-1,-2,-1,1}+\bar{s}_{-1,-1,-2,1}  \end{dmath}
   \begin{dmath}[style={\small}]     s_{-1,-1} \bar{s}_{2,-1}  =   -\frac{1}{4} \zeta _3 \bar{s}_{-1,-1}-\frac{1}{2} \zeta _2 \bar{s}_{-1,2}-\frac{1}{2} \zeta _2 \bar{s}_{2,-1}+\bar{s}_{4,-1}-\bar{s}_{-3,-1,-1}-\bar{s}_{-1,-3,-1}-2 \bar{s}_{2,2,-1}+\bar{s}_{-1,-1,2,-1}+\bar{s}_{-1,2,-1,-1}+\bar{s}_{2,-1,-1,-1}-\ln _2^2 \bar{s}_{-1,-2}+\ln _2^2 \bar{s}_{-1,2}+2 \ln _2^2 \bar{s}_{2,-1}-2 \ln _2 \bar{s}_{-3,-1}-\ln _2 \bar{s}_{-1,-3}+\ln _2 \bar{s}_{-1,3}+\ln _2 \bar{s}_{2,-2}-\ln _2 \bar{s}_{2,2}+2 \ln _2 \bar{s}_{-1,2,-1}+2 \ln _2 \bar{s}_{2,-1,-1}+8 \text{Li}_4\left(\frac{1}{2}\right) \bar{s}_{-1}+\frac{1}{2} \zeta _2 \bar{s}_{-3}-\frac{1}{2} \zeta _3 \bar{s}_{-2}-\frac{111}{40} \zeta _2^2 \bar{s}_{-1}+\frac{3}{4} \zeta _3 \bar{s}_2-2 \zeta _2 \ln _2^2 \bar{s}_{-1}+\frac{1}{2} \zeta _2 \ln _2 \bar{s}_{-2}+3 \zeta _3 \ln _2 \bar{s}_{-1}-\frac{3}{2} \zeta _2 \ln _2 \bar{s}_2+\frac{1}{3} \ln _2^4 \bar{s}_{-1}-\frac{2}{3} \ln _2^3 \bar{s}_{-2}+\frac{2}{3} \ln _2^3 \bar{s}_2-\ln _2^2 \bar{s}_{-3}+\ln _2^2 \bar{s}_3-\ln _2 \bar{s}_{-4}+\ln _2 \bar{s}_4+\frac{1}{4} \zeta _3 s_{-1,-1}-\frac{1}{2} \zeta _2 s_{-1,2}-\frac{1}{2} \zeta _2 s_{2,-1}+s_{4,-1}-2 s_{-3,-1,-1}-s_{-1,-3,-1}-s_{2,2,-1}+s_{-1,2,-1,-1}+2 s_{2,-1,-1,-1}-\ln _2^2 s_{-1,-2}+\ln _2^2 s_{-1,2}-2 \ln _2 s_{-3,-1}-\ln _2 s_{-1,-3}+\ln _2 s_{-1,3}+\ln _2 s_{2,-2}-\ln _2 s_{2,2}+2 \ln _2 s_{-1,2,-1}+2 \ln _2 s_{2,-1,-1}+\frac{753 \zeta _5}{32}-2 \zeta _2 \ln _2^3+3 \zeta _3 \ln _2^2-\frac{169}{20} \zeta _2^2 \ln _2+4 \text{Li}_4\left(\frac{1}{2}\right) s_{-1}-24 \text{Li}_5\left(\frac{1}{2}\right)-\frac{61}{40} \zeta _2^2 s_{-1}+\frac{1}{2} \zeta _2 s_{-3}+\frac{1}{2} \zeta _3 s_{-2}+\frac{3 \zeta _3 s_2}{4}-\zeta _2 s_{-1} \ln _2^2-\frac{1}{2} \zeta _2 s_{-2} \ln _2-\frac{1}{2} \zeta _2 s_2 \ln _2+3 \zeta _3 s_{-1} \ln _2+\frac{1}{6} s_{-1} \ln _2^4+\frac{2}{3} s_{-2} \ln _2^3-\frac{2}{3} s_2 \ln _2^3-s_{-4} \ln _2+s_4 \ln _2+\frac{\ln _2^5}{5}  \end{dmath}

   \begin{dmath}[style={\small}]     s_{-1,-1} \bar{s}_{-1,1,-1}  =   -\frac{\ln _2^5}{4}-\frac{1}{24} s_{-1} \ln _2^4+\frac{1}{24} \bar{s}_{-1} \ln _2^4+\frac{2}{3} s_{-2} \ln _2^3-\frac{2}{3} s_2 \ln _2^3-\frac{4}{3} \zeta _2 \ln _2^3+\frac{2}{3} s_{-1,-1} \ln _2^3-\frac{2}{3} s_{-1,1} \ln _2^3-\frac{2}{3} \bar{s}_{-1,-1} \ln _2^3+\frac{2}{3} \bar{s}_{-1,1} \ln _2^3-\frac{1}{4} s_{-1} \zeta _2 \ln _2^2-\frac{1}{2} \zeta _3 \ln _2^2+\frac{1}{2} \bar{s}_{-3} \ln _2^2-\frac{19}{4} \zeta _2 \bar{s}_{-1} \ln _2^2-\frac{1}{2} \bar{s}_3 \ln _2^2-s_{-1,-2} \ln _2^2+s_{-1,2} \ln _2^2+\frac{1}{2} s_{2,-1} \ln _2^2-\frac{3}{2} s_{2,1} \ln _2^2+\frac{1}{2} \bar{s}_{2,-1} \ln _2^2-\frac{1}{2} \bar{s}_{2,1} \ln _2^2-s_{-1,-1,-1} \ln _2^2+3 s_{-1,-1,1} \ln _2^2-\bar{s}_{-1,-1,-1} \ln _2^2+\bar{s}_{-1,-1,1} \ln _2^2+2 \bar{s}_{-1,1,-1} \ln _2^2+\frac{3}{8} \zeta _2^2 \ln _2-12 \text{Li}_4\left(\frac{1}{2}\right) \ln _2-s_{-4} \ln _2+s_4 \ln _2+\frac{1}{2} s_{-2} \zeta _2 \ln _2+\frac{1}{2} s_2 \zeta _2 \ln _2+\frac{3}{4} s_{-1} \zeta _3 \ln _2+\frac{15}{4} \zeta _3 \bar{s}_{-1} \ln _2+\frac{3}{2} \zeta _2 \bar{s}_2 \ln _2-s_{-2,-2} \ln _2+s_{-2,2} \ln _2-2 s_{-1,-3} \ln _2-\frac{1}{2} \zeta _2 s_{-1,-1} \ln _2-\frac{1}{2} \zeta _2 s_{-1,1} \ln _2+2 s_{-1,3} \ln _2+s_{2,-2} \ln _2-s_{2,2} \ln _2+2 s_{3,-1} \ln _2-\bar{s}_{-1,-3} \ln _2-\frac{5}{2} \zeta _2 \bar{s}_{-1,-1} \ln _2-\frac{3}{2} \zeta _2 \bar{s}_{-1,1} \ln _2+\bar{s}_{-1,3} \ln _2+\bar{s}_{2,-2} \ln _2-\bar{s}_{2,2} \ln _2-2 s_{-2,-1,-1} \ln _2-4 s_{-1,-2,-1} \ln _2-2 s_{-1,-1,-2} \ln _2+2 s_{-1,-1,2} \ln _2+s_{-1,1,-2} \ln _2-s_{-1,1,2} \ln _2-2 s_{2,1,-1} \ln _2-2 \bar{s}_{-1,-2,-1} \ln _2-2 \bar{s}_{-1,-1,-2} \ln _2+2 \bar{s}_{-1,-1,2} \ln _2+\bar{s}_{-1,1,-2} \ln _2-\bar{s}_{-1,1,2} \ln _2-2 \bar{s}_{2,1,-1} \ln _2+4 s_{-1,-1,1,-1} \ln _2+2 s_{-1,1,-1,-1} \ln _2+4 \bar{s}_{-1,-1,1,-1} \ln _2+2 \bar{s}_{-1,1,-1,-1} \ln _2+\frac{1}{5} s_{-1} \zeta _2^2-10 \text{Li}_5\left(\frac{1}{2}\right)-\text{Li}_4\left(\frac{1}{2}\right) s_{-1}-\frac{s_3 \zeta _2}{2}-\frac{3}{4} s_{-2} \zeta _3-\frac{s_2 \zeta _3}{2}+10 \zeta _5-\frac{13}{40} \zeta _2^2 \bar{s}_{-1}+\text{Li}_4\left(\frac{1}{2}\right) \bar{s}_{-1}-\frac{1}{8} \zeta _3 \bar{s}_2-s_{-4,-1}+\frac{1}{2} \zeta _2 s_{-2,-1}+\zeta _2 s_{-1,-2}+\frac{3}{4} \zeta _3 s_{-1,-1}+\frac{3}{4} \zeta _3 s_{-1,1}+\frac{1}{2} \zeta _2 s_{2,1}+\frac{1}{2} \zeta _2 \bar{s}_{-1,-2}+\frac{1}{2} \zeta _3 \bar{s}_{-1,1}+\frac{1}{2} \zeta _2 \bar{s}_{2,1}+s_{-2,2,-1}-\zeta _2 s_{-1,-1,1}-\frac{1}{2} \zeta _2 s_{-1,1,-1}+2 s_{-1,3,-1}+s_{2,-2,-1}+2 s_{3,-1,-1}+\bar{s}_{-3,1,-1}-\zeta _2 \bar{s}_{-1,-1,1}-\frac{1}{2} \zeta _2 \bar{s}_{-1,1,-1}+\bar{s}_{-1,3,-1}+\bar{s}_{2,-2,-1}-2 s_{-2,-1,-1,-1}-3 s_{-1,-2,-1,-1}-2 s_{-1,-1,-2,-1}-s_{-1,1,2,-1}-s_{2,1,-1,-1}-\bar{s}_{-1,-2,-1,-1}-2 \bar{s}_{-1,-1,-2,-1}-\bar{s}_{-1,1,2,-1}-2 \bar{s}_{-1,2,1,-1}-2 \bar{s}_{2,-1,1,-1}-\bar{s}_{2,1,-1,-1}+2 s_{-1,-1,1,-1,-1}+2 s_{-1,1,-1,-1,-1}+3 \bar{s}_{-1,-1,-1,1,-1}+2 \bar{s}_{-1,-1,1,-1,-1}+\bar{s}_{-1,1,-1,-1,-1}  \end{dmath}
   \begin{dmath}[style={\small}]     s_{-1,-1} \bar{s}_{-1,1,1}  =   -\frac{\ln _2^5}{120}-\frac{1}{6} \bar{s}_{-1} \ln _2^4+\frac{1}{3} s_{-2} \ln _2^3-\frac{1}{3} s_2 \ln _2^3-\frac{7}{12} \zeta _2 \ln _2^3-\frac{1}{3} s_{-1,1} \ln _2^3-\frac{2}{3} \bar{s}_{-1,-1} \ln _2^3+\frac{2}{3} \bar{s}_{-1,1} \ln _2^3+\frac{1}{2} s_{-3} \ln _2^2-\frac{1}{2} s_3 \ln _2^2+\frac{11}{16} \zeta _3 \ln _2^2+\frac{1}{4} \zeta _2 \bar{s}_{-1} \ln _2^2+\frac{1}{2} s_{-2,-1} \ln _2^2-\frac{1}{2} s_{-2,1} \ln _2^2+\frac{1}{2} s_{2,-1} \ln _2^2-\frac{1}{2} s_{2,1} \ln _2^2-\frac{1}{2} \bar{s}_{-1,-2} \ln _2^2+\frac{1}{2} \bar{s}_{-1,2} \ln _2^2+\frac{1}{2} \bar{s}_{2,-1} \ln _2^2-\frac{1}{2} \bar{s}_{2,1} \ln _2^2-s_{-1,-1,-1} \ln _2^2+s_{-1,-1,1} \ln _2^2-\frac{1}{2} s_{-1,1,-1} \ln _2^2+\frac{1}{2} s_{-1,1,1} \ln _2^2-\bar{s}_{-1,-1,-1} \ln _2^2+\bar{s}_{-1,-1,1} \ln _2^2+\frac{3}{2} \bar{s}_{-1,1,-1} \ln _2^2+\frac{1}{2} \bar{s}_{-1,1,1} \ln _2^2+\frac{81}{40} \zeta _2^2 \ln _2-3 \text{Li}_4\left(\frac{1}{2}\right) \ln _2-s_{-4} \ln _2+s_4 \ln _2+\frac{9}{8} \zeta _3 \bar{s}_{-1} \ln _2-s_{-2,-2} \ln _2+s_{-2,2} \ln _2-2 s_{-1,-3} \ln _2+\zeta _2 s_{-1,-1} \ln _2+2 s_{-1,3} \ln _2+s_{2,-2} \ln _2-s_{2,2} \ln _2+s_{3,-1} \ln _2-s_{3,1} \ln _2+\zeta _2 \bar{s}_{-1,-1} \ln _2+\zeta _2 \bar{s}_{-1,1} \ln _2-s_{-2,-1,-1} \ln _2+s_{-2,-1,1} \ln _2-2 s_{-1,-2,-1} \ln _2+2 s_{-1,-2,1} \ln _2-2 s_{-1,-1,-2} \ln _2+2 s_{-1,-1,2} \ln _2+s_{-1,1,-2} \ln _2-s_{-1,1,2} \ln _2-s_{2,1,-1} \ln _2+s_{2,1,1} \ln _2-\bar{s}_{-1,-2,-1} \ln _2-\bar{s}_{-1,-2,1} \ln _2-\bar{s}_{2,1,-1} \ln _2-\bar{s}_{2,1,1} \ln _2+2 s_{-1,-1,1,-1} \ln _2-2 s_{-1,-1,1,1} \ln _2+s_{-1,1,-1,-1} \ln _2-s_{-1,1,-1,1} \ln _2+2 \bar{s}_{-1,-1,1,-1} \ln _2+2 \bar{s}_{-1,-1,1,1} \ln _2+\bar{s}_{-1,1,-1,-1} \ln _2+\bar{s}_{-1,1,-1,1} \ln _2-\frac{3}{20} s_{-1} \zeta _2^2+8 \text{Li}_5\left(\frac{1}{2}\right)+\frac{1}{2} s_{-3} \zeta _2+\frac{1}{8} s_{-2} \zeta _3+\frac{s_2 \zeta _3}{4}+\frac{3 \zeta _2 \zeta _3}{16}-\frac{453 \zeta _5}{64}+\frac{27}{40} \zeta _2^2 \bar{s}_{-1}-4 \text{Li}_4\left(\frac{1}{2}\right) \bar{s}_{-1}-\zeta _3 \bar{s}_2-\frac{1}{2} \zeta _2 s_{-2,1}-\frac{11}{8} \zeta _3 s_{-1,-1}-\frac{1}{8} \zeta _3 s_{-1,1}-\zeta _2 s_{-1,2}-\frac{1}{2} \zeta _2 s_{2,-1}+s_{4,-1}-\frac{1}{2} \zeta _2 \bar{s}_{-1,-2}+\frac{9}{8} \zeta _3 \bar{s}_{-1,-1}+\frac{1}{4} \zeta _3 \bar{s}_{-1,1}-\frac{1}{2} \zeta _2 \bar{s}_{-1,2}-\frac{1}{2} \zeta _2 \bar{s}_{2,-1}-\frac{1}{2} \zeta _2 \bar{s}_{2,1}-s_{-3,-1,-1}-s_{-2,-2,-1}-2 s_{-1,-3,-1}+\zeta _2 s_{-1,-1,-1}+\frac{1}{2} \zeta _2 s_{-1,1,1}-s_{2,2,-1}-s_{3,1,-1}+\bar{s}_{-3,1,1}+\zeta _2 \bar{s}_{-1,-1,-1}+\zeta _2 \bar{s}_{-1,-1,1}+\frac{1}{2} \zeta _2 \bar{s}_{-1,1,-1}+\frac{1}{2} \zeta _2 \bar{s}_{-1,1,1}+\bar{s}_{-1,3,1}+\bar{s}_{2,-2,1}+s_{-2,-1,1,-1}+s_{-2,1,-1,-1}+2 s_{-1,-2,1,-1}+2 s_{-1,-1,2,-1}+s_{-1,1,-2,-1}+s_{-1,2,-1,-1}+s_{2,1,1,-1}-\bar{s}_{-1,-2,-1,1}-2 \bar{s}_{-1,-1,-2,1}-\bar{s}_{-1,1,2,1}-2 \bar{s}_{-1,2,1,1}-2 \bar{s}_{2,-1,1,1}-\bar{s}_{2,1,-1,1}-2 s_{-1,-1,1,1,-1}-s_{-1,1,-1,1,-1}-s_{-1,1,1,-1,-1}+3 \bar{s}_{-1,-1,-1,1,1}+2 \bar{s}_{-1,-1,1,-1,1}+\bar{s}_{-1,1,-1,-1,1}  \end{dmath}

   \begin{dmath}[style={\small}]     s_{-1,-1} \bar{s}_{1,-1,-1}  =   \frac{37 \ln _2^5}{120}+\frac{7}{24} s_{-1} \ln _2^4-\frac{5}{8} s_1 \ln _2^4+\frac{1}{6} \bar{s}_{-1} \ln _2^4-\frac{3}{8} \bar{s}_1 \ln _2^4-\frac{1}{12} \zeta _2 \ln _2^3+\frac{1}{6} \bar{s}_{-2} \ln _2^3-\frac{1}{6} \bar{s}_2 \ln _2^3-\frac{1}{6} s_{-1,-1} \ln _2^3+\frac{7}{6} s_{-1,1} \ln _2^3+\frac{1}{6} \bar{s}_{-1,-1} \ln _2^3-\frac{1}{6} \bar{s}_{-1,1} \ln _2^3+\frac{1}{4} s_{-1} \zeta _2 \ln _2^2+\frac{3}{4} s_1 \zeta _2 \ln _2^2+\frac{15}{4} \zeta _3 \ln _2^2+\bar{s}_{-3} \ln _2^2+\frac{3}{4} \zeta _2 \bar{s}_{-1} \ln _2^2+\frac{5}{4} \zeta _2 \bar{s}_1 \ln _2^2-\bar{s}_3 \ln _2^2-2 s_{-2,-1} \ln _2^2-s_{-1,-2} \ln _2^2+s_{-1,2} \ln _2^2+\bar{s}_{-1,-2} \ln _2^2-\bar{s}_{-1,2} \ln _2^2+2 s_{-1,1,-1} \ln _2^2+2 s_{1,-1,-1} \ln _2^2+2 \bar{s}_{1,-1,-1} \ln _2^2+\frac{73}{20} \zeta _2^2 \ln _2+12 \text{Li}_4\left(\frac{1}{2}\right) \ln _2-s_{-4} \ln _2+s_4 \ln _2+s_{-2} \zeta _2 \ln _2-\frac{5}{2} s_{-1} \zeta _3 \ln _2-2 s_1 \zeta _3 \ln _2+\frac{1}{2} \zeta _3 \bar{s}_{-1} \ln _2-2 \zeta _3 \bar{s}_1 \ln _2-2 s_{-2,-2} \ln _2+2 s_{-2,2} \ln _2-s_{-1,-3} \ln _2-\zeta _2 s_{-1,1} \ln _2+s_{-1,3} \ln _2+s_{1,-3} \ln _2-\zeta _2 s_{1,-1} \ln _2-s_{1,3} \ln _2+2 s_{3,-1} \ln _2-\bar{s}_{-2,-2} \ln _2+\bar{s}_{-2,2} \ln _2+\bar{s}_{1,-3} \ln _2-\zeta _2 \bar{s}_{1,-1} \ln _2-\bar{s}_{1,3} \ln _2-4 s_{-2,-1,-1} \ln _2-2 s_{-1,-2,-1} \ln _2+s_{-1,1,-2} \ln _2-s_{-1,1,2} \ln _2+2 s_{1,-1,-2} \ln _2-2 s_{1,-1,2} \ln _2-2 s_{1,2,-1} \ln _2-2 \bar{s}_{-2,-1,-1} \ln _2+\bar{s}_{-1,1,-2} \ln _2-\bar{s}_{-1,1,2} \ln _2+2 \bar{s}_{1,-1,-2} \ln _2-2 \bar{s}_{1,-1,2} \ln _2-2 \bar{s}_{1,2,-1} \ln _2+2 s_{-1,1,-1,-1} \ln _2+4 s_{1,-1,-1,-1} \ln _2+2 \bar{s}_{-1,1,-1,-1} \ln _2+4 \bar{s}_{1,-1,-1,-1} \ln _2+\frac{61}{40} s_{-1} \zeta _2^2+\frac{51}{40} s_1 \zeta _2^2+23 \text{Li}_5\left(\frac{1}{2}\right)-4 \text{Li}_4\left(\frac{1}{2}\right) s_{-1}-4 \text{Li}_4\left(\frac{1}{2}\right) s_1-\frac{s_3 \zeta _2}{2}-\frac{5}{4} s_{-2} \zeta _3-\frac{1437 \zeta _5}{64}-\frac{3}{4} \zeta _3 \bar{s}_{-2}+\frac{13}{40} \zeta _2^2 \bar{s}_{-1}-\text{Li}_4\left(\frac{1}{2}\right) \bar{s}_{-1}+\frac{61}{40} \zeta _2^2 \bar{s}_1-4 \text{Li}_4\left(\frac{1}{2}\right) \bar{s}_1-\frac{1}{4} \zeta _3 \bar{s}_2-s_{-4,-1}+\zeta _2 s_{-2,-1}+\frac{1}{2} \zeta _2 s_{-1,-2}-\frac{1}{4} \zeta _3 s_{-1,-1}+\frac{1}{2} \zeta _3 s_{-1,1}+\frac{5}{4} \zeta _3 s_{1,-1}+\frac{1}{2} \zeta _2 s_{1,2}+\frac{1}{2} \zeta _2 \bar{s}_{-2,-1}+\frac{1}{4} \zeta _3 \bar{s}_{-1,-1}+\frac{3}{4} \zeta _3 \bar{s}_{-1,1}+\frac{5}{4} \zeta _3 \bar{s}_{1,-1}+\frac{1}{2} \zeta _2 \bar{s}_{1,2}+2 s_{-2,2,-1}-\frac{1}{2} \zeta _2 s_{-1,1,-1}+s_{-1,3,-1}+s_{1,-3,-1}-\zeta _2 s_{1,-1,-1}+2 s_{3,-1,-1}+\bar{s}_{-2,2,-1}-\frac{1}{2} \zeta _2 \bar{s}_{-1,1,-1}+\bar{s}_{1,-3,-1}-\zeta _2 \bar{s}_{1,-1,-1}+\bar{s}_{3,-1,-1}-3 s_{-2,-1,-1,-1}-s_{-1,-2,-1,-1}-s_{-1,1,2,-1}-2 s_{1,-1,2,-1}-2 s_{1,2,-1,-1}-2 \bar{s}_{-2,-1,-1,-1}-\bar{s}_{-1,-2,-1,-1}-\bar{s}_{-1,1,2,-1}-2 \bar{s}_{1,-1,2,-1}-2 \bar{s}_{1,2,-1,-1}-\bar{s}_{2,1,-1,-1}+s_{-1,1,-1,-1,-1}+3 s_{1,-1,-1,-1,-1}+\bar{s}_{-1,-1,1,-1,-1}+2 \bar{s}_{-1,1,-1,-1,-1}+3 \bar{s}_{1,-1,-1,-1,-1}  \end{dmath}
   \begin{dmath}[style={\small}]     s_{-1,-1} \bar{s}_{1,-1,1}  =   \frac{11 \ln _2^5}{30}+\frac{5}{24} s_{-1} \ln _2^4-\frac{5}{12} s_1 \ln _2^4+\frac{7}{24} \bar{s}_{-1} \ln _2^4-\frac{1}{24} \bar{s}_1 \ln _2^4+\frac{2}{3} \zeta _2 \ln _2^3+\frac{1}{6} \bar{s}_{-2} \ln _2^3-\frac{1}{6} \bar{s}_2 \ln _2^3-\frac{1}{6} s_{-1,-1} \ln _2^3+\frac{5}{6} s_{-1,1} \ln _2^3+\frac{1}{6} \bar{s}_{-1,-1} \ln _2^3-\frac{1}{6} \bar{s}_{-1,1} \ln _2^3+\frac{1}{2} s_{-3} \ln _2^2-\frac{1}{2} s_3 \ln _2^2+\frac{1}{4} s_{-1} \zeta _2 \ln _2^2+\frac{1}{2} s_1 \zeta _2 \ln _2^2+\frac{49}{16} \zeta _3 \ln _2^2+\frac{1}{2} \bar{s}_{-3} \ln _2^2+\frac{3}{4} \zeta _2 \bar{s}_{-1} \ln _2^2+\frac{9}{4} \zeta _2 \bar{s}_1 \ln _2^2-\frac{1}{2} \bar{s}_3 \ln _2^2-\frac{1}{2} s_{-2,-1} \ln _2^2-\frac{1}{2} s_{-2,1} \ln _2^2-\frac{1}{2} s_{-1,-2} \ln _2^2+\frac{1}{2} s_{-1,2} \ln _2^2-\frac{1}{2} s_{1,-2} \ln _2^2+\frac{1}{2} s_{1,2} \ln _2^2+\frac{1}{2} \bar{s}_{-1,-2} \ln _2^2-\frac{1}{2} \bar{s}_{-1,2} \ln _2^2+s_{-1,1,-1} \ln _2^2+\frac{1}{2} s_{1,-1,-1} \ln _2^2+\frac{1}{2} s_{1,-1,1} \ln _2^2+\frac{3}{2} \bar{s}_{1,-1,-1} \ln _2^2+\frac{1}{2} \bar{s}_{1,-1,1} \ln _2^2-\frac{119}{40} \zeta _2^2 \ln _2+9 \text{Li}_4\left(\frac{1}{2}\right) \ln _2-s_{-4} \ln _2+s_4 \ln _2-\frac{9}{4} s_{-1} \zeta _3 \ln _2-\frac{5}{4} s_1 \zeta _3 \ln _2-\frac{3}{2} \zeta _2 \bar{s}_{-2} \ln _2+\frac{1}{8} \zeta _3 \bar{s}_{-1} \ln _2-\frac{3}{4} \zeta _3 \bar{s}_1 \ln _2-2 s_{-2,-2} \ln _2+2 s_{-2,2} \ln _2-s_{-1,-3} \ln _2-\zeta _2 s_{-1,1} \ln _2+s_{-1,3} \ln _2+s_{1,-3} \ln _2-s_{1,3} \ln _2+s_{3,-1} \ln _2-s_{3,1} \ln _2+\frac{3}{2} \zeta _2 \bar{s}_{-1,1} \ln _2+\frac{7}{2} \zeta _2 \bar{s}_{1,-1} \ln _2-2 s_{-2,-1,-1} \ln _2+2 s_{-2,-1,1} \ln _2-s_{-1,-2,-1} \ln _2+s_{-1,-2,1} \ln _2+s_{-1,1,-2} \ln _2-s_{-1,1,2} \ln _2+2 s_{1,-1,-2} \ln _2-2 s_{1,-1,2} \ln _2-s_{1,2,-1} \ln _2+s_{1,2,1} \ln _2-\bar{s}_{-2,-1,-1} \ln _2-\bar{s}_{-2,-1,1} \ln _2-\bar{s}_{1,2,-1} \ln _2-\bar{s}_{1,2,1} \ln _2+s_{-1,1,-1,-1} \ln _2-s_{-1,1,-1,1} \ln _2+2 s_{1,-1,-1,-1} \ln _2-2 s_{1,-1,-1,1} \ln _2+\bar{s}_{-1,1,-1,-1} \ln _2+\bar{s}_{-1,1,-1,1} \ln _2+2 \bar{s}_{1,-1,-1,-1} \ln _2+2 \bar{s}_{1,-1,-1,1} \ln _2+\frac{51}{40} s_{-1} \zeta _2^2+\frac{9}{20} s_1 \zeta _2^2+3 \text{Li}_5\left(\frac{1}{2}\right)-4 \text{Li}_4\left(\frac{1}{2}\right) s_{-1}-\text{Li}_4\left(\frac{1}{2}\right) s_1+\frac{1}{2} s_{-3} \zeta _2+\frac{3}{8} s_{-2} \zeta _3-\frac{11 \zeta _2 \zeta _3}{16}-\frac{103 \zeta _5}{64}-\frac{1}{2} \zeta _2 \bar{s}_{-3}+\frac{3}{2} \zeta _3 \bar{s}_{-2}-\frac{47}{40} \zeta _2^2 \bar{s}_{-1}+2 \text{Li}_4\left(\frac{1}{2}\right) \bar{s}_{-1}-\frac{51}{40} \zeta _2^2 \bar{s}_1+4 \text{Li}_4\left(\frac{1}{2}\right) \bar{s}_1+\frac{1}{8} \zeta _3 \bar{s}_2+\frac{1}{2} \zeta _2 \bar{s}_3+\frac{1}{2} \zeta _2 s_{-2,-1}-\frac{1}{2} \zeta _2 s_{-2,1}+\frac{1}{8} \zeta _3 s_{-1,-1}+\frac{3}{4} \zeta _3 s_{-1,1}-\frac{1}{2} \zeta _2 s_{-1,2}-\frac{1}{2} \zeta _2 s_{1,-2}-\frac{3}{8} \zeta _3 s_{1,-1}+s_{4,-1}-\zeta _2 \bar{s}_{-2,-1}-\frac{1}{2} \zeta _2 \bar{s}_{-1,-2}-\frac{1}{8} \zeta _3 \bar{s}_{-1,-1}-\frac{3}{2} \zeta _3 \bar{s}_{-1,1}+\frac{1}{2} \zeta _2 \bar{s}_{-1,2}-\frac{9}{4} \zeta _3 \bar{s}_{1,-1}-\zeta _2 \bar{s}_{1,2}-s_{-3,-1,-1}-2 s_{-2,-2,-1}-s_{-1,-3,-1}-\frac{1}{2} \zeta _2 s_{-1,1,-1}-\frac{1}{2} \zeta _2 s_{1,-1,-1}+\frac{1}{2} \zeta _2 s_{1,-1,1}-s_{1,3,-1}-s_{3,1,-1}+\bar{s}_{-2,2,1}+\zeta _2 \bar{s}_{-1,1,-1}+\bar{s}_{1,-3,1}+\frac{3}{2} \zeta _2 \bar{s}_{1,-1,-1}+\frac{1}{2} \zeta _2 \bar{s}_{1,-1,1}+\bar{s}_{3,-1,1}+2 s_{-2,-1,1,-1}+s_{-2,1,-1,-1}+s_{-1,-2,1,-1}+s_{-1,1,-2,-1}+s_{1,-2,-1,-1}+2 s_{1,-1,-2,-1}+s_{1,2,1,-1}-2 \bar{s}_{-2,-1,-1,1}-\bar{s}_{-1,-2,-1,1}-\bar{s}_{-1,1,2,1}-2 \bar{s}_{1,-1,2,1}-2 \bar{s}_{1,2,-1,1}-\bar{s}_{2,1,-1,1}-s_{-1,1,-1,1,-1}-2 s_{1,-1,-1,1,-1}-s_{1,-1,1,-1,-1}+\bar{s}_{-1,-1,1,-1,1}+2 \bar{s}_{-1,1,-1,-1,1}+3 \bar{s}_{1,-1,-1,-1,1}  \end{dmath}
   \begin{dmath}[style={\small}]     s_{-1,-1} \bar{s}_{1,1,-1}  =   \frac{7 \ln _2^5}{40}+\frac{1}{24} s_{-1} \ln _2^4+\frac{3}{8} \bar{s}_{-1} \ln _2^4-\frac{5}{24} \bar{s}_1 \ln _2^4+\frac{2}{3} s_{-2} \ln _2^3-\frac{2}{3} s_2 \ln _2^3-\frac{11}{4} \zeta _2 \ln _2^3+\frac{1}{6} \bar{s}_{-2} \ln _2^3-\frac{1}{6} \bar{s}_2 \ln _2^3-\frac{1}{6} s_{-1,-1} \ln _2^3-\frac{1}{6} s_{-1,1} \ln _2^3-\frac{2}{3} s_{1,-1} \ln _2^3+\frac{2}{3} s_{1,1} \ln _2^3+\frac{1}{6} \bar{s}_{-1,-1} \ln _2^3-\frac{1}{6} \bar{s}_{-1,1} \ln _2^3-\frac{2}{3} \bar{s}_{1,-1} \ln _2^3+\frac{2}{3} \bar{s}_{1,1} \ln _2^3+\frac{1}{4} s_{-1} \zeta _2 \ln _2^2+\frac{27}{8} \zeta _3 \ln _2^2-\frac{1}{2} \bar{s}_{-3} \ln _2^2-\frac{5}{2} \zeta _2 \bar{s}_{-1} \ln _2^2-\frac{9}{4} \zeta _2 \bar{s}_1 \ln _2^2+\frac{1}{2} \bar{s}_3 \ln _2^2-\frac{1}{2} s_{-2,-1} \ln _2^2+\frac{3}{2} s_{-2,1} \ln _2^2-s_{-1,-2} \ln _2^2+s_{-1,2} \ln _2^2+\frac{1}{2} \bar{s}_{-2,-1} \ln _2^2-\frac{1}{2} \bar{s}_{-2,1} \ln _2^2-\frac{1}{2} \bar{s}_{-1,-2} \ln _2^2+\frac{1}{2} \bar{s}_{-1,2} \ln _2^2-\frac{1}{2} \bar{s}_{1,-2} \ln _2^2+\frac{1}{2} \bar{s}_{1,2} \ln _2^2+\frac{1}{2} s_{-1,1,-1} \ln _2^2-\frac{3}{2} s_{-1,1,1} \ln _2^2+\frac{1}{2} s_{1,-1,-1} \ln _2^2-\frac{3}{2} s_{1,-1,1} \ln _2^2-\frac{1}{2} \bar{s}_{-1,1,-1} \ln _2^2+\frac{1}{2} \bar{s}_{-1,1,1} \ln _2^2-\frac{1}{2} \bar{s}_{1,-1,-1} \ln _2^2+\frac{1}{2} \bar{s}_{1,-1,1} \ln _2^2+2 \bar{s}_{1,1,-1} \ln _2^2-\frac{71}{20} \zeta _2^2 \ln _2-s_{-4} \ln _2+s_4 \ln _2-\frac{1}{2} s_{-2} \zeta _2 \ln _2-\frac{1}{2} s_2 \zeta _2 \ln _2-\frac{1}{2} s_{-1} \zeta _3 \ln _2-\frac{1}{4} s_1 \zeta _3 \ln _2+\zeta _2 \bar{s}_{-2} \ln _2+\frac{7}{2} \zeta _3 \bar{s}_{-1} \ln _2+\frac{1}{4} \zeta _3 \bar{s}_1 \ln _2+\frac{1}{2} \zeta _2 \bar{s}_2 \ln _2-2 s_{-3,-1} \ln _2-s_{-2,-2} \ln _2+s_{-2,2} \ln _2-s_{-1,-3} \ln _2+\frac{1}{2} \zeta _2 s_{-1,-1} \ln _2+\frac{1}{2} \zeta _2 s_{-1,1} \ln _2+s_{-1,3} \ln _2+s_{1,-3} \ln _2+\frac{1}{2} \zeta _2 s_{1,-1} \ln _2+\frac{1}{2} \zeta _2 s_{1,1} \ln _2-s_{1,3} \ln _2+s_{2,-2} \ln _2-s_{2,2} \ln _2+\bar{s}_{-2,-2} \ln _2-\bar{s}_{-2,2} \ln _2-\frac{1}{2} \zeta _2 \bar{s}_{-1,-1} \ln _2-\zeta _2 \bar{s}_{-1,1} \ln _2-\bar{s}_{1,-3} \ln _2-\zeta _2 \bar{s}_{1,-1} \ln _2-\frac{3}{2} \zeta _2 \bar{s}_{1,1} \ln _2+\bar{s}_{1,3} \ln _2+2 s_{-2,1,-1} \ln _2+s_{-1,1,-2} \ln _2-s_{-1,1,2} \ln _2+2 s_{-1,2,-1} \ln _2+2 s_{1,-2,-1} \ln _2+s_{1,-1,-2} \ln _2-s_{1,-1,2} \ln _2-s_{1,1,-2} \ln _2+s_{1,1,2} \ln _2+2 s_{2,-1,-1} \ln _2-2 \bar{s}_{-2,1,-1} \ln _2-\bar{s}_{-1,1,-2} \ln _2+\bar{s}_{-1,1,2} \ln _2-2 \bar{s}_{1,-2,-1} \ln _2-\bar{s}_{1,-1,-2} \ln _2+\bar{s}_{1,-1,2} \ln _2+\bar{s}_{1,1,-2} \ln _2-\bar{s}_{1,1,2} \ln _2-2 s_{-1,1,1,-1} \ln _2-2 s_{1,-1,1,-1} \ln _2-2 s_{1,1,-1,-1} \ln _2+2 \bar{s}_{-1,1,1,-1} \ln _2+2 \bar{s}_{1,-1,1,-1} \ln _2+2 \bar{s}_{1,1,-1,-1} \ln _2-\frac{1}{5} s_1 \zeta _2^2-9 \text{Li}_5\left(\frac{1}{2}\right)+\text{Li}_4\left(\frac{1}{2}\right) s_1+\frac{1}{2} s_{-3} \zeta _2+\frac{1}{2} s_{-2} \zeta _3+\frac{3 s_2 \zeta _3}{4}+\frac{285 \zeta _5}{32}-\frac{1}{8} \zeta _3 \bar{s}_{-2}-\frac{8}{5} \zeta _2^2 \bar{s}_{-1}+4 \text{Li}_4\left(\frac{1}{2}\right) \bar{s}_{-1}-\frac{1}{2} \zeta _2 s_{-2,1}-\frac{1}{4} \zeta _3 s_{-1,1}-\frac{1}{2} \zeta _2 s_{-1,2}-\frac{1}{2} \zeta _2 s_{1,-2}-\frac{1}{2} \zeta _3 s_{1,-1}-\frac{3}{4} \zeta _3 s_{1,1}-\frac{1}{2} \zeta _2 s_{2,-1}+s_{4,-1}+\frac{1}{2} \zeta _2 \bar{s}_{-2,1}+\frac{1}{8} \zeta _3 \bar{s}_{-1,1}+\frac{1}{2} \zeta _2 \bar{s}_{1,-2}-\frac{1}{8} \zeta _3 \bar{s}_{1,-1}+\frac{1}{2} \zeta _3 \bar{s}_{1,1}-2 s_{-3,-1,-1}-s_{-2,-2,-1}-s_{-1,-3,-1}+\frac{1}{2} \zeta _2 s_{-1,1,1}+\frac{1}{2} \zeta _2 s_{1,-1,1}+\frac{1}{2} \zeta _2 s_{1,1,-1}-s_{1,3,-1}-s_{2,2,-1}+\bar{s}_{-2,-2,-1}-\frac{1}{2} \zeta _2 \bar{s}_{-1,1,1}-\frac{1}{2} \zeta _2 \bar{s}_{1,-1,1}-\frac{1}{2} \zeta _2 \bar{s}_{1,1,-1}+\bar{s}_{1,3,-1}+\bar{s}_{3,1,-1}+s_{-2,1,-1,-1}+s_{-1,1,-2,-1}+s_{-1,2,-1,-1}+2 s_{1,-2,-1,-1}+s_{1,-1,-2,-1}+s_{1,1,2,-1}+2 s_{2,-1,-1,-1}-\bar{s}_{-2,-1,1,-1}-\bar{s}_{-2,1,-1,-1}-\bar{s}_{-1,-2,1,-1}-\bar{s}_{-1,1,-2,-1}-\bar{s}_{1,-2,-1,-1}-\bar{s}_{1,-1,-2,-1}-\bar{s}_{1,1,2,-1}-\bar{s}_{1,2,1,-1}-\bar{s}_{2,1,1,-1}-s_{-1,1,1,-1,-1}-s_{1,-1,1,-1,-1}-2 s_{1,1,-1,-1,-1}+\bar{s}_{-1,-1,1,1,-1}+\bar{s}_{-1,1,-1,1,-1}+\bar{s}_{-1,1,1,-1,-1}+\bar{s}_{1,-1,-1,1,-1}+\bar{s}_{1,-1,1,-1,-1}+\bar{s}_{1,1,-1,-1,-1}  \end{dmath}
   \begin{dmath}[style={\small}]     s_{-1,-1} \bar{s}_{1,1,1}  =   \frac{\ln _2^5}{6}+\frac{1}{8} s_{-1} \ln _2^4-\frac{1}{8} s_1 \ln _2^4+\frac{1}{6} \bar{s}_{-1} \ln _2^4-\frac{5}{24} \bar{s}_1 \ln _2^4+\frac{1}{3} s_{-2} \ln _2^3-\frac{1}{3} s_2 \ln _2^3-\frac{5}{6} \zeta _2 \ln _2^3+\frac{1}{6} \bar{s}_{-2} \ln _2^3-\frac{1}{6} \bar{s}_2 \ln _2^3-\frac{1}{6} s_{-1,-1} \ln _2^3+\frac{1}{6} s_{-1,1} \ln _2^3-\frac{1}{3} s_{1,-1} \ln _2^3+\frac{1}{3} s_{1,1} \ln _2^3+\frac{1}{6} \bar{s}_{-1,-1} \ln _2^3-\frac{1}{6} \bar{s}_{-1,1} \ln _2^3-\frac{2}{3} \bar{s}_{1,-1} \ln _2^3+\frac{2}{3} \bar{s}_{1,1} \ln _2^3+\frac{1}{2} s_{-3} \ln _2^2-\frac{1}{2} s_3 \ln _2^2-\frac{1}{4} s_{-1} \zeta _2 \ln _2^2+\frac{1}{4} s_1 \zeta _2 \ln _2^2+\frac{35}{16} \zeta _3 \ln _2^2-\frac{1}{2} \zeta _2 \bar{s}_{-1} \ln _2^2+\frac{3}{4} \zeta _2 \bar{s}_1 \ln _2^2-\frac{1}{2} s_{-2,-1} \ln _2^2+\frac{1}{2} s_{-2,1} \ln _2^2-\frac{1}{2} s_{-1,-2} \ln _2^2+\frac{1}{2} s_{-1,2} \ln _2^2-\frac{1}{2} s_{1,-2} \ln _2^2+\frac{1}{2} s_{1,2} \ln _2^2-\frac{1}{2} s_{2,-1} \ln _2^2+\frac{1}{2} s_{2,1} \ln _2^2+\frac{1}{2} \bar{s}_{-2,-1} \ln _2^2-\frac{1}{2} \bar{s}_{-2,1} \ln _2^2-\frac{1}{2} \bar{s}_{1,-2} \ln _2^2+\frac{1}{2} \bar{s}_{1,2} \ln _2^2+\frac{1}{2} s_{-1,1,-1} \ln _2^2-\frac{1}{2} s_{-1,1,1} \ln _2^2+\frac{1}{2} s_{1,-1,-1} \ln _2^2-\frac{1}{2} s_{1,-1,1} \ln _2^2+\frac{1}{2} s_{1,1,-1} \ln _2^2-\frac{1}{2} s_{1,1,1} \ln _2^2-\frac{1}{2} \bar{s}_{-1,1,-1} \ln _2^2+\frac{1}{2} \bar{s}_{-1,1,1} \ln _2^2-\frac{1}{2} \bar{s}_{1,-1,-1} \ln _2^2+\frac{1}{2} \bar{s}_{1,-1,1} \ln _2^2+\frac{3}{2} \bar{s}_{1,1,-1} \ln _2^2+\frac{1}{2} \bar{s}_{1,1,1} \ln _2^2+\frac{4}{5} \zeta _2^2 \ln _2+3 \text{Li}_4\left(\frac{1}{2}\right) \ln _2-s_{-4} \ln _2+s_4 \ln _2-\frac{1}{2} \zeta _2 \bar{s}_{-2} \ln _2+\frac{7}{8} \zeta _3 \bar{s}_{-1} \ln _2+\frac{1}{4} \zeta _3 \bar{s}_1 \ln _2+\frac{1}{2} \zeta _2 \bar{s}_2 \ln _2-s_{-3,-1} \ln _2+s_{-3,1} \ln _2-s_{-2,-2} \ln _2+s_{-2,2} \ln _2-s_{-1,-3} \ln _2+\frac{1}{2} \zeta _2 s_{-1,-1} \ln _2-\frac{1}{2} \zeta _2 s_{-1,1} \ln _2+s_{-1,3} \ln _2+s_{1,-3} \ln _2-s_{1,3} \ln _2+s_{2,-2} \ln _2-s_{2,2} \ln _2-\frac{1}{2} \zeta _2 \bar{s}_{-1,-1} \ln _2+\frac{1}{2} \zeta _2 \bar{s}_{-1,1} \ln _2+\zeta _2 \bar{s}_{1,-1} \ln _2+\zeta _2 \bar{s}_{1,1} \ln _2+s_{-2,1,-1} \ln _2-s_{-2,1,1} \ln _2+s_{-1,1,-2} \ln _2-s_{-1,1,2} \ln _2+s_{-1,2,-1} \ln _2-s_{-1,2,1} \ln _2+s_{1,-2,-1} \ln _2-s_{1,-2,1} \ln _2+s_{1,-1,-2} \ln _2-s_{1,-1,2} \ln _2-s_{1,1,-2} \ln _2+s_{1,1,2} \ln _2+s_{2,-1,-1} \ln _2-s_{2,-1,1} \ln _2-\bar{s}_{-2,1,-1} \ln _2-\bar{s}_{-2,1,1} \ln _2-\bar{s}_{1,-2,-1} \ln _2-\bar{s}_{1,-2,1} \ln _2-s_{-1,1,1,-1} \ln _2+s_{-1,1,1,1} \ln _2-s_{1,-1,1,-1} \ln _2+s_{1,-1,1,1} \ln _2-s_{1,1,-1,-1} \ln _2+s_{1,1,-1,1} \ln _2+\bar{s}_{-1,1,1,-1} \ln _2+\bar{s}_{-1,1,1,1} \ln _2+\bar{s}_{1,-1,1,-1} \ln _2+\bar{s}_{1,-1,1,1} \ln _2+\bar{s}_{1,1,-1,-1} \ln _2+\bar{s}_{1,1,-1,1} \ln _2+\frac{1}{8} s_{-1} \zeta _2^2+\frac{1}{40} s_1 \zeta _2^2+5 \text{Li}_5\left(\frac{1}{2}\right)-\frac{s_3 \zeta _2}{2}-\frac{1}{4} s_{-2} \zeta _3-\frac{s_2 \zeta _3}{8}-\frac{21 \zeta _2 \zeta _3}{16}-\frac{135 \zeta _5}{64}-\frac{1}{8} \zeta _3 \bar{s}_{-2}+\frac{2}{5} \zeta _2^2 \bar{s}_{-1}-\text{Li}_4\left(\frac{1}{2}\right) \bar{s}_{-1}+\frac{1}{8} \zeta _2^2 \bar{s}_1-\frac{7}{8} \zeta _3 \bar{s}_2-s_{-4,-1}+\frac{1}{2} \zeta _2 s_{-2,-1}+\frac{1}{2} \zeta _2 s_{-1,-2}-\frac{7}{8} \zeta _3 s_{-1,-1}+\frac{1}{4} \zeta _3 s_{-1,1}+\frac{1}{4} \zeta _3 s_{1,-1}+\frac{1}{8} \zeta _3 s_{1,1}+\frac{1}{2} \zeta _2 s_{1,2}+\frac{1}{2} \zeta _2 s_{2,1}-\frac{1}{2} \zeta _2 \bar{s}_{-2,-1}-\frac{1}{2} \zeta _2 \bar{s}_{-2,1}+\frac{7}{8} \zeta _3 \bar{s}_{-1,-1}+\frac{1}{8} \zeta _3 \bar{s}_{-1,1}-\frac{1}{2} \zeta _2 \bar{s}_{1,-2}+\frac{1}{8} \zeta _3 \bar{s}_{1,-1}+\frac{1}{4} \zeta _3 \bar{s}_{1,1}-\frac{1}{2} \zeta _2 \bar{s}_{1,2}+s_{-3,1,-1}+s_{-2,2,-1}-\frac{1}{2} \zeta _2 s_{-1,1,-1}+s_{-1,3,-1}+s_{1,-3,-1}-\frac{1}{2} \zeta _2 s_{1,-1,-1}-\frac{1}{2} \zeta _2 s_{1,1,1}+s_{2,-2,-1}+s_{3,-1,-1}+\bar{s}_{-2,-2,1}+\frac{1}{2} \zeta _2 \bar{s}_{-1,1,-1}+\frac{1}{2} \zeta _2 \bar{s}_{-1,1,1}+\frac{1}{2} \zeta _2 \bar{s}_{1,-1,-1}+\frac{1}{2} \zeta _2 \bar{s}_{1,-1,1}+\frac{1}{2} \zeta _2 \bar{s}_{1,1,-1}+\frac{1}{2} \zeta _2 \bar{s}_{1,1,1}+\bar{s}_{1,3,1}+\bar{s}_{3,1,1}-s_{-2,1,1,-1}-s_{-1,1,2,-1}-s_{-1,2,1,-1}-s_{1,-2,1,-1}-s_{1,-1,2,-1}-s_{1,1,-2,-1}-s_{1,2,-1,-1}-s_{2,-1,1,-1}-s_{2,1,-1,-1}-\bar{s}_{-2,-1,1,1}-\bar{s}_{-2,1,-1,1}-\bar{s}_{-1,-2,1,1}-\bar{s}_{-1,1,-2,1}-\bar{s}_{1,-2,-1,1}-\bar{s}_{1,-1,-2,1}-\bar{s}_{1,1,2,1}-\bar{s}_{1,2,1,1}-\bar{s}_{2,1,1,1}+s_{-1,1,1,1,-1}+s_{1,-1,1,1,-1}+s_{1,1,-1,1,-1}+s_{1,1,1,-1,-1}+\bar{s}_{-1,-1,1,1,1}+\bar{s}_{-1,1,-1,1,1}+\bar{s}_{-1,1,1,-1,1}+\bar{s}_{1,-1,-1,1,1}+\bar{s}_{1,-1,1,-1,1}+\bar{s}_{1,1,-1,-1,1}  \end{dmath}
   \begin{dmath}[style={\small}]     s_{-1,-1} \bar{s}_{-1,-2}  =   \zeta _2 \bar{s}_{-1,-2}-4 \zeta _3 \bar{s}_{-1,-1}-\zeta _2 \bar{s}_{-1,2}-\frac{1}{2} \zeta _2 \bar{s}_{2,-1}+\zeta _2 \bar{s}_{-1,-1,-1}+\zeta _2 \ln _2 \bar{s}_{-1,-1}+\bar{s}_{-3,-2}-2 \bar{s}_{-1,2,-2}-2 \bar{s}_{2,-1,-2}+3 \bar{s}_{-1,-1,-1,-2}+\ln _2^2 \bar{s}_{-1,-2}+\ln _2^2 \bar{s}_{-1,2}-\ln _2 \bar{s}_{2,-2}-\ln _2 \bar{s}_{2,2}+2 \ln _2 \bar{s}_{-1,-1,-2}+2 \ln _2 \bar{s}_{-1,-1,2}+8 \text{Li}_4\left(\frac{1}{2}\right) \bar{s}_{-1}+\frac{1}{2} \zeta _2 \bar{s}_{-3}-\frac{29}{20} \zeta _2^2 \bar{s}_{-1}+\frac{3}{2} \zeta _3 \bar{s}_2-\frac{1}{2} \zeta _2 \bar{s}_3-2 \zeta _2 \ln _2^2 \bar{s}_{-1}+\frac{1}{2} \zeta _3 \ln _2 \bar{s}_{-1}+\frac{1}{3} \ln _2^4 \bar{s}_{-1}+\zeta _2 s_{-1,-2}-\frac{1}{2} \zeta _3 s_{-1,-1}-\frac{1}{2} \zeta _2 s_{2,-1}+\zeta _2 s_{-1,-1,-1}+\zeta _2 \ln _2 s_{-1,-1}-s_{-4,-1}+2 s_{-1,3,-1}+s_{2,-2,-1}+s_{3,-1,-1}-s_{-1,-2,-1,-1}-2 s_{-1,-1,-2,-1}+\ln _2^2 s_{-1,-2}-\ln _2^2 s_{-1,2}-2 \ln _2 s_{-1,-3}+2 \ln _2 s_{-1,3}+\ln _2 s_{2,-2}-\ln _2 s_{2,2}-2 \ln _2 s_{-1,-1,-2}+2 \ln _2 s_{-1,-1,2}-\frac{17 \zeta _2 \zeta _3}{8}+\frac{439 \zeta _5}{16}-\frac{2}{3} \zeta _2 \ln _2^3-\frac{3}{2} \zeta _3 \ln _2^2-\frac{7}{2} \zeta _2^2 \ln _2+4 \text{Li}_4\left(\frac{1}{2}\right) s_{-1}-24 \text{Li}_5\left(\frac{1}{2}\right)-8 \text{Li}_4\left(\frac{1}{2}\right) \ln _2-\frac{7}{5} \zeta _2^2 s_{-1}-\frac{\zeta _2 s_3}{2}-\frac{\zeta _3 s_2}{4}-\zeta _2 s_{-1} \ln _2^2+\frac{7}{2} \zeta _3 s_{-1} \ln _2+\frac{1}{6} s_{-1} \ln _2^4+s_{-3} \ln _2^2-s_3 \ln _2^2-s_{-4} \ln _2+s_4 \ln _2-\frac{2 \ln _2^5}{15}  \end{dmath}
   \begin{dmath}[style={\small}]     s_{-1,-1} \bar{s}_{-1,2}  =   3 \zeta _3 \bar{s}_{-1,-1}-\frac{1}{2} \zeta _2 \bar{s}_{2,-1}+\zeta _2 \bar{s}_{-1,-1,-1}+\zeta _2 \ln _2 \bar{s}_{-1,-1}+\bar{s}_{-3,2}-2 \bar{s}_{-1,2,2}-2 \bar{s}_{2,-1,2}+3 \bar{s}_{-1,-1,-1,2}+\ln _2^2 \bar{s}_{-1,-2}+\ln _2^2 \bar{s}_{-1,2}-\ln _2 \bar{s}_{2,-2}-\ln _2 \bar{s}_{2,2}+2 \ln _2 \bar{s}_{-1,-1,-2}+2 \ln _2 \bar{s}_{-1,-1,2}-8 \text{Li}_4\left(\frac{1}{2}\right) \bar{s}_{-1}+\frac{8}{5} \zeta _2^2 \bar{s}_{-1}-2 \zeta _3 \bar{s}_2+2 \zeta _2 \ln _2^2 \bar{s}_{-1}+\frac{1}{2} \zeta _3 \ln _2 \bar{s}_{-1}-\frac{1}{3} \ln _2^4 \bar{s}_{-1}-\frac{3}{2} \zeta _3 s_{-1,-1}-\zeta _2 s_{-1,2}-\frac{1}{2} \zeta _2 s_{2,-1}+\zeta _2 s_{-1,-1,-1}+\zeta _2 \ln _2 s_{-1,-1}+s_{4,-1}-s_{-3,-1,-1}-2 s_{-1,-3,-1}-s_{2,2,-1}+2 s_{-1,-1,2,-1}+s_{-1,2,-1,-1}+\ln _2^2 s_{-1,-2}-\ln _2^2 s_{-1,2}-2 \ln _2 s_{-1,-3}+2 \ln _2 s_{-1,3}+\ln _2 s_{2,-2}-\ln _2 s_{2,2}-2 \ln _2 s_{-1,-1,-2}+2 \ln _2 s_{-1,-1,2}+\frac{\zeta _2 \zeta _3}{8}-\frac{27 \zeta _5}{4}+2 \zeta _2 \ln _2^3-\frac{3}{2} \zeta _3 \ln _2^2+\frac{13}{5} \zeta _2^2 \ln _2+4 \text{Li}_4\left(\frac{1}{2}\right) s_{-1}+8 \text{Li}_5\left(\frac{1}{2}\right)-8 \text{Li}_4\left(\frac{1}{2}\right) \ln _2-\frac{7}{4} \zeta _2^2 s_{-1}+\frac{1}{2} \zeta _2 s_{-3}+\frac{\zeta _3 s_2}{4}-\zeta _2 s_{-1} \ln _2^2+\frac{7}{2} \zeta _3 s_{-1} \ln _2+\frac{1}{6} s_{-1} \ln _2^4+s_{-3} \ln _2^2-s_3 \ln _2^2-s_{-4} \ln _2+s_4 \ln _2-\frac{2 \ln _2^5}{5}  \end{dmath}

   \begin{dmath}[style={\small}]     s_{-1,-1} \bar{s}_{-1,-1,1}  =   \frac{29 \ln _2^5}{120}+\frac{5}{24} s_{-1} \ln _2^4+\frac{7}{24} \bar{s}_{-1} \ln _2^4+\frac{7}{12} \zeta _2 \ln _2^3-s_{-1,-1} \ln _2^3+\frac{1}{2} s_{-3} \ln _2^2-\frac{1}{2} s_3 \ln _2^2-\frac{3}{4} s_{-1} \zeta _2 \ln _2^2+\frac{3}{16} \zeta _3 \ln _2^2-\frac{1}{2} \bar{s}_{-3} \ln _2^2+2 \zeta _2 \bar{s}_{-1} \ln _2^2+\frac{1}{2} \bar{s}_3 \ln _2^2+\frac{1}{2} s_{2,-1} \ln _2^2+\frac{1}{2} s_{2,1} \ln _2^2-\frac{1}{2} \bar{s}_{-1,-2} \ln _2^2+\frac{1}{2} \bar{s}_{-1,2} \ln _2^2-\frac{3}{2} s_{-1,-1,-1} \ln _2^2-\frac{1}{2} s_{-1,-1,1} \ln _2^2+\frac{3}{2} \bar{s}_{-1,-1,-1} \ln _2^2+\frac{1}{2} \bar{s}_{-1,-1,1} \ln _2^2-\frac{189}{40} \zeta _2^2 \ln _2-5 \text{Li}_4\left(\frac{1}{2}\right) \ln _2-s_{-4} \ln _2+s_4 \ln _2+\frac{7}{2} s_{-1} \zeta _3 \ln _2-\frac{5}{8} \zeta _3 \bar{s}_{-1} \ln _2-\frac{3}{2} \zeta _2 \bar{s}_2 \ln _2-s_{-3,-1} \ln _2+s_{-3,1} \ln _2-2 s_{-1,-3} \ln _2+\zeta _2 s_{-1,-1} \ln _2+2 s_{-1,3} \ln _2+2 s_{2,-2} \ln _2-2 s_{2,2} \ln _2+5 \zeta _2 \bar{s}_{-1,-1} \ln _2-3 s_{-1,-1,-2} \ln _2+3 s_{-1,-1,2} \ln _2+2 s_{-1,2,-1} \ln _2-2 s_{-1,2,1} \ln _2+2 s_{2,-1,-1} \ln _2-2 s_{2,-1,1} \ln _2-\bar{s}_{-1,2,-1} \ln _2-\bar{s}_{-1,2,1} \ln _2-\bar{s}_{2,-1,-1} \ln _2-\bar{s}_{2,-1,1} \ln _2-3 s_{-1,-1,-1,-1} \ln _2+3 s_{-1,-1,-1,1} \ln _2+3 \bar{s}_{-1,-1,-1,-1} \ln _2+3 \bar{s}_{-1,-1,-1,1} \ln _2-\frac{69}{40} s_{-1} \zeta _2^2-24 \text{Li}_5\left(\frac{1}{2}\right)+5 \text{Li}_4\left(\frac{1}{2}\right) s_{-1}-\frac{s_3 \zeta _2}{2}-\frac{3 s_2 \zeta _3}{8}-\frac{37 \zeta _2 \zeta _3}{16}+\frac{439 \zeta _5}{16}+\frac{1}{2} \zeta _2 \bar{s}_{-3}-\frac{11}{8} \zeta _2^2 \bar{s}_{-1}+7 \text{Li}_4\left(\frac{1}{2}\right) \bar{s}_{-1}+\frac{13}{8} \zeta _3 \bar{s}_2-\frac{1}{2} \zeta _2 \bar{s}_3-s_{-4,-1}+\zeta _2 s_{-1,-2}-\frac{1}{4} \zeta _3 s_{-1,-1}-\frac{1}{2} \zeta _2 s_{2,-1}+\frac{1}{2} \zeta _2 s_{2,1}+\frac{1}{2} \zeta _2 \bar{s}_{-1,-2}-\frac{31}{8} \zeta _3 \bar{s}_{-1,-1}-\frac{3}{2} \zeta _2 \bar{s}_{-1,2}-\zeta _2 \bar{s}_{2,-1}+s_{-3,1,-1}+\zeta _2 s_{-1,-1,-1}-\frac{1}{2} \zeta _2 s_{-1,-1,1}+2 s_{-1,3,-1}+2 s_{2,-2,-1}+s_{3,-1,-1}+\bar{s}_{-3,-1,1}+\bar{s}_{-1,-3,1}+\frac{5}{2} \zeta _2 \bar{s}_{-1,-1,-1}+\frac{1}{2} \zeta _2 \bar{s}_{-1,-1,1}+\bar{s}_{2,2,1}-s_{-1,-2,-1,-1}-3 s_{-1,-1,-2,-1}-2 s_{-1,2,1,-1}-2 s_{2,-1,1,-1}-s_{2,1,-1,-1}-3 \bar{s}_{-1,-1,2,1}-3 \bar{s}_{-1,2,-1,1}-3 \bar{s}_{2,-1,-1,1}+3 s_{-1,-1,-1,1,-1}+s_{-1,-1,1,-1,-1}+6 \bar{s}_{-1,-1,-1,-1,1}  \end{dmath}

  \begin{dmath}[style={\small}]     s_{-1,1} \bar{s}_{1,2}  =   -\zeta _2 \bar{s}_{-2,1}-2 \zeta _3 \bar{s}_{-1,1}-\frac{1}{2} \zeta _2 \bar{s}_{1,-2}-\frac{29}{8} \zeta _3 \bar{s}_{1,-1}-\frac{1}{2} \zeta _2 \bar{s}_{1,2}+\zeta _2 \bar{s}_{-1,1,1}+\zeta _2 \bar{s}_{1,-1,1}+\frac{3}{2} \zeta _2 \ln _2 \bar{s}_{1,-1}-\bar{s}_{-3,2}+2 \bar{s}_{-2,1,2}+\bar{s}_{-1,2,2}+\bar{s}_{1,-2,2}-2 \bar{s}_{-1,1,1,2}-\bar{s}_{1,-1,1,2}+\frac{1}{2} \ln _2^2 \bar{s}_{1,-2}+\frac{1}{2} \ln _2^2 \bar{s}_{1,2}-3 \text{Li}_4\left(\frac{1}{2}\right) \bar{s}_{-1}+3 \text{Li}_4\left(\frac{1}{2}\right) \bar{s}_1+2 \zeta _3 \bar{s}_{-2}-\frac{1}{8} \zeta _2^2 \bar{s}_{-1}-\frac{43}{40} \zeta _2^2 \bar{s}_1-\frac{1}{4} \zeta _2 \ln _2^2 \bar{s}_{-1}+\frac{1}{4} \zeta _2 \ln _2^2 \bar{s}_1-\frac{1}{8} \ln _2^4 \bar{s}_{-1}+\frac{1}{8} \ln _2^4 \bar{s}_1-\zeta _2 s_{-2,1}+\zeta _3 s_{-1,1}-\zeta _2 s_{-1,2}-\frac{1}{2} \zeta _2 s_{1,-2}-\frac{5}{8} \zeta _3 s_{1,-1}-\frac{1}{2} \zeta _2 s_{1,2}+\zeta _2 s_{-1,1,1}+\zeta _2 s_{1,-1,1}+\frac{3}{2} \zeta _2 \ln _2 s_{1,-1}-s_{-4,1}+s_{-2,2,1}+s_{-1,3,1}+s_{1,-3,1}+s_{3,-1,1}-s_{-1,1,2,1}-s_{1,-1,2,1}-s_{1,2,-1,1}-\frac{1}{2} \ln _2^2 s_{1,-2}+\frac{1}{2} \ln _2^2 s_{1,2}-2 \zeta _2 \zeta _3+\frac{29 \zeta _5}{64}-\frac{1}{2} \zeta _2 \ln _2^3+\frac{15}{8} \zeta _3 \ln _2^2+\frac{3}{40} \zeta _2^2 \ln _2-3 \text{Li}_4\left(\frac{1}{2}\right) s_{-1}+\text{Li}_4\left(\frac{1}{2}\right) s_1+4 \text{Li}_5\left(\frac{1}{2}\right)+\frac{43}{40} \zeta _2^2 s_{-1}-\frac{7}{40} \zeta _2^2 s_1+\frac{1}{2} \zeta _2 s_{-3}+\frac{\zeta _2 s_3}{2}+\frac{5}{8} \zeta _3 s_{-2}-\frac{1}{4} \zeta _2 s_{-1} \ln _2^2+\frac{3}{4} \zeta _2 s_1 \ln _2^2-\frac{3}{2} \zeta _2 s_{-2} \ln _2-\frac{1}{8} s_{-1} \ln _2^4+\frac{1}{24} s_1 \ln _2^4+\frac{1}{2} s_{-3} \ln _2^2-\frac{1}{2} s_3 \ln _2^2-\frac{\ln _2^5}{30}  \end{dmath}

  \begin{dmath}[style={\small}]     s_{1,-1} \bar{s}_{2,1}  =   \frac{\ln _2^5}{12}-\frac{1}{24} s_{-1} \ln _2^4+\frac{1}{8} s_1 \ln _2^4-\frac{1}{24} \bar{s}_{-1} \ln _2^4+\frac{1}{24} \bar{s}_1 \ln _2^4-\frac{1}{6} s_{-2} \ln _2^3+\frac{1}{6} s_2 \ln _2^3-\frac{1}{2} \zeta _2 \ln _2^3+\frac{1}{6} \bar{s}_{-2} \ln _2^3-\frac{1}{6} \bar{s}_2 \ln _2^3+\frac{1}{4} s_{-1} \zeta _2 \ln _2^2-\frac{3}{4} s_1 \zeta _2 \ln _2^2+\frac{9}{8} \zeta _3 \ln _2^2-\frac{1}{2} \bar{s}_{-3} \ln _2^2+\frac{1}{4} \zeta _2 \bar{s}_{-1} \ln _2^2-\frac{1}{4} \zeta _2 \bar{s}_1 \ln _2^2+\frac{1}{2} \bar{s}_3 \ln _2^2-\frac{1}{2} s_{1,-2} \ln _2^2+\frac{1}{2} s_{1,2} \ln _2^2+\frac{1}{2} \bar{s}_{1,-2} \ln _2^2-\frac{1}{2} \bar{s}_{1,2} \ln _2^2-\bar{s}_{2,1} \ln _2^2-\frac{57}{20} \zeta _2^2 \ln _2+2 \text{Li}_4\left(\frac{1}{2}\right) \ln _2+s_{-4} \ln _2-s_4 \ln _2+\frac{3}{2} s_2 \zeta _2 \ln _2-\frac{7}{8} s_{-1} \zeta _3 \ln _2-\frac{7}{8} \zeta _3 \bar{s}_{-1} \ln _2+\frac{9}{4} \zeta _3 \bar{s}_1 \ln _2-2 \zeta _2 \bar{s}_2 \ln _2-s_{1,-3} \ln _2+s_{1,3} \ln _2-s_{2,-2} \ln _2+s_{2,2} \ln _2-s_{3,-1} \ln _2+s_{3,1} \ln _2+\bar{s}_{3,-1} \ln _2+\bar{s}_{3,1} \ln _2+s_{1,2,-1} \ln _2-s_{1,2,1} \ln _2+s_{2,1,-1} \ln _2-s_{2,1,1} \ln _2-\bar{s}_{1,2,-1} \ln _2-\bar{s}_{1,2,1} \ln _2-\bar{s}_{2,1,-1} \ln _2-\bar{s}_{2,1,1} \ln _2+\frac{1}{20} s_{-1} \zeta _2^2-\frac{43}{40} s_1 \zeta _2^2-\text{Li}_4\left(\frac{1}{2}\right) s_{-1}+3 \text{Li}_4\left(\frac{1}{2}\right) s_1-\frac{1}{2} s_{-3} \zeta _2+\frac{3}{4} s_{-2} \zeta _3-\frac{s_2 \zeta _3}{4}+\frac{75 \zeta _5}{64}+\frac{1}{2} \zeta _2 \bar{s}_{-3}-\frac{1}{8} \zeta _3 \bar{s}_{-2}+\frac{1}{20} \zeta _2^2 \bar{s}_{-1}-\text{Li}_4\left(\frac{1}{2}\right) \bar{s}_{-1}-\frac{27}{40} \zeta _2^2 \bar{s}_1+\text{Li}_4\left(\frac{1}{2}\right) \bar{s}_1+\frac{3}{4} \zeta _3 \bar{s}_2+\frac{1}{2} \zeta _2 \bar{s}_3+\frac{1}{2} \zeta _2 s_{1,-2}-\frac{5}{8} \zeta _3 s_{1,-1}+\frac{1}{2} \zeta _2 s_{2,-1}-s_{4,-1}-\bar{s}_{-4,1}-\frac{1}{2} \zeta _2 \bar{s}_{1,-2}-\frac{5}{8} \zeta _3 \bar{s}_{1,-1}-\frac{1}{2} \zeta _2 \bar{s}_{1,2}-\frac{1}{2} \zeta _2 \bar{s}_{2,-1}-\frac{1}{2} \zeta _2 \bar{s}_{2,1}+s_{1,3,-1}+s_{2,2,-1}+2 s_{3,1,-1}+\bar{s}_{-2,2,1}+\bar{s}_{1,-3,1}+\bar{s}_{2,-2,1}+\bar{s}_{3,-1,1}-s_{1,2,1,-1}-2 s_{2,1,1,-1}-\bar{s}_{1,-1,2,1}-\bar{s}_{1,2,-1,1}-\bar{s}_{2,1,-1,1}  \end{dmath}

   \begin{dmath}[style={\small}]     s_{1,-1} \bar{s}_{-1,2}  =   -\frac{\ln _2^5}{12}+\frac{1}{12} s_1 \ln _2^4+\frac{1}{12} \bar{s}_{-1} \ln _2^4+\frac{1}{4} \bar{s}_1 \ln _2^4-\frac{5}{3} \zeta _2 \ln _2^3-\frac{1}{2} s_{-3} \ln _2^2+\frac{1}{2} s_3 \ln _2^2-\frac{1}{4} s_{-1} \zeta _2 \ln _2^2-\frac{1}{4} s_1 \zeta _2 \ln _2^2+\frac{3}{4} \zeta _3 \ln _2^2-\frac{5}{4} \zeta _2 \bar{s}_{-1} \ln _2^2-\frac{3}{4} \zeta _2 \bar{s}_1 \ln _2^2-\frac{1}{2} s_{-1,-2} \ln _2^2+\frac{1}{2} s_{-1,2} \ln _2^2-\frac{1}{2} \bar{s}_{-1,-2} \ln _2^2-\frac{1}{2} \bar{s}_{-1,2} \ln _2^2-\frac{25}{8} \zeta _2^2 \ln _2-6 \text{Li}_4\left(\frac{1}{2}\right) \ln _2+s_{-4} \ln _2-s_4 \ln _2-\frac{3}{2} s_{-2} \zeta _2 \ln _2+\frac{21}{8} s_{-1} \zeta _3 \ln _2+\frac{7}{8} s_1 \zeta _3 \ln _2-\frac{9}{8} \zeta _3 \bar{s}_{-1} \ln _2+\frac{5}{8} \zeta _3 \bar{s}_1 \ln _2+s_{-2,-2} \ln _2-s_{-2,2} \ln _2+s_{-1,-3} \ln _2+\frac{3}{2} \zeta _2 s_{-1,1} \ln _2-s_{-1,3} \ln _2-s_{1,-3} \ln _2+\zeta _2 s_{1,-1} \ln _2+s_{1,3} \ln _2+\bar{s}_{-2,-2} \ln _2+\bar{s}_{-2,2} \ln _2+\frac{1}{2} \zeta _2 \bar{s}_{-1,1} \ln _2-s_{-1,1,-2} \ln _2+s_{-1,1,2} \ln _2-s_{1,-1,-2} \ln _2+s_{1,-1,2} \ln _2-\bar{s}_{-1,1,-2} \ln _2-\bar{s}_{-1,1,2} \ln _2-\bar{s}_{1,-1,-2} \ln _2-\bar{s}_{1,-1,2} \ln _2-\frac{1}{8} s_{-1} \zeta _2^2-\frac{2}{5} s_1 \zeta _2^2-20 \text{Li}_5\left(\frac{1}{2}\right)+2 \text{Li}_4\left(\frac{1}{2}\right) s_1+\frac{s_3 \zeta _2}{2}+\frac{1}{4} s_{-2} \zeta _3-\frac{7 s_2 \zeta _3}{8}-\frac{3 \zeta _2 \zeta _3}{16}+\frac{305 \zeta _5}{16}+2 \zeta _3 \bar{s}_{-2}-\frac{2}{5} \zeta _2^2 \bar{s}_{-1}+2 \text{Li}_4\left(\frac{1}{2}\right) \bar{s}_{-1}-\frac{6}{5} \zeta _2^2 \bar{s}_1+6 \text{Li}_4\left(\frac{1}{2}\right) \bar{s}_1+s_{-4,-1}-\frac{1}{2} \zeta _2 s_{-2,-1}-\frac{1}{2} \zeta _2 s_{-1,-2}+\frac{7}{8} \zeta _3 s_{-1,-1}-\frac{1}{4} \zeta _3 s_{-1,1}-\frac{5}{4} \zeta _3 s_{1,-1}-\frac{1}{2} \zeta _2 s_{1,2}+\frac{1}{2} \zeta _2 \bar{s}_{-2,-1}-\frac{7}{8} \zeta _3 \bar{s}_{-1,-1}-\zeta _3 \bar{s}_{-1,1}-2 \zeta _3 \bar{s}_{1,-1}-\bar{s}_{3,2}-s_{-3,1,-1}-s_{-2,2,-1}+\frac{1}{2} \zeta _2 s_{-1,1,-1}-s_{-1,3,-1}-s_{1,-3,-1}+\frac{1}{2} \zeta _2 s_{1,-1,-1}+2 \bar{s}_{-2,-1,2}+\bar{s}_{-1,-2,2}-\frac{1}{2} \zeta _2 \bar{s}_{-1,1,-1}-\frac{1}{2} \zeta _2 \bar{s}_{1,-1,-1}+\bar{s}_{1,2,2}+s_{-1,1,2,-1}+s_{-1,2,1,-1}+s_{1,-1,2,-1}-\bar{s}_{-1,1,-1,2}-2 \bar{s}_{1,-1,-1,2}  \end{dmath}

   \begin{dmath}[style={\small}]     s_{1,1} \bar{s}_{1,-1,1}  =   -\frac{\ln _2^5}{10}-\frac{1}{8} s_{-1} \ln _2^4+\frac{1}{4} s_1 \ln _2^4-\frac{1}{8} \bar{s}_{-1} \ln _2^4+\frac{1}{12} \bar{s}_1 \ln _2^4+\frac{1}{3} s_{-2} \ln _2^3-\frac{1}{3} s_2 \ln _2^3+\frac{1}{2} \zeta _2 \ln _2^3-\frac{1}{3} s_{1,-1} \ln _2^3+s_{1,1} \ln _2^3-\frac{1}{3} \bar{s}_{1,-1} \ln _2^3+\frac{1}{3} \bar{s}_{1,1} \ln _2^3+\frac{1}{4} s_{-1} \zeta _2 \ln _2^2+2 \zeta _3 \ln _2^2+\frac{1}{2} \bar{s}_{-3} \ln _2^2+\frac{1}{4} \zeta _2 \bar{s}_{-1} \ln _2^2+\frac{3}{2} \zeta _2 \bar{s}_1 \ln _2^2-\frac{1}{2} \bar{s}_3 \ln _2^2-\frac{1}{2} s_{2,-1} \ln _2^2-\frac{1}{2} s_{2,1} \ln _2^2-\bar{s}_{1,-2} \ln _2^2+\bar{s}_{1,2} \ln _2^2-\frac{1}{2} \bar{s}_{2,-1} \ln _2^2+\frac{1}{2} \bar{s}_{2,1} \ln _2^2+s_{1,1,-1} \ln _2^2+s_{1,1,1} \ln _2^2+\bar{s}_{1,1,-1} \ln _2^2-\bar{s}_{1,1,1} \ln _2^2-\frac{51}{40} \zeta _2^2 \ln _2-s_{-2} \zeta _2 \ln _2+s_2 \zeta _2 \ln _2-\frac{1}{2} \zeta _2 \bar{s}_{-2} \ln _2-\zeta _2 \bar{s}_2 \ln _2+\frac{3}{2} \zeta _2 s_{1,-1} \ln _2-\frac{5}{2} \zeta _2 s_{1,1} \ln _2+\frac{3}{2} \zeta _2 \bar{s}_{1,-1} \ln _2+\frac{3}{2} \zeta _2 \bar{s}_{1,1} \ln _2-\frac{1}{8} s_{-1} \zeta _2^2-\frac{23}{10} s_1 \zeta _2^2+4 \text{Li}_5\left(\frac{1}{2}\right)-3 \text{Li}_4\left(\frac{1}{2}\right) s_{-1}+6 \text{Li}_4\left(\frac{1}{2}\right) s_1-s_{-3} \zeta _2+\frac{11}{4} s_{-2} \zeta _3+\frac{s_2 \zeta _3}{4}-2 \zeta _2 \zeta _3+\frac{23 \zeta _5}{64}-\frac{1}{2} \zeta _2 \bar{s}_{-3}+\frac{13}{8} \zeta _3 \bar{s}_{-2}-\frac{1}{8} \zeta _2^2 \bar{s}_{-1}-3 \text{Li}_4\left(\frac{1}{2}\right) \bar{s}_{-1}-\frac{17}{20} \zeta _2^2 \bar{s}_1+2 \text{Li}_4\left(\frac{1}{2}\right) \bar{s}_1+\frac{1}{2} \zeta _2 \bar{s}_3+s_{-4,1}+\zeta _2 s_{-2,1}+2 \zeta _2 s_{1,-2}-\frac{35}{8} \zeta _3 s_{1,-1}-\frac{3}{8} \zeta _3 s_{1,1}+\frac{3}{2} \zeta _2 s_{2,-1}+\frac{1}{2} \zeta _2 s_{2,1}-\frac{11}{8} \zeta _3 \bar{s}_{1,-1}+\frac{1}{8} \zeta _3 \bar{s}_{1,1}-\zeta _2 \bar{s}_{1,2}-\frac{1}{2} \zeta _2 \bar{s}_{2,-1}-\frac{1}{2} \zeta _2 \bar{s}_{2,1}-2 s_{-3,1,1}-s_{-2,2,1}-2 s_{1,-3,1}-\zeta _2 s_{1,-1,1}-3 \zeta _2 s_{1,1,-1}-\zeta _2 s_{1,1,1}-s_{2,-2,1}+\bar{s}_{1,-3,1}+\zeta _2 \bar{s}_{1,-1,1}+\zeta _2 \bar{s}_{1,1,-1}+\zeta _2 \bar{s}_{1,1,1}+\bar{s}_{2,-2,1}+\bar{s}_{3,-1,1}+2 s_{-2,1,1,1}+3 s_{1,-2,1,1}+s_{1,-1,2,1}+2 s_{1,1,-2,1}+s_{2,-1,1,1}-\bar{s}_{1,-2,1,1}-\bar{s}_{1,-1,2,1}-2 \bar{s}_{1,1,-2,1}-2 \bar{s}_{1,2,-1,1}-\bar{s}_{2,-1,1,1}-2 \bar{s}_{2,1,-1,1}-2 s_{1,-1,1,1,1}-2 s_{1,1,-1,1,1}+\bar{s}_{1,-1,1,1,1}+2 \bar{s}_{1,1,-1,1,1}+3 \bar{s}_{1,1,1,-1,1}  \end{dmath}
   
    \begin{dmath}[style={\small}]     s_{-1,-1} \bar{s}_{2,1}  =   -\frac{\ln _2^5}{60}+\frac{1}{8} s_{-1} \ln _2^4-\frac{1}{24} \bar{s}_{-1} \ln _2^4+\frac{1}{2} s_{-2} \ln _2^3-\frac{1}{2} s_2 \ln _2^3+\frac{1}{6} \zeta _2 \ln _2^3-\frac{1}{2} \bar{s}_{-2} \ln _2^3+\frac{1}{2} \bar{s}_2 \ln _2^3+\frac{1}{2} s_{-3} \ln _2^2-\frac{1}{2} s_3 \ln _2^2-\frac{3}{4} s_{-1} \zeta _2 \ln _2^2+\frac{3}{8} \zeta _3 \ln _2^2-\frac{1}{2} \bar{s}_{-3} \ln _2^2+\frac{1}{4} \zeta _2 \bar{s}_{-1} \ln _2^2+\frac{1}{2} \bar{s}_3 \ln _2^2-\frac{1}{2} s_{-1,-2} \ln _2^2+\frac{1}{2} s_{-1,2} \ln _2^2-\frac{1}{2} s_{2,-1} \ln _2^2+\frac{1}{2} s_{2,1} \ln _2^2-\frac{1}{2} \bar{s}_{-1,-2} \ln _2^2+\frac{1}{2} \bar{s}_{-1,2} \ln _2^2+\frac{3}{2} \bar{s}_{2,-1} \ln _2^2+\frac{1}{2} \bar{s}_{2,1} \ln _2^2+\frac{27}{20} \zeta _2^2 \ln _2-s_{-4} \ln _2+s_4 \ln _2-\frac{1}{2} s_{-2} \zeta _2 \ln _2+\frac{1}{2} s_2 \zeta _2 \ln _2+\frac{21}{8} s_{-1} \zeta _3 \ln _2+\frac{1}{2} \zeta _2 \bar{s}_{-2} \ln _2+\frac{3}{8} \zeta _3 \bar{s}_{-1} \ln _2+\frac{3}{2} \zeta _2 \bar{s}_2 \ln _2-s_{-3,-1} \ln _2+s_{-3,1} \ln _2-s_{-1,-3} \ln _2+s_{-1,3} \ln _2+s_{2,-2} \ln _2-s_{2,2} \ln _2-\bar{s}_{-3,-1} \ln _2-\bar{s}_{-3,1} \ln _2+s_{-1,2,-1} \ln _2-s_{-1,2,1} \ln _2+s_{2,-1,-1} \ln _2-s_{2,-1,1} \ln _2+\bar{s}_{-1,2,-1} \ln _2+\bar{s}_{-1,2,1} \ln _2+\bar{s}_{2,-1,-1} \ln _2+\bar{s}_{2,-1,1} \ln _2-\frac{43}{40} s_{-1} \zeta _2^2+2 \text{Li}_5\left(\frac{1}{2}\right)+3 \text{Li}_4\left(\frac{1}{2}\right) s_{-1}-\frac{s_3 \zeta _2}{2}-\frac{3 s_2 \zeta _3}{8}-\frac{21 \zeta _2 \zeta _3}{8}+\frac{111 \zeta _5}{32}-\frac{1}{2} \zeta _2 \bar{s}_{-3}+\frac{27}{40} \zeta _2^2 \bar{s}_{-1}-\text{Li}_4\left(\frac{1}{2}\right) \bar{s}_{-1}-\frac{5}{8} \zeta _3 \bar{s}_2-\frac{1}{2} \zeta _2 \bar{s}_3-s_{-4,-1}+\frac{1}{2} \zeta _2 s_{-1,-2}-\frac{5}{8} \zeta _3 s_{-1,-1}+\frac{1}{2} \zeta _2 s_{2,1}+\frac{1}{2} \zeta _2 \bar{s}_{-1,-2}+\frac{5}{8} \zeta _3 \bar{s}_{-1,-1}+\frac{1}{2} \zeta _2 \bar{s}_{-1,2}+\frac{1}{2} \zeta _2 \bar{s}_{2,-1}+\frac{1}{2} \zeta _2 \bar{s}_{2,1}+\bar{s}_{4,1}+s_{-3,1,-1}+s_{-1,3,-1}+s_{2,-2,-1}+s_{3,-1,-1}-\bar{s}_{-3,-1,1}-\bar{s}_{-1,-3,1}-2 \bar{s}_{2,2,1}-s_{-1,2,1,-1}-s_{2,-1,1,-1}-s_{2,1,-1,-1}+\bar{s}_{-1,-1,2,1}+\bar{s}_{-1,2,-1,1}+\bar{s}_{2,-1,-1,1}  \end{dmath}
   
 \begin{dmath}[style={\small}]     s_{-1,-1} \bar{s}_{-1,-1,-1}  =   \frac{9}{4} \zeta _3 \bar{s}_{-1,-1}+\frac{1}{2} \zeta _2 \bar{s}_{-1,2}+\frac{1}{2} \zeta _2 \bar{s}_{2,-1}-\frac{3}{2} \zeta _2 \bar{s}_{-1,-1,-1}-\zeta _2 \ln _2 \bar{s}_{-1,-1}+\bar{s}_{-3,-1,-1}+\bar{s}_{-1,-3,-1}+\bar{s}_{2,2,-1}-3 \bar{s}_{-1,-1,2,-1}-3 \bar{s}_{-1,2,-1,-1}-3 \bar{s}_{2,-1,-1,-1}+6 \bar{s}_{-1,-1,-1,-1,-1}-\ln _2^2 \bar{s}_{-1,-2}+\ln _2^2 \bar{s}_{-1,2}+2 \ln _2^2 \bar{s}_{-1,-1,-1}+\ln _2 \bar{s}_{-1,-3}-\ln _2 \bar{s}_{-1,3}-\ln _2 \bar{s}_{2,-2}+\ln _2 \bar{s}_{2,2}+3 \ln _2 \bar{s}_{-1,-1,-2}-3 \ln _2 \bar{s}_{-1,-1,2}-2 \ln _2 \bar{s}_{-1,2,-1}-2 \ln _2 \bar{s}_{2,-1,-1}+6 \ln _2 \bar{s}_{-1,-1,-1,-1}-4 \text{Li}_4\left(\frac{1}{2}\right) \bar{s}_{-1}+\frac{21}{20} \zeta _2^2 \bar{s}_{-1}-\zeta _3 \bar{s}_2+\zeta _2 \ln _2^2 \bar{s}_{-1}-\frac{3}{2} \zeta _3 \ln _2 \bar{s}_{-1}-\frac{1}{6} \ln _2^4 \bar{s}_{-1}-\ln _2^2 \bar{s}_{-3}+\ln _2^2 \bar{s}_3-2 \zeta _3 s_{-1,-1}-\zeta _2 s_{-1,2}-\zeta _2 s_{2,-1}+\frac{3}{2} \zeta _2 s_{-1,-1,-1}+2 \zeta _2 \ln _2 s_{-1,-1}+s_{4,-1}-2 s_{-3,-1,-1}-2 s_{-1,-3,-1}-2 s_{2,2,-1}+3 s_{-1,-1,2,-1}+3 s_{-1,2,-1,-1}+3 s_{2,-1,-1,-1}-4 s_{-1,-1,-1,-1,-1}-\frac{4}{3} \ln _2^3 s_{-1,-1}-\ln _2^2 s_{-1,-2}+\ln _2^2 s_{-1,2}+2 \ln _2^2 s_{2,-1}-4 \ln _2^2 s_{-1,-1,-1}-2 \ln _2 s_{-3,-1}-2 \ln _2 s_{-1,-3}+2 \ln _2 s_{-1,3}+2 \ln _2 s_{2,-2}-2 \ln _2 s_{2,2}-3 \ln _2 s_{-1,-1,-2}+3 \ln _2 s_{-1,-1,2}+4 \ln _2 s_{-1,2,-1}+4 \ln _2 s_{2,-1,-1}-6 \ln _2 s_{-1,-1,-1,-1}+\frac{129 \zeta _5}{16}-\frac{2}{3} \zeta _2 \ln _2^3+\frac{5}{2} \zeta _3 \ln _2^2-\frac{7}{2} \zeta _2^2 \ln _2+8 \text{Li}_4\left(\frac{1}{2}\right) s_{-1}-8 \text{Li}_5\left(\frac{1}{2}\right)-\frac{14}{5} \zeta _2^2 s_{-1}+\frac{1}{2} \zeta _2 s_{-3}+\frac{5 \zeta _3 s_2}{4}-\zeta _2 s_{-1} \ln _2^2-\zeta _2 s_2 \ln _2+\frac{9}{2} \zeta _3 s_{-1} \ln _2+\frac{1}{3} s_{-1} \ln _2^4-s_{-4} \ln _2+s_4 \ln _2+\frac{\ln _2^5}{3}  \end{dmath}

   \begin{dmath}[style={\small}]     s_{-1,-1} \bar{s}_{1,2}  =   \frac{11 \ln _2^5}{60}-\frac{1}{8} s_{-1} \ln _2^4-\frac{1}{24} s_1 \ln _2^4-\frac{1}{24} \bar{s}_{-1} \ln _2^4-\frac{1}{8} \bar{s}_1 \ln _2^4-\frac{4}{3} \zeta _2 \ln _2^3+s_{-3} \ln _2^2-s_3 \ln _2^2+\frac{1}{2} s_{-1} \zeta _2 \ln _2^2+\frac{1}{2} s_1 \zeta _2 \ln _2^2+\frac{15}{4} \zeta _3 \ln _2^2-\frac{1}{2} \zeta _2 \bar{s}_{-1} \ln _2^2+\frac{3}{2} \zeta _2 \bar{s}_1 \ln _2^2-s_{1,-2} \ln _2^2+s_{1,2} \ln _2^2+\bar{s}_{1,-2} \ln _2^2+\bar{s}_{1,2} \ln _2^2+\frac{9}{10} \zeta _2^2 \ln _2+6 \text{Li}_4\left(\frac{1}{2}\right) \ln _2-s_{-4} \ln _2+s_4 \ln _2-\frac{21}{8} s_{-1} \zeta _3 \ln _2-\frac{7}{8} s_1 \zeta _3 \ln _2-\frac{1}{2} \zeta _2 \bar{s}_{-2} \ln _2+\frac{9}{8} \zeta _3 \bar{s}_{-1} \ln _2-\frac{5}{8} \zeta _3 \bar{s}_1 \ln _2+\frac{1}{2} \zeta _2 \bar{s}_2 \ln _2-s_{-2,-2} \ln _2+s_{-2,2} \ln _2-s_{-1,-3} \ln _2+\frac{1}{2} \zeta _2 s_{-1,-1} \ln _2-\frac{1}{2} \zeta _2 s_{-1,1} \ln _2+s_{-1,3} \ln _2+s_{1,-3} \ln _2-s_{1,3} \ln _2-\bar{s}_{-2,-2} \ln _2-\bar{s}_{-2,2} \ln _2-\frac{1}{2} \zeta _2 \bar{s}_{-1,-1} \ln _2+\frac{1}{2} \zeta _2 \bar{s}_{-1,1} \ln _2+\zeta _2 \bar{s}_{1,-1} \ln _2+s_{-1,1,-2} \ln _2-s_{-1,1,2} \ln _2+s_{1,-1,-2} \ln _2-s_{1,-1,2} \ln _2+\bar{s}_{-1,1,-2} \ln _2+\bar{s}_{-1,1,2} \ln _2+\bar{s}_{1,-1,-2} \ln _2+\bar{s}_{1,-1,2} \ln _2+\frac{53}{40} s_{-1} \zeta _2^2+\frac{17}{40} s_1 \zeta _2^2+8 \text{Li}_5\left(\frac{1}{2}\right)-3 \text{Li}_4\left(\frac{1}{2}\right) s_{-1}-\text{Li}_4\left(\frac{1}{2}\right) s_1-\frac{s_3 \zeta _2}{2}-\frac{1}{4} s_{-2} \zeta _3+\frac{3 \zeta _2 \zeta _3}{16}-\frac{507 \zeta _5}{64}-\zeta _3 \bar{s}_{-2}+\frac{9}{20} \zeta _2^2 \bar{s}_{-1}-\text{Li}_4\left(\frac{1}{2}\right) \bar{s}_{-1}+\frac{53}{40} \zeta _2^2 \bar{s}_1-3 \text{Li}_4\left(\frac{1}{2}\right) \bar{s}_1-\zeta _3 \bar{s}_2-s_{-4,-1}+\frac{1}{2} \zeta _2 s_{-2,-1}+\frac{1}{2} \zeta _2 s_{-1,-2}-\zeta _3 s_{-1,-1}+\frac{1}{4} \zeta _3 s_{-1,1}+\frac{1}{4} \zeta _3 s_{1,-1}+\frac{1}{2} \zeta _2 s_{1,2}-\frac{1}{2} \zeta _2 \bar{s}_{-2,-1}+\zeta _3 \bar{s}_{-1,-1}+\zeta _3 \bar{s}_{-1,1}+\zeta _3 \bar{s}_{1,-1}+\bar{s}_{3,2}+s_{-2,2,-1}-\frac{1}{2} \zeta _2 s_{-1,1,-1}+s_{-1,3,-1}+s_{1,-3,-1}-\frac{1}{2} \zeta _2 s_{1,-1,-1}+s_{3,-1,-1}-\bar{s}_{-2,-1,2}-\bar{s}_{-1,-2,2}+\frac{1}{2} \zeta _2 \bar{s}_{-1,1,-1}+\frac{1}{2} \zeta _2 \bar{s}_{1,-1,-1}-\bar{s}_{1,2,2}-\bar{s}_{2,1,2}-s_{-1,1,2,-1}-s_{1,-1,2,-1}-s_{1,2,-1,-1}+\bar{s}_{-1,-1,1,2}+\bar{s}_{-1,1,-1,2}+\bar{s}_{1,-1,-1,2}  \end{dmath}

   \begin{dmath}[style={\small}]     s_{1,-1} \bar{s}_{-1,-2}  =   -\frac{4 \ln _2^5}{15}+\frac{1}{12} s_{-1} \ln _2^4+\frac{1}{6} s_1 \ln _2^4-\frac{1}{6} \bar{s}_{-1} \ln _2^4-\frac{1}{6} \bar{s}_1 \ln _2^4+\frac{5}{3} \zeta _2 \ln _2^3-\frac{1}{2} s_{-3} \ln _2^2+\frac{1}{2} s_3 \ln _2^2-\frac{9}{4} s_{-1} \zeta _2 \ln _2^2+\frac{3}{4} s_1 \zeta _2 \ln _2^2+\frac{3}{4} \zeta _3 \ln _2^2+\frac{7}{4} \zeta _2 \bar{s}_{-1} \ln _2^2+\frac{1}{4} \zeta _2 \bar{s}_1 \ln _2^2-\frac{1}{2} s_{-1,-2} \ln _2^2+\frac{1}{2} s_{-1,2} \ln _2^2-\frac{1}{2} \bar{s}_{-1,-2} \ln _2^2-\frac{1}{2} \bar{s}_{-1,2} \ln _2^2-\frac{9}{10} \zeta _2^2 \ln _2-6 \text{Li}_4\left(\frac{1}{2}\right) \ln _2+s_{-4} \ln _2-s_4 \ln _2+\frac{3}{2} s_2 \zeta _2 \ln _2+\frac{21}{8} s_{-1} \zeta _3 \ln _2+\frac{7}{8} s_1 \zeta _3 \ln _2-\frac{9}{8} \zeta _3 \bar{s}_{-1} \ln _2+\frac{5}{8} \zeta _3 \bar{s}_1 \ln _2+s_{-2,-2} \ln _2-s_{-2,2} \ln _2+s_{-1,-3} \ln _2-\frac{3}{2} \zeta _2 s_{-1,-1} \ln _2-s_{-1,3} \ln _2-s_{1,-3} \ln _2+\zeta _2 s_{1,-1} \ln _2+s_{1,3} \ln _2+\bar{s}_{-2,-2} \ln _2+\bar{s}_{-2,2} \ln _2+\frac{3}{2} \zeta _2 \bar{s}_{-1,-1} \ln _2-\zeta _2 \bar{s}_{-1,1} \ln _2-s_{-1,1,-2} \ln _2+s_{-1,1,2} \ln _2-s_{1,-1,-2} \ln _2+s_{1,-1,2} \ln _2-\bar{s}_{-1,1,-2} \ln _2-\bar{s}_{-1,1,2} \ln _2-\bar{s}_{1,-1,-2} \ln _2-\bar{s}_{1,-1,2} \ln _2-\frac{11}{10} s_{-1} \zeta _2^2-\frac{61}{40} s_1 \zeta _2^2+2 \text{Li}_5\left(\frac{1}{2}\right)+2 \text{Li}_4\left(\frac{1}{2}\right) s_{-1}+4 \text{Li}_4\left(\frac{1}{2}\right) s_1-\frac{1}{2} s_{-3} \zeta _2+\frac{5}{8} s_{-2} \zeta _3+\frac{3 \zeta _2 \zeta _3}{16}-\frac{33 \zeta _5}{64}+\frac{1}{2} \zeta _2 \bar{s}_{-3}-\frac{3}{2} \zeta _3 \bar{s}_{-2}+\frac{41}{40} \zeta _2^2 \bar{s}_{-1}-4 \text{Li}_4\left(\frac{1}{2}\right) \bar{s}_{-1}+\frac{17}{40} \zeta _2^2 \bar{s}_1-4 \text{Li}_4\left(\frac{1}{2}\right) \bar{s}_1-\frac{1}{2} \zeta _2 \bar{s}_3-\frac{1}{2} \zeta _2 s_{-2,-1}-\frac{5}{8} \zeta _3 s_{-1,1}+\frac{1}{2} \zeta _2 s_{-1,2}+\frac{1}{2} \zeta _2 s_{1,-2}-\frac{3}{4} \zeta _3 s_{1,-1}-s_{4,-1}+\frac{1}{2} \zeta _2 \bar{s}_{-2,-1}+\frac{1}{2} \zeta _2 \bar{s}_{-1,-2}+\frac{13}{8} \zeta _3 \bar{s}_{-1,1}-\frac{1}{2} \zeta _2 \bar{s}_{-1,2}-\frac{1}{2} \zeta _2 \bar{s}_{1,-2}+\frac{3}{2} \zeta _3 \bar{s}_{1,-1}+\frac{1}{2} \zeta _2 \bar{s}_{1,2}-\bar{s}_{3,-2}+s_{-2,-2,-1}+s_{-1,-3,-1}+\frac{1}{2} \zeta _2 s_{-1,1,-1}+\frac{1}{2} \zeta _2 s_{1,-1,-1}+s_{1,3,-1}+s_{3,1,-1}+2 \bar{s}_{-2,-1,-2}+\bar{s}_{-1,-2,-2}-\frac{1}{2} \zeta _2 \bar{s}_{-1,1,-1}-\frac{1}{2} \zeta _2 \bar{s}_{1,-1,-1}+\bar{s}_{1,2,-2}-s_{-1,-2,1,-1}-s_{-1,1,-2,-1}-s_{1,-1,-2,-1}-\bar{s}_{-1,1,-1,-2}-2 \bar{s}_{1,-1,-1,-2}  \end{dmath}

     \begin{dmath}[style={\small}]     s_{-1,-1} \bar{s}_{1,-2}  =   \frac{\ln _2^5}{5}-\frac{1}{6} s_{-1} \ln _2^4+\frac{1}{6} \bar{s}_1 \ln _2^4-\frac{3}{2} \zeta _2 \ln _2^3+s_{-3} \ln _2^2-s_3 \ln _2^2+\frac{3}{4} s_{-1} \zeta _2 \ln _2^2+\frac{1}{4} s_1 \zeta _2 \ln _2^2+\frac{15}{4} \zeta _3 \ln _2^2-\frac{3}{4} \zeta _2 \bar{s}_{-1} \ln _2^2-\frac{1}{4} \zeta _2 \bar{s}_1 \ln _2^2-s_{1,-2} \ln _2^2+s_{1,2} \ln _2^2+\bar{s}_{1,-2} \ln _2^2+\bar{s}_{1,2} \ln _2^2-\frac{29}{20} \zeta _2^2 \ln _2+6 \text{Li}_4\left(\frac{1}{2}\right) \ln _2-s_{-4} \ln _2+s_4 \ln _2-\frac{21}{8} s_{-1} \zeta _3 \ln _2-\frac{7}{8} s_1 \zeta _3 \ln _2-\frac{1}{2} \zeta _2 \bar{s}_{-2} \ln _2+\frac{9}{8} \zeta _3 \bar{s}_{-1} \ln _2-\frac{5}{8} \zeta _3 \bar{s}_1 \ln _2+\frac{1}{2} \zeta _2 \bar{s}_2 \ln _2-s_{-2,-2} \ln _2+s_{-2,2} \ln _2-s_{-1,-3} \ln _2+\frac{1}{2} \zeta _2 s_{-1,-1} \ln _2-\frac{1}{2} \zeta _2 s_{-1,1} \ln _2+s_{-1,3} \ln _2+s_{1,-3} \ln _2-s_{1,3} \ln _2-\bar{s}_{-2,-2} \ln _2-\bar{s}_{-2,2} \ln _2-\frac{1}{2} \zeta _2 \bar{s}_{-1,-1} \ln _2+\frac{1}{2} \zeta _2 \bar{s}_{-1,1} \ln _2+\zeta _2 \bar{s}_{1,-1} \ln _2+s_{-1,1,-2} \ln _2-s_{-1,1,2} \ln _2+s_{1,-1,-2} \ln _2-s_{1,-1,2} \ln _2+\bar{s}_{-1,1,-2} \ln _2+\bar{s}_{-1,1,2} \ln _2+\bar{s}_{1,-1,-2} \ln _2+\bar{s}_{1,-1,2} \ln _2+\frac{51}{40} s_{-1} \zeta _2^2+\frac{1}{8} s_1 \zeta _2^2+6 \text{Li}_5\left(\frac{1}{2}\right)-4 \text{Li}_4\left(\frac{1}{2}\right) s_{-1}+\frac{1}{2} s_{-3} \zeta _2+\frac{1}{4} s_{-2} \zeta _3-\frac{3 \zeta _2 \zeta _3}{16}-\frac{339 \zeta _5}{64}-\frac{1}{2} \zeta _2 \bar{s}_{-3}+\frac{13}{8} \zeta _3 \bar{s}_{-2}-\frac{29}{40} \zeta _2^2 \bar{s}_{-1}-\frac{51}{40} \zeta _2^2 \bar{s}_1+4 \text{Li}_4\left(\frac{1}{2}\right) \bar{s}_1-\frac{1}{8} \zeta _3 \bar{s}_2+\frac{1}{2} \zeta _2 \bar{s}_3+\frac{1}{2} \zeta _2 s_{-2,-1}-\frac{1}{8} \zeta _3 s_{-1,-1}+\frac{5}{8} \zeta _3 s_{-1,1}-\frac{1}{2} \zeta _2 s_{-1,2}-\frac{1}{2} \zeta _2 s_{1,-2}-\frac{1}{4} \zeta _3 s_{1,-1}+s_{4,-1}-\frac{1}{2} \zeta _2 \bar{s}_{-2,-1}-\frac{1}{2} \zeta _2 \bar{s}_{-1,-2}+\frac{1}{8} \zeta _3 \bar{s}_{-1,-1}-\frac{13}{8} \zeta _3 \bar{s}_{-1,1}+\frac{1}{2} \zeta _2 \bar{s}_{-1,2}+\frac{1}{2} \zeta _2 \bar{s}_{1,-2}-\frac{5}{2} \zeta _3 \bar{s}_{1,-1}-\frac{1}{2} \zeta _2 \bar{s}_{1,2}+\bar{s}_{3,-2}-s_{-3,-1,-1}-s_{-2,-2,-1}-s_{-1,-3,-1}-\frac{1}{2} \zeta _2 s_{-1,1,-1}-\frac{1}{2} \zeta _2 s_{1,-1,-1}-s_{1,3,-1}-\bar{s}_{-2,-1,-2}-\bar{s}_{-1,-2,-2}+\frac{1}{2} \zeta _2 \bar{s}_{-1,1,-1}+\frac{1}{2} \zeta _2 \bar{s}_{1,-1,-1}-\bar{s}_{1,2,-2}-\bar{s}_{2,1,-2}+s_{-1,1,-2,-1}+s_{1,-2,-1,-1}+s_{1,-1,-2,-1}+\bar{s}_{-1,-1,1,-2}+\bar{s}_{-1,1,-1,-2}+\bar{s}_{1,-1,-1,-2}  \end{dmath}

\newpage

  \begin{dmath}[style={\small}]     s_{-1} \bar{s}_{-4}  =   -\bar{s}_{-1,-4}-\frac{7}{20} \zeta _2^2 \bar{s}_{-1}-\frac{1}{2} \zeta _2 \bar{s}_3-\frac{3}{4} \zeta _3 \bar{s}_2-\ln _2 \bar{s}_{-4}-\ln _2 \bar{s}_4+s_{-4,-1}+\frac{75 \zeta _5}{16}-\frac{7}{10} \zeta _2^2 \ln _2-\frac{7}{20} \zeta _2^2 s_{-1}+\frac{\zeta _2 s_3}{2}-\frac{3 \zeta _3 s_2}{4}+s_{-4} \ln _2-s_4 \ln _2 \end{dmath}
  \begin{dmath}[style={\small}]     s_{-1} \bar{s}_4  =   -\bar{s}_{-1,4}-\frac{7}{20} \zeta _2^2 \bar{s}_{-1}-\frac{1}{2} \zeta _2 \bar{s}_{-3}-\frac{3}{4} \zeta _3 \bar{s}_{-2}-\ln _2 \bar{s}_{-4}-\ln _2 \bar{s}_4-s_{4,-1}-\frac{3 \zeta _3 \zeta _2}{4}-\frac{15 \zeta _5}{16}-\frac{7}{10} \zeta _2^2 \ln _2-\frac{7}{20} \zeta _2^2 s_{-1}-\frac{1}{2} \zeta _2 s_{-3}+\frac{3}{4} \zeta _3 s_{-2}+s_{-4} \ln _2-s_4 \ln _2 \end{dmath}
  \begin{dmath}[style={\small}]     s_{-1} \bar{s}_{-3,-1}  =   \bar{s}_{4,-1}-\bar{s}_{-3,-1,-1}-\bar{s}_{-1,-3,-1}-2 \ln _2 \bar{s}_{-3,-1}+\frac{1}{2} \zeta _2 \bar{s}_{-3}+\frac{1}{4} \zeta _3 \bar{s}_{-2}+\frac{1}{8} \zeta _2^2 \bar{s}_{-1}-\frac{1}{2} \zeta _2 \ln _2 \bar{s}_{-2}+\frac{1}{2} \zeta _2 \ln _2 \bar{s}_2-\ln _2^2 \bar{s}_{-3}+\ln _2^2 \bar{s}_3-\ln _2 \bar{s}_{-4}+\ln _2 \bar{s}_4-s_{4,-1}+s_{-3,-1,-1}+2 \ln _2 s_{-3,-1}+\frac{15 \zeta _5}{32}-\frac{7}{4} \zeta _2^2 \ln _2+\frac{1}{8} \zeta _2^2 s_{-1}-\frac{1}{2} \zeta _2 s_{-3}-\frac{1}{4} \zeta _3 s_{-2}+\frac{1}{2} \zeta _2 s_{-2} \ln _2+\frac{1}{2} \zeta _2 s_2 \ln _2+s_{-3} \ln _2^2-s_3 \ln _2^2+s_{-4} \ln _2-s_4 \ln _2 \end{dmath}
  \begin{dmath}[style={\small}]     s_{-1} \bar{s}_{-3,1}  =   \bar{s}_{4,1}-\bar{s}_{-3,-1,1}-\bar{s}_{-1,-3,1}-\ln _2 \bar{s}_{-3,-1}-\ln _2 \bar{s}_{-3,1}+2 \text{Li}_4\left(\frac{1}{2}\right) \bar{s}_{-1}-\frac{1}{2} \zeta _2 \bar{s}_{-3}-\frac{11}{10} \zeta _2^2 \bar{s}_{-1}-\frac{5}{8} \zeta _3 \bar{s}_2-\frac{1}{2} \zeta _2 \bar{s}_3-\frac{1}{2} \zeta _2 \ln _2^2 \bar{s}_{-1}+\frac{7}{4} \zeta _3 \ln _2 \bar{s}_{-1}+\frac{1}{12} \ln _2^4 \bar{s}_{-1}-\frac{1}{2} \ln _2^2 \bar{s}_{-3}+\frac{1}{2} \ln _2^2 \bar{s}_3+s_{-4,-1}-s_{-3,1,-1}+\ln _2 s_{-3,-1}-\ln _2 s_{-3,1}-\frac{17 \zeta _2 \zeta _3}{8}+\frac{225 \zeta _5}{32}-\zeta _2 \ln _2^3+\frac{7}{2} \zeta _3 \ln _2^2-\frac{11}{5} \zeta _2^2 \ln _2+2 \text{Li}_4\left(\frac{1}{2}\right) s_{-1}+4 \text{Li}_4\left(\frac{1}{2}\right) \ln _2-\frac{11}{10} \zeta _2^2 s_{-1}+\frac{\zeta _2 s_3}{2}-\frac{5 \zeta _3 s_2}{8}-\frac{1}{2} \zeta _2 s_{-1} \ln _2^2+\frac{7}{4} \zeta _3 s_{-1} \ln _2+\frac{1}{12} s_{-1} \ln _2^4+\frac{1}{2} s_{-3} \ln _2^2-\frac{1}{2} s_3 \ln _2^2+s_{-4} \ln _2-s_4 \ln _2+\frac{\ln _2^5}{6} \end{dmath}
  \begin{dmath}[style={\small}]     s_{-1} \bar{s}_{2,-2}  =   -\frac{1}{2} \zeta _2 \bar{s}_{2,-1}+\bar{s}_{-3,-2}-\bar{s}_{-1,2,-2}-\bar{s}_{2,-1,-2}-\ln _2 \bar{s}_{2,-2}-\ln _2 \bar{s}_{2,2}+4 \text{Li}_4\left(\frac{1}{2}\right) \bar{s}_{-1}+\frac{1}{2} \zeta _2 \bar{s}_{-3}-\frac{11}{8} \zeta _2^2 \bar{s}_{-1}+\frac{3}{2} \zeta _3 \bar{s}_2-\frac{1}{2} \zeta _2 \bar{s}_3-\zeta _2 \ln _2^2 \bar{s}_{-1}+\frac{1}{2} \zeta _2 \ln _2 \bar{s}_{-2}+\frac{7}{2} \zeta _3 \ln _2 \bar{s}_{-1}-\frac{1}{2} \zeta _2 \ln _2 \bar{s}_2+\frac{1}{6} \ln _2^4 \bar{s}_{-1}+\frac{1}{2} \zeta _2 s_{2,-1}+s_{-4,-1}-s_{2,-2,-1}-\ln _2 s_{2,-2}+\ln _2 s_{2,2}-\frac{25 \zeta _2 \zeta _3}{8}+\frac{235 \zeta _5}{32}-2 \zeta _2 \ln _2^3+7 \zeta _3 \ln _2^2-\frac{11}{4} \zeta _2^2 \ln _2+4 \text{Li}_4\left(\frac{1}{2}\right) s_{-1}+8 \text{Li}_4\left(\frac{1}{2}\right) \ln _2-\frac{11}{8} \zeta _2^2 s_{-1}+\frac{\zeta _2 s_3}{2}-\frac{3 \zeta _3 s_2}{4}-\zeta _2 s_{-1} \ln _2^2-\frac{1}{2} \zeta _2 s_{-2} \ln _2+\frac{1}{2} \zeta _2 s_2 \ln _2+\frac{7}{2} \zeta _3 s_{-1} \ln _2+\frac{1}{6} s_{-1} \ln _2^4+s_{-4} \ln _2-s_4 \ln _2+\frac{\ln _2^5}{3} \end{dmath}
  \begin{dmath}[style={\small}]     s_{-1} \bar{s}_{3,-1}  =   \bar{s}_{-4,-1}-\bar{s}_{-1,3,-1}-\bar{s}_{3,-1,-1}-2 \ln _2 \bar{s}_{3,-1}+\frac{1}{8} \zeta _2^2 \bar{s}_{-1}+\frac{1}{4} \zeta _3 \bar{s}_2+\frac{1}{2} \zeta _2 \bar{s}_3+\frac{1}{2} \zeta _2 \ln _2 \bar{s}_{-2}-\frac{1}{2} \zeta _2 \ln _2 \bar{s}_2+\ln _2^2 \bar{s}_{-3}-\ln _2^2 \bar{s}_3+\ln _2 \bar{s}_{-4}-\ln _2 \bar{s}_4+s_{-4,-1}-s_{3,-1,-1}-2 \ln _2 s_{3,-1}+\frac{3 \zeta _2 \zeta _3}{4}-\frac{85 \zeta _5}{32}+\frac{9}{4} \zeta _2^2 \ln _2+\frac{1}{8} \zeta _2^2 s_{-1}+\frac{\zeta _2 s_3}{2}+\frac{\zeta _3 s_2}{4}-\frac{1}{2} \zeta _2 s_{-2} \ln _2-\frac{1}{2} \zeta _2 s_2 \ln _2+s_{-3} \ln _2^2-s_3 \ln _2^2+s_{-4} \ln _2-s_4 \ln _2 \end{dmath}
  \begin{dmath}[style={\small}]     s_{-1} \bar{s}_{3,1}  =   \bar{s}_{-4,1}-\bar{s}_{-1,3,1}-\bar{s}_{3,-1,1}-\ln _2 \bar{s}_{3,-1}-\ln _2 \bar{s}_{3,1}+2 \text{Li}_4\left(\frac{1}{2}\right) \bar{s}_{-1}-\frac{1}{2} \zeta _2 \bar{s}_{-3}-\frac{5}{8} \zeta _3 \bar{s}_{-2}-\frac{11}{10} \zeta _2^2 \bar{s}_{-1}-\frac{1}{2} \zeta _2 \bar{s}_3-\frac{1}{2} \zeta _2 \ln _2^2 \bar{s}_{-1}+\frac{7}{4} \zeta _3 \ln _2 \bar{s}_{-1}+\frac{1}{12} \ln _2^4 \bar{s}_{-1}+\frac{1}{2} \ln _2^2 \bar{s}_{-3}-\frac{1}{2} \ln _2^2 \bar{s}_3-s_{4,-1}+s_{3,1,-1}-\ln _2 s_{3,-1}+\ln _2 s_{3,1}-\frac{\zeta _2 \zeta _3}{4}+\frac{27 \zeta _5}{8}-\frac{1}{3} \zeta _2 \ln _2^3+\frac{7}{4} \zeta _3 \ln _2^2-\frac{11}{5} \zeta _2^2 \ln _2+2 \text{Li}_4\left(\frac{1}{2}\right) s_{-1}-4 \text{Li}_5\left(\frac{1}{2}\right)-\frac{11}{10} \zeta _2^2 s_{-1}-\frac{1}{2} \zeta _2 s_{-3}+\frac{5}{8} \zeta _3 s_{-2}-\frac{1}{2} \zeta _2 s_{-1} \ln _2^2+\frac{7}{4} \zeta _3 s_{-1} \ln _2+\frac{1}{12} s_{-1} \ln _2^4+\frac{1}{2} s_{-3} \ln _2^2-\frac{1}{2} s_3 \ln _2^2+s_{-4} \ln _2-s_4 \ln _2+\frac{\ln _2^5}{30} \end{dmath}
  \begin{dmath}[style={\small}]     s_{-1} \bar{s}_{-2,-1,-1}  =   \frac{1}{2} \zeta _2 \bar{s}_{-2,-1}+\bar{s}_{-2,2,-1}+\bar{s}_{3,-1,-1}-2 \bar{s}_{-2,-1,-1,-1}-\bar{s}_{-1,-2,-1,-1}-\ln _2 \bar{s}_{-2,-2}+\ln _2 \bar{s}_{-2,2}-2 \ln _2 \bar{s}_{-2,-1,-1}-4 \text{Li}_4\left(\frac{1}{2}\right) \bar{s}_{-1}-\frac{3}{4} \zeta _3 \bar{s}_{-2}+\frac{7}{5} \zeta _2^2 \bar{s}_{-1}-\frac{1}{4} \zeta _3 \bar{s}_2+\zeta _2 \ln _2^2 \bar{s}_{-1}-\frac{7}{2} \zeta _3 \ln _2 \bar{s}_{-1}-\frac{1}{6} \ln _2^4 \bar{s}_{-1}+\frac{2}{3} \ln _2^3 \bar{s}_{-2}-\frac{2}{3} \ln _2^3 \bar{s}_2+\ln _2^2 \bar{s}_{-3}-\ln _2^2 \bar{s}_3-\frac{1}{2} \zeta _2 s_{-2,-1}+s_{-4,-1}-s_{-2,2,-1}-s_{3,-1,-1}+s_{-2,-1,-1,-1}+2 \ln _2^2 s_{-2,-1}+\ln _2 s_{-2,-2}-\ln _2 s_{-2,2}-2 \ln _2 s_{3,-1}+2 \ln _2 s_{-2,-1,-1}+\frac{17 \zeta _2 \zeta _3}{8}-\frac{173 \zeta _5}{16}+\frac{2}{3} \zeta _2 \ln _2^3-\frac{1}{2} \zeta _3 \ln _2^2+\frac{12}{5} \zeta _2^2 \ln _2-4 \text{Li}_4\left(\frac{1}{2}\right) s_{-1}+8 \text{Li}_5\left(\frac{1}{2}\right)+\frac{7}{5} \zeta _2^2 s_{-1}+\frac{\zeta _2 s_3}{2}+\frac{1}{2} \zeta _3 s_{-2}-\frac{\zeta _3 s_2}{4}+\zeta _2 s_{-1} \ln _2^2-\zeta _2 s_{-2} \ln _2-\frac{7}{2} \zeta _3 s_{-1} \ln _2-\frac{1}{6} s_{-1} \ln _2^4+\frac{2}{3} s_{-2} \ln _2^3-\frac{2}{3} s_2 \ln _2^3+s_{-3} \ln _2^2-s_3 \ln _2^2+s_{-4} \ln _2-s_4 \ln _2-\frac{\ln _2^5}{15} \end{dmath}
  \begin{dmath}[style={\small}]     s_{-1} \bar{s}_{-2,-1,1}  =   -\zeta _2 \bar{s}_{-2,-1}+\bar{s}_{-2,2,1}+\bar{s}_{3,-1,1}-2 \bar{s}_{-2,-1,-1,1}-\bar{s}_{-1,-2,-1,1}-\ln _2 \bar{s}_{-2,-1,-1}-\ln _2 \bar{s}_{-2,-1,1}+4 \text{Li}_4\left(\frac{1}{2}\right) \bar{s}_{-1}-\frac{1}{2} \zeta _2 \bar{s}_{-3}+\frac{13}{8} \zeta _3 \bar{s}_{-2}-\frac{5}{4} \zeta _2^2 \bar{s}_{-1}+\frac{1}{2} \zeta _2 \bar{s}_3-\zeta _2 \ln _2^2 \bar{s}_{-1}-2 \zeta _2 \ln _2 \bar{s}_{-2}+\frac{21}{8} \zeta _3 \ln _2 \bar{s}_{-1}+\frac{1}{2} \zeta _2 \ln _2 \bar{s}_2+\frac{1}{6} \ln _2^4 \bar{s}_{-1}+\frac{1}{2} \ln _2^3 \bar{s}_{-2}-\frac{1}{2} \ln _2^3 \bar{s}_2+\frac{1}{2} \ln _2^2 \bar{s}_{-3}-\frac{1}{2} \ln _2^2 \bar{s}_3-\frac{1}{2} \zeta _2 s_{-2,-1}-s_{4,-1}+s_{-2,-2,-1}+s_{3,1,-1}-s_{-2,-1,1,-1}+\ln _2^2 s_{-2,-1}+\ln _2 s_{-2,-2}-\ln _2 s_{-2,2}-\ln _2 s_{3,-1}+\ln _2 s_{3,1}+\ln _2 s_{-2,-1,-1}-\ln _2 s_{-2,-1,1}-\frac{35 \zeta _2 \zeta _3}{16}+\frac{1779 \zeta _5}{64}+\frac{1}{3} \zeta _2 \ln _2^3+\frac{19}{8} \zeta _3 \ln _2^2-\frac{59}{10} \zeta _2^2 \ln _2+4 \text{Li}_4\left(\frac{1}{2}\right) s_{-1}-26 \text{Li}_5\left(\frac{1}{2}\right)-10 \text{Li}_4\left(\frac{1}{2}\right) \ln _2-\frac{5}{4} \zeta _2^2 s_{-1}-\frac{1}{2} \zeta _2 s_{-3}+\frac{5}{8} \zeta _3 s_{-2}-\zeta _2 s_{-1} \ln _2^2-\frac{1}{2} \zeta _2 s_{-2} \ln _2+\frac{1}{2} \zeta _2 s_2 \ln _2+\frac{21}{8} \zeta _3 s_{-1} \ln _2+\frac{1}{6} s_{-1} \ln _2^4+\frac{1}{2} s_{-2} \ln _2^3-\frac{1}{2} s_2 \ln _2^3+\frac{1}{2} s_{-3} \ln _2^2-\frac{1}{2} s_3 \ln _2^2+s_{-4} \ln _2-s_4 \ln _2-\frac{\ln _2^5}{5} \end{dmath}
  \begin{dmath}[style={\small}]     s_{-1} \bar{s}_{-2,1,-1}  =   \frac{1}{2} \zeta _2 \bar{s}_{-2,1}+\bar{s}_{-2,-2,-1}+\bar{s}_{3,1,-1}-\bar{s}_{-2,-1,1,-1}-\bar{s}_{-2,1,-1,-1}-\bar{s}_{-1,-2,1,-1}+\frac{1}{2} \ln _2^2 \bar{s}_{-2,-1}-\frac{1}{2} \ln _2^2 \bar{s}_{-2,1}+\ln _2 \bar{s}_{-2,-2}-\ln _2 \bar{s}_{-2,2}-2 \ln _2 \bar{s}_{-2,1,-1}+3 \text{Li}_4\left(\frac{1}{2}\right) \bar{s}_{-1}-\frac{1}{8} \zeta _3 \bar{s}_{-2}-\frac{6}{5} \zeta _2^2 \bar{s}_{-1}-\frac{3}{4} \zeta _2 \ln _2^2 \bar{s}_{-1}+\zeta _2 \ln _2 \bar{s}_{-2}+\frac{7}{2} \zeta _3 \ln _2 \bar{s}_{-1}+\frac{1}{2} \zeta _2 \ln _2 \bar{s}_2+\frac{1}{8} \ln _2^4 \bar{s}_{-1}-\frac{1}{2} \ln _2^2 \bar{s}_{-3}+\frac{1}{2} \ln _2^2 \bar{s}_3+\frac{1}{2} \zeta _2 s_{-2,1}-s_{4,-1}+s_{-3,-1,-1}+s_{-2,-2,-1}-s_{-2,1,-1,-1}+\frac{1}{2} \ln _2^2 s_{-2,-1}-\frac{3}{2} \ln _2^2 s_{-2,1}+2 \ln _2 s_{-3,-1}+\ln _2 s_{-2,-2}-\ln _2 s_{-2,2}-2 \ln _2 s_{-2,1,-1}+\frac{\zeta _2 \zeta _3}{4}+\frac{477 \zeta _5}{64}+\frac{1}{3} \zeta _2 \ln _2^3+\frac{1}{4} \zeta _3 \ln _2^2-\frac{91}{40} \zeta _2^2 \ln _2+3 \text{Li}_4\left(\frac{1}{2}\right) s_{-1}-8 \text{Li}_5\left(\frac{1}{2}\right)-4 \text{Li}_4\left(\frac{1}{2}\right) \ln _2-\frac{6}{5} \zeta _2^2 s_{-1}-\frac{1}{2} \zeta _2 s_{-3}-\frac{1}{4} \zeta _3 s_{-2}-\frac{3}{4} \zeta _2 s_{-1} \ln _2^2+\frac{1}{2} \zeta _2 s_{-2} \ln _2+\frac{1}{2} \zeta _2 s_2 \ln _2+\frac{7}{2} \zeta _3 s_{-1} \ln _2+\frac{1}{8} s_{-1} \ln _2^4+s_{-3} \ln _2^2-s_3 \ln _2^2+s_{-4} \ln _2-s_4 \ln _2-\frac{\ln _2^5}{10} \end{dmath}
  \begin{dmath}[style={\small}]     s_{-1} \bar{s}_{-2,1,1}  =   -\frac{1}{2} \zeta _2 \bar{s}_{-2,-1}-\frac{1}{2} \zeta _2 \bar{s}_{-2,1}+\bar{s}_{-2,-2,1}+\bar{s}_{3,1,1}-\bar{s}_{-2,-1,1,1}-\bar{s}_{-2,1,-1,1}-\bar{s}_{-1,-2,1,1}+\frac{1}{2} \ln _2^2 \bar{s}_{-2,-1}-\frac{1}{2} \ln _2^2 \bar{s}_{-2,1}-\ln _2 \bar{s}_{-2,1,-1}-\ln _2 \bar{s}_{-2,1,1}-\text{Li}_4\left(\frac{1}{2}\right) \bar{s}_{-1}-\frac{1}{8} \zeta _3 \bar{s}_{-2}+\frac{1}{8} \zeta _2^2 \bar{s}_{-1}-\frac{7}{8} \zeta _3 \bar{s}_2+\frac{1}{4} \zeta _2 \ln _2^2 \bar{s}_{-1}-\frac{1}{2} \zeta _2 \ln _2 \bar{s}_{-2}-\frac{7}{8} \zeta _3 \ln _2 \bar{s}_{-1}+\frac{1}{2} \zeta _2 \ln _2 \bar{s}_2-\frac{1}{24} \ln _2^4 \bar{s}_{-1}+\frac{1}{6} \ln _2^3 \bar{s}_{-2}-\frac{1}{6} \ln _2^3 \bar{s}_2-\frac{1}{2} \zeta _2 s_{-2,-1}+s_{-4,-1}-s_{-3,1,-1}-s_{-2,2,-1}+s_{-2,1,1,-1}+\frac{1}{2} \ln _2^2 s_{-2,-1}-\frac{1}{2} \ln _2^2 s_{-2,1}+\ln _2 s_{-3,-1}-\ln _2 s_{-3,1}+\ln _2 s_{-2,-2}-\ln _2 s_{-2,2}-\ln _2 s_{-2,1,-1}+\ln _2 s_{-2,1,1}-\frac{9 \zeta _2 \zeta _3}{16}+\frac{13 \zeta _5}{16}+\frac{1}{6} \zeta _2 \ln _2^3-\frac{7}{8} \zeta _3 \ln _2^2+\frac{1}{4} \zeta _2^2 \ln _2-\text{Li}_4\left(\frac{1}{2}\right) s_{-1}+2 \text{Li}_5\left(\frac{1}{2}\right)+\frac{1}{8} \zeta _2^2 s_{-1}+\frac{\zeta _2 s_3}{2}+\frac{1}{4} \zeta _3 s_{-2}-\frac{7 \zeta _3 s_2}{8}+\frac{1}{4} \zeta _2 s_{-1} \ln _2^2-\frac{1}{2} \zeta _2 s_{-2} \ln _2+\frac{1}{2} \zeta _2 s_2 \ln _2-\frac{7}{8} \zeta _3 s_{-1} \ln _2-\frac{1}{24} s_{-1} \ln _2^4+\frac{1}{6} s_{-2} \ln _2^3-\frac{1}{6} s_2 \ln _2^3+\frac{1}{2} s_{-3} \ln _2^2-\frac{1}{2} s_3 \ln _2^2+s_{-4} \ln _2-s_4 \ln _2-\frac{\ln _2^5}{60} \end{dmath}
  \begin{dmath}[style={\small}]     s_{-1} \bar{s}_{2,-1,-1}  =   \frac{1}{2} \zeta _2 \bar{s}_{2,-1}+\bar{s}_{-3,-1,-1}+\bar{s}_{2,2,-1}-\bar{s}_{-1,2,-1,-1}-2 \bar{s}_{2,-1,-1,-1}-\ln _2 \bar{s}_{2,-2}+\ln _2 \bar{s}_{2,2}-2 \ln _2 \bar{s}_{2,-1,-1}-4 \text{Li}_4\left(\frac{1}{2}\right) \bar{s}_{-1}-\frac{1}{4} \zeta _3 \bar{s}_{-2}+\frac{7}{5} \zeta _2^2 \bar{s}_{-1}-\frac{3}{4} \zeta _3 \bar{s}_2+\zeta _2 \ln _2^2 \bar{s}_{-1}-\frac{7}{2} \zeta _3 \ln _2 \bar{s}_{-1}-\frac{1}{6} \ln _2^4 \bar{s}_{-1}-\frac{2}{3} \ln _2^3 \bar{s}_{-2}+\frac{2}{3} \ln _2^3 \bar{s}_2-\ln _2^2 \bar{s}_{-3}+\ln _2^2 \bar{s}_3+\frac{1}{2} \zeta _2 s_{2,-1}-s_{4,-1}+s_{-3,-1,-1}+s_{2,2,-1}-s_{2,-1,-1,-1}-2 \ln _2^2 s_{2,-1}+2 \ln _2 s_{-3,-1}-\ln _2 s_{2,-2}+\ln _2 s_{2,2}-2 \ln _2 s_{2,-1,-1}-\frac{5 \zeta _2 \zeta _3}{8}-\frac{369 \zeta _5}{16}-2 \zeta _2 \ln _2^3+\frac{1}{2} \zeta _3 \ln _2^2+\frac{16}{5} \zeta _2^2 \ln _2-4 \text{Li}_4\left(\frac{1}{2}\right) s_{-1}+24 \text{Li}_5\left(\frac{1}{2}\right)+16 \text{Li}_4\left(\frac{1}{2}\right) \ln _2+\frac{7}{5} \zeta _2^2 s_{-1}-\frac{1}{2} \zeta _2 s_{-3}+\frac{1}{4} \zeta _3 s_{-2}-\frac{\zeta _3 s_2}{2}+\zeta _2 s_{-1} \ln _2^2+\zeta _2 s_2 \ln _2-\frac{7}{2} \zeta _3 s_{-1} \ln _2-\frac{1}{6} s_{-1} \ln _2^4+\frac{2}{3} s_{-2} \ln _2^3-\frac{2}{3} s_2 \ln _2^3+s_{-3} \ln _2^2-s_3 \ln _2^2+s_{-4} \ln _2-s_4 \ln _2+\frac{7 \ln _2^5}{15} \end{dmath}
  \begin{dmath}[style={\small}]     s_{-1} \bar{s}_{2,-1,1}  =   -\zeta _2 \bar{s}_{2,-1}+\bar{s}_{-3,-1,1}+\bar{s}_{2,2,1}-\bar{s}_{-1,2,-1,1}-2 \bar{s}_{2,-1,-1,1}-\ln _2 \bar{s}_{2,-1,-1}-\ln _2 \bar{s}_{2,-1,1}+4 \text{Li}_4\left(\frac{1}{2}\right) \bar{s}_{-1}+\frac{1}{2} \zeta _2 \bar{s}_{-3}-\frac{5}{4} \zeta _2^2 \bar{s}_{-1}+\frac{13}{8} \zeta _3 \bar{s}_2-\frac{1}{2} \zeta _2 \bar{s}_3-\zeta _2 \ln _2^2 \bar{s}_{-1}+\frac{1}{2} \zeta _2 \ln _2 \bar{s}_{-2}+\frac{21}{8} \zeta _3 \ln _2 \bar{s}_{-1}-2 \zeta _2 \ln _2 \bar{s}_2+\frac{1}{6} \ln _2^4 \bar{s}_{-1}-\frac{1}{2} \ln _2^3 \bar{s}_{-2}+\frac{1}{2} \ln _2^3 \bar{s}_2-\frac{1}{2} \ln _2^2 \bar{s}_{-3}+\frac{1}{2} \ln _2^2 \bar{s}_3+\frac{1}{2} \zeta _2 s_{2,-1}+s_{-4,-1}-s_{-3,1,-1}-s_{2,-2,-1}+s_{2,-1,1,-1}-\ln _2^2 s_{2,-1}+\ln _2 s_{-3,-1}-\ln _2 s_{-3,1}-\ln _2 s_{2,-2}+\ln _2 s_{2,2}-\ln _2 s_{2,-1,-1}+\ln _2 s_{2,-1,1}-\frac{17 \zeta _2 \zeta _3}{8}+\frac{75 \zeta _5}{16}-\frac{5}{2} \zeta _2 \ln _2^3+\frac{11}{2} \zeta _3 \ln _2^2-\frac{89}{40} \zeta _2^2 \ln _2+4 \text{Li}_4\left(\frac{1}{2}\right) s_{-1}+10 \text{Li}_4\left(\frac{1}{2}\right) \ln _2-\frac{5}{4} \zeta _2^2 s_{-1}+\frac{\zeta _2 s_3}{2}-\frac{5 \zeta _3 s_2}{8}-\zeta _2 s_{-1} \ln _2^2-\frac{1}{2} \zeta _2 s_{-2} \ln _2+\frac{1}{2} \zeta _2 s_2 \ln _2+\frac{21}{8} \zeta _3 s_{-1} \ln _2+\frac{1}{6} s_{-1} \ln _2^4+\frac{1}{2} s_{-2} \ln _2^3-\frac{1}{2} s_2 \ln _2^3+\frac{1}{2} s_{-3} \ln _2^2-\frac{1}{2} s_3 \ln _2^2+s_{-4} \ln _2-s_4 \ln _2+\frac{5 \ln _2^5}{12} \end{dmath}
  \begin{dmath}[style={\small}]     s_{-1} \bar{s}_{2,1,-1}  =   \frac{1}{2} \zeta _2 \bar{s}_{2,1}+\bar{s}_{-3,1,-1}+\bar{s}_{2,-2,-1}-\bar{s}_{-1,2,1,-1}-\bar{s}_{2,-1,1,-1}-\bar{s}_{2,1,-1,-1}+\frac{1}{2} \ln _2^2 \bar{s}_{2,-1}-\frac{1}{2} \ln _2^2 \bar{s}_{2,1}+\ln _2 \bar{s}_{2,-2}-\ln _2 \bar{s}_{2,2}-2 \ln _2 \bar{s}_{2,1,-1}+3 \text{Li}_4\left(\frac{1}{2}\right) \bar{s}_{-1}-\frac{6}{5} \zeta _2^2 \bar{s}_{-1}-\frac{1}{8} \zeta _3 \bar{s}_2-\frac{3}{4} \zeta _2 \ln _2^2 \bar{s}_{-1}+\frac{1}{2} \zeta _2 \ln _2 \bar{s}_{-2}+\frac{7}{2} \zeta _3 \ln _2 \bar{s}_{-1}+\zeta _2 \ln _2 \bar{s}_2+\frac{1}{8} \ln _2^4 \bar{s}_{-1}+\frac{1}{2} \ln _2^2 \bar{s}_{-3}-\frac{1}{2} \ln _2^2 \bar{s}_3-\frac{1}{2} \zeta _2 s_{2,1}+s_{-4,-1}-s_{2,-2,-1}-s_{3,-1,-1}+s_{2,1,-1,-1}-\frac{1}{2} \ln _2^2 s_{2,-1}+\frac{3}{2} \ln _2^2 s_{2,1}-\ln _2 s_{2,-2}+\ln _2 s_{2,2}-2 \ln _2 s_{3,-1}+2 \ln _2 s_{2,1,-1}-\frac{11 \zeta _2 \zeta _3}{8}+\frac{13 \zeta _5}{16}+\frac{1}{6} \zeta _2 \ln _2^3+\frac{19}{8} \zeta _3 \ln _2^2+\frac{3}{5} \zeta _2^2 \ln _2+3 \text{Li}_4\left(\frac{1}{2}\right) s_{-1}+2 \text{Li}_5\left(\frac{1}{2}\right)-\frac{6}{5} \zeta _2^2 s_{-1}+\frac{\zeta _2 s_3}{2}+\frac{\zeta _3 s_2}{4}-\frac{3}{4} \zeta _2 s_{-1} \ln _2^2-\frac{1}{2} \zeta _2 s_{-2} \ln _2-\frac{1}{2} \zeta _2 s_2 \ln _2+\frac{7}{2} \zeta _3 s_{-1} \ln _2+\frac{1}{8} s_{-1} \ln _2^4+s_{-3} \ln _2^2-s_3 \ln _2^2+s_{-4} \ln _2-s_4 \ln _2-\frac{\ln _2^5}{60} \end{dmath}
  \begin{dmath}[style={\small}]     s_{-1} \bar{s}_{2,1,1}  =   -\frac{1}{2} \zeta _2 \bar{s}_{2,-1}-\frac{1}{2} \zeta _2 \bar{s}_{2,1}+\bar{s}_{-3,1,1}+\bar{s}_{2,-2,1}-\bar{s}_{-1,2,1,1}-\bar{s}_{2,-1,1,1}-\bar{s}_{2,1,-1,1}+\frac{1}{2} \ln _2^2 \bar{s}_{2,-1}-\frac{1}{2} \ln _2^2 \bar{s}_{2,1}-\ln _2 \bar{s}_{2,1,-1}-\ln _2 \bar{s}_{2,1,1}-\text{Li}_4\left(\frac{1}{2}\right) \bar{s}_{-1}-\frac{7}{8} \zeta _3 \bar{s}_{-2}+\frac{1}{8} \zeta _2^2 \bar{s}_{-1}-\frac{1}{8} \zeta _3 \bar{s}_2+\frac{1}{4} \zeta _2 \ln _2^2 \bar{s}_{-1}+\frac{1}{2} \zeta _2 \ln _2 \bar{s}_{-2}-\frac{7}{8} \zeta _3 \ln _2 \bar{s}_{-1}-\frac{1}{2} \zeta _2 \ln _2 \bar{s}_2-\frac{1}{24} \ln _2^4 \bar{s}_{-1}-\frac{1}{6} \ln _2^3 \bar{s}_{-2}+\frac{1}{6} \ln _2^3 \bar{s}_2+\frac{1}{2} \zeta _2 s_{2,-1}-s_{4,-1}+s_{2,2,-1}+s_{3,1,-1}-s_{2,1,1,-1}-\frac{1}{2} \ln _2^2 s_{2,-1}+\frac{1}{2} \ln _2^2 s_{2,1}-\ln _2 s_{2,-2}+\ln _2 s_{2,2}-\ln _2 s_{3,-1}+\ln _2 s_{3,1}+\ln _2 s_{2,1,-1}-\ln _2 s_{2,1,1}+\frac{5 \zeta _2 \zeta _3}{8}-\frac{121 \zeta _5}{16}-\frac{1}{2} \zeta _2 \ln _2^3+\frac{7}{8} \zeta _3 \ln _2^2+\frac{1}{4} \zeta _2^2 \ln _2-\text{Li}_4\left(\frac{1}{2}\right) s_{-1}+6 \text{Li}_5\left(\frac{1}{2}\right)+4 \text{Li}_4\left(\frac{1}{2}\right) \ln _2+\frac{1}{8} \zeta _2^2 s_{-1}-\frac{1}{2} \zeta _2 s_{-3}+\frac{7}{8} \zeta _3 s_{-2}-\frac{\zeta _3 s_2}{4}+\frac{1}{4} \zeta _2 s_{-1} \ln _2^2-\frac{1}{2} \zeta _2 s_{-2} \ln _2+\frac{1}{2} \zeta _2 s_2 \ln _2-\frac{7}{8} \zeta _3 s_{-1} \ln _2-\frac{1}{24} s_{-1} \ln _2^4+\frac{1}{6} s_{-2} \ln _2^3-\frac{1}{6} s_2 \ln _2^3+\frac{1}{2} s_{-3} \ln _2^2-\frac{1}{2} s_3 \ln _2^2+s_{-4} \ln _2-s_4 \ln _2+\frac{7 \ln _2^5}{60} \end{dmath}
  \begin{dmath}[style={\small}]     s_{-1} \bar{s}_{-1,1,-1,-1}  =   -\frac{1}{4} \zeta _3 \bar{s}_{-1,-1}-\frac{3}{4} \zeta _3 \bar{s}_{-1,1}+\frac{1}{2} \zeta _2 \bar{s}_{-1,1,-1}+\bar{s}_{-1,-2,-1,-1}+\bar{s}_{-1,1,2,-1}+\bar{s}_{2,1,-1,-1}-2 \bar{s}_{-1,-1,1,-1,-1}-2 \bar{s}_{-1,1,-1,-1,-1}-\frac{1}{6} \ln _2^3 \bar{s}_{-1,-1}+\frac{1}{6} \ln _2^3 \bar{s}_{-1,1}-\ln _2^2 \bar{s}_{-1,-2}+\ln _2^2 \bar{s}_{-1,2}-\ln _2 \bar{s}_{-1,1,-2}+\ln _2 \bar{s}_{-1,1,2}-2 \ln _2 \bar{s}_{-1,1,-1,-1}+\text{Li}_4\left(\frac{1}{2}\right) \bar{s}_{-1}-\frac{9}{20} \zeta _2^2 \bar{s}_{-1}-\zeta _2 \ln _2^2 \bar{s}_{-1}+\frac{1}{24} \ln _2^4 \bar{s}_{-1}+\frac{1}{2} \ln _2^3 \bar{s}_{-2}-\frac{1}{2} \ln _2^3 \bar{s}_2-\frac{1}{2} \zeta _2 s_{-2,-1}-\frac{1}{2} \zeta _2 s_{-1,-2}+\frac{1}{4} \zeta _3 s_{-1,-1}-\frac{1}{2} \zeta _3 s_{-1,1}+\frac{1}{2} \zeta _2 s_{-1,1,-1}+\zeta _2 \ln _2 s_{-1,1}+s_{-4,-1}-s_{-2,2,-1}-s_{-1,3,-1}-s_{3,-1,-1}+s_{-2,-1,-1,-1}+s_{-1,-2,-1,-1}+s_{-1,1,2,-1}-s_{-1,1,-1,-1,-1}+\frac{1}{6} \ln _2^3 s_{-1,-1}-\frac{7}{6} \ln _2^3 s_{-1,1}+2 \ln _2^2 s_{-2,-1}+\ln _2^2 s_{-1,-2}-\ln _2^2 s_{-1,2}-2 \ln _2^2 s_{-1,1,-1}+\ln _2 s_{-2,-2}-\ln _2 s_{-2,2}+\ln _2 s_{-1,-3}-\ln _2 s_{-1,3}-2 \ln _2 s_{3,-1}+2 \ln _2 s_{-2,-1,-1}+2 \ln _2 s_{-1,-2,-1}-\ln _2 s_{-1,1,-2}+\ln _2 s_{-1,1,2}-2 \ln _2 s_{-1,1,-1,-1}+\frac{53 \zeta _5}{8}-\frac{5}{4} \zeta _2 \ln _2^3+3 \zeta _3 \ln _2^2-\frac{97}{40} \zeta _2^2 \ln _2+4 \text{Li}_4\left(\frac{1}{2}\right) s_{-1}-6 \text{Li}_5\left(\frac{1}{2}\right)-\frac{33}{20} \zeta _2^2 s_{-1}+\frac{\zeta _2 s_3}{2}+\frac{1}{2} \zeta _3 s_{-2}-\frac{\zeta _3 s_2}{4}-\frac{1}{2} \zeta _2 s_{-1} \ln _2^2-\zeta _2 s_{-2} \ln _2+3 \zeta _3 s_{-1} \ln _2-\frac{1}{12} s_{-1} \ln _2^4+\frac{2}{3} s_{-2} \ln _2^3-\frac{2}{3} s_2 \ln _2^3+s_{-3} \ln _2^2-s_3 \ln _2^2+s_{-4} \ln _2-s_4 \ln _2+\frac{\ln _2^5}{20} \end{dmath}
  \begin{dmath}[style={\small}]     s_{-1} \bar{s}_{-1,1,-1,1}  =   -\frac{7 \ln _2^5}{120}-\frac{1}{12} \bar{s}_{-1} \ln _2^4+\frac{1}{2} s_{-2} \ln _2^3-\frac{1}{2} s_2 \ln _2^3-\frac{3}{4} \zeta _2 \ln _2^3+\frac{1}{3} \bar{s}_{-2} \ln _2^3-\frac{1}{3} \bar{s}_2 \ln _2^3+\frac{1}{6} s_{-1,-1} \ln _2^3-\frac{5}{6} s_{-1,1} \ln _2^3-\frac{1}{6} \bar{s}_{-1,-1} \ln _2^3+\frac{1}{6} \bar{s}_{-1,1} \ln _2^3+\frac{1}{2} s_{-3} \ln _2^2-\frac{1}{2} s_3 \ln _2^2-\frac{1}{2} s_{-1} \zeta _2 \ln _2^2+\frac{11}{8} \zeta _3 \ln _2^2-\zeta _2 \bar{s}_{-1} \ln _2^2+s_{-2,-1} \ln _2^2+\frac{1}{2} s_{-1,-2} \ln _2^2-\frac{1}{2} s_{-1,2} \ln _2^2-\frac{1}{2} \bar{s}_{-1,-2} \ln _2^2+\frac{1}{2} \bar{s}_{-1,2} \ln _2^2-s_{-1,1,-1} \ln _2^2-\frac{3}{8} \zeta _2^2 \ln _2-3 \text{Li}_4\left(\frac{1}{2}\right) \ln _2+s_{-4} \ln _2-s_4 \ln _2-\frac{1}{2} s_{-2} \zeta _2 \ln _2+\frac{1}{2} s_2 \zeta _2 \ln _2+\frac{19}{8} s_{-1} \zeta _3 \ln _2-\frac{1}{2} \zeta _2 \bar{s}_{-2} \ln _2+\frac{1}{2} \zeta _2 \bar{s}_2 \ln _2+s_{-2,-2} \ln _2-s_{-2,2} \ln _2+s_{-1,-3} \ln _2+\zeta _2 s_{-1,1} \ln _2-s_{-1,3} \ln _2-s_{3,-1} \ln _2+s_{3,1} \ln _2-\frac{3}{2} \zeta _2 \bar{s}_{-1,1} \ln _2+s_{-2,-1,-1} \ln _2-s_{-2,-1,1} \ln _2+s_{-1,-2,-1} \ln _2-s_{-1,-2,1} \ln _2-s_{-1,1,-2} \ln _2+s_{-1,1,2} \ln _2-s_{-1,1,-1,-1} \ln _2+s_{-1,1,-1,1} \ln _2-\bar{s}_{-1,1,-1,-1} \ln _2-\bar{s}_{-1,1,-1,1} \ln _2-\frac{5}{4} s_{-1} \zeta _2^2-6 \text{Li}_5\left(\frac{1}{2}\right)+4 \text{Li}_4\left(\frac{1}{2}\right) s_{-1}-\frac{1}{2} s_{-3} \zeta _2+\frac{5}{8} s_{-2} \zeta _3-\frac{21 \zeta _2 \zeta _3}{16}+\frac{403 \zeta _5}{64}+\frac{1}{8} \zeta _3 \bar{s}_{-2}+\frac{6}{5} \zeta _2^2 \bar{s}_{-1}-2 \text{Li}_4\left(\frac{1}{2}\right) \bar{s}_{-1}-\frac{1}{8} \zeta _3 \bar{s}_2-\frac{1}{2} \zeta _2 s_{-2,-1}-\frac{1}{8} \zeta _3 s_{-1,-1}-\frac{3}{4} \zeta _3 s_{-1,1}+\frac{1}{2} \zeta _2 s_{-1,2}-s_{4,-1}+\frac{1}{2} \zeta _2 \bar{s}_{-1,-2}+\frac{1}{8} \zeta _3 \bar{s}_{-1,-1}+\frac{3}{2} \zeta _3 \bar{s}_{-1,1}-\frac{1}{2} \zeta _2 \bar{s}_{-1,2}+s_{-2,-2,-1}+s_{-1,-3,-1}+\frac{1}{2} \zeta _2 s_{-1,1,-1}+s_{3,1,-1}-\zeta _2 \bar{s}_{-1,1,-1}-s_{-2,-1,1,-1}-s_{-1,-2,1,-1}-s_{-1,1,-2,-1}+\bar{s}_{-1,-2,-1,1}+\bar{s}_{-1,1,2,1}+\bar{s}_{2,1,-1,1}+s_{-1,1,-1,1,-1}-2 \bar{s}_{-1,-1,1,-1,1}-2 \bar{s}_{-1,1,-1,-1,1} \end{dmath}
  \begin{dmath}[style={\small}]     s_{-1} \bar{s}_{-1,1,1,1}  =   -\frac{7 \ln _2^5}{120}+\frac{1}{24} s_{-1} \ln _2^4+\frac{1}{6} s_{-2} \ln _2^3-\frac{1}{6} s_2 \ln _2^3+\frac{1}{6} \zeta _2 \ln _2^3+\frac{1}{6} s_{-1,-1} \ln _2^3-\frac{1}{6} s_{-1,1} \ln _2^3-\frac{1}{6} \bar{s}_{-1,-1} \ln _2^3+\frac{1}{6} \bar{s}_{-1,1} \ln _2^3+\frac{1}{2} s_{-3} \ln _2^2-\frac{1}{2} s_3 \ln _2^2-\frac{1}{4} s_{-1} \zeta _2 \ln _2^2-\frac{7}{16} \zeta _3 \ln _2^2+\frac{1}{2} s_{-2,-1} \ln _2^2-\frac{1}{2} s_{-2,1} \ln _2^2+\frac{1}{2} s_{-1,-2} \ln _2^2-\frac{1}{2} s_{-1,2} \ln _2^2-\frac{1}{2} s_{-1,1,-1} \ln _2^2+\frac{1}{2} s_{-1,1,1} \ln _2^2+\frac{1}{2} \bar{s}_{-1,1,-1} \ln _2^2-\frac{1}{2} \bar{s}_{-1,1,1} \ln _2^2+\frac{3}{20} \zeta _2^2 \ln _2-3 \text{Li}_4\left(\frac{1}{2}\right) \ln _2+s_{-4} \ln _2-s_4 \ln _2-\frac{1}{2} s_{-2} \zeta _2 \ln _2+\frac{1}{2} s_2 \zeta _2 \ln _2+\frac{7}{8} s_{-1} \zeta _3 \ln _2+s_{-3,-1} \ln _2-s_{-3,1} \ln _2+s_{-2,-2} \ln _2-s_{-2,2} \ln _2+s_{-1,-3} \ln _2-\frac{1}{2} \zeta _2 s_{-1,-1} \ln _2+\frac{1}{2} \zeta _2 s_{-1,1} \ln _2-s_{-1,3} \ln _2+\frac{1}{2} \zeta _2 \bar{s}_{-1,-1} \ln _2-\frac{1}{2} \zeta _2 \bar{s}_{-1,1} \ln _2-s_{-2,1,-1} \ln _2+s_{-2,1,1} \ln _2-s_{-1,1,-2} \ln _2+s_{-1,1,2} \ln _2-s_{-1,2,-1} \ln _2+s_{-1,2,1} \ln _2+s_{-1,1,1,-1} \ln _2-s_{-1,1,1,1} \ln _2-\bar{s}_{-1,1,1,-1} \ln _2-\bar{s}_{-1,1,1,1} \ln _2-\frac{1}{8} s_{-1} \zeta _2^2+2 \text{Li}_5\left(\frac{1}{2}\right)-\text{Li}_4\left(\frac{1}{2}\right) s_{-1}+\frac{s_3 \zeta _2}{2}+\frac{1}{4} s_{-2} \zeta _3-\frac{7 s_2 \zeta _3}{8}-\frac{7 \zeta _2 \zeta _3}{16}+\frac{13 \zeta _5}{16}-\frac{2}{5} \zeta _2^2 \bar{s}_{-1}+s_{-4,-1}-\frac{1}{2} \zeta _2 s_{-2,-1}-\frac{1}{2} \zeta _2 s_{-1,-2}+\frac{7}{8} \zeta _3 s_{-1,-1}-\frac{1}{4} \zeta _3 s_{-1,1}-\frac{7}{8} \zeta _3 \bar{s}_{-1,-1}-\frac{1}{8} \zeta _3 \bar{s}_{-1,1}-s_{-3,1,-1}-s_{-2,2,-1}+\frac{1}{2} \zeta _2 s_{-1,1,-1}-s_{-1,3,-1}-\frac{1}{2} \zeta _2 \bar{s}_{-1,1,-1}-\frac{1}{2} \zeta _2 \bar{s}_{-1,1,1}+s_{-2,1,1,-1}+s_{-1,1,2,-1}+s_{-1,2,1,-1}+\bar{s}_{-1,-2,1,1}+\bar{s}_{-1,1,-2,1}+\bar{s}_{2,1,1,1}-s_{-1,1,1,1,-1}-2 \bar{s}_{-1,-1,1,1,1}-\bar{s}_{-1,1,-1,1,1}-\bar{s}_{-1,1,1,-1,1} \end{dmath}
  \begin{dmath}[style={\small}]     s_{-1} \bar{s}_{1,-3}  =   -\frac{3}{4} \zeta _3 \bar{s}_{1,-1}-\frac{1}{2} \zeta _2 \bar{s}_{1,2}+\bar{s}_{-2,-3}-\bar{s}_{-1,1,-3}-\bar{s}_{1,-1,-3}-\ln _2 \bar{s}_{1,-3}-\ln _2 \bar{s}_{1,3}-2 \text{Li}_4\left(\frac{1}{2}\right) \bar{s}_{-1}+2 \text{Li}_4\left(\frac{1}{2}\right) \bar{s}_1+\frac{3}{4} \zeta _3 \bar{s}_{-2}+\frac{3}{4} \zeta _2^2 \bar{s}_{-1}+\frac{3}{10} \zeta _2^2 \bar{s}_1-\frac{3}{4} \zeta _3 \bar{s}_2+\frac{1}{2} \zeta _2 \ln _2^2 \bar{s}_{-1}-\frac{1}{2} \zeta _2 \ln _2^2 \bar{s}_1-\zeta _3 \ln _2 \bar{s}_{-1}+\zeta _3 \ln _2 \bar{s}_1-\frac{1}{12} \ln _2^4 \bar{s}_{-1}+\frac{1}{12} \ln _2^4 \bar{s}_1-\frac{3}{4} \zeta _3 s_{1,-1}-\frac{1}{2} \zeta _2 s_{1,2}+s_{-4,-1}-s_{1,-3,-1}-\ln _2 s_{1,-3}+\ln _2 s_{1,3}+\frac{5 \zeta _2 \zeta _3}{8}-\frac{5 \zeta _5}{32}+\zeta _2 \ln _2^3-\frac{11}{4} \zeta _3 \ln _2^2+\frac{3}{2} \zeta _2^2 \ln _2-2 \text{Li}_4\left(\frac{1}{2}\right) s_{-1}-2 \text{Li}_4\left(\frac{1}{2}\right) s_1-4 \text{Li}_4\left(\frac{1}{2}\right) \ln _2+\frac{3}{4} \zeta _2^2 s_{-1}+\frac{11}{10} \zeta _2^2 s_1+\frac{\zeta _2 s_3}{2}-\frac{3 \zeta _3 s_2}{4}+\frac{1}{2} \zeta _2 s_{-1} \ln _2^2+\frac{1}{2} \zeta _2 s_1 \ln _2^2-\zeta _3 s_{-1} \ln _2-\frac{5}{2} \zeta _3 s_1 \ln _2-\frac{1}{12} s_{-1} \ln _2^4-\frac{1}{12} s_1 \ln _2^4+s_{-4} \ln _2-s_4 \ln _2-\frac{\ln _2^5}{6} \end{dmath}
  \begin{dmath}[style={\small}]     s_{-1} \bar{s}_{1,3}  =   -\frac{1}{2} \zeta _2 \bar{s}_{1,-2}-\frac{3}{4} \zeta _3 \bar{s}_{1,-1}+\bar{s}_{-2,3}-\bar{s}_{-1,1,3}-\bar{s}_{1,-1,3}-\ln _2 \bar{s}_{1,-3}-\ln _2 \bar{s}_{1,3}-\frac{19}{40} \zeta _2^2 \bar{s}_{-1}-\frac{29}{40} \zeta _2^2 \bar{s}_1+\frac{3}{4} \zeta _3 \ln _2 \bar{s}_{-1}-\frac{3}{4} \zeta _3 \ln _2 \bar{s}_1+\frac{1}{2} \zeta _2 s_{1,-2}-\frac{3}{4} \zeta _3 s_{1,-1}-s_{4,-1}+s_{1,3,-1}-\ln _2 s_{1,-3}+\ln _2 s_{1,3}-\frac{7 \zeta _2 \zeta _3}{8}+\frac{47 \zeta _5}{8}+\frac{2}{3} \zeta _2 \ln _2^3-\zeta _3 \ln _2^2-\frac{19}{20} \zeta _2^2 \ln _2-4 \text{Li}_5\left(\frac{1}{2}\right)-4 \text{Li}_4\left(\frac{1}{2}\right) \ln _2-\frac{19}{40} \zeta _2^2 s_{-1}+\frac{1}{8} \zeta _2^2 s_1-\frac{1}{2} \zeta _2 s_{-3}+\frac{3}{4} \zeta _3 s_{-2}+\frac{3}{4} \zeta _3 s_{-1} \ln _2-\frac{3}{4} \zeta _3 s_1 \ln _2+s_{-4} \ln _2-s_4 \ln _2-\frac{2 \ln _2^5}{15} \end{dmath}
  \begin{dmath}[style={\small}]     s_{-1} \bar{s}_{1,-2,-1}  =   \frac{1}{2} \zeta _2 \bar{s}_{1,-2}+\frac{1}{4} \zeta _3 \bar{s}_{1,-1}+\bar{s}_{-2,-2,-1}+\bar{s}_{1,3,-1}-\bar{s}_{-1,1,-2,-1}-\bar{s}_{1,-2,-1,-1}-\bar{s}_{1,-1,-2,-1}-\ln _2^2 \bar{s}_{1,-2}+\ln _2^2 \bar{s}_{1,2}-\ln _2 \bar{s}_{1,-3}+\ln _2 \bar{s}_{1,3}-2 \ln _2 \bar{s}_{1,-2,-1}+4 \text{Li}_4\left(\frac{1}{2}\right) \bar{s}_{-1}-\frac{7}{5} \zeta _2^2 \bar{s}_{-1}+\frac{1}{8} \zeta _2^2 \bar{s}_1-\zeta _2 \ln _2^2 \bar{s}_{-1}-\frac{1}{2} \zeta _2 \ln _2 \bar{s}_{-2}+\frac{19}{8} \zeta _3 \ln _2 \bar{s}_{-1}-\frac{19}{8} \zeta _3 \ln _2 \bar{s}_1+\frac{1}{2} \zeta _2 \ln _2 \bar{s}_2+\frac{1}{6} \ln _2^4 \bar{s}_{-1}+\frac{1}{2} \zeta _2 s_{1,-2}+\frac{1}{4} \zeta _3 s_{1,-1}-s_{4,-1}+s_{-3,-1,-1}+s_{1,3,-1}-s_{1,-2,-1,-1}-\ln _2^2 s_{1,-2}+\ln _2^2 s_{1,2}+2 \ln _2 s_{-3,-1}-\ln _2 s_{1,-3}+\ln _2 s_{1,3}-2 \ln _2 s_{1,-2,-1}+\frac{\zeta _2 \zeta _3}{16}+\frac{369 \zeta _5}{64}-\frac{3}{2} \zeta _2 \ln _2^3+\frac{33}{8} \zeta _3 \ln _2^2-\frac{157}{40} \zeta _2^2 \ln _2+4 \text{Li}_4\left(\frac{1}{2}\right) s_{-1}-4 \text{Li}_4\left(\frac{1}{2}\right) s_1-6 \text{Li}_5\left(\frac{1}{2}\right)+4 \text{Li}_4\left(\frac{1}{2}\right) \ln _2-\frac{7}{5} \zeta _2^2 s_{-1}+\frac{61}{40} \zeta _2^2 s_1-\frac{1}{2} \zeta _2 s_{-3}-\frac{1}{4} \zeta _3 s_{-2}-\zeta _2 s_{-1} \ln _2^2+\zeta _2 s_1 \ln _2^2+\frac{1}{2} \zeta _2 s_{-2} \ln _2+\frac{1}{2} \zeta _2 s_2 \ln _2+\frac{19}{8} \zeta _3 s_{-1} \ln _2-\frac{29}{8} \zeta _3 s_1 \ln _2+\frac{1}{6} s_{-1} \ln _2^4-\frac{1}{6} s_1 \ln _2^4+s_{-3} \ln _2^2-s_3 \ln _2^2+s_{-4} \ln _2-s_4 \ln _2+\frac{13 \ln _2^5}{60} \end{dmath}
  \begin{dmath}[style={\small}]     s_{-1} \bar{s}_{1,-2,1}  =   -\frac{1}{2} \zeta _2 \bar{s}_{1,-2}-\frac{5}{8} \zeta _3 \bar{s}_{1,-1}-\frac{1}{2} \zeta _2 \bar{s}_{1,2}+\bar{s}_{-2,-2,1}+\bar{s}_{1,3,1}-\bar{s}_{-1,1,-2,1}-\bar{s}_{1,-2,-1,1}-\bar{s}_{1,-1,-2,1}-\frac{1}{2} \ln _2^2 \bar{s}_{1,-2}+\frac{1}{2} \ln _2^2 \bar{s}_{1,2}-\ln _2 \bar{s}_{1,-2,-1}-\ln _2 \bar{s}_{1,-2,1}+\frac{5}{8} \zeta _3 \bar{s}_{-2}-\frac{3}{40} \zeta _2^2 \bar{s}_{-1}+\frac{1}{2} \zeta _2^2 \bar{s}_1-\frac{5}{8} \zeta _3 \bar{s}_2+\frac{5}{8} \zeta _3 \ln _2 \bar{s}_{-1}-\frac{5}{8} \zeta _3 \ln _2 \bar{s}_1-\frac{5}{8} \zeta _3 s_{1,-1}-\frac{1}{2} \zeta _2 s_{1,2}+s_{-4,-1}-s_{-3,1,-1}-s_{1,-3,-1}+s_{1,-2,1,-1}-\frac{1}{2} \ln _2^2 s_{1,-2}+\frac{1}{2} \ln _2^2 s_{1,2}+\ln _2 s_{-3,-1}-\ln _2 s_{-3,1}-\ln _2 s_{1,-3}+\ln _2 s_{1,3}-\ln _2 s_{1,-2,-1}+\ln _2 s_{1,-2,1}-\frac{5 \zeta _2 \zeta _3}{4}-\frac{13 \zeta _5}{32}-\frac{2}{3} \zeta _2 \ln _2^3+\frac{19}{8} \zeta _3 \ln _2^2-\frac{3}{20} \zeta _2^2 \ln _2-2 \text{Li}_4\left(\frac{1}{2}\right) s_1+4 \text{Li}_5\left(\frac{1}{2}\right)+4 \text{Li}_4\left(\frac{1}{2}\right) \ln _2-\frac{3}{40} \zeta _2^2 s_{-1}+\frac{41}{40} \zeta _2^2 s_1+\frac{\zeta _2 s_3}{2}-\frac{5 \zeta _3 s_2}{8}+\frac{1}{2} \zeta _2 s_1 \ln _2^2+\frac{5}{8} \zeta _3 s_{-1} \ln _2-\frac{19}{8} \zeta _3 s_1 \ln _2-\frac{1}{12} s_1 \ln _2^4+\frac{1}{2} s_{-3} \ln _2^2-\frac{1}{2} s_3 \ln _2^2+s_{-4} \ln _2-s_4 \ln _2+\frac{2 \ln _2^5}{15} \end{dmath}
  \begin{dmath}[style={\small}]     s_{-1} \bar{s}_{1,-1,-2}  =   -\frac{1}{2} \zeta _2 \bar{s}_{1,-2}+\frac{3}{2} \zeta _3 \bar{s}_{1,-1}+\frac{1}{2} \zeta _2 \bar{s}_{1,2}-\frac{1}{2} \zeta _2 \bar{s}_{1,-1,-1}+\bar{s}_{-2,-1,-2}+\bar{s}_{1,2,-2}-\bar{s}_{-1,1,-1,-2}-2 \bar{s}_{1,-1,-1,-2}-\ln _2 \bar{s}_{1,-1,-2}-\ln _2 \bar{s}_{1,-1,2}-4 \text{Li}_4\left(\frac{1}{2}\right) \bar{s}_1-\frac{7}{20} \zeta _2^2 \bar{s}_{-1}+\frac{17}{40} \zeta _2^2 \bar{s}_1-\frac{1}{4} \zeta _2 \ln _2^2 \bar{s}_{-1}+\frac{5}{4} \zeta _2 \ln _2^2 \bar{s}_1-\frac{1}{2} \zeta _2 \ln _2 \bar{s}_{-2}+\zeta _3 \ln _2 \bar{s}_{-1}-\zeta _3 \ln _2 \bar{s}_1+\frac{1}{2} \zeta _2 \ln _2 \bar{s}_2-\frac{1}{6} \ln _2^4 \bar{s}_1-\frac{1}{2} \zeta _2 s_{-2,-1}+\frac{1}{2} \zeta _2 s_{1,-2}-\frac{3}{4} \zeta _3 s_{1,-1}+\frac{1}{2} \zeta _2 s_{1,-1,-1}+\zeta _2 \ln _2 s_{1,-1}-s_{4,-1}+s_{-2,-2,-1}+s_{1,3,-1}-s_{1,-1,-2,-1}+\ln _2 s_{-2,-2}-\ln _2 s_{-2,2}-\ln _2 s_{1,-3}+\ln _2 s_{1,3}-\ln _2 s_{1,-1,-2}+\ln _2 s_{1,-1,2}-\zeta _2 \zeta _3+\frac{1061 \zeta _5}{64}+\frac{11}{6} \zeta _2 \ln _2^3-\frac{13}{8} \zeta _3 \ln _2^2-\frac{107}{40} \zeta _2^2 \ln _2+4 \text{Li}_4\left(\frac{1}{2}\right) s_1-14 \text{Li}_5\left(\frac{1}{2}\right)-12 \text{Li}_4\left(\frac{1}{2}\right) \ln _2-\frac{7}{20} \zeta _2^2 s_{-1}-\frac{61}{40} \zeta _2^2 s_1-\frac{1}{2} \zeta _2 s_{-3}+\frac{3}{4} \zeta _3 s_{-2}-\frac{1}{4} \zeta _2 s_{-1} \ln _2^2-\frac{1}{4} \zeta _2 s_1 \ln _2^2-\frac{1}{2} \zeta _2 s_{-2} \ln _2+\frac{1}{2} \zeta _2 s_2 \ln _2+\zeta _3 s_{-1} \ln _2+\frac{5}{2} \zeta _3 s_1 \ln _2+\frac{1}{6} s_1 \ln _2^4+s_{-4} \ln _2-s_4 \ln _2-\frac{23 \ln _2^5}{60} \end{dmath}
  \begin{dmath}[style={\small}]     s_{-1} \bar{s}_{1,-1,2}  =   -2 \zeta _3 \bar{s}_{1,-1}-\frac{1}{2} \zeta _2 \bar{s}_{1,-1,-1}+\bar{s}_{-2,-1,2}+\bar{s}_{1,2,2}-\bar{s}_{-1,1,-1,2}-2 \bar{s}_{1,-1,-1,2}-\ln _2 \bar{s}_{1,-1,-2}-\ln _2 \bar{s}_{1,-1,2}-2 \text{Li}_4\left(\frac{1}{2}\right) \bar{s}_{-1}+6 \text{Li}_4\left(\frac{1}{2}\right) \bar{s}_1+\zeta _3 \bar{s}_{-2}+\frac{7}{8} \zeta _2^2 \bar{s}_{-1}-\frac{6}{5} \zeta _2^2 \bar{s}_1-\zeta _3 \bar{s}_2+\frac{1}{4} \zeta _2 \ln _2^2 \bar{s}_{-1}-\frac{5}{4} \zeta _2 \ln _2^2 \bar{s}_1-\frac{1}{2} \zeta _2 \ln _2 \bar{s}_{-2}-\frac{13}{8} \zeta _3 \ln _2 \bar{s}_{-1}+\frac{13}{8} \zeta _3 \ln _2 \bar{s}_1+\frac{1}{2} \zeta _2 \ln _2 \bar{s}_2-\frac{1}{12} \ln _2^4 \bar{s}_{-1}+\frac{1}{4} \ln _2^4 \bar{s}_1-\frac{1}{2} \zeta _2 s_{-2,-1}-\frac{5}{4} \zeta _3 s_{1,-1}-\frac{1}{2} \zeta _2 s_{1,2}+\frac{1}{2} \zeta _2 s_{1,-1,-1}+\zeta _2 \ln _2 s_{1,-1}+s_{-4,-1}-s_{-2,2,-1}-s_{1,-3,-1}+s_{1,-1,2,-1}+\ln _2 s_{-2,-2}-\ln _2 s_{-2,2}-\ln _2 s_{1,-3}+\ln _2 s_{1,3}-\ln _2 s_{1,-1,-2}+\ln _2 s_{1,-1,2}+\frac{29 \zeta _2 \zeta _3}{16}+\frac{535 \zeta _5}{32}+\frac{4}{3} \zeta _2 \ln _2^3-\frac{17}{4} \zeta _3 \ln _2^2-\frac{37}{20} \zeta _2^2 \ln _2-2 \text{Li}_4\left(\frac{1}{2}\right) s_{-1}+2 \text{Li}_4\left(\frac{1}{2}\right) s_1-20 \text{Li}_5\left(\frac{1}{2}\right)-12 \text{Li}_4\left(\frac{1}{2}\right) \ln _2+\frac{7}{8} \zeta _2^2 s_{-1}-\frac{2}{5} \zeta _2^2 s_1+\frac{\zeta _2 s_3}{2}+\frac{1}{4} \zeta _3 s_{-2}-\zeta _3 s_2+\frac{1}{4} \zeta _2 s_{-1} \ln _2^2+\frac{1}{4} \zeta _2 s_1 \ln _2^2-\frac{1}{2} \zeta _2 s_{-2} \ln _2+\frac{1}{2} \zeta _2 s_2 \ln _2-\frac{13}{8} \zeta _3 s_{-1} \ln _2-\frac{1}{8} \zeta _3 s_1 \ln _2-\frac{1}{12} s_{-1} \ln _2^4+\frac{1}{12} s_1 \ln _2^4+s_{-4} \ln _2-s_4 \ln _2-\frac{\ln _2^5}{3} \end{dmath}
  \begin{dmath}[style={\small}]     s_{-1} \bar{s}_{1,1,-2}  =   \frac{1}{2} \zeta _2 \bar{s}_{1,-2}-\frac{1}{8} \zeta _3 \bar{s}_{1,-1}+\frac{13}{8} \zeta _3 \bar{s}_{1,1}-\frac{1}{2} \zeta _2 \bar{s}_{1,2}-\frac{1}{2} \zeta _2 \bar{s}_{1,1,-1}+\frac{1}{2} \zeta _2 \ln _2 \bar{s}_{1,-1}-\frac{1}{2} \zeta _2 \ln _2 \bar{s}_{1,1}+\bar{s}_{-2,1,-2}+\bar{s}_{1,-2,-2}-\bar{s}_{-1,1,1,-2}-\bar{s}_{1,-1,1,-2}-\bar{s}_{1,1,-1,-2}-\ln _2 \bar{s}_{1,1,-2}-\ln _2 \bar{s}_{1,1,2}+\text{Li}_4\left(\frac{1}{2}\right) \bar{s}_{-1}-\text{Li}_4\left(\frac{1}{2}\right) \bar{s}_1+\frac{1}{8} \zeta _3 \bar{s}_{-2}-\frac{3}{20} \zeta _2^2 \bar{s}_{-1}+\frac{31}{40} \zeta _2^2 \bar{s}_1-\frac{1}{8} \zeta _3 \bar{s}_2-\frac{1}{2} \zeta _2 \ln _2^2 \bar{s}_{-1}+\frac{1}{2} \zeta _2 \ln _2^2 \bar{s}_1+\zeta _3 \ln _2 \bar{s}_{-1}-\zeta _3 \ln _2 \bar{s}_1+\frac{1}{24} \ln _2^4 \bar{s}_{-1}-\frac{1}{24} \ln _2^4 \bar{s}_1-\frac{1}{8} \zeta _3 s_{1,-1}+\frac{5}{8} \zeta _3 s_{1,1}-\frac{1}{2} \zeta _2 s_{1,2}+\frac{1}{2} \zeta _2 s_{2,-1}-\frac{1}{2} \zeta _2 s_{1,1,-1}+\frac{1}{2} \zeta _2 \ln _2 s_{1,-1}-\frac{1}{2} \zeta _2 \ln _2 s_{1,1}+s_{-4,-1}-s_{1,-3,-1}-s_{2,-2,-1}+s_{1,1,-2,-1}-\ln _2 s_{1,-3}+\ln _2 s_{1,3}-\ln _2 s_{2,-2}+\ln _2 s_{2,2}+\ln _2 s_{1,1,-2}-\ln _2 s_{1,1,2}-\frac{19 \zeta _2 \zeta _3}{8}+\frac{295 \zeta _5}{64}-\frac{2}{3} \zeta _2 \ln _2^3+\frac{15}{8} \zeta _3 \ln _2^2-\frac{3}{10} \zeta _2^2 \ln _2+\text{Li}_4\left(\frac{1}{2}\right) s_{-1}-3 \text{Li}_4\left(\frac{1}{2}\right) s_1+2 \text{Li}_4\left(\frac{1}{2}\right) \ln _2-\frac{3}{20} \zeta _2^2 s_{-1}+\frac{49}{40} \zeta _2^2 s_1+\frac{\zeta _2 s_3}{2}-\frac{3 \zeta _3 s_2}{4}-\frac{1}{2} \zeta _2 s_{-1} \ln _2^2+\zeta _2 s_1 \ln _2^2-\frac{1}{2} \zeta _2 s_{-2} \ln _2+\frac{1}{2} \zeta _2 s_2 \ln _2+\zeta _3 s_{-1} \ln _2-\frac{11}{4} \zeta _3 s_1 \ln _2+\frac{1}{24} s_{-1} \ln _2^4-\frac{1}{8} s_1 \ln _2^4+s_{-4} \ln _2-s_4 \ln _2+\frac{\ln _2^5}{12} \end{dmath}
  \begin{dmath}[style={\small}]     s_{-1} \bar{s}_{1,1,2}  =   -\zeta _3 \bar{s}_{1,-1}-\zeta _3 \bar{s}_{1,1}-\frac{1}{2} \zeta _2 \bar{s}_{1,1,-1}+\frac{1}{2} \zeta _2 \ln _2 \bar{s}_{1,-1}-\frac{1}{2} \zeta _2 \ln _2 \bar{s}_{1,1}+\bar{s}_{-2,1,2}+\bar{s}_{1,-2,2}-\bar{s}_{-1,1,1,2}-\bar{s}_{1,-1,1,2}-\bar{s}_{1,1,-1,2}-\ln _2 \bar{s}_{1,1,-2}-\ln _2 \bar{s}_{1,1,2}-\text{Li}_4\left(\frac{1}{2}\right) \bar{s}_{-1}+\text{Li}_4\left(\frac{1}{2}\right) \bar{s}_1-\frac{1}{20} \zeta _2^2 \bar{s}_{-1}-\frac{9}{20} \zeta _2^2 \bar{s}_1+\frac{1}{8} \zeta _3 \ln _2 \bar{s}_{-1}-\frac{1}{8} \zeta _3 \ln _2 \bar{s}_1-\frac{1}{24} \ln _2^4 \bar{s}_{-1}+\frac{1}{24} \ln _2^4 \bar{s}_1+\frac{1}{2} \zeta _2 s_{1,-2}-\zeta _3 s_{1,-1}+\frac{1}{4} \zeta _3 s_{1,1}+\frac{1}{2} \zeta _2 s_{2,-1}-\frac{1}{2} \zeta _2 s_{1,1,-1}+\frac{1}{2} \zeta _2 \ln _2 s_{1,-1}-\frac{1}{2} \zeta _2 \ln _2 s_{1,1}-s_{4,-1}+s_{1,3,-1}+s_{2,2,-1}-s_{1,1,2,-1}-\ln _2 s_{1,-3}+\ln _2 s_{1,3}-\ln _2 s_{2,-2}+\ln _2 s_{2,2}+\ln _2 s_{1,1,-2}-\ln _2 s_{1,1,2}+\frac{\zeta _2 \zeta _3}{16}-\frac{241 \zeta _5}{64}-\frac{1}{3} \zeta _2 \ln _2^3+\zeta _3 \ln _2^2-\frac{1}{10} \zeta _2^2 \ln _2-\text{Li}_4\left(\frac{1}{2}\right) s_{-1}-3 \text{Li}_4\left(\frac{1}{2}\right) s_1+4 \text{Li}_5\left(\frac{1}{2}\right)+2 \text{Li}_4\left(\frac{1}{2}\right) \ln _2-\frac{1}{20} \zeta _2^2 s_{-1}+\frac{53}{40} \zeta _2^2 s_1-\frac{1}{2} \zeta _2 s_{-3}+\zeta _3 s_{-2}-\frac{\zeta _3 s_2}{4}+\zeta _2 s_1 \ln _2^2-\frac{1}{2} \zeta _2 s_{-2} \ln _2+\frac{1}{2} \zeta _2 s_2 \ln _2+\frac{1}{8} \zeta _3 s_{-1} \ln _2-\frac{29}{8} \zeta _3 s_1 \ln _2-\frac{1}{24} s_{-1} \ln _2^4-\frac{1}{8} s_1 \ln _2^4+s_{-4} \ln _2-s_4 \ln _2+\frac{\ln _2^5}{20} \end{dmath}
  \begin{dmath}[style={\small}]     s_{-1} \bar{s}_{1,2,-1}  =   \frac{1}{4} \zeta _3 \bar{s}_{1,-1}+\frac{1}{2} \zeta _2 \bar{s}_{1,2}+\bar{s}_{-2,2,-1}+\bar{s}_{1,-3,-1}-\bar{s}_{-1,1,2,-1}-\bar{s}_{1,-1,2,-1}-\bar{s}_{1,2,-1,-1}+\ln _2^2 \bar{s}_{1,-2}-\ln _2^2 \bar{s}_{1,2}+\ln _2 \bar{s}_{1,-3}-\ln _2 \bar{s}_{1,3}-2 \ln _2 \bar{s}_{1,2,-1}-3 \text{Li}_4\left(\frac{1}{2}\right) \bar{s}_{-1}-\text{Li}_4\left(\frac{1}{2}\right) \bar{s}_1-\frac{1}{4} \zeta _3 \bar{s}_{-2}+\frac{6}{5} \zeta _2^2 \bar{s}_{-1}+\frac{7}{40} \zeta _2^2 \bar{s}_1+\frac{1}{4} \zeta _3 \bar{s}_2+\frac{3}{4} \zeta _2 \ln _2^2 \bar{s}_{-1}+\frac{1}{4} \zeta _2 \ln _2^2 \bar{s}_1+\frac{1}{2} \zeta _2 \ln _2 \bar{s}_{-2}-2 \zeta _3 \ln _2 \bar{s}_{-1}+2 \zeta _3 \ln _2 \bar{s}_1-\frac{1}{2} \zeta _2 \ln _2 \bar{s}_2-\frac{1}{8} \ln _2^4 \bar{s}_{-1}-\frac{1}{24} \ln _2^4 \bar{s}_1+\frac{1}{4} \zeta _3 s_{1,-1}-\frac{1}{2} \zeta _2 s_{1,2}+s_{-4,-1}-s_{1,-3,-1}-s_{3,-1,-1}+s_{1,2,-1,-1}-\ln _2^2 s_{1,-2}+\ln _2^2 s_{1,2}-\ln _2 s_{1,-3}+\ln _2 s_{1,3}-2 \ln _2 s_{3,-1}+2 \ln _2 s_{1,2,-1}+\frac{9 \zeta _2 \zeta _3}{8}-\frac{677 \zeta _5}{64}+\frac{2}{3} \zeta _2 \ln _2^3-2 \zeta _3 \ln _2^2+\frac{141}{40} \zeta _2^2 \ln _2-3 \text{Li}_4\left(\frac{1}{2}\right) s_{-1}-3 \text{Li}_4\left(\frac{1}{2}\right) s_1+8 \text{Li}_5\left(\frac{1}{2}\right)+\frac{6}{5} \zeta _2^2 s_{-1}+\frac{43}{40} \zeta _2^2 s_1+\frac{\zeta _2 s_3}{2}+\frac{\zeta _3 s_2}{4}+\frac{3}{4} \zeta _2 s_{-1} \ln _2^2+\frac{3}{4} \zeta _2 s_1 \ln _2^2-\frac{1}{2} \zeta _2 s_{-2} \ln _2-\frac{1}{2} \zeta _2 s_2 \ln _2-2 \zeta _3 s_{-1} \ln _2-2 \zeta _3 s_1 \ln _2-\frac{1}{8} s_{-1} \ln _2^4-\frac{1}{8} s_1 \ln _2^4+s_{-3} \ln _2^2-s_3 \ln _2^2+s_{-4} \ln _2-s_4 \ln _2-\frac{\ln _2^5}{15} \end{dmath}
  \begin{dmath}[style={\small}]     s_{-1} \bar{s}_{1,2,1}  =   -\frac{1}{2} \zeta _2 \bar{s}_{1,-2}-\frac{5}{8} \zeta _3 \bar{s}_{1,-1}-\frac{1}{2} \zeta _2 \bar{s}_{1,2}+\bar{s}_{-2,2,1}+\bar{s}_{1,-3,1}-\bar{s}_{-1,1,2,1}-\bar{s}_{1,-1,2,1}-\bar{s}_{1,2,-1,1}+\frac{1}{2} \ln _2^2 \bar{s}_{1,-2}-\frac{1}{2} \ln _2^2 \bar{s}_{1,2}-\ln _2 \bar{s}_{1,2,-1}-\ln _2 \bar{s}_{1,2,1}-\text{Li}_4\left(\frac{1}{2}\right) \bar{s}_{-1}+\text{Li}_4\left(\frac{1}{2}\right) \bar{s}_1-\frac{1}{40} \zeta _2^2 \bar{s}_{-1}-\frac{27}{40} \zeta _2^2 \bar{s}_1+\frac{1}{4} \zeta _2 \ln _2^2 \bar{s}_{-1}-\frac{1}{4} \zeta _2 \ln _2^2 \bar{s}_1-\frac{1}{4} \zeta _3 \ln _2 \bar{s}_{-1}+\frac{1}{4} \zeta _3 \ln _2 \bar{s}_1-\frac{1}{24} \ln _2^4 \bar{s}_{-1}+\frac{1}{24} \ln _2^4 \bar{s}_1+\frac{1}{2} \zeta _2 s_{1,-2}-\frac{5}{8} \zeta _3 s_{1,-1}-s_{4,-1}+s_{1,3,-1}+s_{3,1,-1}-s_{1,2,1,-1}-\frac{1}{2} \ln _2^2 s_{1,-2}+\frac{1}{2} \ln _2^2 s_{1,2}-\ln _2 s_{1,-3}+\ln _2 s_{1,3}-\ln _2 s_{3,-1}+\ln _2 s_{3,1}+\ln _2 s_{1,2,-1}-\ln _2 s_{1,2,1}-\frac{15 \zeta _2 \zeta _3}{8}+\frac{245 \zeta _5}{64}+\frac{1}{2} \zeta _2 \ln _2^3-\frac{9}{8} \zeta _3 \ln _2^2-\frac{1}{20} \zeta _2^2 \ln _2-\text{Li}_4\left(\frac{1}{2}\right) s_{-1}+3 \text{Li}_4\left(\frac{1}{2}\right) s_1-2 \text{Li}_4\left(\frac{1}{2}\right) \ln _2-\frac{1}{40} \zeta _2^2 s_{-1}-\frac{43}{40} \zeta _2^2 s_1-\frac{1}{2} \zeta _2 s_{-3}+\frac{5}{8} \zeta _3 s_{-2}+\frac{1}{4} \zeta _2 s_{-1} \ln _2^2-\frac{3}{4} \zeta _2 s_1 \ln _2^2-\frac{1}{4} \zeta _3 s_{-1} \ln _2+2 \zeta _3 s_1 \ln _2-\frac{1}{24} s_{-1} \ln _2^4+\frac{1}{8} s_1 \ln _2^4+\frac{1}{2} s_{-3} \ln _2^2-\frac{1}{2} s_3 \ln _2^2+s_{-4} \ln _2-s_4 \ln _2-\frac{\ln _2^5}{12} \end{dmath}
  \begin{dmath}[style={\small}]     s_{-1} \bar{s}_{1,-1,-1,-1}  =   -\frac{43 \ln _2^5}{120}-\frac{1}{12} s_{-1} \ln _2^4-\frac{7}{12} s_1 \ln _2^4-\frac{1}{12} \bar{s}_{-1} \ln _2^4+\frac{1}{12} \bar{s}_1 \ln _2^4+\frac{2}{3} s_{-2} \ln _2^3-\frac{2}{3} s_2 \ln _2^3+\frac{5}{3} \zeta _2 \ln _2^3+\frac{2}{3} \bar{s}_{-2} \ln _2^3-\frac{2}{3} \bar{s}_2 \ln _2^3-\frac{4}{3} s_{1,-1} \ln _2^3+s_{-3} \ln _2^2-s_3 \ln _2^2+\frac{3}{4} s_{-1} \zeta _2 \ln _2^2+\frac{3}{4} s_1 \zeta _2 \ln _2^2-\frac{5}{2} \zeta _3 \ln _2^2+\frac{3}{4} \zeta _2 \bar{s}_{-1} \ln _2^2-\frac{3}{4} \zeta _2 \bar{s}_1 \ln _2^2+2 s_{-2,-1} \ln _2^2-s_{1,-2} \ln _2^2+s_{1,2} \ln _2^2+\bar{s}_{1,-2} \ln _2^2-\bar{s}_{1,2} \ln _2^2-2 s_{1,-1,-1} \ln _2^2+\frac{3}{20} \zeta _2^2 \ln _2-6 \text{Li}_4\left(\frac{1}{2}\right) \ln _2+s_{-4} \ln _2-s_4 \ln _2-s_{-2} \zeta _2 \ln _2-\frac{19}{8} s_{-1} \zeta _3 \ln _2-\frac{3}{8} s_1 \zeta _3 \ln _2-\frac{19}{8} \zeta _3 \bar{s}_{-1} \ln _2+\frac{19}{8} \zeta _3 \bar{s}_1 \ln _2+s_{-2,-2} \ln _2-s_{-2,2} \ln _2-s_{1,-3} \ln _2+\zeta _2 s_{1,-1} \ln _2+s_{1,3} \ln _2-2 s_{3,-1} \ln _2+2 s_{-2,-1,-1} \ln _2-s_{1,-1,-2} \ln _2+s_{1,-1,2} \ln _2+2 s_{1,2,-1} \ln _2-\bar{s}_{1,-1,-2} \ln _2+\bar{s}_{1,-1,2} \ln _2-2 s_{1,-1,-1,-1} \ln _2-2 \bar{s}_{1,-1,-1,-1} \ln _2+\frac{6}{5} s_{-1} \zeta _2^2-\frac{1}{5} s_1 \zeta _2^2-7 \text{Li}_5\left(\frac{1}{2}\right)-3 \text{Li}_4\left(\frac{1}{2}\right) s_{-1}+\text{Li}_4\left(\frac{1}{2}\right) s_1+\frac{s_3 \zeta _2}{2}+\frac{1}{2} s_{-2} \zeta _3-\frac{s_2 \zeta _3}{4}+\frac{27 \zeta _2 \zeta _3}{16}+\frac{119 \zeta _5}{32}+\frac{1}{4} \zeta _3 \bar{s}_{-2}+\frac{6}{5} \zeta _2^2 \bar{s}_{-1}-3 \text{Li}_4\left(\frac{1}{2}\right) \bar{s}_{-1}-\frac{17}{20} \zeta _2^2 \bar{s}_1+3 \text{Li}_4\left(\frac{1}{2}\right) \bar{s}_1-\frac{1}{4} \zeta _3 \bar{s}_2+s_{-4,-1}-\frac{1}{2} \zeta _2 s_{-2,-1}-\frac{3}{4} \zeta _3 s_{1,-1}-\frac{1}{2} \zeta _2 s_{1,2}-\zeta _3 \bar{s}_{1,-1}-s_{-2,2,-1}-s_{1,-3,-1}+\frac{1}{2} \zeta _2 s_{1,-1,-1}-s_{3,-1,-1}+\frac{1}{2} \zeta _2 \bar{s}_{1,-1,-1}+s_{-2,-1,-1,-1}+s_{1,-1,2,-1}+s_{1,2,-1,-1}+\bar{s}_{-2,-1,-1,-1}+\bar{s}_{1,-1,2,-1}+\bar{s}_{1,2,-1,-1}-s_{1,-1,-1,-1,-1}-\bar{s}_{-1,1,-1,-1,-1}-3 \bar{s}_{1,-1,-1,-1,-1} \end{dmath}
  \begin{dmath}[style={\small}]     s_{-1} \bar{s}_{1,-1,-1,1}  =   -\frac{5 \ln _2^5}{12}+\frac{1}{24} s_{-1} \ln _2^4-\frac{3}{8} s_1 \ln _2^4+\frac{1}{24} \bar{s}_{-1} \ln _2^4-\frac{1}{6} \bar{s}_1 \ln _2^4+\frac{1}{2} s_{-2} \ln _2^3-\frac{1}{2} s_2 \ln _2^3+\frac{5}{3} \zeta _2 \ln _2^3+\frac{1}{2} \bar{s}_{-2} \ln _2^3-\frac{1}{2} \bar{s}_2 \ln _2^3-s_{1,-1} \ln _2^3+\frac{1}{2} s_{-3} \ln _2^2-\frac{1}{2} s_3 \ln _2^2-\frac{9}{8} \zeta _3 \ln _2^2+s_{-2,-1} \ln _2^2-\frac{1}{2} s_{1,-2} \ln _2^2+\frac{1}{2} s_{1,2} \ln _2^2+\frac{1}{2} \bar{s}_{1,-2} \ln _2^2-\frac{1}{2} \bar{s}_{1,2} \ln _2^2-s_{1,-1,-1} \ln _2^2-\frac{113}{40} \zeta _2^2 \ln _2-9 \text{Li}_4\left(\frac{1}{2}\right) \ln _2+s_{-4} \ln _2-s_4 \ln _2-\frac{1}{2} s_{-2} \zeta _2 \ln _2+\frac{1}{2} s_2 \zeta _2 \ln _2+\frac{1}{4} s_{-1} \zeta _3 \ln _2+\frac{19}{8} s_1 \zeta _3 \ln _2-\frac{1}{2} \zeta _2 \bar{s}_{-2} \ln _2+\frac{1}{4} \zeta _3 \bar{s}_{-1} \ln _2-\frac{1}{4} \zeta _3 \bar{s}_1 \ln _2+\frac{1}{2} \zeta _2 \bar{s}_2 \ln _2+s_{-2,-2} \ln _2-s_{-2,2} \ln _2-s_{1,-3} \ln _2+\zeta _2 s_{1,-1} \ln _2+s_{1,3} \ln _2-s_{3,-1} \ln _2+s_{3,1} \ln _2-\frac{3}{2} \zeta _2 \bar{s}_{1,-1} \ln _2+s_{-2,-1,-1} \ln _2-s_{-2,-1,1} \ln _2-s_{1,-1,-2} \ln _2+s_{1,-1,2} \ln _2+s_{1,2,-1} \ln _2-s_{1,2,1} \ln _2-s_{1,-1,-1,-1} \ln _2+s_{1,-1,-1,1} \ln _2-\bar{s}_{1,-1,-1,-1} \ln _2-\bar{s}_{1,-1,-1,1} \ln _2-\frac{9}{40} s_{-1} \zeta _2^2-\frac{61}{40} s_1 \zeta _2^2-13 \text{Li}_5\left(\frac{1}{2}\right)+4 \text{Li}_4\left(\frac{1}{2}\right) s_1-\frac{1}{2} s_{-3} \zeta _2+\frac{5}{8} s_{-2} \zeta _3-\frac{3 \zeta _2 \zeta _3}{2}+\frac{259 \zeta _5}{16}-\frac{9}{40} \zeta _2^2 \bar{s}_{-1}+\frac{7}{20} \zeta _2^2 \bar{s}_1-3 \text{Li}_4\left(\frac{1}{2}\right) \bar{s}_1-\frac{1}{2} \zeta _2 s_{-2,-1}+\frac{1}{2} \zeta _2 s_{1,-2}-\frac{5}{8} \zeta _3 s_{1,-1}-s_{4,-1}-\frac{1}{2} \zeta _2 \bar{s}_{1,-2}+\frac{13}{8} \zeta _3 \bar{s}_{1,-1}+\frac{1}{2} \zeta _2 \bar{s}_{1,2}+s_{-2,-2,-1}+\frac{1}{2} \zeta _2 s_{1,-1,-1}+s_{1,3,-1}+s_{3,1,-1}-\zeta _2 \bar{s}_{1,-1,-1}-s_{-2,-1,1,-1}-s_{1,-1,-2,-1}-s_{1,2,1,-1}+\bar{s}_{-2,-1,-1,1}+\bar{s}_{1,-1,2,1}+\bar{s}_{1,2,-1,1}+s_{1,-1,-1,1,-1}-\bar{s}_{-1,1,-1,-1,1}-3 \bar{s}_{1,-1,-1,-1,1} \end{dmath}
  \begin{dmath}[style={\small}]     s_{-1} \bar{s}_{1,-1,1,-1}  =   \frac{\ln _2^5}{40}+\frac{1}{6} s_{-1} \ln _2^4-\frac{5}{24} s_1 \ln _2^4+\frac{1}{6} \bar{s}_{-1} \ln _2^4-\frac{1}{12} \bar{s}_1 \ln _2^4+\frac{5}{12} \zeta _2 \ln _2^3+s_{-3} \ln _2^2-s_3 \ln _2^2-\frac{3}{4} s_{-1} \zeta _2 \ln _2^2+\frac{1}{2} s_1 \zeta _2 \ln _2^2+\frac{35}{16} \zeta _3 \ln _2^2-\frac{3}{4} \zeta _2 \bar{s}_{-1} \ln _2^2+\frac{7}{4} \zeta _2 \bar{s}_1 \ln _2^2+\frac{1}{2} s_{-2,-1} \ln _2^2-\frac{3}{2} s_{-2,1} \ln _2^2-s_{1,-2} \ln _2^2+s_{1,2} \ln _2^2-\frac{1}{2} \bar{s}_{1,-2} \ln _2^2+\frac{1}{2} \bar{s}_{1,2} \ln _2^2-\frac{1}{2} s_{1,-1,-1} \ln _2^2+\frac{3}{2} s_{1,-1,1} \ln _2^2+\frac{1}{2} \bar{s}_{1,-1,-1} \ln _2^2-\frac{1}{2} \bar{s}_{1,-1,1} \ln _2^2-\frac{181}{40} \zeta _2^2 \ln _2+s_{-4} \ln _2-s_4 \ln _2+\frac{1}{2} s_{-2} \zeta _2 \ln _2+\frac{1}{2} s_2 \zeta _2 \ln _2+\frac{5}{2} s_{-1} \zeta _3 \ln _2-\frac{1}{2} s_1 \zeta _3 \ln _2-\frac{1}{2} \zeta _2 \bar{s}_{-2} \ln _2+\frac{5}{2} \zeta _3 \bar{s}_{-1} \ln _2-\frac{5}{2} \zeta _3 \bar{s}_1 \ln _2+\frac{1}{2} \zeta _2 \bar{s}_2 \ln _2+2 s_{-3,-1} \ln _2+s_{-2,-2} \ln _2-s_{-2,2} \ln _2-s_{1,-3} \ln _2+s_{1,3} \ln _2+\frac{3}{2} \zeta _2 \bar{s}_{1,-1} \ln _2-2 s_{-2,1,-1} \ln _2-2 s_{1,-2,-1} \ln _2-s_{1,-1,-2} \ln _2+s_{1,-1,2} \ln _2+\bar{s}_{1,-1,-2} \ln _2-\bar{s}_{1,-1,2} \ln _2+2 s_{1,-1,1,-1} \ln _2-2 \bar{s}_{1,-1,1,-1} \ln _2-\frac{6}{5} s_{-1} \zeta _2^2-11 \text{Li}_5\left(\frac{1}{2}\right)+3 \text{Li}_4\left(\frac{1}{2}\right) s_{-1}-\frac{1}{2} s_{-3} \zeta _2-\frac{1}{4} s_{-2} \zeta _3-\frac{\zeta _2 \zeta _3}{16}+\frac{713 \zeta _5}{64}-\frac{6}{5} \zeta _2^2 \bar{s}_{-1}+3 \text{Li}_4\left(\frac{1}{2}\right) \bar{s}_{-1}+\frac{13}{40} \zeta _2^2 \bar{s}_1-\text{Li}_4\left(\frac{1}{2}\right) \bar{s}_1+\frac{1}{2} \zeta _2 s_{-2,1}+\frac{1}{2} \zeta _2 s_{1,-2}+\frac{1}{4} \zeta _3 s_{1,-1}-s_{4,-1}-\frac{1}{8} \zeta _3 \bar{s}_{1,-1}+s_{-3,-1,-1}+s_{-2,-2,-1}-\frac{1}{2} \zeta _2 s_{1,-1,1}+s_{1,3,-1}+\frac{1}{2} \zeta _2 \bar{s}_{1,-1,1}-s_{-2,1,-1,-1}-s_{1,-2,-1,-1}-s_{1,-1,-2,-1}+\bar{s}_{-2,-1,1,-1}+\bar{s}_{1,-1,-2,-1}+\bar{s}_{1,2,1,-1}+s_{1,-1,1,-1,-1}-\bar{s}_{-1,1,-1,1,-1}-2 \bar{s}_{1,-1,-1,1,-1}-\bar{s}_{1,-1,1,-1,-1} \end{dmath}
  \begin{dmath}[style={\small}]     s_{-1} \bar{s}_{1,-1,1,1}  =   -\frac{\ln _2^5}{30}+\frac{1}{24} s_{-1} \ln _2^4-\frac{1}{4} s_1 \ln _2^4+\frac{1}{24} \bar{s}_{-1} \ln _2^4+\frac{1}{12} \bar{s}_1 \ln _2^4+\frac{1}{6} s_{-2} \ln _2^3-\frac{1}{6} s_2 \ln _2^3+\frac{7}{12} \zeta _2 \ln _2^3+\frac{1}{6} \bar{s}_{-2} \ln _2^3-\frac{1}{6} \bar{s}_2 \ln _2^3-\frac{1}{3} s_{1,-1} \ln _2^3+\frac{1}{2} s_{-3} \ln _2^2-\frac{1}{2} s_3 \ln _2^2+\frac{3}{4} s_1 \zeta _2 \ln _2^2+\frac{3}{16} \zeta _3 \ln _2^2+\frac{1}{2} s_{-2,-1} \ln _2^2-\frac{1}{2} s_{-2,1} \ln _2^2-\frac{1}{2} s_{1,-2} \ln _2^2+\frac{1}{2} s_{1,2} \ln _2^2-\frac{1}{2} s_{1,-1,-1} \ln _2^2+\frac{1}{2} s_{1,-1,1} \ln _2^2+\frac{1}{2} \bar{s}_{1,-1,-1} \ln _2^2-\frac{1}{2} \bar{s}_{1,-1,1} \ln _2^2-\frac{121}{40} \zeta _2^2 \ln _2-\text{Li}_4\left(\frac{1}{2}\right) \ln _2+s_{-4} \ln _2-s_4 \ln _2-\frac{1}{2} s_{-2} \zeta _2 \ln _2+\frac{1}{2} s_2 \zeta _2 \ln _2-\frac{1}{8} s_{-1} \zeta _3 \ln _2-s_1 \zeta _3 \ln _2-\frac{1}{2} \zeta _2 \bar{s}_{-2} \ln _2-\frac{1}{8} \zeta _3 \bar{s}_{-1} \ln _2+\frac{1}{8} \zeta _3 \bar{s}_1 \ln _2+\frac{1}{2} \zeta _2 \bar{s}_2 \ln _2+s_{-3,-1} \ln _2-s_{-3,1} \ln _2+s_{-2,-2} \ln _2-s_{-2,2} \ln _2-s_{1,-3} \ln _2+\zeta _2 s_{1,-1} \ln _2+s_{1,3} \ln _2-s_{-2,1,-1} \ln _2+s_{-2,1,1} \ln _2-s_{1,-2,-1} \ln _2+s_{1,-2,1} \ln _2-s_{1,-1,-2} \ln _2+s_{1,-1,2} \ln _2+s_{1,-1,1,-1} \ln _2-s_{1,-1,1,1} \ln _2-\bar{s}_{1,-1,1,-1} \ln _2-\bar{s}_{1,-1,1,1} \ln _2+\frac{1}{20} s_{-1} \zeta _2^2-\frac{3}{40} s_1 \zeta _2^2-11 \text{Li}_5\left(\frac{1}{2}\right)+\text{Li}_4\left(\frac{1}{2}\right) s_1+\frac{s_3 \zeta _2}{2}+\frac{1}{4} s_{-2} \zeta _3-\frac{7 s_2 \zeta _3}{8}+\frac{3 \zeta _2 \zeta _3}{16}+\frac{367 \zeta _5}{32}+\frac{7}{8} \zeta _3 \bar{s}_{-2}+\frac{1}{20} \zeta _2^2 \bar{s}_{-1}-\frac{3}{5} \zeta _2^2 \bar{s}_1+3 \text{Li}_4\left(\frac{1}{2}\right) \bar{s}_1-\frac{7}{8} \zeta _3 \bar{s}_2+s_{-4,-1}-\frac{1}{2} \zeta _2 s_{-2,-1}-\frac{9}{8} \zeta _3 s_{1,-1}-\frac{1}{2} \zeta _2 s_{1,2}-\zeta _3 \bar{s}_{1,-1}-s_{-3,1,-1}-s_{-2,2,-1}-s_{1,-3,-1}+\frac{1}{2} \zeta _2 s_{1,-1,-1}-\frac{1}{2} \zeta _2 \bar{s}_{1,-1,-1}-\frac{1}{2} \zeta _2 \bar{s}_{1,-1,1}+s_{-2,1,1,-1}+s_{1,-2,1,-1}+s_{1,-1,2,-1}+\bar{s}_{-2,-1,1,1}+\bar{s}_{1,-1,-2,1}+\bar{s}_{1,2,1,1}-s_{1,-1,1,1,-1}-\bar{s}_{-1,1,-1,1,1}-2 \bar{s}_{1,-1,-1,1,1}-\bar{s}_{1,-1,1,-1,1} \end{dmath}
  \begin{dmath}[style={\small}]     s_{-1} \bar{s}_{1,1,-1,-1}  =   \frac{53 \ln _2^5}{120}+\frac{1}{24} s_{-1} \ln _2^4+\frac{1}{8} s_1 \ln _2^4+\frac{1}{24} \bar{s}_{-1} \ln _2^4+\frac{2}{3} s_{-2} \ln _2^3-\frac{2}{3} s_2 \ln _2^3-\frac{35}{12} \zeta _2 \ln _2^3-\frac{1}{2} \bar{s}_{-2} \ln _2^3+\frac{1}{2} \bar{s}_2 \ln _2^3-\frac{1}{6} s_{1,-1} \ln _2^3+\frac{7}{6} s_{1,1} \ln _2^3-\frac{1}{6} \bar{s}_{1,-1} \ln _2^3+\frac{1}{6} \bar{s}_{1,1} \ln _2^3+s_{-3} \ln _2^2-s_3 \ln _2^2-\frac{1}{4} s_{-1} \zeta _2 \ln _2^2+\frac{1}{4} s_1 \zeta _2 \ln _2^2+\frac{39}{8} \zeta _3 \ln _2^2-\frac{1}{4} \zeta _2 \bar{s}_{-1} \ln _2^2-\frac{3}{4} \zeta _2 \bar{s}_1 \ln _2^2-s_{1,-2} \ln _2^2+s_{1,2} \ln _2^2-2 s_{2,-1} \ln _2^2-\bar{s}_{1,-2} \ln _2^2+\bar{s}_{1,2} \ln _2^2+2 s_{1,1,-1} \ln _2^2-\frac{13}{20} \zeta _2^2 \ln _2+10 \text{Li}_4\left(\frac{1}{2}\right) \ln _2+s_{-4} \ln _2-s_4 \ln _2+s_2 \zeta _2 \ln _2-3 s_1 \zeta _3 \ln _2+2 s_{-3,-1} \ln _2-s_{1,-3} \ln _2-\zeta _2 s_{1,1} \ln _2+s_{1,3} \ln _2-s_{2,-2} \ln _2+s_{2,2} \ln _2-2 s_{1,-2,-1} \ln _2+s_{1,1,-2} \ln _2-s_{1,1,2} \ln _2-2 s_{2,-1,-1} \ln _2-\bar{s}_{1,1,-2} \ln _2+\bar{s}_{1,1,2} \ln _2+2 s_{1,1,-1,-1} \ln _2-2 \bar{s}_{1,1,-1,-1} \ln _2-\frac{1}{8} s_{-1} \zeta _2^2+\frac{61}{40} s_1 \zeta _2^2+9 \text{Li}_5\left(\frac{1}{2}\right)-4 \text{Li}_4\left(\frac{1}{2}\right) s_1-\frac{1}{2} s_{-3} \zeta _2+\frac{1}{4} s_{-2} \zeta _3-\frac{s_2 \zeta _3}{2}-\frac{567 \zeta _5}{64}-\frac{1}{8} \zeta _2^2 \bar{s}_{-1}-\frac{13}{40} \zeta _2^2 \bar{s}_1+\text{Li}_4\left(\frac{1}{2}\right) \bar{s}_1+\frac{1}{2} \zeta _2 s_{1,-2}-\frac{1}{4} \zeta _3 s_{1,-1}+\frac{1}{2} \zeta _3 s_{1,1}+\frac{1}{2} \zeta _2 s_{2,-1}-s_{4,-1}-\frac{1}{4} \zeta _3 \bar{s}_{1,-1}-\frac{3}{4} \zeta _3 \bar{s}_{1,1}+s_{-3,-1,-1}-\frac{1}{2} \zeta _2 s_{1,1,-1}+s_{1,3,-1}+s_{2,2,-1}+\frac{1}{2} \zeta _2 \bar{s}_{1,1,-1}-s_{1,-2,-1,-1}-s_{1,1,2,-1}-s_{2,-1,-1,-1}+\bar{s}_{-2,1,-1,-1}+\bar{s}_{1,-2,-1,-1}+\bar{s}_{1,1,2,-1}+s_{1,1,-1,-1,-1}-\bar{s}_{-1,1,1,-1,-1}-\bar{s}_{1,-1,1,-1,-1}-2 \bar{s}_{1,1,-1,-1,-1} \end{dmath}
  \begin{dmath}[style={\small}]     s_{-1} \bar{s}_{1,1,-1,1}  =   \frac{\ln _2^5}{4}+\frac{1}{12} s_{-1} \ln _2^4+\frac{1}{12} s_1 \ln _2^4+\frac{1}{12} \bar{s}_{-1} \ln _2^4-\frac{1}{6} \bar{s}_1 \ln _2^4+\frac{1}{2} s_{-2} \ln _2^3-\frac{1}{2} s_2 \ln _2^3-\frac{9}{4} \zeta _2 \ln _2^3-\frac{1}{3} \bar{s}_{-2} \ln _2^3+\frac{1}{3} \bar{s}_2 \ln _2^3-\frac{1}{6} s_{1,-1} \ln _2^3+\frac{5}{6} s_{1,1} \ln _2^3-\frac{1}{6} \bar{s}_{1,-1} \ln _2^3+\frac{1}{6} \bar{s}_{1,1} \ln _2^3+\frac{1}{2} s_{-3} \ln _2^2-\frac{1}{2} s_3 \ln _2^2-\frac{1}{2} s_{-1} \zeta _2 \ln _2^2+\frac{31}{8} \zeta _3 \ln _2^2-\frac{1}{2} \zeta _2 \bar{s}_{-1} \ln _2^2-\frac{1}{2} \zeta _2 \bar{s}_1 \ln _2^2-\frac{1}{2} s_{1,-2} \ln _2^2+\frac{1}{2} s_{1,2} \ln _2^2-s_{2,-1} \ln _2^2-\frac{1}{2} \bar{s}_{1,-2} \ln _2^2+\frac{1}{2} \bar{s}_{1,2} \ln _2^2+s_{1,1,-1} \ln _2^2+\frac{11}{20} \zeta _2^2 \ln _2+5 \text{Li}_4\left(\frac{1}{2}\right) \ln _2+s_{-4} \ln _2-s_4 \ln _2-\frac{1}{2} s_{-2} \zeta _2 \ln _2+\frac{1}{2} s_2 \zeta _2 \ln _2+\frac{7}{8} s_{-1} \zeta _3 \ln _2-\frac{3}{2} s_1 \zeta _3 \ln _2+\frac{1}{2} \zeta _2 \bar{s}_{-2} \ln _2+\frac{7}{8} \zeta _3 \bar{s}_{-1} \ln _2-\frac{7}{8} \zeta _3 \bar{s}_1 \ln _2-\frac{1}{2} \zeta _2 \bar{s}_2 \ln _2+s_{-3,-1} \ln _2-s_{-3,1} \ln _2-s_{1,-3} \ln _2-\zeta _2 s_{1,1} \ln _2+s_{1,3} \ln _2-s_{2,-2} \ln _2+s_{2,2} \ln _2-\frac{3}{2} \zeta _2 \bar{s}_{1,1} \ln _2-s_{1,-2,-1} \ln _2+s_{1,-2,1} \ln _2+s_{1,1,-2} \ln _2-s_{1,1,2} \ln _2-s_{2,-1,-1} \ln _2+s_{2,-1,1} \ln _2+s_{1,1,-1,-1} \ln _2-s_{1,1,-1,1} \ln _2-\bar{s}_{1,1,-1,-1} \ln _2-\bar{s}_{1,1,-1,1} \ln _2-\frac{3}{20} s_{-1} \zeta _2^2+\frac{11}{10} s_1 \zeta _2^2+5 \text{Li}_5\left(\frac{1}{2}\right)+\text{Li}_4\left(\frac{1}{2}\right) s_{-1}-3 \text{Li}_4\left(\frac{1}{2}\right) s_1+\frac{s_3 \zeta _2}{2}-\frac{5 s_2 \zeta _3}{8}-\frac{21 \zeta _2 \zeta _3}{8}-\frac{5 \zeta _5}{32}-\frac{1}{8} \zeta _3 \bar{s}_{-2}-\frac{3}{20} \zeta _2^2 \bar{s}_{-1}+\text{Li}_4\left(\frac{1}{2}\right) \bar{s}_{-1}+\frac{27}{20} \zeta _2^2 \bar{s}_1-3 \text{Li}_4\left(\frac{1}{2}\right) \bar{s}_1+\frac{1}{8} \zeta _3 \bar{s}_2+s_{-4,-1}+\frac{1}{8} \zeta _3 s_{1,-1}+\frac{3}{4} \zeta _3 s_{1,1}-\frac{1}{2} \zeta _2 s_{1,2}+\frac{1}{2} \zeta _2 s_{2,-1}+\frac{1}{2} \zeta _2 \bar{s}_{1,-2}+\frac{1}{8} \zeta _3 \bar{s}_{1,-1}+\frac{3}{2} \zeta _3 \bar{s}_{1,1}-\frac{1}{2} \zeta _2 \bar{s}_{1,2}-s_{-3,1,-1}-s_{1,-3,-1}-\frac{1}{2} \zeta _2 s_{1,1,-1}-s_{2,-2,-1}-\zeta _2 \bar{s}_{1,1,-1}+s_{1,-2,1,-1}+s_{1,1,-2,-1}+s_{2,-1,1,-1}+\bar{s}_{-2,1,-1,1}+\bar{s}_{1,-2,-1,1}+\bar{s}_{1,1,2,1}-s_{1,1,-1,1,-1}-\bar{s}_{-1,1,1,-1,1}-\bar{s}_{1,-1,1,-1,1}-2 \bar{s}_{1,1,-1,-1,1} \end{dmath}
  \begin{dmath}[style={\small}]     s_{-1} \bar{s}_{1,1,1,-1}  =   -\frac{\ln _2^5}{24}+\frac{1}{24} s_{-1} \ln _2^4-\frac{1}{4} s_1 \ln _2^4+\frac{1}{24} \bar{s}_{-1} \ln _2^4-\frac{1}{12} \bar{s}_1 \ln _2^4+\frac{2}{3} \zeta _2 \ln _2^3+\frac{1}{6} \bar{s}_{-2} \ln _2^3-\frac{1}{6} \bar{s}_2 \ln _2^3-\frac{1}{6} s_{1,-1} \ln _2^3-\frac{1}{6} s_{1,1} \ln _2^3-\frac{1}{6} \bar{s}_{1,-1} \ln _2^3+\frac{1}{6} \bar{s}_{1,1} \ln _2^3+s_{-3} \ln _2^2-s_3 \ln _2^2-\frac{1}{4} s_{-1} \zeta _2 \ln _2^2+\frac{3}{2} s_1 \zeta _2 \ln _2^2-\frac{11}{16} \zeta _3 \ln _2^2-\frac{1}{4} \zeta _2 \bar{s}_{-1} \ln _2^2+\frac{5}{4} \zeta _2 \bar{s}_1 \ln _2^2-s_{1,-2} \ln _2^2+s_{1,2} \ln _2^2-\frac{1}{2} s_{2,-1} \ln _2^2+\frac{3}{2} s_{2,1} \ln _2^2+\frac{1}{2} \bar{s}_{1,-2} \ln _2^2-\frac{1}{2} \bar{s}_{1,2} \ln _2^2+\frac{1}{2} s_{1,1,-1} \ln _2^2-\frac{3}{2} s_{1,1,1} \ln _2^2+\frac{1}{2} \bar{s}_{1,1,-1} \ln _2^2-\frac{1}{2} \bar{s}_{1,1,1} \ln _2^2+\frac{15}{8} \zeta _2^2 \ln _2+s_{-4} \ln _2-s_4 \ln _2-\frac{1}{2} s_{-2} \zeta _2 \ln _2-\frac{1}{2} s_2 \zeta _2 \ln _2+\frac{7}{8} s_{-1} \zeta _3 \ln _2-\frac{25}{8} s_1 \zeta _3 \ln _2+\frac{7}{8} \zeta _3 \bar{s}_{-1} \ln _2-\frac{7}{8} \zeta _3 \bar{s}_1 \ln _2-s_{1,-3} \ln _2+\frac{1}{2} \zeta _2 s_{1,-1} \ln _2+\frac{1}{2} \zeta _2 s_{1,1} \ln _2+s_{1,3} \ln _2-s_{2,-2} \ln _2+s_{2,2} \ln _2-2 s_{3,-1} \ln _2+\frac{1}{2} \zeta _2 \bar{s}_{1,-1} \ln _2+\zeta _2 \bar{s}_{1,1} \ln _2+s_{1,1,-2} \ln _2-s_{1,1,2} \ln _2+2 s_{1,2,-1} \ln _2+2 s_{2,1,-1} \ln _2+\bar{s}_{1,1,-2} \ln _2-\bar{s}_{1,1,2} \ln _2-2 s_{1,1,1,-1} \ln _2-2 \bar{s}_{1,1,1,-1} \ln _2+\frac{6}{5} s_1 \zeta _2^2+5 \text{Li}_5\left(\frac{1}{2}\right)-3 \text{Li}_4\left(\frac{1}{2}\right) s_1+\frac{s_3 \zeta _2}{2}+\frac{s_2 \zeta _3}{4}-5 \zeta _5+\frac{2}{5} \zeta _2^2 \bar{s}_1-\text{Li}_4\left(\frac{1}{2}\right) \bar{s}_1+s_{-4,-1}-\frac{1}{4} \zeta _3 s_{1,1}-\frac{1}{2} \zeta _2 s_{1,2}-\frac{1}{2} \zeta _2 s_{2,1}-\frac{1}{8} \zeta _3 \bar{s}_{1,1}-s_{1,-3,-1}+\frac{1}{2} \zeta _2 s_{1,1,1}-s_{2,-2,-1}-s_{3,-1,-1}+\frac{1}{2} \zeta _2 \bar{s}_{1,1,1}+s_{1,1,-2,-1}+s_{1,2,-1,-1}+s_{2,1,-1,-1}+\bar{s}_{-2,1,1,-1}+\bar{s}_{1,-2,1,-1}+\bar{s}_{1,1,-2,-1}-s_{1,1,1,-1,-1}-\bar{s}_{-1,1,1,1,-1}-\bar{s}_{1,-1,1,1,-1}-\bar{s}_{1,1,-1,1,-1}-\bar{s}_{1,1,1,-1,-1} \end{dmath}
  \begin{dmath}[style={\small}]     s_{-1} \bar{s}_{1,1,1,1}  =   \frac{\ln _2^5}{30}-\frac{1}{24} s_1 \ln _2^4+\frac{1}{6} s_{-2} \ln _2^3-\frac{1}{6} s_2 \ln _2^3-\frac{1}{6} \zeta _2 \ln _2^3-\frac{1}{6} s_{1,-1} \ln _2^3+\frac{1}{6} s_{1,1} \ln _2^3-\frac{1}{6} \bar{s}_{1,-1} \ln _2^3+\frac{1}{6} \bar{s}_{1,1} \ln _2^3+\frac{1}{2} s_{-3} \ln _2^2-\frac{1}{2} s_3 \ln _2^2+\frac{1}{4} s_1 \zeta _2 \ln _2^2+\frac{7}{16} \zeta _3 \ln _2^2-\frac{1}{2} s_{1,-2} \ln _2^2+\frac{1}{2} s_{1,2} \ln _2^2-\frac{1}{2} s_{2,-1} \ln _2^2+\frac{1}{2} s_{2,1} \ln _2^2+\frac{1}{2} s_{1,1,-1} \ln _2^2-\frac{1}{2} s_{1,1,1} \ln _2^2+\frac{1}{2} \bar{s}_{1,1,-1} \ln _2^2-\frac{1}{2} \bar{s}_{1,1,1} \ln _2^2+\text{Li}_4\left(\frac{1}{2}\right) \ln _2+s_{-4} \ln _2-s_4 \ln _2-\frac{1}{2} s_{-2} \zeta _2 \ln _2+\frac{1}{2} s_2 \zeta _2 \ln _2-\frac{7}{8} s_1 \zeta _3 \ln _2-s_{1,-3} \ln _2+\frac{1}{2} \zeta _2 s_{1,-1} \ln _2-\frac{1}{2} \zeta _2 s_{1,1} \ln _2+s_{1,3} \ln _2-s_{2,-2} \ln _2+s_{2,2} \ln _2-s_{3,-1} \ln _2+s_{3,1} \ln _2+\frac{1}{2} \zeta _2 \bar{s}_{1,-1} \ln _2-\frac{1}{2} \zeta _2 \bar{s}_{1,1} \ln _2+s_{1,1,-2} \ln _2-s_{1,1,2} \ln _2+s_{1,2,-1} \ln _2-s_{1,2,1} \ln _2+s_{2,1,-1} \ln _2-s_{2,1,1} \ln _2-s_{1,1,1,-1} \ln _2+s_{1,1,1,1} \ln _2-\bar{s}_{1,1,1,-1} \ln _2-\bar{s}_{1,1,1,1} \ln _2+\frac{1}{8} s_1 \zeta _2^2+3 \text{Li}_5\left(\frac{1}{2}\right)-\text{Li}_4\left(\frac{1}{2}\right) s_{-1}-\frac{1}{2} s_{-3} \zeta _2+\frac{7}{8} s_{-2} \zeta _3-\frac{s_2 \zeta _3}{4}-\frac{7 \zeta _2 \zeta _3}{16}-\frac{59 \zeta _5}{32}-\text{Li}_4\left(\frac{1}{2}\right) \bar{s}_{-1}-\frac{2}{5} \zeta _2^2 \bar{s}_1+\text{Li}_4\left(\frac{1}{2}\right) \bar{s}_1+\frac{1}{2} \zeta _2 s_{1,-2}-\frac{7}{8} \zeta _3 s_{1,-1}+\frac{1}{4} \zeta _3 s_{1,1}+\frac{1}{2} \zeta _2 s_{2,-1}-s_{4,-1}-\frac{7}{8} \zeta _3 \bar{s}_{1,-1}-\frac{1}{8} \zeta _3 \bar{s}_{1,1}-\frac{1}{2} \zeta _2 s_{1,1,-1}+s_{1,3,-1}+s_{2,2,-1}+s_{3,1,-1}-\frac{1}{2} \zeta _2 \bar{s}_{1,1,-1}-\frac{1}{2} \zeta _2 \bar{s}_{1,1,1}-s_{1,1,2,-1}-s_{1,2,1,-1}-s_{2,1,1,-1}+\bar{s}_{-2,1,1,1}+\bar{s}_{1,-2,1,1}+\bar{s}_{1,1,-2,1}+s_{1,1,1,1,-1}-\bar{s}_{-1,1,1,1,1}-\bar{s}_{1,-1,1,1,1}-\bar{s}_{1,1,-1,1,1}-\bar{s}_{1,1,1,-1,1} \end{dmath}
  \begin{dmath}[style={\small}]     s_{-1} \bar{s}_{-1,-3}  =   -\frac{3}{4} \zeta _3 \bar{s}_{-1,-1}-\frac{1}{2} \zeta _2 \bar{s}_{-1,2}+\bar{s}_{2,-3}-2 \bar{s}_{-1,-1,-3}-\ln _2 \bar{s}_{-1,-3}-\ln _2 \bar{s}_{-1,3}-\frac{3}{4} \zeta _3 \bar{s}_{-2}+\frac{21}{20} \zeta _2^2 \bar{s}_{-1}+\frac{3}{4} \zeta _3 \bar{s}_2+\frac{3}{4} \zeta _3 s_{-1,-1}+\frac{1}{2} \zeta _2 s_{-1,2}-s_{4,-1}+s_{-1,-3,-1}+\ln _2 s_{-1,-3}-\ln _2 s_{-1,3}-\frac{9 \zeta _2 \zeta _3}{8}-\frac{15 \zeta _5}{8}+\frac{3}{2} \zeta _3 \ln _2^2+\frac{13}{10} \zeta _2^2 \ln _2-\frac{7}{20} \zeta _2^2 s_{-1}-\frac{1}{2} \zeta _2 s_{-3}+\frac{3}{4} \zeta _3 s_{-2}+\frac{3}{2} \zeta _3 s_{-1} \ln _2+s_{-4} \ln _2-s_4 \ln _2 \end{dmath}
  \begin{dmath}[style={\small}]     s_{-1} \bar{s}_{-1,3}  =   -\frac{1}{2} \zeta _2 \bar{s}_{-1,-2}-\frac{3}{4} \zeta _3 \bar{s}_{-1,-1}+\bar{s}_{2,3}-2 \bar{s}_{-1,-1,3}-\ln _2 \bar{s}_{-1,-3}-\ln _2 \bar{s}_{-1,3}-\frac{6}{5} \zeta _2^2 \bar{s}_{-1}-\frac{1}{2} \zeta _2 s_{-1,-2}+\frac{3}{4} \zeta _3 s_{-1,-1}+s_{-4,-1}-s_{-1,3,-1}+\ln _2 s_{-1,-3}-\ln _2 s_{-1,3}+\frac{3 \zeta _2 \zeta _3}{8}+\frac{75 \zeta _5}{16}+\frac{3}{2} \zeta _3 \ln _2^2-\frac{16}{5} \zeta _2^2 \ln _2-\frac{3}{5} \zeta _2^2 s_{-1}+\frac{\zeta _2 s_3}{2}-\frac{3 \zeta _3 s_2}{4}+\frac{3}{2} \zeta _3 s_{-1} \ln _2+s_{-4} \ln _2-s_4 \ln _2 \end{dmath}
  \begin{dmath}[style={\small}]     s_{-1} \bar{s}_{-1,-2,-1}  =   \frac{1}{2} \zeta _2 \bar{s}_{-1,-2}+\frac{1}{4} \zeta _3 \bar{s}_{-1,-1}+\bar{s}_{-1,3,-1}+\bar{s}_{2,-2,-1}-\bar{s}_{-1,-2,-1,-1}-2 \bar{s}_{-1,-1,-2,-1}-\ln _2^2 \bar{s}_{-1,-2}+\ln _2^2 \bar{s}_{-1,2}-\ln _2 \bar{s}_{-1,-3}+\ln _2 \bar{s}_{-1,3}-2 \ln _2 \bar{s}_{-1,-2,-1}+4 \text{Li}_4\left(\frac{1}{2}\right) \bar{s}_{-1}-\frac{51}{40} \zeta _2^2 \bar{s}_{-1}-\zeta _2 \ln _2^2 \bar{s}_{-1}+\frac{1}{2} \zeta _2 \ln _2 \bar{s}_{-2}-\frac{1}{2} \zeta _2 \ln _2 \bar{s}_2+\frac{1}{6} \ln _2^4 \bar{s}_{-1}-\frac{1}{2} \zeta _2 s_{-1,-2}-\frac{1}{4} \zeta _3 s_{-1,-1}+s_{-4,-1}-s_{-1,3,-1}-s_{3,-1,-1}+s_{-1,-2,-1,-1}+\ln _2^2 s_{-1,-2}-\ln _2^2 s_{-1,2}+\ln _2 s_{-1,-3}-\ln _2 s_{-1,3}-2 \ln _2 s_{3,-1}+2 \ln _2 s_{-1,-2,-1}-\frac{9 \zeta _2 \zeta _3}{8}+\frac{793 \zeta _5}{32}-2 \zeta _2 \ln _2^3+3 \zeta _3 \ln _2^2-\frac{119}{20} \zeta _2^2 \ln _2+8 \text{Li}_4\left(\frac{1}{2}\right) s_{-1}-24 \text{Li}_5\left(\frac{1}{2}\right)-\frac{117}{40} \zeta _2^2 s_{-1}+\frac{\zeta _2 s_3}{2}+\frac{\zeta _3 s_2}{4}-2 \zeta _2 s_{-1} \ln _2^2-\frac{1}{2} \zeta _2 s_{-2} \ln _2-\frac{1}{2} \zeta _2 s_2 \ln _2+6 \zeta _3 s_{-1} \ln _2+\frac{1}{3} s_{-1} \ln _2^4+s_{-3} \ln _2^2-s_3 \ln _2^2+s_{-4} \ln _2-s_4 \ln _2+\frac{\ln _2^5}{5} \end{dmath}
  \begin{dmath}[style={\small}]     s_{-1} \bar{s}_{-1,-2,1}  =   -\frac{1}{2} \zeta _2 \bar{s}_{-1,-2}-\frac{5}{8} \zeta _3 \bar{s}_{-1,-1}-\frac{1}{2} \zeta _2 \bar{s}_{-1,2}+\bar{s}_{-1,3,1}+\bar{s}_{2,-2,1}-\bar{s}_{-1,-2,-1,1}-2 \bar{s}_{-1,-1,-2,1}-\frac{1}{2} \ln _2^2 \bar{s}_{-1,-2}+\frac{1}{2} \ln _2^2 \bar{s}_{-1,2}-\ln _2 \bar{s}_{-1,-2,-1}-\ln _2 \bar{s}_{-1,-2,1}-\frac{5}{8} \zeta _3 \bar{s}_{-2}+\frac{17}{40} \zeta _2^2 \bar{s}_{-1}+\frac{5}{8} \zeta _3 \bar{s}_2+\frac{5}{8} \zeta _3 s_{-1,-1}+\frac{1}{2} \zeta _2 s_{-1,2}-s_{4,-1}+s_{-1,-3,-1}+s_{3,1,-1}-s_{-1,-2,1,-1}+\frac{1}{2} \ln _2^2 s_{-1,-2}-\frac{1}{2} \ln _2^2 s_{-1,2}+\ln _2 s_{-1,-3}-\ln _2 s_{-1,3}-\ln _2 s_{3,-1}+\ln _2 s_{3,1}+\ln _2 s_{-1,-2,-1}-\ln _2 s_{-1,-2,1}-\frac{3 \zeta _2 \zeta _3}{16}-\frac{499 \zeta _5}{64}-\frac{1}{2} \zeta _2 \ln _2^3+\frac{17}{8} \zeta _3 \ln _2^2+\frac{43}{40} \zeta _2^2 \ln _2+2 \text{Li}_4\left(\frac{1}{2}\right) s_{-1}+6 \text{Li}_5\left(\frac{1}{2}\right)+4 \text{Li}_4\left(\frac{1}{2}\right) \ln _2-\frac{11}{10} \zeta _2^2 s_{-1}-\frac{1}{2} \zeta _2 s_{-3}+\frac{5}{8} \zeta _3 s_{-2}-\frac{1}{2} \zeta _2 s_{-1} \ln _2^2+3 \zeta _3 s_{-1} \ln _2+\frac{1}{12} s_{-1} \ln _2^4+\frac{1}{2} s_{-3} \ln _2^2-\frac{1}{2} s_3 \ln _2^2+s_{-4} \ln _2-s_4 \ln _2+\frac{7 \ln _2^5}{60} \end{dmath}
  \begin{dmath}[style={\small}]     s_{-1} \bar{s}_{-1,-1,-2}  =   -\frac{1}{2} \zeta _2 \bar{s}_{-1,-2}+\frac{3}{2} \zeta _3 \bar{s}_{-1,-1}+\frac{1}{2} \zeta _2 \bar{s}_{-1,2}-\frac{1}{2} \zeta _2 \bar{s}_{-1,-1,-1}+\bar{s}_{-1,2,-2}+\bar{s}_{2,-1,-2}-3 \bar{s}_{-1,-1,-1,-2}-\ln _2 \bar{s}_{-1,-1,-2}-\ln _2 \bar{s}_{-1,-1,2}-4 \text{Li}_4\left(\frac{1}{2}\right) \bar{s}_{-1}+\frac{3}{40} \zeta _2^2 \bar{s}_{-1}+\zeta _2 \ln _2^2 \bar{s}_{-1}+\frac{1}{2} \zeta _2 \ln _2 \bar{s}_{-2}-\frac{1}{2} \zeta _2 \ln _2 \bar{s}_2-\frac{1}{6} \ln _2^4 \bar{s}_{-1}-\frac{1}{2} \zeta _2 s_{-1,-2}+\frac{3}{4} \zeta _3 s_{-1,-1}+\frac{1}{2} \zeta _2 s_{2,-1}-\frac{1}{2} \zeta _2 s_{-1,-1,-1}-\zeta _2 \ln _2 s_{-1,-1}+s_{-4,-1}-s_{-1,3,-1}-s_{2,-2,-1}+s_{-1,-1,-2,-1}+\ln _2 s_{-1,-3}-\ln _2 s_{-1,3}-\ln _2 s_{2,-2}+\ln _2 s_{2,2}+\ln _2 s_{-1,-1,-2}-\ln _2 s_{-1,-1,2}+\frac{\zeta _2 \zeta _3}{4}-\frac{643 \zeta _5}{32}-\frac{2}{3} \zeta _2 \ln _2^3+2 \zeta _3 \ln _2^2+\frac{57}{20} \zeta _2^2 \ln _2-4 \text{Li}_4\left(\frac{1}{2}\right) s_{-1}+24 \text{Li}_5\left(\frac{1}{2}\right)+8 \text{Li}_4\left(\frac{1}{2}\right) \ln _2+\frac{47}{40} \zeta _2^2 s_{-1}+\frac{\zeta _2 s_3}{2}-\frac{3 \zeta _3 s_2}{4}-\frac{1}{2} \zeta _2 s_{-2} \ln _2+\frac{1}{2} \zeta _2 s_2 \ln _2-\frac{3}{2} \zeta _3 s_{-1} \ln _2-\frac{1}{6} s_{-1} \ln _2^4+s_{-4} \ln _2-s_4 \ln _2+\frac{2 \ln _2^5}{15} \end{dmath}
  \begin{dmath}[style={\small}]     s_{-1} \bar{s}_{-1,-1,2}  =   -2 \zeta _3 \bar{s}_{-1,-1}-\frac{1}{2} \zeta _2 \bar{s}_{-1,-1,-1}+\bar{s}_{-1,2,2}+\bar{s}_{2,-1,2}-3 \bar{s}_{-1,-1,-1,2}-\ln _2 \bar{s}_{-1,-1,-2}-\ln _2 \bar{s}_{-1,-1,2}+4 \text{Li}_4\left(\frac{1}{2}\right) \bar{s}_{-1}-\zeta _3 \bar{s}_{-2}-\frac{13}{40} \zeta _2^2 \bar{s}_{-1}+\zeta _3 \bar{s}_2-\zeta _2 \ln _2^2 \bar{s}_{-1}+\frac{1}{2} \zeta _2 \ln _2 \bar{s}_{-2}-\frac{1}{2} \zeta _2 \ln _2 \bar{s}_2+\frac{1}{6} \ln _2^4 \bar{s}_{-1}+\frac{5}{4} \zeta _3 s_{-1,-1}+\frac{1}{2} \zeta _2 s_{-1,2}+\frac{1}{2} \zeta _2 s_{2,-1}-\frac{1}{2} \zeta _2 s_{-1,-1,-1}-\zeta _2 \ln _2 s_{-1,-1}-s_{4,-1}+s_{-1,-3,-1}+s_{2,2,-1}-s_{-1,-1,2,-1}+\ln _2 s_{-1,-3}-\ln _2 s_{-1,3}-\ln _2 s_{2,-2}+\ln _2 s_{2,2}+\ln _2 s_{-1,-1,-2}-\ln _2 s_{-1,-1,2}-\frac{5 \zeta _2 \zeta _3}{4}-\frac{291 \zeta _5}{32}-2 \zeta _2 \ln _2^3+2 \zeta _3 \ln _2^2+\frac{41}{20} \zeta _2^2 \ln _2-4 \text{Li}_4\left(\frac{1}{2}\right) s_{-1}+8 \text{Li}_5\left(\frac{1}{2}\right)+8 \text{Li}_4\left(\frac{1}{2}\right) \ln _2+\frac{51}{40} \zeta _2^2 s_{-1}-\frac{1}{2} \zeta _2 s_{-3}+\zeta _3 s_{-2}-\frac{\zeta _3 s_2}{4}-\frac{1}{2} \zeta _2 s_{-2} \ln _2+\frac{1}{2} \zeta _2 s_2 \ln _2-\frac{3}{2} \zeta _3 s_{-1} \ln _2-\frac{1}{6} s_{-1} \ln _2^4+s_{-4} \ln _2-s_4 \ln _2+\frac{4 \ln _2^5}{15} \end{dmath}
  \begin{dmath}[style={\small}]     s_{-1} \bar{s}_{-1,1,-2}  =   \frac{1}{2} \zeta _2 \bar{s}_{-1,-2}-\frac{1}{8} \zeta _3 \bar{s}_{-1,-1}+\frac{13}{8} \zeta _3 \bar{s}_{-1,1}-\frac{1}{2} \zeta _2 \bar{s}_{-1,2}-\frac{1}{2} \zeta _2 \bar{s}_{-1,1,-1}+\frac{1}{2} \zeta _2 \ln _2 \bar{s}_{-1,-1}-\frac{1}{2} \zeta _2 \ln _2 \bar{s}_{-1,1}+\bar{s}_{-1,-2,-2}+\bar{s}_{2,1,-2}-2 \bar{s}_{-1,-1,1,-2}-\bar{s}_{-1,1,-1,-2}-\ln _2 \bar{s}_{-1,1,-2}-\ln _2 \bar{s}_{-1,1,2}-\frac{1}{8} \zeta _3 \bar{s}_{-2}+\frac{5}{8} \zeta _2^2 \bar{s}_{-1}+\frac{1}{8} \zeta _3 \bar{s}_2-\frac{1}{2} \zeta _2 s_{-2,-1}+\frac{1}{8} \zeta _3 s_{-1,-1}-\frac{5}{8} \zeta _3 s_{-1,1}+\frac{1}{2} \zeta _2 s_{-1,2}+\frac{1}{2} \zeta _2 s_{-1,1,-1}-\frac{1}{2} \zeta _2 \ln _2 s_{-1,-1}+\frac{1}{2} \zeta _2 \ln _2 s_{-1,1}-s_{4,-1}+s_{-2,-2,-1}+s_{-1,-3,-1}-s_{-1,1,-2,-1}+\ln _2 s_{-2,-2}-\ln _2 s_{-2,2}+\ln _2 s_{-1,-3}-\ln _2 s_{-1,3}-\ln _2 s_{-1,1,-2}+\ln _2 s_{-1,1,2}-\frac{13 \zeta _2 \zeta _3}{8}+\frac{379 \zeta _5}{64}-\frac{1}{2} \zeta _2 \ln _2^3+\frac{1}{4} \zeta _3 \ln _2^2+\frac{7}{20} \zeta _2^2 \ln _2+4 \text{Li}_4\left(\frac{1}{2}\right) s_{-1}-6 \text{Li}_5\left(\frac{1}{2}\right)-2 \text{Li}_4\left(\frac{1}{2}\right) \ln _2-\frac{11}{8} \zeta _2^2 s_{-1}-\frac{1}{2} \zeta _2 s_{-3}+\frac{3}{4} \zeta _3 s_{-2}-\frac{3}{2} \zeta _2 s_{-1} \ln _2^2-\frac{1}{2} \zeta _2 s_{-2} \ln _2+\frac{1}{2} \zeta _2 s_2 \ln _2+\frac{15}{4} \zeta _3 s_{-1} \ln _2+\frac{1}{6} s_{-1} \ln _2^4+s_{-4} \ln _2-s_4 \ln _2-\frac{\ln _2^5}{30} \end{dmath}
  \begin{dmath}[style={\small}]     s_{-1} \bar{s}_{-1,1,2}  =   -\zeta _3 \bar{s}_{-1,-1}-\zeta _3 \bar{s}_{-1,1}-\frac{1}{2} \zeta _2 \bar{s}_{-1,1,-1}+\frac{1}{2} \zeta _2 \ln _2 \bar{s}_{-1,-1}-\frac{1}{2} \zeta _2 \ln _2 \bar{s}_{-1,1}+\bar{s}_{-1,-2,2}+\bar{s}_{2,1,2}-2 \bar{s}_{-1,-1,1,2}-\bar{s}_{-1,1,-1,2}-\ln _2 \bar{s}_{-1,1,-2}-\ln _2 \bar{s}_{-1,1,2}-\frac{1}{2} \zeta _2^2 \bar{s}_{-1}-\frac{1}{2} \zeta _2 s_{-2,-1}-\frac{1}{2} \zeta _2 s_{-1,-2}+\zeta _3 s_{-1,-1}-\frac{1}{4} \zeta _3 s_{-1,1}+\frac{1}{2} \zeta _2 s_{-1,1,-1}-\frac{1}{2} \zeta _2 \ln _2 s_{-1,-1}+\frac{1}{2} \zeta _2 \ln _2 s_{-1,1}+s_{-4,-1}-s_{-2,2,-1}-s_{-1,3,-1}+s_{-1,1,2,-1}+\ln _2 s_{-2,-2}-\ln _2 s_{-2,2}+\ln _2 s_{-1,-3}-\ln _2 s_{-1,3}-\ln _2 s_{-1,1,-2}+\ln _2 s_{-1,1,2}-\frac{\zeta _2 \zeta _3}{8}-\frac{23 \zeta _5}{32}+\frac{1}{3} \zeta _2 \ln _2^3+\frac{1}{4} \zeta _3 \ln _2^2-\frac{11}{40} \zeta _2^2 \ln _2+2 \text{Li}_4\left(\frac{1}{2}\right) s_{-1}+4 \text{Li}_5\left(\frac{1}{2}\right)-2 \text{Li}_4\left(\frac{1}{2}\right) \ln _2-\frac{11}{8} \zeta _2^2 s_{-1}+\frac{\zeta _2 s_3}{2}+\frac{1}{4} \zeta _3 s_{-2}-\zeta _3 s_2-\zeta _2 s_{-1} \ln _2^2-\frac{1}{2} \zeta _2 s_{-2} \ln _2+\frac{1}{2} \zeta _2 s_2 \ln _2+\frac{15}{4} \zeta _3 s_{-1} \ln _2+\frac{1}{12} s_{-1} \ln _2^4+s_{-4} \ln _2-s_4 \ln _2-\frac{7 \ln _2^5}{60} \end{dmath}
  \begin{dmath}[style={\small}]     s_{-1} \bar{s}_{-1,2,-1}  =   \frac{1}{4} \zeta _3 \bar{s}_{-1,-1}+\frac{1}{2} \zeta _2 \bar{s}_{-1,2}+\bar{s}_{-1,-3,-1}+\bar{s}_{2,2,-1}-2 \bar{s}_{-1,-1,2,-1}-\bar{s}_{-1,2,-1,-1}+\ln _2^2 \bar{s}_{-1,-2}-\ln _2^2 \bar{s}_{-1,2}+\ln _2 \bar{s}_{-1,-3}-\ln _2 \bar{s}_{-1,3}-2 \ln _2 \bar{s}_{-1,2,-1}-4 \text{Li}_4\left(\frac{1}{2}\right) \bar{s}_{-1}+\frac{1}{4} \zeta _3 \bar{s}_{-2}+\frac{11}{8} \zeta _2^2 \bar{s}_{-1}-\frac{1}{4} \zeta _3 \bar{s}_2+\zeta _2 \ln _2^2 \bar{s}_{-1}-\frac{1}{2} \zeta _2 \ln _2 \bar{s}_{-2}+\frac{1}{2} \zeta _2 \ln _2 \bar{s}_2-\frac{1}{6} \ln _2^4 \bar{s}_{-1}-\frac{1}{4} \zeta _3 s_{-1,-1}+\frac{1}{2} \zeta _2 s_{-1,2}-s_{4,-1}+s_{-3,-1,-1}+s_{-1,-3,-1}-s_{-1,2,-1,-1}+\ln _2^2 s_{-1,-2}-\ln _2^2 s_{-1,2}+2 \ln _2 s_{-3,-1}+\ln _2 s_{-1,-3}-\ln _2 s_{-1,3}-2 \ln _2 s_{-1,2,-1}+\frac{7 \zeta _2 \zeta _3}{8}+\frac{231 \zeta _5}{32}+\frac{10}{3} \zeta _2 \ln _2^3-4 \zeta _3 \ln _2^2+\frac{7}{20} \zeta _2^2 \ln _2-8 \text{Li}_5\left(\frac{1}{2}\right)-16 \text{Li}_4\left(\frac{1}{2}\right) \ln _2+\frac{1}{8} \zeta _2^2 s_{-1}-\frac{1}{2} \zeta _2 s_{-3}-\frac{1}{4} \zeta _3 s_{-2}+\frac{1}{2} \zeta _2 s_{-2} \ln _2+\frac{1}{2} \zeta _2 s_2 \ln _2+s_{-3} \ln _2^2-s_3 \ln _2^2+s_{-4} \ln _2-s_4 \ln _2-\frac{3 \ln _2^5}{5} \end{dmath}
  \begin{dmath}[style={\small}]     s_{-1} \bar{s}_{-1,2,1}  =   -\frac{1}{2} \zeta _2 \bar{s}_{-1,-2}-\frac{5}{8} \zeta _3 \bar{s}_{-1,-1}-\frac{1}{2} \zeta _2 \bar{s}_{-1,2}+\bar{s}_{-1,-3,1}+\bar{s}_{2,2,1}-2 \bar{s}_{-1,-1,2,1}-\bar{s}_{-1,2,-1,1}+\frac{1}{2} \ln _2^2 \bar{s}_{-1,-2}-\frac{1}{2} \ln _2^2 \bar{s}_{-1,2}-\ln _2 \bar{s}_{-1,2,-1}-\ln _2 \bar{s}_{-1,2,1}-\frac{7}{10} \zeta _2^2 \bar{s}_{-1}-\frac{1}{2} \zeta _2 s_{-1,-2}+\frac{5}{8} \zeta _3 s_{-1,-1}+s_{-4,-1}-s_{-3,1,-1}-s_{-1,3,-1}+s_{-1,2,1,-1}+\frac{1}{2} \ln _2^2 s_{-1,-2}-\frac{1}{2} \ln _2^2 s_{-1,2}+\ln _2 s_{-3,-1}-\ln _2 s_{-3,1}+\ln _2 s_{-1,-3}-\ln _2 s_{-1,3}-\ln _2 s_{-1,2,-1}+\ln _2 s_{-1,2,1}+\frac{7 \zeta _2 \zeta _3}{8}+\frac{173 \zeta _5}{32}+\frac{7}{6} \zeta _2 \ln _2^3-\frac{9}{4} \zeta _3 \ln _2^2-\frac{47}{40} \zeta _2^2 \ln _2-4 \text{Li}_4\left(\frac{1}{2}\right) s_{-1}-4 \text{Li}_5\left(\frac{1}{2}\right)-6 \text{Li}_4\left(\frac{1}{2}\right) \ln _2+\frac{21}{20} \zeta _2^2 s_{-1}+\frac{\zeta _2 s_3}{2}-\frac{5 \zeta _3 s_2}{8}+\zeta _2 s_{-1} \ln _2^2-\frac{9}{4} \zeta _3 s_{-1} \ln _2-\frac{1}{6} s_{-1} \ln _2^4+\frac{1}{2} s_{-3} \ln _2^2-\frac{1}{2} s_3 \ln _2^2+s_{-4} \ln _2-s_4 \ln _2-\frac{13 \ln _2^5}{60} \end{dmath}
  \begin{dmath}[style={\small}]     s_{-1} \bar{s}_{-1,-1,-1,-1}  =   -\zeta _3 \bar{s}_{-1,-1}+\frac{1}{2} \zeta _2 \bar{s}_{-1,-1,-1}+\bar{s}_{-1,-1,2,-1}+\bar{s}_{-1,2,-1,-1}+\bar{s}_{2,-1,-1,-1}-4 \bar{s}_{-1,-1,-1,-1,-1}+\ln _2^2 \bar{s}_{-1,-2}-\ln _2^2 \bar{s}_{-1,2}-\ln _2 \bar{s}_{-1,-1,-2}+\ln _2 \bar{s}_{-1,-1,2}-2 \ln _2 \bar{s}_{-1,-1,-1,-1}-\frac{1}{4} \zeta _3 \bar{s}_{-2}+\frac{7}{20} \zeta _2^2 \bar{s}_{-1}+\frac{1}{4} \zeta _3 \bar{s}_2-\frac{2}{3} \ln _2^3 \bar{s}_{-2}+\frac{2}{3} \ln _2^3 \bar{s}_2+\frac{3}{4} \zeta _3 s_{-1,-1}+\frac{1}{2} \zeta _2 s_{-1,2}+\frac{1}{2} \zeta _2 s_{2,-1}-\frac{1}{2} \zeta _2 s_{-1,-1,-1}-\zeta _2 \ln _2 s_{-1,-1}-s_{4,-1}+s_{-3,-1,-1}+s_{-1,-3,-1}+s_{2,2,-1}-s_{-1,-1,2,-1}-s_{-1,2,-1,-1}-s_{2,-1,-1,-1}+s_{-1,-1,-1,-1,-1}+\frac{4}{3} \ln _2^3 s_{-1,-1}+\ln _2^2 s_{-1,-2}-\ln _2^2 s_{-1,2}-2 \ln _2^2 s_{2,-1}+2 \ln _2^2 s_{-1,-1,-1}+2 \ln _2 s_{-3,-1}+\ln _2 s_{-1,-3}-\ln _2 s_{-1,3}-\ln _2 s_{2,-2}+\ln _2 s_{2,2}+\ln _2 s_{-1,-1,-2}-\ln _2 s_{-1,-1,2}-2 \ln _2 s_{-1,2,-1}-2 \ln _2 s_{2,-1,-1}+2 \ln _2 s_{-1,-1,-1,-1}-\frac{\zeta _2 \zeta _3}{4}-\frac{33 \zeta _5}{4}-2 \zeta _3 \ln _2^2+\frac{14}{5} \zeta _2^2 \ln _2-4 \text{Li}_4\left(\frac{1}{2}\right) s_{-1}+8 \text{Li}_5\left(\frac{1}{2}\right)+\frac{7}{5} \zeta _2^2 s_{-1}-\frac{1}{2} \zeta _2 s_{-3}+\frac{1}{4} \zeta _3 s_{-2}-\frac{\zeta _3 s_2}{2}+\zeta _2 s_2 \ln _2-2 \zeta _3 s_{-1} \ln _2+\frac{1}{2} s_{-1} \ln _2^4+\frac{2}{3} s_{-2} \ln _2^3-\frac{2}{3} s_2 \ln _2^3+s_{-3} \ln _2^2-s_3 \ln _2^2+s_{-4} \ln _2-s_4 \ln _2+\frac{\ln _2^5}{5} \end{dmath}
  \begin{dmath}[style={\small}]     s_{-1} \bar{s}_{-1,-1,-1,1}  =   -\frac{1}{2} \zeta _2 \bar{s}_{-1,-2}+\frac{13}{8} \zeta _3 \bar{s}_{-1,-1}+\frac{1}{2} \zeta _2 \bar{s}_{-1,2}-\zeta _2 \bar{s}_{-1,-1,-1}-\frac{3}{2} \zeta _2 \ln _2 \bar{s}_{-1,-1}+\bar{s}_{-1,-1,2,1}+\bar{s}_{-1,2,-1,1}+\bar{s}_{2,-1,-1,1}-4 \bar{s}_{-1,-1,-1,-1,1}+\frac{1}{2} \ln _2^2 \bar{s}_{-1,-2}-\frac{1}{2} \ln _2^2 \bar{s}_{-1,2}-\ln _2 \bar{s}_{-1,-1,-1,-1}-\ln _2 \bar{s}_{-1,-1,-1,1}-3 \text{Li}_4\left(\frac{1}{2}\right) \bar{s}_{-1}+\frac{1}{8} \zeta _2^2 \bar{s}_{-1}+\frac{1}{2} \zeta _2 \ln _2 \bar{s}_{-2}-\frac{1}{2} \zeta _2 \ln _2 \bar{s}_2-\frac{1}{8} \ln _2^4 \bar{s}_{-1}-\frac{1}{2} \ln _2^3 \bar{s}_{-2}+\frac{1}{2} \ln _2^3 \bar{s}_2-\frac{1}{2} \zeta _2 s_{-1,-2}+\frac{5}{8} \zeta _3 s_{-1,-1}+\frac{1}{2} \zeta _2 s_{2,-1}-\frac{1}{2} \zeta _2 s_{-1,-1,-1}-\zeta _2 \ln _2 s_{-1,-1}+s_{-4,-1}-s_{-3,1,-1}-s_{-1,3,-1}-s_{2,-2,-1}+s_{-1,-1,-2,-1}+s_{-1,2,1,-1}+s_{2,-1,1,-1}-s_{-1,-1,-1,1,-1}+\ln _2^3 s_{-1,-1}+\frac{1}{2} \ln _2^2 s_{-1,-2}-\frac{1}{2} \ln _2^2 s_{-1,2}-\ln _2^2 s_{2,-1}+\ln _2^2 s_{-1,-1,-1}+\ln _2 s_{-3,-1}-\ln _2 s_{-3,1}+\ln _2 s_{-1,-3}-\ln _2 s_{-1,3}-\ln _2 s_{2,-2}+\ln _2 s_{2,2}+\ln _2 s_{-1,-1,-2}-\ln _2 s_{-1,-1,2}-\ln _2 s_{-1,2,-1}+\ln _2 s_{-1,2,1}-\ln _2 s_{2,-1,-1}+\ln _2 s_{2,-1,1}+\ln _2 s_{-1,-1,-1,-1}-\ln _2 s_{-1,-1,-1,1}+\frac{7 \zeta _2 \zeta _3}{8}-\frac{51 \zeta _5}{4}-\frac{1}{2} \zeta _2 \ln _2^3-\frac{3}{4} \zeta _3 \ln _2^2+\frac{99}{40} \zeta _2^2 \ln _2-4 \text{Li}_4\left(\frac{1}{2}\right) s_{-1}+14 \text{Li}_5\left(\frac{1}{2}\right)+3 \text{Li}_4\left(\frac{1}{2}\right) \ln _2+\frac{13}{10} \zeta _2^2 s_{-1}+\frac{\zeta _2 s_3}{2}-\frac{5 \zeta _3 s_2}{8}-\frac{1}{2} \zeta _2 s_{-2} \ln _2+\frac{1}{2} \zeta _2 s_2 \ln _2-\frac{17}{8} \zeta _3 s_{-1} \ln _2+\frac{5}{12} s_{-1} \ln _2^4+\frac{1}{2} s_{-2} \ln _2^3-\frac{1}{2} s_2 \ln _2^3+\frac{1}{2} s_{-3} \ln _2^2-\frac{1}{2} s_3 \ln _2^2+s_{-4} \ln _2-s_4 \ln _2+\frac{31 \ln _2^5}{120} \end{dmath}
  \begin{dmath}[style={\small}]     s_{-1} \bar{s}_{-1,-1,1,-1}  =   -\frac{1}{8} \zeta _3 \bar{s}_{-1,-1}+\frac{1}{2} \zeta _2 \bar{s}_{-1,-1,1}+\frac{3}{2} \zeta _2 \ln _2 \bar{s}_{-1,-1}+\bar{s}_{-1,-1,-2,-1}+\bar{s}_{-1,2,1,-1}+\bar{s}_{2,-1,1,-1}-3 \bar{s}_{-1,-1,-1,1,-1}-\bar{s}_{-1,-1,1,-1,-1}-\frac{1}{2} \ln _2^2 \bar{s}_{-1,-2}+\frac{1}{2} \ln _2^2 \bar{s}_{-1,2}+\frac{1}{2} \ln _2^2 \bar{s}_{-1,-1,-1}-\frac{1}{2} \ln _2^2 \bar{s}_{-1,-1,1}+\ln _2 \bar{s}_{-1,-1,-2}-\ln _2 \bar{s}_{-1,-1,2}-2 \ln _2 \bar{s}_{-1,-1,1,-1}+2 \text{Li}_4\left(\frac{1}{2}\right) \bar{s}_{-1}-\frac{7}{8} \zeta _2^2 \bar{s}_{-1}+\zeta _2 \ln _2^2 \bar{s}_{-1}+\frac{1}{2} \zeta _2 \ln _2 \bar{s}_{-2}-\frac{1}{2} \zeta _2 \ln _2 \bar{s}_2+\frac{1}{12} \ln _2^4 \bar{s}_{-1}-\frac{1}{2} \zeta _2 s_{-1,-2}-\frac{1}{4} \zeta _3 s_{-1,-1}-\frac{1}{2} \zeta _2 s_{2,1}+\frac{1}{2} \zeta _2 s_{-1,-1,1}+s_{-4,-1}-s_{-1,3,-1}-s_{2,-2,-1}-s_{3,-1,-1}+s_{-1,-2,-1,-1}+s_{-1,-1,-2,-1}+s_{2,1,-1,-1}-s_{-1,-1,1,-1,-1}+\ln _2^2 s_{-1,-2}-\ln _2^2 s_{-1,2}-\frac{1}{2} \ln _2^2 s_{2,-1}+\frac{3}{2} \ln _2^2 s_{2,1}+\frac{1}{2} \ln _2^2 s_{-1,-1,-1}-\frac{3}{2} \ln _2^2 s_{-1,-1,1}+\ln _2 s_{-1,-3}-\ln _2 s_{-1,3}-\ln _2 s_{2,-2}+\ln _2 s_{2,2}-2 \ln _2 s_{3,-1}+2 \ln _2 s_{-1,-2,-1}+\ln _2 s_{-1,-1,-2}-\ln _2 s_{-1,-1,2}+2 \ln _2 s_{2,1,-1}-2 \ln _2 s_{-1,-1,1,-1}-\frac{21 \zeta _2 \zeta _3}{16}+\frac{13 \zeta _5}{16}-\frac{13}{12} \zeta _2 \ln _2^3+\frac{49}{16} \zeta _3 \ln _2^2-\frac{9}{10} \zeta _2^2 \ln _2+3 \text{Li}_4\left(\frac{1}{2}\right) s_{-1}+2 \text{Li}_5\left(\frac{1}{2}\right)+6 \text{Li}_4\left(\frac{1}{2}\right) \ln _2-\frac{6}{5} \zeta _2^2 s_{-1}+\frac{\zeta _2 s_3}{2}+\frac{\zeta _3 s_2}{4}-\frac{5}{4} \zeta _2 s_{-1} \ln _2^2-\frac{1}{2} \zeta _2 s_{-2} \ln _2-\frac{1}{2} \zeta _2 s_2 \ln _2+3 \zeta _3 s_{-1} \ln _2+\frac{3}{8} s_{-1} \ln _2^4+s_{-3} \ln _2^2-s_3 \ln _2^2+s_{-4} \ln _2-s_4 \ln _2+\frac{2 \ln _2^5}{5} \end{dmath}
  \begin{dmath}[style={\small}]     s_{-1} \bar{s}_{-1,-1,1,1}  =   \frac{17 \ln _2^5}{40}+\frac{7}{24} s_{-1} \ln _2^4+\frac{1}{8} \bar{s}_{-1} \ln _2^4+\frac{1}{6} s_{-2} \ln _2^3-\frac{1}{6} s_2 \ln _2^3-\frac{17}{12} \zeta _2 \ln _2^3-\frac{1}{6} \bar{s}_{-2} \ln _2^3+\frac{1}{6} \bar{s}_2 \ln _2^3+\frac{1}{3} s_{-1,-1} \ln _2^3+\frac{1}{2} s_{-3} \ln _2^2-\frac{1}{2} s_3 \ln _2^2-\frac{3}{4} s_{-1} \zeta _2 \ln _2^2+\frac{39}{16} \zeta _3 \ln _2^2+\frac{1}{2} s_{-1,-2} \ln _2^2-\frac{1}{2} s_{-1,2} \ln _2^2-\frac{1}{2} s_{2,-1} \ln _2^2+\frac{1}{2} s_{2,1} \ln _2^2+\frac{1}{2} s_{-1,-1,-1} \ln _2^2-\frac{1}{2} s_{-1,-1,1} \ln _2^2+\frac{1}{2} \bar{s}_{-1,-1,-1} \ln _2^2-\frac{1}{2} \bar{s}_{-1,-1,1} \ln _2^2+\frac{3}{4} \zeta _2^2 \ln _2+7 \text{Li}_4\left(\frac{1}{2}\right) \ln _2+s_{-4} \ln _2-s_4 \ln _2-\frac{1}{2} s_{-2} \zeta _2 \ln _2+\frac{1}{2} s_2 \zeta _2 \ln _2+\frac{7}{8} s_{-1} \zeta _3 \ln _2+\frac{1}{2} \zeta _2 \bar{s}_{-2} \ln _2-\frac{1}{2} \zeta _2 \bar{s}_2 \ln _2+s_{-1,-3} \ln _2-\zeta _2 s_{-1,-1} \ln _2-s_{-1,3} \ln _2-s_{2,-2} \ln _2+s_{2,2} \ln _2-s_{3,-1} \ln _2+s_{3,1} \ln _2+s_{-1,-2,-1} \ln _2-s_{-1,-2,1} \ln _2+s_{-1,-1,-2} \ln _2-s_{-1,-1,2} \ln _2+s_{2,1,-1} \ln _2-s_{2,1,1} \ln _2-s_{-1,-1,1,-1} \ln _2+s_{-1,-1,1,1} \ln _2-\bar{s}_{-1,-1,1,-1} \ln _2-\bar{s}_{-1,-1,1,1} \ln _2+\frac{1}{8} s_{-1} \zeta _2^2+6 \text{Li}_5\left(\frac{1}{2}\right)-\text{Li}_4\left(\frac{1}{2}\right) s_{-1}-\frac{1}{2} s_{-3} \zeta _2+\frac{7}{8} s_{-2} \zeta _3-\frac{s_2 \zeta _3}{4}-\frac{\zeta _2 \zeta _3}{4}-\frac{505 \zeta _5}{64}-\frac{7}{8} \zeta _3 \bar{s}_{-2}-\frac{11}{20} \zeta _2^2 \bar{s}_{-1}+3 \text{Li}_4\left(\frac{1}{2}\right) \bar{s}_{-1}+\frac{7}{8} \zeta _3 \bar{s}_2+\frac{9}{8} \zeta _3 s_{-1,-1}+\frac{1}{2} \zeta _2 s_{-1,2}+\frac{1}{2} \zeta _2 s_{2,-1}-s_{4,-1}-\zeta _3 \bar{s}_{-1,-1}+s_{-1,-3,-1}-\frac{1}{2} \zeta _2 s_{-1,-1,-1}+s_{2,2,-1}+s_{3,1,-1}-\frac{1}{2} \zeta _2 \bar{s}_{-1,-1,-1}-\frac{1}{2} \zeta _2 \bar{s}_{-1,-1,1}-s_{-1,-2,1,-1}-s_{-1,-1,2,-1}-s_{2,1,1,-1}+\bar{s}_{-1,-1,-2,1}+\bar{s}_{-1,2,1,1}+\bar{s}_{2,-1,1,1}+s_{-1,-1,1,1,-1}-3 \bar{s}_{-1,-1,-1,1,1}-\bar{s}_{-1,-1,1,-1,1} \end{dmath}
  \begin{dmath}[style={\small}]     s_{-1} \bar{s}_{-1,1,1,-1}  =   -\frac{\ln _2^5}{4}+\frac{7}{24} s_{-1} \ln _2^4-\frac{1}{24} \bar{s}_{-1} \ln _2^4+\frac{7}{6} \zeta _2 \ln _2^3-\frac{1}{6} \bar{s}_{-2} \ln _2^3+\frac{1}{6} \bar{s}_2 \ln _2^3+\frac{1}{6} s_{-1,-1} \ln _2^3+\frac{1}{6} s_{-1,1} \ln _2^3-\frac{1}{6} \bar{s}_{-1,-1} \ln _2^3+\frac{1}{6} \bar{s}_{-1,1} \ln _2^3+s_{-3} \ln _2^2-s_3 \ln _2^2-\frac{7}{4} s_{-1} \zeta _2 \ln _2^2-\frac{31}{16} \zeta _3 \ln _2^2+\zeta _2 \bar{s}_{-1} \ln _2^2+\frac{1}{2} s_{-2,-1} \ln _2^2-\frac{3}{2} s_{-2,1} \ln _2^2+s_{-1,-2} \ln _2^2-s_{-1,2} \ln _2^2+\frac{1}{2} \bar{s}_{-1,-2} \ln _2^2-\frac{1}{2} \bar{s}_{-1,2} \ln _2^2-\frac{1}{2} s_{-1,1,-1} \ln _2^2+\frac{3}{2} s_{-1,1,1} \ln _2^2+\frac{1}{2} \bar{s}_{-1,1,-1} \ln _2^2-\frac{1}{2} \bar{s}_{-1,1,1} \ln _2^2-\frac{1}{8} \zeta _2^2 \ln _2-10 \text{Li}_4\left(\frac{1}{2}\right) \ln _2+s_{-4} \ln _2-s_4 \ln _2+\frac{1}{2} s_{-2} \zeta _2 \ln _2+\frac{1}{2} s_2 \zeta _2 \ln _2+4 s_{-1} \zeta _3 \ln _2+2 s_{-3,-1} \ln _2+s_{-2,-2} \ln _2-s_{-2,2} \ln _2+s_{-1,-3} \ln _2-\frac{1}{2} \zeta _2 s_{-1,-1} \ln _2-\frac{1}{2} \zeta _2 s_{-1,1} \ln _2-s_{-1,3} \ln _2+\frac{1}{2} \zeta _2 \bar{s}_{-1,-1} \ln _2+\zeta _2 \bar{s}_{-1,1} \ln _2-2 s_{-2,1,-1} \ln _2-s_{-1,1,-2} \ln _2+s_{-1,1,2} \ln _2-2 s_{-1,2,-1} \ln _2+\bar{s}_{-1,1,-2} \ln _2-\bar{s}_{-1,1,2} \ln _2+2 s_{-1,1,1,-1} \ln _2-2 \bar{s}_{-1,1,1,-1} \ln _2-\frac{6}{5} s_{-1} \zeta _2^2-8 \text{Li}_5\left(\frac{1}{2}\right)+3 \text{Li}_4\left(\frac{1}{2}\right) s_{-1}-\frac{1}{2} s_{-3} \zeta _2-\frac{1}{4} s_{-2} \zeta _3+\frac{7 \zeta _2 \zeta _3}{16}+\frac{57 \zeta _5}{8}+\frac{2}{5} \zeta _2^2 \bar{s}_{-1}-\text{Li}_4\left(\frac{1}{2}\right) \bar{s}_{-1}+\frac{1}{2} \zeta _2 s_{-2,1}+\frac{1}{4} \zeta _3 s_{-1,1}+\frac{1}{2} \zeta _2 s_{-1,2}-s_{4,-1}-\frac{1}{8} \zeta _3 \bar{s}_{-1,1}+s_{-3,-1,-1}+s_{-2,-2,-1}+s_{-1,-3,-1}-\frac{1}{2} \zeta _2 s_{-1,1,1}+\frac{1}{2} \zeta _2 \bar{s}_{-1,1,1}-s_{-2,1,-1,-1}-s_{-1,1,-2,-1}-s_{-1,2,-1,-1}+\bar{s}_{-1,-2,1,-1}+\bar{s}_{-1,1,-2,-1}+\bar{s}_{2,1,1,-1}+s_{-1,1,1,-1,-1}-2 \bar{s}_{-1,-1,1,1,-1}-\bar{s}_{-1,1,-1,1,-1}-\bar{s}_{-1,1,1,-1,-1} \end{dmath}
  \begin{dmath}[style={\small}]     s_{-1} \bar{s}_{-2,-2}  =   -\frac{1}{2} \zeta _2 \bar{s}_{-2,-1}+\bar{s}_{3,-2}-\bar{s}_{-2,-1,-2}-\bar{s}_{-1,-2,-2}-\ln _2 \bar{s}_{-2,-2}-\ln _2 \bar{s}_{-2,2}+4 \text{Li}_4\left(\frac{1}{2}\right) \bar{s}_{-1}-\frac{1}{2} \zeta _2 \bar{s}_{-3}+\frac{3}{2} \zeta _3 \bar{s}_{-2}-\frac{11}{8} \zeta _2^2 \bar{s}_{-1}+\frac{1}{2} \zeta _2 \bar{s}_3-\zeta _2 \ln _2^2 \bar{s}_{-1}-\frac{1}{2} \zeta _2 \ln _2 \bar{s}_{-2}+\frac{7}{2} \zeta _3 \ln _2 \bar{s}_{-1}+\frac{1}{2} \zeta _2 \ln _2 \bar{s}_2+\frac{1}{6} \ln _2^4 \bar{s}_{-1}-\frac{1}{2} \zeta _2 s_{-2,-1}-s_{4,-1}+s_{-2,-2,-1}+\ln _2 s_{-2,-2}-\ln _2 s_{-2,2}-\frac{15 \zeta _2 \zeta _3}{8}+\frac{547 \zeta _5}{32}+\frac{2}{3} \zeta _2 \ln _2^3-\frac{11}{4} \zeta _2^2 \ln _2+4 \text{Li}_4\left(\frac{1}{2}\right) s_{-1}-16 \text{Li}_5\left(\frac{1}{2}\right)-8 \text{Li}_4\left(\frac{1}{2}\right) \ln _2-\frac{11}{8} \zeta _2^2 s_{-1}-\frac{1}{2} \zeta _2 s_{-3}+\frac{3}{4} \zeta _3 s_{-2}-\zeta _2 s_{-1} \ln _2^2-\frac{1}{2} \zeta _2 s_{-2} \ln _2+\frac{1}{2} \zeta _2 s_2 \ln _2+\frac{7}{2} \zeta _3 s_{-1} \ln _2+\frac{1}{6} s_{-1} \ln _2^4+s_{-4} \ln _2-s_4 \ln _2-\frac{\ln _2^5}{5} \end{dmath}
  \begin{dmath}[style={\small}]     s_{-1} \bar{s}_{-2,2}  =   -\frac{1}{2} \zeta _2 \bar{s}_{-2,-1}+\bar{s}_{3,2}-\bar{s}_{-2,-1,2}-\bar{s}_{-1,-2,2}-\ln _2 \bar{s}_{-2,-2}-\ln _2 \bar{s}_{-2,2}-4 \text{Li}_4\left(\frac{1}{2}\right) \bar{s}_{-1}-\zeta _3 \bar{s}_{-2}+\frac{51}{40} \zeta _2^2 \bar{s}_{-1}-\zeta _3 \bar{s}_2+\zeta _2 \ln _2^2 \bar{s}_{-1}-\frac{1}{2} \zeta _2 \ln _2 \bar{s}_{-2}-\frac{7}{2} \zeta _3 \ln _2 \bar{s}_{-1}+\frac{1}{2} \zeta _2 \ln _2 \bar{s}_2-\frac{1}{6} \ln _2^4 \bar{s}_{-1}-\frac{1}{2} \zeta _2 s_{-2,-1}+s_{-4,-1}-s_{-2,2,-1}+\ln _2 s_{-2,-2}-\ln _2 s_{-2,2}+\frac{19 \zeta _2 \zeta _3}{8}-\frac{75 \zeta _5}{32}+2 \zeta _2 \ln _2^3-7 \zeta _3 \ln _2^2+\frac{51}{20} \zeta _2^2 \ln _2-4 \text{Li}_4\left(\frac{1}{2}\right) s_{-1}-8 \text{Li}_4\left(\frac{1}{2}\right) \ln _2+\frac{51}{40} \zeta _2^2 s_{-1}+\frac{\zeta _2 s_3}{2}+\frac{1}{4} \zeta _3 s_{-2}-\zeta _3 s_2+\zeta _2 s_{-1} \ln _2^2-\frac{1}{2} \zeta _2 s_{-2} \ln _2+\frac{1}{2} \zeta _2 s_2 \ln _2-\frac{7}{2} \zeta _3 s_{-1} \ln _2-\frac{1}{6} s_{-1} \ln _2^4+s_{-4} \ln _2-s_4 \ln _2-\frac{\ln _2^5}{3} \end{dmath}
  \begin{dmath}[style={\small}]     s_{-1} \bar{s}_{2,2}  =   -\frac{1}{2} \zeta _2 \bar{s}_{2,-1}+\bar{s}_{-3,2}-\bar{s}_{-1,2,2}-\bar{s}_{2,-1,2}-\ln _2 \bar{s}_{2,-2}-\ln _2 \bar{s}_{2,2}-4 \text{Li}_4\left(\frac{1}{2}\right) \bar{s}_{-1}-\zeta _3 \bar{s}_{-2}+\frac{51}{40} \zeta _2^2 \bar{s}_{-1}-\zeta _3 \bar{s}_2+\zeta _2 \ln _2^2 \bar{s}_{-1}+\frac{1}{2} \zeta _2 \ln _2 \bar{s}_{-2}-\frac{7}{2} \zeta _3 \ln _2 \bar{s}_{-1}-\frac{1}{2} \zeta _2 \ln _2 \bar{s}_2-\frac{1}{6} \ln _2^4 \bar{s}_{-1}+\frac{1}{2} \zeta _2 s_{2,-1}-s_{4,-1}+s_{2,2,-1}-\ln _2 s_{2,-2}+\ln _2 s_{2,2}-\frac{3 \zeta _2 \zeta _3}{8}-\frac{507 \zeta _5}{32}-\frac{2}{3} \zeta _2 \ln _2^3+\frac{51}{20} \zeta _2^2 \ln _2-4 \text{Li}_4\left(\frac{1}{2}\right) s_{-1}+16 \text{Li}_5\left(\frac{1}{2}\right)+8 \text{Li}_4\left(\frac{1}{2}\right) \ln _2+\frac{51}{40} \zeta _2^2 s_{-1}-\frac{1}{2} \zeta _2 s_{-3}+\zeta _3 s_{-2}-\frac{\zeta _3 s_2}{4}+\zeta _2 s_{-1} \ln _2^2-\frac{1}{2} \zeta _2 s_{-2} \ln _2+\frac{1}{2} \zeta _2 s_2 \ln _2-\frac{7}{2} \zeta _3 s_{-1} \ln _2-\frac{1}{6} s_{-1} \ln _2^4+s_{-4} \ln _2-s_4 \ln _2+\frac{\ln _2^5}{5} \end{dmath}
  \begin{dmath}[style={\small}]    \label{s1scm4} s_1 \bar{s}_{-4}  =   \bar{s}_{1,-4}-\frac{3}{4} \zeta _2^2 \bar{s}_{-1}+\frac{7}{20} \zeta _2^2 \bar{s}_1-\zeta _2 \bar{s}_{-3}-\zeta _3 \bar{s}_{-2}+s_{-4,1}-\frac{3 \zeta _3 \zeta _2}{2}+\frac{15 \zeta _5}{16}-\frac{3}{2} \zeta _2^2 \ln _2-\frac{3}{4} \zeta _2^2 s_{-1}-\frac{7}{20} \zeta _2^2 s_1-\zeta _2 s_{-3}+\zeta _3 s_{-2} \end{dmath}
  \begin{dmath}[style={\small}]     s_1 \bar{s}_4  =   \bar{s}_{1,4}-\frac{2}{5} \zeta _2^2 \bar{s}_1-\zeta _2 \bar{s}_3-\zeta _3 \bar{s}_2-s_{4,1}+5 \zeta _5+\frac{2}{5} \zeta _2^2 s_1+\zeta _2 s_3-\zeta _3 s_2 \end{dmath}
  \begin{dmath}[style={\small}]     s_1 \bar{s}_{-3,-1}  =   -\bar{s}_{-4,-1}+\bar{s}_{-3,1,-1}+\bar{s}_{1,-3,-1}-\ln _2 \bar{s}_{-3,-1}+\ln _2 \bar{s}_{-3,1}+2 \text{Li}_4\left(\frac{1}{2}\right) \bar{s}_1-\frac{3}{5} \zeta _2^2 \bar{s}_1+\frac{5}{8} \zeta _3 \bar{s}_2-\frac{1}{2} \zeta _2 \bar{s}_3-\frac{1}{2} \zeta _2 \ln _2^2 \bar{s}_1+\zeta _2 \ln _2 \bar{s}_{-2}+\frac{7}{4} \zeta _3 \ln _2 \bar{s}_{-1}-\zeta _2 \ln _2 \bar{s}_2+\frac{1}{12} \ln _2^4 \bar{s}_1-\frac{1}{2} \ln _2^2 \bar{s}_{-3}+\frac{1}{2} \ln _2^2 \bar{s}_3-\ln _2 \bar{s}_{-4}+\ln _2 \bar{s}_4-s_{4,1}+s_{-3,-1,1}+\ln _2 s_{-3,-1}+\ln _2 s_{-3,1}+\frac{3 \zeta _2 \zeta _3}{8}-\frac{35 \zeta _5}{16}+\zeta _2 \ln _2^3+\frac{69}{20} \zeta _2^2 \ln _2-2 \text{Li}_4\left(\frac{1}{2}\right) s_1-4 \text{Li}_4\left(\frac{1}{2}\right) \ln _2+\frac{3}{5} \zeta _2^2 s_1+\frac{1}{2} \zeta _2 s_{-3}+\frac{\zeta _2 s_3}{2}+\frac{5 \zeta _3 s_2}{8}+\frac{1}{2} \zeta _2 s_1 \ln _2^2-\zeta _2 s_{-2} \ln _2-\zeta _2 s_2 \ln _2+\frac{7}{4} \zeta _3 s_{-1} \ln _2-\frac{1}{12} s_1 \ln _2^4+\frac{1}{2} s_{-3} \ln _2^2-\frac{1}{2} s_3 \ln _2^2-\frac{\ln _2^5}{6} \end{dmath}
  \begin{dmath}[style={\small}]     s_1 \bar{s}_{-3,1}  =   -\bar{s}_{-4,1}+\bar{s}_{-3,1,1}+\bar{s}_{1,-3,1}+2 \text{Li}_4\left(\frac{1}{2}\right) \bar{s}_{-1}-2 \text{Li}_4\left(\frac{1}{2}\right) \bar{s}_1+\zeta _2 \bar{s}_{-3}-2 \zeta _3 \bar{s}_{-2}-\frac{8}{5} \zeta _2^2 \bar{s}_{-1}+\frac{11}{10} \zeta _2^2 \bar{s}_1-\frac{1}{2} \zeta _2 \ln _2^2 \bar{s}_{-1}+\frac{1}{2} \zeta _2 \ln _2^2 \bar{s}_1+\frac{7}{4} \zeta _3 \ln _2 \bar{s}_{-1}-\frac{7}{4} \zeta _3 \ln _2 \bar{s}_1+\frac{1}{12} \ln _2^4 \bar{s}_{-1}-\frac{1}{12} \ln _2^4 \bar{s}_1+s_{-4,1}-s_{-3,1,1}+\frac{7 \zeta _2 \zeta _3}{8}+\frac{39 \zeta _5}{16}-\frac{1}{3} \zeta _2 \ln _2^3+\frac{7}{4} \zeta _3 \ln _2^2-\frac{16}{5} \zeta _2^2 \ln _2+2 \text{Li}_4\left(\frac{1}{2}\right) s_{-1}+2 \text{Li}_4\left(\frac{1}{2}\right) s_1-4 \text{Li}_5\left(\frac{1}{2}\right)-\frac{8}{5} \zeta _2^2 s_{-1}-\frac{11}{10} \zeta _2^2 s_1-\zeta _2 s_{-3}+2 \zeta _3 s_{-2}-\frac{1}{2} \zeta _2 s_{-1} \ln _2^2-\frac{1}{2} \zeta _2 s_1 \ln _2^2+\frac{7}{4} \zeta _3 s_{-1} \ln _2+\frac{7}{4} \zeta _3 s_1 \ln _2+\frac{1}{12} s_{-1} \ln _2^4+\frac{1}{12} s_1 \ln _2^4+\frac{\ln _2^5}{30} \end{dmath}
  \begin{dmath}[style={\small}] \label{s1sc2m2}    s_1 \bar{s}_{2,-2}  =   -\frac{3}{2} \zeta _2 \bar{s}_{2,-1}+\frac{1}{2} \zeta _2 \bar{s}_{2,1}-\bar{s}_{3,-2}+\bar{s}_{1,2,-2}+\bar{s}_{2,1,-2}+4 \text{Li}_4\left(\frac{1}{2}\right) \bar{s}_{-1}-4 \text{Li}_4\left(\frac{1}{2}\right) \bar{s}_1+\frac{1}{2} \zeta _2 \bar{s}_{-3}-\frac{13}{8} \zeta _3 \bar{s}_{-2}-\frac{17}{10} \zeta _2^2 \bar{s}_{-1}+\frac{17}{8} \zeta _2^2 \bar{s}_1+\frac{1}{8} \zeta _3 \bar{s}_2-\frac{1}{2} \zeta _2 \bar{s}_3-\zeta _2 \ln _2^2 \bar{s}_{-1}+\zeta _2 \ln _2^2 \bar{s}_1+\frac{3}{2} \zeta _2 \ln _2 \bar{s}_{-2}+\frac{7}{2} \zeta _3 \ln _2 \bar{s}_{-1}-\frac{7}{2} \zeta _3 \ln _2 \bar{s}_1-\frac{3}{2} \zeta _2 \ln _2 \bar{s}_2+\frac{1}{6} \ln _2^4 \bar{s}_{-1}-\frac{1}{6} \ln _2^4 \bar{s}_1+\frac{3}{2} \zeta _2 s_{2,-1}+\frac{1}{2} \zeta _2 s_{2,1}+s_{-4,1}-s_{2,-2,1}-\frac{\zeta _2 \zeta _3}{8}+\frac{83 \zeta _5}{16}-\frac{2}{3} \zeta _2 \ln _2^3+\frac{7}{2} \zeta _3 \ln _2^2-\frac{17}{5} \zeta _2^2 \ln _2+4 \text{Li}_4\left(\frac{1}{2}\right) s_{-1}+4 \text{Li}_4\left(\frac{1}{2}\right) s_1-8 \text{Li}_5\left(\frac{1}{2}\right)-\frac{17}{10} \zeta _2^2 s_{-1}-\frac{17}{8} \zeta _2^2 s_1-\zeta _2 s_{-3}+\frac{13}{8} \zeta _3 s_{-2}-\frac{5 \zeta _3 s_2}{8}-\zeta _2 s_{-1} \ln _2^2-\zeta _2 s_1 \ln _2^2-\frac{3}{2} \zeta _2 s_{-2} \ln _2+\frac{3}{2} \zeta _2 s_2 \ln _2+\frac{7}{2} \zeta _3 s_{-1} \ln _2+\frac{7}{2} \zeta _3 s_1 \ln _2+\frac{1}{6} s_{-1} \ln _2^4+\frac{1}{6} s_1 \ln _2^4+\frac{\ln _2^5}{15} \end{dmath}
  \begin{dmath}[style={\small}]     s_1 \bar{s}_{3,-1}  =   -\bar{s}_{4,-1}+\bar{s}_{1,3,-1}+\bar{s}_{3,1,-1}-\ln _2 \bar{s}_{3,-1}+\ln _2 \bar{s}_{3,1}+2 \text{Li}_4\left(\frac{1}{2}\right) \bar{s}_{-1}-\frac{1}{2} \zeta _2 \bar{s}_{-3}+\frac{5}{8} \zeta _3 \bar{s}_{-2}-\frac{19}{40} \zeta _2^2 \bar{s}_{-1}-\frac{1}{8} \zeta _2^2 \bar{s}_1-\frac{1}{2} \zeta _2 \ln _2^2 \bar{s}_{-1}-\zeta _2 \ln _2 \bar{s}_{-2}+\frac{7}{4} \zeta _3 \ln _2 \bar{s}_1+\zeta _2 \ln _2 \bar{s}_2+\frac{1}{12} \ln _2^4 \bar{s}_{-1}+\frac{1}{2} \ln _2^2 \bar{s}_{-3}-\frac{1}{2} \ln _2^2 \bar{s}_3+\ln _2 \bar{s}_{-4}-\ln _2 \bar{s}_4+s_{-4,1}-s_{3,-1,1}-\ln _2 s_{3,-1}-\ln _2 s_{3,1}-\frac{9 \zeta _2 \zeta _3}{8}+\frac{163 \zeta _5}{16}-\frac{2}{3} \zeta _2 \ln _2^3-\frac{22}{5} \zeta _2^2 \ln _2+2 \text{Li}_4\left(\frac{1}{2}\right) s_{-1}-8 \text{Li}_5\left(\frac{1}{2}\right)-\frac{19}{40} \zeta _2^2 s_{-1}+\frac{1}{8} \zeta _2^2 s_1-\frac{1}{2} \zeta _2 s_{-3}-\frac{\zeta _2 s_3}{2}-\frac{5}{8} \zeta _3 s_{-2}-\frac{1}{2} \zeta _2 s_{-1} \ln _2^2+\zeta _2 s_{-2} \ln _2+\zeta _2 s_2 \ln _2-\frac{7}{4} \zeta _3 s_1 \ln _2+\frac{1}{12} s_{-1} \ln _2^4+\frac{1}{2} s_{-3} \ln _2^2-\frac{1}{2} s_3 \ln _2^2+\frac{\ln _2^5}{15} \end{dmath}
  \begin{dmath}[style={\small}]     s_1 \bar{s}_{3,1}  =   -\bar{s}_{4,1}+\bar{s}_{1,3,1}+\bar{s}_{3,1,1}-\frac{1}{2} \zeta _2^2 \bar{s}_1+\zeta _2 \bar{s}_3-2 \zeta _3 \bar{s}_2-s_{4,1}+s_{3,1,1}-2 \zeta _3 \zeta _2+\frac{15 \zeta _5}{2}+\frac{1}{2} \zeta _2^2 s_1+\zeta _2 s_3-2 \zeta _3 s_2 \end{dmath}
  \begin{dmath}[style={\small}]     s_1 \bar{s}_{-2,-1,-1}  =   -\frac{1}{2} \zeta _2 \bar{s}_{-2,1}-\bar{s}_{-3,-1,-1}-\bar{s}_{-2,-2,-1}+\bar{s}_{-2,-1,1,-1}+\bar{s}_{-2,1,-1,-1}+\bar{s}_{1,-2,-1,-1}+\frac{1}{2} \ln _2^2 \bar{s}_{-2,-1}-\frac{1}{2} \ln _2^2 \bar{s}_{-2,1}-\ln _2 \bar{s}_{-2,-2}+\ln _2 \bar{s}_{-2,2}-\ln _2 \bar{s}_{-2,-1,-1}+\ln _2 \bar{s}_{-2,-1,1}-5 \text{Li}_4\left(\frac{1}{2}\right) \bar{s}_{-1}+4 \text{Li}_4\left(\frac{1}{2}\right) \bar{s}_1+\zeta _3 \bar{s}_{-2}+\frac{11}{8} \zeta _2^2 \bar{s}_{-1}-\frac{7}{5} \zeta _2^2 \bar{s}_1-\frac{1}{4} \zeta _2 \ln _2^2 \bar{s}_{-1}+\frac{1}{2} \zeta _2 \ln _2^2 \bar{s}_1-\frac{1}{2} \zeta _2 \ln _2 \bar{s}_{-2}-\frac{7}{2} \zeta _3 \ln _2 \bar{s}_{-1}+\frac{21}{8} \zeta _3 \ln _2 \bar{s}_1+\frac{1}{2} \zeta _2 \ln _2 \bar{s}_2-\frac{5}{24} \ln _2^4 \bar{s}_{-1}+\frac{1}{6} \ln _2^4 \bar{s}_1+\frac{1}{2} \ln _2^3 \bar{s}_{-2}-\frac{1}{2} \ln _2^3 \bar{s}_2+\ln _2^2 \bar{s}_{-3}-\ln _2^2 \bar{s}_3+\frac{1}{2} \zeta _2 s_{-2,-1}+\frac{1}{2} \zeta _2 s_{-2,1}+s_{-4,1}-s_{-2,2,1}-s_{3,-1,1}+s_{-2,-1,-1,1}+\frac{3}{2} \ln _2^2 s_{-2,-1}+\frac{1}{2} \ln _2^2 s_{-2,1}-\ln _2 s_{3,-1}-\ln _2 s_{3,1}+\ln _2 s_{-2,-1,-1}+\ln _2 s_{-2,-1,1}-\frac{7 \zeta _2 \zeta _3}{8}-\frac{849 \zeta _5}{64}-\frac{25}{6} \zeta _2 \ln _2^3+\frac{5}{4} \zeta _3 \ln _2^2+\frac{7}{5} \zeta _2^2 \ln _2-5 \text{Li}_4\left(\frac{1}{2}\right) s_{-1}-4 \text{Li}_4\left(\frac{1}{2}\right) s_1+16 \text{Li}_5\left(\frac{1}{2}\right)+10 \text{Li}_4\left(\frac{1}{2}\right) \ln _2+\frac{11}{8} \zeta _2^2 s_{-1}+\frac{7}{5} \zeta _2^2 s_1-\frac{1}{2} \zeta _2 s_{-3}-\frac{\zeta _2 s_3}{2}-\frac{5}{8} \zeta _3 s_{-2}-\frac{1}{4} \zeta _2 s_{-1} \ln _2^2-\frac{1}{2} \zeta _2 s_1 \ln _2^2+\frac{3}{2} \zeta _2 s_{-2} \ln _2+\frac{1}{2} \zeta _2 s_2 \ln _2-\frac{7}{2} \zeta _3 s_{-1} \ln _2-\frac{21}{8} \zeta _3 s_1 \ln _2-\frac{5}{24} s_{-1} \ln _2^4-\frac{1}{6} s_1 \ln _2^4+\frac{1}{2} s_{-2} \ln _2^3-\frac{1}{2} s_2 \ln _2^3+\frac{1}{2} s_{-3} \ln _2^2-\frac{1}{2} s_3 \ln _2^2+\frac{17 \ln _2^5}{60} \end{dmath}
  \begin{dmath}[style={\small}]     s_1 \bar{s}_{-2,-1,1}  =   \frac{1}{2} \zeta _2 \bar{s}_{-2,-1}+\frac{1}{2} \zeta _2 \bar{s}_{-2,1}-\bar{s}_{-3,-1,1}-\bar{s}_{-2,-2,1}+\bar{s}_{-2,-1,1,1}+\bar{s}_{-2,1,-1,1}+\bar{s}_{1,-2,-1,1}+\frac{1}{2} \ln _2^2 \bar{s}_{-2,-1}-\frac{1}{2} \ln _2^2 \bar{s}_{-2,1}+3 \text{Li}_4\left(\frac{1}{2}\right) \bar{s}_1-\frac{1}{2} \zeta _2 \bar{s}_{-3}+\frac{1}{8} \zeta _3 \bar{s}_{-2}+\frac{3}{4} \zeta _2^2 \bar{s}_{-1}-\frac{39}{40} \zeta _2^2 \bar{s}_1-\frac{7}{4} \zeta _3 \bar{s}_2+\frac{1}{2} \zeta _2 \bar{s}_3-\frac{3}{4} \zeta _2 \ln _2^2 \bar{s}_{-1}+\frac{1}{2} \zeta _2 \ln _2 \bar{s}_{-2}+\zeta _2 \ln _2 \bar{s}_2+\frac{1}{8} \ln _2^4 \bar{s}_1+\frac{1}{3} \ln _2^3 \bar{s}_{-2}-\frac{1}{3} \ln _2^3 \bar{s}_2+\frac{1}{2} \ln _2^2 \bar{s}_{-3}-\frac{1}{2} \ln _2^2 \bar{s}_3-\frac{3}{2} \zeta _2 s_{-2,-1}-\frac{1}{2} \zeta _2 s_{-2,1}-s_{4,1}+s_{-2,-2,1}+s_{3,1,1}-s_{-2,-1,1,1}+\frac{1}{2} \ln _2^2 s_{-2,-1}+\frac{1}{2} \ln _2^2 s_{-2,1}-\frac{5 \zeta _2 \zeta _3}{2}+\frac{599 \zeta _5}{64}-2 \zeta _2 \ln _2^3+\frac{3}{2} \zeta _3 \ln _2^2+\frac{33}{40} \zeta _2^2 \ln _2-3 \text{Li}_4\left(\frac{1}{2}\right) s_1-6 \text{Li}_5\left(\frac{1}{2}\right)+\frac{3}{4} \zeta _2^2 s_{-1}+\frac{39}{40} \zeta _2^2 s_1+\zeta _2 s_3-\frac{1}{4} \zeta _3 s_{-2}-\frac{7 \zeta _3 s_2}{4}-\frac{3}{4} \zeta _2 s_{-1} \ln _2^2-\zeta _2 s_{-2} \ln _2+\zeta _2 s_2 \ln _2-\frac{1}{8} s_1 \ln _2^4+\frac{1}{3} s_{-2} \ln _2^3-\frac{1}{3} s_2 \ln _2^3+\frac{\ln _2^5}{20} \end{dmath}
  \begin{dmath}[style={\small}]     s_1 \bar{s}_{-2,1,-1}  =   -\frac{1}{2} \zeta _2 \bar{s}_{-2,-1}-\bar{s}_{-3,1,-1}-\bar{s}_{-2,2,-1}+2 \bar{s}_{-2,1,1,-1}+\bar{s}_{1,-2,1,-1}+\ln _2 \bar{s}_{-2,-2}-\ln _2 \bar{s}_{-2,2}-\ln _2 \bar{s}_{-2,1,-1}+\ln _2 \bar{s}_{-2,1,1}+\frac{1}{4} \zeta _3 \bar{s}_{-2}-\frac{3}{40} \zeta _2^2 \bar{s}_1-\frac{1}{8} \zeta _3 \bar{s}_2+\frac{3}{4} \zeta _2 \ln _2^2 \bar{s}_{-1}-\frac{3}{4} \zeta _2 \ln _2^2 \bar{s}_1-\frac{3}{2} \zeta _2 \ln _2 \bar{s}_{-2}+\frac{21}{8} \zeta _3 \ln _2 \bar{s}_{-1}-\frac{1}{6} \ln _2^3 \bar{s}_{-2}+\frac{1}{6} \ln _2^3 \bar{s}_2-\frac{1}{2} \ln _2^2 \bar{s}_{-3}+\frac{1}{2} \ln _2^2 \bar{s}_3-\frac{1}{2} \zeta _2 s_{-2,-1}-\frac{1}{2} \zeta _2 s_{-2,1}-s_{4,1}+s_{-3,-1,1}+s_{-2,-2,1}-s_{-2,1,-1,1}-\ln _2^2 s_{-2,1}+\ln _2 s_{-3,-1}+\ln _2 s_{-3,1}-\ln _2 s_{-2,1,-1}-\ln _2 s_{-2,1,1}-\frac{17 \zeta _2 \zeta _3}{16}-\frac{181 \zeta _5}{32}+\frac{19}{6} \zeta _2 \ln _2^3-\frac{5}{8} \zeta _3 \ln _2^2+\frac{13}{4} \zeta _2^2 \ln _2+8 \text{Li}_5\left(\frac{1}{2}\right)-4 \text{Li}_4\left(\frac{1}{2}\right) \ln _2+\frac{3}{40} \zeta _2^2 s_1+\frac{1}{2} \zeta _2 s_{-3}+\frac{\zeta _2 s_3}{2}+\frac{3}{4} \zeta _3 s_{-2}-\frac{\zeta _3 s_2}{8}+\frac{3}{4} \zeta _2 s_{-1} \ln _2^2+\frac{3}{4} \zeta _2 s_1 \ln _2^2-2 \zeta _2 s_{-2} \ln _2+\frac{21}{8} \zeta _3 s_{-1} \ln _2-\frac{1}{6} s_{-2} \ln _2^3+\frac{1}{6} s_2 \ln _2^3+\frac{1}{2} s_{-3} \ln _2^2-\frac{1}{2} s_3 \ln _2^2-\frac{7 \ln _2^5}{30} \end{dmath}
  \begin{dmath}[style={\small}]     s_1 \bar{s}_{-2,1,1}  =   \zeta _2 \bar{s}_{-2,1}-\bar{s}_{-3,1,1}-\bar{s}_{-2,2,1}+2 \bar{s}_{-2,1,1,1}+\bar{s}_{1,-2,1,1}-\text{Li}_4\left(\frac{1}{2}\right) \bar{s}_{-1}+\text{Li}_4\left(\frac{1}{2}\right) \bar{s}_1+\zeta _3 \bar{s}_{-2}-\frac{43}{40} \zeta _2^2 \bar{s}_{-1}-\frac{1}{8} \zeta _2^2 \bar{s}_1+\frac{1}{4} \zeta _2 \ln _2^2 \bar{s}_{-1}-\frac{1}{4} \zeta _2 \ln _2^2 \bar{s}_1-\frac{7}{8} \zeta _3 \ln _2 \bar{s}_{-1}+\frac{7}{8} \zeta _3 \ln _2 \bar{s}_1-\frac{1}{24} \ln _2^4 \bar{s}_{-1}+\frac{1}{24} \ln _2^4 \bar{s}_1+\zeta _2 s_{-2,1}+s_{-4,1}-s_{-3,1,1}-s_{-2,2,1}+s_{-2,1,1,1}+\frac{\zeta _2 \zeta _3}{16}+\frac{7 \zeta _5}{4}+\frac{1}{6} \zeta _2 \ln _2^3-\frac{7}{8} \zeta _3 \ln _2^2-\frac{43}{20} \zeta _2^2 \ln _2-\text{Li}_4\left(\frac{1}{2}\right) s_{-1}-\text{Li}_4\left(\frac{1}{2}\right) s_1+2 \text{Li}_5\left(\frac{1}{2}\right)-\frac{43}{40} \zeta _2^2 s_{-1}+\frac{1}{8} \zeta _2^2 s_1-\zeta _2 s_{-3}+2 \zeta _3 s_{-2}+\frac{1}{4} \zeta _2 s_{-1} \ln _2^2+\frac{1}{4} \zeta _2 s_1 \ln _2^2-\frac{7}{8} \zeta _3 s_{-1} \ln _2-\frac{7}{8} \zeta _3 s_1 \ln _2-\frac{1}{24} s_{-1} \ln _2^4-\frac{1}{24} s_1 \ln _2^4-\frac{\ln _2^5}{60} \end{dmath}
  \begin{dmath}[style={\small}]     s_1 \bar{s}_{2,-1,-1}  =   -\frac{1}{2} \zeta _2 \bar{s}_{2,1}-\bar{s}_{2,-2,-1}-\bar{s}_{3,-1,-1}+\bar{s}_{1,2,-1,-1}+\bar{s}_{2,-1,1,-1}+\bar{s}_{2,1,-1,-1}+\frac{1}{2} \ln _2^2 \bar{s}_{2,-1}-\frac{1}{2} \ln _2^2 \bar{s}_{2,1}-\ln _2 \bar{s}_{2,-2}+\ln _2 \bar{s}_{2,2}-\ln _2 \bar{s}_{2,-1,-1}+\ln _2 \bar{s}_{2,-1,1}-\text{Li}_4\left(\frac{1}{2}\right) \bar{s}_1-\frac{1}{40} \zeta _2^2 \bar{s}_1+\zeta _3 \bar{s}_2+\frac{3}{2} \zeta _2 \ln _2^2 \bar{s}_{-1}-\frac{5}{4} \zeta _2 \ln _2^2 \bar{s}_1+\frac{1}{2} \zeta _2 \ln _2 \bar{s}_{-2}-\frac{7}{8} \zeta _3 \ln _2 \bar{s}_{-1}-\frac{1}{2} \zeta _2 \ln _2 \bar{s}_2-\frac{1}{24} \ln _2^4 \bar{s}_1-\frac{1}{2} \ln _2^3 \bar{s}_{-2}+\frac{1}{2} \ln _2^3 \bar{s}_2-\ln _2^2 \bar{s}_{-3}+\ln _2^2 \bar{s}_3-\frac{1}{2} \zeta _2 s_{2,-1}-\frac{1}{2} \zeta _2 s_{2,1}-s_{4,1}+s_{-3,-1,1}+s_{2,2,1}-s_{2,-1,-1,1}-\frac{3}{2} \ln _2^2 s_{2,-1}-\frac{1}{2} \ln _2^2 s_{2,1}+\ln _2 s_{-3,-1}+\ln _2 s_{-3,1}-\ln _2 s_{2,-1,-1}-\ln _2 s_{2,-1,1}+2 \zeta _2 \zeta _3-\frac{765 \zeta _5}{64}+\frac{7}{3} \zeta _2 \ln _2^3+\frac{11}{8} \zeta _3 \ln _2^2+\frac{27}{20} \zeta _2^2 \ln _2+\text{Li}_4\left(\frac{1}{2}\right) s_1+10 \text{Li}_5\left(\frac{1}{2}\right)+6 \text{Li}_4\left(\frac{1}{2}\right) \ln _2+\frac{1}{40} \zeta _2^2 s_1+\frac{1}{2} \zeta _2 s_{-3}+\frac{\zeta _2 s_3}{2}+\frac{5 \zeta _3 s_2}{8}+\frac{3}{2} \zeta _2 s_{-1} \ln _2^2+\frac{5}{4} \zeta _2 s_1 \ln _2^2-\frac{1}{2} \zeta _2 s_{-2} \ln _2-\frac{3}{2} \zeta _2 s_2 \ln _2-\frac{7}{8} \zeta _3 s_{-1} \ln _2+\frac{1}{24} s_1 \ln _2^4+\frac{1}{2} s_{-2} \ln _2^3-\frac{1}{2} s_2 \ln _2^3+\frac{1}{2} s_{-3} \ln _2^2-\frac{1}{2} s_3 \ln _2^2+\frac{\ln _2^5}{6} \end{dmath}
  \begin{dmath}[style={\small}]     s_1 \bar{s}_{2,-1,1}  =   \frac{1}{2} \zeta _2 \bar{s}_{2,-1}+\frac{1}{2} \zeta _2 \bar{s}_{2,1}-\bar{s}_{2,-2,1}-\bar{s}_{3,-1,1}+\bar{s}_{1,2,-1,1}+\bar{s}_{2,-1,1,1}+\bar{s}_{2,1,-1,1}+\frac{1}{2} \ln _2^2 \bar{s}_{2,-1}-\frac{1}{2} \ln _2^2 \bar{s}_{2,1}+7 \text{Li}_4\left(\frac{1}{2}\right) \bar{s}_{-1}-4 \text{Li}_4\left(\frac{1}{2}\right) \bar{s}_1+\frac{1}{2} \zeta _2 \bar{s}_{-3}-\frac{7}{4} \zeta _3 \bar{s}_{-2}-\frac{89}{40} \zeta _2^2 \bar{s}_{-1}+2 \zeta _2^2 \bar{s}_1+\frac{1}{8} \zeta _3 \bar{s}_2-\frac{1}{2} \zeta _2 \bar{s}_3-\zeta _2 \ln _2^2 \bar{s}_{-1}+\frac{1}{4} \zeta _2 \ln _2^2 \bar{s}_1+\zeta _2 \ln _2 \bar{s}_{-2}+\frac{21}{8} \zeta _3 \ln _2 \bar{s}_{-1}-\frac{21}{8} \zeta _3 \ln _2 \bar{s}_1+\frac{1}{2} \zeta _2 \ln _2 \bar{s}_2+\frac{7}{24} \ln _2^4 \bar{s}_{-1}-\frac{1}{6} \ln _2^4 \bar{s}_1-\frac{1}{3} \ln _2^3 \bar{s}_{-2}+\frac{1}{3} \ln _2^3 \bar{s}_2-\frac{1}{2} \ln _2^2 \bar{s}_{-3}+\frac{1}{2} \ln _2^2 \bar{s}_3+\frac{3}{2} \zeta _2 s_{2,-1}+\frac{1}{2} \zeta _2 s_{2,1}+s_{-4,1}-s_{-3,1,1}-s_{2,-2,1}+s_{2,-1,1,1}-\frac{1}{2} \ln _2^2 s_{2,-1}-\frac{1}{2} \ln _2^2 s_{2,1}-\frac{17 \zeta _2 \zeta _3}{16}+\frac{307 \zeta _5}{16}-\frac{1}{3} \zeta _2 \ln _2^3+\frac{15}{4} \zeta _3 \ln _2^2-\frac{127}{20} \zeta _2^2 \ln _2+7 \text{Li}_4\left(\frac{1}{2}\right) s_{-1}+4 \text{Li}_4\left(\frac{1}{2}\right) s_1-22 \text{Li}_5\left(\frac{1}{2}\right)-\frac{89}{40} \zeta _2^2 s_{-1}-2 \zeta _2^2 s_1-\zeta _2 s_{-3}+\frac{7}{4} \zeta _3 s_{-2}+\frac{\zeta _3 s_2}{4}-\zeta _2 s_{-1} \ln _2^2-\frac{1}{4} \zeta _2 s_1 \ln _2^2-\zeta _2 s_{-2} \ln _2+\zeta _2 s_2 \ln _2+\frac{21}{8} \zeta _3 s_{-1} \ln _2+\frac{21}{8} \zeta _3 s_1 \ln _2+\frac{7}{24} s_{-1} \ln _2^4+\frac{1}{6} s_1 \ln _2^4+\frac{1}{3} s_{-2} \ln _2^3-\frac{1}{3} s_2 \ln _2^3+\frac{11 \ln _2^5}{60} \end{dmath}
  \begin{dmath}[style={\small}]     s_1 \bar{s}_{2,1,-1}  =   -\frac{1}{2} \zeta _2 \bar{s}_{2,-1}-\bar{s}_{2,2,-1}-\bar{s}_{3,1,-1}+\bar{s}_{1,2,1,-1}+2 \bar{s}_{2,1,1,-1}+\ln _2 \bar{s}_{2,-2}-\ln _2 \bar{s}_{2,2}-\ln _2 \bar{s}_{2,1,-1}+\ln _2 \bar{s}_{2,1,1}+3 \text{Li}_4\left(\frac{1}{2}\right) \bar{s}_{-1}-3 \text{Li}_4\left(\frac{1}{2}\right) \bar{s}_1-\frac{1}{8} \zeta _3 \bar{s}_{-2}-\frac{51}{40} \zeta _2^2 \bar{s}_{-1}+\frac{6}{5} \zeta _2^2 \bar{s}_1+\frac{1}{4} \zeta _3 \bar{s}_2-\frac{3}{2} \zeta _2 \ln _2^2 \bar{s}_{-1}+\frac{3}{2} \zeta _2 \ln _2^2 \bar{s}_1+\frac{7}{2} \zeta _3 \ln _2 \bar{s}_{-1}-\frac{7}{8} \zeta _3 \ln _2 \bar{s}_1-\frac{3}{2} \zeta _2 \ln _2 \bar{s}_2+\frac{1}{8} \ln _2^4 \bar{s}_{-1}-\frac{1}{8} \ln _2^4 \bar{s}_1+\frac{1}{6} \ln _2^3 \bar{s}_{-2}-\frac{1}{6} \ln _2^3 \bar{s}_2+\frac{1}{2} \ln _2^2 \bar{s}_{-3}-\frac{1}{2} \ln _2^2 \bar{s}_3+\frac{1}{2} \zeta _2 s_{2,-1}+\frac{1}{2} \zeta _2 s_{2,1}+s_{-4,1}-s_{2,-2,1}-s_{3,-1,1}+s_{2,1,-1,1}+\ln _2^2 s_{2,1}-\ln _2 s_{3,-1}-\ln _2 s_{3,1}+\ln _2 s_{2,1,-1}+\ln _2 s_{2,1,1}-\frac{7 \zeta _2 \zeta _3}{16}+\frac{577 \zeta _5}{64}-\frac{7}{6} \zeta _2 \ln _2^3+\frac{5}{8} \zeta _3 \ln _2^2-\frac{129}{40} \zeta _2^2 \ln _2+3 \text{Li}_4\left(\frac{1}{2}\right) s_{-1}+3 \text{Li}_4\left(\frac{1}{2}\right) s_1-8 \text{Li}_5\left(\frac{1}{2}\right)-4 \text{Li}_4\left(\frac{1}{2}\right) \ln _2-\frac{51}{40} \zeta _2^2 s_{-1}-\frac{6}{5} \zeta _2^2 s_1-\frac{1}{2} \zeta _2 s_{-3}-\frac{\zeta _2 s_3}{2}+\frac{1}{8} \zeta _3 s_{-2}-\frac{3 \zeta _3 s_2}{4}-\frac{3}{2} \zeta _2 s_{-1} \ln _2^2-\frac{3}{2} \zeta _2 s_1 \ln _2^2+2 \zeta _2 s_2 \ln _2+\frac{7}{2} \zeta _3 s_{-1} \ln _2+\frac{7}{8} \zeta _3 s_1 \ln _2+\frac{1}{8} s_{-1} \ln _2^4+\frac{1}{8} s_1 \ln _2^4-\frac{1}{6} s_{-2} \ln _2^3+\frac{1}{6} s_2 \ln _2^3+\frac{1}{2} s_{-3} \ln _2^2-\frac{1}{2} s_3 \ln _2^2-\frac{\ln _2^5}{10} \end{dmath}
  \begin{dmath}[style={\small}]     s_1 \bar{s}_{2,1,1}  =   \zeta _2 \bar{s}_{2,1}-\bar{s}_{2,2,1}-\bar{s}_{3,1,1}+\bar{s}_{1,2,1,1}+2 \bar{s}_{2,1,1,1}-\frac{6}{5} \zeta _2^2 \bar{s}_1+\zeta _3 \bar{s}_2-\zeta _2 s_{2,1}-s_{4,1}+s_{2,2,1}+s_{3,1,1}-s_{2,1,1,1}-\zeta _3 \zeta _2+5 \zeta _5+\frac{6}{5} \zeta _2^2 s_1+\zeta _2 s_3-2 \zeta _3 s_2 \end{dmath}
  \begin{dmath}[style={\small}]     s_1 \bar{s}_{-1,1,-1,-1}  =   \frac{\ln _2^5}{12}-\frac{1}{3} s_{-1} \ln _2^4+\frac{1}{24} s_1 \ln _2^4-\frac{1}{24} \bar{s}_1 \ln _2^4+\frac{1}{2} s_{-2} \ln _2^3-\frac{1}{2} s_2 \ln _2^3-\frac{17}{12} \zeta _2 \ln _2^3+\frac{1}{2} \bar{s}_{-2} \ln _2^3-\frac{1}{2} \bar{s}_2 \ln _2^3-s_{-1,1} \ln _2^3+\frac{1}{2} s_{-3} \ln _2^2-\frac{1}{2} s_3 \ln _2^2-2 s_{-1} \zeta _2 \ln _2^2-\frac{1}{4} s_1 \zeta _2 \ln _2^2+\frac{41}{16} \zeta _3 \ln _2^2+\frac{3}{4} \zeta _2 \bar{s}_{-1} \ln _2^2+\frac{1}{4} \zeta _2 \bar{s}_1 \ln _2^2+\frac{3}{2} s_{-2,-1} \ln _2^2+\frac{1}{2} s_{-2,1} \ln _2^2+\frac{1}{2} s_{-1,-2} \ln _2^2-\frac{1}{2} s_{-1,2} \ln _2^2-\bar{s}_{-1,-2} \ln _2^2+\bar{s}_{-1,2} \ln _2^2-\frac{3}{2} s_{-1,1,-1} \ln _2^2-\frac{1}{2} s_{-1,1,1} \ln _2^2+\frac{1}{2} \bar{s}_{-1,1,-1} \ln _2^2-\frac{1}{2} \bar{s}_{-1,1,1} \ln _2^2-\frac{31}{40} \zeta _2^2 \ln _2+3 \text{Li}_4\left(\frac{1}{2}\right) \ln _2+\frac{3}{2} s_{-2} \zeta _2 \ln _2+\frac{1}{2} s_2 \zeta _2 \ln _2+\frac{1}{2} s_{-1} \zeta _3 \ln _2-\frac{1}{8} s_1 \zeta _3 \ln _2-\frac{1}{8} \zeta _3 \bar{s}_{-1} \ln _2+\frac{1}{8} \zeta _3 \bar{s}_1 \ln _2-\frac{1}{2} \zeta _2 s_{-1,-1} \ln _2-\frac{3}{2} \zeta _2 s_{-1,1} \ln _2-s_{3,-1} \ln _2-s_{3,1} \ln _2+\frac{1}{2} \zeta _2 \bar{s}_{-1,-1} \ln _2-\frac{1}{2} \zeta _2 \bar{s}_{-1,1} \ln _2+s_{-2,-1,-1} \ln _2+s_{-2,-1,1} \ln _2+s_{-1,-2,-1} \ln _2+s_{-1,-2,1} \ln _2-\bar{s}_{-1,1,-2} \ln _2+\bar{s}_{-1,1,2} \ln _2-s_{-1,1,-1,-1} \ln _2-s_{-1,1,-1,1} \ln _2-\bar{s}_{-1,1,-1,-1} \ln _2+\bar{s}_{-1,1,-1,1} \ln _2-\frac{3}{20} s_{-1} \zeta _2^2-\frac{1}{8} s_1 \zeta _2^2+3 \text{Li}_5\left(\frac{1}{2}\right)-\text{Li}_4\left(\frac{1}{2}\right) s_{-1}-\frac{1}{2} s_{-3} \zeta _2-\frac{s_3 \zeta _2}{2}-\frac{5}{8} s_{-2} \zeta _3-\frac{7 \zeta _2 \zeta _3}{16}-\frac{53 \zeta _5}{32}+\frac{13}{40} \zeta _2^2 \bar{s}_{-1}-\text{Li}_4\left(\frac{1}{2}\right) \bar{s}_{-1}+\frac{1}{8} \zeta _2^2 \bar{s}_1+s_{-4,1}+\frac{1}{2} \zeta _2 s_{-2,-1}+\frac{1}{2} \zeta _2 s_{-2,1}+\frac{1}{2} \zeta _2 s_{-1,-2}+\frac{5}{8} \zeta _3 s_{-1,1}+\frac{1}{2} \zeta _2 s_{-1,2}+\zeta _3 \bar{s}_{-1,1}-s_{-2,2,1}-\frac{1}{2} \zeta _2 s_{-1,1,-1}-\frac{1}{2} \zeta _2 s_{-1,1,1}-s_{-1,3,1}-s_{3,-1,1}-\frac{1}{2} \zeta _2 \bar{s}_{-1,1,1}+s_{-2,-1,-1,1}+s_{-1,-2,-1,1}+s_{-1,1,2,1}-\bar{s}_{-2,1,-1,-1}-\bar{s}_{-1,1,-2,-1}-\bar{s}_{-1,2,-1,-1}-s_{-1,1,-1,-1,1}+\bar{s}_{-1,1,-1,1,-1}+2 \bar{s}_{-1,1,1,-1,-1}+\bar{s}_{1,-1,1,-1,-1} \end{dmath}
  \begin{dmath}[style={\small}]     s_1 \bar{s}_{-1,1,-1,1}  =   \frac{\ln _2^5}{120}-\frac{1}{3} s_{-1} \ln _2^4+\frac{1}{24} s_1 \ln _2^4+\frac{1}{8} \bar{s}_{-1} \ln _2^4-\frac{1}{24} \bar{s}_1 \ln _2^4+\frac{1}{3} s_{-2} \ln _2^3-\frac{1}{3} s_2 \ln _2^3-\frac{1}{4} \zeta _2 \ln _2^3+\frac{1}{3} \bar{s}_{-2} \ln _2^3-\frac{1}{3} \bar{s}_2 \ln _2^3-\frac{2}{3} s_{-1,1} \ln _2^3-\frac{1}{4} s_1 \zeta _2 \ln _2^2+\frac{15}{16} \zeta _3 \ln _2^2+\frac{3}{4} \zeta _2 \bar{s}_{-1} \ln _2^2+\frac{1}{4} \zeta _2 \bar{s}_1 \ln _2^2+\frac{1}{2} s_{-2,-1} \ln _2^2+\frac{1}{2} s_{-2,1} \ln _2^2-\frac{1}{2} \bar{s}_{-1,-2} \ln _2^2+\frac{1}{2} \bar{s}_{-1,2} \ln _2^2-\frac{1}{2} s_{-1,1,-1} \ln _2^2-\frac{1}{2} s_{-1,1,1} \ln _2^2+\frac{1}{2} \bar{s}_{-1,1,-1} \ln _2^2-\frac{1}{2} \bar{s}_{-1,1,1} \ln _2^2-\frac{17}{40} \zeta _2^2 \ln _2-s_{-2} \zeta _2 \ln _2+s_2 \zeta _2 \ln _2-\frac{1}{8} s_{-1} \zeta _3 \ln _2-\frac{1}{8} s_1 \zeta _3 \ln _2-\frac{1}{2} \zeta _2 \bar{s}_{-2} \ln _2-\frac{1}{8} \zeta _3 \bar{s}_{-1} \ln _2+\frac{1}{8} \zeta _3 \bar{s}_1 \ln _2+\frac{1}{2} \zeta _2 \bar{s}_2 \ln _2-\frac{1}{2} \zeta _2 s_{-1,-1} \ln _2+\frac{3}{2} \zeta _2 s_{-1,1} \ln _2+\frac{1}{2} \zeta _2 \bar{s}_{-1,-1} \ln _2+\zeta _2 \bar{s}_{-1,1} \ln _2+\frac{27}{20} s_{-1} \zeta _2^2+\frac{3}{8} s_1 \zeta _2^2-\text{Li}_5\left(\frac{1}{2}\right)-3 \text{Li}_4\left(\frac{1}{2}\right) s_{-1}+s_3 \zeta _2-\frac{1}{4} s_{-2} \zeta _3-\frac{7 s_2 \zeta _3}{4}-\frac{11 \zeta _2 \zeta _3}{8}+\frac{289 \zeta _5}{64}+\frac{1}{8} \zeta _3 \bar{s}_{-2}-\frac{33}{40} \zeta _2^2 \bar{s}_{-1}+2 \text{Li}_4\left(\frac{1}{2}\right) \bar{s}_{-1}-\frac{3}{8} \zeta _2^2 \bar{s}_1-\frac{1}{8} \zeta _3 \bar{s}_2-\frac{3}{2} \zeta _2 s_{-2,-1}-\frac{1}{2} \zeta _2 s_{-2,1}-\zeta _2 s_{-1,-2}+\frac{13}{8} \zeta _3 s_{-1,-1}+\frac{1}{8} \zeta _3 s_{-1,1}-s_{4,1}+\frac{1}{2} \zeta _2 \bar{s}_{-1,-2}-\frac{13}{8} \zeta _3 \bar{s}_{-1,-1}-\frac{1}{2} \zeta _2 \bar{s}_{-1,2}+s_{-2,-2,1}+s_{-1,-3,1}+\frac{3}{2} \zeta _2 s_{-1,1,-1}+\frac{1}{2} \zeta _2 s_{-1,1,1}+s_{3,1,1}+\frac{1}{2} \zeta _2 \bar{s}_{-1,1,-1}+\frac{1}{2} \zeta _2 \bar{s}_{-1,1,1}-s_{-2,-1,1,1}-s_{-1,-2,1,1}-s_{-1,1,-2,1}-\bar{s}_{-2,1,-1,1}-\bar{s}_{-1,1,-2,1}-\bar{s}_{-1,2,-1,1}+s_{-1,1,-1,1,1}+\bar{s}_{-1,1,-1,1,1}+2 \bar{s}_{-1,1,1,-1,1}+\bar{s}_{1,-1,1,-1,1} \end{dmath}
  \begin{dmath}[style={\small}]     s_1 \bar{s}_{-1,1,1,1}  =   \zeta _3 \bar{s}_{-1,1}+\zeta _2 \bar{s}_{-1,1,1}-\bar{s}_{-2,1,1,1}-\bar{s}_{-1,1,2,1}-\bar{s}_{-1,2,1,1}+3 \bar{s}_{-1,1,1,1,1}+\bar{s}_{1,-1,1,1,1}-\text{Li}_4\left(\frac{1}{2}\right) \bar{s}_{-1}+\text{Li}_4\left(\frac{1}{2}\right) \bar{s}_1+\frac{2}{5} \zeta _2^2 \bar{s}_{-1}+\zeta _2 s_{-2,1}-2 \zeta _3 s_{-1,1}+\zeta _2 s_{-1,2}-\zeta _2 s_{-1,1,1}+s_{-4,1}-s_{-3,1,1}-s_{-2,2,1}-s_{-1,3,1}+s_{-2,1,1,1}+s_{-1,1,2,1}+s_{-1,2,1,1}-s_{-1,1,1,1,1}+\frac{7 \zeta _2 \zeta _3}{8}-\frac{127 \zeta _5}{32}-\frac{3}{20} \zeta _2^2 \ln _2-\text{Li}_4\left(\frac{1}{2}\right) s_{-1}-\text{Li}_4\left(\frac{1}{2}\right) s_1+5 \text{Li}_5\left(\frac{1}{2}\right)-\frac{6}{5} \zeta _2^2 s_{-1}-\zeta _2 s_{-3}+2 \zeta _3 s_{-2}+\frac{\ln _2^5}{40} \end{dmath}
  \begin{dmath}[style={\small}]     s_1 \bar{s}_{1,-3}  =   -\zeta _2 \bar{s}_{1,-2}-\frac{7}{4} \zeta _3 \bar{s}_{1,-1}+\frac{3}{4} \zeta _3 \bar{s}_{1,1}-\bar{s}_{2,-3}+2 \bar{s}_{1,1,-3}-4 \text{Li}_4\left(\frac{1}{2}\right) \bar{s}_{-1}+4 \text{Li}_4\left(\frac{1}{2}\right) \bar{s}_1+\frac{3}{4} \zeta _3 \bar{s}_{-2}+\frac{19}{20} \zeta _2^2 \bar{s}_{-1}-2 \zeta _2^2 \bar{s}_1-\frac{3}{4} \zeta _3 \bar{s}_2+\zeta _2 \ln _2^2 \bar{s}_{-1}-\zeta _2 \ln _2^2 \bar{s}_1-\frac{7}{4} \zeta _3 \ln _2 \bar{s}_{-1}+\frac{7}{4} \zeta _3 \ln _2 \bar{s}_1-\frac{1}{6} \ln _2^4 \bar{s}_{-1}+\frac{1}{6} \ln _2^4 \bar{s}_1+\zeta _2 s_{1,-2}-\frac{7}{4} \zeta _3 s_{1,-1}-\frac{3}{4} \zeta _3 s_{1,1}+s_{-4,1}-s_{1,-3,1}-\frac{25 \zeta _2 \zeta _3}{8}-\frac{63 \zeta _5}{16}+\frac{2}{3} \zeta _2 \ln _2^3-\frac{7}{4} \zeta _3 \ln _2^2+\frac{19}{10} \zeta _2^2 \ln _2-4 \text{Li}_4\left(\frac{1}{2}\right) s_{-1}+8 \text{Li}_5\left(\frac{1}{2}\right)+\frac{19}{20} \zeta _2^2 s_{-1}+\frac{3}{20} \zeta _2^2 s_1-\zeta _2 s_{-3}+\zeta _3 s_{-2}+\zeta _2 s_{-1} \ln _2^2-\frac{7}{4} \zeta _3 s_{-1} \ln _2-\frac{7}{4} \zeta _3 s_1 \ln _2-\frac{1}{6} s_{-1} \ln _2^4-\frac{\ln _2^5}{15} \end{dmath}
  \begin{dmath}[style={\small}]     s_1 \bar{s}_{1,3}  =   -\zeta _2 \bar{s}_{1,2}-\zeta _3 \bar{s}_{1,1}-\bar{s}_{2,3}+2 \bar{s}_{1,1,3}+\frac{6}{5} \zeta _2^2 \bar{s}_1-\zeta _2 s_{1,2}+\zeta _3 s_{1,1}-s_{4,1}+s_{1,3,1}-\zeta _3 \zeta _2+5 \zeta _5+\frac{2}{5} \zeta _2^2 s_1+\zeta _2 s_3-\zeta _3 s_2 \end{dmath}
  \begin{dmath}[style={\small}]     s_1 \bar{s}_{1,-2,-1}  =   \frac{5}{8} \zeta _3 \bar{s}_{1,1}-\frac{1}{2} \zeta _2 \bar{s}_{1,2}+\frac{3}{2} \zeta _2 \ln _2 \bar{s}_{1,-1}-\frac{3}{2} \zeta _2 \ln _2 \bar{s}_{1,1}-\bar{s}_{1,-3,-1}-\bar{s}_{2,-2,-1}+\bar{s}_{1,-2,1,-1}+2 \bar{s}_{1,1,-2,-1}-\frac{1}{2} \ln _2^2 \bar{s}_{1,-2}+\frac{1}{2} \ln _2^2 \bar{s}_{1,2}-\ln _2 \bar{s}_{1,-3}+\ln _2 \bar{s}_{1,3}-\ln _2 \bar{s}_{1,-2,-1}+\ln _2 \bar{s}_{1,-2,1}-4 \text{Li}_4\left(\frac{1}{2}\right) \bar{s}_1+\frac{51}{40} \zeta _2^2 \bar{s}_1-\frac{3}{2} \zeta _2 \ln _2^2 \bar{s}_{-1}+\frac{5}{2} \zeta _2 \ln _2^2 \bar{s}_1-\frac{1}{2} \zeta _2 \ln _2 \bar{s}_{-2}+\frac{7}{4} \zeta _3 \ln _2 \bar{s}_{-1}-\frac{7}{4} \zeta _3 \ln _2 \bar{s}_1+\frac{1}{2} \zeta _2 \ln _2 \bar{s}_2-\frac{1}{6} \ln _2^4 \bar{s}_1-\frac{1}{2} \zeta _2 s_{1,-2}-\frac{5}{8} \zeta _3 s_{1,1}-\frac{1}{2} \zeta _2 s_{1,2}+\frac{3}{2} \zeta _2 \ln _2 s_{1,-1}+\frac{3}{2} \zeta _2 \ln _2 s_{1,1}-s_{4,1}+s_{-3,-1,1}+s_{1,3,1}-s_{1,-2,-1,1}-\frac{1}{2} \ln _2^2 s_{1,-2}+\frac{1}{2} \ln _2^2 s_{1,2}+\ln _2 s_{-3,-1}+\ln _2 s_{-3,1}-\ln _2 s_{1,-2,-1}-\ln _2 s_{1,-2,1}-\frac{39 \zeta _2 \zeta _3}{16}-\frac{9 \zeta _5}{16}-\frac{7}{6} \zeta _2 \ln _2^3+\frac{7}{4} \zeta _3 \ln _2^2+\frac{33}{10} \zeta _2^2 \ln _2-2 \text{Li}_4\left(\frac{1}{2}\right) s_1+4 \text{Li}_5\left(\frac{1}{2}\right)+\frac{3}{5} \zeta _2^2 s_1+\frac{1}{2} \zeta _2 s_{-3}+\frac{\zeta _2 s_3}{2}+\frac{5 \zeta _3 s_2}{8}-\frac{3}{2} \zeta _2 s_{-1} \ln _2^2+2 \zeta _2 s_1 \ln _2^2-\zeta _2 s_{-2} \ln _2-\zeta _2 s_2 \ln _2+\frac{7}{4} \zeta _3 s_{-1} \ln _2-\frac{5}{4} \zeta _3 s_1 \ln _2-\frac{1}{12} s_1 \ln _2^4+\frac{1}{2} s_{-3} \ln _2^2-\frac{1}{2} s_3 \ln _2^2-\frac{\ln _2^5}{30} \end{dmath}
  \begin{dmath}[style={\small}]     s_1 \bar{s}_{1,-2,1}  =   \zeta _2 \bar{s}_{1,-2}-\frac{21}{8} \zeta _3 \bar{s}_{1,-1}+\frac{5}{8} \zeta _3 \bar{s}_{1,1}-\bar{s}_{1,-3,1}-\bar{s}_{2,-2,1}+\bar{s}_{1,-2,1,1}+2 \bar{s}_{1,1,-2,1}-3 \text{Li}_4\left(\frac{1}{2}\right) \bar{s}_{-1}+3 \text{Li}_4\left(\frac{1}{2}\right) \bar{s}_1+\frac{5}{8} \zeta _3 \bar{s}_{-2}+\frac{3}{20} \zeta _2^2 \bar{s}_{-1}-\frac{23}{40} \zeta _2^2 \bar{s}_1-\frac{5}{8} \zeta _3 \bar{s}_2+\frac{3}{4} \zeta _2 \ln _2^2 \bar{s}_{-1}-\frac{3}{4} \zeta _2 \ln _2^2 \bar{s}_1-\frac{1}{8} \ln _2^4 \bar{s}_{-1}+\frac{1}{8} \ln _2^4 \bar{s}_1+\zeta _2 s_{1,-2}-\frac{21}{8} \zeta _3 s_{1,-1}-\frac{5}{8} \zeta _3 s_{1,1}+s_{-4,1}-s_{-3,1,1}-s_{1,-3,1}+s_{1,-2,1,1}-\frac{7 \zeta _2 \zeta _3}{4}-\frac{107 \zeta _5}{32}+\frac{1}{2} \zeta _2 \ln _2^3+\frac{3}{10} \zeta _2^2 \ln _2-3 \text{Li}_4\left(\frac{1}{2}\right) s_{-1}+3 \text{Li}_4\left(\frac{1}{2}\right) s_1+6 \text{Li}_5\left(\frac{1}{2}\right)+\frac{3}{20} \zeta _2^2 s_{-1}-\frac{4}{5} \zeta _2^2 s_1-\zeta _2 s_{-3}+2 \zeta _3 s_{-2}+\frac{3}{4} \zeta _2 s_{-1} \ln _2^2-\frac{3}{4} \zeta _2 s_1 \ln _2^2-\frac{1}{8} s_{-1} \ln _2^4+\frac{1}{8} s_1 \ln _2^4-\frac{\ln _2^5}{20} \end{dmath}
  \begin{dmath}[style={\small}]     s_1 \bar{s}_{1,-1,-2}  =   -\frac{1}{2} \zeta _2 \bar{s}_{1,-2}+\frac{1}{8} \zeta _3 \bar{s}_{1,-1}-\frac{13}{8} \zeta _3 \bar{s}_{1,1}+\frac{1}{2} \zeta _2 \bar{s}_{1,2}-\frac{3}{2} \zeta _2 \bar{s}_{1,-1,-1}+\frac{1}{2} \zeta _2 \bar{s}_{1,-1,1}-\zeta _2 \ln _2 \bar{s}_{1,-1}+\zeta _2 \ln _2 \bar{s}_{1,1}-\bar{s}_{1,-2,-2}-\bar{s}_{2,-1,-2}+\bar{s}_{1,-1,1,-2}+2 \bar{s}_{1,1,-1,-2}+4 \text{Li}_4\left(\frac{1}{2}\right) \bar{s}_1+\frac{1}{4} \zeta _2^2 \bar{s}_{-1}-\frac{13}{40} \zeta _2^2 \bar{s}_1-\zeta _2 \ln _2^2 \bar{s}_1-\frac{1}{2} \zeta _2 \ln _2 \bar{s}_{-2}+\frac{1}{2} \zeta _2 \ln _2 \bar{s}_2+\frac{1}{6} \ln _2^4 \bar{s}_1-\frac{3}{2} \zeta _2 s_{-2,-1}-\frac{1}{2} \zeta _2 s_{-2,1}-\frac{5}{8} \zeta _3 s_{1,-1}+\frac{13}{8} \zeta _3 s_{1,1}-\zeta _2 s_{1,2}+\frac{3}{2} \zeta _2 s_{1,-1,-1}+\frac{1}{2} \zeta _2 s_{1,-1,1}+2 \zeta _2 \ln _2 s_{1,-1}-\zeta _2 \ln _2 s_{1,1}-s_{4,1}+s_{-2,-2,1}+s_{1,3,1}-s_{1,-1,-2,1}+\frac{5 \zeta _2 \zeta _3}{16}+\frac{89 \zeta _5}{16}-\frac{1}{6} \zeta _2 \ln _2^3-\frac{7}{10} \zeta _2^2 \ln _2-4 \text{Li}_5\left(\frac{1}{2}\right)+\frac{1}{4} \zeta _2^2 s_{-1}+\frac{13}{40} \zeta _2^2 s_1+\zeta _2 s_3+\frac{5}{8} \zeta _3 s_{-2}-\frac{13 \zeta _3 s_2}{8}+\frac{1}{2} \zeta _2 s_1 \ln _2^2-\frac{3}{2} \zeta _2 s_{-2} \ln _2+\frac{3}{2} \zeta _2 s_2 \ln _2+\frac{\ln _2^5}{30} \end{dmath}
  \begin{dmath}[style={\small}]     s_1 \bar{s}_{1,-1,2}  =   \zeta _3 \bar{s}_{1,-1}+\zeta _3 \bar{s}_{1,1}-\zeta _2 \bar{s}_{1,-1,1}+\frac{1}{2} \zeta _2 \ln _2 \bar{s}_{1,-1}-\frac{1}{2} \zeta _2 \ln _2 \bar{s}_{1,1}-\bar{s}_{1,-2,2}-\bar{s}_{2,-1,2}+\bar{s}_{1,-1,1,2}+2 \bar{s}_{1,1,-1,2}-5 \text{Li}_4\left(\frac{1}{2}\right) \bar{s}_{-1}+\text{Li}_4\left(\frac{1}{2}\right) \bar{s}_1+\zeta _3 \bar{s}_{-2}+\frac{5}{4} \zeta _2^2 \bar{s}_{-1}-\frac{37}{40} \zeta _2^2 \bar{s}_1-\zeta _3 \bar{s}_2+\frac{1}{2} \zeta _2 \ln _2^2 \bar{s}_{-1}+\frac{1}{2} \zeta _2 \ln _2^2 \bar{s}_1-\frac{1}{2} \zeta _2 \ln _2 \bar{s}_{-2}-\frac{21}{8} \zeta _3 \ln _2 \bar{s}_{-1}+\frac{21}{8} \zeta _3 \ln _2 \bar{s}_1+\frac{1}{2} \zeta _2 \ln _2 \bar{s}_2-\frac{5}{24} \ln _2^4 \bar{s}_{-1}+\frac{1}{24} \ln _2^4 \bar{s}_1+\zeta _2 s_{-2,1}+\zeta _2 s_{1,-2}-2 \zeta _3 s_{1,-1}-\zeta _3 s_{1,1}-\zeta _2 s_{1,-1,1}+\frac{1}{2} \zeta _2 \ln _2 s_{1,-1}+\frac{1}{2} \zeta _2 \ln _2 s_{1,1}+s_{-4,1}-s_{-2,2,1}-s_{1,-3,1}+s_{1,-1,2,1}-\frac{49 \zeta _2 \zeta _3}{16}-\frac{503 \zeta _5}{64}+\frac{2}{3} \zeta _2 \ln _2^3-\frac{21}{8} \zeta _3 \ln _2^2+\frac{139}{40} \zeta _2^2 \ln _2-5 \text{Li}_4\left(\frac{1}{2}\right) s_{-1}+\text{Li}_4\left(\frac{1}{2}\right) s_1+12 \text{Li}_5\left(\frac{1}{2}\right)+\frac{5}{4} \zeta _2^2 s_{-1}-\frac{1}{5} \zeta _2^2 s_1-\zeta _2 s_{-3}+\zeta _3 s_{-2}+\frac{1}{2} \zeta _2 s_{-1} \ln _2^2+\zeta _2 s_1 \ln _2^2-\frac{21}{8} \zeta _3 s_{-1} \ln _2-\frac{21}{8} \zeta _3 s_1 \ln _2-\frac{5}{24} s_{-1} \ln _2^4+\frac{1}{24} s_1 \ln _2^4-\frac{\ln _2^5}{10} \end{dmath}
  \begin{dmath}[style={\small}]     s_1 \bar{s}_{1,1,-2}  =   \frac{1}{2} \zeta _2 \bar{s}_{1,-2}-\frac{7}{4} \zeta _3 \bar{s}_{1,-1}+\frac{1}{4} \zeta _3 \bar{s}_{1,1}-\frac{1}{2} \zeta _2 \bar{s}_{1,2}-\frac{3}{2} \zeta _2 \bar{s}_{1,1,-1}+\frac{1}{2} \zeta _2 \bar{s}_{1,1,1}+\frac{3}{2} \zeta _2 \ln _2 \bar{s}_{1,-1}-\frac{3}{2} \zeta _2 \ln _2 \bar{s}_{1,1}-\bar{s}_{1,2,-2}-\bar{s}_{2,1,-2}+3 \bar{s}_{1,1,1,-2}-\text{Li}_4\left(\frac{1}{2}\right) \bar{s}_{-1}+\text{Li}_4\left(\frac{1}{2}\right) \bar{s}_1+\frac{1}{8} \zeta _3 \bar{s}_{-2}+\frac{1}{20} \zeta _2^2 \bar{s}_{-1}-\frac{27}{40} \zeta _2^2 \bar{s}_1-\frac{1}{8} \zeta _3 \bar{s}_2-\frac{1}{2} \zeta _2 \ln _2^2 \bar{s}_{-1}+\frac{1}{2} \zeta _2 \ln _2^2 \bar{s}_1+\frac{7}{8} \zeta _3 \ln _2 \bar{s}_{-1}-\frac{7}{8} \zeta _3 \ln _2 \bar{s}_1-\frac{1}{24} \ln _2^4 \bar{s}_{-1}+\frac{1}{24} \ln _2^4 \bar{s}_1+\zeta _2 s_{1,-2}-\frac{7}{4} \zeta _3 s_{1,-1}+\frac{1}{2} \zeta _3 s_{1,1}+\frac{3}{2} \zeta _2 s_{2,-1}+\frac{1}{2} \zeta _2 s_{2,1}-\frac{3}{2} \zeta _2 s_{1,1,-1}-\frac{1}{2} \zeta _2 s_{1,1,1}+\frac{3}{2} \zeta _2 \ln _2 s_{1,-1}-\frac{3}{2} \zeta _2 \ln _2 s_{1,1}+s_{-4,1}-s_{1,-3,1}-s_{2,-2,1}+s_{1,1,-2,1}-\frac{9 \zeta _2 \zeta _3}{4}-\frac{19 \zeta _5}{32}-\frac{1}{3} \zeta _2 \ln _2^3+\frac{7}{8} \zeta _3 \ln _2^2+\frac{1}{10} \zeta _2^2 \ln _2-\text{Li}_4\left(\frac{1}{2}\right) s_{-1}+3 \text{Li}_4\left(\frac{1}{2}\right) s_1+2 \text{Li}_5\left(\frac{1}{2}\right)+\frac{1}{20} \zeta _2^2 s_{-1}-\frac{57}{40} \zeta _2^2 s_1-\zeta _2 s_{-3}+\frac{13}{8} \zeta _3 s_{-2}-\frac{5 \zeta _3 s_2}{8}-\frac{1}{2} \zeta _2 s_{-1} \ln _2^2-\frac{3}{2} \zeta _2 s_{-2} \ln _2+\frac{3}{2} \zeta _2 s_2 \ln _2+\frac{7}{8} \zeta _3 s_{-1} \ln _2+\frac{7}{8} \zeta _3 s_1 \ln _2-\frac{1}{24} s_{-1} \ln _2^4+\frac{1}{8} s_1 \ln _2^4-\frac{\ln _2^5}{60} \end{dmath}
  \begin{dmath}[style={\small}]     s_1 \bar{s}_{1,1,2}  =   -\zeta _2 \bar{s}_{1,1,1}+2 \zeta _3 \bar{s}_{1,1}-\bar{s}_{1,2,2}-\bar{s}_{2,1,2}+3 \bar{s}_{1,1,1,2}+\frac{1}{2} \zeta _2^2 \bar{s}_1-\zeta _2 s_{1,2}-\zeta _2 s_{2,1}+\zeta _2 s_{1,1,1}+\zeta _3 s_{1,1}-s_{4,1}+s_{1,3,1}+s_{2,2,1}-s_{1,1,2,1}-2 \zeta _3 \zeta _2+\frac{15 \zeta _5}{2}+\frac{7}{10} \zeta _2^2 s_1+\zeta _2 s_3-\zeta _3 s_2 \end{dmath}
  \begin{dmath}[style={\small}]     s_1 \bar{s}_{1,2,-1}  =   -\frac{1}{2} \zeta _2 \bar{s}_{1,-2}+\frac{7}{8} \zeta _3 \bar{s}_{1,-1}-\frac{1}{4} \zeta _3 \bar{s}_{1,1}-\frac{3}{2} \zeta _2 \ln _2 \bar{s}_{1,-1}+\frac{3}{2} \zeta _2 \ln _2 \bar{s}_{1,1}-\bar{s}_{1,3,-1}-\bar{s}_{2,2,-1}+2 \bar{s}_{1,1,2,-1}+\bar{s}_{1,2,1,-1}+\frac{1}{2} \ln _2^2 \bar{s}_{1,-2}-\frac{1}{2} \ln _2^2 \bar{s}_{1,2}+\ln _2 \bar{s}_{1,-3}-\ln _2 \bar{s}_{1,3}-\ln _2 \bar{s}_{1,2,-1}+\ln _2 \bar{s}_{1,2,1}+\text{Li}_4\left(\frac{1}{2}\right) \bar{s}_{-1}+3 \text{Li}_4\left(\frac{1}{2}\right) \bar{s}_1-\frac{1}{4} \zeta _3 \bar{s}_{-2}-\frac{1}{20} \zeta _2^2 \bar{s}_{-1}-\frac{53}{40} \zeta _2^2 \bar{s}_1+\frac{1}{4} \zeta _3 \bar{s}_2+\frac{5}{4} \zeta _2 \ln _2^2 \bar{s}_{-1}-\frac{9}{4} \zeta _2 \ln _2^2 \bar{s}_1+\frac{1}{2} \zeta _2 \ln _2 \bar{s}_{-2}-\frac{7}{4} \zeta _3 \ln _2 \bar{s}_{-1}+\frac{7}{4} \zeta _3 \ln _2 \bar{s}_1-\frac{1}{2} \zeta _2 \ln _2 \bar{s}_2+\frac{1}{24} \ln _2^4 \bar{s}_{-1}+\frac{1}{8} \ln _2^4 \bar{s}_1+\frac{1}{2} \zeta _2 s_{1,-2}+\frac{7}{8} \zeta _3 s_{1,-1}+\frac{1}{4} \zeta _3 s_{1,1}+\frac{1}{2} \zeta _2 s_{1,2}-\frac{3}{2} \zeta _2 \ln _2 s_{1,-1}-\frac{3}{2} \zeta _2 \ln _2 s_{1,1}+s_{-4,1}-s_{1,-3,1}-s_{3,-1,1}+s_{1,2,-1,1}-\frac{1}{2} \ln _2^2 s_{1,-2}+\frac{1}{2} \ln _2^2 s_{1,2}-\ln _2 s_{3,-1}-\ln _2 s_{3,1}+\ln _2 s_{1,2,-1}+\ln _2 s_{1,2,1}+\frac{\zeta _2 \zeta _3}{4}+\frac{251 \zeta _5}{64}+\frac{2}{3} \zeta _2 \ln _2^3-\frac{7}{8} \zeta _3 \ln _2^2-\frac{17}{5} \zeta _2^2 \ln _2+\text{Li}_4\left(\frac{1}{2}\right) s_{-1}-5 \text{Li}_4\left(\frac{1}{2}\right) s_1-4 \text{Li}_5\left(\frac{1}{2}\right)+2 \text{Li}_4\left(\frac{1}{2}\right) \ln _2-\frac{1}{20} \zeta _2^2 s_{-1}+\frac{37}{20} \zeta _2^2 s_1-\frac{1}{2} \zeta _2 s_{-3}-\frac{\zeta _2 s_3}{2}-\frac{5}{8} \zeta _3 s_{-2}+\frac{5}{4} \zeta _2 s_{-1} \ln _2^2-\frac{1}{4} \zeta _2 s_1 \ln _2^2+\zeta _2 s_{-2} \ln _2+\zeta _2 s_2 \ln _2-\frac{7}{4} \zeta _3 s_{-1} \ln _2-4 \zeta _3 s_1 \ln _2+\frac{1}{24} s_{-1} \ln _2^4-\frac{5}{24} s_1 \ln _2^4+\frac{1}{2} s_{-3} \ln _2^2-\frac{1}{2} s_3 \ln _2^2+\frac{7 \ln _2^5}{60} \end{dmath}
  \begin{dmath}[style={\small}]     s_1 \bar{s}_{1,2,1}  =   \zeta _2 \bar{s}_{1,2}-2 \zeta _3 \bar{s}_{1,1}-\bar{s}_{1,3,1}-\bar{s}_{2,2,1}+2 \bar{s}_{1,1,2,1}+\bar{s}_{1,2,1,1}+\frac{7}{10} \zeta _2^2 \bar{s}_1-\zeta _2 s_{1,2}+2 \zeta _3 s_{1,1}-s_{4,1}+s_{1,3,1}+s_{3,1,1}-s_{1,2,1,1}+3 \zeta _3 \zeta _2-\frac{5 \zeta _5}{2}+\frac{1}{2} \zeta _2^2 s_1+\zeta _2 s_3-2 \zeta _3 s_2 \end{dmath}
  \begin{dmath}[style={\small}]     s_1 \bar{s}_{1,-1,-1,-1}  =   -\frac{17 \ln _2^5}{120}-\frac{1}{8} s_{-1} \ln _2^4-\frac{3}{4} s_1 \ln _2^4-\frac{1}{8} \bar{s}_{-1} \ln _2^4+\frac{1}{8} \bar{s}_1 \ln _2^4+\frac{1}{2} s_{-2} \ln _2^3-\frac{1}{2} s_2 \ln _2^3-\frac{3}{2} \zeta _2 \ln _2^3+\frac{2}{3} \bar{s}_{-2} \ln _2^3-\frac{2}{3} \bar{s}_2 \ln _2^3-\frac{7}{6} s_{1,-1} \ln _2^3-\frac{1}{6} s_{1,1} \ln _2^3-\frac{1}{6} \bar{s}_{1,-1} \ln _2^3+\frac{1}{6} \bar{s}_{1,1} \ln _2^3+\frac{1}{2} s_{-3} \ln _2^2-\frac{1}{2} s_3 \ln _2^2+\frac{1}{2} s_{-1} \zeta _2 \ln _2^2-\frac{3}{4} s_1 \zeta _2 \ln _2^2-\frac{17}{16} \zeta _3 \ln _2^2+\frac{1}{2} \zeta _2 \bar{s}_{-1} \ln _2^2-\frac{1}{2} \zeta _2 \bar{s}_1 \ln _2^2+\frac{3}{2} s_{-2,-1} \ln _2^2+\frac{1}{2} s_{-2,1} \ln _2^2-\frac{1}{2} s_{1,-2} \ln _2^2+\frac{1}{2} s_{1,2} \ln _2^2+\bar{s}_{1,-2} \ln _2^2-\bar{s}_{1,2} \ln _2^2-\frac{3}{2} s_{1,-1,-1} \ln _2^2-\frac{1}{2} s_{1,-1,1} \ln _2^2+\frac{1}{2} \bar{s}_{1,-1,-1} \ln _2^2-\frac{1}{2} \bar{s}_{1,-1,1} \ln _2^2+\frac{31}{20} \zeta _2^2 \ln _2+3 \text{Li}_4\left(\frac{1}{2}\right) \ln _2+\frac{3}{2} s_{-2} \zeta _2 \ln _2+\frac{1}{2} s_2 \zeta _2 \ln _2-\frac{21}{8} s_{-1} \zeta _3 \ln _2-\frac{11}{4} s_1 \zeta _3 \ln _2-\frac{21}{8} \zeta _3 \bar{s}_{-1} \ln _2+\frac{21}{8} \zeta _3 \bar{s}_1 \ln _2-\frac{3}{2} \zeta _2 s_{1,-1} \ln _2-\frac{1}{2} \zeta _2 s_{1,1} \ln _2-s_{3,-1} \ln _2-s_{3,1} \ln _2-\frac{1}{2} \zeta _2 \bar{s}_{1,-1} \ln _2+\frac{1}{2} \zeta _2 \bar{s}_{1,1} \ln _2+s_{-2,-1,-1} \ln _2+s_{-2,-1,1} \ln _2+s_{1,2,-1} \ln _2+s_{1,2,1} \ln _2-\bar{s}_{1,-1,-2} \ln _2+\bar{s}_{1,-1,2} \ln _2-s_{1,-1,-1,-1} \ln _2-s_{1,-1,-1,1} \ln _2-\bar{s}_{1,-1,-1,-1} \ln _2+\bar{s}_{1,-1,-1,1} \ln _2+\frac{39}{40} s_{-1} \zeta _2^2+\frac{7}{5} s_1 \zeta _2^2+10 \text{Li}_5\left(\frac{1}{2}\right)-3 \text{Li}_4\left(\frac{1}{2}\right) s_{-1}-4 \text{Li}_4\left(\frac{1}{2}\right) s_1-\frac{1}{2} s_{-3} \zeta _2-\frac{s_3 \zeta _2}{2}-\frac{5}{8} s_{-2} \zeta _3-\frac{\zeta _2 \zeta _3}{2}-\frac{589 \zeta _5}{64}+\frac{1}{4} \zeta _3 \bar{s}_{-2}+\frac{39}{40} \zeta _2^2 \bar{s}_{-1}-3 \text{Li}_4\left(\frac{1}{2}\right) \bar{s}_{-1}-\frac{53}{40} \zeta _2^2 \bar{s}_1+3 \text{Li}_4\left(\frac{1}{2}\right) \bar{s}_1-\frac{1}{4} \zeta _3 \bar{s}_2+s_{-4,1}+\frac{1}{2} \zeta _2 s_{-2,-1}+\frac{1}{2} \zeta _2 s_{-2,1}+\frac{1}{2} \zeta _2 s_{1,-2}+\frac{3}{8} \zeta _3 s_{1,-1}-\frac{1}{4} \zeta _3 s_{1,1}+\frac{1}{2} \zeta _2 s_{1,2}+\frac{3}{4} \zeta _3 \bar{s}_{1,-1}+\frac{1}{4} \zeta _3 \bar{s}_{1,1}-s_{-2,2,1}-s_{1,-3,1}-\frac{1}{2} \zeta _2 s_{1,-1,-1}-\frac{1}{2} \zeta _2 s_{1,-1,1}-s_{3,-1,1}-\frac{1}{2} \zeta _2 \bar{s}_{1,-1,1}+s_{-2,-1,-1,1}+s_{1,-1,2,1}+s_{1,2,-1,1}-\bar{s}_{1,-2,-1,-1}-\bar{s}_{1,-1,-2,-1}-\bar{s}_{2,-1,-1,-1}-s_{1,-1,-1,-1,1}+\bar{s}_{1,-1,-1,1,-1}+\bar{s}_{1,-1,1,-1,-1}+2 \bar{s}_{1,1,-1,-1,-1} \end{dmath}
  \begin{dmath}[style={\small}]     s_1 \bar{s}_{1,-1,-1,1}  =   -\frac{\ln _2^5}{6}-\frac{5}{8} s_1 \ln _2^4+\frac{1}{8} \bar{s}_1 \ln _2^4+\frac{1}{3} s_{-2} \ln _2^3-\frac{1}{3} s_2 \ln _2^3-\frac{5}{6} \zeta _2 \ln _2^3+\frac{1}{2} \bar{s}_{-2} \ln _2^3-\frac{1}{2} \bar{s}_2 \ln _2^3-\frac{5}{6} s_{1,-1} \ln _2^3-\frac{1}{6} s_{1,1} \ln _2^3-\frac{1}{6} \bar{s}_{1,-1} \ln _2^3+\frac{1}{6} \bar{s}_{1,1} \ln _2^3-\frac{1}{4} s_{-1} \zeta _2 \ln _2^2+\frac{1}{2} s_1 \zeta _2 \ln _2^2+\frac{5}{16} \zeta _3 \ln _2^2-\frac{1}{4} \zeta _2 \bar{s}_{-1} \ln _2^2+\frac{1}{4} \zeta _2 \bar{s}_1 \ln _2^2+\frac{1}{2} s_{-2,-1} \ln _2^2+\frac{1}{2} s_{-2,1} \ln _2^2+\frac{1}{2} \bar{s}_{1,-2} \ln _2^2-\frac{1}{2} \bar{s}_{1,2} \ln _2^2-\frac{1}{2} s_{1,-1,-1} \ln _2^2-\frac{1}{2} s_{1,-1,1} \ln _2^2+\frac{1}{2} \bar{s}_{1,-1,-1} \ln _2^2-\frac{1}{2} \bar{s}_{1,-1,1} \ln _2^2+\frac{5}{8} \zeta _2^2 \ln _2-s_{-2} \zeta _2 \ln _2+s_2 \zeta _2 \ln _2-\frac{1}{2} \zeta _2 \bar{s}_{-2} \ln _2+\frac{1}{2} \zeta _2 \bar{s}_2 \ln _2+\frac{3}{2} \zeta _2 s_{1,-1} \ln _2-\frac{1}{2} \zeta _2 s_{1,1} \ln _2+\zeta _2 \bar{s}_{1,-1} \ln _2+\frac{1}{2} \zeta _2 \bar{s}_{1,1} \ln _2+\frac{1}{4} s_{-1} \zeta _2^2+\frac{39}{40} s_1 \zeta _2^2-3 \text{Li}_4\left(\frac{1}{2}\right) s_1+s_3 \zeta _2-\frac{1}{4} s_{-2} \zeta _3-\frac{7 s_2 \zeta _3}{4}+\frac{37 \zeta _2 \zeta _3}{16}-\frac{145 \zeta _5}{64}+\frac{1}{4} \zeta _2^2 \bar{s}_{-1}-\frac{3}{8} \zeta _2^2 \bar{s}_1+3 \text{Li}_4\left(\frac{1}{2}\right) \bar{s}_1-\frac{3}{2} \zeta _2 s_{-2,-1}-\frac{1}{2} \zeta _2 s_{-2,1}+\frac{1}{4} \zeta _3 s_{1,-1}+\frac{7}{4} \zeta _3 s_{1,1}-\zeta _2 s_{1,2}-s_{4,1}-\frac{1}{2} \zeta _2 \bar{s}_{1,-2}+\frac{1}{8} \zeta _3 \bar{s}_{1,-1}-\frac{7}{4} \zeta _3 \bar{s}_{1,1}+\frac{1}{2} \zeta _2 \bar{s}_{1,2}+s_{-2,-2,1}+\frac{3}{2} \zeta _2 s_{1,-1,-1}+\frac{1}{2} \zeta _2 s_{1,-1,1}+s_{1,3,1}+s_{3,1,1}+\frac{1}{2} \zeta _2 \bar{s}_{1,-1,-1}+\frac{1}{2} \zeta _2 \bar{s}_{1,-1,1}-s_{-2,-1,1,1}-s_{1,-1,-2,1}-s_{1,2,1,1}-\bar{s}_{1,-2,-1,1}-\bar{s}_{1,-1,-2,1}-\bar{s}_{2,-1,-1,1}+s_{1,-1,-1,1,1}+\bar{s}_{1,-1,-1,1,1}+\bar{s}_{1,-1,1,-1,1}+2 \bar{s}_{1,1,-1,-1,1} \end{dmath}
  \begin{dmath}[style={\small}]     s_1 \bar{s}_{1,-1,1,-1}  =   -\frac{13 \ln _2^5}{120}-\frac{1}{6} s_1 \ln _2^4-\frac{1}{12} \bar{s}_1 \ln _2^4-\frac{1}{6} s_{-2} \ln _2^3+\frac{1}{6} s_2 \ln _2^3-\frac{1}{12} \zeta _2 \ln _2^3+\frac{1}{6} s_{1,-1} \ln _2^3-\frac{1}{6} s_{1,1} \ln _2^3-\frac{1}{6} \bar{s}_{1,-1} \ln _2^3+\frac{1}{6} \bar{s}_{1,1} \ln _2^3+\frac{1}{2} s_{-3} \ln _2^2-\frac{1}{2} s_3 \ln _2^2-\frac{1}{4} s_{-1} \zeta _2 \ln _2^2+\frac{9}{4} s_1 \zeta _2 \ln _2^2+\zeta _3 \ln _2^2-\frac{1}{4} \zeta _2 \bar{s}_{-1} \ln _2^2-\frac{3}{4} \zeta _2 \bar{s}_1 \ln _2^2-s_{-2,1} \ln _2^2-\frac{1}{2} s_{1,-2} \ln _2^2+\frac{1}{2} s_{1,2} \ln _2^2-\frac{1}{2} \bar{s}_{1,-2} \ln _2^2+\frac{1}{2} \bar{s}_{1,2} \ln _2^2+s_{1,-1,1} \ln _2^2+\frac{25}{8} \zeta _2^2 \ln _2+\text{Li}_4\left(\frac{1}{2}\right) \ln _2-2 s_{-2} \zeta _2 \ln _2+\frac{13}{8} s_{-1} \zeta _3 \ln _2-\frac{1}{2} s_1 \zeta _3 \ln _2-\frac{1}{2} \zeta _2 \bar{s}_{-2} \ln _2+\frac{13}{8} \zeta _3 \bar{s}_{-1} \ln _2-\frac{13}{8} \zeta _3 \bar{s}_1 \ln _2+\frac{1}{2} \zeta _2 \bar{s}_2 \ln _2+s_{-3,-1} \ln _2+s_{-3,1} \ln _2+\frac{5}{2} \zeta _2 s_{1,-1} \ln _2+\frac{1}{2} \zeta _2 s_{1,1} \ln _2-\zeta _2 \bar{s}_{1,-1} \ln _2-\frac{1}{2} \zeta _2 \bar{s}_{1,1} \ln _2-s_{-2,1,-1} \ln _2-s_{-2,1,1} \ln _2-s_{1,-2,-1} \ln _2-s_{1,-2,1} \ln _2+\bar{s}_{1,-1,-2} \ln _2-\bar{s}_{1,-1,2} \ln _2+s_{1,-1,1,-1} \ln _2+s_{1,-1,1,1} \ln _2-\bar{s}_{1,-1,1,-1} \ln _2+\bar{s}_{1,-1,1,1} \ln _2+\frac{3}{40} s_1 \zeta _2^2+8 \text{Li}_5\left(\frac{1}{2}\right)+\frac{1}{2} s_{-3} \zeta _2+\frac{s_3 \zeta _2}{2}+\frac{3}{4} s_{-2} \zeta _3-\frac{s_2 \zeta _3}{8}-\frac{9 \zeta _2 \zeta _3}{8}-\frac{181 \zeta _5}{32}+\frac{7}{8} \zeta _2^2 \bar{s}_1-2 \text{Li}_4\left(\frac{1}{2}\right) \bar{s}_1-\frac{1}{2} \zeta _2 s_{-2,-1}-\frac{1}{2} \zeta _2 s_{-2,1}-\frac{1}{2} \zeta _2 s_{1,-2}-\frac{3}{4} \zeta _3 s_{1,-1}+\frac{1}{8} \zeta _3 s_{1,1}-\frac{1}{2} \zeta _2 s_{1,2}-s_{4,1}+\frac{1}{4} \zeta _3 \bar{s}_{1,-1}-\frac{1}{8} \zeta _3 \bar{s}_{1,1}+s_{-3,-1,1}+s_{-2,-2,1}+\frac{1}{2} \zeta _2 s_{1,-1,-1}+\frac{1}{2} \zeta _2 s_{1,-1,1}+s_{1,3,1}-\frac{1}{2} \zeta _2 \bar{s}_{1,-1,-1}-s_{-2,1,-1,1}-s_{1,-2,-1,1}-s_{1,-1,-2,1}-\bar{s}_{1,-2,1,-1}-\bar{s}_{1,-1,2,-1}-\bar{s}_{2,-1,1,-1}+s_{1,-1,1,-1,1}+2 \bar{s}_{1,-1,1,1,-1}+2 \bar{s}_{1,1,-1,1,-1} \end{dmath}
  \begin{dmath}[style={\small}]     s_1 \bar{s}_{1,-1,1,1}  =   \frac{1}{8} \zeta _3 \bar{s}_{1,-1}+\frac{7}{8} \zeta _3 \bar{s}_{1,1}+\zeta _2 \bar{s}_{1,-1,1}+\frac{1}{2} \zeta _2 \ln _2 \bar{s}_{1,-1}-\frac{1}{2} \zeta _2 \ln _2 \bar{s}_{1,1}-\bar{s}_{1,-2,1,1}-\bar{s}_{1,-1,2,1}-\bar{s}_{2,-1,1,1}+2 \bar{s}_{1,-1,1,1,1}+2 \bar{s}_{1,1,-1,1,1}-\frac{1}{6} \ln _2^3 \bar{s}_{1,-1}+\frac{1}{6} \ln _2^3 \bar{s}_{1,1}-3 \text{Li}_4\left(\frac{1}{2}\right) \bar{s}_{-1}+\frac{7}{8} \zeta _3 \bar{s}_{-2}+\frac{1}{20} \zeta _2^2 \bar{s}_{-1}+\frac{1}{2} \zeta _2^2 \bar{s}_1-\frac{7}{8} \zeta _3 \bar{s}_2+\frac{1}{2} \zeta _2 \ln _2^2 \bar{s}_{-1}-\frac{1}{2} \zeta _2 \ln _2^2 \bar{s}_1-\frac{1}{2} \zeta _2 \ln _2 \bar{s}_{-2}-\zeta _3 \ln _2 \bar{s}_{-1}+\zeta _3 \ln _2 \bar{s}_1+\frac{1}{2} \zeta _2 \ln _2 \bar{s}_2-\frac{1}{8} \ln _2^4 \bar{s}_{-1}+\frac{1}{6} \ln _2^3 \bar{s}_{-2}-\frac{1}{6} \ln _2^3 \bar{s}_2+\zeta _2 s_{-2,1}+\zeta _2 s_{1,-2}-\frac{23}{8} \zeta _3 s_{1,-1}-\frac{7}{8} \zeta _3 s_{1,1}-\zeta _2 s_{1,-1,1}+\frac{1}{2} \zeta _2 \ln _2 s_{1,-1}+\frac{1}{2} \zeta _2 \ln _2 s_{1,1}+s_{-4,1}-s_{-3,1,1}-s_{-2,2,1}-s_{1,-3,1}+s_{-2,1,1,1}+s_{1,-2,1,1}+s_{1,-1,2,1}-s_{1,-1,1,1,1}-\frac{1}{6} \ln _2^3 s_{1,-1}-\frac{1}{6} \ln _2^3 s_{1,1}-\frac{9 \zeta _2 \zeta _3}{4}-\frac{281 \zeta _5}{64}+\frac{1}{12} \zeta _2 \ln _2^3-\zeta _3 \ln _2^2+\frac{13}{8} \zeta _2^2 \ln _2-3 \text{Li}_4\left(\frac{1}{2}\right) s_{-1}+3 \text{Li}_4\left(\frac{1}{2}\right) s_1+8 \text{Li}_5\left(\frac{1}{2}\right)+\frac{1}{20} \zeta _2^2 s_{-1}-\frac{23}{20} \zeta _2^2 s_1-\zeta _2 s_{-3}+2 \zeta _3 s_{-2}+\frac{1}{2} \zeta _2 s_{-1} \ln _2^2+\frac{1}{2} \zeta _2 s_1 \ln _2^2-\zeta _3 s_{-1} \ln _2-\zeta _3 s_1 \ln _2-\frac{1}{8} s_{-1} \ln _2^4-\frac{1}{8} s_1 \ln _2^4-\frac{\ln _2^5}{6} \end{dmath}
  \begin{dmath}[style={\small}]     s_1 \bar{s}_{1,1,-1,-1}  =   \frac{31 \ln _2^5}{120}+\frac{3}{8} s_1 \ln _2^4-\frac{1}{24} \bar{s}_1 \ln _2^4+\frac{1}{2} s_{-2} \ln _2^3-\frac{1}{2} s_2 \ln _2^3+\frac{7}{12} \zeta _2 \ln _2^3-\frac{1}{2} \bar{s}_{-2} \ln _2^3+\frac{1}{2} \bar{s}_2 \ln _2^3+s_{1,1} \ln _2^3+\frac{1}{2} s_{-3} \ln _2^2-\frac{1}{2} s_3 \ln _2^2+\frac{7}{4} s_1 \zeta _2 \ln _2^2+\frac{57}{16} \zeta _3 \ln _2^2+\zeta _2 \bar{s}_1 \ln _2^2-\frac{1}{2} s_{1,-2} \ln _2^2+\frac{1}{2} s_{1,2} \ln _2^2-\frac{3}{2} s_{2,-1} \ln _2^2-\frac{1}{2} s_{2,1} \ln _2^2-\bar{s}_{1,-2} \ln _2^2+\bar{s}_{1,2} \ln _2^2+\frac{3}{2} s_{1,1,-1} \ln _2^2+\frac{1}{2} s_{1,1,1} \ln _2^2+\frac{1}{2} \bar{s}_{1,1,-1} \ln _2^2-\frac{1}{2} \bar{s}_{1,1,1} \ln _2^2+\frac{3}{20} \zeta _2^2 \ln _2+5 \text{Li}_4\left(\frac{1}{2}\right) \ln _2-\frac{1}{2} s_{-2} \zeta _2 \ln _2-\frac{3}{2} s_2 \zeta _2 \ln _2+\frac{1}{8} s_{-1} \zeta _3 \ln _2-\frac{1}{2} s_1 \zeta _3 \ln _2+\frac{1}{8} \zeta _3 \bar{s}_{-1} \ln _2-\frac{1}{8} \zeta _3 \bar{s}_1 \ln _2+s_{-3,-1} \ln _2+s_{-3,1} \ln _2+\frac{1}{2} \zeta _2 s_{1,-1} \ln _2+\frac{3}{2} \zeta _2 s_{1,1} \ln _2+\frac{1}{2} \zeta _2 \bar{s}_{1,-1} \ln _2-\frac{1}{2} \zeta _2 \bar{s}_{1,1} \ln _2-s_{1,-2,-1} \ln _2-s_{1,-2,1} \ln _2-s_{2,-1,-1} \ln _2-s_{2,-1,1} \ln _2-\bar{s}_{1,1,-2} \ln _2+\bar{s}_{1,1,2} \ln _2+s_{1,1,-1,-1} \ln _2+s_{1,1,-1,1} \ln _2-\bar{s}_{1,1,-1,-1} \ln _2+\bar{s}_{1,1,-1,1} \ln _2+\frac{1}{40} s_1 \zeta _2^2+4 \text{Li}_5\left(\frac{1}{2}\right)+\text{Li}_4\left(\frac{1}{2}\right) s_1+\frac{1}{2} s_{-3} \zeta _2+\frac{s_3 \zeta _2}{2}+\frac{5 s_2 \zeta _3}{8}-\frac{21 \zeta _2 \zeta _3}{16}-\frac{21 \zeta _5}{64}+\frac{9}{20} \zeta _2^2 \bar{s}_1-\text{Li}_4\left(\frac{1}{2}\right) \bar{s}_1-\frac{1}{2} \zeta _2 s_{1,-2}-\frac{5}{8} \zeta _3 s_{1,1}-\frac{1}{2} \zeta _2 s_{1,2}-\frac{1}{2} \zeta _2 s_{2,-1}-\frac{1}{2} \zeta _2 s_{2,1}-s_{4,1}+\zeta _3 \bar{s}_{1,1}+s_{-3,-1,1}+\frac{1}{2} \zeta _2 s_{1,1,-1}+\frac{1}{2} \zeta _2 s_{1,1,1}+s_{1,3,1}+s_{2,2,1}-\frac{1}{2} \zeta _2 \bar{s}_{1,1,1}-s_{1,-2,-1,1}-s_{1,1,2,1}-s_{2,-1,-1,1}-\bar{s}_{1,1,-2,-1}-\bar{s}_{1,2,-1,-1}-\bar{s}_{2,1,-1,-1}+s_{1,1,-1,-1,1}+\bar{s}_{1,1,-1,1,-1}+3 \bar{s}_{1,1,1,-1,-1} \end{dmath}
  \begin{dmath}[style={\small}]     s_1 \bar{s}_{1,1,-1,1}  =   \frac{2 \ln _2^5}{15}+\frac{1}{24} s_{-1} \ln _2^4+\frac{5}{12} s_1 \ln _2^4+\frac{1}{24} \bar{s}_{-1} \ln _2^4+\frac{1}{24} \bar{s}_1 \ln _2^4+\frac{1}{3} s_{-2} \ln _2^3-\frac{1}{3} s_2 \ln _2^3+\frac{7}{12} \zeta _2 \ln _2^3-\frac{1}{3} \bar{s}_{-2} \ln _2^3+\frac{1}{3} \bar{s}_2 \ln _2^3+\frac{2}{3} s_{1,1} \ln _2^3-\frac{1}{4} s_{-1} \zeta _2 \ln _2^2-\frac{1}{2} s_1 \zeta _2 \ln _2^2+\frac{41}{16} \zeta _3 \ln _2^2-\frac{1}{4} \zeta _2 \bar{s}_{-1} \ln _2^2+\frac{5}{4} \zeta _2 \bar{s}_1 \ln _2^2-\frac{1}{2} s_{2,-1} \ln _2^2-\frac{1}{2} s_{2,1} \ln _2^2-\frac{1}{2} \bar{s}_{1,-2} \ln _2^2+\frac{1}{2} \bar{s}_{1,2} \ln _2^2+\frac{1}{2} s_{1,1,-1} \ln _2^2+\frac{1}{2} s_{1,1,1} \ln _2^2+\frac{1}{2} \bar{s}_{1,1,-1} \ln _2^2-\frac{1}{2} \bar{s}_{1,1,1} \ln _2^2-\frac{71}{20} \zeta _2^2 \ln _2-s_{-2} \zeta _2 \ln _2+s_2 \zeta _2 \ln _2+s_{-1} \zeta _3 \ln _2+s_1 \zeta _3 \ln _2+\frac{1}{2} \zeta _2 \bar{s}_{-2} \ln _2+\zeta _3 \bar{s}_{-1} \ln _2-\zeta _3 \bar{s}_1 \ln _2-\frac{1}{2} \zeta _2 \bar{s}_2 \ln _2+\frac{1}{2} \zeta _2 s_{1,-1} \ln _2-\frac{3}{2} \zeta _2 s_{1,1} \ln _2+\frac{1}{2} \zeta _2 \bar{s}_{1,-1} \ln _2+\zeta _2 \bar{s}_{1,1} \ln _2-\frac{21}{40} s_{-1} \zeta _2^2-\frac{3}{2} s_1 \zeta _2^2-8 \text{Li}_5\left(\frac{1}{2}\right)+\text{Li}_4\left(\frac{1}{2}\right) s_{-1}+4 \text{Li}_4\left(\frac{1}{2}\right) s_1-s_{-3} \zeta _2+\frac{7}{4} s_{-2} \zeta _3+\frac{s_2 \zeta _3}{4}-\frac{27 \zeta _2 \zeta _3}{16}+\frac{137 \zeta _5}{16}-\frac{1}{8} \zeta _3 \bar{s}_{-2}-\frac{21}{40} \zeta _2^2 \bar{s}_{-1}+\text{Li}_4\left(\frac{1}{2}\right) \bar{s}_{-1}-\frac{27}{40} \zeta _2^2 \bar{s}_1+\text{Li}_4\left(\frac{1}{2}\right) \bar{s}_1+\frac{1}{8} \zeta _3 \bar{s}_2+s_{-4,1}+\zeta _2 s_{1,-2}-\frac{13}{8} \zeta _3 s_{1,-1}-\frac{1}{8} \zeta _3 s_{1,1}+\frac{3}{2} \zeta _2 s_{2,-1}+\frac{1}{2} \zeta _2 s_{2,1}+\frac{1}{2} \zeta _2 \bar{s}_{1,-2}-\frac{13}{8} \zeta _3 \bar{s}_{1,-1}-\frac{1}{2} \zeta _2 \bar{s}_{1,2}-s_{-3,1,1}-s_{1,-3,1}-\frac{3}{2} \zeta _2 s_{1,1,-1}-\frac{1}{2} \zeta _2 s_{1,1,1}-s_{2,-2,1}+\frac{1}{2} \zeta _2 \bar{s}_{1,1,-1}+\frac{1}{2} \zeta _2 \bar{s}_{1,1,1}+s_{1,-2,1,1}+s_{1,1,-2,1}+s_{2,-1,1,1}-\bar{s}_{1,1,-2,1}-\bar{s}_{1,2,-1,1}-\bar{s}_{2,1,-1,1}-s_{1,1,-1,1,1}+\bar{s}_{1,1,-1,1,1}+3 \bar{s}_{1,1,1,-1,1} \end{dmath}
  \begin{dmath}[style={\small}]     s_1 \bar{s}_{1,1,1,-1}  =   -\frac{\ln _2^5}{24}+\frac{1}{24} s_{-1} \ln _2^4-\frac{1}{8} s_1 \ln _2^4+\frac{1}{24} \bar{s}_{-1} \ln _2^4-\frac{1}{6} s_{-2} \ln _2^3+\frac{1}{6} s_2 \ln _2^3-\frac{1}{2} \zeta _2 \ln _2^3+\frac{1}{6} \bar{s}_{-2} \ln _2^3-\frac{1}{6} \bar{s}_2 \ln _2^3-\frac{1}{3} s_{1,1} \ln _2^3+\frac{1}{2} s_{-3} \ln _2^2-\frac{1}{2} s_3 \ln _2^2-\frac{1}{4} s_{-1} \zeta _2 \ln _2^2-\frac{3}{4} s_1 \zeta _2 \ln _2^2-\frac{9}{8} \zeta _3 \ln _2^2-\frac{1}{4} \zeta _2 \bar{s}_{-1} \ln _2^2-\frac{3}{4} \zeta _2 \bar{s}_1 \ln _2^2-\frac{1}{2} s_{1,-2} \ln _2^2+\frac{1}{2} s_{1,2} \ln _2^2+s_{2,1} \ln _2^2+\frac{1}{2} \bar{s}_{1,-2} \ln _2^2-\frac{1}{2} \bar{s}_{1,2} \ln _2^2-s_{1,1,1} \ln _2^2-\frac{83}{40} \zeta _2^2 \ln _2-\text{Li}_4\left(\frac{1}{2}\right) \ln _2+2 s_2 \zeta _2 \ln _2+\frac{7}{8} s_{-1} \zeta _3 \ln _2-\frac{9}{4} s_1 \zeta _3 \ln _2+\frac{7}{8} \zeta _3 \bar{s}_{-1} \ln _2-\frac{7}{8} \zeta _3 \bar{s}_1 \ln _2-2 \zeta _2 s_{1,1} \ln _2-s_{3,-1} \ln _2-s_{3,1} \ln _2-\frac{3}{2} \zeta _2 \bar{s}_{1,1} \ln _2+s_{1,2,-1} \ln _2+s_{1,2,1} \ln _2+s_{2,1,-1} \ln _2+s_{2,1,1} \ln _2+\bar{s}_{1,1,-2} \ln _2-\bar{s}_{1,1,2} \ln _2-s_{1,1,1,-1} \ln _2-s_{1,1,1,1} \ln _2-\bar{s}_{1,1,1,-1} \ln _2+\bar{s}_{1,1,1,1} \ln _2-\frac{2}{5} s_{-1} \zeta _2^2+\frac{13}{40} s_1 \zeta _2^2-2 \text{Li}_5\left(\frac{1}{2}\right)+\text{Li}_4\left(\frac{1}{2}\right) s_{-1}-\text{Li}_4\left(\frac{1}{2}\right) s_1-\frac{1}{2} s_{-3} \zeta _2-\frac{s_3 \zeta _2}{2}+\frac{1}{8} s_{-2} \zeta _3-\frac{3 s_2 \zeta _3}{4}-\frac{7 \zeta _2 \zeta _3}{16}+\frac{23 \zeta _5}{8}-\frac{2}{5} \zeta _2^2 \bar{s}_{-1}+\text{Li}_4\left(\frac{1}{2}\right) \bar{s}_{-1}+s_{-4,1}+\frac{1}{2} \zeta _2 s_{1,-2}-\frac{1}{8} \zeta _3 s_{1,-1}+\frac{3}{4} \zeta _3 s_{1,1}+\frac{1}{2} \zeta _2 s_{1,2}+\frac{1}{2} \zeta _2 s_{2,-1}+\frac{1}{2} \zeta _2 s_{2,1}-\frac{1}{8} \zeta _3 \bar{s}_{1,-1}+\frac{1}{4} \zeta _3 \bar{s}_{1,1}-s_{1,-3,1}-\frac{1}{2} \zeta _2 s_{1,1,-1}-\frac{1}{2} \zeta _2 s_{1,1,1}-s_{2,-2,1}-s_{3,-1,1}-\frac{1}{2} \zeta _2 \bar{s}_{1,1,-1}+s_{1,1,-2,1}+s_{1,2,-1,1}+s_{2,1,-1,1}-\bar{s}_{1,1,2,-1}-\bar{s}_{1,2,1,-1}-\bar{s}_{2,1,1,-1}-s_{1,1,1,-1,1}+4 \bar{s}_{1,1,1,1,-1} \end{dmath}
  \begin{dmath}[style={\small}]     s_1 \bar{s}_{1,1,1,1}  =   \zeta _2 \bar{s}_{1,1,1}+\zeta _3 \bar{s}_{1,1}-\bar{s}_{1,1,2,1}-\bar{s}_{1,2,1,1}-\bar{s}_{2,1,1,1}+4 \bar{s}_{1,1,1,1,1}+\frac{2}{5} \zeta _2^2 \bar{s}_1-\zeta _2 s_{1,2}-\zeta _2 s_{2,1}+\zeta _2 s_{1,1,1}+2 \zeta _3 s_{1,1}-s_{4,1}+s_{1,3,1}+s_{2,2,1}+s_{3,1,1}-s_{1,1,2,1}-s_{1,2,1,1}-s_{2,1,1,1}+s_{1,1,1,1,1}+5 \zeta _5+\frac{6}{5} \zeta _2^2 s_1+\zeta _2 s_3-2 \zeta _3 s_2 \end{dmath}
  \begin{dmath}[style={\small}]     s_1 \bar{s}_{-1,-3}  =   -\zeta _2 \bar{s}_{-1,-2}-\frac{7}{4} \zeta _3 \bar{s}_{-1,-1}+\frac{3}{4} \zeta _3 \bar{s}_{-1,1}-\bar{s}_{-2,-3}+\bar{s}_{-1,1,-3}+\bar{s}_{1,-1,-3}+2 \text{Li}_4\left(\frac{1}{2}\right) \bar{s}_{-1}-2 \text{Li}_4\left(\frac{1}{2}\right) \bar{s}_1-\frac{3}{4} \zeta _3 \bar{s}_{-2}-\frac{5}{4} \zeta _2^2 \bar{s}_{-1}+\frac{1}{5} \zeta _2^2 \bar{s}_1+\frac{3}{4} \zeta _3 \bar{s}_2-\frac{1}{2} \zeta _2 \ln _2^2 \bar{s}_{-1}+\frac{1}{2} \zeta _2 \ln _2^2 \bar{s}_1+\frac{3}{4} \zeta _3 \ln _2 \bar{s}_{-1}-\frac{3}{4} \zeta _3 \ln _2 \bar{s}_1+\frac{1}{12} \ln _2^4 \bar{s}_{-1}-\frac{1}{12} \ln _2^4 \bar{s}_1-\zeta _2 s_{-1,-2}+\frac{7}{4} \zeta _3 s_{-1,-1}+\frac{3}{4} \zeta _3 s_{-1,1}-s_{4,1}+s_{-1,-3,1}+\frac{21 \zeta _2 \zeta _3}{8}+\frac{5 \zeta _5}{32}+\frac{3}{4} \zeta _3 \ln _2^2-\frac{9}{4} \zeta _2^2 \ln _2-2 \text{Li}_4\left(\frac{1}{2}\right) s_{-1}+2 \text{Li}_4\left(\frac{1}{2}\right) s_1+\frac{3}{5} \zeta _2^2 s_{-1}-\frac{1}{5} \zeta _2^2 s_1+\zeta _2 s_3-\zeta _3 s_2+\frac{1}{2} \zeta _2 s_{-1} \ln _2^2-\frac{1}{2} \zeta _2 s_1 \ln _2^2+\frac{3}{4} \zeta _3 s_{-1} \ln _2+\frac{3}{4} \zeta _3 s_1 \ln _2-\frac{1}{12} s_{-1} \ln _2^4+\frac{1}{12} s_1 \ln _2^4 \end{dmath}
  \begin{dmath}[style={\small}]     s_1 \bar{s}_{-1,3}  =   -\zeta _3 \bar{s}_{-1,1}-\zeta _2 \bar{s}_{-1,2}-\bar{s}_{-2,3}+\bar{s}_{-1,1,3}+\bar{s}_{1,-1,3}+\frac{29}{40} \zeta _2^2 \bar{s}_{-1}+\frac{19}{40} \zeta _2^2 \bar{s}_1+\frac{3}{4} \zeta _3 \ln _2 \bar{s}_{-1}-\frac{3}{4} \zeta _3 \ln _2 \bar{s}_1-\zeta _3 s_{-1,1}+\zeta _2 s_{-1,2}+s_{-4,1}-s_{-1,3,1}-\frac{3 \zeta _2 \zeta _3}{8}-\frac{47 \zeta _5}{8}+\frac{1}{3} \zeta _2 \ln _2^3+\frac{3}{4} \zeta _3 \ln _2^2+\frac{17}{10} \zeta _2^2 \ln _2+4 \text{Li}_5\left(\frac{1}{2}\right)-\frac{7}{8} \zeta _2^2 s_{-1}-\frac{19}{40} \zeta _2^2 s_1-\zeta _2 s_{-3}+\zeta _3 s_{-2}+\frac{3}{4} \zeta _3 s_{-1} \ln _2+\frac{3}{4} \zeta _3 s_1 \ln _2-\frac{\ln _2^5}{30} \end{dmath}
  \begin{dmath}[style={\small}]     s_1 \bar{s}_{-1,-2,-1}  =   \frac{5}{8} \zeta _3 \bar{s}_{-1,1}-\frac{1}{2} \zeta _2 \bar{s}_{-1,2}+\frac{3}{2} \zeta _2 \ln _2 \bar{s}_{-1,-1}-\frac{3}{2} \zeta _2 \ln _2 \bar{s}_{-1,1}-\bar{s}_{-2,-2,-1}-\bar{s}_{-1,-3,-1}+\bar{s}_{-1,-2,1,-1}+\bar{s}_{-1,1,-2,-1}+\bar{s}_{1,-1,-2,-1}-\frac{1}{2} \ln _2^2 \bar{s}_{-1,-2}+\frac{1}{2} \ln _2^2 \bar{s}_{-1,2}-\ln _2 \bar{s}_{-1,-3}+\ln _2 \bar{s}_{-1,3}-\ln _2 \bar{s}_{-1,-2,-1}+\ln _2 \bar{s}_{-1,-2,1}-4 \text{Li}_4\left(\frac{1}{2}\right) \bar{s}_1-\frac{1}{8} \zeta _2^2 \bar{s}_{-1}+\frac{7}{5} \zeta _2^2 \bar{s}_1+\frac{3}{2} \zeta _2 \ln _2^2 \bar{s}_{-1}-\frac{1}{2} \zeta _2 \ln _2^2 \bar{s}_1+\frac{1}{2} \zeta _2 \ln _2 \bar{s}_{-2}+\frac{5}{8} \zeta _3 \ln _2 \bar{s}_{-1}-\frac{5}{8} \zeta _3 \ln _2 \bar{s}_1-\frac{1}{2} \zeta _2 \ln _2 \bar{s}_2-\frac{1}{6} \ln _2^4 \bar{s}_1+\frac{1}{2} \zeta _2 s_{-1,-2}+\frac{5}{8} \zeta _3 s_{-1,1}+\frac{1}{2} \zeta _2 s_{-1,2}-\frac{3}{2} \zeta _2 \ln _2 s_{-1,-1}-\frac{3}{2} \zeta _2 \ln _2 s_{-1,1}+s_{-4,1}-s_{-1,3,1}-s_{3,-1,1}+s_{-1,-2,-1,1}+\frac{1}{2} \ln _2^2 s_{-1,-2}-\frac{1}{2} \ln _2^2 s_{-1,2}-\ln _2 s_{3,-1}-\ln _2 s_{3,1}+\ln _2 s_{-1,-2,-1}+\ln _2 s_{-1,-2,1}-\frac{3 \zeta _2 \zeta _3}{8}+\frac{685 \zeta _5}{64}+\frac{5}{3} \zeta _2 \ln _2^3+\frac{3}{2} \zeta _3 \ln _2^2-\frac{151}{40} \zeta _2^2 \ln _2+6 \text{Li}_4\left(\frac{1}{2}\right) s_{-1}+4 \text{Li}_4\left(\frac{1}{2}\right) s_1-10 \text{Li}_5\left(\frac{1}{2}\right)-4 \text{Li}_4\left(\frac{1}{2}\right) \ln _2-2 \zeta _2^2 s_{-1}-\frac{7}{5} \zeta _2^2 s_1-\frac{1}{2} \zeta _2 s_{-3}-\frac{\zeta _2 s_3}{2}-\frac{5}{8} \zeta _3 s_{-2}-3 \zeta _2 s_{-1} \ln _2^2+\frac{1}{2} \zeta _2 s_1 \ln _2^2+\zeta _2 s_{-2} \ln _2+\zeta _2 s_2 \ln _2+\frac{29}{8} \zeta _3 s_{-1} \ln _2+\frac{5}{8} \zeta _3 s_1 \ln _2+\frac{1}{4} s_{-1} \ln _2^4+\frac{1}{6} s_1 \ln _2^4+\frac{1}{2} s_{-3} \ln _2^2-\frac{1}{2} s_3 \ln _2^2-\frac{\ln _2^5}{12} \end{dmath}
  \begin{dmath}[style={\small}]     s_1 \bar{s}_{-1,-2,1}  =   \zeta _2 \bar{s}_{-1,-2}-\frac{21}{8} \zeta _3 \bar{s}_{-1,-1}+\frac{5}{8} \zeta _3 \bar{s}_{-1,1}-\bar{s}_{-2,-2,1}-\bar{s}_{-1,-3,1}+\bar{s}_{-1,-2,1,1}+\bar{s}_{-1,1,-2,1}+\bar{s}_{1,-1,-2,1}+3 \text{Li}_4\left(\frac{1}{2}\right) \bar{s}_{-1}-3 \text{Li}_4\left(\frac{1}{2}\right) \bar{s}_1-\frac{5}{8} \zeta _3 \bar{s}_{-2}-\frac{13}{20} \zeta _2^2 \bar{s}_{-1}+\frac{9}{40} \zeta _2^2 \bar{s}_1+\frac{5}{8} \zeta _3 \bar{s}_2-\frac{3}{4} \zeta _2 \ln _2^2 \bar{s}_{-1}+\frac{3}{4} \zeta _2 \ln _2^2 \bar{s}_1+\frac{5}{8} \zeta _3 \ln _2 \bar{s}_{-1}-\frac{5}{8} \zeta _3 \ln _2 \bar{s}_1+\frac{1}{8} \ln _2^4 \bar{s}_{-1}-\frac{1}{8} \ln _2^4 \bar{s}_1-\zeta _2 s_{-1,-2}+\frac{21}{8} \zeta _3 s_{-1,-1}+\frac{5}{8} \zeta _3 s_{-1,1}-s_{4,1}+s_{-1,-3,1}+s_{3,1,1}-s_{-1,-2,1,1}-\frac{11 \zeta _2 \zeta _3}{16}+\frac{305 \zeta _5}{64}+\frac{5}{8} \zeta _3 \ln _2^2-\frac{9}{8} \zeta _2^2 \ln _2-3 \text{Li}_4\left(\frac{1}{2}\right) s_{-1}+3 \text{Li}_4\left(\frac{1}{2}\right) s_1+\frac{29}{40} \zeta _2^2 s_{-1}-\frac{9}{40} \zeta _2^2 s_1+\zeta _2 s_3-2 \zeta _3 s_2+\frac{3}{4} \zeta _2 s_{-1} \ln _2^2-\frac{3}{4} \zeta _2 s_1 \ln _2^2+\frac{5}{8} \zeta _3 s_{-1} \ln _2+\frac{5}{8} \zeta _3 s_1 \ln _2-\frac{1}{8} s_{-1} \ln _2^4+\frac{1}{8} s_1 \ln _2^4 \end{dmath}
  \begin{dmath}[style={\small}]     s_1 \bar{s}_{-1,-1,-2}  =   -\frac{1}{2} \zeta _2 \bar{s}_{-1,-2}+\frac{1}{8} \zeta _3 \bar{s}_{-1,-1}-\frac{13}{8} \zeta _3 \bar{s}_{-1,1}+\frac{1}{2} \zeta _2 \bar{s}_{-1,2}-\frac{3}{2} \zeta _2 \bar{s}_{-1,-1,-1}+\frac{1}{2} \zeta _2 \bar{s}_{-1,-1,1}-\zeta _2 \ln _2 \bar{s}_{-1,-1}+\zeta _2 \ln _2 \bar{s}_{-1,1}-\bar{s}_{-2,-1,-2}-\bar{s}_{-1,-2,-2}+\bar{s}_{-1,-1,1,-2}+\bar{s}_{-1,1,-1,-2}+\bar{s}_{1,-1,-1,-2}+4 \text{Li}_4\left(\frac{1}{2}\right) \bar{s}_{-1}-\frac{27}{40} \zeta _2^2 \bar{s}_{-1}+\frac{3}{5} \zeta _2^2 \bar{s}_1-\frac{5}{4} \zeta _2 \ln _2^2 \bar{s}_{-1}+\frac{1}{4} \zeta _2 \ln _2^2 \bar{s}_1+\frac{1}{2} \zeta _2 \ln _2 \bar{s}_{-2}+\zeta _3 \ln _2 \bar{s}_{-1}-\zeta _3 \ln _2 \bar{s}_1-\frac{1}{2} \zeta _2 \ln _2 \bar{s}_2+\frac{1}{6} \ln _2^4 \bar{s}_{-1}+\frac{5}{8} \zeta _3 s_{-1,-1}-\frac{13}{8} \zeta _3 s_{-1,1}+\zeta _2 s_{-1,2}+\frac{3}{2} \zeta _2 s_{2,-1}+\frac{1}{2} \zeta _2 s_{2,1}-\frac{3}{2} \zeta _2 s_{-1,-1,-1}-\frac{1}{2} \zeta _2 s_{-1,-1,1}-2 \zeta _2 \ln _2 s_{-1,-1}+\zeta _2 \ln _2 s_{-1,1}+s_{-4,1}-s_{-1,3,1}-s_{2,-2,1}+s_{-1,-1,-2,1}-\frac{\zeta _2 \zeta _3}{16}-\frac{7 \zeta _5}{64}-\frac{7}{6} \zeta _2 \ln _2^3+\zeta _3 \ln _2^2-\frac{9}{40} \zeta _2^2 \ln _2-2 \text{Li}_5\left(\frac{1}{2}\right)-\frac{27}{40} \zeta _2^2 s_{-1}-\frac{3}{5} \zeta _2^2 s_1-\zeta _2 s_{-3}+\frac{13}{8} \zeta _3 s_{-2}-\frac{5 \zeta _3 s_2}{8}-\frac{3}{4} \zeta _2 s_{-1} \ln _2^2-\frac{1}{4} \zeta _2 s_1 \ln _2^2-\frac{3}{2} \zeta _2 s_{-2} \ln _2+\frac{3}{2} \zeta _2 s_2 \ln _2+\zeta _3 s_{-1} \ln _2+\zeta _3 s_1 \ln _2+\frac{\ln _2^5}{60} \end{dmath}
  \begin{dmath}[style={\small}]     s_1 \bar{s}_{-1,-1,2}  =   \zeta _3 \bar{s}_{-1,-1}+\zeta _3 \bar{s}_{-1,1}-\zeta _2 \bar{s}_{-1,-1,1}+\frac{1}{2} \zeta _2 \ln _2 \bar{s}_{-1,-1}-\frac{1}{2} \zeta _2 \ln _2 \bar{s}_{-1,1}-\bar{s}_{-2,-1,2}-\bar{s}_{-1,-2,2}+\bar{s}_{-1,-1,1,2}+\bar{s}_{-1,1,-1,2}+\bar{s}_{1,-1,-1,2}-\text{Li}_4\left(\frac{1}{2}\right) \bar{s}_{-1}-3 \text{Li}_4\left(\frac{1}{2}\right) \bar{s}_1-\zeta _3 \bar{s}_{-2}-\frac{1}{20} \zeta _2^2 \bar{s}_{-1}+\frac{3}{8} \zeta _2^2 \bar{s}_1+\zeta _3 \bar{s}_2+\frac{3}{4} \zeta _2 \ln _2^2 \bar{s}_{-1}+\frac{1}{4} \zeta _2 \ln _2^2 \bar{s}_1+\frac{1}{2} \zeta _2 \ln _2 \bar{s}_{-2}+\zeta _3 \ln _2 \bar{s}_{-1}-\zeta _3 \ln _2 \bar{s}_1-\frac{1}{2} \zeta _2 \ln _2 \bar{s}_2-\frac{1}{24} \ln _2^4 \bar{s}_{-1}-\frac{1}{8} \ln _2^4 \bar{s}_1-\zeta _2 s_{-1,-2}+2 \zeta _3 s_{-1,-1}+\zeta _3 s_{-1,1}-\zeta _2 s_{2,1}+\zeta _2 s_{-1,-1,1}-\frac{1}{2} \zeta _2 \ln _2 s_{-1,-1}-\frac{1}{2} \zeta _2 \ln _2 s_{-1,1}-s_{4,1}+s_{-1,-3,1}+s_{2,2,1}-s_{-1,-1,2,1}+\frac{25 \zeta _2 \zeta _3}{8}-\frac{667 \zeta _5}{64}+\frac{1}{6} \zeta _2 \ln _2^3+\zeta _3 \ln _2^2+\frac{41}{40} \zeta _2^2 \ln _2-3 \text{Li}_4\left(\frac{1}{2}\right) s_{-1}+3 \text{Li}_4\left(\frac{1}{2}\right) s_1+8 \text{Li}_5\left(\frac{1}{2}\right)+\frac{43}{40} \zeta _2^2 s_{-1}-\frac{3}{8} \zeta _2^2 s_1+\zeta _2 s_3-\zeta _3 s_2-\frac{3}{4} \zeta _2 s_{-1} \ln _2^2-\frac{1}{4} \zeta _2 s_1 \ln _2^2+\zeta _3 s_{-1} \ln _2+\zeta _3 s_1 \ln _2-\frac{1}{8} s_{-1} \ln _2^4+\frac{1}{8} s_1 \ln _2^4-\frac{\ln _2^5}{15} \end{dmath}
  \begin{dmath}[style={\small}]     s_1 \bar{s}_{-1,1,-2}  =   \frac{1}{2} \zeta _2 \bar{s}_{-1,-2}-\frac{7}{4} \zeta _3 \bar{s}_{-1,-1}+\frac{1}{4} \zeta _3 \bar{s}_{-1,1}-\frac{1}{2} \zeta _2 \bar{s}_{-1,2}-\frac{3}{2} \zeta _2 \bar{s}_{-1,1,-1}+\frac{1}{2} \zeta _2 \bar{s}_{-1,1,1}+\frac{3}{2} \zeta _2 \ln _2 \bar{s}_{-1,-1}-\frac{3}{2} \zeta _2 \ln _2 \bar{s}_{-1,1}-\bar{s}_{-2,1,-2}-\bar{s}_{-1,2,-2}+2 \bar{s}_{-1,1,1,-2}+\bar{s}_{1,-1,1,-2}+2 \text{Li}_4\left(\frac{1}{2}\right) \bar{s}_{-1}-2 \text{Li}_4\left(\frac{1}{2}\right) \bar{s}_1-\frac{1}{8} \zeta _3 \bar{s}_{-2}-\frac{33}{40} \zeta _2^2 \bar{s}_{-1}+\frac{1}{5} \zeta _2^2 \bar{s}_1+\frac{1}{8} \zeta _3 \bar{s}_2+\frac{1}{8} \zeta _3 \ln _2 \bar{s}_{-1}-\frac{1}{8} \zeta _3 \ln _2 \bar{s}_1+\frac{1}{12} \ln _2^4 \bar{s}_{-1}-\frac{1}{12} \ln _2^4 \bar{s}_1-\frac{3}{2} \zeta _2 s_{-2,-1}-\frac{1}{2} \zeta _2 s_{-2,1}-\zeta _2 s_{-1,-2}+\frac{7}{4} \zeta _3 s_{-1,-1}-\frac{1}{2} \zeta _3 s_{-1,1}+\frac{3}{2} \zeta _2 s_{-1,1,-1}+\frac{1}{2} \zeta _2 s_{-1,1,1}-\frac{3}{2} \zeta _2 \ln _2 s_{-1,-1}+\frac{3}{2} \zeta _2 \ln _2 s_{-1,1}-s_{4,1}+s_{-2,-2,1}+s_{-1,-3,1}-s_{-1,1,-2,1}-\zeta _2 \zeta _3-\frac{\zeta _5}{4}+\frac{1}{8} \zeta _3 \ln _2^2+\frac{9}{20} \zeta _2^2 \ln _2-2 \text{Li}_4\left(\frac{1}{2}\right) s_{-1}+2 \text{Li}_4\left(\frac{1}{2}\right) s_1+4 \text{Li}_5\left(\frac{1}{2}\right)+\frac{51}{40} \zeta _2^2 s_{-1}-\frac{1}{5} \zeta _2^2 s_1+\zeta _2 s_3+\frac{5}{8} \zeta _3 s_{-2}-\frac{13 \zeta _3 s_2}{8}-\frac{1}{2} \zeta _2 s_{-1} \ln _2^2-\frac{3}{2} \zeta _2 s_{-2} \ln _2+\frac{3}{2} \zeta _2 s_2 \ln _2+\frac{1}{8} \zeta _3 s_{-1} \ln _2+\frac{1}{8} \zeta _3 s_1 \ln _2-\frac{1}{12} s_{-1} \ln _2^4+\frac{1}{12} s_1 \ln _2^4-\frac{\ln _2^5}{30} \end{dmath}
  \begin{dmath}[style={\small}]     s_1 \bar{s}_{-1,1,2}  =   2 \zeta _3 \bar{s}_{-1,1}-\zeta _2 \bar{s}_{-1,1,1}-\bar{s}_{-2,1,2}-\bar{s}_{-1,2,2}+2 \bar{s}_{-1,1,1,2}+\bar{s}_{1,-1,1,2}-\text{Li}_4\left(\frac{1}{2}\right) \bar{s}_{-1}+\text{Li}_4\left(\frac{1}{2}\right) \bar{s}_1+\frac{9}{20} \zeta _2^2 \bar{s}_{-1}+\frac{1}{20} \zeta _2^2 \bar{s}_1+\frac{1}{8} \zeta _3 \ln _2 \bar{s}_{-1}-\frac{1}{8} \zeta _3 \ln _2 \bar{s}_1-\frac{1}{24} \ln _2^4 \bar{s}_{-1}+\frac{1}{24} \ln _2^4 \bar{s}_1+\zeta _2 s_{-2,1}-\zeta _3 s_{-1,1}+\zeta _2 s_{-1,2}-\zeta _2 s_{-1,1,1}+s_{-4,1}-s_{-2,2,1}-s_{-1,3,1}+s_{-1,1,2,1}-\frac{13 \zeta _2 \zeta _3}{16}-\frac{131 \zeta _5}{64}-\frac{1}{2} \zeta _2 \ln _2^3+\frac{1}{8} \zeta _3 \ln _2^2+\frac{33}{40} \zeta _2^2 \ln _2-\text{Li}_4\left(\frac{1}{2}\right) s_{-1}-\text{Li}_4\left(\frac{1}{2}\right) s_1+4 \text{Li}_5\left(\frac{1}{2}\right)-\frac{3}{4} \zeta _2^2 s_{-1}-\frac{1}{20} \zeta _2^2 s_1-\zeta _2 s_{-3}+\zeta _3 s_{-2}+\frac{1}{8} \zeta _3 s_{-1} \ln _2+\frac{1}{8} \zeta _3 s_1 \ln _2-\frac{1}{24} s_{-1} \ln _2^4-\frac{1}{24} s_1 \ln _2^4-\frac{\ln _2^5}{30} \end{dmath}
  \begin{dmath}[style={\small}]     s_1 \bar{s}_{-1,2,-1}  =   -\frac{1}{2} \zeta _2 \bar{s}_{-1,-2}+\frac{7}{8} \zeta _3 \bar{s}_{-1,-1}-\frac{1}{4} \zeta _3 \bar{s}_{-1,1}-\frac{3}{2} \zeta _2 \ln _2 \bar{s}_{-1,-1}+\frac{3}{2} \zeta _2 \ln _2 \bar{s}_{-1,1}-\bar{s}_{-2,2,-1}-\bar{s}_{-1,3,-1}+\bar{s}_{-1,1,2,-1}+\bar{s}_{-1,2,1,-1}+\bar{s}_{1,-1,2,-1}+\frac{1}{2} \ln _2^2 \bar{s}_{-1,-2}-\frac{1}{2} \ln _2^2 \bar{s}_{-1,2}+\ln _2 \bar{s}_{-1,-3}-\ln _2 \bar{s}_{-1,3}-\ln _2 \bar{s}_{-1,2,-1}+\ln _2 \bar{s}_{-1,2,1}+4 \text{Li}_4\left(\frac{1}{2}\right) \bar{s}_1+\frac{1}{4} \zeta _3 \bar{s}_{-2}-\frac{1}{8} \zeta _2^2 \bar{s}_{-1}-\frac{5}{4} \zeta _2^2 \bar{s}_1-\frac{1}{4} \zeta _3 \bar{s}_2-\frac{3}{2} \zeta _2 \ln _2^2 \bar{s}_{-1}+\frac{1}{2} \zeta _2 \ln _2^2 \bar{s}_1-\frac{1}{2} \zeta _2 \ln _2 \bar{s}_{-2}-\frac{1}{4} \zeta _3 \ln _2 \bar{s}_{-1}+\frac{1}{4} \zeta _3 \ln _2 \bar{s}_1+\frac{1}{2} \zeta _2 \ln _2 \bar{s}_2+\frac{1}{6} \ln _2^4 \bar{s}_1-\frac{1}{2} \zeta _2 s_{-1,-2}-\frac{7}{8} \zeta _3 s_{-1,-1}-\frac{1}{4} \zeta _3 s_{-1,1}-\frac{1}{2} \zeta _2 s_{-1,2}+\frac{3}{2} \zeta _2 \ln _2 s_{-1,-1}+\frac{3}{2} \zeta _2 \ln _2 s_{-1,1}-s_{4,1}+s_{-3,-1,1}+s_{-1,-3,1}-s_{-1,2,-1,1}+\frac{1}{2} \ln _2^2 s_{-1,-2}-\frac{1}{2} \ln _2^2 s_{-1,2}+\ln _2 s_{-3,-1}+\ln _2 s_{-3,1}-\ln _2 s_{-1,2,-1}-\ln _2 s_{-1,2,1}+\frac{7 \zeta _2 \zeta _3}{8}+\frac{135 \zeta _5}{8}-\frac{2}{3} \zeta _2 \ln _2^3-2 \zeta _3 \ln _2^2-\frac{61}{40} \zeta _2^2 \ln _2+2 \text{Li}_4\left(\frac{1}{2}\right) s_{-1}-4 \text{Li}_4\left(\frac{1}{2}\right) s_1-20 \text{Li}_5\left(\frac{1}{2}\right)-10 \text{Li}_4\left(\frac{1}{2}\right) \ln _2-\frac{13}{20} \zeta _2^2 s_{-1}+\frac{5}{4} \zeta _2^2 s_1+\frac{1}{2} \zeta _2 s_{-3}+\frac{\zeta _2 s_3}{2}+\frac{5 \zeta _3 s_2}{8}+\zeta _2 s_{-1} \ln _2^2-\frac{1}{2} \zeta _2 s_1 \ln _2^2-\zeta _2 s_{-2} \ln _2-\zeta _2 s_2 \ln _2+2 \zeta _3 s_{-1} \ln _2-\frac{1}{4} \zeta _3 s_1 \ln _2+\frac{1}{12} s_{-1} \ln _2^4-\frac{1}{6} s_1 \ln _2^4+\frac{1}{2} s_{-3} \ln _2^2-\frac{1}{2} s_3 \ln _2^2-\frac{\ln _2^5}{4} \end{dmath}
  \begin{dmath}[style={\small}]     s_1 \bar{s}_{-1,2,1}  =   -2 \zeta _3 \bar{s}_{-1,1}+\zeta _2 \bar{s}_{-1,2}-\bar{s}_{-2,2,1}-\bar{s}_{-1,3,1}+\bar{s}_{-1,1,2,1}+\bar{s}_{-1,2,1,1}+\bar{s}_{1,-1,2,1}-\text{Li}_4\left(\frac{1}{2}\right) \bar{s}_{-1}+\text{Li}_4\left(\frac{1}{2}\right) \bar{s}_1+\frac{27}{40} \zeta _2^2 \bar{s}_{-1}+\frac{1}{40} \zeta _2^2 \bar{s}_1+\frac{1}{4} \zeta _2 \ln _2^2 \bar{s}_{-1}-\frac{1}{4} \zeta _2 \ln _2^2 \bar{s}_1-\frac{1}{4} \zeta _3 \ln _2 \bar{s}_{-1}+\frac{1}{4} \zeta _3 \ln _2 \bar{s}_1-\frac{1}{24} \ln _2^4 \bar{s}_{-1}+\frac{1}{24} \ln _2^4 \bar{s}_1-2 \zeta _3 s_{-1,1}+\zeta _2 s_{-1,2}+s_{-4,1}-s_{-3,1,1}-s_{-1,3,1}+s_{-1,2,1,1}+\frac{9 \zeta _2 \zeta _3}{8}-\frac{617 \zeta _5}{64}+\frac{2}{3} \zeta _2 \ln _2^3-\frac{1}{4} \zeta _3 \ln _2^2+\frac{61}{40} \zeta _2^2 \ln _2-\text{Li}_4\left(\frac{1}{2}\right) s_{-1}-\text{Li}_4\left(\frac{1}{2}\right) s_1+8 \text{Li}_5\left(\frac{1}{2}\right)-\frac{21}{40} \zeta _2^2 s_{-1}-\frac{1}{40} \zeta _2^2 s_1-\zeta _2 s_{-3}+2 \zeta _3 s_{-2}+\frac{1}{4} \zeta _2 s_{-1} \ln _2^2+\frac{1}{4} \zeta _2 s_1 \ln _2^2-\frac{1}{4} \zeta _3 s_{-1} \ln _2-\frac{1}{4} \zeta _3 s_1 \ln _2-\frac{1}{24} s_{-1} \ln _2^4-\frac{1}{24} s_1 \ln _2^4-\frac{\ln _2^5}{15} \end{dmath}
  \begin{dmath}[style={\small}]     s_1 \bar{s}_{-1,-1,-1,-1}  =   \frac{\ln _2^5}{6}+\frac{2}{3} s_{-1} \ln _2^4+\frac{1}{24} s_1 \ln _2^4+\frac{1}{24} \bar{s}_{-1} \ln _2^4-\frac{1}{24} \bar{s}_1 \ln _2^4+\frac{1}{2} s_{-2} \ln _2^3-\frac{1}{2} s_2 \ln _2^3+\frac{3}{2} \zeta _2 \ln _2^3-\frac{2}{3} \bar{s}_{-2} \ln _2^3+\frac{2}{3} \bar{s}_2 \ln _2^3+\frac{7}{6} s_{-1,-1} \ln _2^3+\frac{1}{6} s_{-1,1} \ln _2^3-\frac{1}{6} \bar{s}_{-1,-1} \ln _2^3+\frac{1}{6} \bar{s}_{-1,1} \ln _2^3+\frac{1}{2} s_{-3} \ln _2^2-\frac{1}{2} s_3 \ln _2^2+\frac{3}{2} s_{-1} \zeta _2 \ln _2^2+\frac{1}{4} s_1 \zeta _2 \ln _2^2-\frac{15}{16} \zeta _3 \ln _2^2+\frac{1}{4} \zeta _2 \bar{s}_{-1} \ln _2^2-\frac{1}{4} \zeta _2 \bar{s}_1 \ln _2^2+\frac{1}{2} s_{-1,-2} \ln _2^2-\frac{1}{2} s_{-1,2} \ln _2^2-\frac{3}{2} s_{2,-1} \ln _2^2-\frac{1}{2} s_{2,1} \ln _2^2+\bar{s}_{-1,-2} \ln _2^2-\bar{s}_{-1,2} \ln _2^2+\frac{3}{2} s_{-1,-1,-1} \ln _2^2+\frac{1}{2} s_{-1,-1,1} \ln _2^2+\frac{1}{2} \bar{s}_{-1,-1,-1} \ln _2^2-\frac{1}{2} \bar{s}_{-1,-1,1} \ln _2^2-\frac{2}{5} \zeta _2^2 \ln _2-3 \text{Li}_4\left(\frac{1}{2}\right) \ln _2-\frac{1}{2} s_{-2} \zeta _2 \ln _2-\frac{3}{2} s_2 \zeta _2 \ln _2+\frac{3}{8} s_{-1} \zeta _3 \ln _2+\frac{1}{4} s_1 \zeta _3 \ln _2+\frac{1}{4} \zeta _3 \bar{s}_{-1} \ln _2-\frac{1}{4} \zeta _3 \bar{s}_1 \ln _2+s_{-3,-1} \ln _2+s_{-3,1} \ln _2+\frac{3}{2} \zeta _2 s_{-1,-1} \ln _2+\frac{1}{2} \zeta _2 s_{-1,1} \ln _2-\frac{1}{2} \zeta _2 \bar{s}_{-1,-1} \ln _2+\frac{1}{2} \zeta _2 \bar{s}_{-1,1} \ln _2-s_{-1,2,-1} \ln _2-s_{-1,2,1} \ln _2-s_{2,-1,-1} \ln _2-s_{2,-1,1} \ln _2-\bar{s}_{-1,-1,-2} \ln _2+\bar{s}_{-1,-1,2} \ln _2+s_{-1,-1,-1,-1} \ln _2+s_{-1,-1,-1,1} \ln _2-\bar{s}_{-1,-1,-1,-1} \ln _2+\bar{s}_{-1,-1,-1,1} \ln _2-\frac{1}{5} s_{-1} \zeta _2^2+\frac{9}{40} s_1 \zeta _2^2-5 \text{Li}_5\left(\frac{1}{2}\right)+\text{Li}_4\left(\frac{1}{2}\right) s_{-1}+\frac{1}{2} s_{-3} \zeta _2+\frac{s_3 \zeta _2}{2}+\frac{5 s_2 \zeta _3}{8}+\frac{25 \zeta _2 \zeta _3}{16}+\frac{165 \zeta _5}{64}-\frac{1}{4} \zeta _3 \bar{s}_{-2}-\frac{1}{8} \zeta _2^2 \bar{s}_{-1}-\frac{9}{40} \zeta _2^2 \bar{s}_1+\frac{1}{4} \zeta _3 \bar{s}_2-\frac{1}{2} \zeta _2 s_{-1,-2}-\frac{3}{8} \zeta _3 s_{-1,-1}+\frac{1}{4} \zeta _3 s_{-1,1}-\frac{1}{2} \zeta _2 s_{-1,2}-\frac{1}{2} \zeta _2 s_{2,-1}-\frac{1}{2} \zeta _2 s_{2,1}-s_{4,1}+\frac{3}{4} \zeta _3 \bar{s}_{-1,-1}+\frac{1}{4} \zeta _3 \bar{s}_{-1,1}+s_{-3,-1,1}+s_{-1,-3,1}+\frac{1}{2} \zeta _2 s_{-1,-1,-1}+\frac{1}{2} \zeta _2 s_{-1,-1,1}+s_{2,2,1}-\frac{1}{2} \zeta _2 \bar{s}_{-1,-1,1}-s_{-1,-1,2,1}-s_{-1,2,-1,1}-s_{2,-1,-1,1}-\bar{s}_{-2,-1,-1,-1}-\bar{s}_{-1,-2,-1,-1}-\bar{s}_{-1,-1,-2,-1}+s_{-1,-1,-1,-1,1}+\bar{s}_{-1,-1,-1,1,-1}+\bar{s}_{-1,-1,1,-1,-1}+\bar{s}_{-1,1,-1,-1,-1}+\bar{s}_{1,-1,-1,-1,-1} \end{dmath}
  \begin{dmath}[style={\small}]     s_1 \bar{s}_{-1,-1,-1,1}  =   \frac{7 \ln _2^5}{24}+\frac{2}{3} s_{-1} \ln _2^4+\frac{1}{24} s_1 \ln _2^4+\frac{1}{6} \bar{s}_{-1} \ln _2^4-\frac{1}{24} \bar{s}_1 \ln _2^4+\frac{1}{3} s_{-2} \ln _2^3-\frac{1}{3} s_2 \ln _2^3+\frac{1}{3} \zeta _2 \ln _2^3-\frac{1}{2} \bar{s}_{-2} \ln _2^3+\frac{1}{2} \bar{s}_2 \ln _2^3+\frac{5}{6} s_{-1,-1} \ln _2^3+\frac{1}{6} s_{-1,1} \ln _2^3-\frac{1}{6} \bar{s}_{-1,-1} \ln _2^3+\frac{1}{6} \bar{s}_{-1,1} \ln _2^3-\frac{1}{2} s_{-1} \zeta _2 \ln _2^2+\frac{1}{4} s_1 \zeta _2 \ln _2^2+\frac{5}{16} \zeta _3 \ln _2^2+\frac{1}{4} \zeta _2 \bar{s}_{-1} \ln _2^2-\frac{1}{4} \zeta _2 \bar{s}_1 \ln _2^2-\frac{1}{2} s_{2,-1} \ln _2^2-\frac{1}{2} s_{2,1} \ln _2^2+\frac{1}{2} \bar{s}_{-1,-2} \ln _2^2-\frac{1}{2} \bar{s}_{-1,2} \ln _2^2+\frac{1}{2} s_{-1,-1,-1} \ln _2^2+\frac{1}{2} s_{-1,-1,1} \ln _2^2+\frac{1}{2} \bar{s}_{-1,-1,-1} \ln _2^2-\frac{1}{2} \bar{s}_{-1,-1,1} \ln _2^2-\frac{15}{8} \zeta _2^2 \ln _2-s_{-2} \zeta _2 \ln _2+s_2 \zeta _2 \ln _2+\frac{1}{4} s_{-1} \zeta _3 \ln _2+\frac{1}{4} s_1 \zeta _3 \ln _2+\frac{1}{2} \zeta _2 \bar{s}_{-2} \ln _2+\frac{1}{4} \zeta _3 \bar{s}_{-1} \ln _2-\frac{1}{4} \zeta _3 \bar{s}_1 \ln _2-\frac{1}{2} \zeta _2 \bar{s}_2 \ln _2-\frac{3}{2} \zeta _2 s_{-1,-1} \ln _2+\frac{1}{2} \zeta _2 s_{-1,1} \ln _2+\zeta _2 \bar{s}_{-1,-1} \ln _2+\frac{1}{2} \zeta _2 \bar{s}_{-1,1} \ln _2-\frac{6}{5} s_{-1} \zeta _2^2-\frac{19}{40} s_1 \zeta _2^2-7 \text{Li}_5\left(\frac{1}{2}\right)+3 \text{Li}_4\left(\frac{1}{2}\right) s_{-1}-s_{-3} \zeta _2+\frac{7}{4} s_{-2} \zeta _3+\frac{s_2 \zeta _3}{4}+\frac{\zeta _2 \zeta _3}{16}+\frac{319 \zeta _5}{64}-\frac{3}{5} \zeta _2^2 \bar{s}_{-1}+3 \text{Li}_4\left(\frac{1}{2}\right) \bar{s}_{-1}+\frac{19}{40} \zeta _2^2 \bar{s}_1+s_{-4,1}-\frac{1}{4} \zeta _3 s_{-1,-1}-\frac{7}{4} \zeta _3 s_{-1,1}+\zeta _2 s_{-1,2}+\frac{3}{2} \zeta _2 s_{2,-1}+\frac{1}{2} \zeta _2 s_{2,1}-\frac{1}{2} \zeta _2 \bar{s}_{-1,-2}+\frac{1}{8} \zeta _3 \bar{s}_{-1,-1}-\frac{7}{4} \zeta _3 \bar{s}_{-1,1}+\frac{1}{2} \zeta _2 \bar{s}_{-1,2}-s_{-3,1,1}-\frac{3}{2} \zeta _2 s_{-1,-1,-1}-\frac{1}{2} \zeta _2 s_{-1,-1,1}-s_{-1,3,1}-s_{2,-2,1}+\frac{1}{2} \zeta _2 \bar{s}_{-1,-1,-1}+\frac{1}{2} \zeta _2 \bar{s}_{-1,-1,1}+s_{-1,-1,-2,1}+s_{-1,2,1,1}+s_{2,-1,1,1}-\bar{s}_{-2,-1,-1,1}-\bar{s}_{-1,-2,-1,1}-\bar{s}_{-1,-1,-2,1}-s_{-1,-1,-1,1,1}+\bar{s}_{-1,-1,-1,1,1}+\bar{s}_{-1,-1,1,-1,1}+\bar{s}_{-1,1,-1,-1,1}+\bar{s}_{1,-1,-1,-1,1} \end{dmath}
  \begin{dmath}[style={\small}]     s_1 \bar{s}_{-1,-1,1,-1}  =   \frac{3 \ln _2^5}{20}+\frac{1}{3} s_{-1} \ln _2^4+\frac{1}{6} s_1 \ln _2^4+\frac{1}{12} \bar{s}_{-1} \ln _2^4-\frac{1}{6} \bar{s}_1 \ln _2^4-\frac{1}{6} s_{-2} \ln _2^3+\frac{1}{6} s_2 \ln _2^3-\frac{19}{12} \zeta _2 \ln _2^3-\frac{1}{6} s_{-1,-1} \ln _2^3+\frac{1}{6} s_{-1,1} \ln _2^3-\frac{1}{6} \bar{s}_{-1,-1} \ln _2^3+\frac{1}{6} \bar{s}_{-1,1} \ln _2^3+\frac{1}{2} s_{-3} \ln _2^2-\frac{1}{2} s_3 \ln _2^2-3 s_{-1} \zeta _2 \ln _2^2-\frac{1}{2} s_1 \zeta _2 \ln _2^2+\frac{3}{2} \zeta _3 \ln _2^2-\frac{3}{2} \zeta _2 \bar{s}_{-1} \ln _2^2+\frac{1}{2} \zeta _2 \bar{s}_1 \ln _2^2+\frac{1}{2} s_{-1,-2} \ln _2^2-\frac{1}{2} s_{-1,2} \ln _2^2+s_{2,1} \ln _2^2-\frac{1}{2} \bar{s}_{-1,-2} \ln _2^2+\frac{1}{2} \bar{s}_{-1,2} \ln _2^2-s_{-1,-1,1} \ln _2^2-\frac{11}{5} \zeta _2^2 \ln _2-\text{Li}_4\left(\frac{1}{2}\right) \ln _2+2 s_2 \zeta _2 \ln _2+3 s_{-1} \zeta _3 \ln _2+\frac{7}{8} s_1 \zeta _3 \ln _2+\frac{1}{2} \zeta _2 \bar{s}_{-2} \ln _2+\frac{7}{8} \zeta _3 \bar{s}_{-1} \ln _2-\frac{7}{8} \zeta _3 \bar{s}_1 \ln _2-\frac{1}{2} \zeta _2 \bar{s}_2 \ln _2-\frac{5}{2} \zeta _2 s_{-1,-1} \ln _2-\frac{1}{2} \zeta _2 s_{-1,1} \ln _2-s_{3,-1} \ln _2-s_{3,1} \ln _2-\zeta _2 \bar{s}_{-1,-1} \ln _2-\frac{1}{2} \zeta _2 \bar{s}_{-1,1} \ln _2+s_{-1,-2,-1} \ln _2+s_{-1,-2,1} \ln _2+s_{2,1,-1} \ln _2+s_{2,1,1} \ln _2+\bar{s}_{-1,-1,-2} \ln _2-\bar{s}_{-1,-1,2} \ln _2-s_{-1,-1,1,-1} \ln _2-s_{-1,-1,1,1} \ln _2-\bar{s}_{-1,-1,1,-1} \ln _2+\bar{s}_{-1,-1,1,1} \ln _2-\frac{51}{40} s_{-1} \zeta _2^2-\frac{6}{5} s_1 \zeta _2^2-5 \text{Li}_5\left(\frac{1}{2}\right)+3 \text{Li}_4\left(\frac{1}{2}\right) s_{-1}+3 \text{Li}_4\left(\frac{1}{2}\right) s_1-\frac{1}{2} s_{-3} \zeta _2-\frac{s_3 \zeta _2}{2}+\frac{1}{8} s_{-2} \zeta _3-\frac{3 s_2 \zeta _3}{4}-\frac{\zeta _2 \zeta _3}{4}+\frac{341 \zeta _5}{64}-\frac{13}{40} \zeta _2^2 \bar{s}_{-1}+\text{Li}_4\left(\frac{1}{2}\right) \bar{s}_{-1}+\frac{6}{5} \zeta _2^2 \bar{s}_1-3 \text{Li}_4\left(\frac{1}{2}\right) \bar{s}_1+s_{-4,1}+\frac{1}{2} \zeta _2 s_{-1,-2}+\frac{3}{4} \zeta _3 s_{-1,-1}-\frac{1}{8} \zeta _3 s_{-1,1}+\frac{1}{2} \zeta _2 s_{-1,2}+\frac{1}{2} \zeta _2 s_{2,-1}+\frac{1}{2} \zeta _2 s_{2,1}+\frac{1}{4} \zeta _3 \bar{s}_{-1,-1}-\frac{1}{8} \zeta _3 \bar{s}_{-1,1}-\frac{1}{2} \zeta _2 s_{-1,-1,-1}-\frac{1}{2} \zeta _2 s_{-1,-1,1}-s_{-1,3,1}-s_{2,-2,1}-s_{3,-1,1}-\frac{1}{2} \zeta _2 \bar{s}_{-1,-1,-1}+s_{-1,-2,-1,1}+s_{-1,-1,-2,1}+s_{2,1,-1,1}-\bar{s}_{-2,-1,1,-1}-\bar{s}_{-1,-2,1,-1}-\bar{s}_{-1,-1,2,-1}-s_{-1,-1,1,-1,1}+2 \bar{s}_{-1,-1,1,1,-1}+\bar{s}_{-1,1,-1,1,-1}+\bar{s}_{1,-1,-1,1,-1} \end{dmath}
  \begin{dmath}[style={\small}]     s_1 \bar{s}_{-1,-1,1,1}  =   \frac{1}{8} \zeta _3 \bar{s}_{-1,-1}+\frac{7}{8} \zeta _3 \bar{s}_{-1,1}+\zeta _2 \bar{s}_{-1,-1,1}+\frac{1}{2} \zeta _2 \ln _2 \bar{s}_{-1,-1}-\frac{1}{2} \zeta _2 \ln _2 \bar{s}_{-1,1}-\bar{s}_{-2,-1,1,1}-\bar{s}_{-1,-2,1,1}-\bar{s}_{-1,-1,2,1}+2 \bar{s}_{-1,-1,1,1,1}+\bar{s}_{-1,1,-1,1,1}+\bar{s}_{1,-1,-1,1,1}-\frac{1}{6} \ln _2^3 \bar{s}_{-1,-1}+\frac{1}{6} \ln _2^3 \bar{s}_{-1,1}-3 \text{Li}_4\left(\frac{1}{2}\right) \bar{s}_1-\frac{7}{8} \zeta _3 \bar{s}_{-2}+\frac{11}{20} \zeta _2^2 \bar{s}_{-1}+\frac{7}{8} \zeta _3 \bar{s}_2-\frac{1}{2} \zeta _2 \ln _2^2 \bar{s}_{-1}+\frac{1}{2} \zeta _2 \ln _2^2 \bar{s}_1+\frac{1}{2} \zeta _2 \ln _2 \bar{s}_{-2}+\frac{7}{8} \zeta _3 \ln _2 \bar{s}_{-1}-\frac{7}{8} \zeta _3 \ln _2 \bar{s}_1-\frac{1}{2} \zeta _2 \ln _2 \bar{s}_2+\frac{1}{24} \ln _2^4 \bar{s}_{-1}-\frac{1}{6} \ln _2^4 \bar{s}_1-\frac{1}{6} \ln _2^3 \bar{s}_{-2}+\frac{1}{6} \ln _2^3 \bar{s}_2-\zeta _2 s_{-1,-2}+\frac{23}{8} \zeta _3 s_{-1,-1}+\frac{7}{8} \zeta _3 s_{-1,1}-\zeta _2 s_{2,1}+\zeta _2 s_{-1,-1,1}-\frac{1}{2} \zeta _2 \ln _2 s_{-1,-1}-\frac{1}{2} \zeta _2 \ln _2 s_{-1,1}-s_{4,1}+s_{-1,-3,1}+s_{2,2,1}+s_{3,1,1}-s_{-1,-2,1,1}-s_{-1,-1,2,1}-s_{2,1,1,1}+s_{-1,-1,1,1,1}+\frac{1}{6} \ln _2^3 s_{-1,-1}+\frac{1}{6} \ln _2^3 s_{-1,1}-\frac{7 \zeta _2 \zeta _3}{16}-\frac{13 \zeta _5}{16}-\frac{7}{12} \zeta _2 \ln _2^3+\frac{7}{8} \zeta _3 \ln _2^2+\frac{23}{20} \zeta _2^2 \ln _2-3 \text{Li}_4\left(\frac{1}{2}\right) s_{-1}+3 \text{Li}_4\left(\frac{1}{2}\right) s_1+3 \text{Li}_5\left(\frac{1}{2}\right)+\frac{6}{5} \zeta _2^2 s_{-1}+\zeta _2 s_3-2 \zeta _3 s_2-\frac{1}{2} \zeta _2 s_{-1} \ln _2^2-\frac{1}{2} \zeta _2 s_1 \ln _2^2+\frac{7}{8} \zeta _3 s_{-1} \ln _2+\frac{7}{8} \zeta _3 s_1 \ln _2+\frac{1}{6} s_{-1} \ln _2^4+\frac{1}{6} s_1 \ln _2^4+\frac{17 \ln _2^5}{120} \end{dmath}
  \begin{dmath}[style={\small}]    s_1 \bar{s}_{-1,1,1,-1}  =   -\frac{1}{8} \zeta _3 \bar{s}_{-1,-1}+\frac{1}{4} \zeta _3 \bar{s}_{-1,1}-\frac{1}{2} \zeta _2 \bar{s}_{-1,1,-1}-\frac{3}{2} \zeta _2 \ln _2 \bar{s}_{-1,1}-\bar{s}_{-2,1,1,-1}-\bar{s}_{-1,1,2,-1}-\bar{s}_{-1,2,1,-1}+3 \bar{s}_{-1,1,1,1,-1}+\bar{s}_{1,-1,1,1,-1}+\frac{1}{2} \ln _2^2 \bar{s}_{-1,-2}-\frac{1}{2} \ln _2^2 \bar{s}_{-1,2}+\ln _2 \bar{s}_{-1,1,-2}-\ln _2 \bar{s}_{-1,1,2}-\ln _2 \bar{s}_{-1,1,1,-1}+\ln _2 \bar{s}_{-1,1,1,1}+\text{Li}_4\left(\frac{1}{2}\right) \bar{s}_1-\frac{2}{5} \zeta _2^2 \bar{s}_1-\zeta _2 \ln _2^2 \bar{s}_{-1}+\frac{1}{24} \ln _2^4 \bar{s}_{-1}-\frac{1}{6} \ln _2^3 \bar{s}_{-2}+\frac{1}{6} \ln _2^3 \bar{s}_2-\frac{1}{2} \zeta _2 s_{-2,-1}-\frac{1}{2} \zeta _2 s_{-2,1}-\frac{1}{2} \zeta _2 s_{-1,-2}+\frac{1}{8} \zeta _3 s_{-1,-1}-\frac{3}{4} \zeta _3 s_{-1,1}-\frac{1}{2} \zeta _2 s_{-1,2}+\frac{1}{2} \zeta _2 s_{-1,1,-1}+\frac{1}{2} \zeta _2 s_{-1,1,1}+2 \zeta _2 \ln _2 s_{-1,1}-s_{4,1}+s_{-3,-1,1}+s_{-2,-2,1}+s_{-1,-3,1}-s_{-2,1,-1,1}-s_{-1,1,-2,1}-s_{-1,2,-1,1}+s_{-1,1,1,-1,1}+\frac{1}{3} \ln _2^3 s_{-1,1}-\ln _2^2 s_{-2,1}+\frac{1}{2} \ln _2^2 s_{-1,-2}-\frac{1}{2} \ln _2^2 s_{-1,2}+\ln _2^2 s_{-1,1,1}+\ln _2 s_{-3,-1}+\ln _2 s_{-3,1}-\ln _2 s_{-2,1,-1}-\ln _2 s_{-2,1,1}-\ln _2 s_{-1,2,-1}-\ln _2 s_{-1,2,1}+\ln _2 s_{-1,1,1,-1}+\ln _2 s_{-1,1,1,1}+5 \zeta _5+\zeta _2 \ln _2^3-\frac{3}{2} \zeta _3 \ln _2^2+\frac{13}{40} \zeta _2^2 \ln _2+\text{Li}_4\left(\frac{1}{2}\right) s_{-1}-\text{Li}_4\left(\frac{1}{2}\right) s_1-5 \text{Li}_5\left(\frac{1}{2}\right)-7 \text{Li}_4\left(\frac{1}{2}\right) \ln _2-\frac{13}{40} \zeta _2^2 s_{-1}+\frac{2}{5} \zeta _2^2 s_1+\frac{1}{2} \zeta _2 s_{-3}+\frac{\zeta _2 s_3}{2}+\frac{3}{4} \zeta _3 s_{-2}-\frac{\zeta _3 s_2}{8}+\frac{1}{2} \zeta _2 s_{-1} \ln _2^2-2 \zeta _2 s_{-2} \ln _2+\frac{25}{8} \zeta _3 s_{-1} \ln _2+\frac{1}{6} s_{-1} \ln _2^4-\frac{1}{6} s_{-2} \ln _2^3+\frac{1}{6} s_2 \ln _2^3+\frac{1}{2} s_{-3} \ln _2^2-\frac{1}{2} s_3 \ln _2^2-\frac{\ln _2^5}{6} \end{dmath}
  \begin{dmath}[style={\small}]    s_1 \bar{s}_{-2,-2}  =   -\frac{3}{2} \zeta _2 \bar{s}_{-2,-1}+\frac{1}{2} \zeta _2 \bar{s}_{-2,1}-\bar{s}_{-3,-2}+\bar{s}_{-2,1,-2}+\bar{s}_{1,-2,-2}+\frac{3}{4} \zeta _2^2 \bar{s}_{-1}-\frac{13}{40} \zeta _2^2 \bar{s}_1-\frac{1}{2} \zeta _2 \bar{s}_{-3}+\frac{1}{2} \zeta _2 \bar{s}_3+\frac{1}{8} \zeta _3 \bar{s}_{-2}-\frac{13}{8} \zeta _3 \bar{s}_2-\frac{3}{2} \zeta _2 \ln _2 \bar{s}_{-2}+\frac{3}{2} \zeta _2 \ln _2 \bar{s}_2-\frac{3}{2} \zeta _2 s_{-2,-1}-\frac{1}{2} \zeta _2 s_{-2,1}-s_{4,1}+s_{-2,-2,1}-\frac{15 \zeta _3 \zeta _2}{4}+\frac{115 \zeta _5}{16}+\frac{3}{2} \zeta _2^2 \ln _2+\frac{3}{4} \zeta _2^2 s_{-1}+\frac{13}{40} \zeta _2^2 s_1+\zeta _2 s_3+\frac{5}{8} \zeta _3 s_{-2}-\frac{13 \zeta _3 s_2}{8}-\frac{3}{2} \zeta _2 s_{-2} \ln _2+\frac{3}{2} \zeta _2 s_2 \ln _2 \end{dmath}
  
  \begin{dmath}[style={\small}] \label{s1scm22}   s_1 \bar{s}_{-2,2}  =   -\zeta _2 \bar{s}_{-2,1}-\bar{s}_{-3,2}+\bar{s}_{-2,1,2}+\bar{s}_{1,-2,2}-4 \text{Li}_4\left(\frac{1}{2}\right) \bar{s}_{-1}+4 \text{Li}_4\left(\frac{1}{2}\right) \bar{s}_1+2 \zeta _3 \bar{s}_{-2}+\frac{23}{40} \zeta _2^2 \bar{s}_{-1}-\frac{51}{40} \zeta _2^2 \bar{s}_1+\zeta _2 \ln _2^2 \bar{s}_{-1}-\zeta _2 \ln _2^2 \bar{s}_1-\frac{7}{2} \zeta _3 \ln _2 \bar{s}_{-1}+\frac{7}{2} \zeta _3 \ln _2 \bar{s}_1-\frac{1}{6} \ln _2^4 \bar{s}_{-1}+\frac{1}{6} \ln _2^4 \bar{s}_1+\zeta _2 s_{-2,1}+s_{-4,1}-s_{-2,2,1}-\frac{5 \zeta _2 \zeta _3}{2}-\frac{51 \zeta _5}{32}+\frac{2}{3} \zeta _2 \ln _2^3-\frac{7}{2} \zeta _3 \ln _2^2+\frac{23}{20} \zeta _2^2 \ln _2-4 \text{Li}_4\left(\frac{1}{2}\right) s_{-1}-4 \text{Li}_4\left(\frac{1}{2}\right) s_1+8 \text{Li}_5\left(\frac{1}{2}\right)+\frac{23}{40} \zeta _2^2 s_{-1}+\frac{51}{40} \zeta _2^2 s_1-\zeta _2 s_{-3}+\zeta _3 s_{-2}+\zeta _2 s_{-1} \ln _2^2+\zeta _2 s_1 \ln _2^2-\frac{7}{2} \zeta _3 s_{-1} \ln _2-\frac{7}{2} \zeta _3 s_1 \ln _2-\frac{1}{6} s_{-1} \ln _2^4-\frac{1}{6} s_1 \ln _2^4-\frac{\ln _2^5}{15} \end{dmath}
  \begin{dmath}[style={\small}] \label{s1sc22}   s_1 \bar{s}_{2,2}  =   -\zeta _2 \bar{s}_{2,1}-\bar{s}_{3,2}+\bar{s}_{1,2,2}+\bar{s}_{2,1,2}-\frac{7}{10} \zeta _2^2 \bar{s}_1+2 \zeta _3 \bar{s}_2-\zeta _2 s_{2,1}-s_{4,1}+s_{2,2,1}+3 \zeta _3 \zeta _2-\frac{5 \zeta _5}{2}+\frac{7}{10} \zeta _2^2 s_1+\zeta _2 s_3-\zeta _3 s_2 \end{dmath}



\end{document}